\documentclass[10pt]{book}

\usepackage[
    includehead=true,
    includefoot=false,
    papersize={156mm,234mm},
    top=13mm,
    bottom=13mm,
    left=17mm, 
    right=17mm, 
    twoside,
    pdftex
]
{geometry}
\usepackage{ifthen}
\usepackage{fancyhdr}
\usepackage{setspace}

\makeatletter
\newboolean{THESISFM@hassubtitle}\setboolean{THESISFM@hassubtitle}{false}
\newboolean{THESISFM@hasthesisepigraph}\setboolean{THESISFM@hasthesisepigraph}{false}
\newboolean{THESISFM@hascopromotor}\setboolean{THESISFM@hascopromotor}{false}
\newboolean{THESISFM@hascolofon}\setboolean{THESISFM@hascolofon}{false}

\newcommand{\THESISFM@subtitle}{}
\newcommand{\subtitle}[1]
{
    \setboolean{THESISFM@hassubtitle}{true}
    \renewcommand{\THESISFM@subtitle}{#1}
}
\newcommand{\THESISFM@rectormagnificus}{}
\newcommand{\rectormagnificus}[1]
{
    \renewcommand{\THESISFM@rectormagnificus}{#1}
}
\newcommand{\THESISFM@thetime}{}
\newcommand{\thetime}[1]
{
    \renewcommand{\THESISFM@thetime}{#1}
}
\newcommand{\THESISFM@theday}{}
\newcommand{\theday}[1]
{
    \renewcommand{\THESISFM@theday}{#1}
}
\newcommand{\THESISFM@placeofbirth}{}
\newcommand{\placeofbirth}[1]
{
    \renewcommand{\THESISFM@placeofbirth}{#1}
}
\newcommand{\THESISFM@thesisepigraph}{}
\newcommand{\thesisepigraph}[1]
{
    \setboolean{THESISFM@hasthesisepigraph}{true}
    \renewcommand{\THESISFM@thesisepigraph}{#1}
}
\newcommand{\THESISFM@promotor}{}
\newcommand{\promotor}[1]
{
    \renewcommand{\THESISFM@promotor}{#1}
}
\newcommand{\THESISFM@copromotor}{}
\newcommand{\copromotor}[1]
{
    \setboolean{THESISFM@hascopromotor}{true}
    \renewcommand{\THESISFM@copromotor}{#1}
}
\newcommand{\THESISFM@othercomitteemembers}{}
\newcommand{\othercomitteemembers}[1]
{
    \renewcommand{\THESISFM@othercomitteemembers}{#1}
}
\newcommand{\THESISFM@colofon}{}
\newcommand{\thesiscolofon}[1]
{
    \setboolean{THESISFM@hascolofon}{true}
    \renewcommand{\THESISFM@colofon}{#1}
}

\newcommand{\THESISFM@department}
{
    Faculteit der Natuurwetenschappen, Wiskunde en Informatica
}
\newcommand{\department}[1]{\renewcommand{\THESISFM@department}{#1}}

\newcommand{\makethesisfrontpage}
{
    \newpage\pagestyle{empty}
    \vspace*{\fill}
    \begin{center}%
    \textsc{\Huge \@title} \par
    \ifTHESISFM@hassubtitle
    \vskip 12pt
    \textsc{\Large \THESISFM@subtitle}\par
    \fi
    \end{center}
    \vspace{\fill}
}

\newcommand{\makethesiscolofonnew}
{
    \newpage\pagestyle{empty}
    This work has been accomplished at the Institute for Theoretical
    Physics (ITFA) of the University of Amsterdam (UvA) and is financially
    supported by a Spinoza grant of the Netherlands Organisation for
    Scientific Research (NWO). 
    \\
\begin{figure}[h!]
\begin{center} 
\includegraphics[width=.1\textwidth]{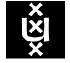}
\qquad \qquad
\includegraphics[width=.15\textwidth]{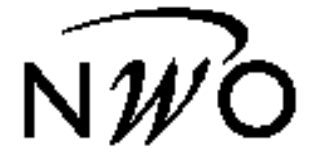}
\end{center}
\end{figure}
    \vspace*{\fill}\\
    Cover illustration: Lotte Hollands \\
    Cover design: The DocWorkers \\
    Lay-out: Lotte Hollands, typeset using \LaTeX \\
    \vskip .5cm%
\hspace*{-3mm}
\begin{tabular}{ll}
    ISBN &	978 90 8555 020 4\\
    NUR	& 924 \\
\end{tabular}
   \vskip .2cm%
\copyright~ L. Hollands~/~Pallas Publications --- Amsterdam University
Press,~2009\\
  \vskip 0.1cm%
All rights reserved. Without limiting the rights under copyright reserved above, no part of this book may be reproduced, stored in or introduced into a retrieval system, or transmitted, in any form or by any means (electronic, mechanical, photocopying, recording or otherwise) without the written permission of both the copyright owner and the author of the book.
}

\newcommand{\makethesiscolofon}
{
    \newpage\pagestyle{empty}
    \THESISFM@colofon
}

\newcommand{\makethesisannouncepage}
{
    \cleardoublepage\pagestyle{empty}
    \vspace*{\fill}
    \begin{center}%
    \textsc{\Huge \@title}\par
    \ifTHESISFM@hassubtitle
        \vskip 12pt
        \textsc{\Large \THESISFM@subtitle }
    \fi
    \vskip 2cm%
    \textsc{\LARGE Academisch Proefschrift} \par\vskip 1cm
    ter verkrijging van de graad van doctor\par\vskip 1em
    aan de Universiteit van Amsterdam\par\vskip 1em
    op gezag van de Rector Magnificus\par\vskip 1em
    \THESISFM@rectormagnificus\par\vskip 1em
    ten overstaan van een door het college voor promoties\par\vskip 1em
    ingestelde commissie,\par\vskip 1em in het openbaar te verdedigen
    in de 
    Agnietenkapel \par\vskip 1em
    op \THESISFM@theday{} \@date,  te \THESISFM@thetime{} uur\par\vskip 1cm
    door \par\vskip 1em
    \textsc{\@author}\par\vskip 1em
    geboren te \THESISFM@placeofbirth
    \end{center}
    \vspace{\fill}
}

\newcommand{\makethesiscomitteepage}
{
    \newpage\pagestyle{empty}
    \begin{center}\textsc{\Large Promotiecommissie}\end{center}\par\vskip 2em
    {Promotor}\par\vskip 1em
    \hspace{2cm}\THESISFM@promotor\par\vskip 1em
    \ifTHESISFM@hascopromotor
        \textsc{Co-Promotor}\par\vskip 1em
        \hspace{2cm}\THESISFM@copromotor\par\vskip 1em
    \fi
    {Overige leden}\par\vskip 1em
    
    \@for \THESISFM@comitteemember:=\THESISFM@othercomitteemembers \do 
    {\hspace{2cm}\THESISFM@comitteemember\vskip 1em}
    
    {\THESISFM@department}\par
    
    \vspace*{\fill}
    \THESISFM@colofon
}

\newcommand{\makethesisepigraphpage}
{
    \ifTHESISFM@hasthesisepigraph
        \newpage\pagestyle{empty}
        \vspace*{\fill}
        \THESISFM@thesisepigraph
        \vspace{\fill}
        \newpage\pagestyle{empty}
        \vspace*{\fill}
    \fi
}

\newcommand{\thesisfont}[1]
{
    \renewcommand{\rmdefault}{#1}
    \renewcommand{\sfdefault}{#1}
    \renewcommand{\ttdefault}{#1}
}
\thesisfont{bch}

\newcommand{\THESIS@chapterstyle}{\centering\Huge\it\bfseries}
\newcommand{\thesischapterstyle}[1]{\renewcommand{\THESIS@chapterstyle}{#1}}

\def\@makechapterhead#1%
{%
    \vspace*{50\p@}%
    {
        \parindent \z@ \raggedright \normalfont
        \ifnum \c@secnumdepth >\m@ne
            \if@mainmatter
                \THESIS@chapterstyle \huge \@chapapp\space \thechapter
                \par\nobreak
                \vskip 20\p@
            \fi
        \fi
        \interlinepenalty\@M
        \THESIS@chapterstyle #1\par
        \vskip 40\p@
    }
}

\def\@makeschapterhead#1%
{%
    \vspace*{50\p@}%
    {
        \parindent \z@ \raggedright
        \normalfont
        \interlinepenalty\@M
        \THESIS@chapterstyle  #1\par\nobreak
        \vskip 40\p@
    }
}

\newcommand{\THESIS@sectiondecorationleft}{}
\newcommand{\THESIS@sectiondecorationright}{}
\newcommand{\thesissectiondecoration}[2]
{
    \renewcommand{\THESIS@sectiondecorationleft}{#1}
    \renewcommand{\THESIS@sectiondecorationright}{#2}
}
\renewcommand{\@seccntformat}[1]
{
    \THESIS@sectiondecorationleft\@nameuse{the#1}\THESIS@sectiondecorationright\quad
}

\newcommand{\THESIS@sectionstyle}{\bfseries}
\newcommand{\thesissectionstyle}[1]{\renewcommand{\THESIS@sectionstyle}{#1}}
\newcommand{\THESIS@sectionfont}{\normalfont}
\newcommand{\thesissectionfont}[1]{\renewcommand{\THESIS@sectionfont}{#1}}
\newcommand{\THESIS@sectionsize}{\Large}
\newcommand{\THESIS@subsectionsize}{\large}
\newcommand{\THESIS@subsubsectionsize}{\normalsize}
\newcommand{\thesissectionsizes}[3]
{
    \renewcommand{\THESIS@sectionsize}{#1}
    \renewcommand{\THESIS@subsectionsize}{#2}
    \renewcommand{\THESIS@subsubsectionsize}{#3}
}
\renewcommand\section{\@startsection {section}{1}{\z@}%
             {-3.5ex \@plus -1ex \@minus -.2ex}%
             {2.3ex \@plus.2ex}%
             {\THESIS@sectionfont\THESIS@sectionsize\THESIS@sectionstyle}}
\renewcommand\subsection{\@startsection{subsection}{2}{\z@}%
             {-2.75ex\@plus -1ex \@minus -1ex}
             {.75ex \@plus .2ex \@minus -.25ex}
             {\THESIS@sectionfont\THESIS@subsectionsize\THESIS@sectionstyle}}
\renewcommand\subsubsection{\@startsection{subsubsection}{3}{\z@}%
             {-2.25ex\@plus -1ex \@minus -1ex}
             {.75ex \@plus .2ex \@minus -.25ex}
             {\THESIS@sectionfont\THESIS@subsubsectionsize\THESIS@sectionstyle}}

\newcommand{\THESIS@paragraphstyle}{\bfseries}
\newcommand{\thesisparagraphstyle}[1]{\renewcommand{\THESIS@paragraphstyle}{#1}}
\newcommand{\THESIS@paragraphfont}{\normalfont}
\newcommand{\thesisparagraphfont}[1]{\renewcommand{\THESIS@paragraphfont}{#1}}
\newcommand{\THESIS@paragraphsize}{\normalsize}
\newcommand{\THESIS@subparagraphsize}{\normalsize}
\newcommand{\thesisparagraphsizes}[2]
{
    \renewcommand{\THESIS@paragraphsize}{#1}
    \renewcommand{\THESIS@subparagraphsize}{#2}
}
\renewcommand\paragraph{\@startsection{paragraph}{4}{\z@}%
             {3.25ex \@plus1ex \@minus.2ex}%
             {-1em}%
             {\THESIS@paragraphfont\THESIS@paragraphsize\THESIS@paragraphstyle}}
\renewcommand\subparagraph{\@startsection{subparagraph}{5}{\parindent}%
             {3.25ex \@plus1ex \@minus .2ex}%
             {-1em}%
             {\THESIS@paragraphfont\THESIS@subparagraphsize\THESIS@paragraphstyle}}

\newcommand{\maketitlefinish}
{
    \setcounter{footnote}{0}%
    \global\let\maketitle\relax
    \global\let\@author\@empty
    \global\let\@date\@empty
    \global\let\@title\@empty
    \global\let\THESIS@subtitle\@empty
    \global\let\THESIS@rectormagnificus\@empty
    \global\let\THESIS@theday\@empty
    \global\let\THESIS@thetime\@empty
    \global\let\THESIS@placeofbirth\@empty
    \global\let\THESIS@promotor\@empty
    \global\let\THESIS@copromotor\@empty
    \global\let\title\relax
    \global\let\subtitle\relax
    \global\let\rectormagnificus\relax
    \global\let\theday\relax
    \global\let\thetime\relax
    \global\let\placeofbirth\relax
    \global\let\promotor\relax
    \global\let\copromotor\relax
    \global\let\author\relax
    \global\let\date\relax
    \global\let\and\relax
}

\if@titlepage
\renewcommand{\maketitle}
{
    \makethesisfrontpage
    \makethesiscolofonnew
    \makethesisannouncepage
    \makethesiscomitteepage
    \makethesisepigraphpage
    \maketitlefinish
    
    \newpage\pagestyle{empty}
    \begin{center}\textsc{\Large Publications}\end{center}\par\vskip 2em
     \par\vskip .5em

\begin{center}
This thesis is based on the following publications:
\par\vskip -0.5em

\begin{minipage}[t]{0.9\textwidth}
  \vspace*{0.2cm}
\begin{center} 
 \item
  R.~Dijkgraaf, L.~Hollands, P. Su{\l}kowski and C.~Vafa,\\
  \emph{Supersymmetric Gauge Theories, Intersecting Branes and Free Fermions},\\
  \textsf{arXiv/0709.4446 [hep-th]}, JHEP 02 (2008) 106.
\end{center}
\end{minipage}

\begin{minipage}[t]{0.9\textwidth}
  \vspace*{0.5cm}
\begin{center} 
  L.~Hollands, J.~Marsano, K.~Papadodimas and  M.~Shigemori,\\ 
  \emph{Nonsupersymmetric Flux Vacua and Perturbed N=2 Systems},\\
  \textsf{arXiv/0804.4006 [hep-th]}, JHEP 10 (2008) 102.
\end{center}
\end{minipage}

\begin{minipage}[t]{0.9\textwidth} 
 \vspace*{0.5cm}
\begin{center} 
  R.~Dijkgraaf, L.~Hollands and P. Su{\l}kowski, \\
  \emph{Quantum Curves and D-Modules},\\ \textsf{arXiv/0810.4157
    [hep-th]}, JHEP 11 (2009) 047.
 \end{center}
\end{minipage}

\begin{minipage}[t]{0.9\textwidth} 
 \vspace*{0.5cm}
\begin{center} 
  M. Cheng and L.~Hollands, \\
  \emph{A Geometric Derivation of the Dyon Wall-Crossing Group},\\
  \textsf{arXiv/0901.1758 [hep-th]}, JHEP 04 (2009) 067.  
\end{center}
\end{minipage}

\end{center}

    \newpage\pagestyle{empty}
}
\fi

\newcommand{\contents}{
    \pagestyle{empty}
    \tableofcontents
    \cleardoublepage\pagestyle{thesisheadings}
}

\fancypagestyle{plain}
{
    \fancyhead{}

}
\fancypagestyle{thesisheadings}
{
    \fancyhead{}\fancyfoot{}
    \fancyhead[CE]{\nouppercase{\it\leftmark}}
    \fancyhead[CO]{\nouppercase{\it\rightmark}}
    \fancyhead[RO]{\thepage}
    \fancyhead[LE]{\thepage}

}
\makeatletter
\def\cleardoublepage{\clearpage\if@twoside \ifodd\c@page\else
  \hbox{}
  \vspace*{\fill}
  \thispagestyle{empty}
  \newpage
  \if@twocolumn\hbox{}\newpage\fi\fi\fi}
\makeatother

\usepackage{dsfont}
\usepackage{youngtab}
\makeatother

\newcommand{\bea}{\begin{eqnarray}}
\newcommand{\eea}{\end{eqnarray}}
\newcommand{\be}{\begin{equation}}
\newcommand{\ee}{\end{equation}}

\newcommand{\bem}{\begin{pmatrix}}
\newcommand{\eem}{\end{pmatrix}}

\newcommand{\del}{\partial}
\newcommand{\delbar}{\overline{\partial}}

\newcommand{\hf}{\frac{1}{2}}

\newcommand{\Z}{{\mathbb Z}}
\newcommand{\R}{{\mathbb R}}
\newcommand{\C}{{\mathbb C}}

\renewcommand{\P}{{\mathbb P}}
\newcommand{\M}{{\cal M}}
\newcommand{\Q}{{\mathbb Q}}

\def\Tr{{\rm Tr}}

\def\g{\gamma} 
\def\G{\Gamma}
\def\e{\epsilon}

\def\la{\lambda}
\def\bl{\bar\lambda}

\def\n{\nu}
 
\def\s{\sigma} 
\def\bt{\bar\tau}
\def\f{\phi} 
 
\def\w{\omega}
\def\W{\Omega} 
 
\def\z{\zeta}
\def\S{\varSigma}
\def\pa{\partial}

\def\re{\mathrm{Re}}  
\def\im{\mathrm{Im}}

\def\lieg{{\mathfrak g}}

\newcommand{\cB}{{\cal B }}            
\newcommand{\cM}{{\cal M }}            
\newcommand{\cF}{{\cal F }}            

\newcommand{\cK}{{\cal K }} 
\newcommand{\cE}{{\cal E }}                  
\newcommand{\cL}{{\cal L }}                
\newcommand{\cO}{{\cal O }}            
\newcommand{\cH}{{\cal H }}            
\newcommand{\cA}{{\cal A }}            

\newcommand{\cN}{{\cal N }}            
\newcommand{\cY}{{\cal Y }}    
\newcommand{\cD}{{\cal D }} 
\newcommand{\cV}{{\cal V }}    
\newcommand{\cW}{{\cal W }}
\newcommand{\cZ}{{\cal Z }}
   
\newcommand{\B}{{\cal B}} 
\newcommand{\lotjesd}{{\partial}}

\def\wt{\widetilde}

\usepackage[english,dutch]{babel}
\usepackage{amsmath,amssymb}
\usepackage[dvips]{graphicx}
\usepackage{pstricks}

\usepackage{makeidx}
\makeindex

\newcommand{\Index}[1]{\index{#1}\emph{#1}}

\usepackage{caption}

\usepackage[pageanchor]{hyperref}
\usepackage{cleveref}
\usepackage{epsf}
\usepackage{hyperref}
\usepackage{amssymb,amsmath,dsfont,theorem,multirow}

\thesischapterstyle{\centering\Huge\it\bfseries}
\thesissectionstyle{\bfseries}
\thesissectionsizes{\Large}{\large}{\normalsize}
\thesiscolofon{\vspace*{\fill}{
    \itshape\small
    \begin{minipage}[b]{\textwidth}
     
    \end{minipage}    
}
}
\title{Topological Strings and Quantum Curves}

\author{Lotte Hollands}
\placeofbirth{Maasbree}

\theday{donderdag}
\date{3 september}
\thetime{14.00}

\rectormagnificus{prof. dr. D.C. van den Boom}
\promotor{prof. dr. R.H. Dijkgraaf}
\othercomitteemembers{
prof. dr. J. de Boer,
prof. dr. A.O. Klemm,
prof. dr. E.M. Opdam,
dr. H.B. Posthuma,
dr. S.J.G. Vandoren,
prof. dr. E.P. Verlinde,

}
\department{Faculteit der Natuurwetenschappen, Wiskunde en
Informatica}

\begin{document}

    \selectlanguage{english}
    \frontmatter
        \maketitle
        \contents
    \mainmatter
        
\chapter{Introduction}

The twentieth century has seen the birth of two influential theories
of physics. On very small scales quantum mechanics is very successful,
whereas general relativity rules our universe on large
scales. Unfortunately, in regimes at small scale where gravity is
nevertheless non-negligible---such as black holes or the big
bang---neither of these two descriptions suffice. String theory is the
best candidate for a unified theoretical description of nature to
date.  

However, strings live at even smaller scales than those governed by
quantum mechanics. Hence string theory necessarily involves levels of
energy that are so high that they cannot be simulated in a
laboratory. Therefore, we need another way to arrive at valid
predictions. Although string theory thus rests on physical arguments,
it turns out that this framework carries rich mathematical
structure. This means that a broad array of mathematical techniques
can be deployed to explore string theory. 
Vice versa, physics can also benefit mathematics. For example, 
different physical perspectives can relate previously unconnected
topics in mathematics.  

In this thesis we are motivated by this fertile interaction. We both
use string theory to find new directions in mathematics, and employ
mathematics to discover novel structures in string theory. Let us
illustrate this with an example.



\section{Fermions on Riemann surfaces}

This section briefly introduces the main ingredients of this
thesis. We will meet all in much greater detail in the following
chapters. 

\begin{figure}[h]
\begin{center} 
\includegraphics[width=5cm]{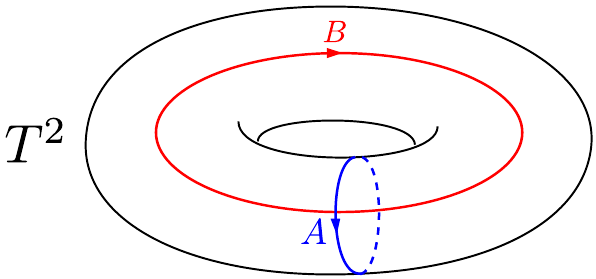}
\caption{A compact Riemann surface with just one hole in it is called a 2-torus $T^2$. We refer to its two 1-cycles as the $A$ and the $B$-cycle. }\label{fig:torus}
\end{center}
\end{figure}

So-called Riemann surfaces play a prominent role in many of the
fruitful interactions between mathematics and theoretical physics that
have developed in the second half of the twentieth century. Riemann
surfaces are smooth two-dimensional curved surfaces that have a number
of holes, which is called their genus $g$. Fig.~\ref{fig:torus} shows
an example.

Riemann surfaces additionally come equipped with a complex structure,
so that any small region of the surface resembles the complex plane
$\mathbb{C}$. An illustrative class of Riemann surfaces is defined by
equations of the form
\begin{align*}
\Sigma: \quad  F(x,y) = 0,  \quad \textrm{where} \quad x, \, y \in \mathbb{C}.
\end{align*}
Here, $F$ can for instance be a polynomial in the complex variables
$x$ and $y$. A simple example is a (hyper)elliptic curve defined by  
\begin{align*}
y^2 = p(x),  
\end{align*}
where $p(x)$ is a polynomial. The curve is elliptic if $p$ has degree
3 or 4, in which case its genus is $g=1$. For higher degrees of $p$ it
is hyperelliptic and has genus $g > 1$.

\begin{figure}[h]
\begin{center} 
\includegraphics[width=5cm]{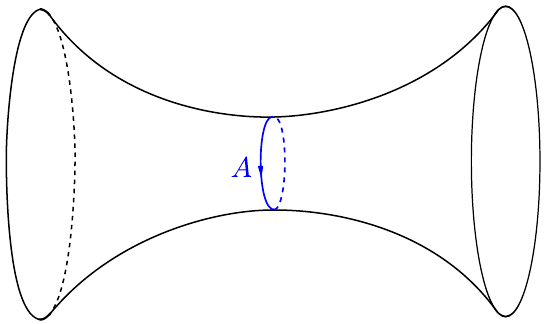}
\caption{A simple example of a non-compact Riemann surface is defined by the equation $x^2 + y^2 = 1$ in the complex plane $\C^2$.}\label{fig:curveconifold}
\end{center}
\end{figure}

The complex structure of a compact Riemann surface can be conveniently
characterized by a period matrix.
This is a symmetric square matrix $\tau_{ij}$ of complex numbers,
whose rank equals the genus of the surface and whose imaginary part is
strictly positive. The complex structure of a 2-torus, illustrated in
Fig.~\ref{fig:torus}, is for example determined by one complex number
$\tau$ that takes values in the upper half plane.

The period matrix encodes the contour integrals of the $g$ independent
holomorphic 1-forms over the 1-cycles on the surface. To this end one
picks a canonical basis of $A$-cycles and $B$-cycles, where the only
non-trivial intersection is $A_1 \cap B_j = \delta_{ij}$. The 1-forms
$\omega_i$ may then be normalized by integrating them over the
$A$-cycles, so that their integrals over the $B$-cycles determine the
period matrix:
%
\begin{align*}
 \int_{A_i} \omega_j = \delta_{ij}, \quad \int_{B_i} \omega_j =  \tau_{ij}. 
\end{align*}

\subsubsection{Conformal field theory}

Riemann surfaces play a dominant role in the study of conformal field
theories (CFT's). Quantum field theories can be defined on a
space-time background $M$, which is usually a Riemannian manifold with
a metric $g$. The quantum field theory is called conformal when it is
invariant under arbitrary rescaling of the metric $g$. A CFT
therefore depends only on the conformal class of the metric $g$.

The simplest quantum field theories study a bosonic scalar field
$\phi$ on the background $M$. The contribution to the action for a
massless scalar is 
\begin{align*}
S_{\textrm{boson}} = \int_M \sqrt{g} \, g^{mn} \, \partial_{m} \phi \, \partial_{n} \phi,   
\end{align*}
yielding the Klein-Gordon equation $\partial_{m} \partial^{m} \phi =0$
(for zero mass) as equation of motion. Note that, on the level of the
classical action, this theory is clearly conformal in 2 dimensions;
this still holds for the full quantum theory. Since a 2-dimensional
conformal structure uniquely determines a complex structure, the free
boson defines a CFT on any Riemann surface $\Sigma$.

Let us compactify the scalar field $\phi$ on a circle $S^1$, so that
it has winding modes along the 1-cycles of the Riemann surface.  The
classical part of the CFT partition function is determined by the
solutions to the equation of motion. The holomorphic contribution to
the classical partition function is well-known to be encoded in a
Riemann theta function
\begin{align*}
 \theta \left(\tau, \nu \right) = \sum_{p \in \mathbb{Z}^g} e^{2 \pi i \left( \frac{1}{2} p^t \tau p + p^t \nu \right) },   
\end{align*}
where the integers $p$ represent the momenta of $\phi$ that flow
through the $A$-cycles of the 2-dimensional geometry.

Another basic example of a CFT is generated by a fermionic field
$\psi$ on a Riemann surface. Let us consider a chiral fermion
$\psi(z)$. Mathematically, this field transforms on the Riemann
surface as a $(1/2,0)$-form, whence
\begin{align*}
\psi(z) \sqrt{dz} = \psi(z') \sqrt{dz'},
\end{align*}
where $z$ and $z'$ are two complex coordinates.
Such a chiral fermion contributes to the 2-dimensional action as
\begin{align*}
S_{\textrm{fermion}} = \int_{\Sigma} \psi(z) \, \overline{\partial}_A \psi(z),
\end{align*}
where $\partial_A = \partial + A$, and $A$ is a connection 1-form on
$\Sigma$. This action determines the Dirac equation
$\overline{\partial}_A \psi(z) = 0$ as its equation of motion. 
The total partition function of the fermion field $\psi(z)$ is
computed as the determinant 
\begin{align*}
Z_{\textrm{fermion}} =  \textrm{det} (\overline{\partial}_ A).
\end{align*}
Remarkably, this partition function is also proportional to a Riemann
theta function. The fact that this is not just a coincidence, but a
reflection of a deeper symmetry between 2-dimensional bosons and
chiral fermions, is known as the boson-fermion correspondence.


\subsubsection{Integrable hierarchy}

Fermions on Riemann surfaces are also familiar from the perspective of
integrable systems. Traditionally, integrable systems are Hamiltonian
systems  
\begin{align*}
\dot{x}_i = \{ x_i, H \} = \frac{\partial H}{ \partial p_i}, \qquad \dot{p}_i = \{
p_i, H\} = \frac{\partial H}{ \partial x_i},  
\end{align*}
with coordinates $x_i$, momenta $p_i$ and a Hamiltonian $H$, for which
there exists an equal amount of integrals of motion $I_i$ such that 
\begin{align*}
 \dot{I}_i = \{ I_i, H\} = 0, \quad  \{I_i, I_j \} =0. 
\end{align*}
The left equation requires the integrals $I_i$ to be constants
of motion whereas the right one forces them to commute
among each other. 
\index{integrable system}

\begin{figure}[h]
\begin{center} 
\includegraphics[width=5cm]{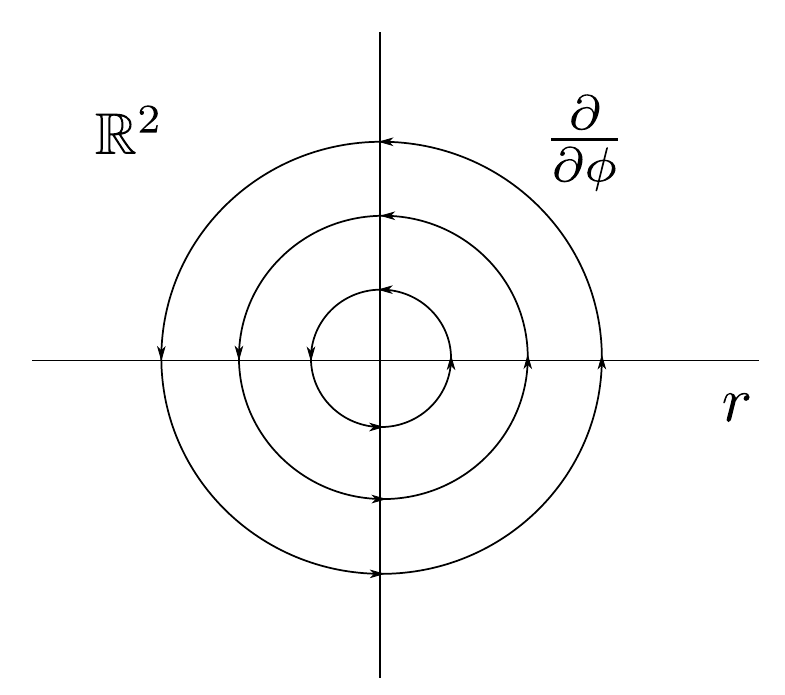}
\caption{The two-dimensional plane $\R^2$ seen as an integrable system.}\label{fig:simpleintsystem}
\end{center}
\end{figure}

A simple example is the real two-dimensional plane $\mathbb{R}^2$ with
polar coordinates $r$ and $\phi$, see
Fig.~\ref{fig:simpleintsystem}. Take the Hamiltonian $H$ to be the
radius $r$. Then $H=r$ is itself an integral of motion, so that the
system is integrable. Notice that $H$ is constant on the flow generated
by the differential $\partial/\partial \phi$.

A characteristic example of an integrable system that is intimately
related to Riemann surfaces and theta-functions is the Korteweg de
Vries (KdV) hierarchy. Although this system was already studied by
Korteweg and de Vries at the end of the 19th century as a non-linear
partial differential equation
\begin{align*}
 \frac{\partial u}{\partial t} =  \frac{\partial^3 u}{\partial x^3} + 6u \frac{\partial u}{\partial x},
\end{align*}
it was only realized later that it is part of a very rich geometric
and algebraic structure. It is impossible to do justice to this
beautiful story in this short introduction. Instead, let us just touch on
the aspects that are relevant for this thesis. 

Geometrically, a special class of solutions to the KdV differential equation
yields linear flows over a $2g$-dimensional torus that is associated to
a Riemann surface $\Sigma$. This torus is called the Jacobian. Its
complex structure is determined by the period matrix $\tau_{ij}$ of
$\Sigma$.  The Riemann surface $\Sigma$ is called the spectral curve
of the KdV hierarchy.

The KdV spectral curve is an elliptic curve
\begin{align*}
Q^2 = P^3 - g_2 P - g_3,  
\end{align*}
where $g_2$ and $g_3$ are the Weierstrass invariants. The coordinates $P$ and $Q$ on this elliptic curve can best be described as commuting differential operators
\begin{align*}
 [P,Q]=0, 
\end{align*}
that arise naturally when the KdV differential equation is written in the form of a Hamiltonian system. 

As a side remark we notice that the KdV system is closely related to
the Hitchin integrable system, which studies certain holomorphic
bundles over a complex curve $C$. In this integrable system, too, the
dynamics can be expressed in terms of a linear flow on the Jacobian
associated to a spectral cover $\Sigma$ of $C$. 

We go one step beyond the geometrical structure of the KdV system when
introducing chiral fermions on the spectral curve $\Sigma$. Their 
partition function  $\det(\overline{\partial})$ transforms as a
section of the determinant line bundle over the moduli space of the
integrable hierarchy. It is known as the tau-function. 
The tau-function is proportional to the theta-function associated to
the period matrix $\tau_{ij}$.  

\subsubsection{Gauge theory and random matrices}

Similar integrable structures have been found in wide variety of physical
theories recently. An important ingredient of this thesis is the
appearance of an auxiliary Riemann surface in 4-dimensional gauge theories. The
basic field in a gauge theory is a gauge field $A$, which is
mathematically a connection 1-form of a principal $G$-bundle over the
4-manifold 
$M$. Let us take $G = U(1)$. In that situation the classical equations
of motion reproduce the Maxwell equations. Once more, the holomorphic
part of the gauge theory partition function is essentially a
theta-function 
\begin{align*}
\int DA~ e^{\int_M \frac{1}{2} \tau F_+ \wedge F_+} \sim \sum_{p_+} e^{\frac{1}{2} \tau p_+^2} ,
\end{align*}
where $\tau$ is the complex coupling constant of the gauge theory, and $p_+$ are the fluxes of the self-dual field strength $F_+$ through the 2-cycles of $M$. It
turns out to be useful to think of $\tau$ as the complex structure
parameter of an auxiliary elliptic curve. In other gauge theories
a similar auxiliary curve is known to play a significant role as
well. We will explain this in detail in the main body of this thesis.

Another interesting application is the theory of random matrices. This
is a 0-dimensional quantum field theory based on a Hermitean
$N$-by-$N$-matrix $X$. The simplest matrix model is the so-called
Gaussian one, whose action reads  
\begin{align*}
S_{\textrm{matrix}} = -  \frac{1}{2 \lambda} \textrm{Tr} X^2,
\end{align*}
where $\lambda$ is a coupling constant. Other matrix models are
obtained by adding higher order interactions to the matrix model
potential $W(X) = 
X^2/2$. 
Such a matrix model can be
conveniently evaluated by diagonalizing the
matrices $X$. This reduces the path integral to a standard
integral over eigenvalues $x_i$, yet adds the extra term  
\begin{align*}
 \frac{1}{N^2} \sum_{i,j} \log |x_i - x_j| 
\end{align*}
to the matrix model action. Depending on the values of the parameters
$N$ and $\lambda$ the eigenvalues will either localize on the minima
of the potential $W(X)$ or, oppositely, spread over a larger
interval. When one lets 
$N$ tend to infinity while keeping the 't Hooft parameter $\mu = N \lambda$
fixed, the eigenvalues can be seen to form a smooth distribution. For
instance, in the Gaussian matrix model, the density of eigenvalues is
given by the Wigner-Dyson semi-circle 
\begin{align*}
\rho(x) \sim \sqrt{4 \mu -x^2},
\end{align*}
leading to the algebraic curve
\begin{align*}
 x^2 + y^2 = 4 \mu. 
\end{align*}
Likewise, the eigenvalue distribution of a more general matrix model
takes the form of a hyperelliptic curve in the limit $N \to \infty$
with fixed 't Hooft parameter. Many properties of the matrix
model are captured in this spectral curve. 
 
\subsubsection{Topological string theory}

String theory provides a unifying framework to discuss all these
models. In particular, topological string theory studies embeddings of a Riemann surface $C$---which shouldn't be
confused with the spectral curve---into 6-dimensional target manifolds $X$ of
a certain kind. These target manifolds are 
called Calabi-Yau manifolds. The Riemann surface $C$ is the
worldvolume swept out in time by a 1-dimensional string. The
topological string partition function is a series expansion 
\begin{align*}
Z_{\textrm{top}}(\lambda) = \exp \left( \sum_g \lambda^{2g-2} \mathcal{F}_g \right),  
\end{align*}
where $\lambda$ is called the topological string coupling
constant. Each $\mathcal{F}_g$ contains the contribution of curves $C$
of genus $g$ to the partition function.    

Calabi-Yau manifolds are not well understood in general, and studying
topological string theory is very difficult task. 
Nevertheless, certain classes of non-compact Calabi-Yau manifolds are much
easier to analyse, and contain precisely the relevant backgrounds
to study the above integrable structures. 

Consider an equation of the form 
\begin{align*}
X_{\Sigma}: \quad  uv - F(x,y) = 0, 
\end{align*}
where $u$ and $v$ are $\mathbb{C}$-coordinates and $$ \Sigma: \quad F(x,y)=0$$ 
defines the spectral curve $\Sigma$ in the complex $(x,y)$-plane.
The 6-dimensional manifold $X_{\Sigma}$
may be regarded as a $\mathbb{C}^*$-fibration over the $(x,y)$-plane that
degenerates over the Riemann surface $F(x,y)=0$.

\begin{figure}[h]
\begin{center} 
\includegraphics[width=4.8 cm]{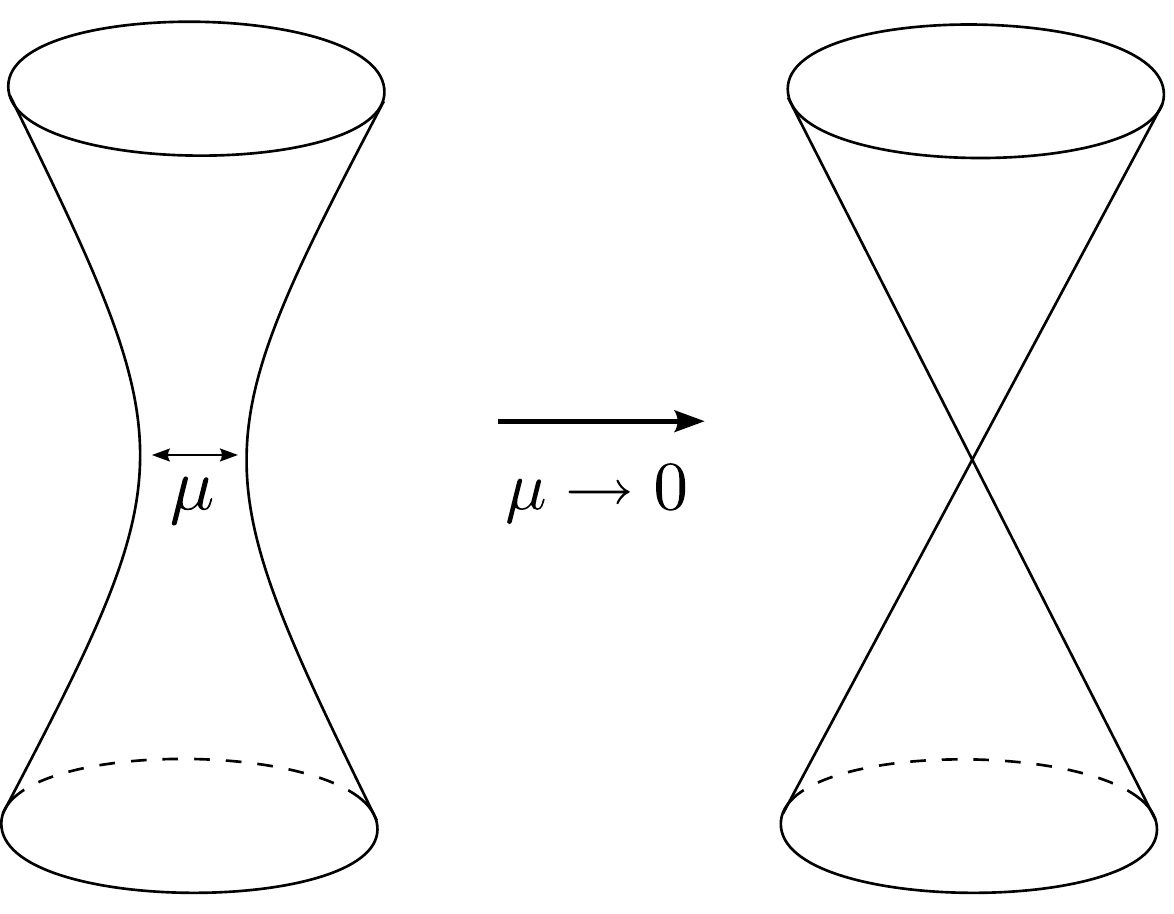}
\caption{A local Calabi-Yau threefold defined by an equation of the
  form $uv - F(x,y) = 0$ can be viewed as a $\C^*$-fibration over the
  base parametrized by $x$ and $y$. The fibers are hyperboloids over
  general points in the base, but degenerate when $\mu = F(x,y) \to
  0$. }\label{fig:hyperboloid}  
\end{center}
\end{figure}

Indeed, the fiber over a point $(x,y)$ is defined by $uv =
F(x,y)$. Over a generic point in the base this fiber is a
hyperboloid. However, when $F(x,y) \to 0$ the hyperboloid degenerates to a cone. 
This is illustrated in Fig.~\ref{fig:hyperboloid}. The
zero locus of the polynomial $F(x,y)$ thus 
determines the degeneration locus of the fibration. 
We therefore refer to
the variety $X_{\Sigma}$ as a non-compact Calabi-Yau threefold modeled
on a Riemann surface.  

When the Riemann surface $\Sigma$ equals the auxiliary Riemann surface that emerges in the study of gauge theories, the topological string partition function is known to capture important properties of the gauge theory. Furthermore, when it equals the spectral curve in the matrix model, it is known to capture the full matrix model partition function. 

\subsubsection{Quantum curves and $\cD$-modules}

However, topological string computations go well beyond computing a fermion determinant on $\Sigma$. In fact, it is known that the fermion determinant appears as the genus 1 part $\mathcal{F}_1$ of the free energy. We call this the semi-classical part of the free energy, as it remains finite when $\lambda \to 0$. Similarly, $\mathcal{F}_0$ is the classical contribution that becomes very large in this limit, and all higher $\mathcal{F}_g$'s encode the quantum corrections. In other words, we may interpret the loop parameter $\lambda$ as a quantization of the semi-classical tau-function. But what does this mean in terms of the underlying integrable system?
Answering this question is one of the goals in this thesis. 

In both the gauge theory and the matrix model several hints have been
obtained. In the context of gauge theories the higher loop corrections
$\mathcal{F}_g$ are known to correspond to couplings of the gauge
theory to gravity. Mathematically, these appear by making the theory
equivariant with respect to an $SU(2)$-action on the underlying
4-dimensional background. N. Nekrasov and A. Okounkov have shown that
the partition function of this theory is the tau-function of an
integrable hierarchy.  
 
In the theory of random matrices, $\lambda$-corrections correspond to
finite $N$ corrections. In this theory, too, it has been found that the
partition function behaves as the tau-function of an integrable
hierarchy. However, like in the gauge theory examples, there is no
interpretation of the full partition function as a fermion
determinant. 
Instead, a new perspective arises, in which
the spectral curve is be replaced by a non-commutative spectral
curve. In matrix models that relate to the KdV integrable system,  
it is found that the KdV differential operators $P$ and $Q$ do not
commute anymore when the quantum corrections are taken into account:  
\begin{align*}
 [P,Q]= \lambda. 
\end{align*}
Similar hints have been found in the study of topological string theory on general local Calabi-Yau's modeled on a Riemann surface.


This thesis introduces a new physical perspective to study topological
string theory, clarifying the interpretation of the parameter
$\lambda$ as a ``non-commutativity parameter''. We show that 
$\lambda$ quantizes the Riemann surface $\Sigma$, and that fermions on
the curve are sections of a so-called
$\mathcal{D}$-module instead. This attributes Fourier-like
transformation properties to the fermions on $\Sigma$:
\begin{align*}
 \psi(y) = \int dy\, e^{xy/\lambda} \, \psi(x). 
\end{align*}

\section{Outline of this thesis}\label{sec:overview}

\Cref{ch:intro} starts with an introduction to the geometry of Calabi-Yau threefolds. They serve as an important class of backgrounds in string theory, that can be used to make contact with the 4-dimensional world we perceive. Specifically, we explain the idea of a Calabi-Yau compactification, and introduce certain non-compact Calabi-Yau backgrounds that appear in many guises in this thesis. 


Topological string theory on these non-compact Calabi-Yau backgrounds is of physical relevance in the description of 4-dimensional supersymmetric gauge theories and supersymmetric black holes. For example, certain 
holomorphic corrections to supersymmetric gauge theories are known to have an elegant description in terms of an auxiliary Riemann surface $\Sigma$. A compactification of string theory on a local Calabi-Yau that is modeled on this Riemann surface ``geometrically
engineers'' the corresponding gauge theory in four
dimensions. Topological string amplitudes capture the holomorphic
corrections.

\begin{figure}[h]
\begin{center} 
 \includegraphics[width=\linewidth]{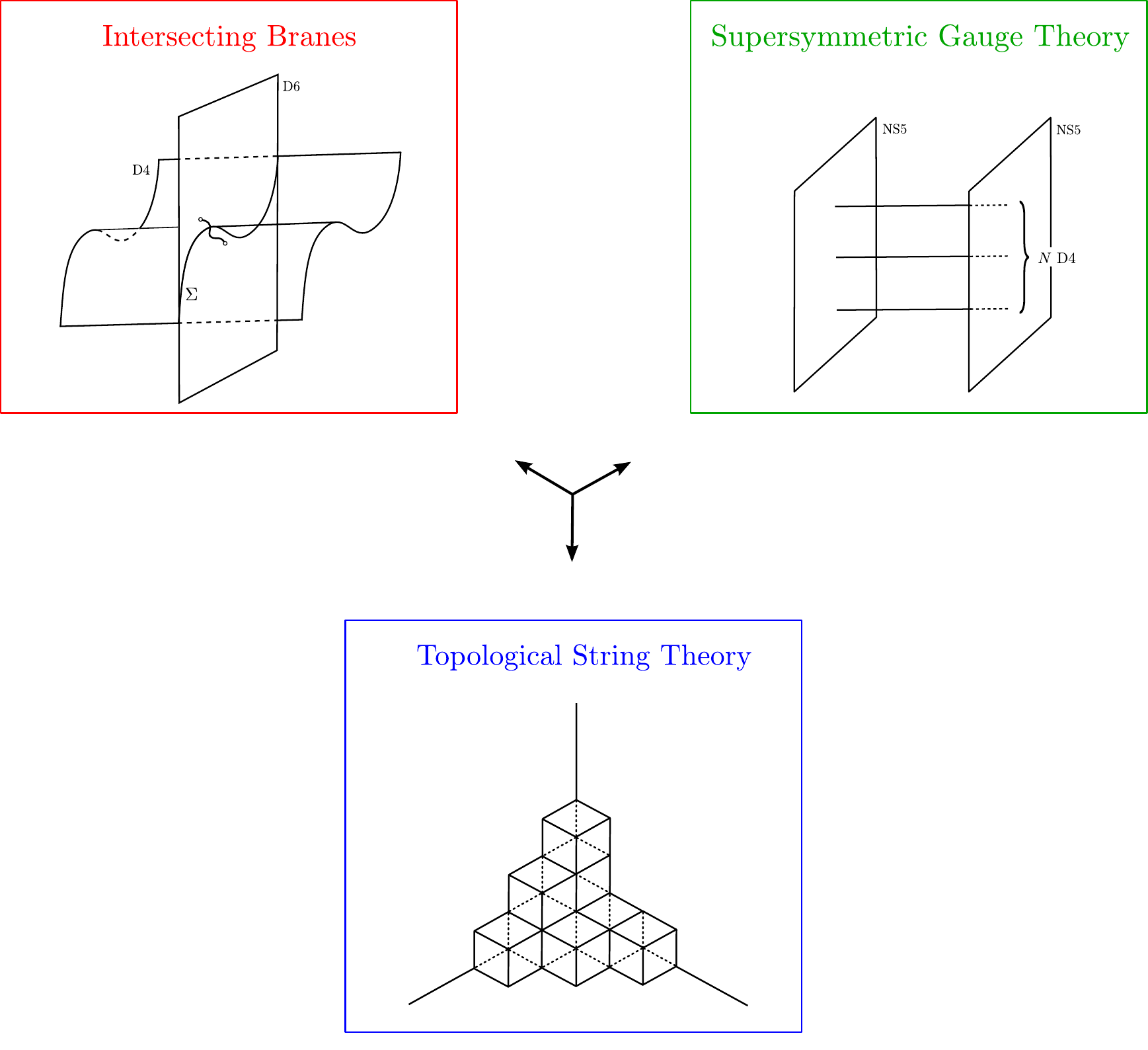}
\caption{The web of dualities relates 4-dimensional supersymmetric
  gauge theory to topological string theory and to an intersecting brane
  configuration.}\label{fig:pictorialweb}
\end{center}
\end{figure}
\begin{figure}[b]
\begin{center} 
 \includegraphics[width=\linewidth]{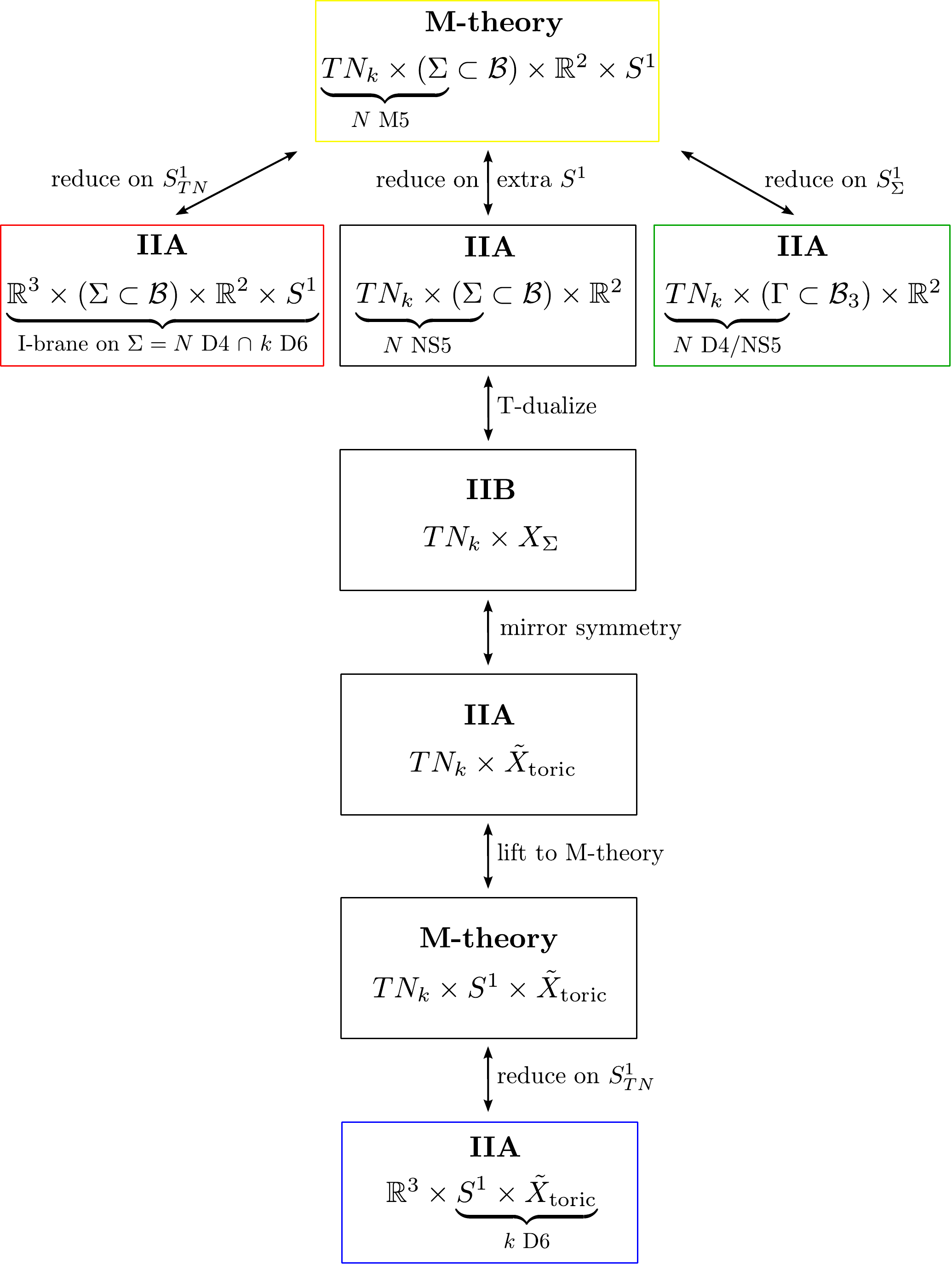}
\caption{Web of dualities.}\label{fig:webofdualities}
\end{center}
\end{figure}

In Chapter~\ref{chapter2} and \ref{chapter3+4} we elucidate these relations by introducing a web of dualities in string theory. This web
is given in full detail in Fig.~\ref{fig:webofdualities}. Most important are its outer boxes, that are illustrated pictorially in Fig.~\ref{fig:pictorialweb}. They correspond to string theory embeddings of the theories we mentioned above. The upper right box is a string theory embedding of  supersymmetric gauge theory, whereas the lowest box is a string theory embedding of topological string theory. 

The main objective of the duality
web is to relate both string frames to the upper left box, that
describes a string theory configuration of \emph{intersecting D-branes}. The most relevant feature of this intersecting brane configuration is that the D-branes intersect over a Riemann surface $\Sigma$, which is the same Riemann surface that underlies the gauge theory and appears in topological string theory. We explain that
the duality chain gives a dual description of topological string theory in terms of a 2-dimensional quantum field theory of free fermions that live on the 2-dimensional intersecting brane, the so-called I-brane, wrapping~$\Sigma$. 

Chapter~\ref{chapter2} studies 4-dimensional
gauge theories that preserve a maximal amount \hyphenation {a-mount}
of supersymmetry. They
are called $\cN=4$ supersymmetric Yang-Mills theories.  Instead of
analyzing this theory on flat $\R^4$, we consider more general ALE
backgrounds.  These non-compact manifolds are
resolved singularities of the type $$\C^2/\Gamma,$$ where $\Gamma$ is a
finite subgroup of $SU(2)$ that acts linearly on $\C^2$.
In the mid-nineties it has been discovered that the gauge theory partition
function on an ALE space computes 2-dimensional CFT characters.
In Chapter~\ref{chapter2} we introduce a string theoretic set-up that
explains this duality between 4-dimensional gauge theories and
2-dimensional CFT's from a higher standpoint. 

In Chapter~\ref{chapter3+4}  we extend
the duality between supersymmetric gauge theories and intersecting
brane configurations to the full duality web in
Fig.~\ref{fig:webofdualities}. We start out with 4-dimensional $\cN=2$
gauge theories, that preserve half of the supersymmetry of the
above $\cN=4$ theories, and line out their relation to the
intersecting brane configuration and to local
Calabi-Yau compactifications. 
Furthermore, we relate several types of  objects that play
an important role in the duality sequence, and propose a partition
function for the intersecting brane configuration. 


The I-brane configuration emphasizes the key role of the
auxiliary Riemann surface $\Sigma$ in 4-dimensional gauge theory and topological string theory. All the
relevant physical modes are localized on the intersecting brane that
wraps $\Sigma$. These modes turn out to be free fermions.  
In Chapter~\ref{chapter5} and~\ref{chapter6} we describe the resulting
2-dimensional quantum field on $\Sigma$ in the language of integrable
hierarchies.

We connect local Calabi-Yau compactifications 
to the Kadomtsev-Petviashvili, or shortly, KP
integrable hierarchy, which is closely related to 2-dimensional free
fermion conformal field theories. To any semi-classical free fermion
system on a curve $\Sigma$ it associates a solution of the integrable
hierarchy. This is well-known as a Krichever solution. 

%
%

In Chapter~\ref{chapter5} we explain how to interpret the topological
string partition function on a local Calabi-Yau $X_{\Sigma}$ as a
quantum deformation of a Krichever solution. We discover that the
curve $\Sigma$ should be replaced by a \emph{quantum curve}, 
defined as a differential operator $P(x)$ obtained by quantizing $y
= \la \partial/\partial x$. Mathematically, we are led to a novel
description of topological string theory in terms of $\cD$-modules. 

Chapter~\ref{chapter6} illustrates this formalism
with several examples related to matrix models and gauge theory. In
all these examples there is a canonical way to quantize the curve as a
differential operator. Moreover, we can simply read off the known
partition functions from the associated $\cD$-modules. This tests our proposal.



The topological string partition function has several interpretations
in terms of topological invariants on a Calabi-Yau
threefold. Moreover, its usual expansion in the topological string
coupling constant $\la$ is only valid in a certain regime of the
background Calabi-Yau parameters. Although the topological invariants
stay invariant under small deformations in these background
parameters, there are so-called \emph{walls of marginal stability},
where the index of these invariants may jump. This
phenomenon has recently attracted much attention among both physicists and
mathematicians. 


Chapter~\ref{chapter7} studies wall-crossing in string theory
compactifications that preserve $\cN=4$ supersymmetry. The invariants
in this theory that are sensitive to wall-crossing are called quarter
BPS states. They have been studied extensively in the
last years. 
In Chapter~\ref{chapter7} we work out in detail
the relation of these supersymmetric states to a genus 2 surface
$\Sigma$. Moreover, we find that wall-crossing in this theory has a
simple and elegant interpretation in terms of the genus 2 surface.




Finally, Chapter~\ref{chapter8} focuses on supersymmetry breaking in
Calabi-Yau compactifications. Since supersymmetry is not an exact
symmetry of nature, we need to go 
beyond Calabi-Yau compactifications to find a more accurate description of
nature. One of the possibilities is that supersymmetry is broken at a
lower scale than the compactification scale. In Chapter~\ref{chapter8}
we study such a supersymmetry breaking mechanism for $\cN=2$
supersymmetric gauge theories. These gauge theories are related to
type II string compactifications on a local Calabi-Yau
$X_{\Sigma}$. By turning on non-standard fluxes (in the form of 
generalized gauge fields) on the underlying Riemann
surface $\Sigma$ we find a potential on the moduli space of the
theory. We show that this potential generically has non-supersymmetric
minima.   

 \emph{Chapter~\ref{chapter2} and
  \ref{chapter3+4} explain the 
ideas of the publication \cite{Dijkgraaf:2007sw}, whereas
Chapter~\ref{chapter5} and \ref{chapter6} contain the results of
\cite{Dijkgraaf:2008fh}. The first part of Chapter~\ref{chapter7}
includes examples from \cite{Dijkgraaf:2007sw}, whereas the second part
is based on the article \cite{Cheng:2009hm}. Finally,
Chapter~\ref{chapter8} is a 
slightly shortened version of the work \cite{Hollands:2008cs}.}  

\newpage

\vspace*{2cm}

\chapter{Calabi-Yau Geometry}\label{ch:intro}

Although string theory lives on 10-dimensional backgrounds, it is
possible to make contact with the 4-dimensional world we perceive by
shrinking 6 out of the 10 dimensions to very small scales. Remarkably,
this also yields many important examples of interesting interactions
between physics and mathematics.  The simplest way to compactify a
10-dimensional string background to 4 dimensions is to start with a
compactification of the form $\R^{4} \times X,$ where $X$ is a compact
6-dimensional Riemannian manifold and $\R^{4}$ represents Minkowski
space-time.

\begin{figure}[h]
\begin{center} 
\includegraphics[width=5.8cm]{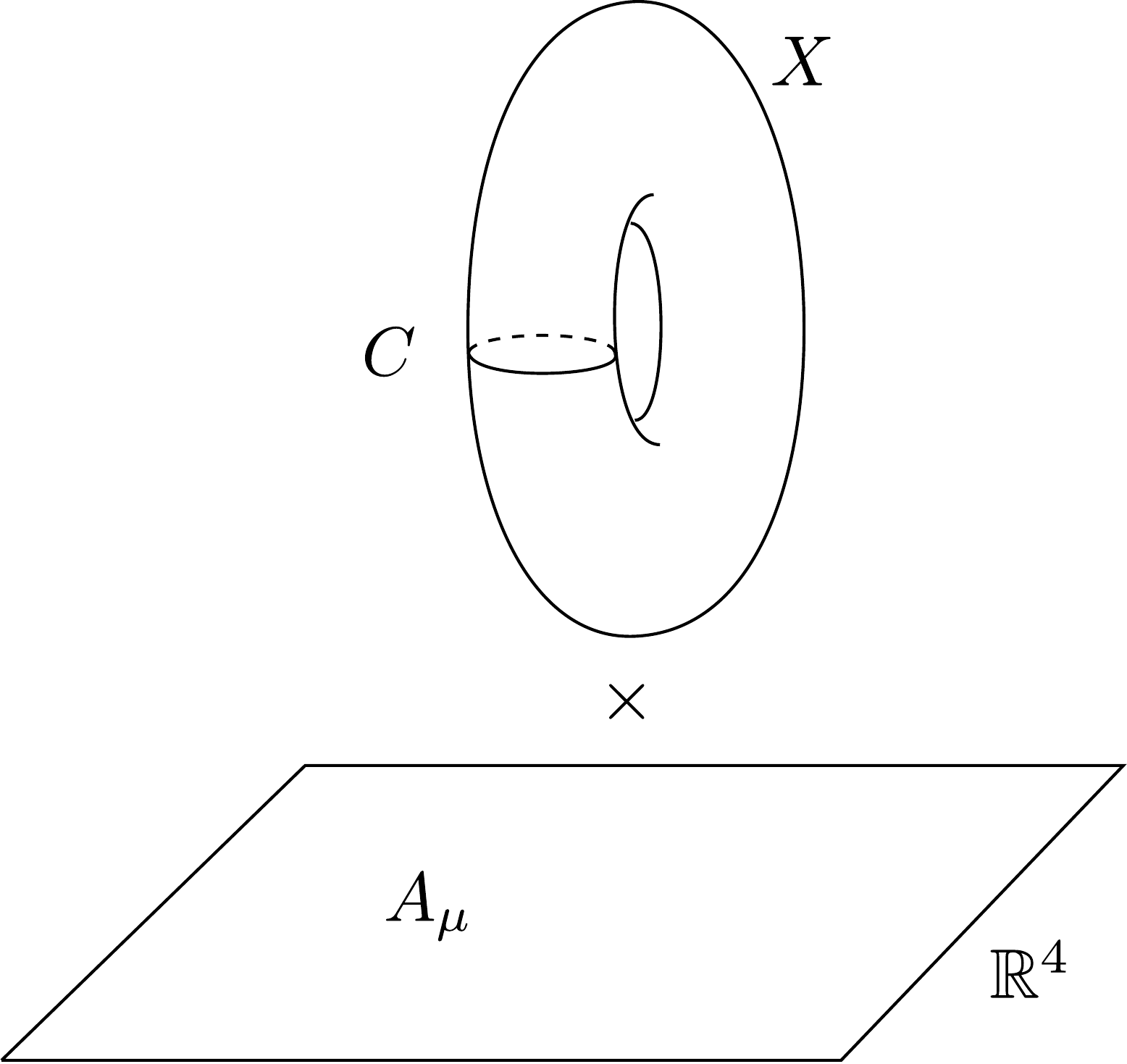}
\caption{Compactifying string theory over an internal space
  $X$ geometrizes 4-dimensional physics. Here we represented the
  6-dimensional internal space as a 2-torus. The Yang-Mills gauge field
  $A_{\mu}$ on $\R^4$ is obtained by integrating a 10-dimensional
  RR 4-form over a 3-cycle $C$.}\label{fig:calabiyaucomp}
\end{center}
\end{figure}

Apart from the metric, string theory is based on several more
quantum fields. Amongst them are generalized Ramond-Ramond (RR) gauge fields, the analogues of the famous 
Yang-Mills field in four dimensions. In a compactification
to four dimensions these fields can be integrated over any cycle of
the internal manifold $X$. This results in e.g. scalars and gauge
fields, which are the building blocks of the standard model. 
This is illustrated in Fig.~\ref{fig:calabiyaucomp}.
Four-dimensional symmetries, like gauge symmetries,
can thus be directly related to geometrical symmetries of the internal
space. This geometrization of physics is a main theme in string
theory, and we will see many examples in this thesis. 

Although the topological properties of the compactification manifold $X$
are restricted by the model that we try to engineer in four
dimensions, other moduli of $X$, such as its size and shape, are a priori
allowed to fluctuate. This leads to the so-called landscape of string
theory, that parametrizes all the possible vacua. A way to truncate
the possibilities into a discrete number of vacua is to
introduce extra fluxes of some higher dimensional gauge fields. We
will come back to this at the end of this thesis, in Chapter~\ref{chapter8}.      

The best studied compactifications are those that preserve
supersymmetry in four dimensions. This yields severe constraints on
the internal manifold. Not only should it be provided with a complex
structure, but also its metric should be of a special form. They are
known as \Index{Calabi-Yau} manifolds. Remarkably, this class
of manifolds is also very rich from a mathematical
perspective.


This chapter is meant to acquaint the reader with Calabi-Yau
manifolds. The aim of \Cref{sec:SYZ} is to describe
 \emph{compact} real 6-dimensional Calabi-Yau manifolds in terms of
real 3-dimensional 
geometry. \Cref{sec:Fermatquintic} illustrates this with the prime example of a compact
Calabi-Yau manifold, the Fermat quintic. We find that its
3-dimensional representation is characterized by a Riemann
surface. In \Cref{sec:localCY} we explain how this places the
simpler \emph{local} or \emph{non-compact} Calabi-Yau
compactifications into context. These are studied in the main body of this
thesis.   



\section{The Strominger-Yau-Zaslow conjecture}\label{sec:SYZ} 

Calabi-Yau manifolds $X$ are complex Riemannian manifolds with some
additional structures. These structures can be characterized in a few equivalent
ways. 
One approach is to combine the metric $g$ and the complex
structure $J$ into a 2-form $k = g \circ J$. For Calabi-Yau manifolds
this 2-form needs to be closed $dk=0$. It is called the \Index{K\"ahler form}
and equips the Calabi-Yau with a K\"aher structure.

Moreover, a
Calabi-Yau manifold must have a trivializable canonical bundle
$K_X = \bigwedge^3 T^*X$, where $T^*X$ is the holomorphic cotangent
space of $X$. The canonical bundle is a line bundle over $X$. When it
is trivializable the Calabi-Yau manifold contains a non-vanishing
\Index{holomorphic 3-form} $\Omega$.   
Together, the K\"ahler form $k$ and the holomorphic 3-form $\Omega$
determine a Calabi-Yau manifold. 

The only non-trivial cohomology
of a Calabi-Yau $X$ is 
contained in $H^{1,1}(X)$ and $H^{2,1}(X)$. These cohomology classes
 parametrize its moduli. The elements in $H^{1,1}(X)$ can
change the K\"ahler structure of the Calabi-Yau infinitesimally, and
are therefore called \emph{K\"ahler moduli}. In string theory we
usually complexify these moduli. On the other hand,
$H^{2,1}(X)$ parametrizes 
the \emph{complex structure  moduli} of the Calabi-Yau. It is related to
the choice of the holomorphic 3-form $\Omega$, since contracting
$(2,1)$-forms with $\Omega$ 
determines an isomorphism with the 
cohomology class $H^1_{\bar{\lotjesd}}(X)$, that is well-known to
characterize infinitesimal complex structure deformations.

Celebrated work of E.~Calabi and S.-T. Yau shows that the K\"ahler
metric $g$ can be tuned  within its cohomology class to a unique K\"ahler
metric that satisfies $R_{mn}=0$. Calabi-Yau manifolds can therefore
also be characterized by a unique Ricci-flat K\"ahler metric.

A tantalizing question is how to visualize a Calabi-Yau
threefold. The above definition in terms of a K\"ahler form $k$ and a
holomorphic 3-form $\Omega$ is rather abstract. Is there a more concrete way
to picture a Calabi-Yau manifold?

\subsubsection{The SYZ conjecture}

From the string theory perspective there are many relations
between different string theory set-ups. These are known as 
\emph{dualities}\index{duality}, and relate the different incarnations
of string 
theory (type I, type IIA/B, heterotic, M-theory and F-theory) and
different background parameters. One of the famous dualities in string
theory connects type IIB string theory compactified on some Calabi-Yau
$X$ with type IIA string theory compactified on another Calabi-Yau
$\tilde{X}$, where both Calabi-Yau manifolds are related by swapping their K\"ahler
and complex structure parameters. This duality is called \Index{mirror
  symmetry}. It suggests that any Calabi-Yau threefold $X$
has a mirror $\tilde{X}$ such that 
\begin{align*}
  H^{1,1}(X) \cong H^{2,1}(\tilde{X}) \quad \textrm{and} \quad H^{2,1}(X)
  \cong H^{1,1}(\tilde{X}). 
\end{align*}
Although the mirror conjecture doesn't hold exactly as it is stated above, many mirror pairs have been found and much intuition has been
obtained about the underlying Calabi-Yau geometry. 

One of the examples that illustrates this best is
known as the \Index{Strominger-Yau-Zaslow (SYZ) conjecture}
\cite{Strominger:1996it}. The starting 
point for this argument is mirror symmetry in type II string
theory. Without explaining in detail what type II 
string theory is, we just note that an important role in this theory
is played by so-called D-branes. These branes can be considered
mathematically as submanifolds of the Calabi-Yau manifold with certain
$U(1)$ bundles on them. 

The submanifolds are even (odd) dimensional in type
IIA (type IIB) string theory, and they are restricted by
supersymmetry. Even-dimensional 
cycles need to be holomorphic, whereas 3-dimensional cycles have to be
of \emph{special Lagrangian} type.  (Note that these are the only
odd-dimensional cycles in case the first Betti number $b_1$ of the
Calabi-Yau is zero.) A special 
Lagrangian 3-cycle $C$ is defined by the requirement that
\begin{align*}
k|_{C}=0 \quad \textrm{as well as} \quad \textrm{Im}\,\Omega|_{C}=0.
\end{align*}
Both the holomorphic and the special Lagrangian cycles minimize the
volume in their homology class, and are therefore stable against small
deformations. This turns them into supersymmetric cycles.  

Mirror symmetry not only suggests that the backgrounds $X$ and
$\tilde{X}$ are related, but that type IIA string theory
compactified on $\tilde{X}$ is equivalent to type IIB string theory on
$X$. This implies in particular that moduli spaces of
supersymmetric D-branes should be isomorphic. 
In type IIB theory in the background $\R^4 \times X$ the
relevant D-branes are mathematically characterized
by special Lagrangian submanifolds in $X$ with flat
$U(1)$-bundles on them. 
In type IIA theory
there is one particularly simple type of D-branes:
D0-branes that just wrap a point in $\tilde{X}$. Their moduli space is simply
the space $\tilde{X}$ itself. 
Since mirror symmetry
maps D0-branes into D3-branes that wrap special Lagrangian cycles in
$X$, the moduli space of these D3-branes, i.e. the moduli space of 
special Lagrangian submanifolds with a flat $U(1)$ bundle on it, should be
isomorphic to $\tilde{X}$.

\begin{figure}[h]
\begin{center} 
\includegraphics[width=8.5cm]{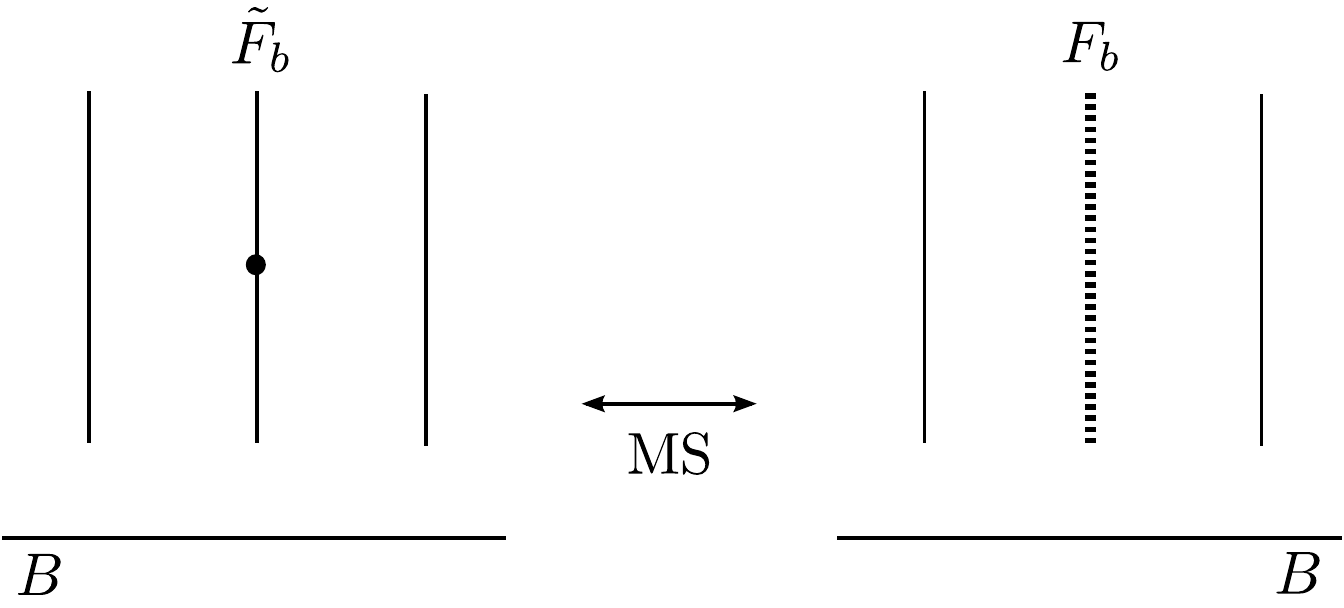}
\caption{Schematic illustration of an SYZ fibration of $\tilde{X}$
  (on the left) and of $X$ (on the right). Mirror symmetry (MS)
  acts by mapping each point in $\tilde{X}$ to a special Lagrangian
  cycle of $X$. In the SYZ fibration of $\tilde{X}$ this point is part
  of a fiber $\tilde{F}_b$ over some $b \in B$. Mirror symmetry sends
  it to the total fiber $F_b$, over $b \in B$, of the SYZ fibration of
  $X$.}\label{fig:SYZ} 
\end{center}
\end{figure}

Special Lagrangian deformations of a
special Lagrangian submanifold $C$ are well-known to be characterized by the
first cohomology group $H^1(C,\R)$. On the other hand, flat
$U(1)$-connections moduli gauge equivalence are parametrized by
$H^1(C,\R)/H^1(C,\Z)$. The first Betti number of $C$ therefore
has to match the complex dimension of $\tilde{X}$, which is 3. So
mirror symmetry implies that $\tilde{X}$ must admit a 3-dimensional fibration
of 3-tori $T^3 = (S^1)^3$ over a 3-dimensional base space that
parametrizes the special Lagrangian submanifolds in $X$. 
Of course, the same argument proves the converse. 
We thus conclude that both $X$ and $\tilde{X}$ are the total space of a special
Lagrangian fibration of 3-tori over some real three-dimensional base $B$. 

Such a special Lagrangian fibration is called an \emph{SYZ
  fibration}. It is illustrated in Fig.~\ref{fig:SYZ}. Notice that
generically this fibration is not 
smooth over the whole base $B$. We refer to the locus $\Gamma \subset
B$ where the fibers degenerate as the singular, degenerate or
discriminant locus of the SYZ fibration. Smooth
fibers $F_b$ and $\tilde{F}_b$, over some point $b \in B$,  can be
argued to be dual in the sense 
that $H^1(F_b, \R/\Z) = \tilde{F}_b$ and vice versa, since flat $U(1)$
connections on a torus are parametrized by the dual torus.

\subsubsection{Large complex structure}

Mirror symmetry, as well as the SYZ conjecture, is not
exactly true in the way as stated above. 
More precisely, it should be considered as a symmetry
only around certain special points in the K\"ahler and complex
structure moduli space. In the complex structure moduli space these
are the large complex structure limit points, and in the K\"ahler
moduli space the large K\"ahler structure limit points. Mirror symmetry will
exchange a Calabi-Yau threefold near a large complex structure point
with its mirror near a large K\"ahler structure limit
point. Moreover, only in these regions the SYZ conjecture may be
expected to be valid. 

Topological aspects of type II string theory compactifications can be
fully described in terms of either the K\"ahler structure deformations
or the complex structure deformations of the internal Calabi-Yau
manifold. In the following we will focus solely on describing the
K\"ahler structure of an SYZ fibered Calabi-Yau 
threefold. Starting with this motivation we can freely go
to a large complex structure point on the complex structure moduli
space.

\begin{figure}[h]
\begin{center} 
\includegraphics[width=8cm]{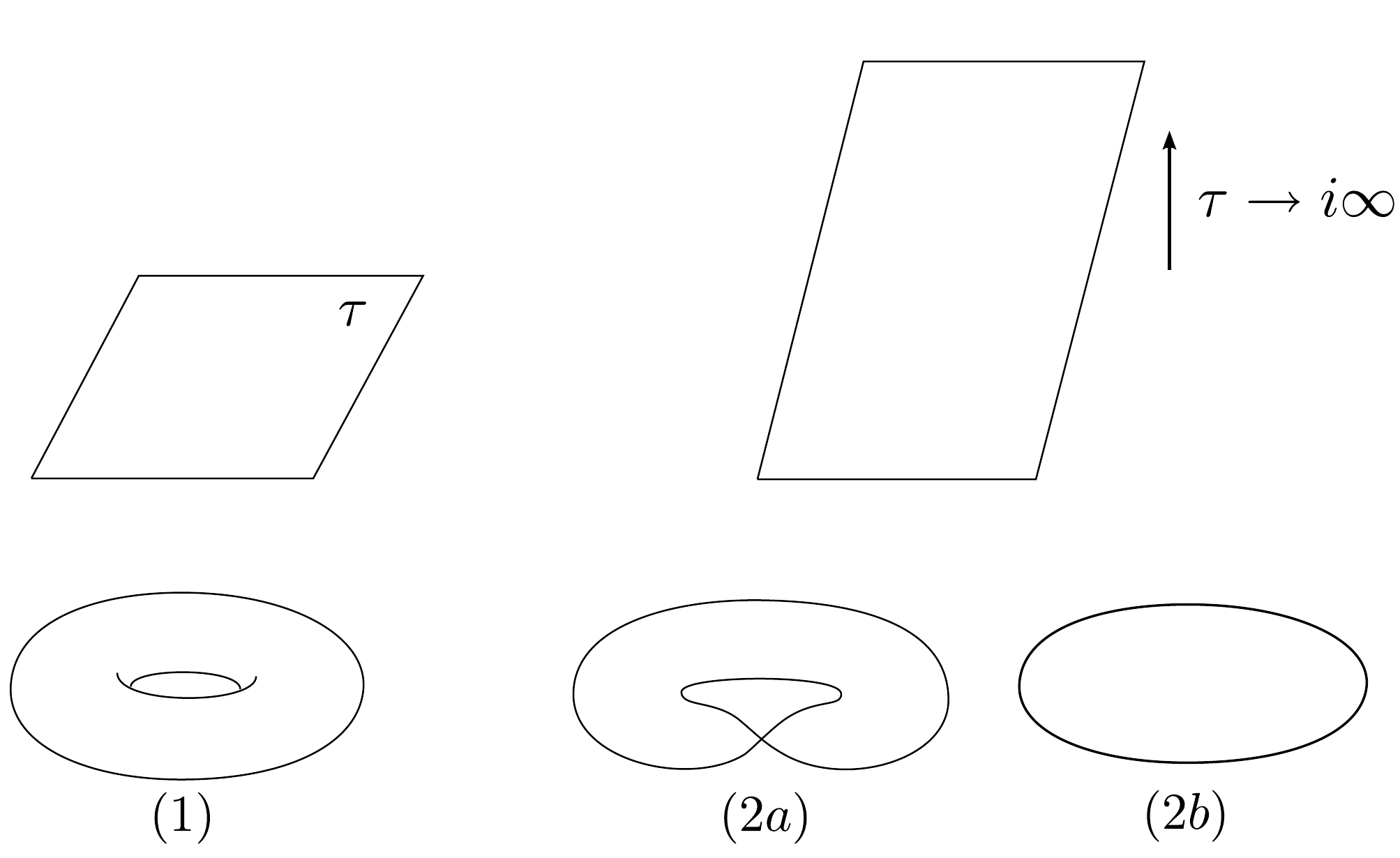}
\caption{In picture $(1)$ we see a regular torus with complex
  structure $\tau$: below the torus itself and above its fundamental
  domain (that is obtained by removing the
  A and B-cycle from the torus). In the large complex structure limit $\tau$ is sent to 
  $i \infty$. In picture $(2a)$ this degeneration is illustrated
  topologically and in $(2b)$ metrically: when we hold the
  volume fixed,  the torus  collapses to a circle.}\label{fig:SYZtorus} 
\end{center}
\end{figure}

How should we think of such a large complex structure? Let us
illustrate this with a 2-torus $T^2$. This torus is characterized by
one K\"ahler parameter $t$, which is the volume of the torus, and one
complex structure parameter $\tau$. When performing a mirror symmetry,
these two parameters are exchanged. When we send the complex structure
to infinity $\tau \mapsto  i \infty$, the torus degenerates into a
nodal torus. See Fig.~\ref{fig:SYZtorus}. However, in this limit the
volume also blows up. When we 
hold this volume fixed, by rescaling the Ricci flat metric $ds^2 =
\frac{t}{\tau_2} dz \otimes d \bar{z}$, the torus will collapse to a
line. This line should be viewed as the base of the SYZ fibration.

For a general Calabi-Yau threefold a similar picture is thought to be
true. When approaching a large complex structure point, the
Calabi-Yau threefold should have a description as a special Lagrangian
fibration over a 3-sphere whose $T^3$-fibers get very
small. The metric on these fibers is expected to become flat. Moreover,
the singular locus of the fibration is thought to 
be of codimension two in the base in that limit, so that it
defines a 1-dimensional graph on the base of the SYZ
fibration. 

In recent years much progress has been made in the mathematical study
of the SYZ conjecture. Comprehensive reviews can be found in e.g.  
\cite{Thomas:2005xf,gross-2008,Joyce-slag}. Especially the
topological approach of M.~Gross and others,
with recent connections to tropical geometry, seems 
very promising.   
The main idea in this program is to characterize the SYZ fibration by
a natural affine structure on its base $B$. 







This affine structure is found as follows.  Once we
pick a zero-section of the (in particular) Lagrangian fibration, we
can contract the inverse symplectic form with a one-form on the base
$B$ to find a vector field along the fiber. An integral affine
structure can then be defined on $B$ by an integral affine function
$f$ whose derivative $df$ defines a time-one flow along the fiber that
equals the identity. The affine structure may be visualized as a
lattice on the base $B$. 

Although this affine structure will be trivial in
a smooth open region of the fibration, it contains important
information near the singular locus. Especially the monodromy of the
$T^3$-fibers is encoded in it. In the large complex
structure limit all non-trivial symplectic information of
the SYZ fibration is thought to be  
captured in the monodromy of the $T^3$-fibers around the discriminant
locus of the fibration \cite{Gross:1999hc,gross-2008}. 
The SYZ conjecture thus leads to a beautiful picture of how
a Calabi-Yau threefold may be visualized as an affine real 3-dimensional
space with an additional integral monodromy structure around an affine
1-dimensional degeneration locus (see also
\cite{kontsevich-2004}).

One of the concrete results that have been accomplished is an SYZ description
of elliptic K3-surfaces  \cite{gross-2000} and Calabi-Yau
hypersurfaces in toric varieties (in a series of papers by W.-D. Ruan
starting with \cite{ruan}). Let us illustrate this 
with the prime example of a compact Calabi-Yau threefold, the Fermat
quintic. This is also the first Calabi-Yau for which a mirror pair was found
\cite{Greene:1990ud}. On a first reading it is not necessary to go
through all of the formulas in this example; looking at the pictures
should be sufficient.


\section{The Fermat quintic}\label{sec:Fermatquintic}

The Fermat quintic is defined by an equation of degree 5 in
projective space $\P^4$:
\begin{align*}
X_{\mu}: \quad \sum_{k=1}^5 z_k^5 - 5 \mu \prod_{k=1}^5 z_k = 0,
\end{align*}
where $[z] =[z_1:\ldots:z_5]$ are projective coordinates on
$\P^4$. Here $\mu$ denotes the complex structure of the 
threefold. 

Pulling back the K\"ahler form of $\P^4$ provides the Fermat quintic
with a K\"ahler structure. Moreover, the so-called adjunction formula
shows that $X_{\mu}$ has a trivial canonical bundle. So the Fermat
quintic $X_{\mu}$ is a Calabi-Yau threefold. Note as well that $\P^4$
admits an action of 
$T^4$ that is parametrized by five angles $\theta_k$ with
$\sum_k \theta_k =0$:
\begin{align*}
T^4:\quad [z_1:\ldots:z_5] \mapsto  [e^{i \theta_1}
  z_1:\ldots: e^{i \theta_5} z_5]  
\end{align*}
Such a variety is called toric. The quintic is thus embedded as a
hypersurface in a toric variety. 


\begin{figure}[h!]
\begin{center} \label{fig:SYZbasequintic}
\includegraphics[width=7cm]{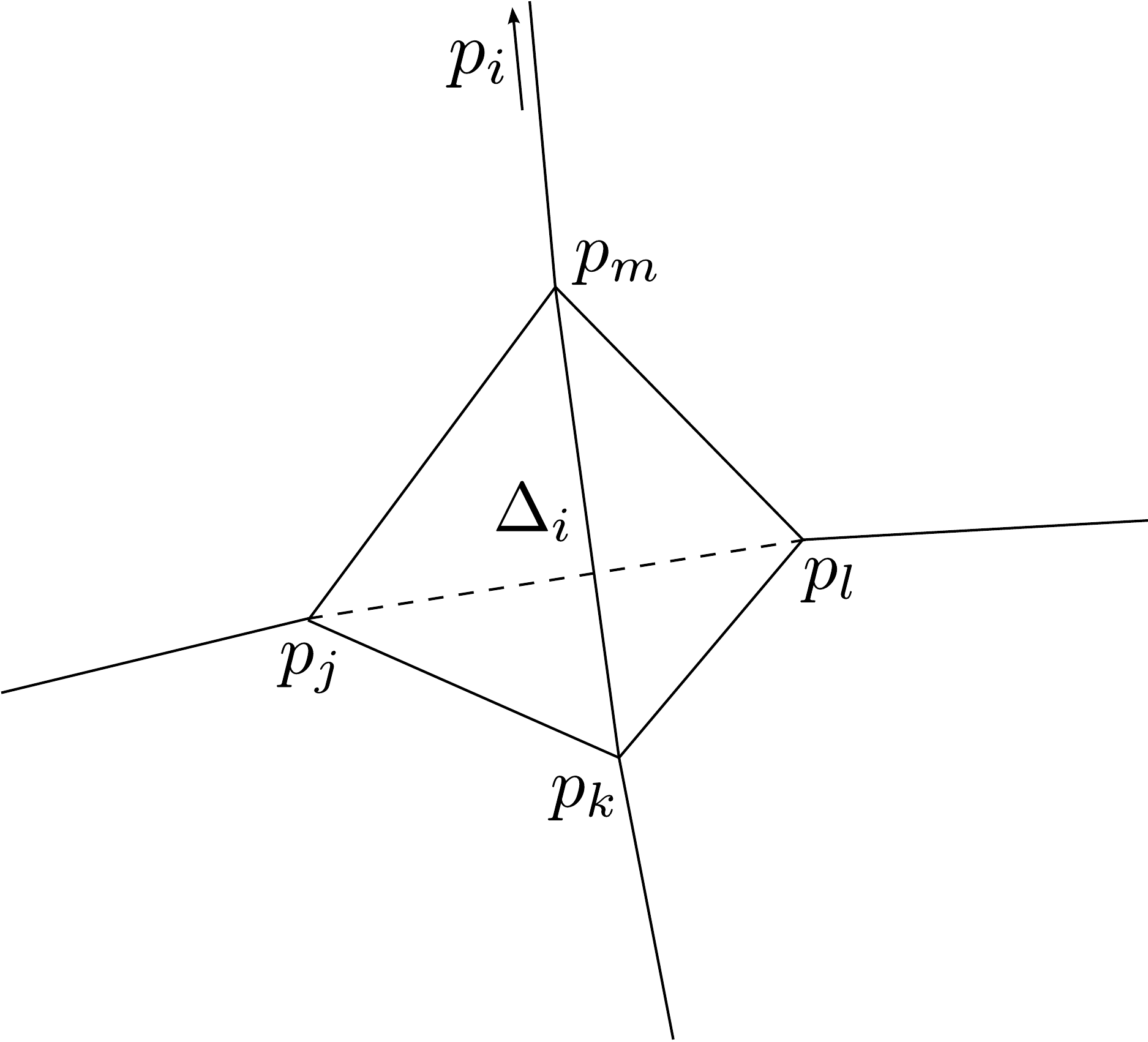}
\caption{The base of the SYZ fibration of the Fermat
  quintic $X_{\mu}$ is the boundary $\lotjesd \Delta$ of a 4-simplex
  $\Delta$. To be able to draw this 
  base we have placed the vertex $p_i$ at infinity. The tetrahedron
  $\Delta_i$ is the projection of the subset $\{z_i=0\} \subset
  \P^4$. The boundary $\lotjesd \Delta$ consists of 5 such tetrahedrons.}
\end{center}
\end{figure}

When we take $\mu \to \infty$ we reach the large complex structure
limit point 
\begin{align*}
X_{\infty}: \quad \prod_{k=1}^5 z_k = 0.  
\end{align*}
This is just a union of five $\P^3$'s, that each inherit a
$T^3$-action from the above toric action on $\P^4$. From this
observation it simply follows that $X_{\infty}$ can be seen as a
SYZ fibration. We can make this explicit by considering
the fibration $\pi: \mathbb{P}^4 \to \mathbb{R}^4$ given
by 
\begin{align*}
 \pi([z]) =   \sum_{k=1}^5\frac{  |z_k|^2 }{\sum_{l=1}^5 |z_l|^2} p_k, 
\end{align*}
where the $p_k$ are five generic points in $\mathbb{R}^4$. The image
of $F$ is a 4-simplex $\Delta$ spanned by the five points $p_k$ in
$\R^4$ and $X_{\infty}$ is naturally fibered over the
boundary $\partial \Delta$ with generic fiber being $T^3$. The base of
the SYZ fibration is thus topologically a 3-sphere $S^3$. Similar to the
large complex structure limit of a 2-torus, the Ricci-flat metric on $X_{\infty}$
is degenerate and the SYZ fibers are very small.

Let us introduce some notation to refer to different patches
of $\P^4$ and $\Delta$. Call $D_{i_1, \ldots, i_n}$ the closed part of
$\P^4$ where $z_{i_1}$ up to $z_{i_n}$ vanish, and denote its
projection to $\R^4$ by $\Delta_{i_1, \ldots, i_n}$. The ten faces of the
boundary $\lotjesd \Delta$ are thus labeled by $\Delta_{ij}$, with $1 \le
i, j \le 5$.

\begin{figure}[h!]
\begin{center} 
 \includegraphics[width=6.3cm]{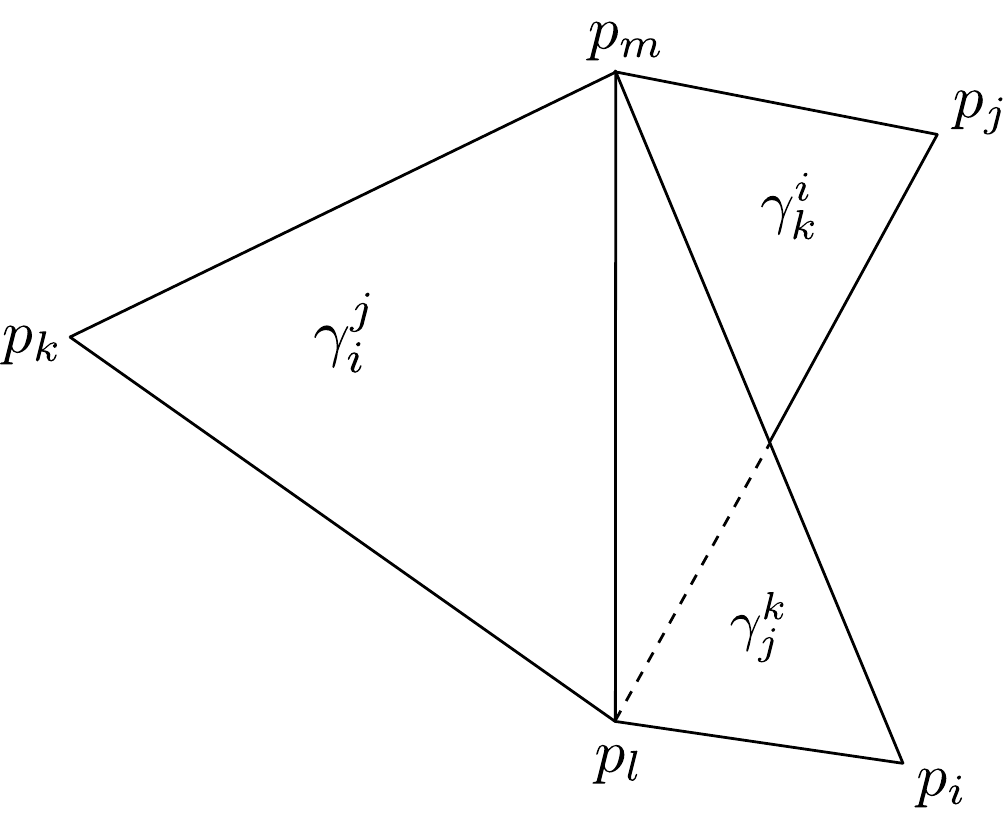}
\caption{The boundary of each tetrahedron is a union of four
  triangles. The $T^3$-fibration of the Fermat quintic degenerates over these 
  triangles. This figure illustrates which cycle in the $T^3$-fiber 
  vanishes over which triangle in $\lotjesd \Delta$: $\gamma^j_i$
  degenerates over the triangle $\Delta_{ij}$, $\gamma_j^k$ over
  $\Delta_{jk}$, etc.}\label{fig:degSYZbasequintic}    
\end{center}
\end{figure}

In the following it is also useful to introduce a notation for the
$S^1$-cycles that are fibered over the base $\Delta$. So define the circles
\begin{align*}
\gamma^k_{i} = \{ \, z_i = 0,\, |\frac{z_k}{z_j}| = a_1,\,
\frac{z_l}{z_j} = a_2,\, \frac{z_m}{z_j} = a_3 \, \},
\end{align*}
where the indices $\{i, j, k, l, m \}$ are a permutation of
$\{1, \ldots, 5 \}$ and the numbers $a_1$, $a_2$, $a_3$ are determined
by choosing a base 
point $[z]$ on the circle. Since a different choice for the index $j$
leaves the circle
invariant, it is not included as a label in the name. Which circle
shrinks to zero-size over each cell of the boundary $\lotjesd \Delta$ is
summarized  in  
Fig.~\ref{fig:degSYZbasequintic}. Notice that the only non-vanishing 
circle in the fibers over the triple intersection $\Delta_{ijk}$ is
the circle  $\gamma^l_i$ (which may also be denoted as
$\gamma^l_j$, $\gamma^l_k$, $ - \gamma^m_i$, $- \gamma^m_j$ or
$-\gamma^m_k$).


W.-D. Ruan works out how to generate a Lagrangian
fibration for any $X_{\mu}$ where $\mu$ is large  \cite{ruan}. His idea is to 
use a gradient flow that deforms the Lagrangian fibration of
$X_{\infty}$ into a Lagrangian fibration of $X_{\mu}$. Let us
consider the region $D_i$ and choose
$z_j \neq 0$ for some $j \neq i$. In this patch $x_k = z_k/z_j$ are
local coordinates. It turns out that the flow of the vector field 
\begin{align*}
V = \textrm{Re} \Big(\frac{ \sum_{k \neq i,j} x_k^5 +1}{\prod_{k \neq i,j}
  x_k} \frac{\partial}{\partial x_i} \Big)
\end{align*}
produces a Lagrangian fibration of $X_{\mu}$ over $\lotjesd \Delta$.


\begin{figure}[h]
\begin{center} 
\includegraphics[width=5.8cm]{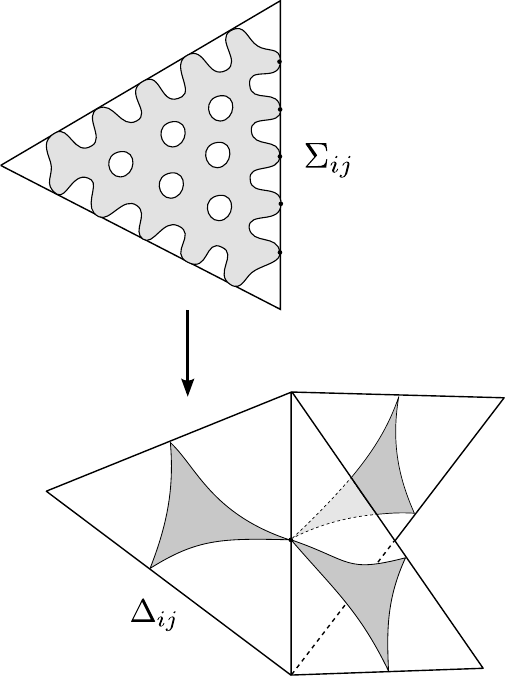}
\caption{The shaded area in this figure illustrates the discriminant
  locus of the Lagrangian fibration of the quintic $X_{\mu}$, for
  large $\mu$. In the total space it has the shape of a genus 6 Riemann
  surface $\Sigma_{ij}$ over each 2-cell  $\Delta_{ij}$. The five 
  dots on the Riemann surface project to a single dot on the base. }\label{fig:degfiberquintic}  
\end{center}
\end{figure}

All the smooth points of $X_{\infty}$, i.e. those for which only one
of the coordinates $z_i$ vanishes, will be transformed into regular
points of the Lagrangian fibration. Only the points in the
intersection of $X_{\mu}$ with the singular locus of $X_{\infty}$
won't move with the flow. These can be shown to form the
complete singular locus of the Lagrangian fibration of $X_{\mu}$. Let us
call this singular locus  $\Sigma$ and denote 
$$\Sigma_{ij} = D_{ij} \cap \Sigma = \{ [z] \, | \, z_i = z_j = 0,\,
z_k^5 + z_l^5 + z_m^5 = 0 \, \}.$$ 
The singular locus $\Sigma_{ij}$ is thus a projective curve which has
genus 6.  It intersects $D_{ijk}$ at the five points. 
The image of $\Sigma_{ij}$ under the projection $\pi$ is
a deformed triangle in $\Delta_{ij}$ that intersects the boundary
lines of $\Delta_{ij}$ once. This is illustrated in
Fig.~\ref{fig:degfiberquintic}.

In the neighborhood of an inverse image of the intersection point
$\frac{1}{2}(p_l + p_m)$ in $D_{ij}$ the coordinate $z_k$ gets very
small. This implies that if we write $z_k = r_k e^{ i \phi_k}$, it is
the circle parametrized by $\phi_k$ that wraps the leg of the pair of
pants in the limit $r_k \to 0$. In the notation we introduced before
this circle is $\gamma_j^k = \gamma_i^k$.

\begin{figure}[h]
\begin{center} 
 \includegraphics[width=7.5cm]{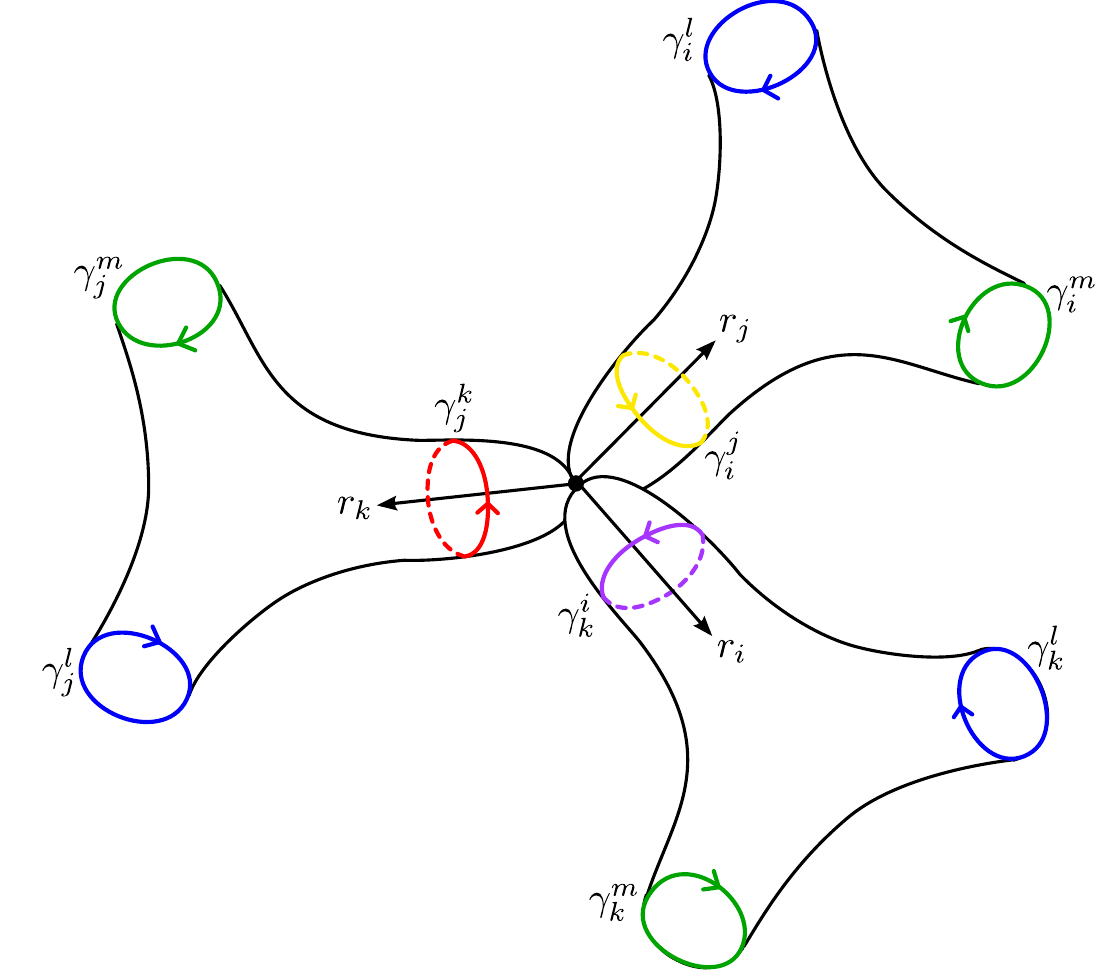}
\caption{This picture shows one plus-vertex surrounded by three
  minus-vertices. It illustrates the discriminant locus in the
  neighborhood of an inverse image of the center of $\Delta_{ijk}$,
  marked by a dot in the center of the plus-vertex. At this point  the three
  genus 6 surfaces $\Sigma_{ij}$, $\Sigma_{jk}$ and $\Sigma_{ki}$ meet
  transversely. Notice that e.g. 
  $\gamma_j^k + 
  \gamma_j^l + \gamma_j^m =0$ and $\gamma_i^j + \gamma_j^k +
  \gamma_k^i =0$.    
}\label{fig:degfiberquintic2} 
\end{center}
\end{figure}

The study of the cycles in the fibration reveals the structure of the
singular locus. It is built out of two kinds of 3-vertices. We call a
3-vertex whose center lies at an edge $D_{ijk}$ a plus-vertex and a
3-vertex that lies in the interior of some 2-cell $D_{ij}$ a
minus-vertex. The plus-vertex is 
described by a different degenerating cycle at each of the three
legs. Together they sum up to zero. The minus-vertex is
characterized by a single vanishing cycle. The precise topological
picture is shown in Fig.~\ref{fig:degfiberquintic2}.
 

So in the large complex structure limit $\mu \to \infty$ the
quintic has an elegant structure in terms of ten transversely
intersecting genus 6 Riemann surfaces or equivalently in terms of 50
plus-vertices and 250 minus-vertices.


\subsubsection{Two kinds of vertices}

The structure of the singular locus of the Fermat quintic in the above
example, ten genus 6 Riemann surfaces that intersect each other
transversely, is very elegant. It is remarkable that it may be
described by just two types of 3-vertices. This immediately raises the
question whether this is accidental or a generic feature of SYZ
fibrations of Calabi-Yau threefolds. In fact, M.~Gross shows that
under reasonable assumptions there are just a few possibilities
for the topological structure in the neighborhood of the discriminant
locus \cite{Gross:1999hc} (see also \cite{gross-2008}). Indeed only
two types of trivalent vertices can 
occur. Both are characterized by the monodromy that acts on three generators
$\gamma_i$ of the homology $H_1(F,\mathbb{Z})$ of the $T^3$-fiber $F$
when we transport them around each single leg of the vertex. Let 
us summarize this monodromy in three matrices $M_{\alpha}$ such that  
$$ \gamma_i \mapsto \gamma_j (M_{\alpha})_{ji}$$ 
when we encircle the $\alpha$th leg of the vertex.

The first type of trivalent vertex is described by the three monodromy matrices 
\begin{align*}
M_+:\quad \left(
\begin{array}{ccc} 1 & 0 & 1 \\ 0 & 1 & 0 \\ 0 & 0 & 1 \end{array} \right),\,
\left(
\begin{array}{ccc} 1 & 0 & 0 \\ 0 & 1 & 1 \\ 0 & 0 & 1 \end{array} \right),\,
\left(
\begin{array}{ccc} 1 & 0 & -1 \\ 0 & 1 & -1 \\ 0 & 0 & ~1 \end{array}
\right).
\end{align*}
This vertex is characterized by a a distinguished $T^2
\subset T^3$ that is generated by $\gamma_1$ and $\gamma_2$ and stays
invariant under the monodromy. Only the element $\gamma_3$ picks up a
different cycle at each leg. This latter cycle must therefore be a
vanishing cycle at the corresponding leg.

\begin{figure}[h]
\begin{center} 
\includegraphics[width=4cm]{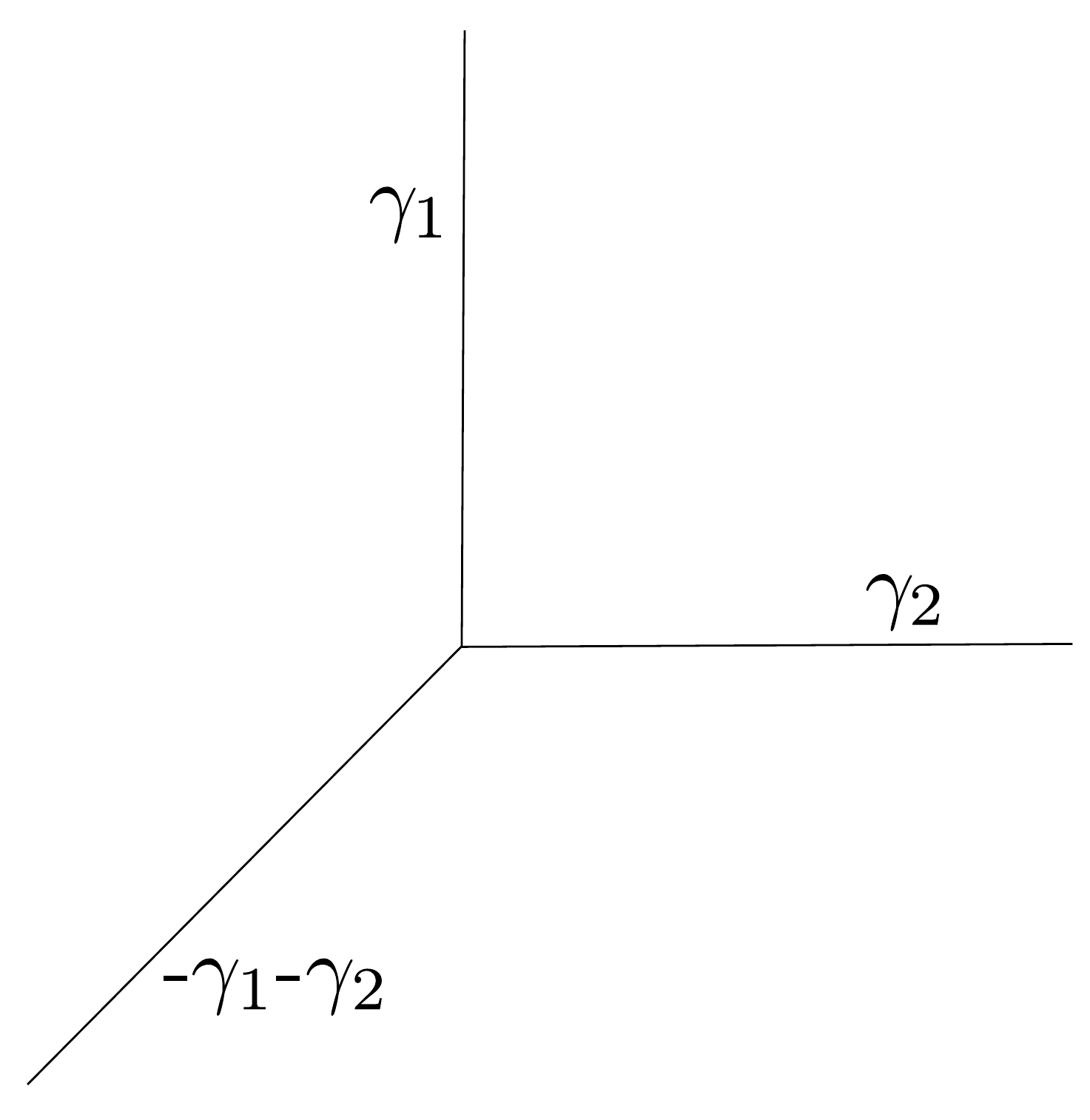}
\caption{The plus-vertex is illustrated as a 1-dimensional
  graph. Its legs are labeled by the cycle in $T^2 \subset T^3$ that
  vanishes there. The cycle $\gamma_3$ picks up the monodromy.}\label{fig:plusvertex} 
\end{center}
\end{figure}

The legs of this vertex can thus be labeled by the
cycles $\gamma_1$, $\gamma_2$ and $-\gamma_1-\gamma_2$ respectively,
so that the vertex is conveniently represented as in
Fig.~\ref{fig:plusvertex}. Note that this is precisely the topological
structure of the plus-vertex in the Lagrangian fibration   
of the Fermat quintic. \index{topological vertex}        

\begin{figure}[h]
\begin{center} 
 \includegraphics[width=3.8cm]{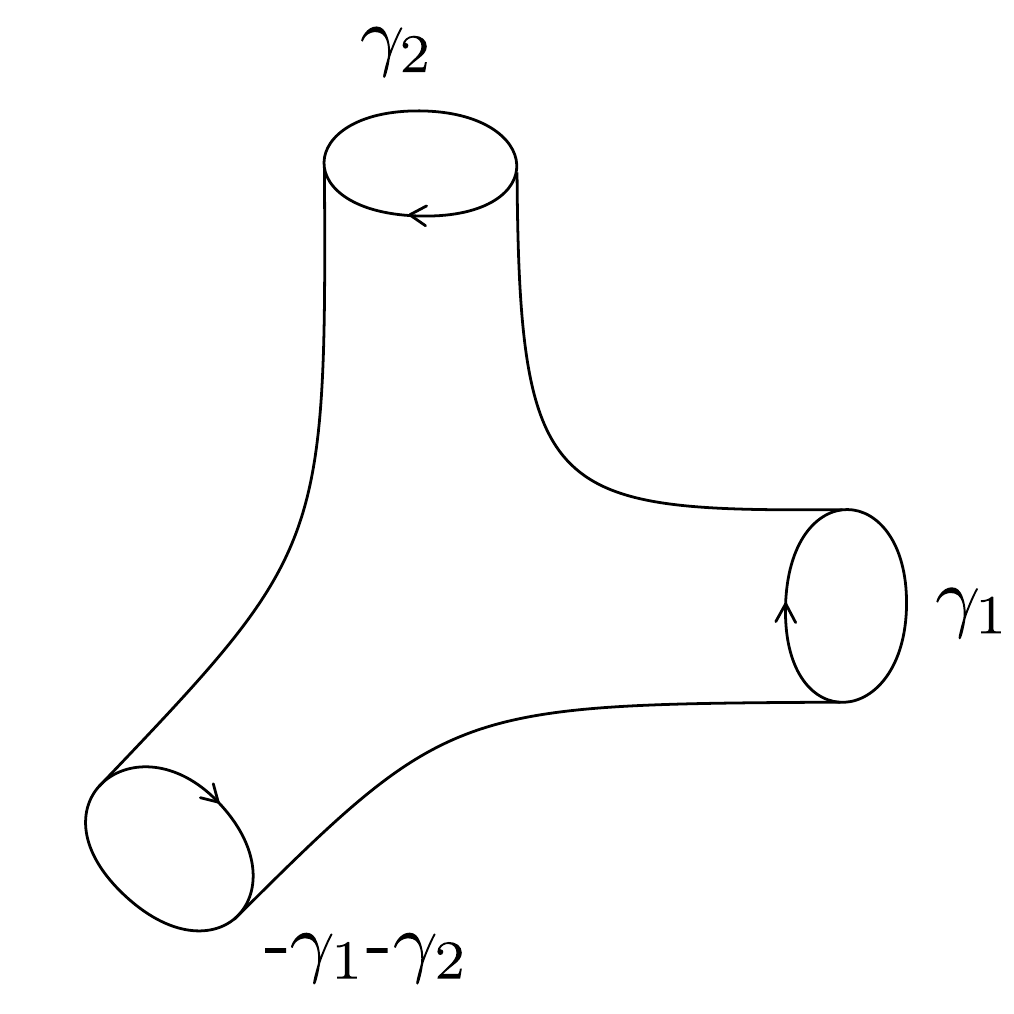}
\caption{The minus-vertex is illustrated as a pair of pants. It is
  characterized by a single vanishing cycle $\gamma_3$. Furthermore,
  its legs are labeled by the cycle that wraps it. }\label{fig:minusvertex} 
\end{center}
\end{figure}

The monodromy of the second type of vertex is summarized by
the matrices
\begin{align*}
M_-: \quad \left(
\begin{array}{ccc} ~1 & 0 & 0 \\ ~0 & 1 & 0 \\ -1 & 0 & 1 \end{array} \right),\,
\left(
\begin{array}{ccc} 1 & ~0 & 0 \\ 0 & ~1 & 0 \\ 0 & -1 & 1 \end{array} \right),\,
\left(
\begin{array}{ccc} 1 & 0 & 0 \\ 0 & 1 & 0 \\ 1 & 1 & 1 \end{array}
\right).
\end{align*}
In contrast to the plus-vertex these monodromy matrices single out a unique
1-cycle $\gamma_3$ that degenerates at all three legs of the
vertex. Instead of labeling the vertex by this vanishing cycle, it is
now more convenient to label the legs with the non-vanishing cycle
that does not pick up any monodromy. Like in the example of
the Fermat quintic these cycles topologically form a pair of
pants. This is illustrated in Fig.~\ref{fig:minusvertex}. 
\index{pair of pants}


Notice that both sets of monodromy matrices are related by simple
duality $(M_{+})^{-t} = M_-$. Since the mirror of an SYZ fibered Calabi-Yau may be obtained
by dualizing the $T^3$-fibration, the above vertices must be related
by mirror symmetry as well. 
Obviously these vertices will become important when we describe string
compactifications on Calabi-Yau threefolds.

\section{Local Calabi-Yau threefolds}\label{sec:localCY}

Topological string theory captures topological aspects of type II
string theory compactifications; K\"ahler aspects of type IIA
compactifications and complex structure aspects of type IIB 
compactification. Mirror symmetry relates them. The topological string
partition function $Z_{\textrm{top}}$ can be written as a generating
function of either symplectic or complex structure invariants of the
underlying Calabi-Yau 
manifold. It contains for example a
series in the number of genus zero curves that are embedded in the
Calabi-Yau threefold. Using mirror symmetry it is possible to go well
beyond the classical computations of these numbers. A famous result
is the computation of the whole series of these genus zero invariants
for the Fermat quintic: 2875 different lines, 609250 conics, 317206375
cubics, etc. \cite{Candelas:1990rm}.    

It is much more difficult to find the complete topological
string partition function, which also contains information about
higher genus curves in the Calabi-Yau threefold. The state of the art
for the quintic is the computation of these invariant up to $g=51$
\cite{Huang:2006hq}. Although this is an impressive result, it is far
from computing the total partition function. In contrast, the
all-genus partition function has been found
 for a simpler type of Calabi-Yau manifolds, that are
 non-compact. What kind of spaces are these? And why is it
 so much easier to compute their partition function?  
 
The simplest Calabi-Yau threefold is plain $\C^3$ with
complex coordinates $z_i$. It admits a K\"ahler form $$k = \sum_{i=1}^3
dz_i \wedge d\bar{z}_i,$$ and a non-vanishing holomorphic 3-form $$\Omega
= dz_1 \wedge dz_2 \wedge dz_3.$$ A special Lagrangian $T^2 \times \R$
fibration of $\C^3$ over $\R^3$ has been known for a long time
\cite{Harvey:1982xk}. It is defined by the map
\begin{align*}
(z_1,z_2,z_3) \, \mapsto \, 
(\, \textrm{Im} \,z_1 z_2 z_3,\, |z_1|^2 - |z_2|^2,\, |z_1|^2 - |z_3|^2\,).
\end{align*}
Notice that its degeneration locus is a 3-vertex with legs $(0,t,0)$,
$(0,0,t)$ and $(0,-t,-t)$, for $t \in \R_{\ge 0}$. Over all these legs
some cycle in the $T^2$-fiber shrinks to zero-size. We can name these
cycles $\gamma_1$, $\gamma_2$ and $-\gamma_1 - \gamma_2$ since they add
up to zero. The degenerate fiber over each leg is a pinched cylinder
times $S^1$. It is thus the noncompact cycle ($\cong \R$) in the fiber
that picks
up monodromy when we move around one of the toric legs. This 3-vertex
clearly has the same topological structure as the plus-vertex.    

Many more non-compact Calabi-Yau's can be constructed by
gluing $\C^3$-pieces together. In fact, these constitute all 
non-compact toric Calabi-Yau threefolds. Their 
degeneration locus can be drawn in $\R^2$ as a 1-dimensional trivalent
graph. The fact that their singular graph is simply planar, as opposed to
actually 3-dimensional (as for the Fermat quintic), makes the
computation of the partition function on such non-compact toric
Calabi-Yau's much simpler. 
String theorists have managed to find
the full topological partition function $Z_{\textrm{top}}$ 
\cite{Aganagic:2003db} (using a
duality with a 3-dimensional topological theory, called
Chern-Simons theory \cite{Gopakumar:1998ki}). The recipe to compute
the partition function involves cutting the graph in basic
3-vertices. To generalize this for compact Calabi-Yau's it seems one would need
to find a way to glue the partition function for a plus-vertex with
that of a minus-vertex.  
\index{toric Calabi-Yau threefold}

Since the partition function is fully known, topological string
theory on these toric manifolds is the ideal playground to learn more
about its underlying structure.
This has revealed many interesting mathematical and physical
connections, for example  
to several algebraic invariants such as Donaldson-Thomas invariants
\cite{mnop1, mnop2,Szendroi:2007nu} and Gopakumar-Vafa invariants
\cite{Gopakumar:1998ii,Gopakumar:1998jq},  to knot theory
\cite{Labastida:2000yw, Gukov:2004hz, Gukov:2007tf, Dijkgraaf:2009sb}, and
to a duality with crystal melting
\cite{Okounkov:2003sp,dijkgraaf-2009-811, Ooguri:2008yb}. 

\begin{figure}[h]
\begin{center} 
\includegraphics[width=9cm]{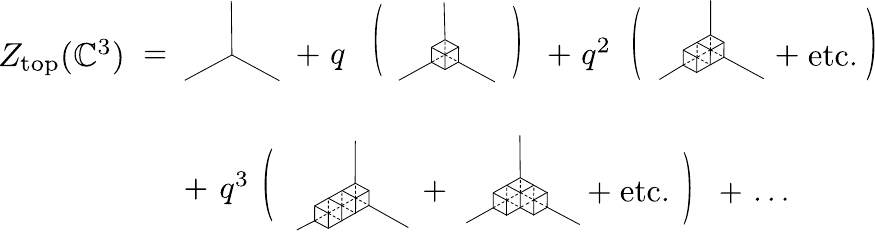}
\caption{Interpretation of the first terms in the expansion of
  $Z_{\textrm{top}}(\C^3)$ in terms of a three-dimensional crystal in
  the positive octant of $\R^3$.}\label{fig:crystal} 
\end{center}
\end{figure}

To illustrate the last duality, 
let us write down the plain $\C^3$ partition function: 
\begin{align*}
 Z_{\textrm{top}}(\C^3) = \prod_{n>0} \frac{1}{(1-q^n)^n} = 1+ q + 3
 q^2 + 6 q^3 + \ldots.
\end{align*}
This $q$-expansion is well-known to be generating function of 3-dimensional
partitions; it is called the MacMahon function \cite{Macmahon}. The
3-dimensional 
partition can be
visualized as boxes that are stacked in the positive octant 
of $\R^3$. Three of the sides of each box must either touch the
walls or another box. This is pictured in Fig.~\ref{fig:crystal}.

Since $q = e^{\la}$, where $\la$ is the coupling
constant of topological string theory, the boxes naturally have length
$\la$. Whereas the regime with $\la$ finite is described by a
discrete quantum structure, in the limit $\la \to 0$ surprisingly a
smooth Calabi-Yau geometry emerges. In a duality with statistical
mechanics this corresponds to the shape of a melting crystal. These
observations have led to 
deep insights in the quantum description of space and time
\cite{Iqbal:2003ds, Ooguri:2009ri}. 

Remarkably, it has been shown that the emergent smooth geometry of the
crystal can be 
identified with the mirror of $\C^3$. How does this limit shape look like? 
Using local mirror symmetry the equation for the mirror of $\C^3$ was
found in \cite{Hori:2000kt}. It is given by
\begin{align*}
uv -  x - y +1 = 0, 
\end{align*}
where $u,v \in \C$ and $x,\, y \in \C^*$. 
Remember that the topological structure of the mirror of a plus-vertex
should be that of a minus-vertex. Viewing the mirror as a
$(u,v)$-fibration over a complex plane spanned by $x$ and $y$ confirms
this: 

The degeneration locus of the fibration equals the zero-locus $x+y-1=0$. Parametrizing this curve by $x$, it is easily seen that
this is a 2-sphere with three punctures at the points $x = 0, \, 1$
and $\infty$.  We can equivalently represent this curve as a pair of
pants, by cutting off a disc at each boundary $|\tilde{x}|=1$, where $\tilde{x}$ 
is a local coordinate that vanishes at the corresponding
puncture. This realizes the mirror of $\C^3$ topologically as the
minus-vertex in Fig. \ref{fig:minusvertex}.

Mirrors of general non-compact toric manifolds are of the same form
\begin{align*}
X_{\Sigma}: \quad uv - H(x,y) =0,
\end{align*}
where the equation $H(x,y)=0$ now defines a generic Riemann surface $\Sigma$
embedded in $(\C^*)^2$. This surface is a thickening of the
1-dimensional degeneration graph $\Gamma$  of its mirror. Its
non-vanishing holomorphic 3-form is proportional to
\begin{align*}
 \Omega =  \frac{du}{u} \wedge dx \wedge dy. 
\end{align*}
These geometries allow a Ricci-flat metric
that is conical at infinity
\cite{calabi,tianyaulocalCY1,tianyaulocalCY2,tianyaulocalCY3}. We
refer to the threefold $X_{\Sigma}$ as the 
\Index{local Calabi-Yau threefold modeled on $\Sigma$}.
The curve $\Sigma$ plays a central role in this thesis. We study
several set-ups in string theory whose common denominator is the
relevance of the Riemann surface $\Sigma$. In particular, we study the
melting crystal 
picture from the mirror perspective. One of our main results is a simple
representation of 
topological string theory in terms a quantum Riemann surface, that
reduces to the smooth Riemann surface $\Sigma$ in the limit $\la 
\to 0$.

\chapter{I-brane Perspective on Vafa-Witten Theory and WZW~Models}\label{chapter2}

In the last decades enormous progress has been made in analyzing
4-dimensional supersymmetric gauge theories. Partition functions and
correlation 
functions of some theories have been computed, spectra of BPS
operators have been discovered and many other structures have been
revealed. Most fascinating to us is that many exact results
can be related to two-dimensional geometries.   

Since 4-dimensional supersymmetric gauge theories appear in several
contexts in string theory, much of this progress is strongly
influenced by string theory. Often, string theory tools
can be used to compute important quantities in supersymmetric 
gauge theories. Moreover, in many cases string theory provides
a key understanding of new results. For example, when
auxiliary structures in the gauge theory can be realized geometrically
in string theory and when symmetries in the gauge theory can be understood as
stringy dualities.  

In this chapter we study a remarkable correspondence
between 4-dimensional gauge theories and 2-dimensional conformal field
theories. This correspondence connects a ``twisted'' version of
supersymmetric Yang-Mills theory to a so-called Wess-Zumino-Witten
model. In particular, generating functions of $SU(N)$ gauge
instantons on the 4-manifold $\C^2/\Z_k$ are related to characters of the affine
Kac-Moody algebra $\widehat{su}(k)$ at level $N$.      
This connection was originally discovered by Nakajima \cite{nakajima},
and further analyzed by Vafa and Witten \cite{Vafa:1994tf}. The goal
of this chapter is to make it more transparent. Once
again, we find that string theory offers the right perspective.

We have strived to make this chapter self-contained by starting in
\Cref{sec:gaugetheories} with a short 
introduction in gauge and string theory. We review how
supersymmetric gauge theories show up as low energy world-volume
theories on D-branes and how they naturally get
twisted. Twisting emphasizes the role of topological
contributions to the theory. Furthermore, we 
introduce fundamental string dualities as T-duality and S-duality. 
  
In \Cref{sec:N=4YM} we introduce Vafa-Witten theory as an
example of a twisted 4-dimensional gauge theory, and study
it on non-compact 4-manifolds that are asymptotically
Euclidean. In \Cref{sec:stringrealization} we show that Vafa-Witten
theory on such a 4-manifold is embedded in string theory as a D4-D6
brane intersection over 
a torus $T^2$.  We refer to the intersecting brane wrapping $T^2$ as
the \Index{I-brane}. Since the open 4-6 strings introduce chiral 
fermions on the I-brane, we find a duality between Vafa-Witten theory
and a CFT of free fermions on $T^2$.  

In \Cref{sec:NVW} we show that the full I-brane partition
function is simply given by a fermionic character, and reduces to
the Nakajima-Vafa-Witten results after taking a decoupling
limit. The I-brane thus elucidates the Nakajima-Vafa-Witten correspondence
from a string theoretic perspective.  Moreover, we gain more
insights in level-rank duality and the McKay correspondence from this
stringy point of view.

\section{Instantons and branes}\label{sec:gaugetheories}

A four-dimensional gauge theory with gauge group $G$ on
a Euclidean 4-manifold $M$ is mathematically formulated in terms of a
$G$-bundle $E \to M$. A gauge field $A$ is part of a local
connection $D = d + A$ of this bundle, whose curvature is the
electro-magnetic field strength 
\begin{align*}
F = dA + A \wedge A. 
\end{align*}
If we denote the electro-magnetic gauge coupling by $e$ and call $*$ 
the Hodge star operator in four dimensions, the Yang-Mills path
integral is
\begin{align*}
Z =  \int_{\cA/G} DA ~\exp \left(-
  \frac{1}{e^2} \int_M d^4x~ \Tr F \wedge *F \right),
\end{align*}
This path integral, over the moduli space of connections $\cA$ modulo
gauge invariance, defines quantum corrections to the classical
Yang-Mills equation $D * F  = 0$.
When the gauge group is abelian, $G = U(1)$, the equation of
motion plus Bianchi identity combine into the familiar Maxwell equations 
\begin{align*}
d * F = 0 , \qquad
d F = 0.
\end{align*}

\subsubsection{Topological terms}\index{topological term}

Topologically non-trivial configurations of the gauge field are
measured by characteristic classes. If $G$ is connected and simply-connected the
gauge bundle $E$ is characterized topologically by the 
\emph{instanton charge}\index{gauge instanton}
\begin{align}\label{eqn:4dinstanton}
 ch_2(F) = \Tr \left[ \frac{F \wedge F}{8 \pi^2} \right] \in
 H^4(M, \Z). 
\end{align}
Instanton configuration are included in the Yang-Mills formalism by
adding a topological term to the Yang-Mills Lagrangian
\begin{align}\label{eqn:fullYMaction}
\cL = -\frac{1}{e^2}  F \wedge * F +
\frac{i\theta}{8 \pi^2} F \wedge F  
\end{align}
Note that this doesn't change the equations of motion. The path
integral is invariant under $\theta \to \theta + 2 
\pi$, and the parameter $\theta$ is therefore called the
$\theta$-angle. The total Yang-Mills Lagrangian can be rewritten as 
\begin{align}\label{YMactionthetaform}
\cL = \frac{i \tau}{4 \pi}  F_+ \wedge F_+ + \frac{i \bar{\tau}}{4
  \pi}  F_- \wedge F_- , 
\end{align}
where $F_{\pm} = \frac{1}{2} \left( F \pm  *F \right)$ are the
(anti-)selfdual field strengths while 
\begin{align}\label{eqn:tau}
\tau = \frac{\theta}{2 \pi} + \frac{4 \pi i}{e^2}
\end{align}
is the \emph{complexified gauge coupling constant}
\index{coupling constant $\tau$}.  
When $G$ is not simply-connected magnetic fluxes on 2-cycles in $M$
are detected by the first Chern class
\begin{align}\label{magneticflux}
 c_1(F) = \Tr \left[ \frac{F}{2 \pi} \right] \in H^2(M,\Z).
\end{align}


\subsubsection{Electro-magnetic duality}\label{sec:e-mduality}

\begin{figure}[h]
\centering
\includegraphics[width=7.5cm]{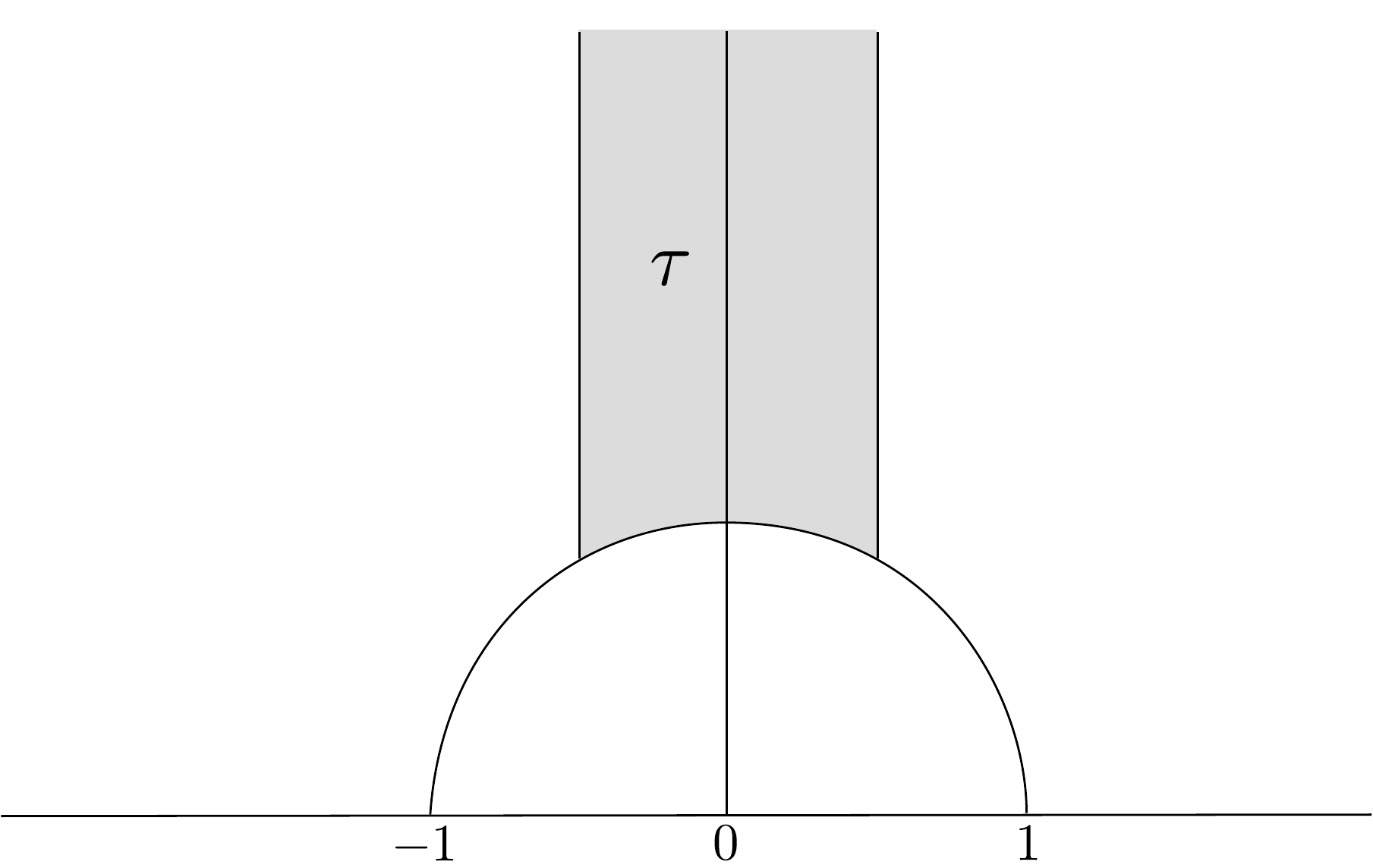}
\caption{The fundamental domain of the modular group $Sl(2,\Z)$ in the
upper half plane.}\label{upphalfplane}
\end{figure}

The Maxwell equations are clearly invariant under the transformation
$F \leftrightarrow *F$
that exchanges the electric and the magnetic field. 
To see that this is even a symmetry at the quantum level, we introduce a
Lagrange multiplier field $A_D$ in the $U(1)$ Yang-Mills path integral that explicitly imposes $dF=0$:
\begin{align*}
\int DA D A_D \, \exp  \int_M \left( \frac{i \tau}{4 \pi}  F_+ \wedge F_+ +
  \frac{i \bar{\tau}}{4 \pi}  F_- \wedge F_- + \frac{1}{2\pi}  F \wedge * dA_D  \right).  
\end{align*}
Integrating out $A$ yields the dual path integral  
\begin{align*}
\int D A_D \, \exp  \int_M  \left(\frac{ i }{ 4 \pi \tau}  F^D_+ \wedge F^D_+ +
  \frac{i }{4 \pi \bar{\tau}}  F^D_- \wedge F^D_-  \right). 
\end{align*}
So electric-magnetic duality is  a
strong-weak coupling 
duality, that sends the complexified gauge coupling $\tau \mapsto
-1/\tau$.
Moreover, this argument
suggests an important role for the \Index{modular group} $Sl(2,\Z)$. 
This group acts on $\tau$ as 
\begin{align*}
 \tau \mapsto \frac{a \tau + b}{c \tau + d}, \quad   \textrm{for} \quad
\left( \begin{array}{cc} a & b \\ c & d \end{array} \right) \in Sl(2,\Z).
\end{align*}
and is generated by 
\begin{align*}
 S = \left( \begin{array}{cc} 0 & 1 \\ -1 & 0 \end{array} \right)
 \quad  \mbox{and} \quad  T = \left( \begin{array}{cc} 1 & 1 \\ 0 & 1 \end{array} \right).
\end{align*}
Hence, $S$ is the generator of electro-magnetic duality (later
also called \Index{S-duality}) and $T$ the generator of shifts in the
$\theta$-angle. 
The gauge coupling $\tau$ is thus part of the 
fundamental domain of $Sl(2,\Z)$ in the upper-half plane, as shown in
Fig.~\ref{upphalfplane}.  

Montonen and Olive \cite{Montonen:1977sn} where pioneers in conjecturing
that electro-magnetic duality is an exact non-abelian symmetry, that exchanges the
opposite roles of electric and magnetic particles in 4-dimensional gauge
theories. This involves replacing the gauge group $G$ by
the dual group $\hat{G}$ (whose weight lattice is dual to that of
$G$). The first important tests of S-duality have been performed in
supersymmetric gauge theories.   

For $U(1)$ theories the partition function $Z^{U(1)}$ can be
explicitly computed   
\cite{Witten:1995gf,Verlinde:1995mz}. The classical contribution to
the partition function is given by integral fluxes $p \in H^2(M,\Z)$, as in
equation~(\ref{magneticflux}), to the Langrangian
(\ref{YMactionthetaform}). Decomposing the flux $p$ into a self-dual
and anti-selfdual contribution yields the generalized theta-function
\begin{align}
\theta_{\G}(q,\bar{q}) = \sum_{(p_+,p_-) \in \G} q^{\frac{1}{2} p_+^2}
\bar{q}^{\frac{1}{2}p_-^2}  
\end{align}
with $q = \exp( 2 \pi i \tau)$, whereas $\Gamma = H^2(M,\Z)$ is the
intersection lattice of $M$ and $p^2 = \int_M p \wedge p$. The total
$U(1)$ partition function is 
found by adding quantum corrections to the above result, which are
captured by some determinants \cite{Witten:1995gf}. Instead of
transforming as a modular invariant, $Z^{U(1)}$ transforms as a 
\Index{modular form} under $Sl(2,\Z)$-transformation of $\tau$
\begin{align*}
 Z^{U(1)}\left( \frac{a \tau + b}{c \tau + d} \right) = (c \tau
 +d)^{u/2} (c \overline{\tau} +d)^{v/2}  Z^{U(1)}\left( \tau \right),
 \quad \left( \begin{array}{cc} a & b \\c & d \end{array} \right) \in
 Sl(2, \Z).
\end{align*}
We will come back
on this in \Cref{sec:N=4YM}.

\subsection{Supersymmetry}\label{sec:susygaugetheories}
\index{supersymmetry}


In \emph{supersymmetric} theories quantum corrections are much better under
control, so that much more can be learned about non-perturbative properties of
the theory. We will soon discuss such elegant results, but let us first
introduce supersymmetric gauge theories.  

The field content of the simplest supersymmetric gauge theories just 
consists of a bosonic gauge field $A$ and a fermionic gaugino field 
$\la$. Supersymmetry relates the gauge field $A$ to
its superpartner $\chi$. In any supersymmetric theory the number of
physical bosonic degrees of freedom must be the same as the number of
physical fermionic degrees of freedom. This constraints supersymmetric
Yang-Mills theories to dimension $d\le10$. 

The Lagrangian of a minimal supersymmetric gauge theory is 
\begin{align*}
\mathcal{L} = -\frac{1}{4} \Tr(F_{\mu \nu} F^{\mu \nu}) +
\frac{i}{2} \bar{\chi} \Gamma^{\mu} D_{\mu} \chi,  
\end{align*}
and supersymmetry variations of the fields $A$ and $\chi$ are generated by a
spinor $\epsilon$
\begin{align}\label{eqn:susytrans}
\delta A_{\mu}  = \frac{i}{2} \bar{\epsilon} \Gamma_{\mu} \chi, \qquad
\delta \chi  = \frac{1}{4} F_{\mu \nu} \Gamma^{\mu \nu} \epsilon.
\end{align}
The number of supersymmetries equals the number of components of
$\epsilon$.
 
Dimensionally reducing the above minimal $\cN=1$ susy gauge theories
to lower-dimensional space-times yield $\cN=2,4$ and possibly $\cN=8$
susy gauge theories. Their supersymmetry variations are
determined by an extended supersymmetry algebra. 
In four dimensions this is a unique extension of the Poincar\'e
algebra generated by the supercharges $Q_{\alpha}^A$ and $Q_{A
  \dot{\alpha}}$, with $A \in \{1,
\ldots, \cN\}$ and $\alpha, \dot{\alpha} \in \{1,2\}$ are indices in the
4-dimensional spin
group $\mathfrak{su}(2)_L \times \mathfrak{su}(2)_R$. Non-vanishing anti-commutation relations are given by
\begin{align*}
 \{ Q^A_{\alpha}, \overline{Q}_{B \dot{\beta}} \} &= 2(\sigma^{\mu})_{\alpha
   \dot{\beta}} P_{\mu} \delta^A_B \\
\{ Q^A_{\alpha}, Q^B_{\beta} \} &= \epsilon_{\alpha \beta} \cZ^{AB} \\
\{ \overline{Q}_{A \dot{\alpha}}, \overline{Q}_{B \dot{\beta}} \} &= \epsilon_{\dot{\alpha} \dot{\beta}} \cZ^{\dag}_{AB}  
\end{align*}
where $\cZ^{AB}$ and its Hermitean conjugate are the central
charges. The automorphism group of this algebra, that acts on the
supercharges, is known as the R-symmetry group. \index{R-symmetry} 
\index{central charge $\cZ$}

\subsubsection{BPS states}

A special role in extended supersymmetric theories is played by
supersymmetric \emph{BPS states}\index{BPS state} \cite{Witten:1978mh}. They
are annihilated by a some of the supersymmetry generators,
e.g. quarter BPS states satisfy    
\begin{align*}
 Q^A | \textrm{BPS} \rangle = 0,   
\end{align*}
for $1/4$ $\cN$ indices $ A \in \{1, \ldots, \cN\}$. BPS states saturate the
bound $M^2 \le |\cZ|^2$ and form ``small'' representations of the
above supersymmetry algebra. This implies that supersymmetry protects
them against quantum corrections: a small deformation won't
just change the dimension of the representation.      

\subsubsection{Twisting}\label{sec:tqft}

Supersymmetric Yang-Mills requires a covariantly constant
spinor $\epsilon$ in the rigid supersymmetry variations
(\ref{eqn:susytrans}). Since these are impossible to find on a generic
manifold $M$, the concept of \emph{twisting}\index{topological twist} has been
invented. Twisting makes use of the fact that supersymmetric gauge
theories are invariant under a non-trivial internal symmetry, the R-symmetry group. By
choosing a homomorphism from the space-time symmetry group into this
internal global symmetry 
group, the spinor representations change and often contain a
representation that transforms as a scalar under the new Lorentz
group. 

Such an odd scalar $Q_T$ can be argued to obey $Q_T^2 = 0$. It is
 a topological supercharge that turns the
theory into a \emph{cohomological} quantum field theory. Observables
$\cO$ can be identified with the cohomology generated by $Q_T$, and
correlation functions are independent of continuous deformations of
the metric  
\begin{align}
\frac{\lotjesd}{\lotjesd g_{\mu \nu}}  \langle \cO_1 \cdots \cO_k \rangle = 0  
\end{align}
These correlation functions can thus be computed by going to short
distances. This yields techniques to study the dynamics of these
theories non-perturbatively. 
For these topological theories it is sometimes possible to compute the
partition function and other correlators.
Witten initiated twisting in the context
of $\cN=2$ supersymmetric Yang-Mills \cite{Witten:1994ev}. He showed
that correlators in the so-called Donaldson-Witten twist compute
the famous Donaldson invariants.

A general theme in cohomological field theories is
\emph{localization}. Unlike in general physical theories, in these
topological theories the saddle point
approximation is actually exact. The path integral only receives
contributions from fixed point locus $\cM$ of the scalar
supercharge $Q_T$. Since the kinetic part of the action (that contains
all metric-dependent terms) is $Q_T$-exact, the only non-trivial
contribution to the path integral is given by topological terms:
\begin{align*}
Z_{\textrm{cohTFT}}=  \int D \mathcal{X} ~\exp \left(- \frac{1}{e^2}
  S_{\textrm{kin}}(\mathcal{X} ) + 
   S_{\textrm{top}}(\mathcal{X} ) \right) \to   \int_{\cM} D
 \mathcal{X}  ~\exp \left(  
   S_{\textrm{top}}(\mathcal{X} ) \right).
\end{align*}%
Here $\mathcal{X} $ represents a general field content.
An elegant example in this respect is
2-dimensional gauge theory \cite{Witten:1992xu}. Extensive reviews
of localization are \cite{Blau:1995rs,Cordes:1994fc}. We will
encounter localization on quite a few occasions, starting with
Vafa-Witten theory in \Cref{sec:N=4YM}. 

\subsection{Extended objects}\label{sec:branes}

Whereas Yang-Mills theory is formulated in terms of a single
gauge potential $A$, string theory is equipped with a whole set of
higher-form gauge fields.  Instead of coupling to electro-magnetic
particles they couple to \Index{extended objects}, such as D-branes. This is
analogous to the coupling of a particle of electric charge $q$ to the
Maxwell gauge field $A$
$$
q \int_{\cW} A = q \int_{\cW} A_{\mu} \frac{\lotjesd x^{\mu}}{\lotjesd t} dt,
$$
where $\cW$ is the worldline of the particle. Notice that we
need to pull-back the space-time gauge field $A$ in the first term before
integrating it over the worldline. We often don't write down the
pull-back explicitly to simplify notation.
D-branes and other extended objects appear all over this
thesis. Let us therefore give a very brief account of the properties
that are relevant for us. 

\subsubsection{Couplings and branes in type
  II}
\index{D-brane}\index{NS5-brane}
\index{fundamental string}\index{string coupling constant}
\index{string length}

Gauge potentials in type II theory either belong to the so-called RR or the
NS-NS sector. The RR potentials couple to D-branes, whereas the NS-NS
potential couples to the fundamental string (which is often denoted by F1)
and the NS5-brane. Let us discuss these sectors in a little more
detail. 

The only NS-NS gauge field is the 2-form $B$. The $B$-field \index{B-field}
plays a crucial role in \Cref{chapter5}. Aside from the $B$-field
the NS-NS sector contains the dilaton field $\phi$ and the
space-time metric $g_{mn}$. Together these NS-NS fields combine into the
sigma model \index{sigma model} action
\begin{align}\label{eqn:worldsheetBfield}
S_{\sigma\textrm{-model}} = - \frac{1}{2 \pi \alpha'} \int_{\Sigma} g_{mn} d x^m \wedge *
 d x^n + iB_{mn}  d x^m \wedge d x^n  + \alpha' \phi R.
\end{align}
This action describes a string that wraps the Riemann surface $\Sigma$
and is embedded in a space-time with coordinates $x^m$.
In particular, it follows that the $B$-field couples to a
(fundamental) string F1. 

Remember that the 1-form $dx^m$ refers to the pull-back
$\partial_{\alpha} x^m d \sigma^{\alpha}$  to the worldsheet $\Sigma$
with coordinates 
$\sigma^{\alpha}$.
Furthermore, the symbol $*$ stands for the 2-dimensio-nal worldsheet
Hodge star operator  and $R$ is the worldsheet curvature 2-form. 

This formula requires some more explanation though. The symbol $\sqrt{\alpha'}=
l_s$ sets the string length, since $\alpha'$ is inversely proportional to
its tension.  Since the
Ricci scalar of a Riemann surface equals its Euler 
number, the last term in the action contributes $2g-2$ powers of
$$
g_s = e^{\phi}
$$
to a stringy $g$-loop diagram; $g_s$ is therefore
called the string coupling constant.  

The extended object to which the B-field couples
magnetically is called the NS5-brane. It can wrap any
6-dimensional geometry in the full 10-dimensional string
background, but its presence will deform the transverse geometry. In
the transverse directions the dilaton field $\phi$ is non-constant, and there
is a flux $H = dB$ of the $B$-field through the boundary of the 
transerve 4-dimensional space. The tension of a NS5-brane is proportional to
$1/g_s^2$ so that it is a very heavy object when $g_s \to
0$. Moreover, unlike for D-branes
open strings cannot end on it. This makes it quite a mysterious object. 

For the RR-sector it makes a difference whether we are in type IIA
or in type IIB theory: type IIA contains all odd-degree RR forms and
type IIB the even ones. In particular, the gauge potentials $C_{1}$
and $C_4$ are known as the graviphoton fields 
\index{graviphoton field} for type IIA and type 
IIB, respectively, and the RR potential $C_{0}$ is the axion field. 
Any RR potential $C_{p+1}$ couples
electrically to a \emph{D$p$-brane}. This is a $p$-dimensional extended
object that sweeps out a $(p+1)$-dimensional
world-volume $\Sigma_{p+1}$. Type IIA thus contains
D$p$-branes with $p$ even, whereas in type IIB $p$ is
odd.

The electric D$p$-brane coupling to $C_{p+1}$  introduces the term 
\begin{align}\label{eqn:Dbranecoupling}
T_p \int_{\Sigma_{p+1}} C_{p+1} 
\end{align}
in the 10-dimensional string theory action, where 
$ 1/T_p = (2 \pi)^p  \sqrt{\alpha'}^{p+1} g_s$
is the inverse tension of the
D$p$-brane.
 Magnetically, the RR-potential $C_{p+1}$ couples to a D$(6-p)$-brane 
that wraps a $(7-p)$-dimensional submanifold $\Sigma_{7-p}$.
%

\subsubsection{Calibrated cycles}\index{calibration}

D-branes are solitonic states as
their tension $T_p$ is proportional to $1/g_s$. To be stable
against decay the brane 
needs to wrap a submanifold that preserves some supersymmetries. Geometrically,
such configurations are defined by a \emph{calibration} \cite{harveylawson}. A
calibration form is a closed form $\Phi$ such that $\Phi \le \textrm{vol}$ at
any point of the background. A submanifold $\Sigma$ that is calibrated
satisfies 
\begin{align*}
 \int_{\Sigma} \Phi = \int_{\Sigma} \textrm{vol}, 
\end{align*}
and minimizes the volume in its holomogy class. On a K\"ahler manifold
a calibration is given by the K\"ahler form $t$, and the calibrated
submanifolds are complex submanifolds. On a Calabi-Yau threefold the
holomorphic threeform $\Omega$ provides a calibration, whose
calibrated submanifolds are special Lagrangians. Calibrated
submanifolds support covariantly constant spinors, and therefore
preserve some supersymmetry. D-branes wrapping them are supersymmetric
BPS states.  

\subsubsection{Worldvolume theory}

D-branes have a perturbative description in terms of open strings that
end on them. 
The massless modes of these open strings recombine in
a Yang-Mills gauge field $A$. When the D-brane worldvolume is flat the
corresponding field theory on the $p$-dimensional brane is a reduction
of $\mathcal{N}=1$ susy Yang-Mills from 10 dimensions to $p+1$. The
$9-p$ scalar fields in this theory correspond to the transverse
D-brane excitations. When $N$ D-branes coincide the wordvolume theory
is a $U(N)$ supersymmetric Yang-Mills theory.  

For more general calibrated submanifolds the low energy gauge theory
is a twisted topological gauge theory \cite{Bershadsky:1995qy}, 
which we introduced in 
\Cref{sec:tqft}. Which particular twist is realized, can be
argued by determining the normal bundel to the submanifold. Sections
of the normal bundel fix the transverse bosonic excitations of
the gauge theory, and should correspond to the bosonic field content
of the twisted theory. 

\subsubsection{I-branes and bound states}
\index{I-brane}\index{bound state}\index{Chern-Simons coupling}

Branes can intersect each others such that they preserve some amount
of supersymmetry. This is called an \emph{I-brane} configuration. In such a
set-up there are more degrees of freedom than the ones (we described
above) that reside on the individual branes. These extra degrees of
freedom are given by the modes of open strings that stretch between
the branes. In stringy constructions of the standard model on a set
of branes they often provide the chiral fermions. 

Chiral fermions are intimately connected with quantum anomalies, and
brane intersections likewise. To cure all possible anomalies in an
I-brane system, a topological Chern-Simons term has to be added to the string
action 
\begin{align}\label{eq:CSterms}
S_{CS} = T_p \int_{\Sigma_{p+1}} \Tr \, \exp \left(\frac{F}{2\pi} \right) \wedge
\sum_i C_i \wedge \sqrt{ \hat{A}(R) }.
\end{align}
This term is derived through a so-called anomaly inflow analysis
\cite{Green:1996dd, Cheung:1997az}. The
last piece contains the $A$-roof genus for the 10-dimensional
curvature 2-form $R$ pulled back to $\Sigma_{p+1}$. It may be expanded as
\begin{align*}
\hat{A}(R) =  1 - \frac{p_1(R)}{24} + \frac{7 p_1(R)^2- 4 p_2(R)}{5760} +
\ldots,
\end{align*}
where $p_k(R)$ is a Pontryagin class. 
For example, the Chern-Simons term~(\ref{eq:CSterms})  includes a
factor 
$$ 
T_p \int_{\Sigma_{p+1}} \Tr\, \left( \frac{F}{2 \pi} \right) \wedge C_{p-1}
$$
when a gauge field $F$ on the worldvolume $\Sigma_{p+1}$ is turned on. It describes an
induced D$(p-2)$ brane 
wrapping the Poincare dual of $[F/2 \pi]$ in $\Sigma_{p+1}$. 

Vice versa, a bound state of a D$(p-2)$-brane on a D$p$-brane may be
interpreted as turning on a field strength $F$ on the D$p$-brane. 
Analogously, instantons (\ref{eqn:4dinstanton}) in a 4-dimensional
gauge theory, say of rank zero and second Chern class $n$, 
 have an interpretation in type IIA
theory  bound states of $n$ D0-branes on a D4-brane.  
More generally, topological excitations in the worldvolume theory of a
brane are often caused by other extended objects that end on it
\cite{Strominger:1995ac}.  


\subsection{String dualities}

The different appearances of string theory, type I, II, heterotic and
M-theory, are connected through a zoo of dualities. Let us briefly
introduce T-duality and S-duality in type II. There are many more
dualities, some of which we will meet on our way.

\subsubsection{T-duality}\index{T-duality}

T-duality originates in the worldsheet description of type II theory
in terms of open and closed strings. T-duality on an $S^1$ in the
background interchanges the Dirichlet and Neumann boundary conditions
of the open strings on that circle, and thereby maps branes that wrap
this $S^1$ into branes that don't wrap it (and vice versa). It thus
interchanges type IIA and type IIB theory. 

Similar to electro-magnetic duality (see 
\Cref{sec:e-mduality}), T-duality follows from a path integral argument 
\cite{Rocek:1991ps}. The sigma model action for a fundamental string
is based on the term 
\begin{align*}
 - \frac{1}{2 \pi \alpha'} \int_{\Sigma}  g_{mn} d x^m \wedge * dx^n,  
\end{align*}
in equation~(\ref{eqn:worldsheetBfield}) when we forget the $B$-field for simplicity. Let us suppose that the
metric is diagonal in the coordinate $x$ that parametrizes the
T-duality circle $S^1$. Then we can add a Lagrange multiplier field
$dy$ to the relevant part of the action 
\begin{align*}
  \int D x D y \exp  \int \left( -\frac{1}{2 \pi \alpha'} dx \wedge
   * dx + \frac{i}{\pi} dx \wedge * d y \right). 
\end{align*}
On the one hand the Lagrange multiplier field $d y$ forces $d (dx) = 0$,
which locally says that $dx$ is exact. On the other hand integrating
out $dx$ yields
\begin{align*}
 \int D y \exp  \int \left(-  \frac{\alpha'}{2 \pi} d y
   \wedge * d y \right) 
\end{align*}
So T-duality exchanges
$$
\alpha' \leftrightarrow \frac{1}{\alpha'}
$$
and is therefore a strong-weak coupling on the worldsheet. 
More precisely, one should also take into account the B-field coupling
(\ref{eqn:worldsheetBfield}), which is related to non-diagonal terms
is the space-time metric. This leads to the well-known Buscher
rules \cite{Buscher:1987qj}

Since the
differential $dx$ may be identified with a component of the gauge field $A$ on the
brane, and $dy$ with a normal 1-form to the brane, the reduction
of supersymmetric Yang-Mills from ten to $10-d$ dimensions 
can  be understood as applying T-duality $d$ times.

\subsubsection{S-duality}\index{S-duality}

Since $\cN=4$ supersymmetric Yang-Mills is realized as the low-energy
effective theory on a D3-brane wrapping $\R^{4}$. Since electro-magnetic
duality in this theory is an exact symmetry, it should have a string
theoretic realization in type IIB theory as well. Indeed it
does, and this symmetry is known as S-duality. In type IIB theory S-duality
is a (space-time) strong-weak coupling duality that 
maps 
$
g_s \leftrightarrow 1/g_s.
$
Analogous to Yang-Mills theory the complete symmetry group is $Sl(2,\Z)$, where
the complex coupling constant $\tau$ (\ref{eqn:tau}) is realized as
$$
\tau = C_0 + i e^{- \phi}.
$$
Since the ratio of the tensions of the fundamental string F1 and the
D1-brane is equal to $g_s$, S-duality exchanges these objects as well
as the $B$-field and the $C_2$-field they couple to. Likewise, it
exchanges the NS5-brane with the D5-brane.

\subsubsection{M-theory}\index{M-theory}\index{M-brane}

Type IIA is not invariant under S-duality. Instead, in the strong
coupling limit another dimension of size $g_s l_s$ opens up. This
eleven dimensional theory is called M-theory. The field content of
M-theory contains a 3-form gauge field $C_3$ with 4-form flux $G_4=
dC_3$ that couples to an M2-brane. It magnetic dual is an M5-brane. 
Other extended objects include the KK-modes and Taub-NUT space. We
will introduce them in more detail later. For now, let us just note
that a reduction over the M-theory circle consistently reproduces all the fields
and objects of type IIA theory. (An extensive review can for instance be
found in \cite{Obers:1998fb}.)




\section{Vafa-Witten twist on ALE spaces}\label{sec:N=4YM}

The maximal amount of supersymmetry in 4-dimensional gauge theories is
$\cN=4$ supersymmetry. This gauge theory preserves so many supercharges
that it has a few very special properties. Its beta function is argued
to vanish non-perturbatively, making the theory exactly finite and
conformally invariant \cite{Seiberg:1988ur}. 
It is also, not unrelated, the only 4-dimensional gauge theory
where electro-magnetic and $Sl(2,\Z)$ duality are conjectured to hold
at all energy scales.

In this section we apply the techniques of the previous section to study a
twisted version of $\cN=4$ supersymmetric Yang-Mills theory in four
dimensions. It is called the Vafa-Witten twist.
We study Vafa-Witten theory on ALE (asympto-tically
locally Euclidean) spaces, which are defined as hyper-K\"ahler
resolutions of the singularity 
\begin{align*}
 \C^2/\Gamma, 
\end{align*}
where $\Gamma$ is a finite subgroup of $SU(2)$.
ALE spaces are intimately connected to ADE Lie algebras.
 
The Vafa-Witten twist is an example of a topological gauge theory. It
localizes on anti-selfdual instantons that are defined by the
vanishing of the selfdual component $F_+$ of the field strength. The
Vafa-Witten partition function is therefore a generating function that counts
anti-selfdual instantons. 
On an ALE space this partition function turns
out to compute the character of an affine ADE algebra. 

This section starts off with the Vafa-Witten twist and ALE spaces. 
We discuss Vafa-Witten theory on ALE spaces, and embed the gauge theory 
into string theory as a worldvolume theory on top of D4-branes. This is the
first step in finding a deeper explanation for the duality between
$\cN=4$ supersymmetric gauge theories and 2-dimensional conformal
field theories.




\subsection{Vafa-Witten twist}\index{Vafa-Witten theory}

C.~Vafa and E.~Witten have performed an important S-duality
check of $\cN=4$ supersymmetric gauge theory by computing a twisted partition function  on certain 4-manifolds $M$ 
\cite{Vafa:1994tf}.

In total three inequivalent twists of $\cN=4$ Yang-Mills theory are possible. These are
characterized by an embedding of the rotation group $SO(4) \cong SU(2)_L
\times SU(2)_R$ in the R-symmetry group $SU(4)_R$ of the
supersymmetric gauge theory. The 
Vafa-Witten twist considers the branching $$SU(4)_R \to SU(2)_A \times
SU(2)_B$$ and twists either $SU(2)_L$ or $SU(2)_R$ with $SU(2)_A$. Both
twists are related by changing the orientation of the 4-fold $M$ and
at the same time changing $\tau$ with $\bar{\tau}$. Let us choose the
left-twist here. This results in a bosonic field content consisting of a
gauge field $A$, an anti-selfdual 2-form and three scalars.


The twisted theory is a cohomological gauge theory with $\cN_T=2$
equivariant topological supercharges $Q_{\pm}$, whose Lagrangian can be
written in the form
\begin{align}\label{eqn:SWaction}
 \cL &\,=\,  \frac{i \tau}{4 \pi} \,  F \wedge  F 
\,-\, \frac{2}{e^2} \, F_- \wedge * F_- \,+\,
 \ldots \notag \\ 
& \,=\, \frac{i \tau}{4 \pi} \, F \wedge F \, +\, Q_+ Q_- \,\cF , 
\end{align}
where $\cF$ is called the action potential \cite{Dijkgraaf:1996tz}. In
the spirit of our 
discussion in \Cref{sec:tqft} this implies that the path integral
localizes onto the 
critical points of the potential $\cF$ modulo gauge equivalence. On K\"ahler
manifolds 
this critical locus is
characterized by the vanishing of the anti-selfdual 2-form and the three
scalars, whereas the gauge field obeys 
\begin{align*}
F_{-}=0. 
\end{align*}
The Vafa-Witten twist thus localizes to the instanton moduli space 
\begin{align*}
 \M = \cW/G, \quad \cW= \{A:  F_-(A) = 0 \}. 
\end{align*}
of selfdual connections.\footnote{Alternatively, the localization
  to $F_-$ follows from the field equations. Since the field content
  of the right twist involves only anti-selfdual (instead of
  selfdual) fields, setting the
  fermion variations to zero forces $F_-=0$. Likewise, performing the left
  twist corresponds to changing $\tau$ with $\bar{\tau}$ as well
  changing $F_-$ with $F_+$ in the Lagrangian
  (\ref{eqn:SWaction}). Ultimately, our conventions in 
  equation~(\ref{eqn:fullYMaction}) and
  (\ref{YMactionthetaform}) imply that the selfdual
  instantons receive contributions in $\tau$. This choice is
  non-standard in the Vafa-Witten literature, but it fits better with the
  content of this thesis. 
}
The moduli space $\cM$ naturally decomposes in connected components $\cM_n$
that are labeled by the instanton number
\begin{align*}
n =  \int_M  \Tr \left[ \frac{F \wedge F}{8 \pi^2} \right]. 
\end{align*}
The Vafa-Witten partition function computes the Euler characteristic
of these components (without $\pm$ signs). 
Up to possible holomorphic anomalies it is a holomorphic
function of ${\tau}$ with a Fourier-expansion of the type
\begin{align} \label{eqn:noncompleteVWFourier}
 Z^G(\tau) \sim \sum_n d(n) q^{n},   
\end{align}
where  the numbers $d(n)$ represent the Euler
characteristic of $\cM_n$, whereas $q^n = \exp (2 \pi i n \tau)$ denotes
the contribution of the instantons to the topological term in the
Lagrangian~(\ref{eqn:SWaction}). The numbers $d(n)$ are integers when
$G$ is connected and simply connected.

%

%
%

\subsubsection{S-duality and modular forms}
\index{modular form}\index{S-duality}

The Vafa-Witten partition function only transforms nicely under
S-duality once local curvature corrections in the Euler characteristic
$\chi$ and the signature $\sigma$ of $M$ have been added to the action
\cite{Vafa:1994tf}. Notice that this is justified since they do not
change the untwisted 
theory on $\R^4$. In particular, these additional terms introduce an
extra exponent in 
(\ref{eqn:noncompleteVWFourier}) 
\begin{align*}
 Z^G(\tau) = q^{-c/24} \sum_n d(n) q^{n},
\end{align*}
where $c$ is a number depending on $\chi$ and $\sigma$. The resulting
Vafa-Witten partition function conjecturally transforms as a \emph{modular form}
of weight $w = - \chi(M)$ that exchanges $G$ with its 
dual $\hat{G}$  
\begin{align*}
 Z^G \left(- 1/\tau \right) \sim \tau^{w/2} Z^{\hat{G}}(\tau).
\end{align*}

Since $\hat{G}$ is often not simply connected (for
example the dual of $G = SU(2)$ is $\hat{G}=SO(3)= SU(2)/\Z_2$) one
has to take into account magnetic fluxes $v \in H^2(M,
\pi_1(\hat{G}))$. The components of the partition function $Z_v$ mix under the
S-duality transformation $\tau \to - 1/\tau$. Furthermore, 
since for such $\hat{G}$ instantons numbers are not integer, the
vector valued partition function $Z_{v}$ will only be covariant 
under a subgroup of $Sl(2, \Z)$. 

Characters of affine Lie algebras are examples of such vector valued
modular forms. We will soon introduce them and see that they indeed
appear as Vafa-Witten partition functions.       



\subsubsection{Unitary gauge group and Jacobi forms}\index{Jacobi form}

In the following we will be especially interested in
Vafa-Witten theory with gauge group $U(N)$. This gauge group is not
simply-connected, since it contains an abelian subgroup $U(1) \subset
U(N)$. Instanton bundles are therefore not only characterized
topologically by their second Chern class $ch_2$, but also carry
abelian fluxes measured by the first Chern class $c_1$.    

The $U(N)$ Yang-Mills partition function gets extra contributions from
these magnetic fluxes. The path integral can be performed by first
taking care of the $U(1)$ part of the field strength. This gives a
contribution in the form of a Siegel theta function, precisely
as explained in \Cref{sec:e-mduality}.  

We can make these fluxes more explicit by
introducing a topological coupling $v \in H^2(M,\Z)$ in the original Yang-Mills
Lagrangian:
\begin{align}\label{eqn:abelianpart}
\cL \,=\,  \frac{i \tau}{4 \pi} \,\Tr\, F_+ \wedge F_+ \,+\, v \wedge
\Tr\,F_+ \,+\, \textrm{c.c}..
\end{align}
Here we define complex conjugation c.c. not only to change $\tau$
and $v$ into their anti-holomorphic conjugates, but also to map the
selfdual part $F_+$ of the field strength to the anti-selfdual part
$F_-$.  The $v$-dependence of the partition function is
entirely captured by the $U(1)$ factor of the field strength. It
results in a Siegel theta-function of signature $(b_2^+,b_2^-)$   
\begin{align}
\theta_{\Gamma}(\tau,\bar{\tau};v, \bar{v}) = \sum_{p \in \G} e^{ i \pi \left(\tau p_+^2 - \bar\tau p_-^2
\right)} e^{ 2 \pi i  \left(
  v \cdot p_+ - \bar{v} \cdot p_-\right)}. \label{eqn:U(1)piece}
\end{align}
Here $p$ and $v$ are elements of $\G = H^2(M, \Z)$, so that $v \cdot p$ (and 
likewise $p^2$) refers to the intersection product $\int_M v \wedge
p$.


In this chapter we focus on non-compact hyper-K\"ahler manifolds whose
Betti number $b^+_2 = 0$. We change their orientation to find a
non-trivial Vafa-Witten partition function. The $U(1)$ contribution to
their partition function is then purely holomorphic.

Because of S-duality the total $U(N)$ partition function $Z(v,\tau)$ is
expected to be given by a \emph{Jacobi form} 
determined by the geometry $M$. That is, for 
\begin{align*}
\left(\begin{array}{cc}
a & b \\
c & d \\
\end{array}\right) \in SL(2,\Z)& \quad \textrm{and} \quad
n,m \in H^2(M,\Z) \cong \Z^{b_2}, 
\end{align*}
it should have the
transformation properties
\begin{align*}
Z\left({v \over c\tau + d},{a\tau +b \over c\tau +d} \right) 
& = (c\tau + d)^{w/2} e^{ \frac{2\pi i \kappa c v^2}{c\tau
    +d}}  Z(v,\tau)
 \\
Z(v + n\tau + m, \tau) &= e^{ -2\pi i \kappa (n^2\tau + 2n
  \cdot v)}
Z(v,\tau), 
\end{align*}
where $w$ is the weight and $\kappa$ is the index of the Jacobi form. 
Using the localization to instantons, the partition function has a
Fourier expansion of the form
\begin{align}\label{eqn:vw-jacobi}
Z(v,\tau) = \sum_{m\in H^2(M),n\geq 0} d(m,n)\, y^m q^{n-c/24},
\end{align}
where $y = e^{2\pi i v}$, $q = e^{2\pi i \tau}$ and $c = N \chi$.
The coefficients $d(n,m)$ are roughly computed as the Euler number of
the moduli space of $U(N)$ instantons on $M$ with total instanton
numbers $c_1=m$ and $ch_2=n$.

\subsection{M5-brane interpretation}\index{M-brane}


In string theory $U(N)$ Vafa-Witten theory is embedded as the topological
subsector of $\cN=4$ super Yang-Mills theory on $N$ D4-branes that
wrap a holomorphic 4-cycle $M \subset X$ in the IIA background
\begin{align}
\textrm{IIA}: \quad X \times \R^4.   
\end{align}
Topological excitations in the gauge theory amount to bound states of D0 and D2-branes on the D4-brane. 
 
Let us now consider the 5-dimensional gauge theory on a Euclidean D4-brane
wrapping $M \times S^1$. The partition function of this theory is
given by a trace over its Hilbert space, whose components are labeled by the number $n = ch_2(F)$ of D0-branes and the number $m=c_1(F)$ of D2-branes. 
The coefficients $d(n,m)$  in the Fourier expansion
(\ref{eqn:vw-jacobi}) thus have a direct interpretation as
computing BPS invariants: the number of bound states of $n$ D0-branes
and $m$ D2-branes on the D4-brane. For this reason
they are believed to be integers in general. We can compute $d(m,n)$ as the
index\footnote{Here and in the subsequent sections we assume that the two
  fermion zero modes associated to the center of mass movements of the
  D4-brane have been absorbed.}
$$
d(m,n) = \Tr (-1)^F \in \Z,
$$ 
in the subsector of field configurations on $M$ of given instanton
numbers $m,n$.

From the string theory point of view the modular invariance of $Z$ is
explained naturally by lifting the D4-brane to M-theory, where it
becomes an M5-brane on the product manifold
$$
M \times T^2.
$$ 
The world-volume theory of the M5-brane is (in the low-energy limit)
the rather mysterious 6-dimensional $U(N)$ conformal field theory
with $(0,2)$ supersymmetry. The complexified gauge coupling $\tau$ can
now be interpreted as the modulus of the elliptic curve $T^2$, while
the Wilson loops of the 3-form potential $C_3$ along this curve are
related to the couplings $v$, as we explain in more detail in
\Cref{ssec-D4D6}.  With this interpretation the action of modular
group $SL(2,\Z)$ on $v$ and $\tau$ is the obvious geometric one.

Instead of compactifying over $T^2$, we can also consider a
compactification over $M$. We then find a 2-dimensional $(0,8)$ CFT on the
two-torus, whose moduli space consists of the solutions to the
Vafa-Witten field equations on $M$. This duality motivates the
appearance of CFT characters in Vafa-Witten theory. In 
\Cref{sec:stringrealization} we will reach a deeper understanding. 
  





\subsubsection{$\R^4$ -- Example}\index{affine character}

The simplest example is $U(1)$ Vafa-Witten theory on $\R^4$ corresponding to a
single D4-brane on $\R^4$. Point-like instantons in this theory
correspond to bound states with D0-branes and yield a non-trivial partition
function  
%
$$ 
Z(\tau, v) = {\theta_3(v,\tau) \over \eta(\tau)} = {\sum_{p\in\Z} e^{\pi
    i \tau p^2 + 2\pi i v p} \over q^{1/24} \prod_{n=1}^{\infty} (1-q^n)}.
$$
The Dedekind eta function $\eta(\tau)$ can be rewritten as
a generating function 
$$ 
\frac{1}{\eta(\tau)} = q^{-1/24} \frac{1}{\prod_{n>0} \left(1-q^n
  \right)} =  q^{-1/24}  \sum_{n \ge 0} p(n) q^n, 
$$ 
of the number $p(n)$ of partitions $(n_1, \ldots, n_k)$ of
$n$. We can identity each such a partition with a bosonic
state 
$$
\alpha_{-n_1} \cdots \alpha_{-n_k} |p \rangle
$$
in the Fock space $\cH_p$ of a chiral boson $\phi(x)$ with mode
expansion 
$$\del \phi(x) = \sum_{n \in \Z} \alpha_n x^{-n-1}.$$
The state $|p \rangle$
is the vacuum whose Fermi level is raised by $p$ units. The partition
function $Z(\tau,v)$ is exactly reproduced by  the $\widehat{u}(1)$ character
$$
Z(\tau,v)= \Tr_{\cH_p} \left( y^{J_0} q^{L_0 - c/24} \right) =
\chi^{\widehat{u}(1)}(\tau,v),  
$$
where $L_0 = \frac{1}{2} \alpha_0^2 + \sum_{n>0} \alpha_{-n} \alpha_n$
measures the energy of the states and $J_0 = \alpha_{0}$ 
the $U(1)$ charge. The instanton zero point energy $c=1$ now corresponds
to the central charge for a single free boson.

\subsection{Vafa-Witten theory on ALE spaces}\label{sec:VafaWittenALEspace}

So far we motivated that the Vafa-Witten partition function transforms
under S-duality in a modular way. Furthermore, we 
have seen a simple example with $M = \R^4$ where
the partition function equals a CFT-character. In this
section we will see that this relation is more generally true for ALE
spaces. 

In the forthcoming sections we introduce quite a few notions
from the theory of affine Lie algebras $\widehat{\lieg}$ and their
appearance in WZW (Wess-Zumino-Witten)
models. The classic reference for this subject is \cite{franscesco}.

\subsubsection{ALE spaces and geometric McKay
  correspondence}\label{sec:ALEspace}
\index{ALE space}\index{McKay correspondence}

An ALE space $M_{\G}$ is a
non-compact hyper-K\"ahler surface. It is obtained 
by resolving the singularity at the origin of $\C^2/\Gamma$, 
$$
M_\G \to \C^2/\G,
$$ 
where $\Gamma$ is a finite subgroup of $SU(2)$ that acts linearly on
$\C^2$. These Kleinian singularities are classified into three
families: the cyclic groups $A_{k}$, the dihedral groups $D_k$ and 
the symmetries of the platonic solids $E_k$. For example, an
$A_{k-1}$ singularity is generated by the element 
$$(z,w) \mapsto (e^{2 \pi i/k}z, e^{-2
  \pi i/k}w).$$  
of the cyclic subgroup $\Gamma = \Z_{k}$.

\begin{figure}[h]
\centering
\includegraphics[width=11cm]{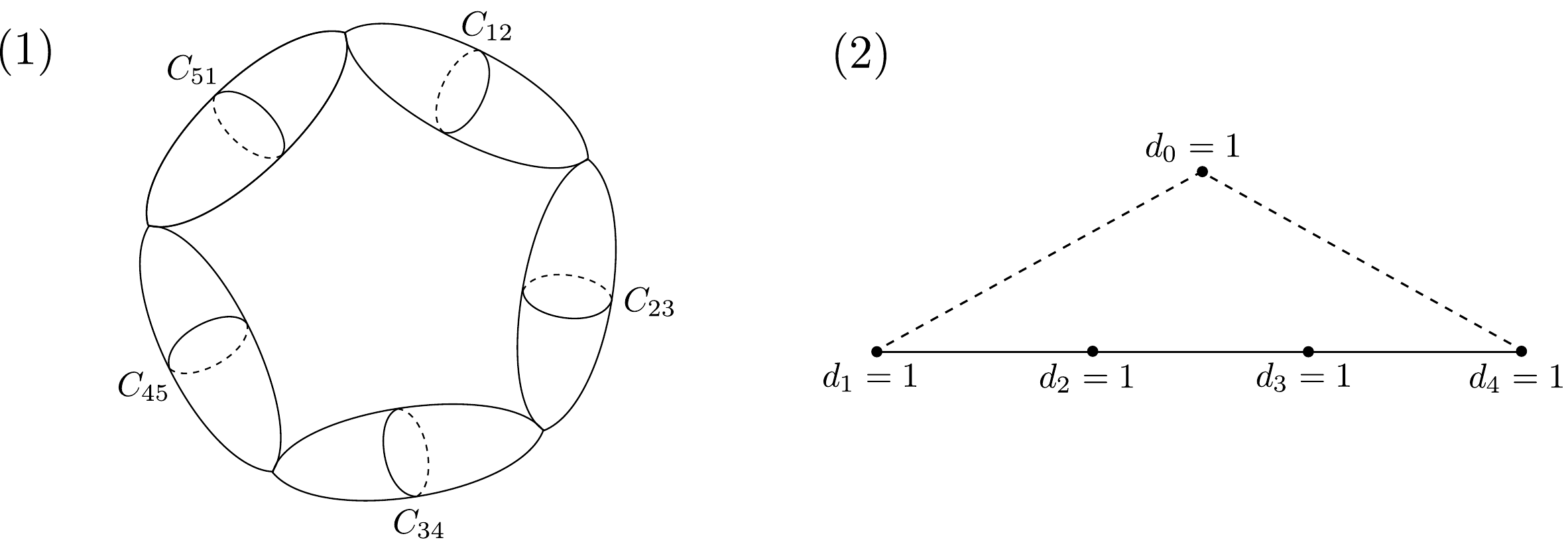}
\caption{The left picture $(1)$ illustrates an $A_4$-singularity, which is a
  hyper-K\"ahler resolution of $\C^2/\Z_5$. Its homology is generated
  by 4 independent 2-cycles. They have self-intersection number $-2$
  and intersect once with their neighbours. This Kleinian singularity
  is therefore dual to the Dynkin diagram of the Lie algebra
  $\mathfrak{su}(5)$, which is illustrated on the right in picture
  $(2)$. The dotted lines complete this diagram into
  the Dynkin diagram 
  of the extended Lie algebra $\widehat{\mathfrak{su}}(5)$. The labels
  are the dual Dynkin indices of the simple roots. }\label{fig:mckay}
\end{figure}

A hyper-K\"ahler resolution replaces the singularity at the origin
with a number of 2-spheres. The (oriented)
intersection product 
$$(S^2_i, S^2_j) \mapsto S^2_i \cup S^2_j$$
puts a lattice structure on
the second homology. This turns out to be
determined by the Cartan matrix of the corresponding ADE Lie algebra
$\lieg$, so that there is a bijection between a basis of 2-cycles and
a choice of simple roots. 
 $A_{k-1}$ singularities correspond to the Lie group $SU(k)$,
 $D_k$ singularities lead to $SO(2k)$ and $E_k$ ones
 are related to one of the exceptional Lie groups $E_6$, $E_7$ or $E_8$.

The Dynkin diagram of each Lie algebra is thus realized geometrically
in terms of intersections of 2-cyles in the resolution of the
corresponding Kleinian 
singularity. This is the famous \emph{geometric McKay 
correspondence} \cite{mckay,mckay-slodowy}. We will encounter its
string theoretic interpretation in the next chapter.

In this thesis we mainly consider $A_{k-1}$ surface
singularities, for which $\G=\Z_{k}$. Let us work out this example
in some more detail. A 
resolved $A_{k-1}$ singularity $M_k$ is defined by an equation of the form: 
\begin{align*}
W_{{k}} =   \prod_{i=1}^{k} (z- a_i) + u^2 + v^2 = 0, \qquad
\mbox{for}~z,\,u,\,v \in \C. 
\end{align*}
More precisely, this equation defines a family of $A_{k-1}$
spaces that are parametrized by $k$ complex numbers $a_i$. 
For any configuration with
$a_1 \neq \ldots \neq a_{k}$ the surface $M_k$ is smooth. 

The 4-manifold $M_k$ can be thought of $S^1 \times \R$-fibration over
the complex plane $\C$, where the fiber is defined by the equation
$u^2 + v^2 = \mu = - \prod_{i=1}^{k} (z- a_i)$  over a point $z \in
\C$. Notice, however, that the size $\mu$ of the circle $S^1$ becomes infinite
when $z \to  \infty$.  

Over each of the points $z=a_i$ the fiber circle vanishes. Hence, non-trivial
2-cycles $C_{ij}$ in the 4-manifold can be constructed as $S^1$-fibrations over
some line segment  $[a_i, a_j]$ in the $z$-plane. In fact, the second
homology of the 4-manifold $M_{k}$ is spanned by $k-1$ of these
two-spheres, say $C_{i (i+1)}$ for $1 \le i \le k-1$. This is
illustrated in in Fig. \ref{fig:mckay}. 

Since the $(k-1) \times (k-1)$ intersection
form on $H_2(M_{k})$ in this basis
\begin{align*}
 \bem -2 & 1  & 0 & \cdots  \\
1 & -2 & 1 & \cdots  \\
0 & 1 & -2 & \cdots \\
\vdots & \vdots& \vdots & \ddots  
\eem
\end{align*}
coincides with
the Cartan form 
of the Lie algebra $\mathfrak{su}(k)$ (up to an overall minus-sign),
the 2-cycles $C_{i  (i+1)}$ generate the root lattice of $A_{k-1}$.

%
%

\subsubsection{McKay-Nakajima correspondence}
\index{affine character}\index{WZW model}

Let us now study Vafa-Witten theory on these ALE spaces. 
First of all we have to address the fact that the 4-manifold $M_\G$
is non-compact, so that we have to fix boundary conditions for the
gauge field. The boundary at infinity is given by the Lens space
$S^3/\G$ and here the $U(N)$ gauge field should approach a flat
connection. Up to gauge equivalence this flat connection is labeled
by an $N$-dimensional representation of the quotient group $\Gamma$,
that is, an element
$$
\rho \in {\rm Hom}\left(\G,U(N)\right).
$$
If $\rho_i$ label the irreducible representations of $\Gamma$ (with
$\rho_0$ the trivial representation), then 
$\rho$ can be decomposed as
$$
\rho= \bigoplus_{i=0}^r N_i \rho_i,
$$
where the multiplicities $N_i$ are non-negative integers satisfying
the restriction
$$
\sum_{i=0}^r N_i d_i = N, \qquad d_i = \dim \rho_i.
$$
and $r$ is the rank of the gauge group. 
Now the classic \emph{McKay correspondence} (without the adjective
``geometric'') 
relates  the irreducible representations
$\rho_i$ of the finite subgroup $\G$ to the nodes of the
Dynkin diagram of the corresponding affine extension
$\widehat{\lieg}$, such that the
dimensions $d_i$ of these irreps can be identified with the dual
Dynkin indices (see Fig.~\ref{fig:mckay}). 

Furthermore, the non-negative integers $N_i$
label a dominant weight of the affine algebra $\widehat{\lieg}$ whose
level is equal to $N$. 
Through the McKay correspondence each $N$-dimensional representation
$\rho$ of $\G$ thus determines an integrable highest-weight representation
of $\widehat{\lieg}_N$ at level $N$. We will denote this
(infinite-dimensional) Lie algebra representation as $V_\rho$.
For $\Gamma=\Z_k$, which is the case that we will mostly
concentrate on, flat connections on $S^3/\Z_k$ get identified
with integrable representations of $\widehat{su}(k)_N$. In this particular
case all Dynkin indices satisfy $d_i=1$.

With $\rho$ labeling the boundary conditions of the gauge field at
infinity, we will get a vector-valued partition function
$Z_\rho(v,\tau)$. Formally the $U(N)$ gauge theory partition function on the
ALE manifold again has an expansion
$$
Z_\rho(v,\tau) = \sum_{n,m} d(m,n) y^m q^{h_\rho+n-c/24},
$$ 
where $c=N k$ with $k$ the regularized Euler number of the $A_{k-1}$
manifold \cite{Vafa:1994tf}.  The usual instanton numbers given by the
second Chern class $n=ch_2$ in the exponent are now shifted by a
rational number $h_{\rho}$, which is related to the Chern-Simons
invariant of the flat connection $\rho$. 
As we explain in \Cref{ssec-mckay}, $h_{\rho}$ gets mapped to
the conformal dimension 
of
the corresponding integrable weight in the affine Lie algebra
$\widehat{\lieg}$ related to $\Gamma$ by the McKay correspondence.
S-duality will act non-trivially on the boundary conditions $\rho$,
and therefore $Z_{\rho}(v,\tau)$ will be a vector-valued Jacobi form
\cite{Vafa:1994tf}.

For these ALE spaces the instanton computations can be explicitly
performed, because there exists a generalized ADHM construction in
which the instanton moduli space is represented as a quiver
variety. The physical intuition underlying this formalism has been
justified by the beautiful mathematical work of H.~Nakajima
\cite{nakajima,Nakajima:1995ka}, who has proved that on the middle
dimensional cohomology of the instanton moduli space one can actually
realize the action of the affine Kac-Moody algebra $\widehat{\mathfrak
  g}_N$ in terms of geometric operations. In fact, this work leads to
the identification
$$
Z_{\rho}(v,\tau) = \Tr_{\strut V_\rho}\!\left(y^{J_0} q^{L_0-c/24} \right) =
\chi_\rho(v,\tau),
$$ 
with $V_\rho$ the integrable highest-weight representation of
$\widehat{\lieg}_N$ and $\chi_\rho$ its affine character. Here $c$ is 
the appropriate central charge of the corresponding WZW model. A
remarkable fact is that, in the case of a $U(N)$ gauge theory on a
$\Z_k$ singularity, we find an action of $\widehat{su}(k)_N$ and not
of the gauge group $SU(N)$. 
 This is a important example of level-rank duality of affine Lie algebras. This setup has
been   studied from various perspectives in for instance
\cite{Bianchi:1996zj,Fucito:2004ry,Fucito:2006kn}.

Interestingly, I.~Frenkel has suggested
\cite{frenkel} that, if one works equivariantly for the action of the
gauge group $SU(N)$ at infinity (we ignore the $U(1)$ part for the
moment), there would similarly be an action of the $\widehat{su}(N)_k$
affine Lie algebra.  Physically this means ``ungauging'' the
$SU(N)$ at infinity.  In other words, we consider making the $SU(N)$ into a
global symmetry instead of a gauge symmetry at the boundary.  This
suggestion has recently been confirmed in \cite{licata}. So, depending
on how we deal with the theory at infinity, there are reasons to
expect both affine symmetry structures to appear and have a combined
action of the Lie algebra
$$
\widehat{su}(N)_K \times \widehat{su}(k)_N.
$$ 
We will now turn to a dual string theory realization, where this
structure indeed becomes transparent.






\section{Free fermion realization}\label{sec:stringrealization}

In this section we discover a string theoretic set-up to study the
correspondence between Vafa-Witten theory on ALE spaces and the
holomorphic part of a WZW model. We find that Vafa-Witten theory is
dual to a system of 
intersecting D4 and D6-branes on a torus $T^2$.

\subsection{Taub-NUT geometry}\label{sec:taubnut}
\index{Taub-NUT space}

To study Vafa-Witten theory on ALE spaces within string theory, we use a
trick that proved to be very effectively in relating 4d and 5d black
holes \cite{Gaiotto:2005gf, Gaiotto:2005xt, Shih:2005uc,
  Dijkgraaf:2006um, Bena:2006qm} and is in line with 
the duality between ALE spaces and 5-brane geometries \cite{Ooguri:1995wj}.
We will replace the local $A_{k-1}$
singularity with a Taub-NUT geometry. This can be best understood as
an $S^1$ compactification of the singularity. The $TN_k$ geometry is a
hyper-K\"ahler manifold with metric \cite{Ruback:1986ag, Sen:1997js},
$$
ds^2_{TN} = R^2 \left[ {1\over V}(d\chi
+ \alpha)^2 + V d{\vec x}^2 \right],
$$
with $\chi \in S^1$ (with period $4\pi$) and ${\vec x}\in \R^3$. Here
the function $V$ and 1-form $\alpha$ are determined as
$$ 
V(\vec{x}) = {1 + \sum_{a=1}^k {1\over |\vec{x} - {\vec x}_a|}}, \qquad
d \alpha = *_3 \,\, dV.
$$ 
Just like a local $A_{k-1}$ singularity, the Taub-NUT manifold may be
seen as a circle fibration 
$$
\begin{matrix}
S^1_{TN} & \to & TN_k \\
& &  \downarrow     \\[2mm]
& & \R^3
\end{matrix}
$$
where the size of the $S^1_{TN}$ shrinks at the points
$\vec{x}_1,\ldots,\vec{x}_k \in \R^3$, whose positions are the hyperk\"ahler
moduli of the space. The main difference with the (resolved) $A_{k-1}$
singularity is that the Taub-NUT fiber stays of finite size $R$ at
infinity. 

The total Taub-NUT manifold is perfectly
smooth. At infinity it approximates the cylinder $\R^3 \times S^1_{TN}$,
but is non-trivially fibered 
over the $S^2$ at infinity as a monopole bundle of charge (first
Chern class) $k$
$$
\int_{S^2} d\alpha = 2\pi k. 
$$
In the core, where we can ignore the constant $1$ that appears in the
expression for the potential $V(\vec{x})$, the Taub-NUT geometry can be
approximated by the (resolved) $A_{k-1}$ singularity.

The manifold $TN_k$ has non-trivial 2-cycles $C_{a,b} \cong S^2$ that
are fibered over the line segments joining the locations $\vec{x}_a$ and
$\vec{x}_b$ in $\R^3$. Only $k-1$ of these cycles are homologically
independent. As a basis we can pick the cycles
$$
C_a:=C_{a,a+1}, \qquad a=1,\ldots,k-1.
$$
The intersection matrix of these 2-cycles gives the Cartan matrix of
$A_{k-1}$.

From a dual perspective, there are $k$ independent normalizable
harmonic 2-forms $\w_a$ on $TN_k$, that can be chosen to be localized
around the centers or NUTs $\vec{x}_a$. With 
$$
V_a={1\over |\vec{x}-\vec{x}_a|}, \qquad
d\alpha_a=*dV_a,
$$
they are given as
$$
\w_a = d \eta_a,\qquad \eta_a = \alpha_a-{V_a \over V}(d\chi +\alpha).
$$
Furthermore, these 2-forms satisfy
$$
\int_{TN} \w_a \wedge \w_b = 16 \pi^2 \delta_{ab},
$$
and are dual to the cycles $C_{a,b}$
$$
\int_{C_{a,b}} \w_c = 4 \pi (\delta_{ac} - \delta_{bc}).
$$
A special role is played by the sum of these harmonic 2-forms
\be \w_{TN} = \sum_a \w_a.
\label{omega}
\ee
This is the unique normalizable harmonic 2-form that is invariant under
the tri-holomorphic $U(1)$ isometry of $TN$. The form $\w_{TN}$ has zero
pairings with all the cycles $C_{ab}$.  In the ``decompactification
limit'', where $TN_k$ gets replaced by $A_{k-1}$, the linear
combination $\w_{TN}$ becomes non-normalizable, while the $k-1$ two-forms
orthogonal to it survive.  

We will make convenient use of the following elegant interpretation of
the two-form $\w_{TN}$.  Consider the $U(1)$ action on the $TN_k$
manifold that rotates 
the $S^1_{TN}$ fiber. It is generated by a Killing vector field $\xi$. Let
$\eta_{TN}$ be the corresponding dual one-form given as
$
(\eta_{TN})_{\mu} = g_{\mu\nu} \xi^\nu,
$
where we used the $TN$-metric to convert the vector field to a
one-form. Up to an overall rescaling this gives
\be
\label{eta}
\eta_{TN} = {1\over V}(d\chi + \alpha).
\ee
In terms of this one-form, $\w_{TN}$ is given by
$\w_{TN} = d \eta_{TN}.$

\subsection{The D4-D6 system}  \label{ssec-realization}\index{I-brane}

Our strategy will be that, since we consider the twisted
partition function of the topological field theory, the answer will be
formally independent of the radius $R$ of the Taub-NUT geometry. So we
can take both the limit $R \to \infty$, where we recover the result
for the ALE space $\C^2/\Z_k$, and the limit $R \to 0$, where the
problem becomes essentially 3-dimensional.

Now, there are some subtleties with this argument, since a priori the
partition function of the gauge theory on the $TN$ manifold is
\emph{not} identical to that of the ALE space. In particular there are
new topological configurations of the gauge field that can
contribute. These can be thought of as monopoles going around the
$S^1$ at infinity. We will come back to this subtle point
later.\footnote{Recently, instantons on Taub-NUT spaces have been
  studied extensively in \cite{Cherkis:2008ip, Witten:2009xu,
    Cherkis:2009jm}. In particular, \cite{Witten:2009at} gives a closely related description of the duality between $\cN=4$ supersymmetric gauge theory on Taub-NUT space and WZW models from the perspective of an M5-brane wrapping $\R \times S^1 \times TN$. It is called a geometric Langlands duality for surfaces.}


In type IIA string theory, the partition function of the $\cN=4$ SYM
theory on the $TN_k$ manifold can be obtained by considering a
compactification of the form
$$
\hbox{(IIA)}\quad TN \times S^1 \times \R^5,
$$
and wrapping $N$ D4-branes on $TN \times S^1$. This is a special case of 
the situation presented in the box on the right-hand side in
Fig.~\ref{fig:webofdualities}, 
with $\Gamma=S^1$, $\cB_3=S^1\times\R^2$, and $S^1$ decompactified.
In the decoupling limit the partition function of this set of D-branes will
reproduce the Vafa-Witten partition function on $TN_k$. This partition
function can be also written as an index
$$
Z(v,\tau) = \Tr \left((-1)^F
e^{-\beta H}  e^{i n \theta}e^{2\pi i m v} \right) 
$$
where $\beta=2\pi R_9$ is the circumference of the ``9th dimension''
$S^1$, and $m=c_1$, $n=ch_2$ are the Chern characters of the gauge
bundle on the $TN_k$ space. Here we can think of the theta angle $\theta$
as the Wilson loop for the graviphoton field $C_1$ along the
$S^1$. Similarly $v$ is the Wilson loop for $C_3$. The gauge coupling
of the 4d gauge theory is now identified as
$$
{1\over g^2} = { \beta \over g_s \ell_s}.
$$ 
Because only BPS configurations contribute in this index, again only
the holomorphic combination $\tau$ (\ref{eqn:tau}) will appear.

We can now further lift this configuration to M-theory
with an additional $S^1$ of size  $R_{11} = g_s l_s$, where we
obtain the compactification
$$
\hbox{(M)}\quad TN \times T^2 \times \R^5,
$$ 
now with $N$ M5-branes wrapping the 6-manifold $TN_k \times T^2$. 
This corresponds to the top box in Fig.~\ref{fig:webofdualities}, with
$\Sigma=T^2$. 
As we remarked earlier, after this lift the coupling constant $\tau$ is
interpreted as the geometric modulus of the elliptic curve $T^2$.
In particular its imaginary part is given by the ratio $R_9/R_{11}$.
Dimensionally reducing the 6-dimensional $U(N)$ theory on the
M5-brane world-volume over the Taub-NUT space gives a 2-dimensional
$(0,8)$ superconformal field theory, in which the gauge theory
partition function is computed as the elliptic genus
$$
Z = \Tr \left((-1)^F y^{J_0} q^{L_0-c/24} \right).
$$

In order to further analyze this system we switch to yet another
duality frame by compactifying back to Type IIA theory, but now
along the $S^1$ fiber in the Taub-NUT geometry. This is the familiar
9-11 exchange. In this fashion we end up with a IIA compactification
on
$$
\hbox{(IIA)}\quad \R^3 \times T^2 \times \R^5,
$$
with $N$ D4-branes wrapping $\R^3 \times T^2$. However, because the
circle fibration of the $TN$ space has singular points, we have to
include D6-branes as well. In fact, there will be $k$ D6-branes that wrap $T^2
\times \R^5$ and are localized at the points
$\vec{x}_1,\ldots,\vec{x}_k$ in the $\R^3$. 
This situation is represented in the box on the left-hand side in Fig.~\ref{fig:webofdualities}.

\begin{figure}[h]
\begin{center}   \label{fig1}
\includegraphics[width=9cm]{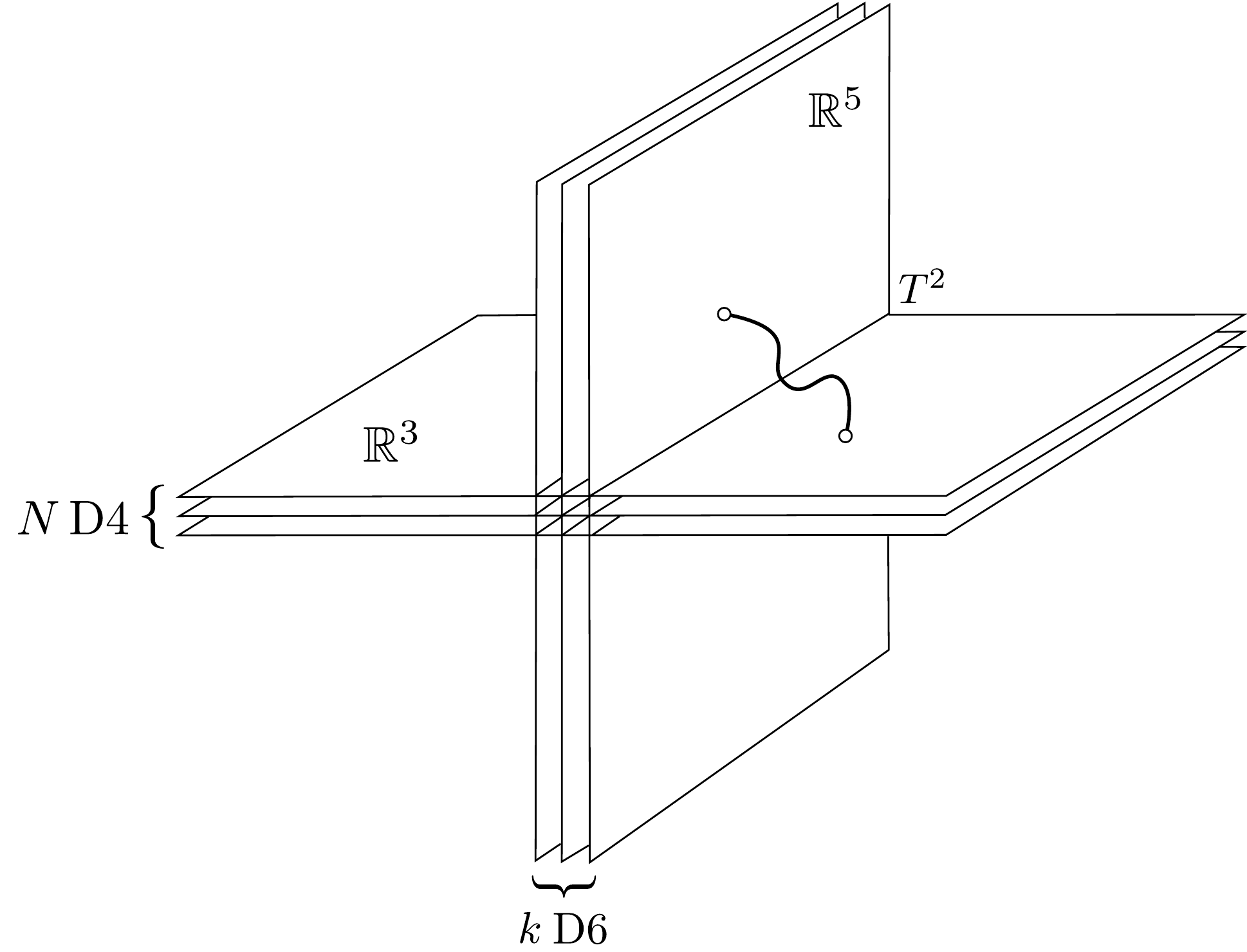}
\label{fig:I-brane}
\caption{Configuration of
  intersecting D4 and D6-branes with one of the 4-6 open strings that
  gives rise to a chiral fermion localized on the I-brane.}
\end{center} 
\end{figure}

Summarizing, we get a system of $N$ D4-branes and $k$ D6-branes intersecting
along the $T^2$. This intersection locus is called the I-brane. It
is pictured in Fig.~\ref{fig:I-brane}.  We will now study this I-brane
system in greater detail.

\subsection{Free fermions}  \label{ssec-D4D6}\index{free fermions}

A collection of D4-branes and D6-branes that intersect along two
(flat) dimensions is a supersymmetric configuration. One way to see
this is that after some T-dualities, it can be related to a D0-D8 or
D1-D9 system. The supersymmetry in this case is of type $(0,8)$. The
massless modes of the 4-6 open strings stretching between the D4 and
D6 branes reside entirely in the Ramond sector. All modes in the NS
sector are massive. These massless modes are well-known to be chiral
fermions on the 2-dimensional I-brane \cite{Green:1996dd,
  Bachas:1997kn, Hung:2006nn}. If we have
$N$ D4-branes and $k$ D6-branes, the chiral fermions
$$
\psi_{i,\overline a}(z),\,\, \psi^\dagger_{\overline\imath, a}(z), \qquad
i=1,\ldots,N,\ a=1,\ldots,k
$$ 
transform in the bifundamental representations $(N,\overline{k})$ and
$(\overline{N},k)$ of $U(N) \times U(k)$. Since we are computing an
index, we can take the $\alpha' \to 0$ limit, in which all massive
modes decouple. In this limit we are just 
left with the chiral fermions. Their action is necessarily free and given
by 
$$ S = \int d^2z\ \psi^\dagger \bar\partial_{A + \wt A}\psi,
$$ 
where $A$ and $\wt A$ are the restrictions to the I-brane $T^2$ of the
$U(N)$ and $U(k)$ gauge fields, that live on the worldvolumes of the
D4-branes and the D6-branes respectively. (Here we have absorbed the overall  
coupling constant). 

Under the two $U(1)$'s the fermions have charge $(+1,-1)$. Therefore
the overall (diagonal) $U(1)$ decouples and the fermions
effectively  couple to the gauge group
$$
U(1) \times SU(N) \times SU(k),
$$ 
where the remaining $U(1)$ is the anti-diagonal. At this point we ignore  
certain discrete identifications under the $\Z_N$ and $\Z_k$ centers,
that we will return to later.

\subsubsection{Zero modes}

A special role is played by the zero-modes of the D-brane gauge
fields. In the supersymmetric configuration we can have both a
non-trivial flat $U(N)$ and $U(k)$ gauge field turned on along the
$T^2$.  We will denote these moduli as $u_i$ and $v_a$
respectively.
The partition function of the chiral fermions on the I-brane will be a
function $Z(u,v,\tau)$ of both the flat connections $u,v$ and the
modulus $\tau$. It will transform as a (generalized) Jacobi-form under
the action of $SL(2,\Z)$ on the two-torus.

The couplings $u$ and $v$ have straightforward identifications in the
$\cN=4$ gauge theory on the $TN$ space. First of all, the parameters $u_i$
are Wilson loops along the circle of the $D4$ compactified on $TN
\times S^1$, and so in the 4-dimensional theory they just describe
the values of the scalar fields on the Higgs moduli space. That is,
they parametrize the positions $u_i$ of the $N$ D4-branes along the
$S^1$. Clearly, we are not interested in describing these kind of
configurations where the gauge group $U(N)$ gets broken to $U(1)^N$
(or some intermediate case). Therefore we will in general put
$u=0$. 

The parameters $v_a$ are the Wilson lines on the D6-branes and are
directly related to fluxes along the non-trivial two-cycles of $TN_k$
and (in the limiting case) on the $A_{k-1}$ geometry. To see this, let
us briefly review how the world-volume fields of the D6-branes are
related to the $TN$ geometry in the M-theory compactification. 

The positions of the NUTs $\vec{x}_a$ of the $TN$ manifold
are given by the vev's of the three scalar Higgs fields of the 6+1
dimensional gauge theory on the D6-brane.  In a similar fashion the
$U(1)$ gauge fields $\wt A_a$ on the D6-branes are obtained from the
3-form $C_3$~field in M-theory. More precisely, if $\w_a$ are the $k$
harmonic two-forms on $TN_k$ introduced in
\Cref{sec:taubnut}, we have a 
decomposition
\be
\label{C}
C_3 = \sum_a \w_a \wedge \wt A_a.
\ee
We recall that the forms $\w_a$ are localized around the centers
$\vec{x}_a$ of the $TN$ geometry (the fixed points of the circle
action). So in this fashion the bulk $C_3$~field gets replaced by $k$
$U(1)$~brane fields $\wt A_a$. This relation also holds for a single
D6-brane, because the two-form $\w$ is normalizable in the
$TN_1$ geometry. Relation (\ref{C}) holds in particular for a flat
connection, in which case we get the M-theory background
$$
C_3 = \sum_a v_a\, \w_a \wedge dz + c.c.
$$
Reducing this 3-form down to the type IIA configuration on $TN \times
S^1$ gives a mixture of NS $B$ fields and RR $C_3$ fields on the Taub-NUT 
geometry. Finally, in the $\cN=4$ gauge theory this translates (for an
instanton background) into a topological coupling
$$
\int v \wedge \Tr\, F_+  + \bar{v} \wedge \Tr\, F_-, 
$$
with $v$ the harmonic two-form
$$
v = \sum_a  v_a \w_a.
$$
The existence of this coupling can also be seen by recalling that
the M5-brane action contains the term $\int H \wedge C_3$. On the
manifold $M \times T^2$ the tensor field strength $H$ reduces as $H =
F_+\wedge d\bar{z} + F_- \wedge dz$ and similarly one has $C_3 = v
\wedge dz + \bar{v} \wedge d \bar{z}$, which gives the above
result. If one thinks of the gauge theory in terms of a D3-brane, the
couplings $v, \bar{v}$ are the fluxes of the complexified 2-form
combination $B_{RR} + \tau B_{NS}$.

\subsubsection{Chiral anomaly}\index{Chern-Simons coupling}
\index{chiral anomaly}

We should address another point: the chiral fermions on the I-brane
are obviously anomalous. Under a gauge transformation of, say, the
$U(N)$ gauge field
$$
\delta A = D\xi,
$$
the effective action of the fermions transforms as
$$
k \int_{T^2}\Tr(\xi F_A).
$$ 
A similar story holds for the $U(k)$ gauge symmetry. Nonetheless, the
overall theory including both the chiral fermions on the I-brane and
the gauge fields in the bulk of the D-branes is consistent, due to the
coupling between both systems. The consistency is ensured by
Chern-Simons terms (\ref{eq:CSterms}) in the D-brane actions, which
cancel the anomaly 
through the process of anomaly inflow \cite{Green:1996dd, Itzhaki:2005tu}. For
example, on the D4-brane there is a term coupling to the RR 2-form
(graviphoton) field strength $G_2$:
\be
\label{CS}
I_{CS} = {1 \over 2 \pi} \int_{T^2 \times \R^3} G_2 \wedge CS(A),
\ee
with Chern-Simons term 
$$
CS(A) = \Tr\big(AdA + {2\over 3} A\wedge A \wedge A \big).
$$
Because of the presence of the D6-branes, the 2-form $G_2$ is no
longer closed, but satisfies instead
$$
dG_2 = 2 \pi k \cdot \delta_{T^2}.
$$ 
Therefore under a gauge transformation $\delta A = D\xi$ the D4-brane
action gives the required compensating term
$$
\delta I_{CS} = {1 \over 2 \pi} \int G_2 \wedge d\, \Tr(\xi F_A) =  -k \int_{T^2} \ \Tr(\xi F_A),
$$
which makes the whole system gauge invariant.

\section{Nakajima-Vafa-Witten correspondence}\label{sec:NVW}

So far we have obtained a configuration of $N$ D4-branes and $k$
D6-branes that intersect transversely along a 2-torus. Moreover, the
massless modes of the 4-6 open strings combine into $Nk$ free fermions
on this 2-torus. This already relates $SU(N)$ Vafa-Witten theory on an
$A_k$-singularity to a 2-dimensional conformal field theory of free
fermions. However, the I-brane system contains more
information than just the Vafa-Witten partition function. In this
section we analyze the I-brane system and extract the
Nakajima-Vafa-Witten correspondence 
from the I-brane partition function.


\subsection{Conformal embeddings and level-rank
  duality}\index{level-rank duality}

The system of intersecting branes gives an elegant
realization of the level-rank duality
$$
\widehat{su}(N)_k \ \leftrightarrow \ \widehat{su}(k)_N
$$ 
that is well-known in CFT and 3d topological field theory. The
analysis has been conducted in \cite{Itzhaki:2005tu} for a system of D5-D5
branes, which is of course T-dual to the D4-D6 system that we consider
here. Hence we can follow this analysis to a large extent.

The system of $Nk$ free fermions has central charge $c=Nk$ and gives a
realization of the $\widehat{u}(Nk)_1$ affine symmetry at level
one. In terms of affine Kac-Moody Lie groups we have the embedding
\be
\label{conf}
\widehat{u}(1)_{Nk} \times \widehat{su}(N)_k \times \widehat{su}(k)_N \subset \widehat{u}(Nk)_1.
\ee
This is a conformal embedding, in the sense that the central charges
of the WZW models on both sides are equal. Indeed, using that the
central charge of $\widehat{su}(N)_k$ is
$$
c_{N,k} = \frac{k(N^2-1)}{k+N},
$$
it is easily checked that
$$
1 + c_{N,k} + c_{k,N} = Nk.
$$ 
The generators for these commuting subalgebras are bilinears
constructed out of the fermions $\psi_{i,a}$ and their conjugates
$\psi_{i,a}^\dagger$.  In terms of these fields one can define the
currents of the $\widehat{u}(N)_k$ and $\widehat{u}(k)_N$ subalgebras
as respectively
$$
J_{j \overline k}(z) = \sum_a \psi_{j}{}^a \psi^\dagger_{\overline k a}, 
$$
and
$$
J_{\overline a b}(z) = \sum_j \psi_{j,\overline a} \psi^\dagger_{b}{}^j.
$$

Now it is exactly the conformal embedding (\ref{conf}) that gives the
most elegant explanation of level-rank duality. This correspondence
should be considered as the affine version of the well-known
Schur-Weyl duality for finite-dimensional Lie groups.  Let us recall
that the latter is obtained by considering the (commuting)
actions of the unitary group and symmetric group
$$
U(N) \times S_k \subset U(Nk)
$$ 
on the vector space $\C^{Nk}$, regarded as the $k$-th tensor product
of the fundamental representation $\C^N$.  Schur-Weyl duality is
the statement that the corresponding  group algebras are maximally
commuting in ${\rm End}\left((\C^N)^{\otimes k}\right)$, in the sense
that the two algebras are each other's commutants. 
Under these actions one obtains the decomposition
$$
\C^{Nk} = \bigoplus_\rho V_\rho\otimes \widetilde{V}_\rho,
$$
with $V_\rho$ and $\widetilde{V}_\rho$ irreducible representations of
$u(N)$ and $S_k$ respectively. Here $\rho$ runs over all partitions of
$k$ with at most $N$ parts. This duality gives the famous pairing
between the representation theory of the unitary group and the
symmetric group.

In the affine case we have a similar situation, where we now take the
$k$th tensor product of the $N$ free fermion Fock spaces, viewed as
the fundamental representation of $\widehat{u}(N)_1$. The symmetric
group $S_k$ gets replaced by $\widehat{u}(k)_N$ (which reminds one of
constructions in D-branes and matrix string theory, where the symmetry
group appears as the Weyl group of a non-Abelian symmetry). The affine
Lie algebras
$$
\widehat{u}(1)_{Nk} \times \widehat{su}(N)_k \times \widehat{su}(k)_N 
$$
again have the property that they form maximally commuting subalgebras
within $\widehat{u}(Nk)_1$. The total Fock space $\cF^{\otimes Nk}$ of $Nk$ free
fermions now decomposes under the embedding (\ref{conf}) as
\be
\label{decom}
\cF^{\otimes Nk} = \bigoplus_{\rho} U_{\|\rho\|} \otimes V_{\rho} \otimes
{\wt V}_{\wt\rho}.
\ee
Here $U_{\|\rho\|}$, $V_{\rho}$ and ${\wt V}_{\wt\rho}$ denote irreducible
integrable representations of $\widehat{u}(1)_{Nk}$, $\widehat{su}(k)_N$, 
and $\widehat{su}(N)_k$ respectively.

The precise formula for the decomposition (\ref{decom}) is a bit
complicated, in particular due to the role of the overall $U(1)$
symmetry, and is given in detail in \Cref{app-decompose}.  But
roughly it can be understood as follows: the irreducible
representations of $\widehat{u}(N)_k$ are given by Young diagrams that
fit into a box of size $N \times k$. Similarly, the representations of
$\widehat{u}(k)_N$ fit in a reflected box of size $k \times N$. In
this fashion level-rank duality relates a representation $V_\rho$ of
$\widehat{u}(N)_k$ to the representation ${\wt V}_{\wt\rho}$ of
$\widehat{u}(k)_N$ labeled by the transposed Young diagram. If we
factor out the $\widehat{u}(1)_{Nk}$ action, we get a representation
of charge $\|\rho\|$, which is related to the total number of boxes
$|\rho|$ in $\rho$ (or equivalently $\wt\rho$).

At the level of the partition function we have a similar decomposition
into characters. To write this in more generality it is useful to add
the Cartan generators. That is, we consider the characters for
$\widehat{u}(N)_k$ that are given by
$$ 
\chi_\rho^{\widehat{u}(N)_k} (u,\tau) = \Tr_{\strut V_\rho} \left( e^{2\pi i u_j
  J_0^j} q^{L_0-c_{N,k}/24} \right),
$$
and similarly for $\widehat{u}(k)_N$ we have
$$
\chi_{\wt\rho}^{\widehat{u}(k)_N} (v,\tau) = \Tr_{\strut \wt V_{\wt\rho}} 
\left( e^{2\pi i v_a J_0^a} q^{L_0-c_{k,N}/24} \right).
$$ 
Here the diagonal currents
$$
J_0^j = \oint \frac{dz}{2\pi i} J_{jj}(z),\qquad  
J_0^a = \oint \frac{dz}{2\pi i} J_{aa}(z)
$$ 
generate the Cartan tori $U(1)^N \subset U(N)$ and $U(1)^k \subset U(k)$.

Including the Wilson lines $u$ and $v$ for the $U(N)$ and $U(k)$ gauge
fields, the partition function of the I-brane system is given by the
character of the fermion Fock space 
\begin{align}
&Z_I(u,v,\tau) = 
\Tr_{\strut \mathcal{F}}\Big(e^{2\pi i (u_jJ^j_0 + v_a J^a_0)}\, q^{L_0 - \frac{Nk}{24}}
\Big) \\
&= q^{-\frac{Nk}{24}}\prod_{\scriptstyle j=1, \ldots,N \atop 
\scriptstyle a=1,\ldots,k} \prod_{n \geq 0}\Big(1+e^{2\pi i (u_j+v_a)}
q^{n+1/2} \Big)  
\Big(1+e^{- 2\pi i (u_j+v_a)} q^{n+1/2} \Big). \nonumber
\end{align}
Writing the decomposition (\ref{decom}) in terms of characters
gives 
\begin{align*}
&Z_I(u,v,\tau) \\ &= \sum_{[\rho]\subset
  \mathcal{Y}_{N-1,k}} \sum_{j=0}^{N-1} \sum_{a=0}^{k-1}
\chi^{\widehat{u}(1)_{Nk}}_{|\rho|+jk+aN}(N |u| + k
|v|,\tau)\ \chi^{\widehat{su}(N)_k}_{\sigma^j_N(\rho)} (\overline{u},\tau)
\chi^{\widehat{su}(k)_N}_{\sigma^a_k(\wt \rho)}
(\overline{v},\tau),
\end{align*}
where the Young diagrams $\rho\in \cY_{N-1,k}$ of size $(N-1)\times k$
represent $\widehat{su}(N)_k$ integrable representations and $\sigma$
denote generators of the outer automorphism groups $\Z_N$ and $\Z_k$
that connect the centers of $SU(N)$ and $SU(k)$ to the $U(1)$
factor (see again \Cref{app-decompose} for notation and more details).

\subsection{Deriving the McKay-Nakajima
  correspondence}   \label{ssec-mckay}\index{McKay correspondence}

In the intersecting D-brane configuration both the D4-branes and the
D6-branes are non-compact. So, we can choose both the $U(N)$ and
$U(k)$ gauge groups to be non-dynamical and freeze the background
gauge fields $A$ and $\wt A$. In fact, this set-up is entirely
symmetric between the two gauge systems, which makes level-rank
duality transparent. 

However, in order to make contact with the $\cN=4$ gauge theory
computation, we will have to break this symmetry. Clearly, we want the
$U(N)$ gauge field to be dynamical --- our starting point was to
compute the partition function of the $U(N)$ Yang-Mills theory. The
$U(k)$ symmetry should however {\it not} be dynamical, since we want
to freeze the geometry of the Taub-NUT manifold. So, to derive the
gauge theory result, we will have to integrate out the $U(N)$ gauge
field $A$ on the I-brane. Particular attention has to be payed to the
$U(1)$ factor in the CFT on the I-brane. We will argue that in this
string theory set-up we should not take that to be dynamical.

Therefore we are dealing with a partially gauged CFT or coset theory
$$
\widehat{u}(Nk)_1/\widehat{su}(N)_k.
$$
In particular the $\widehat{su}(N)_k$ WZW model will be replaced by the
corresponding $G/G$ model. Gauging the model will reduce the
characters. (Note that this only makes sense if the Coulomb parameters
$u$ are set to zero. If not, we can only gauge the residual gauge
symmetry, which leads to fractionalization and a product structure.)
In the gauged WZW model, which is a topological field theory, only the
ground state remains in each irreducible integrable representation. So
we have a reduction
$$
\chi^{\widehat{su}(N)_k}_\rho(\overline{u},\tau)\ \to\ q^{h_{\rho}-c/24},
$$ 
with $h_\rho$ the conformal dimension of the ground state
representation $\rho$. Note that the choice of $\rho$ corresponds
exactly to the boundary condition for the gauge theory on the
$A_{k-1}$ manifold. We will explain this fact, that is crucial to the
McKay correspondence, in a moment.

Gauging the full I-brane theory and restricting to
the sector $\rho$ finally gives
$$
Z_I(u,v,\tau)\ \to\ Z^{N,k}_\rho(v,\tau) = q^{h_{\rho}-c/24}
\sum_{a=0}^{k-1}
\chi^{\widehat{u}(1)_{Nk}}_{|\rho|+aN}(k|v|,\tau)\ 
\chi^{\widehat{su}(k)_N}_{\sigma^a_k(\wt\rho)}(\overline{v},\tau).
$$ 
Up to the $\chi^{\widehat{u}(1)_{Nk}}$ factor, this reproduces the
results presented in \cite{Vafa:1994tf, nakajima} for ALE spaces, which
involve just $\widehat{su}(k)_N$ characters. This extra factor is
is due to additional monopoles mentioned in 
\Cref{ssec-realization}. They are related to the finite radius $S^1$
at infinity of the Taub-NUT space and are absent in case of ALE
geometries. 

In fact, the extra $U(1)$ factor can already be seen at the classical level,
because the extra normalizable harmonic two-form $\omega$ in
(\ref{omega}) disappears in the decompactification limit where
$TN_k$ degenerates into $A_{k-1}$. The lattice $H^2(TN_k,\Z)$ is isomorphic 
to $\Z^k$ with the standard inner product and contains the root
lattice $A_{k-1}$ as a sublattice given by $\sum_I n_I=0$. Note also
that the lattice $\Z^k$ is not even, which explains why the I-brane
partition function has a fermionic character and only transforms under
a subgroup of $SL(2,\Z)$ that leaves invariant the spin structure on
$T^2$.

\subsubsection{Relating the boundary conditions}

By relating the original 4-dimensional gauge theory to the
intersecting brane picture one can in fact derive the McKay
correspondence directly. Moreover we can understand the appearance of
characters of the WZW models (for both the $SU(N)$ and the $SU(k)$
symmetry) in a more natural way in this set-up.  Recall that the
$SU(N)$ gauge theory on the $A_{k-1}$ singularity or $TN_k$ manifold is
specified by a boundary state. This state is given by picking a flat
connection on the boundary that is topologically $S^3/\Z_k$.  If we
think of this system in radial quantization near the boundary, where
we consider a wave function for the time evolution along
$$
S^3/\Z_k \times \R,
$$ we have a Hilbert space with one state $|\rho\rangle$ for each
$N$-dimensional representation
$$
\rho:\ \Z_k \to U(N).
$$ 

After the duality to the I-brane system, we are dealing with a
5-dimensional $SU(N)$ gauge theory on $\R^3 \times T^2$, with 
$k$ D6-branes intersecting it along $\{p\}\times T^2$ where $p$ is
(say) the origin of $\R^3$. Here the boundary of the D4-brane system
is $S^2 \times T^2$.  In other words, near the boundary the space-time
geometry looks like $\R\times S^2\times T^2$. We now ask ourselves what
specifies the boundary states for this theory.  Since we need a finite
energy condition, this is equivalent to considering the IR limit of the
theory.
In M-theory the $S^1$-bundle over 
$S^2$ carries a first Chern class $k$, which translates into the flux
of the graviphoton field strength 
$$
\int_{S^2} G_2 = 2\pi k.
$$
Therefore the term 
$$
\int_{S^2\times T^2\times \R} G_2\wedge CS(A),
$$
living on the D4 brane, leads upon reduction on $S^2$ (as is done in
\cite{Itzhaki:2005tu}) to the term
$$
I_{CS} = 2\pi k \int_{T^2 \times \R}CS(A).
$$ 
Hence we have learned that the boundary condition for the D4-brane
requires specifying a state of the $SU(N)$ Chern-Simons theory at
level $k$ living on $T^2$.  The Hilbert space for Chern-Simons theory
on $T^2$ is well-known to have a state for each integrable
representation of the $\widehat{u}(N)_k$ WZW model, which up to the
level-rank duality described in the previous section, gives the McKay
correspondence.  

In fact, the full level-rank duality can be brought to
life. Just as we discussed for the $N$ D4-branes, a $SU(k)$ gauge theory 
lives on the $k$ D6-branes on $T^2\times \R^5$.  The boundary of
the space is $S^4\times T^2$. Furthermore, taking into account that the $N$
D4-branes source the $G_4$ RR flux through $S^4$, we get, as in the
above, a $SU(k)$ Chern-Simons theory at level $N$ living on
$T^2\times \R$.  Therefore the boundary condition should be specified by a
state in the Hilbert space of the $SU(k)$ Chern-Simons theory on
$T^2$.  So we see three distinct ways to specify the boundary
conditions: as a representation of $\Z_k$ in $SU(N)$, as a character
of $SU(N)$ at level $k$, and as a character of $SU(k)$ at level $N$.
Thus we have learned that, quite independently of the fermionic realization,
there should be an equivalence between these objects.

To make the map more clearly we could try to show that the choice of the
flat connection of the $SU(N)$ theory on $S^3/ \Z_k$ gets mapped to
the characters that we have discussed in the dual intersecting brane
picture.  To accomplish this, recall that the original $SU(N)$ action
on the $A_{k-1}$ space leads to a boundary term (modulo an integer
multiple of $2\pi i \tau$) given by the Chern-Simons invariant
$$
 {\tau\over 4\pi i}\int_{A_{k-1}} \Tr\,F \wedge F =
{\tau\over 4\pi i} \int_{S^3/ \Z_k} CS(A).
$$ 
Restricting to a particular flat connection on $S^3/ \Z_k$ yields
the value of the classical Chern-Simons action.

If we show that
$$
S(\rho) = {1\over 8\pi^2} \int_{S^3/ \Z_k} CS(A)
$$ 
for the flat connection $\rho$ on $S^3/\Z_k$ gets mapped to the
conformal dimension $h_\rho$ of the corresponding state of the quantum
Chern-Simons theory on $T^2$, we would have completed a direct check of the map, 
because the gauge coupling constant $\tau$ above is nothing but the 
modulus of the torus in the dual description.

To see how this works, let us first consider the abelian case of
$N=1$. In that case the flat connection $\rho$ is given by a phase
$e^{2\pi i n/k}$ with $n \in \Z/k\Z$. The corresponding CS term gives
$$
S^{U(1)}(\rho)= {n^2 \over 2k}.
$$
This is the conformal dimension of a primary state of the
$U(1)$ WZW model at level $k$. 

A general $U(N)$ connection can always be diagonalized to $U(1)^N$,
which therefore gives integers $n_1,\ldots,n_N \in \Z/k\Z$. The
Chern-Simons action is therefore given by
$$
S^{U(N)}(\rho) = \sum_{i=1}^N {n_i^2 \over 2k}.
$$ 
On the other hand, a conformal dimension of a primary state in the
corresponding WZW model is given by
$$
h_{\rho} = \frac{C_2(\rho)}{2(k+N)},
$$
where $\rho$ is an irreducible integrable $\widehat{u}(N)_k$
weight. Such a weight can be encoded in a Young diagram with at most
$N$ rows of lengths $R_i$. There is a natural change of basis $n_i =
R_i + \rho^{\textrm{Weyl}}_i$ where we shift by the Weyl vector
$\rho^{\textrm{Weyl}}$. If we decompose $U(N)$ into $SU(N)$ and $U(1)$, the
basis $n_i$ cannot be longer than $k$, which relates to the condition $n_i
\in \Z_k$ on the Chern-Simons side. In this basis the second Casimir $C_2$ takes a
simple form. Therefore the conformal dimension becomes
$$
h_{\rho} = -\frac{N(N^2-1)}{24(k+N)} + \frac{1}{2(k+N)} \sum_{i=1}^N n_i^2.
$$
The constant term combines nicely with the central charge contribution
$-c_{N,k}/24$ to give an overall constant $(N^2-1)/24$. Apart from
this term we see that $h_{\rho}$ indeed matches the expression for
$S^{U(N)}(\rho)$ given above, up to the usual quantum shift $k \to
k+N$.

According to the McKay correspondence one might expect to find a
relation between representations of $\Z_k$ and $\widehat{u}(k)_N$
integrable weights. Instead, we have just shown how $\widehat{u}(N)_k$
weights $\rho$ arise. Nonetheless, one can relate integrable weights
of those algebras by a transposition of the corresponding Young
diagrams. Then the conformal dimensions of $\widehat{u}(k)_N$ weights
$\wt\rho$ are determined by the relation \cite{Naculich:1990pa}
$$
h_{\rho}+h_{\wt\rho} = \frac{|\rho|}{2} - \frac{|\rho|^2}{2Nk},
$$
which is a consequence of the level-rank duality described in
\Cref{app-decompose}.  The above chain of arguments connects $\Z_k$ 
representations and $\widehat{u}(k)_N$ integrable weights, thereby
realizing the McKay correspondence.

\subsection{Orientifolds and $SO/Sp$ gauge groups}

In this chapter we have considered a system of $N$ D4-branes and $k$
D6-branes intersecting along a torus, whose low energy theory is
described by $U(N)$ and $U(k)$ gauge theories on each stack of branes,
together with bifundamental fermions. We can reduce this system to
orthogonal or symplectic gauge groups in a standard way by adding an
orientifold plane. This construction can also be lifted to
M-theory. Let us recall that D6-branes in our system originated from
a Taub-NUT solution in M-theory. The O6-orientifold can also be
understood from M-theory perspective, and it corresponds to the
Atiyah-Hitchin space \cite{Sen:1997kz}. Combining both ingredients, it is
possible to construct the M-theory background for a collection of
D6-branes with an O6-plane. The details of this construction are
explained in \cite{Sen:1997kz}.

Let us see what are the consequences of introducing the orientifold
into our I-brane system. We start with a stack of $k$ D6-branes. To
get orthogonal or symplectic gauge groups one should add an
orientifold O6-plane parallel to D6-branes \cite{Gimon:1996rq}, which
induces an orientifold projection $\Omega$ which acts on the
Chan-Paton factors via a matrix $\gamma_{\Omega}$. Let us recall there
are in fact two species O6${}^{\pm}$ of such an orientifold. As the
$\Omega$ must square to identity, this requires 
$$
\gamma_{\Omega}^t = \pm \gamma_{\Omega},
$$ 
with the $\pm$ sign corresponding to O6$^{\pm}$-plane, which gives
respectively $SO(k)$ and $Sp(2k)$ gauge group. In the former case $k$
can be even or odd; $k$ odd requires having \emph{half-branes}, fixed
to the orientifold plane (as explained {\it e.g.}~in
\cite{Evans:1997hk}).

Let us add now $N$ D4-branes intersecting D6 along two directions. The
presence of O6${}^{\pm}$-plane induces appropriate reduction of the D4
gauge group as well. The easiest way to argue what gauge group arises
is as follows. We can perform a T-duality along three directions to get a
system of D1-D9-branes, now with a spacetime-filling O9-plane. This is
analogous to the D5-D9-09 system in \cite{Gimon:1996rq}, in which case
the gauge groups on both stacks of branes must be different (either
orthogonal on D5-branes and symplectic on D9-branes, or the other
way round); the derivation of this fact is a consequence of having 4
possible mixed Neumann-Dirichlet boundary conditions for open strings
stretched between branes. On the contrary, for D1-D9-O9 system
there is twice as many possible mixed boundary conditions, which in
consequence leads to the same gauge group on both stack of branes. By
T-duality we also expect to get the same gauge groups in D4-D6 system
under orientifold projection. 

Let us explain now that the appearance of the same type of gauge
groups is consistent with character decompositions resulting from
consistent conformal embeddings or the existence of the so-called dual
pairs of affine Lie algebras related to systems of free fermions.  We
have already come across one such consistent embedding in (\ref{conf})
for $\widehat{u}(Nk)_1$. A dual pair of affine algebras in this case
is $(\widehat{su}(N)_k,\, \widehat{su}(k)_N)$. These two algebras are
related by the level-rank duality discussed in \Cref{app-decompose}.
As proved in \cite{jimbo-miwa,Hasegawa}, all 
other consistent dual pairs are necessarily of one of the following
forms
$$ (\widehat{sp}(2N)_k ,\, \widehat{sp}(2k)_N), $$
$$ (\widehat{so}(2N+1)_{2k+1} ,\, \widehat{so}(2k+1)_{2N+1}), $$
$$ (\widehat{so}(2N)_{2k+1} ,\, \widehat{so}(2k+1)_{2N}),  $$
$$ (\widehat{so}(2N)_{2k} ,\, \widehat{so}(2k)_{2N}). $$ 
Corresponding expressions in terms of characters, analogous to
(\ref{uNk-decompose}), are also given in \cite{Hasegawa}. The crucial
point is that both elements of those pair involve algebras of the same
type, which confirms and agrees with the string theoretic orientifold
analysis above.

Finally we wish to stress that the appearance of $U$, $Sp$ and $SO$
gauge groups which we considered so far in this paper is related to
the fact that their respective affine Lie algebras can be realized in
terms of free fermions, which arise on the I-brane from our
perspective.  It turns out there are other Lie groups $G$ whose affine
algebras have free fermion realization.  There is a finite number of
them, and fermionic realizations can be found only if there exists a
symmetric space of the form $G'/G$ for some other group $G'$
\cite{Goddard:1985jp}. It is an interesting question whether I-brane
configurations can be engineered in string theory that
support fermions realizing all these affine algebras.

From a geometric point of view we can remark the following. For ALE
singularities of $A$-type and $D$-type a non-compact dimension can be
compactified on a $S^1$ to give Taub-NUT geometries. For exceptional
groups such manifolds do not exist. But one can compactify \emph{two}
directions on a $T^2$ to give an elliptic fibration. In this setting
exotic singularities can appear as well. Such construction have a direct
analogue in type IIB string theory where they correspond to a
collection of $(p,q)$ 7-branes
\cite{Morrison:1996na, Morrison:1996pp}. The I-brane is
now generalized to the intersection of $N$ D3-branes with this
non-abelian 7-brane configuration \cite{Harvey:2007ab}. However, there
is in general 
no regime where all the 7-branes are weakly coupled, so it is not
straightforward to write down the I-brane system.



\chapter{Topological Strings, Free Fermions and Gauge Theory}\label{chapter3+4}

In Chapter~\ref{chapter2}  we found a stringy explanation for the appearance of
CFT characters in $\cN=4$ supersymmetric gauge theories. We discovered
that these characters emerge from a free fermion system living on
a torus $T^2$. 

Also in $\cN=2$ supersymmetric gauge theories important exact
quantities have turned out to be expressible in terms of an effective
Riemann surface or complex curve $\Sigma$. In this chapter we find
that the I-brane configuration can be generalized to this $\cN=2$
setting by replacing the torus $T^2$ with the more general
2-dimensional topology $\Sigma$. Moreover, we find an 
extension of the duality 
chain to the complete web of dualities in
Fig.~\ref{fig:webofdualities}. Interestingly, this makes it possible
to compare local Calabi-Yau compactifications with intersecting brane
configurations. This sheds new light on the presence of free
fermions in those theories.

The theme of this chapter is the web of dualities in
Fig.~\ref{fig:webofdualities}. The three keywords ``topological
strings'', ``free fermions'' and ``gauge theory'' refer to the three
cornerstones of the duality web. In all of these frames a holomorphic curve
$\Sigma$ plays a central role. The duality web relates the curves 
in all three settings, thereby giving a more fundamental understanding of the
appearance of curves in $\cN=2$ theories.  
The goal of this chapter is to introduce the three corners of
the web and their relations. This yields a fruitful dual perspective on $\cN=2$
supersymmetric gauge theory as well as topological string theory. 

An instructive example of an $\cN=2$ supersymmetric gauge theory is
the celebrated Seiberg-Witten theory. In \Cref{sec:SWcurves} we 
summarize how the low energy behaviour of $SU(N)$
supersymmetric Yang-Mills is encoded in a Riemann surface
$\Sigma_{SW}$ of genus $N-1$, which is widely known as the
\emph{Seiberg-Witten curve}.     
In \Cref{sec:firstdualitychain} we show that it is dual to an I-brane
configuration of D4 and D6 branes that intersect at the Seiberg-Witten curve
$\Sigma_{SW}$. This set-up is easily generalized to more general
$\cN=2$ gauge theories. 

In \Cref{sec:ge} we study non-compact Calabi-Yau threefolds that are
modeled on a Riemann surface. Such 6-dimensional backgrounds
\emph{geometrically engineer} a supersymmetric gauge theory in the
four transverse dimensions. In \Cref{dualitysequenceII} we 
relate the I-brane configuration to such Calabi-Yau compactifications
in a second chain of dualities.      

This far we haven't discussed topological invariants in these duality
frames. This is subject of \Cref{sec:CYcomp}. We review the most
relevant aspects of Calabi-Yau compactifications and the way
topological string theory enters. 
In \Cref{sec:BPSstatesfreefermions} we introduce the several types of
topological invariants that the topological string captures, and we
show how they enter the web of dualities. Moreover, we discuss the
relation of these invariants to the free fermions on the I-brane. 
As an application we write down a partition function that counts bound
states of D0-D2-D4 branes on a D6 brane and argue that this computes
the I-brane  
partition function.     

Let us emphasize that novel results
in this chapter may be found in 
\Cref{sec:firstdualitychain}, \Cref{dualitysequenceII},
\Cref{sec:Ibranepartfunc} and \Cref{fermionsBPS}.


\section{Curves in $\cN=2$ gauge theories}\label{sec:SWcurves}   
\index{Seiberg-Witten theory}\index{$\cN=2$ gauge theory}

$\cN=2$ supersymmetric gauge theories (unlike their $\cN=4$ relatives)
are 
sensitive to quantum corrections and thus not conformally invariant. In
particular, the $SU(N)$ theory is 
\emph{asymptotically free}: its complex gauge coupling constant $\tau$
depends on the energy scale $\mu$ such that $g_{YM}(\mu)$ decreases at high
energies. This dependence can be argued to be of the form \cite{Seiberg:1988ur} 
\begin{align}\label{eqn:efftau}\index{coupling constant $\tau$}
\tau_{\textrm{eff}}(\mu) = \tau_{\textrm{clas}} + \frac{i}{\pi} \log \frac{\mu^2}{\Lambda^2} + \sum_{k=1}^{\infty} c_k \left( \frac{\Lambda}{\mu} \right)^{4k}   
\end{align}
for some to be determined constants $c_k$, where $\Lambda$ is the
scale at which the gauge coupling becomes strong. The second term on
the right-hand side is the only perturbative contribution, which
follows from a one-loop computation, and the third term captures all
possible instanton contributions. 

Surprisingly, N.~Seiberg and E.~Witten discovered that an elegant
geometrical story is hidden behind the coefficients $c_k$
\cite{Seiberg:1994rs}.  They realized that many properties of
$\cN=2$ supersymmetric gauge theories have a geometrical
interpretation in terms of an auxiliary Riemann surface, which is now
called the Seiberg-Witten curve.
One of the successes of string theory is the physical embedding of the
Seiberg-Witten curve in a 10- or 11-dimensional geometry. This has
deepened the insight in supersymmetric gauge theories considerably. 

In this first section of this chapter we explain how the
Seiberg-Witten curve comes 
about, and which information it holds about the underlying gauge
theory. Moreover, we explain its embedding in string theory as the
rightmost diagram in the web of dualities in
Fig.~\ref{fig:webofdualities}. All these preliminaries are needed to
get to the main result of this section: the duality of
$\cN=2$ supersymmetric gauge theories with intersecting brane
configurations of D4 and D6-branes wrapping the gauge theory curve
$\Sigma$.

\subsection{Low energy effective description}\label{sec:clasmodspace}

Let us start with the basics. Since  $\cN=2$ super Yang-Mills on
$\R^4$  is a
reduction of $\cN=1$ super Yang-Mills in six dimensions, it follows
immediately that its field content consists of a gauge field
$A_{\mu}$, a complex scalar field $\phi$ and two Weyl spinors
$\la_{\pm}$. The last three fields transform in the adjoint
representation of the gauge group.  
The bosonic part of the action follows likewise from this reduction
%
\begin{align}\label{eqn:n=2action}
 \cL = -\frac{1}{e^2} \Tr \left( F \wedge *F + 2 D \phi \wedge
 * D \phi^{\dag}  + [\phi, \phi^{\dag}]^2 \right), 
\end{align}
where we could have added the topological term $\frac{i\theta}{8 \pi^2} \Tr (F \wedge
F)$. Supersymmetric vacua are therefore found as solutions of  
\begin{align*}
 V(\phi) = \Tr [\phi, \phi^{\dag}]^2 = 0,
\end{align*}
i.e. $\phi$ and $\phi^{\dag}$ have to commute. Notice that this gives a continuum
set of solutions, since $\phi$ has an expansion in the Cartan generators
$\{ h_i \}$ of the gauge group 
\begin{align*}
  \phi = \sum a_i h_i \qquad a_i \in \C.
\end{align*}
The gauge group is thus generically broken to a number of
$U(1)$-factors. Dividing out the residual Weyl symmetry, for $SU(N)$ we find a
moduli space $\cM_c$ of classical vacua that is parametrized by the
symmetric polynomials 
\begin{align*}
u_k = \Tr \phi^k  
\end{align*}
in the parameters $a_i$. 

Classically, there are singularities in this moduli space where
$W$-particles become massless and the gauge symmetry is partially 
restored. To understand the theory fully, it is important to find out
what happens to these singularities quantum-mechanically. This
information is contained in the quantum metric on the moduli space,
which is part of the \emph{low energy effective action}.

\subsubsection{Quantum moduli
  space}\label{sec:prepotential}\index{Seiberg-Witten moduli space}


Let us explain this in some detail. The abelian low energy effective action is
very much restricted by supersymmetry  
\begin{align}\label{eqn:n=2f-term}
\cL =  \textrm{Im} \int d^4 \theta~ \Tr \cF_0(\mathbf{\Psi}^i),   
\end{align}
where $\cF_0$ is any holomorphic function in the $\cN=2$ abelian vector
superfields $\mathbf{\Psi}^i$.  The holomorphic function $\cF_0$ is 
known as the \emph{prepotential}, whereas the supermultiplets
$\mathbf{\Psi}^i$ form a
representation of the $\cN=2$ supersymmetry algebra and contain the
$U(1)$ fields $A^i_{\mu}$, $\phi^i$ and $\la^i_{\pm}$ as physical degrees of
freedom. 

In $\cN=1$ language $\mathbf{\Psi}$ is decomposed into two $\cN=1$
chiral multiplets $\mathbf{\Phi}$, containing the scalar field $\phi$
and $\la_-$, and 
$\mathbf{W}_{\alpha}$, which can be expanded in terms of $\la_+$ and
the field strength $F_{\mu \nu}$. This results in the well-known low
energy Lagrangian  
\begin{align}\label{eqn:n2lagrangian}
\cL =   \int d^2 \theta d^2 \bar{\theta}~ \cK(\mathbf{\Phi}^k,
\overline{\mathbf{\Phi}^k}) 
   + \int d^2 \theta ~
  \tau_{ij}(\mathbf{\Phi}^k)
  \mathbf{W}^i_{\alpha} \mathbf{W}^{\alpha j}, 
\end{align}
with 
\begin{align*}
\cK(\mathbf{\Phi}^k, \overline{\mathbf{\Phi}^k}) =  \textrm{Im} \left[
  \frac{\lotjesd
  \cF_0(\mathbf{\Phi}^k)}{\lotjesd \mathbf{\Phi}^i}
\overline{\mathbf{\Phi}^i}\right]  \quad \textrm{and} \quad 
\tau_{i j}(\mathbf{\Phi}^i) =   \frac{\lotjesd^2 \cF_0(\mathbf{\Phi}^k)}{\partial
    \mathbf{\Phi}^i \partial \mathbf{\Phi}^j} 
\end{align*}

Important is that the first term (the so-called \Index{D-term}) in this
Lagrangian determines a 
K\"ahler metric $g_{i \bar{j}}$ on the quantum vacuum moduli space $\cM_q$. Indeed,
when written in terms of components, we find a sigma model action 
$\cL = g_{i \bar{j}} \partial \phi^{i} \overline{\partial \phi^{j}} +
\ldots $ for the scalar fields $\phi^i$ with K\"ahler metric  
\begin{align}\label{eqn:gaugekahlerpot}
g_{i \bar{j}}(\phi^k, \overline{\phi^k}) =  \frac{\lotjesd^2 \cK(\phi^k,
  \overline{\phi^{k}})}{\lotjesd \phi^i 
\lotjesd \overline{\phi^j}} = \textrm{Im} \left(\frac{\lotjesd^2
  \cF_0(\phi^k)}{\lotjesd \phi^i \lotjesd \phi^j} \right).
\end{align}

Furthermore, the second term in the $\cN=2$ Lagrangian (the
\Index{F-term}) yields the familiar
Yang-Mills action for the field strengths $F^i_{\mu \nu}$ with gauge coupling
constants $\tau_{ij}$. It captures the holomorphic dependence of the theory. 

For the $SU(2)$ theory, when $\phi = a \sigma_3$, the quantum prepotential $\cF_0$
has an expansion   
\begin{align}\label{eqn:fullprepotential}\index{prepotential}
\cF_0 = \frac{1}{2} \tau_0 a^2 + \frac{i}{2 \pi} a^2 \log \frac{a^2}{\Lambda^2} + \sum_{k=1}^{\infty} \cF_{0,k} \left( \frac{\Lambda}{a} \right)^{4k} a^2,  
\end{align}
whose second derivative determines the effective gauge coupling $\tau_{\textrm{eff}}(a)$ in
equation~(\ref{eqn:efftau}). But this expression cannot be valid all over
the moduli space: the resulting metric is harmonic, and thus cannot
have a minimum, while it should be positive definite. So $\tau_{\textrm{eff}}$
must have singularities and $a$ cannot be a global
coordinate on the quantum moduli space $\cM_q$. Instead, one needs
another local description in the strong coupling regions on the moduli
space.

\begin{figure}[h!]
\begin{center} 
\includegraphics[width=9cm]{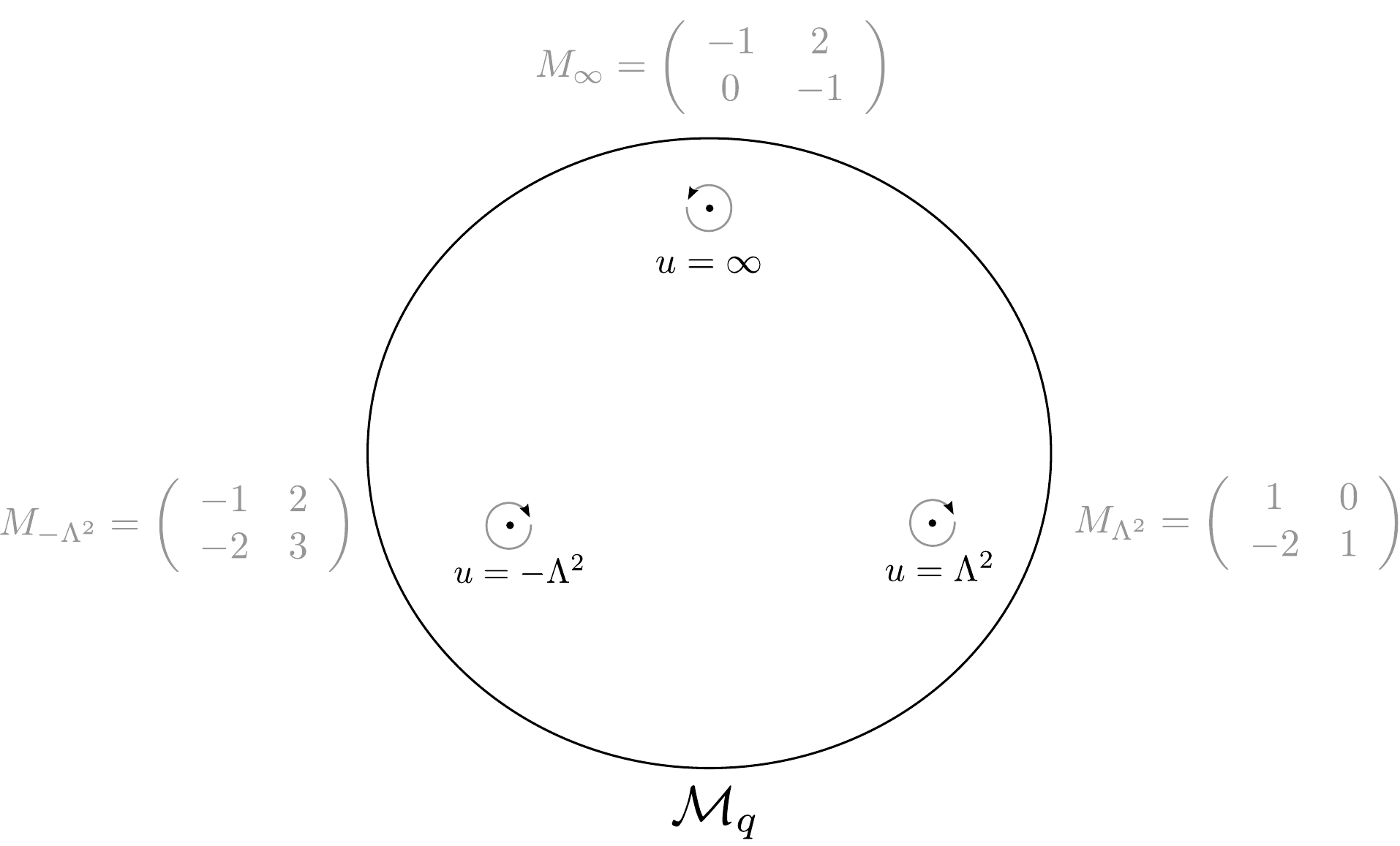}
\caption{The quantum moduli space $\cM_q$ for the $SU(2)$ Seiberg-Witten
  theory is a 2-sphere with three singularities at $u = \infty$ and $u
= \pm \Lambda^2$. The low energy effective theory is described by an
$Sp(1,\Z) = Sl(2,\Z)$-bundle over $\cM_q$ with monodromies $M_{ \infty,
  \pm \Lambda^2}$ around the
singularities.}\label{fig:quantummodspace} 
\end{center}
\end{figure}

 Let us introduce the magnetically dual coordinate 
\begin{align*}
 a_D = \frac{\lotjesd \cF_0}{\lotjesd a}.
\end{align*}
The idea of Seiberg and Witten is that the tuple $(a_D, a)$ should be
considered as a holomorphic \emph{section} of a $Sl(2,\Z)= Sp(1,\Z)$-bundle over the
moduli space $\cM_q$. Indeed, the metric on $\cM$ may be rewritten as 
\begin{align}\label{eqn:SWsymplmetric}
 ds^2  =\textrm{Im} \tau_{\textrm{eff}}~ da \otimes
        d \bar{a} 
 = \textrm{Im}~d a_{D} \otimes        d \overline{a}, \quad
 \textrm{with}\quad \tau_{\textrm{eff}} = \left( \frac{\lotjesd a_D}{\lotjesd a}  \right)
\end{align}
Since this tuple experiences a monodromy around the 
singularities of $\cM_q$
\begin{align*}
   \bem a_D \\ a \eem \to M
 \bem a_D \\ a \eem, 
\end{align*}
just finding these monodromies defines a
Riemann-Hilbert problem whose solution determines the quantum metric.  
And this turns out to be feasible.  Except for the monodromy
$M_{\infty}$ around $u=\infty$, Seiberg and Witten
find two other quantum singularities at $u = \pm \Lambda^2$ with
monodromy matrices $M_{\pm \Lambda}$. They are shown in
Fig.~\ref{fig:quantummodspace}.

\subsection{Seiberg-Witten curve}\label{sec:SWcurve}
\index{Seiberg-Witten curve} 

\begin{figure}[h]
\centering
\includegraphics[width=\textwidth]{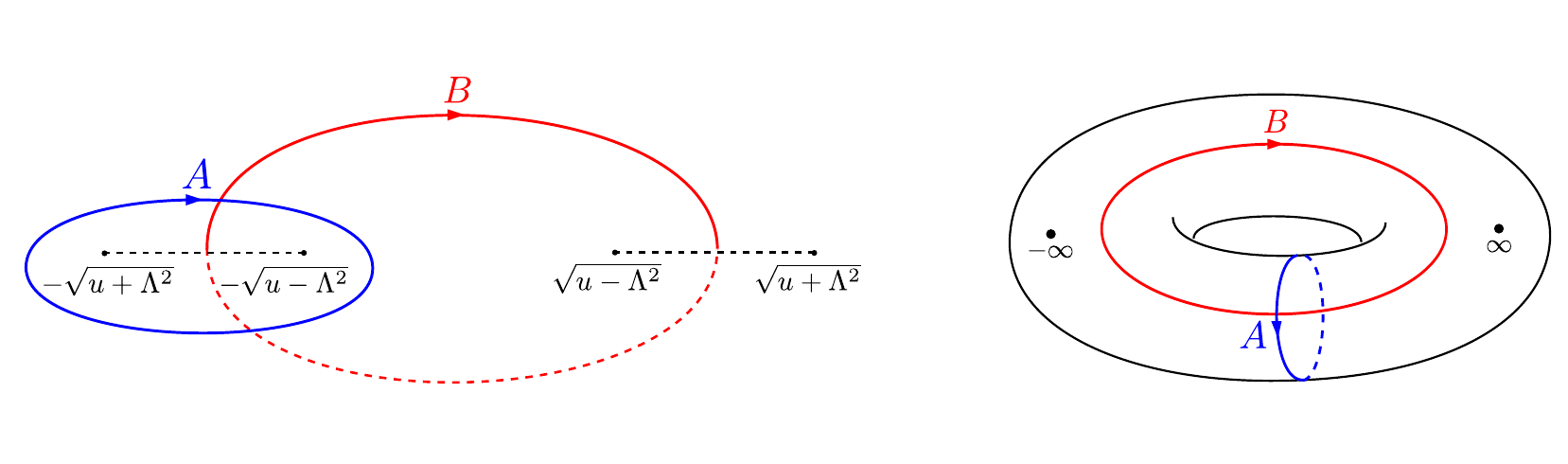}
\caption{On the left we see a hyperelliptic representation of the
  $SU(2)$ Seiberg-Witten curve defined in
  equation~(\ref{eqn:SU(2)SWcurve}), together with a choice of A and 
  B-cycle. On the right it is compactified by adding two points at
  infinity.}\label{fig:SU(2)SWcurve} 
\end{figure}

Mathematically, this solution has another interesting
characterization.  Notice that the metric in
equation~(\ref{eqn:SWsymplmetric}) equals that of an elliptic
curve at each point $a \in \cM_q$.
Moreover, the monodromies $M_{\infty, \pm \Lambda^2}$ altogether generate a subgroup
$\Gamma_0(2) \subset Sl(2, \Z)$, which is exactly the moduli space for an
elliptic curve $\Sigma_{SW}$. The singularity structure suggests that
the moduli space is parametrized by the family
\begin{align}\label{eqn:SU(2)SWcurve}
\Sigma_{SW}(u): \quad y^2 = (x^2 - u)^2 -  \Lambda^4
\end{align}
of Seiberg-Witten curves.
This family of elliptic curves, illustrated in
Fig.~\ref{fig:SU(2)SWcurve}, has four branch points in 
the $x$-plane, which can be connected by two cuts running from $\pm
\sqrt{u - \Lambda^2}$ to $\pm \sqrt{u + \Lambda^2}$.


\vspace*{1mm}

At the points $u \in \{ \infty, \pm \Lambda^2 \}$ some of these branch
points come together so that the elliptic curve degenerates. Precisely
which 1-cycle degenerates at a quantum singularity, labeled by the monodromy
matrix $M$, can be found by solving the equation 
\begin{align*}
  \bem p & q \eem \, M  = 0. 
\end{align*}
The 1-cycle $p B + q A$  vanishes at the corresponding
singularity of $\cM_q$. 

Remember that the period
matrix $\tau$ of an elliptic curve is defined by  
\begin{align*}
 \tau = \frac{ \int_B \omega}{\int_A \omega}, 
\end{align*}
where $\omega$ is a holomorphic 1-form on $\Sigma_{SW}$, and satisfies
$\textrm{Im} \tau >0$. This suggests the identifications
\begin{align*}
 \partial_u a_D = \int_B \omega, \quad \partial_u a = \int_A \omega
\end{align*}
and leads to the introduction of a meromorphic Seiberg-Witten form
$\eta_{SW}$ that obeys
\begin{align*}
 \partial_u \eta_{SW} = \omega. 
\end{align*}
The metric on the quantum moduli space $\cM_q$ can thus be given a
geometric meaning in terms of an auxiliary Riemann surface
$\Sigma_{SW}$ together
with a meromorphic 1-form $\eta_{SW}$.

\subsubsection{Monodromies and BPS particles}
\index{BPS particle}

Physically, a quantum singularity hints that certain BPS particles
becomes massless. In $\cN=2$ supersymmetric Yang-Mills BPS particles are
characterized by their electro-magnetic charge $\gamma = (p,\, q)$. BPS particles with both types of charge are called BPS dyons. \index{BPS dyon}
Their central charge $Z$ is of the form 
\begin{align}\label{eqn:SWBPSmass}
 Z = q a + p a_D = q \int_A \eta_{SW} + p \int_B \eta_{SW},
\end{align}
where on the righthand-side we have written the parameters $a$ and
$a_D$ in terms of the geometric variables $\eta_{SW}$ and the 1-cycles
on the Seiberg-Witten curve. This formula implies that BPS dyons have
a geometric interpretation as wrapping a combination of A and B-cycles
of the Seiberg-Witten curve. This is shown in
Fig.~\ref{fig:SU(2)SWcurve}: magnetic particles wrap the
B-cycle of the curve, whereas electric particles  wrap the
A-cycle.   

The monodromies $M_{\infty, \pm \Lambda^2}$ can therefore indeed be
explained in terms of BPS particles that become massless. Magnetic
monopoles  of charge $(1,0)$ become very light in the neighbourhood of
$u =\Lambda^2$, since they are associated with the vanishing of the
B-cycle. On the other hand, compared to the electric gauge bosons $W^{\pm}$ of charge
$\pm(0,1)$, they become very heavy in the weak-coupling region
of the moduli space.  
This has led to an understanding of confinement in $\cN=2$
supersymmetric gauge theories \cite{Seiberg:1994rs}.

Determining the full spectrum of the $SU(2)$ Seiberg-Witten theory is
more subtle. For example, because of the monodromy around the three
singular points, the BPS charges are not determined uniquely. A
careful analysis \cite{Ferrari:1996sv} reveals that there is a contour 
on $\cM_q$ going through the singular points $u = \pm \Lambda^2$,
where BPS dyons may decay into other BPS dyons. This contour separates
the strong and the weak coupling region and leads to a consistent BPS
spectrum.

\subsubsection{$U(N)$ Seiberg-Witten curve}
\index{M5 brane}



The discussion in \Cref{sec:clasmodspace} on the classical
moduli space $\cM_c$ can easily be extended to other gauge groups. For
gauge group $U(N)$ the singularity structure 
on $\cM_c$ is encoded in the characteristic polynomial
\begin{align*}
 P_N(x,\phi) = \det [x \mathds{1} -\phi],
\end{align*}
that defines coordinates $u_k = \Tr \phi^k$ on the moduli
space. When two or more $a_i$'s assume the same value, the gauge 
group is classically partially restored.
 
On the quantum level  the low energy theory is captured by a
section
\begin{align*}
 (a_i, a_{D,i}) \in \G (\cM_q,\cH) 
\end{align*}
of an $Sp(N,\Z)$-bundle $\cH$ over $\cM_q$. This section is related to
the genus $N-1$ hyperelliptic curve
\begin{align*}
 \Sigma_{SW}: \quad y^2 = P_N(x,u_k)^2-\Lambda^{2N}, 
\end{align*}
where $u_k$ now stands for the quantum vacuum expectation value (vev) $u_k =
\langle
\Tr \phi^k \rangle$ of the scalar field $\phi$. The extra constraint
$u_1 = 0$ defines the Seiberg-Witten curve for gauge group $SU(N)$.
%
%
The curve $\Sigma_{SW}$ can be represented by a two-sheeted
$x$-plane with $N$ cuts. Whenever two branch points coincide a quantum
singularity arises where some BPS dyon becomes massless.

\begin{figure}[h]
\begin{center} 
\includegraphics[width=7cm]{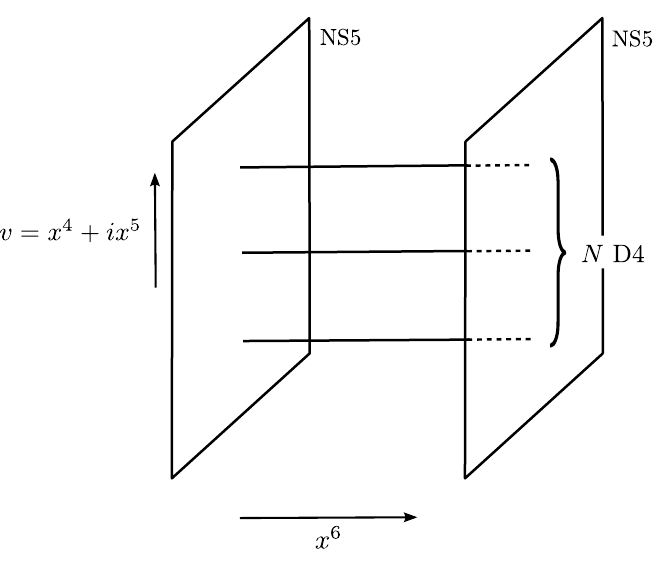}
\caption{Configuration of $N$ D4-branes stretched between two
  NS5-branes that realizes the $SU(N)$ Seiberg-Witten gauge theory on the D4
  worldvolume. This picture is only valid in the small $g_s$
  limit where the effect of the ending of the D4-branes on the NS5
  branes may be neglected.}\label{fig:D4NS5config}  
\end{center}
\end{figure}

By applying a coordinate
transformation $y = P_N(x;u_k) + \Lambda^N t$ we find 
\begin{align}\label{eqn:swcurve}
\Sigma_{SW}: \quad \Lambda^N (t + t^{-1}) + 2 P_N(v;u_k) =0.
\end{align}
The Seiberg-Witten differential now takes a
particularly simple form 
\begin{align}
 \eta_{SW} = v(t,u_k) \frac{ dt}{t}. 
\end{align}
Exactly this representation of the Seiberg-Witten curve and
differential plays a crucial 
role in the following. It relates $\cN=2$ gauge theories to integrable
systems and makes it possible to embed them in string theory as
configurations of D4 and NS5 branes. 
Note that the section $(a_i, a_{D,i})$ is recovered by the period
integrals 
\begin{align*}
  a_i = \int_{A_i} \eta_{SW}, \quad  a_{D,i} = \int_{B_i} \eta_{SW}.
\end{align*}

\subsection{Brane webs}\label{sec:Wittenbrane} 
\index{D4 brane} \index{NS5 brane}

In string theory $\cN=2$ supersymmetric gauge theories can be studied on a
configuration of D4, NS5, and D6 branes \cite{Witten:1997sc}. Pure
 $U(N)$ (and $SU(N)$) Seiberg-Witten theory are embedded in type IIA theory on
$\R^{10}$ as a combination of two NS5-branes with $N$ D4-branes stretched between
them, see Fig.~\ref{fig:D4NS5config}. The NS5-branes are located at some fixed
classical value of $x^6$ and are parametrized by $x^0, \ldots, x^5$, while the
D4-branes are parametrized by $x^0, \ldots, x^3$ and stretch between the two
NS5-branes in the $x^6$-direction. Furthermore, we define a complex
coordinate $v = x^4 + i x^5$. 
Since the coordinate $x^6$ has to fulfill the Laplace equation
$\nabla^2 x^6=0$ on the NS5-worldvolume, the NS5-branes are actually
curved logarithmically at infinity
\begin{align}\label{eqn:boundns5}
 x_6 \sim \pm N \log |v|, \quad v \to \infty.
\end{align}
The D4-branes cause dimples on the NS5-worldvolume.

The $N$ abelian gauge fields on the $N$ D4-branes realize a low energy
description of $\cN=2$ super Yang-Mills in the Minkowski directions $x^0,
\ldots, x^3$. The vev's for the complex scalar $\phi$ correspond to
the positions of the D4-branes on the NS5-brane, and the effective gauge
coupling is given by
\begin{align*}
 \frac{1}{g^2_{\textrm{eff}}(v)} = \frac{L(v)}{\la}, 
\end{align*}
where $L(v)$ is the $x^6$-distance between both NS5-branes at position
$v$. Hence $v$  plays the role of mass scale in four dimensions. The
corresponding logarithmic behaviour of the gauge coupling constant
agrees with its one-loop correction (see equation (\ref{eqn:efftau})).

Since the gauge coupling constant is naturally complexified, it is 
 natural to introduce a new complex coordinate $s = x^6 + ix^{10}$ and 
lift the configuration to M-theory, where $x^{10}$ parametrizes the
M-theory circle 
$S^1$. In M-theory the $N$ D4-branes and the two NS5-branes lift
to a single M5-brane. Supersymmetry forces this M5-brane to wrap a
holomorphic cycle in the complex 2-dimensional surface spanned by
$v$ and $t = e^{-s}$. Moreover, the classical IIA geometry forces this
Riemann surface to have genus $g=N-1$. 

By decomposing the self-dual 3-form field strength $T$ on the M5-brane  
\begin{align*}
 T = F \wedge \Lambda + * F \wedge * \Lambda,
\end{align*}
into a 2-form $F$ on $\R^4$ and a 1-form $\Lambda$ on the Seiberg-Witten curve,
we recover the abelian gauge field strengths $F^i$ in the 4-dimensional theory.
The five-brane kinetic energy $\int T \wedge * T$ reduces to the
4-dimensional effective gauge Lagrangian 
\begin{align}\label{eqn:effgaugelag}
\cL = \tau_{ij} F_+^i \wedge  F_+^j,  
\end{align}
where $\tau_{ij}$ is the period matrix of the M-theory curve \cite{Witten:1997sc}. So 
choosing the Seiberg-Witten curve
\begin{align*}
\Sigma_{SW}: \quad \Lambda^N (t + t^{-1}) + 2 P_N(v;u_k) =0.
\end{align*}
as M-theory curve, consistent with the boundary conditions
(\ref{eqn:boundns5}), indeed engineers the $U(N)$ (or $SU(N)$) gauge
theory dynamics in 4 dimensions (depending on whether $u_1 =0$).

\subsection{I-brane configuration}\label{sec:firstdualitychain}
\index{free fermions}\index{I-brane}


This brings us to the most important paragraph in this section. 
We can generalize the I-brane configuration in Figure~\ref{fig1} to gauge
theories with $\cN=2$ supersymmetry. Like for the $\cN=4$ gauge
theories in Chapter~\ref{chapter2}, we study  $\cN=2$ theories on more
general Taub-NUT backgrounds $TN_k$. Remember that 
$TN_1$ is related to $\R^4$ in the limit that the Taub-NUT circle
becomes very large. 

\subsubsection{The first duality chain}

The M5-brane configuration in \Cref{sec:Wittenbrane} is in many ways
the most elegant starting point to study supersymmetric
gauge theories. For a pure $\cN=2$ susy Yang-Mills theory it wraps the
Seiberg-Witten curve $\Sigma_{SW}$ we met in equation (\ref{eqn:swcurve}). We
call $\cB$ the 2-dimensional complex surface in which this curve is
embedded. So let us start with the M-theory compactification
$$
\hbox{(M)}\quad TN \times \B \times \wt\R^3,
$$ 
corresponding to the top box in Fig.~\ref{fig:webofdualities} (with
$S^1$ decompactified). 
Here we have denoted the three non-compact directions as $\wt\R^3$
to distinguish them from the $\R^3$ in the base of $TN$.
We further
pick $\B$ to be a flat complex surface that is topologically a $T^4$ or
some decompactification of it.  
That is, in the most general case $\B$ will be a product
$$
\B = E \times E'
$$ 
of two elliptic curves. But more often we will consider the
degenerations $\B = \C^* \times \C^*$ and $\B = \C \times \C$, or any
mixed combination. (In the relation with integrable hierarchies the
cases $\C,$ $\C^*$, and $E$ correspond to rational, trigonometric, and
elliptic solutions respectively.) We will denote the affine
coordinates on $\B$ as $(x,y) \in \B$. The complex surface $\B$ has a
(2,0) holomorphic form
$$
\omega = dx \wedge dy.\\
$$
We will now pick a holomorphic curve $\Sigma$ inside $\B$ given by an equation
$$
\Sigma:\ H(x,y)=0,
$$
and wrap a single M5-brane over $TN \times \Sigma$.
Because $\Sigma$ is holomorphically embedded this is a
configuration with $\cN=2$ supersymmetry in four dimensions.

There are two obvious reductions to type IIA string theory
depending on whether we take the $S^1$ inside $\B$,
or an $S^1$ in the Taub-NUT fibration.
In the first case we will compactify $\B$ along a $S^1$ down
to a three dimensional base $\cB_3$. The curve $\Sigma$, and therefore
also the M5-brane, will partially wrap this $S^1$. Consequently, we
arrive at a configuration of NS5-branes and D4-branes that are
spanned between them \cite{Witten:1997sc}. In the classical situation
discussed by Witten 
we take $\B = \C \times \C^*$ and end up with a IIA string theory on
$$ 
\hbox{(IIA)}\quad TN \times \R^6
$$ 
with a set of parallel NS 5-branes with D4-branes ending on them,
exactly as we discussed in \Cref{sec:Wittenbrane}.

\begin{figure}[h]
\begin{center}
 \includegraphics[width=6.5cm]{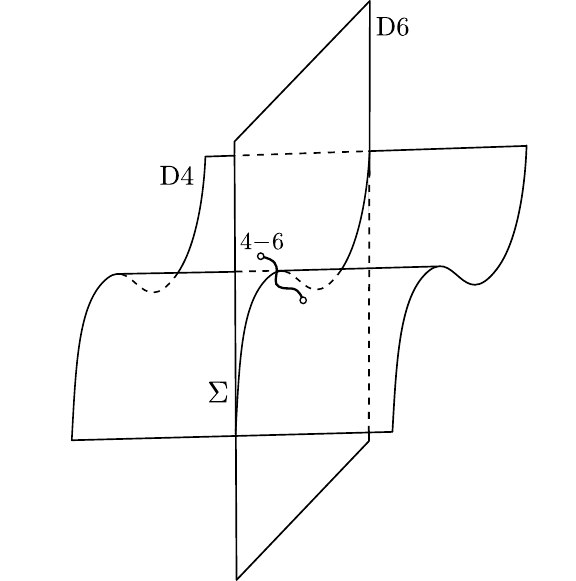}
\caption{A more general configuration of
  D4 and D6-branes where the intersection locus is an affine holomorphic curve $\Sigma$.}\label{fig:curve}
\end{center} 
\end{figure}

In the dual interpretation we switch to the other
duality frame, by compactifying to Type IIA theory
along the $S^1$ fiber in the Taub-NUT geometry. This is the familiar
9-11 exchange. In this fashion we end up with a IIA compactification
on
$$
\hbox{(IIA)}\quad \R^3 \times \B \times\wt \R^3.
$$
with $N$ D4-branes wrapping $\R^3 \times \Sigma$. However, because the
circle fibration of the $TN$ space has singular points, we have to
include D6-branes as well. In fact, there will be $k$ D6-branes that wrap $\cB
\times \wt\R^3$ and are localized at the points
$\vec{x}_1,\ldots,\vec{x}_k$ in the $\wt\R^3$. 
This situation is represented in the box on the left-hand side in Fig.~\ref{fig:webofdualities} and  illustrated in Fig.~\ref{fig:curve}.
Summarizing, we get a system of $N$ D4-branes and $k$ D6-branes intersecting
along the holomorphic curve $\Sigma$. As before we refer to this intersection
locus as the I-brane.

\subsubsection{Comparing partition functions}

In this generalized geometry we should consider the free fermion
system on a higher genus Riemann surface with action
$$
I = \int_\Sigma \psi^\dagger \overline\partial \psi.
$$

Let us compare the I-brane with the gauge theory
computation. In the gauge theory we are computing two
contributions. Firstly, there is a
gauge coupling matrix $\tau_{ij}$ of the $U(1)^g$ fields $F^i$
$$
\int_{TN} \frac{i}{4 \pi} \tau_{ij} F^i_+ \wedge F^j_+ +
v_i \wedge F^i_+
$$ 
for a genus $g$ curve $\Sigma$. Compared to
equation~(\ref{eqn:effgaugelag}) we added magnetic couplings $v_i$,
like in equation (\ref{eqn:abelianpart}). On the $TN$ geometry the
gauge field strengths $F^i$ have fluxes in 
the lattice 
$$
[F^i/2\pi] = p^i \in H^2(TN,\Z).
$$ 
Since the cohomology lattice $H^2(TN,\Z) \cong \Z^k$, these fluxes are
labeled by integers $p^i_a$ with $i=1,\ldots, g$ and $a=1,\ldots,k$.

Secondly, there is a gravitational coupling $\cF_i$ that appears in
the term (we discuss this term in detail in \Cref{sec:CYcomp} and
Chapter~\ref{chapter5})
$$
\int_{TN} \cF_1(\tau)\ \Tr\, R_+ \wedge R_+.
$$
Since the regularized Euler number of $TN_k$ equals $k$, 
combining these two terms yields the partition function
\be
\label{Z-bos}
Z_{\textrm{gauge}} = \sum_{p^i_a \in \Z} e^{\pi i p^i_a \tau_{ij} p^{j,a} +
  2\pi i v_i^a p^i_a} e^{k \cF_1}.
\ee

In the I-brane model the partition function $Z_{\textrm{gauge}}$ is nothing but
the determinant of the chiral Dirac operator acting on $k$ free
fermions living on the ``spectral curve'' $\Sigma_g$, coupled to the
flat $U(k)$ connection $v$ on corresponding rank $k$ vector bundle
${\cal E} \to \Sigma$. So we are led in a very direct way to
\be
\label{Z-fer}
Z_{\textrm{gauge}}
 = \det \overline\partial_{\cal E}.
\ee
The results (\ref{Z-bos}) and (\ref{Z-fer}) are just the usual
bosonization formula, where the fermion determinant is equivalent to
 a sum over the lattice of momenta together with a boson
determinant. Here we use the identification
$$
\cF_1 = - \frac{1}{2} \log \det \Delta_\Sigma.
$$

To complete this map we need to show why the $p^i$ are identified with
fermion currents on the Riemann surface through the corresponding
cycle, but this is relatively clear. Consider a cycle $A_i$ on
the Riemann surface and a disc ending on this cycle (which can
always be done as $\Sigma$ is contractible in the full CY).  Then the
statement that $F^i$ is turned on corresponds to the fact that the
integral of the corresponding flux over this disc is not zero.  Since
the fermions are charged under the $U(k)$ gauge group, this means that
they pick up a phase as they go along this cycle on the Riemann
surface (the Aharanov-Bohm effect).  Thus the holonomy of the fermions
correlates with the $p^i$.  Later (in \Cref{sec:fermioncharge}),
we provide an alternative view of the fluxes $p^i$:
they also correspond to D4-branes, wrapping 4-cycles of the Calabi-Yau and
bound to the D6-brane.

\section{Geometric engineering}\label{sec:ge}
\index{geometric engineering}

In \Cref{sec:N=4YM} we considered $\cN=4$ supersymmetric
Yang-Mills in the background of an ALE space $\C^2/\Gamma$, and
discovered an interesting relationship with two dimensional conformal
field theory. We explained this by embedding the gauge theory in type IIA theory
on a D4-brane wrapping the ALE space. In the ten dimensions
of string theory, however, much more is possible.  

Remember
 that ALE spaces $\C^2/\Gamma$ are characterized by their
vanishing two-cycles (see \Cref{sec:ALEspace}), whose
intersection matrix realizes the Dynkin 
diagram for the corresponding ADE Lie group. Resolving the singularity
gives a finite volume $\epsilon$ to the vanishing cycles. Let us call
the dual two-forms $\omega_i$. The RR gauge field $C_3$ reduces to a
set of $r$ abelian gauge fields $A^i$  
\begin{align}
 C_3 = \sum_{i=1}^r A^i \omega_i. 
\end{align}
When $\epsilon \to 0$ the gauge symmetry is enhanced to the
corresponding ADE gauge, similarly (and in fact dual) to when D-branes
approach each other \cite{Bershadsky:1995sp}. This realizes the geometric
McKay correspondence in string theory.  

Wrapping
a D2-brane over any of these two-cycles yields a BPS particle in the
six transverse directions to the ALE space, whose mass is proportional to
the volume $\epsilon$ of the two-cycle it is wrapping. 
Since these 6-dimensional BPS particles transform as vectors and are charged 
under the Cartan of the ADE-group, they are the $W$-bosons corresponding to the
breaking of the ADE gauge group to
its Cartan subalgebra.

A simple example is given by the ALE-fibration
\begin{align*}
 (z - a)(z+ a) + u^2 + v^2 =0. 
\end{align*}
Its only two-cycle is spanned in between $z=- a$ and $z =
a$. Two-branes can wrap this two-cycle with two possible
orientations, generating a $W^+$ and $W^-$-boson with masses
proportional to $a$. So this engineers a broken $SU(2)$ gauge group
for a generic value for $a$ that is enhanced to $SU(2)$ when $a
=0$.
Note that this matches with the classical Seiberg-Witten
moduli space $\cM_c$ (see \Cref{sec:clasmodspace}). Indeed,
the ALE fibration only breaks half of the supersymmetries, which
amounts to $\cN=4$ supersymmetry in four dimensions. As we pointed out before,
in $\cN=4$ theories classical results are exact.  

We can turn this example into a string theoretic setting that studies
4-dimen-sional $\cN=2$ Yang-Mills by fibering the ALE space
over a genus zero curve. The genus zero curve breaks the supersymmetry
from $\cN=4$ to $\cN=2$, without introducing extra particles. 
This idea of looking for a string theory set-up that engineers a
particular supersymmetric gauge theory in string theory is called
geometric engineering
\cite{Katz:1996xe,Katz:1996fh,Katz:1996th,Katz:1997eq}. 

In \Cref{sec:revisitSW} we will see how the results of
Seiberg and Witten 
can be elegantly embedded in string theory in the language of string
compactifications.

\subsection{Non-compact Calabi-Yau threefolds}\label{sec:toricgeometry} 

Let us return to the local Calabi-Yau threefolds that we introduced in
Chapter~\ref{ch:intro}, and give some explicit examples that we will
meet later-on in this thesis. We start
with one of the simplest Calabi-Yau threefolds that is modeled on an affine
curve $\Sigma$: the deformed conifold.

\subsubsection{Deformed conifold}\index{conifold}
\index{local Calabi-Yau modeled on a curve $\Sigma$}

The deformed conifold $X_{\mu}$ is defined by the equation
\begin{align*}
X_{\mu}: \quad xy - uv = \mu, \quad (x,y,u,v) \in \C^4. 
\end{align*}
More precisely, the parameter $\mu \in \C$ parametrizes a family of Calabi-Yau
threefolds $X_{\mu}$ that becomes singular at $\mu=0$. This
singularity is called the conifold singularity.  The non-vanishing holomorphic
three-form $\Omega$ equals  
\begin{align}
 \Omega = \frac{du}{u} \wedge dx \wedge dy. 
\end{align}

The threefold $X_{\mu}$ just contains one compact cycle: a 3-cycle with topology
$S^3$ that shrinks to zero-size when $\mu \to 0$. This is particularly easy to
see after a change of variables  
\begin{align*}
X_{\mu}: \quad z^2 + w^2 + \tilde{u}^2 + \tilde{v}^2 = \mu, \quad (z,w,\tilde{u},\tilde{v}) \in \C^4. 
\end{align*}
So ${\mu}$ parametrizes a family of $T^* S^3$'s.
%

\begin{figure}[h!]
\begin{center}
 \includegraphics[width=7.4cm]{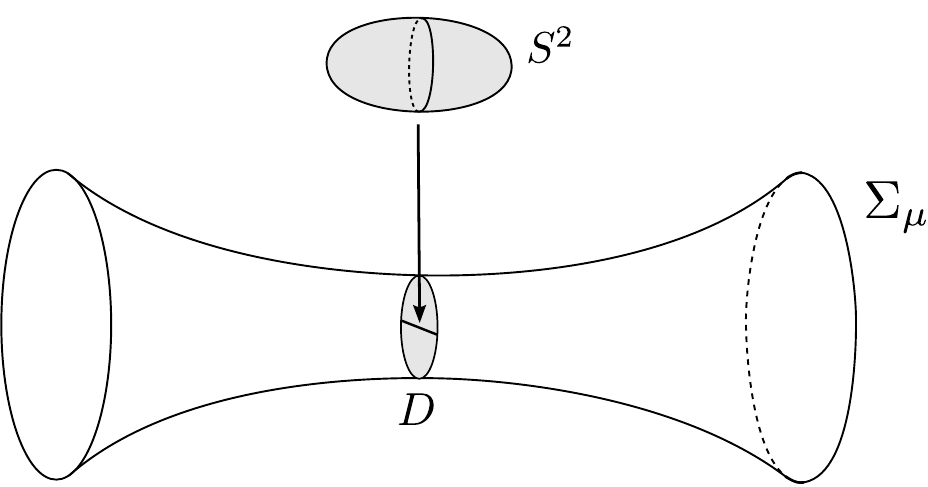}
\caption{The deformed conifold is a non-compact Calabi-Yau threefold
  that is modeled on the curve $\Sigma_{\mu}$ defined by $z^2 + w^2 =
  \mu$. Its only compact 3-cycle is  $S^1$-fibered over the disk
  $D$: Over each line segment in $D$ with endpoints on $\partial D$
  the $S^1$ fibration forms a 2-sphere. Moving the line segment over
  the disk $D$ shows that the compact 3-cycle is homeomorphic to
  $S^3$.}\label{fig:defconifold} 
\end{center} 
\end{figure}

When we view $X_{\mu}$ as a $(u,v)$-fibration over the complex plane
spanned by $z$ and $w$, its degeneration locus is
\begin{align*}
\Sigma_{\mu}: \quad z^2 + w^2 = \mu
\end{align*}
in the $(z,w)$-plane. The $S^3$-cycle in $X_{\mu}$ may then be viewed
as an $S^1$-fibration over a disk $D$ in the $(z,w)$-plane, that is
bounded by the curve $\Sigma_{\mu}$. Since the $(u,v)$-fibration degenerates at the
locus $z^2 + w^2 = \mu$ the resulting 3-cycle has topology $S^3$. This
is illustrated in Fig.~\ref{fig:defconifold}.

Since the $S^3$-cycle is special Lagrangian it is a supersymmetric
cycle. In type IIB we can wrap a D3-brane around it and find a vector
BPS particle in the 
transverse four dimensions, whose mass is proportional to the complex
structure modulus $X$. With Cauchy's theorem we can reduce this
formula to the 1-cycle $\lotjesd D$ of 
$\Sigma_{\mu}$ 
\begin{align}\label{eqn:redOmegatau}
 \int_{S^3} \Omega = \int_{D} dz \wedge dw  = \int_{\lotjesd D} \eta,
\end{align}
where $\eta= w dz$ is a meromorphic 1-form on the $(z,w)$ plane. The
4-dimensional BPS particle couples to a $U(1)$ gauge field 
\begin{align*}
A_{\mu} = \int_{S^3} C^4,  
\end{align*}
that is obtained as a reduction of the RR 4-form over
the $S^3$-cycle. 

This bijective correspondence between 3-cycles in a local Calabi-Yau
threefold $X_{\Sigma}$ modeled on a curve $\Sigma$ and 1-cycles on
$\Sigma$ holds in general. The 3-cycles may be constructed by filling
in a disk $D$ whose boundary $\lotjesd D$ is a 1-cycle on
$\Sigma$. If one of the variables in the complex surface $\cB$ is
$\C^*$-valued, the disk $D$ will be punctured. In such a situation
differences of 1-cycles have to be considered. We will see an example
of this shortly.



\subsubsection{Resolved conifold and toric Calabi-Yau's}\label{sec:resconifold}
\index{toric Calabi-Yau}

Instead of deforming the conifold singularity we can also resolve
the singularity. This is described by $\C^4$ parametrized by $(x,y,u,v)$
together with the identification 
\begin{align*}
 (x,y,z,w) \sim (k^{-1} x, k^{-1} y, k z, kw), \qquad k \in \C^*.  
\end{align*}
The first two complex coordinates parametrize a sphere $\C\P^1$ and
the last two coordinates two line-bundle over it. Altogether this gives
the total space of the line bundles $\cO(-1) \oplus \cO(-1) \to
\C\P^1$. The resolved conifold is an example of a local toric
Calabi-Yau, just like $\C^3$ in Chapter~\ref{ch:intro}.


\begin{figure}[h!]
\begin{center}
 \includegraphics[width=4.6cm]{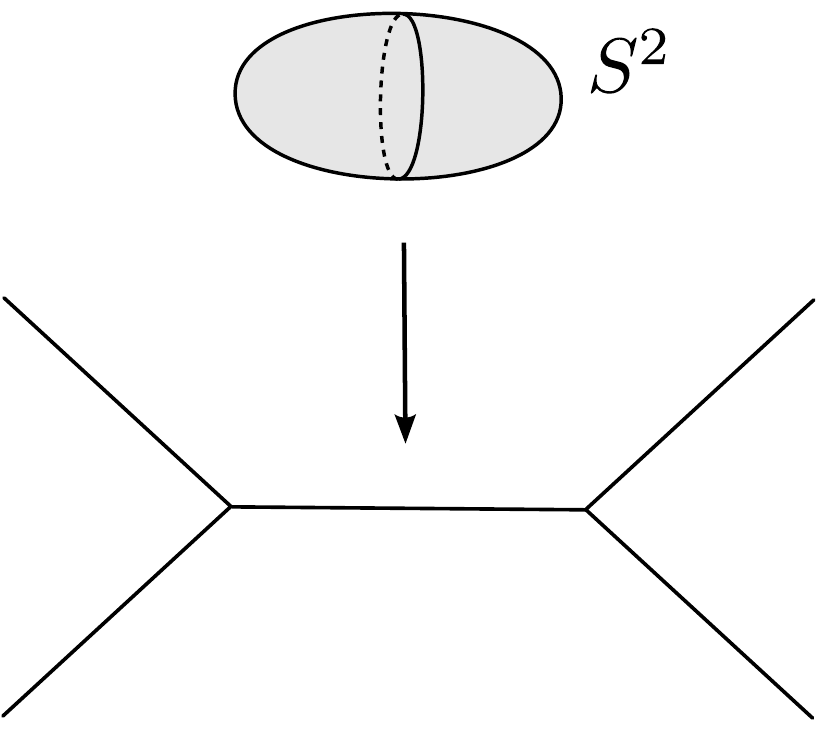}
\caption{The resolved conifold is a non-compact toric Calabi-Yau
  threefold. The $T^2 \times \R$-fibration degenerates over a
  trivalent graph consisting of two 3-vertices. The $S^1$-fibration
  over the inner leg of this graph forms a 2-sphere.}\label{fig:resconifold}
\end{center} 
\end{figure}

Like any local toric Calabi-Yau, the resolved conifold can be obtained
by glueing a few copies of $\C^3$ such that its singular locus 
is a linear trivalent graph in $\R^2$. The toric diagram of the
resolved conifold is shown in Fig~\ref{fig:resconifold}. It consists
of two copies of $\C^3$ with coordinates $(x,z,w)$ and $(y,z,w)$
respectively. Since both 1-cycles in $T^2$ shrink at the vertices, the
middle toric leg represents the $\C\P^1$-cycle.

\subsection{Geometrically engineering Seiberg-Witten
  theory}\label{sec:revisitSW}\index{Seiberg-Witten theory} 

Supersymmetric gauge theories with any gauge group and matter content
may be engineered by a local Calabi-Yau compactification. Pure $\cN=2$
Seiberg-Witten theory with an ADE gauge group is embedded string
theory as a local K3-fibration over $\C\P^1$. Zooming in on the
ALE-singularities of the K3 reveals an ADE gauge group in four
dimensions, which is broken by the Higgs mechanism when some of the
two-cycles gain a non-zero mass.  

\index{geometric engineering}

\begin{figure}[h]
\begin{center}
 \includegraphics[width=3.9cm]{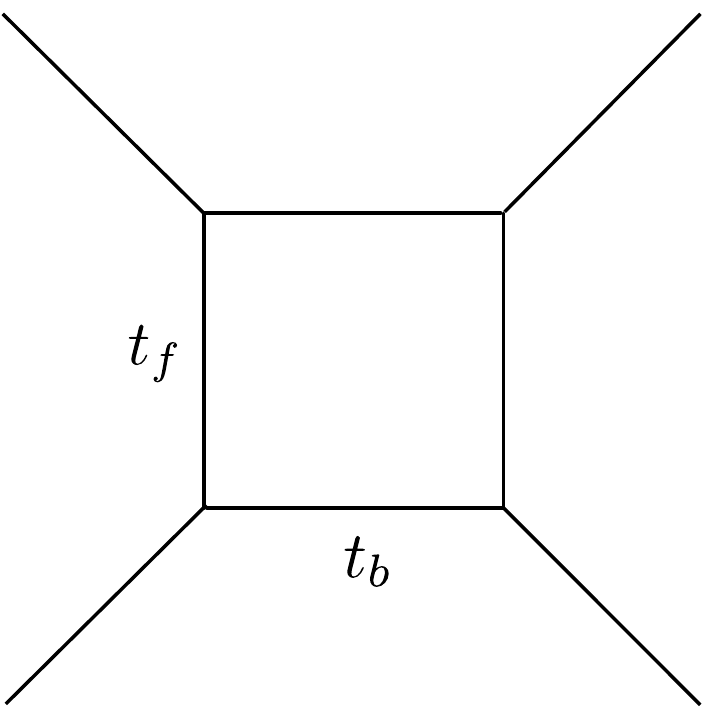}
\caption{$SU(2)$ Seiberg-Witten theory can be engineered as a local
  Calabi-Yau compactification, based on a local Hirzebruch
  surface. The 4-dimensional field theory results are recovered when
  taking the limit $t_b \to \infty$, as in
  equation~(\ref{eq:fieldtheorylimit}).}\label{fig:localhirzebruch} 
\end{center} 
\end{figure}

Let us consider $SU(2)$ Seiberg-Witten theory in some detail. This may be
engineered in type IIA by any local Hirzebruch surface, which is the
total space of 
the canonical bundle over a Hirzebruch surface. For example, take the
simplest Hirzebruch surface  $\C\P^1_b \times \C\P^1_f$, where
$\C\P^1_b$ denotes the base sphere, and $\C\P^1_f$ is the only
two-cycle in the resolved $A_1$-singularity. This non-compact
Calabi-Yau manifold is toric, and its toric diagram is shown in Fig.~\ref{fig:localhirzebruch}. The $W_{\pm}$ bosons correspond to $D2$-branes that are wrapped
around the $\C\P^1_f$ with opposite orientations.

To go to the field theory limit we should take the string scale to
infinity, while keeping the masses of the $W$-bosons fixed. This
corresponds to letting the size $t_b$ of $\C\P^1_b$ to infinity, while
taking the size $t_f$ of $\C\P^1_f$ proportional to the mass $a$ of
the $W$-bosons, as in 
\begin{align}\label{eq:fieldtheorylimit}
 \exp \left(-t_b \right) = \exp \left( -1/g_{YM}^2 \right) = \left( \frac{\beta \Lambda}{2}
 \right)^4, \quad
 \exp \left( -t_f \right) = \exp \left( -\beta a \right)  
\end{align}
when $\beta \to 0$. Without this decoupling limit we end up with a
5-dimensional theory on $S^1 \times \R^4$, where the size of the
$S^1$ is given by $\beta$. 

The Seiberg-Witten prepotential $\cF_0$ is reproduced by stringy instanton corrections to the toric Calabi-Yau geometry. In this geometry the instantons may wrap the base $n$ times and the fiber $m$ times. This gives a contribution to the so-called type IIA prepotential (\ref{eqn:instantoncontr}), as we explain in much more detail in \Cref{sec:CYcomp}. In the 4-dimensional field theory limit only the fiber worldsheet instantons remain. They recombine into the Seiberg-Witten prepotential (\ref{eqn:fullprepotential}). 

The mirror map translates this type IIA configuration into a type IIB
configuration where we can see the Seiberg-Witten curve explicitly in
the limit $\beta \to 0$. We find a non-compact Calabi-Yau $X_{SW}$
based on the Seiberg-Witten curve (\ref{eqn:swcurve}) 
\begin{align}
H_{SW}(t,v)= \Lambda (t + t^{-1}) - 2(v^2-a^2) = 0.
\end{align}
Recall that $t \in \C^*$ while $v \in \C$. The holomorphic three-form
thus reduces to the Seiberg-Witten form  
\begin{align}
\eta_{SW} = v \frac{dt}{t}.  
\end{align}

\begin{figure}[h]
 \begin{center}
\includegraphics[width=9.5cm]{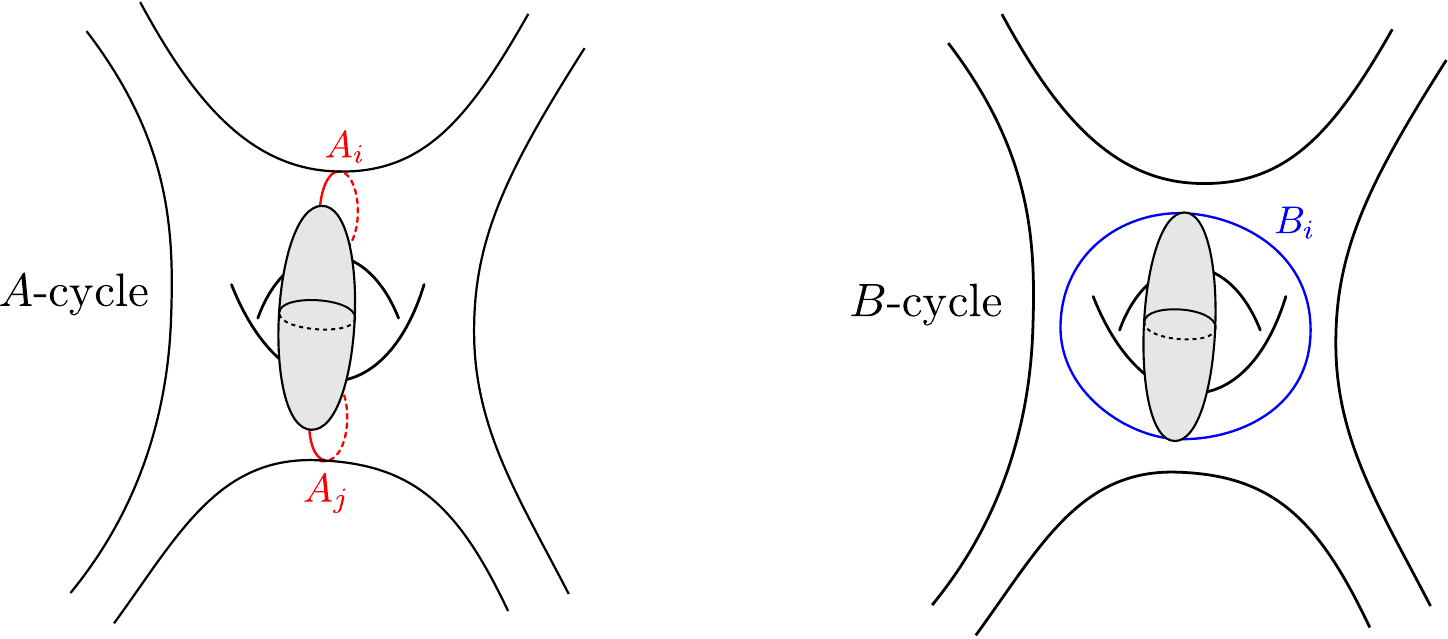}  
\caption{The
 relation between 3-cycles in the Calabi-Yau $X_{SW}$ and 1-cycles on
 the Riemann surface $\Sigma_{SW}$ for the Seiberg-Witten geometry. For
 the $A$-cycle, by fibering $S^1$ over the line segment whose
 endpoints are at a point on $A_i$ and a point on $-A_j$, one obtains
 $S^2$.  By moving the endpoints over $A_i$ and $-A_j$, one obtains
 $S^2\times S^1$.  For the $B$-cycle, similarly moving the $S^2$
 ending on $B_i$, one obtains $S^3$.  \label{SW3cycles}}
 \end{center}
\end{figure}

$A$-cycles on $\Sigma_{SW}$ are not contractible
on the $(t,v)$-plane. Instead, compact $A$-cycles in the
noncompact Calabi-Yau threefold will reduce to a difference of $A$-cycles
on $\Sigma_{SW}$. Indeed, notice that a point on the 1-cycle $A^i$ and
one on another 1-cycle $-A^j$, with opposite orientation, are connected
by a $\C\P^1$ in the Calabi-Yau. The resulting 3-cycle therefore has the
topology of $S^2 \times S^1$. For the $B$-cycles this subtlety does not
arise, and compact $B$-cycles in the Calabi-Yau have $S^3$ topology and
reduce to compact 1-cycles connecting the two hyperelliptic planes. See
Figure \ref{SW3cycles}.

\subsubsection{Self-dual strings}\label{SW:selfdualstring}

Wrapping D3-branes around a 3-cyle $\G= q A + p B$ of
$X_{SW}$, introduces a BPS dyon in four dimensions whose mass is the
absolute value of
\begin{align*}
 Z \sim q \int_A \Omega + p \int_B \Omega. 
\end{align*}
This reproduces the mass-formula (\ref{eqn:SWBPSmass}). On the
Seiberg-Witten curve $\Sigma_{SW}$ the D3-branes reduce to strings,
whose tension is proportional to to the size of the 2-cycle above
it. These strings are not fundamental strings, but instead
non-critical. They live in six dimensions and couple to the $B$-field
that is part of a 6-dimensional tensor multiplet. Since the field
strength of $B$ is self-dual, they are called \emph{selfdual strings}
\cite{Klemm:1996bj}.

The crucial difference between $\cN=2$ and $\cN=4$ gauge theories is
that the metric on the $SU(2)$ Seiberg-Witten curve is not just the
usual flat metric, as it is on the $\cN=4$ torus. The Seiberg-Witten
metric can be derived  
from the metric on the D3-branes. Since these branes wrap supersymmetric
3-cycles they are calibrated by $\textrm{Re }\Omega$. The
pull-back of $\Omega$ to their worldvolume is proportional to the
volume element on the brane \cite{Becker:1995kb}. On the
space part of the self-dual string, parametrized by $x$,  we thus find
\begin{align*}
 \frac{\lotjesd}{\lotjesd x} \left( \eta^t_{SW}(t) \frac{\lotjesd \log t}{\lotjesd x} \right) = 0,  
\end{align*}
where $\eta_{SW} = \eta^t_{SW} dt/t$. This is the geodesic equation for a
metric
\begin{align}
 g_{t \bar{t}} =   \eta^t_{SW} \overline{\eta^t}_{SW}. 
\end{align}
Since $\eta_{SW}$ is meromorphic, this metric has poles. Its geodesics
are therefore curved, so that not all BPS particles are stable. Studying
these BPS trajectories gives a powerful method to determine 
quantum BPS dyon spectra. In particular this leads to the correct $SU(2)$
Seiberg-Witten spectrum \cite{Klemm:1996bj}.  
 
Notice that when we scale the Seiberg-Witten form $\eta_{SW}$ as 
\begin{align}
 \eta_{SW} \to \frac{\eta_{SW}}{\la} 
\end{align}
and take the limit $\la \to 0$, the Seiberg-Witten metric reduces to
the usual $\cN=4$ flat metric. This is a preview on \Cref{sec:CYcomp}, 
where we study gravitational corrections to the prepotential
$\cF_0$ that scale the prepotential to $\cF_0/\la^2$. In the limit $\la \to 0$ we
recover $\cN=4$ results out of the $\cN=2$ data.

\subsection{Completing the web of dualities}\label{dualitysequenceII}
\index{Calabi-Yau compactification}

Now that we have seen the intimate connection between non-compact Calabi-Yau
compactifications and $\cN=2$ supersymmetric gauge theories, it is
time to link them to our first chain of dualities in
\Cref{sec:firstdualitychain}.
We accomplish this by going through another
chain of dualities, represented by the vertical sequence of boxes in
Fig.~\ref{fig:webofdualities}. 

\subsubsection{The second duality chain}

Let us consider a slightly more general
compactification of M-theory, namely
$$ 
\hbox{(M)}\quad TN \times \B \times \R^2 \times S^1,
$$ 
corresponding to the top box in Fig.~\ref{fig:webofdualities}.
The extra $S^1$ does not really influence the earlier results, since
the D6-branes remain non-compact and therefore the gauge theory that
they support stays non-dynamical. Hence the I-brane configuration remains
the same. In the other compactification with an ensemble of NS5-branes and D4-branes we can now perform a T-duality on $S^1$ to give
a web of $(p,q)$ 5-branes in Type IIB, which is another familiar and
convenient realization of the $\cN=2$ system.

However, in this situation there is an obvious third possible compactification to
type IIA, by just reducing on the extra $S^1$ that we have
introduced. This will give IIA on
$$
\hbox{(IIA)}\quad TN \times \B \times \R^2
$$ 
with $N$ NS5-branes wrapping $TN \times \Sigma$. We haven't gained
much in this step, but now we can T-dualize the NS5-brane away to
remain with a purely geometric situation \cite{Ooguri:1995wj,Klemm:1996bj}. 
In general a T-duality {\it transverse} to a set of $N$ NS5-branes
produces a local $A_{N-1}$ singularity of the form
\be
uv = z^N.
\label{tdual}
\ee 
In the case of a single 5-brane $N=1$ this gives an $A_0$
``singularity''.  We recall that world-sheet instanton effects are
important to understand this very non-trivial duality
\cite{Ooguri:1995wj,Tong:2002rq,Harvey:2005ab}.

Applying this T-duality in the present set-up gives us a type IIB
compactification of the form
$$
\hbox{(IIB)}\quad TN \times X,
$$
where $X$ is a non-compact Calabi-Yau geometry of the form
$$
X:\ uv + H(x,y)=0,
$$ 
modeled on the affine Riemann surface $\Sigma$ that is defined by $H(x,y)=0$.
 This is just an application of 
(\ref{tdual}) in the case $N=1$, where $z$ is the local coordinate
transverse to the curve $\Sigma$.

Let us emphasize that the resulting web~\ref{fig:webofdualities} can
be used to study any non-compact Calabi-Yau threefold $X$ that is modeled on a
Riemann surface $\Sigma$. In particular, the Calabi-Yau doesn't need to have a
toric mirror. Examples of such Calabi-Yau's are the Dijkgraaf-Vafa
geometries that we will encounter in Chapter~\ref{chapter6} as well as
Chapter~\ref{chapter8}. 

Moreover, notice that we restricted ourselves to $N=1$, corresponding
to a single M5-brane in the upper node of the duality
web~\ref{fig:webofdualities}. There is no problem in generalizing this
to arbitrary number $N$ of M5-branes though, we just find non-compact
Calabi-Yau threefolds whose 
fiber over the affine curve $\Sigma$ is given by a more complicated
local singularity.

\section{Topological invariants of Calabi-Yau threefolds}\label{sec:CYcomp}
\index{Calabi-Yau compactification}\index{sigma model}

So far we paid attention to some geometrical aspects of the Calabi-Yau
background itself in a Calabi-Yau compactification. For associating
topological invariants to Calabi-Yau threefolds, we need to go one step deeper
and study the moduli spaces of the fields in a Calabi-Yau
compactification. That is the theme of this section.   

A string compactification is a generalization of the
old idea of Kaluza and Klein \cite{kaluza,klein} to unify
electromagnetics and gravity in four dimensions by introducing an
extra dimension. A reduction of the 5-dimensional metric over a
circle $S^1$ yields the 4-dimensional metric $g_{\mu \nu}$  and gauge field
$ A_{\mu} = \int_{S^1} g_{\mu 5}$.

In this thesis we will mostly consider
compactifications of type II to four dimensions, with a
Calabi-Yau threefold $X$ as compactification manifold. The low
energy effective theory in four dimensions is determined by reducing
the massless fields of type II theory. These are given by a metric
$g_{MN}$, the $B$-field $B_{MN}$ and the dilaton $\phi$, together with the
R-R gauge potentials and their superpartners. 

String compactifications are characterized by the amount of
supersymmetry that is 
preserved. This is proportional to the number of covariantly constant 
spinors on the compactification manifold, or equivalently, to the amount of
reduced holonomy. For example, complex Calabi-Yau threefolds with
$SU(3)$ holonomy preserve a quarter of the 10-dimensional
supersymmetry. This results in a 4-dimensional theory with $\cN=2$
supersymmetry. Supersymmetry dictates that the resulting massless
fields can be combined into 4-dimensional $\cN=2$
multiplets. E.g. the 4-dimensional metric $g_{\mu \nu}$ combines
with the graviphoton and two gravitinos in a gravity multiplet, and
yields for example the 4-dimensional Einstein gravity Lagrangian 
\begin{align*}
\int d^{4} x ~\sqrt{-g} R. 
\end{align*}

But there are more multiplets that play a role in the resulting
4-dimensional low energy effective theory. The scalar components of
these multiplets all have an interpretation in terms of the internal
Calabi-Yau metric $g_{mn}$. As this internal metric may vary 
over $x \in \R^4$, though preserving the Calabi-Yau condition $R_{mn}=0$, their
moduli appear in the 4-dimension theory as fields $\phi(x)$. A
reduction of the 10-dimensional Einstein gravity Lagrangian shows
that they are part of a 4-dimensional sigma model action  
\begin{align}\label{eq:sigmamodel}
\int d^4x~ G_{\alpha \beta}(\phi) d \phi^{\alpha} \wedge * d
\phi^{\beta},
\end{align}
where $G_{\alpha \beta}(\phi)$ is the metric on the moduli space of the
Calabi-Yau. Now we already learned in Chapter~\ref{ch:intro} that this
metric splits into two pieces, the first describing $h^{2,1}$ complex
structure moduli $X^i$ and the second $h^{1,1}$ complexified K\"ahler
moduli $t^j$, which combine the K\"ahler metric
$g$ and the B-field.   

In the language of $\cN=2$ supersymmetry the type IIB low energy effective
theory is captured by a $\mathcal{N}=2$ supergravity theory in 4d coupled to
$h^{2,1}$ vector multiplets, whose scalar components
parametrize the complex structure deformations of $X$,  and to $h^{1,1}+1$
hypermultiplets, whose scalar components describe
deformations of the complexified K\"ahler moduli of $X$ plus
the axion-dilaton $C + i e^{-\phi}$. In the context of IIA
compactifications these   
identifications are reversed: there are $h^{1,1}$ vector multiplets
and $h^{2,1}+1$ hypermultiplets. 

As we will see in detail in a moment the metric on the scalar moduli
spaces is captured by classical geometry, up to such quantum
corrections in $g_s$ in full string theory and in $\alpha'$ at
world-sheet level. 
The fact that these two sets of multiplets are
decoupled in the 4-dimensional theory is important in constraining
these corrections. 
For example, the metric on the IIB moduli
space $\cM_V$ doesn't receive any string loop corrections  nor
instanton corrections, since the dilaton as
well as the K\"ahler moduli (which control the size of the Calabi-Yau)
are part of the hypermultiplets. On the
other hand, the metric on the IIA moduli space $\cM_V$ is sensitive to
$\alpha'$ corrections.

\subsection{Special geometry}\label{sec:specialgeometry}
\index{special geometry}\index{vector moduli space}\index{prepotential}

Supersymmetry imposes strong constraints on the geometry of the
moduli spaces of the scalar fields. It forces the metric $G_{i\bar{j}}$ on
any $\cN=2$ vector multiplet moduli space $\cM_V$ to be of the special K\"ahler
form  \cite{deWit:1984px,Strominger:1990pd} 
\begin{align}\label{eqn:specialkahlermetric}
G_{i \bar{j}} = \lotjesd_i \lotjesd_{\bar{j}} \cK,  
\end{align}
with K\"ahler potential $\cK$. This K\"ahler potential can be
expressed in terms of a 
prepotential $\cF_0$, similar to our discussion in
\Cref{sec:prepotential}. Let 
us first show how the prepotential $\cF_0$ is related to the Calabi-Yau
geometry in both type IIA and IIB compactifications, and in
\Cref{sec:decouplegravity} we explain the 
relation to the quantum moduli space of $\cN=2$ supersymetric gauge
theories.

\subsubsection{Type IIB vector moduli space}

In a type IIB Calabi-Yau compactification the K\"ahler potential on $\cM_V$ is given by
\begin{align}\label{eqn:IIBkahlerpot}
\cK &= - \log i \left(  \int_X \Omega \wedge \overline{\Omega} \right).
\end{align}
Here $\Omega$ is the nonvanishing holomorphic three-form of $X$, locally
determined up to multiplication by a nonvanishing holomorphic
function. The transformation $\Omega \to e^{f(z)} \Omega$ 
changes the K\"ahler potential into  
\begin{align*}
\cK(z, \bar{z}) \to \cK(z, \bar{z}) - f(z) - \overline{f(z)},  
\end{align*}
so that the K\"ahler metric (\ref{eqn:specialkahlermetric}) indeed
remains invariant. 
Notice
that (\ref{eqn:IIBkahlerpot}) corresponds precisely to  the
Weil-Petersson metric on the moduli 
space of complex structures of the Calabi-Yau manifold $X$. 

The fact
that the K\"ahler potential $\cK$ is captured by a single function
$\cF_0$ is known as special geometry. 
To see this let us start by choosing a basis of three-cycles $\{ A_i, B_j \}$
of $X$, with $i,j = 0,1, \ldots, h^{2,1}$, such that
\begin{align*}
 A^i \cap B_j = \delta^i_j, \quad A^i \cap A^j = B_i \cap B_j =0. 
\end{align*}
As such a basis is preserved under the symplectic group
$Sp(b^{3},\Z)$\footnote{Recall that $Sp(N,\mathbb{Z})$ is the group 
consisting of $2N \times 2N$ integer-valued matrices
\begin{align*}
g=  \bem a & b \\ c & d \eem \quad \textrm{such that}
\quad g^t J g =J \quad \textrm{with}
\quad J = \bem 0 & -I \\ I & 0 \eem,
\end{align*}
where $a$, $b$, $c$ and $d$ are $N \times N$ matrices.}, 
it is called a symplectic basis of $X$.


The complex structure periods of
$X$ with respect to this basis are defined as  
\begin{align}
X^i = \int_{A^i} \Omega, \quad \cF_{0,i} = \int_{B_i} \Omega.
\end{align} 
The tuple $(X^i, \cF_{0,i})$ forms a local section of an $Sp(b^3,
\Z)$-bundle over the moduli space $\M_V$, whereas the quotients 
\begin{align*}
 Z^i = \frac{X^i}{X^0}, \qquad i = 1, \ldots, h^{2,1} 
\end{align*}
yield a set of local coordinates on $\cM_V$. 

Since the variation of a $(3,0)$-form just contains a $(3,0)$ piece
and a $(2,1)$ piece,
%
$ \int_X \Omega \wedge \partial_i \Omega = 0.$
This implies that the functions $\cF_i(X)$ are in fact first derivatives 
\begin{align*}
\cF_{0,i}(X) = \partial_i \cF_0(X), \quad \cF_0(X^i) = \frac{1}{2} X^i \cF_{0,i}(X)
\end{align*}
of a homogeneous function $\cF_0$ of degree 2 in the complex moduli $X =
\{ X^i \}$. This is the prepotential. Geometrically, this means that
the image of the period 
map, which sends 
\begin{align*}
 \Omega \mapsto (X^i, \cF_{0,i}),  
\end{align*}
is a Lagrangian submanifold of $H^3(X, \C)$. 

The Riemann bilinear identity
$\int_X \alpha \wedge \beta = \sum_i \left(
\int_{A^i} \alpha \int_{B_i} \beta - \int_{A^i} \beta \int_{B_i}
\alpha \right)$
immediately shows that
the K\"ahler potential is fully determined by the prepotential~$\cF_0$:
\begin{align}\label{eqn:kahler+prepotential}
  \cK(X,\overline{X}) =  - \log \,i  \left(   X^i \overline{\cF}_{0,i}  -
    \overline{X}^i \cF_{0,i}
  \right). 
\end{align}

\subsubsection{Decoupling gravity}\label{sec:decouplegravity}

In general, to make contact with the 4-dimensional world around us,
we would like to consider compactifications with a relatively small
Calabi-Yau. The Planck scale in four dimensions is then proportional to
the size of the threefold (and inversely related to its curvature). 

However, when we are purely interested in the gauge dynamics of the
reduced theory, we want to \emph{decouple} gravity in 4d. This is
commonly done by zooming in on the part of the complex moduli space
where the Calabi-Yau develops a singularity. In such a region of the
moduli space a particular set of cycles becomes very small. (Think of
the singularities on the quantum
$SU(2)$ Seiberg-Witten moduli space $\cM_q$ where 1-cycles degenerate.)  

In four dimensions this results in a new
lower energy scale, governing the dynamics of the fields that
correspond to the singularity. Zooming in close to the singularity, we
can forget about the rest of the Calabi-Yau so that we are
effectively studying non-compact Calabi-Yau
compactifications. Depending on the type of the compactification,
wrapping branes around vanishing 2, 3 and 4-cycles yield massless BPS vector
and hyperparticles, that can engineer any gauge symmetry and matter
content that one is interested in. 

In this non-compact limit the local special geometry of $\cN=2$ supergravity
reduces to rigid special geometry, relevant for the low energy
dynamics of $\cN=2$ gauge theories (more details
e.g. \cite{Seiberg:1994rs, Craps:1997gp, Billo:1998yr}). In
particular, the K\"ahler  
potential reduces to    
\begin{align}
\cK = i \int_X \Omega \wedge \overline{\Omega}. 
\end{align}
so that the metric is given by 
\begin{align}
 g_{i \bar{j}} =  \textrm{Im} \left( \frac{\lotjesd^2 \cF_0}{\lotjesd X^i
     \lotjesd X^j} \right). 
\end{align}
This is indeed in agreement with the Seiberg-Witten expression
(\ref{eqn:gaugekahlerpot}).

\subsubsection{Normalizable and non-normalizable modes}

An important point is the distinction between normalizable and
non-normalizable complex structure moduli in the case of noncompact
Calabi-Yau manifolds. To be more precise let us consider the local
Calabi-Yau modeled on a Riemann surface $\Sigma$. The coefficients
of the polynomial $H(x,y)$ 
characterize the complex structure of $X$, so that they are the
complex structure moduli of $X$. However not all of them are
dynamical.  Some of them control the complex structure of 3-cycles
which are localized in the ``interior'' of the singularity and are
dynamical, while others describe how the singularity is embedded in
the bigger Calabi-Yau and become frozen when we take the decoupling
limit.
 
To determine if a specific complex structure modulus $X$
is dynamical or not, one has to compute the corresponding K\"ahler metric
\begin{equation}
  g_{X\overline{X}} = \partial_X \overline{\partial_X}K = i 
\partial_X \overline{\partial_X}\int \Omega \wedge \overline{ \Omega}.
\label{metriccomplex}
\end{equation}
If this expression is finite, then the modulus $X$ is dynamical,
otherwise it is decoupled and should be treated as a parameter of the
theory.  We will refer to the first set of moduli as
\emph{normalizable} and to the second as \emph{non-normalizable}.

\subsubsection{Type IIA vector moduli space}

Similar to how the complex structure moduli enter in the IIB vector
moduli space, the K\"ahler moduli parametrize the IIA vector moduli
space. It is convenient to enlarge the moduli space to all
complex even-dimensional forms  on the Calalabi-Yau threefold $\widetilde{X}$. The K\"ahler periods of $\widetilde{X}$ are then found
with respect to the element
\begin{align}
 \Omega = \exp(t) \in H^{2*}(\widetilde{X}, \C),
\end{align}
where $t = B + iJ $ denote the complexified K\"ahler form on the
Calabi-Yau. When integrated over a
basis of two-cycles $C^i$  this yields the K\"aher parameters
%
$ t^i = \int_{C^i} t$, for $ i = 1, \ldots, h^{1,1}$. 
%
More precisely, these are the normalized complex K\"ahler parameters $t^i =
X^i/X^0$. 

The K\"ahler potential is given by
\begin{align*}
 \cK = - \log \, i \left(  \int \Omega \wedge \tilde{\overline{\Omega}} \right),
\end{align*}
where $\tilde{~}$ above $\overline{\Omega}$ reminds us that we should
include a minus sign for components in $H^{1,1}$ and $H^{3,3}$ (to
define a Hermitean inner product). The K\"ahler potential can then be
written in terms of a holomorphic function $\cF_{0,\textrm{clas}}$ as in
equation~(\ref{eqn:kahler+prepotential}). It equals the
triple intersection number  
\begin{align*}
 \cF_{0,\textrm{clas}}(t) = \frac{1}{6} \int_{\widetilde{X}} t \wedge t \wedge t.
\end{align*}
in the normalized coordinates $t$. 
As we already suggested in our notation, in a type IIA
compactification this only determines the full 
prepotential $\cF_0$ up to non-perturbative corrections in $\alpha'$. 
 
Both type II vector multiplet moduli spaces are related by
\Index{mirror symmetry}, which exchanges a type IIB compactification
on a Calabi-Yau threefold $X$, with Hodge numbers $h^{1,1}$ and
$h^{2,1}$, with a type IIA compactification on its mirror
$\widetilde{X}$, whose Hodge numbers are flipped.
There is a precise mirror map between
the complex structure moduli $Z^i$ on $X$ and the K\"ahler moduli
$t^i$ on its mirror $\widetilde{X}$, that determines the full IIA
prepotential in the large radius limit to be of the form 
\begin{align}\label{eqn:instantoncontr} 
\cF_0(t) = \frac{1}{6} \int_{\widetilde{X}} t \wedge t \wedge t 
 -i \frac{\chi}{2(2 \pi)^3} \zeta(3)
 +  \sum_{d \in H_2(\tilde{X})} \textrm{GW}_{0}(d)~
  e^{-\frac{1}{\alpha'} \sum_i d_i t^i} 
\end{align}
where $\textrm{GW}_0(d)$ are the so-called genus zero
Gromov-Witten invariants. 
They count the number of worldsheet instantons in the K\"ahler class
$d = \sum_i d_i C^i$. Gromov-Witten invariants are one type of
topological invariants on a Calabi-Yau threefold. They give a
mathematical definition of what is called the A-model in topological string
theory. This is the topic of \Cref{sec:topstringtheory}.

%
%


\subsection{Topological string theory}\label{sec:topstringtheory}

Topological strings were introduced by Witten in \cite{Witten:1988xj} as a
twisted version of a supersymmetric 2-dimensional sigma model  
\begin{align*}
Z_{\sigma}(X;t) = \int D x \, D g_{C} ~\exp \left( - t
  \int_{C}\, g_{mn} d x^{m} \wedge * d  x^{n} + \ldots \right),  
\end{align*}
that integrates all maps $x$ of a worldsheet $C$ into the
target manifold $X$ with metric $g_{mn}$.\footnote{Be careful not to
  confuse the worldsheet $C$ in the sigma-model with the target
  space curve $\Sigma$ in the local Calabi-Yau $X_{\Sigma}$.} The maps are weighted by the
volume of $x(C)$.  
The extra integral over all 
metrics $g_{C}$ on $C$ couples the sigma model to gravity.
When $X$
is a Calabi-Yau threefold the sigma model is conformally invariant and
$\cN=(2,2)$ supersymmetry on the worldsheet is preserved.  
 
Similarly as in \Cref{sec:tqft}, the sigma model is twisted by
choosing an embedding of the worldsheet Lorentz group $U(1)$ into the
R-symmetry group $U(1)_R$ of the supersymmetry algebra. Essentially
this gives two possibilities, which are called the A-model and the
B-model. Precisely as in \Cref{sec:tqft} both models are characterized
by their BRST operators, which make sure the amplitudes are
independent of the metric $g_{mn}$ of $X$. In fact, one finds that the A-model
only depends on the $h^{1,1}$ K\"ahler moduli of $X$, while the B-model is
just based on the $h^{2,1}$ complex structure moduli of $X$. Mirror
symmetry equates the B-model on the Calabi-Yau $X$ and the A-model on
its mirror $\tilde{X}$.    

There are many excellent reviews on topological string theory and mirror symmetry, see for instance \cite{Vonk:2005yv, Neitzke:2004ni, notesKlemm, horimirrorsymmetry, coxkatz}.

\subsection{Gravitational corrections}\label{sec:gravcorr}
\index{gravitational coupling}

How do topological
string amplitudes pop up in the physical string?   
In \Cref{sec:SWcurves} we wrote down the most general
Lagrangian (\ref{eqn:n=2f-term}) one can build out of any number of
$\cN=2$ vector multiplets $\mathbf{X}^i$ (which contain the scalars $X^i$ as
lowest components). A Calabi-Yau compactification furthermore produces an
$\cN=2$ gravity multiplet, whose components may be recombined in the
so-called Weyl multiplet $\cW_{\mu \nu}$. This multiplet is expanded 
as
\begin{align*}
\cW_{\mu \nu} = F^+_{\mu \nu} - R^+_{\mu \nu \la \rho} \theta \sigma^{\la \rho} \theta + \ldots,  
\end{align*}
where $F^+$ is the self-dual part of graviphoton field strength and
$R^+$ the self-dual Riemann tensor. In the 4-dimensional low energy
effective action $\cW_{\mu \nu}$ appears together with the $\mathbf{Z}^i$'s in
F-terms of the form    
\begin{align*}
\int d^4x \int d^4 \theta ~ \cF_g(\mathbf{X}^i) ( \cW^2)^g,  
\end{align*}
which clearly generalize the Lagrangian (\ref{eqn:n=2f-term}) that only
captures the prepotential
$\cF_0$. Expanding the above Lagrangian 
in components yields the terms 
\begin{align}
\int d^4 x ~\cF_g(X^i)\, R^2_+ \,F_+^{2g-2}.  
\end{align}
In other words, the $F_g$'s are coefficients in a gravitational
correction to the amplitude for the scattering of $2g-2$ graviphotons.  
In \cite{Bershadsky:1993cx,Antoniadis:1993ze} it was shown that these
coefficients may be computed in topological string theory. The vector
multiplets in a IIB compactification parametrize complex structure
moduli, so that the above F-terms compute B-model free energies $\cF_g$,
whereas in a type IIA compactification the F-terms generate A-model
invariants.  

These gravitational couplings are crucial in the study of
supersymmetric black holes
\cite{Gopakumar:1998ii,Gopakumar:1998jq,Katz:1999xq,Ooguri:2004zv,Vafa:2004qa,Aganagic:2004js}. But
even when we decouple gravity by zooming in on the singularity of a 
Calabi-Yau (the topic of the next paragraph), they appear as meaningful
holomorphic corrections to the effective 
4-dimensional gauge theory \cite{Dijkgraaf:2002fc,Dijkgraaf:2002vw,
Dijkgraaf:2002dh}.

\subsubsection{A-model and Gromov-Witten invariants}

The A-model enumerates degree $d$ holomorphic maps $x: C_g \to
\tilde{X}$ from a 
genus $g$ worldsheet $C_g$ into the Calabi-Yau $\tilde{X}$. These
are captured by the 
\emph{Gromov-Witten invariants}\index{Gromov-Witten invariant} 
\begin{align*}
 GW_{g}(d) =  \int_{\overline{\cM}_{g}(\tilde{X},d)]^{\textrm{vir}}} 1 \in \Q,
\end{align*}
that ``count'' the number of points in their compactified moduli space
$\overline{\cM}_{g}(\tilde{X},d)$ (which is rigorously defined in \cite{behrend, li-tian}). In a large radius
limit the partition function of the A-model has a genus expansion  
\begin{align}\label{eqn:freeenergy}
Z_{GW}(t;\la) = \exp \left( \cF(t;\la) \right);  \qquad \cF (t;  \la) = \sum_{g \ge 0} \la^{2g-2} \cF_{g}(t),
\end{align}
where we introduced the topological string coupling constant $\la$
that keeps track of the worldsheet genus $g$, and where we absorbed the
coupling constant $t$ in the complexified K\"ahler class  $t$. The 
$\cF_g$'s can be expressed in terms of the Gromov-Witten invariants
\begin{align}
 \cF_g(t) = \sum_d \textrm{GW}_g(d) e^{d \cdot t} + \left\{ \begin{array}{cc}
     \frac{1}{6 } \int_X t\wedge t \wedge t & (g=0) \\ \frac{1}{24}
     \int_{\tilde{X}} c_2(\tilde{X}) \wedge 
     t & (g=1)\end{array} \right.,   
\end{align}
with the notation $d \cdot t = \int_{d} x^*(t)$. 
In particular, the genus zero contribution $\cF_0$ equals the
prepotential~(\ref{eqn:instantoncontr}).

\subsubsection{A-model on toric Calabi-Yau threefolds}\label{sec:topvertex}

In \cite{Aganagic:2003db} the A-model on a non-compact toric
Calabi-Yau threefold is solved by taking advantage of the large $N$
open-closed string duality between Chern-Simons theory on $S^3$ and the A-model
on the resolved conifold \cite{Gopakumar:1998ki}. It was discovered
that the all-genus free energy can be  
computed by splitting the toric target manifold in $\C^3$-patches, as
discussed in \Cref{sec:toricgeometry}. On each $\C^3$-patch
the worldsheet instantons localize onto the trivalent vertex in the
base of the $T^2 \times \R$-fibration, where at least
one of the cycles of the $T^2$-fibration degenerates. In fact,
the topological string amplitude on such a patch is an open string
amplitude. 

The boundary conditions are specified by so-called
A-branes that sit at the end points of the trivalent graphs. Supersymmetry
conditions determine these A-branes to wrap special Lagrangian cycles
in the Calabi-Yau threefold. In non-compact toric Calabi-Yau's such
cycles end on the degeneration locus and wrap the $T^2$-fiber
together with a half-line in the base transverse to the trivalent
graph. Since one of the $T^2$-cycles degenerates at the
graph the topology of the A-brane is $S^1 \times \C$. It can only
move up and down a toric edge. Open strings may wrap the $S^1$ of
each A-brane an arbitrary number of times.  

In total there can be three stacks of A-branes at the  toric legs
of the trivalent vertex, which the open string wraps. With 
the Frobenius relation this winding information can be rewritten in
terms of representations of the Lie group $U(\infty)$. The open topological
string amplitude on each $\C^3$-patch yields the \Index{topological vertex}
expression 
\begin{align}
C_{R_1 R_2 R_3} =  \sum_{Q_1, Q_3} N_{Q_1 Q_3^t}^{R_1 R_3^t} q^{\kappa_{R_2}/2 +
    \kappa_{R_3}/2} \frac{W_{R_2^t Q_1}W_{R_2 Q_3^t}}{W_{R_2 0}},
\end{align}
where the $R_i$ are $U(\infty)$ representations, the numbers $N$ are
integers that count the number of ways that the representations $Q_1$
and $Q_3^t$ go into $R_1$ and $R_3^t$, and the $W$'s are link
invariants for the $U(\infty)$ Hopf link. For more details we
refer to reference \cite{Aganagic:2003db}. 
 
The full closed topological string amplitude is found by multiplying the
topological vertices. When two $\C^3$-patches are glued
along a toric edge, one has to sum over the attached representation
$R$ with weight $\exp \left(-t l(R) \right)$, where $t$ is the
K\"ahler parameter of the toric edge and $l(R)$ the number of boxes in
the Young diagram corresponding to $R$.  
We will see examples of such computations in Chapter~\ref{chapter6}
and Chapter \ref{chapter7}.

\subsubsection{B-model and Kodaira-Spencer theory}

The free energy of the B-model has a similar genus expansion as in
(\ref{eqn:freeenergy}) in a large complex structure limit, that is
related to the A-model expansion by 
the mirror map. But in contrast to the A-model, the B-model is best
understood from a target space 
point of view, where it is described as a
\emph{Kodaira-Spencer}\index{Kodaira-Spencer theory} path
integral over complex structure deformations of
$X$ \cite{Bershadsky:1993cx} 
\begin{align*}
 Z = \int D \phi \exp \left( \int \lotjesd \phi \wedge \overline{\lotjesd} \phi +
   \la A^3 \right).
\end{align*}
Here, the complex structure field $A = X + \lotjesd \phi \in H^{2,1}$ is
expanded around a background complex structure
specified by $Z \in H^{2,1}(X)$, so that $\phi \in H^{1,1}(X)$ describes  
cohomologically trivial complex structure fluctuations. The
interaction term schematically denoted by $A^3$ is more carefully
equal to $\Omega \cdot (A' \wedge A') \wedge A$, where $A' \in
H^{0,1}(T_X)$ is uniquely determined $\Omega \cdot A' = A$. 

This
partition function quantizes complex structures as its equation of
motion is the Kodaira-Spencer equation
\begin{align*}
 \overline{\lotjesd} A' + \frac{1}{2}[A',A']=0, 
\end{align*}
which is condition for a complex structure $\overline{\lotjesd} + A' \cdot \lotjesd$ to
be integral. 
At tree level the free energy is captured by special geometry in terms
of the periods 
\begin{align*}
 X^i = \int_{A^i} \Omega, \qquad \frac{\lotjesd \cF_0}{ \lotjesd X^i} = \int_{B_i}
\Omega 
\end{align*}
as described in \Cref{sec:specialgeometry}. At one
loop level the free energy computes the logarithm of the
\emph{analytic (Ray-Singer) torsion} of $X$ 
\cite{Bershadsky:1993cx}  
\begin{align}\label{eqn:Ray-Singer}
 \cF_1 = \sum_{p,q=0}^3 \frac{1}{2} pq (-)^{p+q} \log \mbox{det}' \Delta_{p,q},  
\end{align}
where the prime $'$ refers to subtracting the zero modes. Higher
$\cF_g$'s don't have such a neat interpretation; they just give quantum
corrections to special geometry.  

\subsubsection{B-model on local Calabi-Yau modeled on a curve}

\index{integrable hierarchy}
\index{free fermions}
\index{$\cW_{1+\infty}$-algebra}
\index{complex structure deformations}
\index{topological B-brane}
\index{complex symplectic plane}
\index{Ward identity}

Kodaira-Spencer theory on a local Calabi-Yau $X_{\Sigma}$ reduces to a
2-dimensional  
quantum field theory on the curve $\Sigma$ that
quantizes the complex structure deformations of $\Sigma$
\cite{Aganagic:2003db, Aganagic:2003qj}. These complex structure variations are
controlled by a bosonic chiral field $\phi$ such that 
\begin{align}\label{eqn:1formboson}
 \eta = \del \phi(x), 
\end{align}

Let us assume for a moment that the complex
structure of $\Sigma$ can only be changed at its $k$ asymptotic
infinities (so $\Sigma$ is a genus zero curve), and that these regions are
described by a local coordinate $x_i$ such that $x_i \to \infty$ at
the boundary. Near each boundary $\lotjesd \Sigma_i$ the chiral boson $\phi$ admits a mode
expansion
\begin{align*}
 \lotjesd \phi(x) = \sum_{n \in \Z} \alpha_n x_i^{-n-1}, \qquad [\alpha_n.
 \alpha_m] = n \la^2 \delta_{n+m,0}  
\end{align*}
(When $x_i$ is a periodic coordinate, the series should be expanded in $e^{nx_i}$.)
 
Denote the Hilbert space of this free boson by $\cH$. The B-model free
energy sweeps out a state 
\begin{align}
|\cW \rangle \in \cH^{\otimes k}. 
\end{align}
At each boundary $\lotjesd \Sigma_i$ we can introduce a coherent state
\begin{align*}
 | t^i \rangle = \exp \left( \sum_{n>0} t^i_n \alpha_{-n} \right) |0 \rangle, 
\end{align*}
where the $t^i_n$'s represent the complex structure deformations at
the corresponding asymptotic infinity. Roughly, $t^i_n$ corresponds to
complex structure variations
$y_i \sim x_i^{n-1}$ at the $i$th boundary when $n>0$. It follows that
the B-model 
partition function can be written as the correlator   
\begin{align}
  \exp \cF(t^1, \ldots, t^k) = \langle t^1 | \otimes \ldots \otimes
 \langle t^k |\cW \rangle. 
\end{align}

In \cite{Aganagic:2003qj} it is discovered that there is a large
symmetry group that acts on this theory, whose broken symmetries
generate an infinite sequence of Ward identities that are strong
enough to 
determine the free energy $\cF$ as a function of the deformation
parameters $t^i_n$. In three complex dimensions these
symmetries are the global 
diffeomorphisms that preserve $\Omega$. In two complex dimensions they
reduce to diffeomorphisms that preserve the complex symplectic form $dy \wedge
dx$ on $\cB$. Obviously, these symmetries are broken by
$\Sigma$. This gives $\phi$ an interpretation as Goldstone boson.  

Geometrically, the Lie algebra of infinitesimal tranformations is
generated by Hamiltonian vector fields
\begin{align*}
 \delta x_i = \frac{\lotjesd f(y_i,x_i)}{\lotjesd y_i}, \quad \delta y_i = - \frac{\lotjesd
   f(y_i,x_i)}{\lotjesd x_i},  
\end{align*}
with the Hamiltonian $f(y,x)$ being any polynomial in $y$ and
$x$. This is known as the $\cW_{1+\infty}$-algebra. On the quantum
level any change of local coordinates $\delta x_i = \epsilon(x_i)$ is
implemented by the operator 
\begin{align*}
\oint \epsilon(x_i) T(x_i) dx_i  
\end{align*}
acting on the Hilbert space, where $T(x)= (\del \phi)^2/2$ is the
energy momentum tensor of the free boson CFT. Similarly, when $f(y_i,x_i)=
y_i^n x_i^m$ at the $i$th asymptotic infinity, we find a quantum
generator 
\begin{align*}
 W^{n+1}_{m}(x_i) \sim \oint_{\lotjesd \Sigma_i} x_i^m \frac{(\lotjesd \phi)^{n+1}}{n+1}.  
\end{align*}
These generators are the modes of the $\cW$-current 
$W^{n+1}(x) \sim
(\del \phi)^{n+1}/(n+1)$, that includes the Virasoro current for $n=1$.  

To write down the Ward identities, corresponding to the broken
$\cW_{1+\infty}$-symme-try, we first need to see how the
complex structure deformations in the various asymptotic regions are
related. It turns out that this is highly non-trivial, and can be best
understood in terms of the topological B-branes in the geometry $X_{\Sigma}$.  
 
Topological B-branes wrap holomorphic cycles of the Calabi-Yau
geometry. In the local geometry $X_{\sigma}$ complex 1-dimensional
branes are parametrized by $\Sigma$. These branes wrap the non-compact
fiber $uv=0$ over the curve $\Sigma$. More precisely, such a fiber is
degenerated into two parts $v=0$ and $u=0$, which are wrapped by a
brane and an anti-brane resp. Inserting such a brane at position $p$
on an asymptotic
infinity deforms the meromorphic 1-form $\eta$ by a small amount
%
$ \oint_p \eta = \la,$
%
where the contour encircles the point $p$ on $\Sigma$. Since we
identified $\eta$ with $\del \phi$ in~(\ref{eqn:1formboson}), we find the
equality 
\begin{align*}
\langle \cdots \oint_p \del \phi(x_i)~ \psi(p) \cdots \rangle = \la
~\langle \cdots \psi(p) \cdots \rangle.  
\end{align*}
Now the OPE 
%
$ :\psi^*(x_i) \psi(p):~\sim  1/(x_i-p),$ when $ x_i \to p,$ 
 shows that the B-brane corresponds to the fermion $\psi$ by the
familiar bosonization rules 
\begin{align*}
  \quad :\psi(x_i) \psi^*(x_i): ~= \del \phi(x_i)/\la, \quad \psi(x_i) = \exp
  \left( \phi(x_i)/\la \right). 
\end{align*}
Similarly, an anti-brane corresponds to its conjugate $\psi^* = \exp
\left(- \phi/\la \right)$. 

This has some very interesting
consequences. It is argued in  \cite{Aganagic:2003qj} that the brane
partition function 
%
 $\Psi(x_i) = \langle \psi(x_i) \rangle, $
%
for a B-brane inserted at a point $x_i$ on the $i$th asymptotic patch,
transforms to another asymptotic region as a \emph{wave
  function} (similar to 
the closed partition function in \cite{Witten:1993ed}, see also
\cite{KashaniPoor:2006nc,Bouchard:2007ys}). As a result the 
fermion fields in both patches are related in a Fourier-like way
\begin{align}\label{eqn:fourierfermion}
  \psi(x_j) = \int dx_i~ e^{S(x_i,x_j)/\la} \psi(x_i),
\end{align}
where $S$ is a gauge transformation that relates the symplectic
coordinates $(y,x)$ in both patches  
\begin{align*}
 y_i dx_i - y_j d x_j = dS(x_i,x_j). 
\end{align*}

Notice that when the local coordinates are related by a symplectic
transformation 
$Sp(1, \Z) \cong  SL(2,\Z)$, the exponent $S(x_i,x_j)$ is a homogeneous
quadratic function in $x_i$ and $x_j$, so that the wave-function
transforms in the metaplectic representation.  

%
%

%

Putting everything together we find the Ward identities \cite{Aganagic:2003qj}
\begin{align*}
 \oint_{\lotjesd \Sigma_i} \psi^*(x_i) x_i^m  y_i^n &\psi(x_i) | \cW \rangle =
 \notag \\
& -
 \sum_{j \neq i} \oint_{C_j} \psi(x_j)^* x_i(x_j,y_j)^m
 y_i(x_j,y_j)^n \psi(x_j) | \cW \rangle. 
\end{align*}
By bosonizing the fermionic fields and contracting the above state with a
coherent state $\langle t|$, this yields a set of differential
equations that can be solved recursively in the worldsheet genus $g$. 

Let us recapitulate: Topological B-branes that end on $\Sigma$ behave
as free fermions on $\Sigma$. However, their transformation properties
from patch to patch are rather unusual. In this thesis we examine
these free fermions from the I-brane perspective. In
\Cref{fermionsBPS} we relate the B-model fermions to the ones on the
I-brane, whereas Chapter~\ref{chapter5} is devoted to explaining
their characteristics from the I-brane point of view.



\section{BPS states and free
  fermions}\label{sec:BPSstatesfreefermions}  

It is time to take another look at the web of dualities in
Fig.~\ref{fig:webofdualities}. The lowest 4 boxes in the vertical
sequence of the web are all related to Calabi-Yau compactification
down to a 4-dimensional Taub-NUT manifold. The middle ones correspond
to topological string theory, in the sense that the holomorphic
F-terms on the Taub-NUT compute topological string amplitudes. The
lowest ones are also familiar as string theory realizations of other
types of topological invariants on Calabi-Yau threefolds, called
Gopakumar-Vafa (GV) invariants and Donaldson-Thomas (DT) invariants. We are
particularly interested in the DT perspective, as this leads to a
natural I-brane partition function, see \Cref{sec:Ibranepartfunc}.

\subsection{BPS states in the 9-11 flip}

Let us start by going through the 9-11 flip in the vertical sequence
once again. This time we pay attention to the topological invariants
that are associated to these duality frames \cite{Dijkgraaf:2006um}. 

\subsubsection{Donaldson-Thomas theory and D0-D2-D6 bound
  states}\label{sec:DT}  

Instead of studying holomorphic embeddings $\phi$ of a worldsheet
$\Sigma_g$ into the Calabi-Yau threefold $\tilde{X}$, one can also think of
its image $\phi_*([\Sigma_g]) \subset H_2(\tilde{X})$ as the zero-locus of a
finite set of holomorphic functions on $\tilde{X}$. Mathematically
this is rigorously 
defined in terms of rank 1 ideal sheaves $\cE$. On a Calabi-Yau
$\tilde{X}$ ideal 
sheaves $\cE$ with $ch_2 = d$ and $c_3=n$ can be counted as the
\emph{Donaldson-Thomas invariants} \index{Donaldson-Thomas invariant}
\begin{align*}
 \textrm{DT}_{n,d} = \int_{[\cE_n(\tilde{X},d)]^{\textrm{vir}}} 1 \in \Z,  
\end{align*}
since their moduli space $[\cE_n(\tilde{X},d)]^{\textrm{vir}}$ has
virtual dimension 
zero \cite{donaldson-thomas,dt-thomas}. Their generating function is
\begin{align}\label{eqn:DTpartfunction}
 Z_{\textrm{DT}}(t, e^{\la}) = \sum_{d,n} \textrm{DT}_{d,n} e^{-d \cdot t} e^{n \la} 
\end{align}
The degree zero Gromov-Witten and Donaldson-Thomas partition functions
are conjectured to be related as $Z_0^{\textrm{DT}} (t, e^{\la}) = \left(
  Z_0^{\textrm{GW}} (t, \la) \right)^2$,
whereas their reduced parts $Z' = Z/Z_0$ are conjectured
mathematically to be equal 
\begin{align*}
 {Z}_{\textrm{DT}}'(t, e^{\la}) = {Z}_{\textrm{GW}}'(t,\la). 
\end{align*}
This has been
verified in some important cases, such as toric Calabi-Yau
threefolds \cite{maulik-2003,maulik-2004,maulik-2008}.  

Physically, these invariants can be thought of as capturing D0-D2 bound states
on a single D6 brane.
In the maximally supersymmetric
$U(1)$ worldvolume theory on the D6-brane these bound states
correspond to non-trivial topological sectors of the rank 1 gauge
bundle $\cE$: the D2 charge $d$ equals $ch_2(E)$, while
the D0 charge $n$ equals $ch_3(E)$. So $\textrm{DT}_{d,n}$ should correspond to
the number of bound states of a D6 brane with D2-branes that are
wrapped around a cycle $d \in H_2(\tilde{X})$ and $n$ added D0-branes. 

When we wrap the D6-brane around a Calabi-Yau threefold $\tilde{X}$ the
worldvolume gauge theory is naturally twisted. On $\tilde{X} \times S^1$ its
partition function is given by the Witten index
\begin{align*}
 Tr[(-1)^F e^{-\beta H}] 
\end{align*}
where $\beta$ is the radius of $S^1$. Since the theory is topological
we can take the limit $\beta \to 0$ to find a 6-dimensional
maximally supersymmetric $U(1)$ gauge theory. The D0-D2 bound states 
are captured by the topological terms 
\begin{align}\label{eqn:topologicalterms}
 S = \la \int_{\tilde{X}} c_3(\cE) + \int_{\tilde{X}} t \wedge ch_2(\cE), 
\end{align}
where $t$ denotes the K\"ahler class of $\tilde{X}$.  Since the only relevant
contributions in this topological gauge theory are given by the
above topological terms, 
we find the Donaldson-Thomas partition function (\ref{eqn:DTpartfunction}).
%
%
The mathematical relationship between GW and DT invariants corresponds
physically to a gauge-gravity correspondence between the
6-dimensional $U(1)$ gauge theory and topological string theory
\cite{Iqbal:2003ds}.  


In string theory this configuration is embedded in the type IIA background
\begin{align*}
\textrm{IIA}: \quad \tilde{X} \times \R^3 \times S^1_{\beta} 
\end{align*}
with a single D6-brane wrapped on $\tilde{X} \times S^1_{\beta}$ bound to D2 
and D0-branes. 

Notice that in our duality web the Taub-NUT space $TN$ may have an arbitrary
number $k$ of nuts. In the remainder of this thesis we restrict to
$k=1$ just because that makes it easier to write down generating
functions for BPS states. It is in
fact a striking question how to use the I-brane web of dualities to
write down a partition 
function that holds for any $k$.\footnote{The obstacle in writing down higher-rank Donaldson-Thomas invariants is the definition of a nice moduli space. Lately, this has been studied in \cite{joyce-2008}. It would be interesting to 
  find out how this generalized Donaldson-Thomas theory is related to our web of dualities.}

\subsubsection{The 9-11 flip and Gopakumar-Vafa invariants}

As the D6-brane dissolves into a non-trivial Taub-NUT
geometry, this configuration lifts to M-theory as 
\begin{align}\label{eqn:GVmtheory}
\textrm{M}: \quad \tilde{X} \times TN \times S^1_{\beta}, 
\end{align}
accompanied by M2-branes and KK momenta along the Taub-NUT circle
$S^1_{TN}$. Note that instead of reducing over $S^1_{TN}$ to go down
to IIA, it is also possible to shrink the 
thermal circle $S^1_{\beta}$. This yields the background 
$ \tilde{X} \times TN$ with fundamental strings wrapping 2-cycles of the
Calabi-Yau $X$. 
Since the 5-dimensional angular momentum 
around $S^1_{\beta}$ couples to the graviphoton field, the Witten
index will in this configuration compute the A-model partition
function $Z_{\textrm{GW}}(t,\la)$ \cite{Dijkgraaf:2006um}.

Moreover, in \cite{Dijkgraaf:2006um} it is argued that this
$S^1_{TN}$-$S^1_{\beta}$ flip (9-11 flip)  encodes
 yet another interpretation of the topological string partition
function in terms of  the \emph{Gopakumar-Vafa invariants}.
\index{Gopakumar-Vafa invariant} These integer invariants 
\begin{align*}
\textrm{GV}^m_{d} \in \Z
\end{align*} 
count the number of $M2$-branes wrapped over a 2-cycle $d$ in the
Calabi-Yau $\tilde{X}$ in a compactification of M-theory on
$\tilde{X}$ to five dimensions
\cite{Gopakumar:1998ii,Gopakumar:1998jq,Katz:1999xq}.\footnote{Mathematically
  it is not so easy to define 
  Gopakumar-Vafa invariants rigorously. Recent attempts can be found
  in \cite{katz-2006, pandharipande-2007}.}  
The integer $m$ stands for the left spin  
content of the corresponding 5-dimensional BPS particle. More
accurately, its spin $(m_L,m_R) = (2j_L^3,2j_R^3)$ takes value in the
5-dimensional rotation group $SO(4) = SU(2)_L \times SU(2)_R$, but to
find a proper complex structure
invariant one should sum over the right spin content. Compactifying
the fifth dimension gives the BPS particles an extra angular momentum. 

Let us go back to the M-theory set-up (\ref{eqn:GVmtheory}).
In the strong coupling limit in which the radius $R$ of the Taub-NUT
circle $S^1_{TN}$ is 
large, we effectively zoom in on the center of the Taub-NUT. Near this
center the geometry cannot be distinguished from $\R^4$, so that the
M2-branes form a free gas of spinning BPS particles in $\R^4$. Their
spin receives a internal contribution $m$ from the $U(1) \subset
SU(2)_L$ isometry 
of the Taub-NUT plus an orbital contribution $n$ that originates from
the KK-momenta along the $S^1_{TN}$.
This motivates the Gopakumar-Vafa partition function
\begin{align}\label{eqn:GVpartitionfunction}
 Z_{\textrm{GV}}(t, \la) = \prod_{d \in H_2} \prod_{m \in \Z} \prod_{k=-m}^m
\prod_{n=1}^{\infty} (1-e^{\la(k+n)} e^{t \cdot d})^{(-1)^{m+1}n \, \textrm{GV}^{m}_d},
\end{align}
as a second quantized product over
the spinning BPS states
\cite{Dijkgraaf:2006um,Hollowood:2003cv,Gopakumar:1998ii}. 
The equivalence of the GV partition function  $Z_{\textrm{GV}}(t,
\la)$ to the GW partition function
$Z_{\textrm{GW}}(t,\la)$ conjectures the integrality of the Gopakumar-Vafa
invariants $\textrm{GV}^m_{d}$.  

So although the Gromov-Witten invariants are not integer, and thus
cannot be made sense of as counting BPS states in string theory,
their generating function does allow a reformulation in terms of the
integer-valued Donaldson-Thomas invariants and Gopakumar-Vafa
invariants, which do have a BPS-interpretation in string theory and
are connected by this 9-11 flip.

\subsubsection{Degree zero example}

Let us give a simple example of these dualities (more can be found in
e.g. \cite{notesKlemm,katz-2004}). In Gromov-Witten theory the  
constant map contributions are captured by the generating function
\begin{align*}
 \textrm{GW}_g(0) = \frac{(-1)^g \chi(\tilde{X})}{2} \int_{\overline{\cM}_g} c_{g-1}^3 =
 \frac{(-1)^g \chi(\tilde{X})}{2} \frac{|B_{2g}||B_{2g-2}|}{2g(2g-2)(2g-2)!} 
\end{align*}
where $c_{g-1}^3$ is
$(g-1)$th Chern class of the Hodge bundle over $\cM_g$ (the moduli
space of Riemann surfaces of genus $g$)  and $B_n$ are
the Bernouilli numbers. The
corresponding partition function $Z_{\textrm{GW},0}$ can be
rewritten as a Gopakumar-Vafa sum
\begin{align*}
Z_{\textrm{GV},0}  
  = \prod_{n \ge 0} \left( 1-e^{\la n} \right)^{-n \chi (\tilde{X})} 
= \prod_{n \ge 0} \left( 1-e^{\la n} \right)^{n \textrm{GV}_{0,0}}.
\end{align*}
This  
generating function is a power of the MacMahon function 
\begin{align*}
M(q) = \prod_{n \ge   1} (1-q^n)^{-n},
\end{align*}
which counts 3-dimensional
partitions, and as such appears in the localization of the
6-dimensional gauge theory and the computation of the
Donaldson-Thomas invariants \cite{Iqbal:2003ds,
  Okounkov:2003sp, Cirafici:2008sn}.  


\subsection{Fermions and BPS states}\label{fermionsBPS}

Let us compare the different types of objects that
play a role in the remainder of the duality web. In the I-brane frame
these are the 
free fermions themselves and the fluxes through the $a$-cycles of the
Riemann surface that they couple to. In the B-model these are BPS
states that are obtained by wrapping D3-branes around $A$-cycles and
the fluxes through these 3-cycles. These B-model invariants can be
translated to analogous charges in the Gopakumar-Vafa M-theory setting
or the Donaldson-Thomas frame in type IIA.  Let us look at this closer.

\subsubsection{Relating the I-brane fermions to BPS states}
\index{free fermions}  \index{BPS state}

Perhaps it is clarifying to compare the chiral fermions, that appear
so naturally in the intersecting brane system, to the usual BPS
states. This is most naturally done in the M-theory picture, where we
consider an M5-brane with topology $TN_k \times \Sigma$.

First of all, the chiral fermions are given by the open fundamental
strings that stretch between D4-branes and D6-branes. On the Riemann
surface $\Sigma$ they wrap an $A$-cycle, which is their time trajectory, times an
interval $I$. At one end  of the interval the open string ends on a
D4-brane, whereas on the other end it ends on a D6-brane. 
These fundamental strings are lifted to open M2-branes in M-theory. Since the
Taub-NUT circle shrinks at the D6-endpoint, the topology of the
membranes is $S^1 \times D^2$, where $S^1$ is the time trajectory and
$D^2$ is a 2-dimensional disc whose boundary $\lotjesd D^2$ lies 
on the M5-brane. This boundary encircles the $S^1_{TN}$ of the Taub-NUT 
geometry. 
Another way to see this is that the BPS mass of these open M2-brane
states is given by 
$$
\cZ = \int_{D^2} \w_{TN} = \oint_{\lotjesd D^2} \eta_{TN},
$$
where $\eta_{TN}$ is the one-form (\ref{eta}) on $TN$. This mass
goes to zero exactly when the M2-brane approaches one of the
NUTs of the Taub-NUT geometry. There the chiral fermions appear. 

\begin{figure}[h]
\begin{center}
\includegraphics[width=8cm]{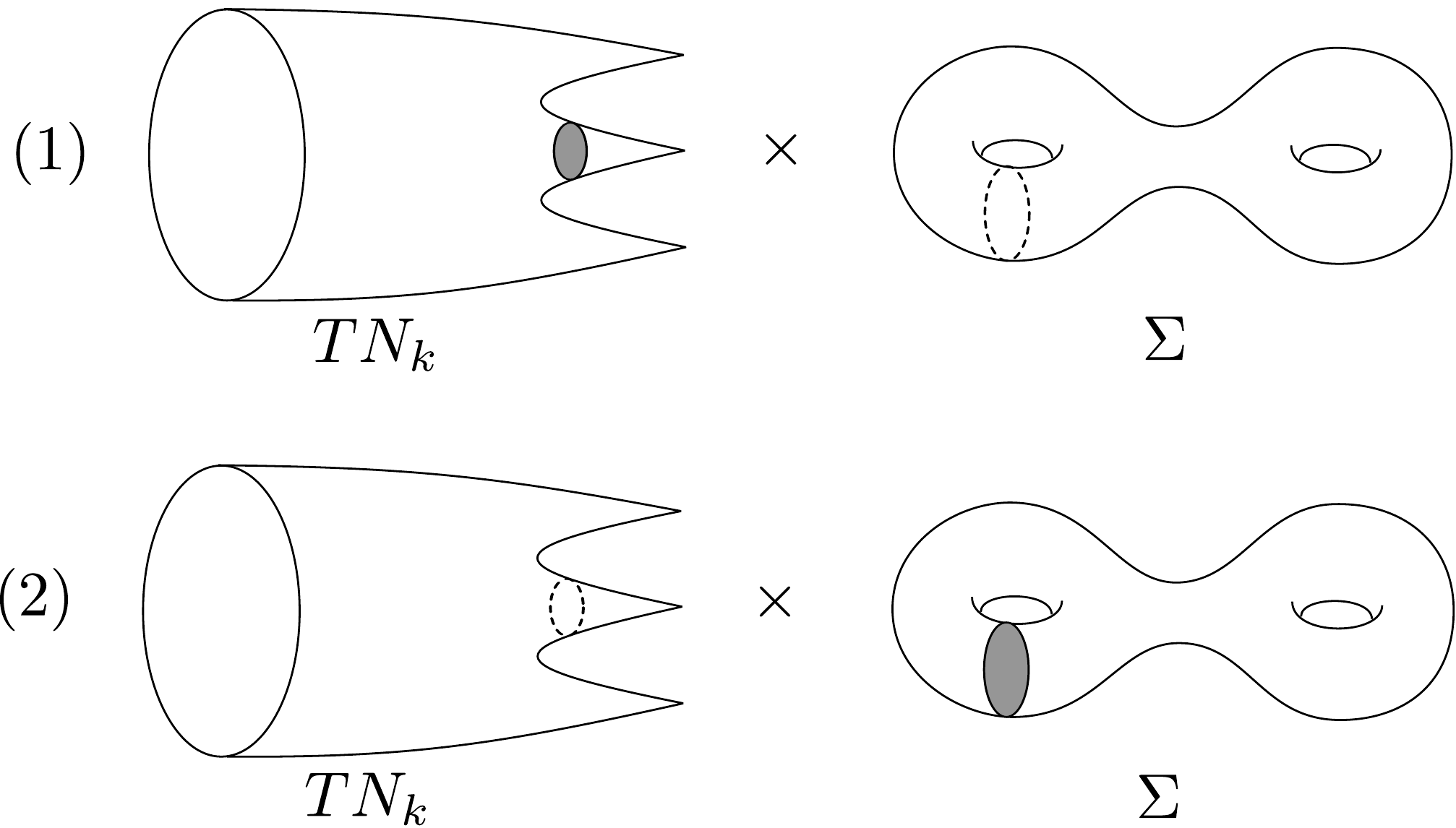}
\caption{Two kinds of open M2-brane
  instantons that contribute to a M5-brane with geometry $TN_k \times
  \Sigma$: $(1)$ the free fermions, massless at the NUTs of $TN_k$;
  $(2)$ the usual BPS states that become massless when the Riemann
  surface $\Sigma$ pinches.}\label{fig:openM2}
\end{center} 
\end{figure}

So, if we compute a fermion one-loop diagram, this is
represented in M-theory in terms of open M2-brane world-volumes with
geometry $D^2 \times S^1$ and boundary $T^2=S^1 \times S^1$ on the
M5-brane. Note that the $S^1$ on the $TN$ factor is filled in by a
disc. This is illustrated in Fig.~\ref{fig:openM2}.

On the other hand we have the traditional BPS states in Calabi-Yau
compactifications, that we discussed in \Cref{sec:ge}. In the
IIB compactification on $TN_k  
\times X$ these will be given by D3-branes that wrap special
Lagrangian 3-cycles in $X$. Let us consider the D3-branes that wrap
the A-cycles of the local Calabi-Yau, and thus fill in a disc $D$ that ends
on the Riemann surface $\Sigma$.  After the dualities that map this
configuration to an M5-brane in M-theory, these states become open
M2-branes with the topology of a disc as well. But now the boundary $S^1$ of
these discs will lie on the surface $\Sigma$. The time trajectory will
be along the space-time $TN_k$. More invariantly, we have again
M2-branes with world-volumes $D^2 \times S^1$ and boundary $S^1 \times
S^1$, but now the $S^1$ on the Riemann surface $\Sigma$ is filled
in. This makes sense, since the mass of the BPS states, given by
$$
\cZ= \oint y dx,
$$
goes to zero exactly when the surface $\Sigma$ is pinched or forms a
long neck. These states are the M-theory interpretation of the
well-known massless monopoles of Seiberg and Witten
\cite{Witten:1997sc,Henningson:1997hy,Lambert:1998wc}.

In a full quantum theory of the M5-brane, both kinds of open M2-branes
should contribute to the partition function. In fact, the boundaries
of these M2-branes are the celebrated self-dual strings (see
\Cref{SW:selfdualstring}) that should
describe the M5-brane world-volume theory
\cite{Duff:1993ye,Klemm:1996bj}.  Clearly the corresponding massless
states are contributing in different regimes. In that sense the
relation between the free fermions and the usual BPS states can be
considered as a strong-weak coupling duality.

\subsubsection{Fermion charges on the I-brane}\label{sec:fermioncharge}

It might be good as well to follow the fermion numbers $p^1,\ldots,p^g$
(through the handles of the Riemann surface) and the dual $U(1)$
holonomies $v_1,\ldots,v_g$ (that couple to these fluxes) through this chain
of dualities. We pick $k=1$ for simplicity. First we remark that we
have to choose a basis of $a$-cycles on $\Sigma$ to define these
quantities.

In the IIB compactification on $TN \times X$ the quantum numbers $p^i$
appear as fluxes of the RR field $G_5$ through a basis of 3-cycles of
the Calabi-Yau $X$.\footnote{Indeed, the fermion flux originates from
  a 2-form field strength on the D6-brane. This lifts in M-theory to a
  4-form flux $G_4$ and arises as a $G_5$-flux in type IIB. We will
  explain these last steps in more detail in \Cref{sec:chasinglambda}.}
In reduction of $G_5$ to the 4-dimensional low-energy
theory on $TN$ this gives the $U(1)$ gauge fields $F^i$ of the vector
multiplets that appear in the gauge coupling
$$
\int_{TN} \tau_{ij} F^i_+ \wedge F^j_+,
$$
analogous to the reduction in equation (\ref{eqn:effgaugelag}).

After mapping this configuration to IIA theory on $\R^3 \times S^1
\times \wt X$ with a D6-brane wrapping $S^1 \times \wt X$, the flux
$p$ is carried by the $U(1)$ gauge field strength $F$ on the
world-volume of the D6-brane, $[F^i / 2\pi] = p^i \in H^2(\wt X,\Z)$.
This can be interpreted as a bound state of a single D6-brane with $p$
D4-branes. In a similar fashion the Wilson lines $v_i$ are mapped to
potentials for the D4-branes. So in this duality frame we can simply
shift the fermion number by adding D4-branes.

\subsubsection{B-model branes compared to I-brane fermions}
\index{B-model brane}

Another natural question to ask is whether these fermions in the topological
B-model, see \Cref{sec:topstringtheory}, are the same as the physical
fermions we encountered on the 
I-brane. An
important property of the fermions in the context of topological
vertex is that the insertions of the operators $\psi(x_i)$, that
create fermions (which is of course not the same as the quanta of the
corresponding field), change the geometry of the curve. The fermions
will produce extra poles in the meromorphic differential $ydx$ which
encodes the embedding of the curve as $H(x,y)=0$.  We have the
identification $y=\partial\phi(x)$, and by bosonization the operator
product with a fermion insertion gives a single pole at each fermion
insertion
$$
\partial\phi(x) \cdot \psi(x_i) \sim {1\over x-x_i} \psi(x_i).
$$

In the superstring such a correlator 
$$
\langle \psi(x_1) \cdots \psi(x_n)\rangle
$$
of fermion creation operators corresponds to the insertion of
D5-branes in type IIB compactification. These 5-branes all have 
topology $\R^4 \times \C$, where in the Calabi-Yau $uv + H(x,y)$ they
are located at specific points $x_i$ of the curve $H(x,y)=0$ along the
line $v=0$ (so they are parametrized by the remaining coordinate
$u$).   Having this extra pole for $y$ means that the Riemann surface
has extra tubes attached to it at $x=x_i$.

If we T-dualize this geometry to replace the Calabi-Yau $X$ by an NS5-brane
wrapping $\R^4 \times \Sigma$, the D5-branes, which are all transverse
to $\Sigma$, will become D4-branes. So we get an NS5-brane with a
bunch of D4-branes attached, that all end on the NS5-brane. This
configuration can be lifted to M-theory to give a single irreducible
M5-brane, now with ``spikes'' at the positions $x_1,\ldots,x_n$. So we
indeed see that the two kinds of fermions (or at least their sources)
are directly related and have the same effect on the geometry of the
Riemann surface.

\subsection{I-brane partition function}\label{sec:Ibranepartfunc}
\index{I-brane partition function}
\index{dual partition function}

In Type II compactifications on Calabi-Yau geometries four-dimensional
F-terms can be computed using topological string
theory techniques. We now want to see how these kind of computations
can be mapped to the I-brane model.

Starting point will be the end of the chain of dualities in
Figure~\ref{fig:webofdualities}: the IIA compactification on $\R^3
\times S^1 \times 
\wt X$ with $\wt X$ a (non-compact) Calabi-Yau geometry, where we wrap
a D6-brane along $S^1 \times \wt X$. This is the set-up of
Donaldson-Thomas theory. With the right background values of the
moduli turned on, the topological string partition function can be 
reproduced as an index that counts BPS states
degeneracies of D-branes in
this configuration. 

More precisely, the topological string theory partition function in
the A-model naturally splits in a classical and a quantum (or
instanton) contribution
\be
\label{Z-top}
Z_{\textrm{top}}(t,\la)  = \exp \left(- \frac{t^3}{6 \la^2} -
  \frac{1}{24} t \wedge 
c_2\right) Z_{\textrm{qu}}(t,\la).
\ee
Here $t \in H^2(\wt X)$ is the complexified K\"ahler class, $\la$ the
topological string coupling constant, and $c_2=c_2(\wt X)$. The
classical contributions should be read as integrals over the
Calabi-Yau $\tilde{X}$. The quantum
contribution is decomposed as
$$
Z_{\textrm{qu}}(t,\la) = \exp \left( \sum_{g\geq 0} \la^{2g-2} \cF_g(t) \right),
$$
with the genus $g$ free energy expressed in terms of the Gromov-Witten
invariants of degree $d$ as
\begin{align}
\cF_g(t) = \sum_d \textrm{GW}_g(d) e^{d \cdot t}.
\end{align}

We can now use the fact that $Z_{\textrm{qu}}$ has a dual interpretation as the
Donaldson-Thomas partition function counting D0-D2-D6 bound states
(here we ignore a subtlety with the degree zero maps)
\be
\label{DT}
Z_{\textrm{qu}}(t,\la) =  \sum_{n,d} \textrm{DT}(n,d) \, e^{-n\la} e^{d \cdot t}.
\ee
In this sum $n \in H_0(\wt X,\Z)\cong \Z$ and $d \in H_2(\wt X,\Z)$
are the numbers of D0-branes and D2-branes. As before $d \cdot t$
stands for $\int_{d} t$. The integers $\textrm{DT}(n,d)$
are the Donaldson-Thomas invariants of the ideal sheaves with these
characteristic classes. From the BPS counting perspective it is also
natural to add the exponential cubic prefactor in (\ref{Z-top}), since this
is nothing but the tension of a single D6-brane (including the
geometrically induced D2-brane charge). Remember  that this tension is
measured by integrating $\Omega = \exp t$ over the submanifold.

In type IIA string theory set-up the complex parameters $\la$ and
$t$ can be expressed in terms of geometric moduli of the $S^1$ and the
Calabi-Yau $\wt X$, and the Wilson loops of the flat RR fields $C_1$
and $C_3$. In particular, we can write
\be
\label{lambda}
{\la} = \frac{\beta}{\ell_s g_s} + i \theta, 
\ee
with
$$
\beta = 2\pi R_9 = \oint_{S^1} ds
$$ 
the length of the Euclidean time circle $S^1$, and $\theta$ the Wilson
loop
$$ \theta = \oint_{S^1} C_1.
$$ 
That is, $\lambda$ can be written as the holonomy of the complexified
one-form $ds/g_s+ i C_1$ (in string units). An important remark is
that, as expressed in equation (\ref{DT}), for BPS states $\theta$ and
$\beta$ only appear in the {\it holomorphic} combination
(\ref{lambda}). In the same way the parameter $t$ is given by the
integral of the complex 3-form
$$ t = \oint_{S^1} \frac{k \wedge ds}{g_s} + i C_3.
$$ 

It is rather trivial to also include the coupling to
D4-branes in this BPS sum. As explained before, such a bound state of
$p$ D4-branes to 
a D6-brane is given by a flux of the $U(1)$ gauge field on the
D6-brane.  We can think of this as a non-trivial first Chern class of
the line bundle over the D6-brane that wraps $\wt X$, so that the
D4-brane has charge $p = \textrm{Tr} [F/2 \pi]$. Tensoring with
the extra line bundle will not change the BPS degeneracy, since the
moduli space of such twisted sheaves is isomorphic to that of the
untwisted one. The only part that changes are the induced D0 and D2
charges, that shift as 
\begin{eqnarray}
d & \to & d - \frac{1}{2}p^2 - \frac{1}{24}c_2\\
n & \to & n + d \wedge p + \frac{1}{6}p^3 + \frac{1}{24} p \wedge c_2.
\nonumber
\end{eqnarray}
Here we also wrote the charges $d \in H^4(\tilde{X})$ and $n \in
H^6(\tilde{X})$ as de Rham forms by 
using Poincar\'e 
duality. The induced charges follow from applying equation
(\ref{eq:CSterms}). For example, the 
D2-brane charge $d$ gets
a contribution 
from the couplings $1/2 \, \textrm{Tr}\,[F/2 \pi]^2$ and $c_2/24$ on the
D6-brane worldvolume. 

So, if we also include a sum over the number of D4-branes, weighted by
a potential $v \in H^4(\tilde{X})$, we get a generalized partition
function  
\begin{align}\label{fermionpartfunction}
&Z(v,t,\la) = \nonumber \\
& = \, \sum_{p \in H^2(\wt X,\Z)}e^{p (v-t^2/2\la)} e^{-(p^2/2+c_2/24)t}e^{-(p^3/6 + p
  c_2/24)\la}e^{-t^3/6\la^2}\ Z_{\textrm{qu}}(t+p\la,\la) \nonumber \\
 & =\,  \sum_{p \in H^2(\wt X,\Z)} e^{pv}Z_{\textrm{top}}(t+p\la,\la).
\end{align}
Apart from adding the D6-brane tension $-t^3/6\la^2$, we have also
added the tension $- \int_{p} t^2/2\la$ of the D4-branes.  
Remember that the D4-branes translate into fermion fluxes on the
I-brane. From this fermionic (I-brane) point of view it is natural to
sum of these fermion numbers. The structure  
(\ref{fermionpartfunction}) is therefore the object that we want to
identify with the  
I-brane partition function and that should be computable in terms of
free fermions. 
It should be remarked that the partition function
(\ref{fermionpartfunction}) was also found in \cite{Dijkgraaf:2002ac},
where a dual 
object was studied: a NS5-brane wrapping $\wt X$. Also in
\cite{Nekrasov:2003rj}, a partition function of this type was
considered and directly related to fermionic expressions. Moreover,
recently it was even conjectured to capture non-perturbative aspects
of the topological string \cite{Eynard:2008he}. We will comment more
on this in \Cref{sec:ch6discussion}.   

Let us finish this chapter by noting that the above expression for
$Z(v,t,\la)$ has an interesting limit for $\la \to 0$, where only
genus zero and one contribute. In that case we have
\be
\label{pert}
 Z_{\textrm{top}}(t+p\la) \sim \exp\left[\frac{1}{\la^2}\cF_0(t) + \frac{1}{
     \la} p^i \partial_i \mathcal{F}_0 + \frac{1}{ 2} p^i p^j \tau_{ij}
   + \cF_1(t) + 
\cO(\la) \right].
\ee
If we now subtract the singular terms (which have a straightforward
interpretation as we shall see in a moment), we are left with the
familiar $\la=0$ answer
$$
Z(v,t) = \sum_p e^{\frac{1}{2} p\cdot \tau \cdot p + p \cdot v}
e^{\cF_1}.
$$
%



\chapter{Quantum Integrable Hierarchies}\label{chapter5}

The string
theory embedding of topological string theory in terms of a Calabi-Yau
compactification to a Taub-NUT space, suggests a  duality of the
topological string partition function and the I-brane partition
function. We thus expect that the topological string
partition function on a non-compact Calabi-Yau threefold $X_{\Sigma}$, that
is modeled on a holomorphic curve $\Sigma$, has an interpretation in terms of
\emph{physical} chiral fermions on $\Sigma$. However, we
discover that these fermions don't just transforms as sections of the
square-root of the canonical bundle on $\Sigma$ (just like the
topological B-branes in the B-model). Instead, the topological string
coupling constant translates into a non-trivial flux on $\Sigma$ that
quantizes the curve into a non-commutative object. Fermions should in
this context be mathematically interpreted as elements of a holomorphic
$\cD$-module supported on $\Sigma$. The goal of 
this chapter is to explain the topological string partition function
as a tau-function of a quantum or non-commutative integrable hierarchy.     

In \Cref{sec:semi-clas} we start with (semi-classical) Krichever solutions of the Kadomtsev-Petviashvili (KP) hierarchy. We explain how the semi-classical contribution $\cF_1$ to the free energy can be interpreted in terms of such a Krichever solution. In \Cref{sec:b-field} we study the effect of
the topological string coupling constant $\la$ on the I-brane. We
introduce $\cD$-modules and show in detail how the resulting I-brane
configuration should be understood mathematically in terms of a quantum
integrable hierarchy. In \Cref{sec:quantumpartfunction} we use our
framework to assign a fermionic state to a non-commutative curve
corresponding to $\Sigma$. In analogy with the semi-classical recipe
the determinant of such a state yields a tau-function. We
conjecture that this \emph{quantum} tau-function is equal to the all-genus 
topological string partition function. The advantage of this line of
thought is that the total topological string partition function on a
local Calabi-Yau threefold acquires a very simple interpretation as a fermion
determinant on an underlying non-commutative curve. In
Chapter~\ref{chapter6} we work out some insightful examples.

\section{Semi-classical integrable hierarchies}\label{sec:semi-clas}

An important class of conformal field theories is that of free chiral 
fermions on a Riemann surface $\Sigma$. Already in the mid-eighties it was
discovered that these CFT's are 
intimitely related to a certain integrable system, well-known as the
Kadomtsev-Petviashvili (KP) hierarchy.  
The CFT partition function appears as a so-called
tau-function of this integrable hierarchy. It computes the 
determinant of the Cauchy-Riemann operator $\bar{\partial}$ on
$\Sigma$. Let us explain this in more detail.


\subsection{KP integrable hierarchy}\label{sec:KP}
\index{KP integrable hierarchy}
\index{integrable hierarchy}


The KP integrable system originates from the non-linear partial
differential equation 
\begin{align}\label{KdVequation}
 \frac{\lotjesd u}{\lotjesd t} =  \frac{\lotjesd^3
   u}{\lotjesd x^3} + 6 u \frac{\lotjesd u}{\lotjesd x}.  
\end{align}
This equation was written down around 1900 by Kortweg and de Vries as
a description of non-linear waves in shallow water. To see how this
differential equation is included in an integrable hierarchy, let us
write down a pseudo-differential operator 
\begin{align*}
 L = \lotjesd+ \sum_{i=0}^{\infty} u_i(x) \lotjesd^{-i}, 
\end{align*}
where $\lotjesd = \lotjesd/\lotjesd x$. This operator $L$ is known as a Lax operator. The
negative powers in its expansion are defined through the Leibnitz rule
\begin{align*}
\partial^i f = \sum_{k=0}^{\infty} { i \choose k} ( \partial^k f
) \partial^{i-k}, \qquad \mathrm{for}~i \in \Z
\end{align*}
with ${i \choose k} = i \cdot \ldots \cdot (i-k+1)
/ k!$. So $\lotjesd^{-1}$ should be interpreted as partial integration.
The Lax operator may depend on an infinite set of times $x=t_1, t_2,
\ldots$ and satisfies the KP flow equations 
\begin{align}\label{eqn:KP}
 \frac{\lotjesd L}{\lotjesd t_n} = [H_n, L], \qquad \mathrm{for}~n\ge 1
\end{align}
where the Hamiltonians  $H_n = (L^{n})_+$ are powers of the Lax
operator $L$. The subscript in $(L^n)_+$ denotes the restriction to the positive
powers of the pseudo-differential operator $L^n$. The Hamiltonians obey the
Zakharov-Shabat relations 
\begin{align*}
\frac{\lotjesd }{\lotjesd t_n} H_m - \frac{\lotjesd }{\lotjesd t_m} H_n  = [H_n, H_m].   
\end{align*}
%
They make sure that the defining KP
equations (\ref{eqn:KP}) line out a commuting flow. 

We can just as well write down the KP integrable hierarchy (\ref{eqn:KP})
for a power $P = L^p$ of the Lax operator $L$. When $P$ is actually a
differential operator, the resulting integrable hierarchy is of
so-called KdV type. For 
\begin{align}\label{eqn:kdvop}
 P = \lotjesd^2 + u(x), 
\end{align}
we easily recover the original KdV equation (\ref{KdVequation}). 

\begin{figure}[h]
\begin{center}   
\includegraphics[width=\textwidth]{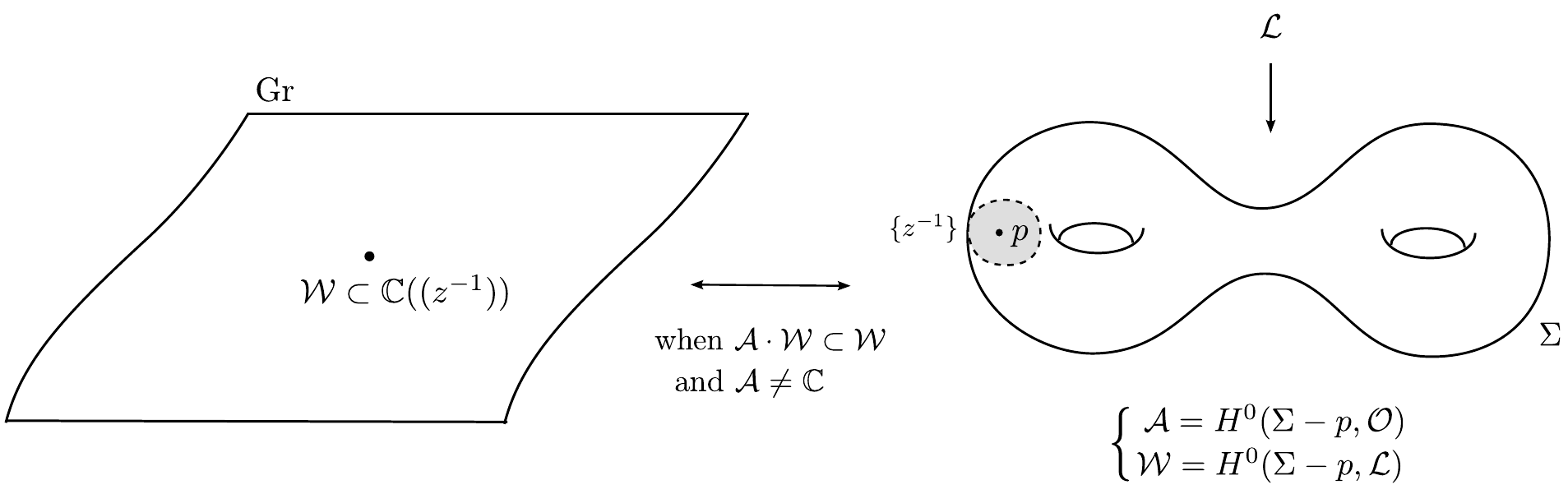}
\caption{The Krichever correspondence assigns a geometric
  configuration $(\Sigma -p, z^{-1}, \cL)$ to a
  dense subset of 
  points $\cW$ in the Grassmannian $Gr$, where $z^{-1}$ is a
  local coordinate in the neighborhood $p$ on the curve $\Sigma$. Such
  a geometric 
  interpretation only exists when the stabilizer $\cA$ of $\cW$ is
  non-trivial. Under the Krichever map it turns into the algebra of
  holomorphic functions on 
  $\Sigma - p$. }\label{fig:KPKrichever}
\end{center} 
\end{figure}

In the last decades many complementary perspectives on the KP integrable
hierachy have been developed (starting with
\cite{sato1981,Date:1982tj,segalwilson}). In this section we are
mostly interested  
in its geometric interpretation in terms of (possible singular)
Riemann surfaces. This relation, discovered by Krichever and
illustrated in Fig.~\ref{fig:KPKrichever}, comes about
as follows. Look at the algebra of differential operators that commute
with a given differential operator $P$. This turns out to be a
commutative algebra that we denote by $\cA_P$. In the KdV example
$\cA_{P}$ is a polynomial ring generated by the degree two KdV operator  
$P$ from equation (\ref{eqn:kdvop}) and a degree three operator $Q$  
\begin{align*}
\cA_P =  \C[P,Q]/(Q^2 = P^3 - g_2 P - g_3),   
\end{align*}
in terms of the Weierstrass invariants $g_2$ and $g_3$. Note that this
forms the algebra of functions on an elliptic curve. Any (pointed) curve that
is obtained via  a commuting ring of differential operators is
called a \emph{Krichever} curve. It is parametrized by the
eigenvalues of the commuting operators in $\cA_P$, and therefore also
referred to as a \emph{spectral curve} of the KP system corresponding
to $P$.

According to the above map many differential operators $P$ will correspond
to the same spectral curve $\Sigma$: in fact, for most of them the
algebra $\cA_p$ will be generated by $P$ only, so that the Krichever
curve is just a pointed projective plane. However, when we fix a line
bundle\footnote{For singular curve we need to consider rank 1 torsion free
sheafs.} $\cL$  on the spectral 
curve $\Sigma$, together with a point $p \in \Sigma$ and 
trivializations of $\Sigma$ and $\cL$ in the neighbourhood of $p$,
there \'is a natural correspondence with a subset of solutions to the KP
hierarchy. This is called the \emph{Krichever correspondence}.    

So let us start with a spectral curve $\Sigma$ together with a
line bundle $\cL$, that comes equipped with a connection $A$. Both from
the physical and mathematical perspective it is 
natural to consider a chiral fermion field $\psi(z)$ on $\Sigma$ that
couples to the gauge field $A$. The free fermion partition function
computes the 
determinant of the twisted Dirac operator $\overline\partial_A$
coupled to the line bundle $\cL$.

It has been known for a long time that these chiral determinants are closely
related to integrable hierarchies of KP-type. 
If one picks a point $p \in \Sigma$ on
the curve together with a local trivialization $e^t$ of the line bundle
around $p$, the ratio with respect to a reference connection $A_0$
$$
\tau(t) = {\det \overline\partial_{A} \over \det \overline\partial_{A_0}}
$$
equals a so-called \Index{tau-function} of the KP integrable hierarchy.
This determinant has many interesting properties, {\it e.g.}\ the
dependence of this determinant on the connection $A$ is captured by
a Jacobi theta-function
%
%

How is the above tau-function related to the KP hierarchy~(\ref{eqn:KP})
that we started with? This becomes clear when we study the
system from a Hamiltonian perspective. In this formulation
the partition function $Z(S^1)$ is an element of a Hilbert space $\cH$
assigned to a circle $S^1$ surrounding $p$.  
This Hilbert space consists of functionals on all possible boundary
conditions of the fields at the circle around $p$. In this example
that Hilbert space is just the Fock space $\cF$ that is constructed
out of the modes $\psi_n$ and $\psi_n^*$ of the fermion field
$\psi(z)$. So we find a state
\begin{align*}
| \cW \rangle \in \cF.
\end{align*}
In fact, we can describe the state $|\cW \rangle$ explicitly. It is
the semi-infinite wedge product of a basis that spans the space $\cW$ of
holomorphic sections   
$$
\cW=H^0(\Sigma-p, {\cal L}).
$$
These are all meromorphic sections of the line bundle $\cL$ that are
only allowed to have poles at $p$. Notice that $\cW$ is a module for
the algebra $\cA = H^0(\Sigma-p, \cO)$ consisting of holomorphic
functions on $\Sigma-p$. 

Let us choose a coordinate $z^{-1}$ in a neighbourhood around
$p$. Since a holomorphic section of $\cL$ on $\Sigma - p$ can at most
have a (finite order) pole at $p$, the subspace $\cW$ is a subset of
$\C((z^{-1})) = \C[z] \oplus \C[[z^{-1}]]$. The set of all subspaces
$\cW$ of $\C((z^{-1}))$ whose projection to $\C[z]$ has a finite index $\mu$,
with respect to the projection map $\C((z^{-1})) \mapsto \C[z]$, has
the structure of a Grassmannian $Gr(\mu)$.  The Krichever map assigns
a geometric set-up to an element $\cW$ of the Grassmannian that has a 
a non-trivial ($\cA \neq \C)$ stabilizer\footnote{More carefully, the
  rank of $\cW$ 
  over $\cA$ determines the rank of the bundle $\cL$ over $\Sigma$.}
\begin{align*}
 \cA \cdot \cW \subset \cW 
\end{align*}
The infinite wedge products $|\cW \rangle$ form a line bundle over the
Grassmannian. Pulling this line bundle back to the family of configurations
$(\Sigma, P, \cL)$ reconstructs the determinant line bundle 
corresponding to the $\delbar$-operator on $\Sigma$. 

The big-cell of the index zero Grassmannian is the subset that is
comparable to the vacuum $\C[z^{-1}]$ (both the kernel and the
cokernel of the projection map have dimension zero).
To compute the
tau-function on the big cell, we first associate a
coherent state 
$|t\rangle$ to the local trivialisation around $P$, corresponding to
the chosen boundary conditions. Then, combining the above ingredients, the
tau-function can be written as 
\be
\tau(t) = \langle t |\cW \rangle.
\label{taufunction}
\ee
%

It is even possible to reconstruct the differential
operator $P$. Notice that any subspace
$\cW$ in the big cell contains a unique element of the form  
\begin{align*}
 \eta(z,t) = 1 + \sum_{i=1}^{\infty} a_i(t) z^{-i}. 
\end{align*}
such that the \emph{Baker function} 
\begin{align*}
 \psi(t,z) = \eta(z,t) g(z), \quad g(z)=  \exp \left( \sum_n t_n z^n \right) 
\end{align*}
is an element of $\cW$ for all times $t$. Substituting $z
\leftrightarrow \partial$ in the expression for $\eta$ defines a
pseudo-differential operator $K(\partial, t)$, such that $P = K
\lotjesd^p K^{-1}$ solves the differential equation 
\begin{align*}
 P \psi(t,z) = z^p \psi(t,z). 
\end{align*}
We conclude that this differential operator $P$ is fixed by the state
$\cW$ and thus by the geometrical configuration $(\Sigma, P, L)$
together with trivializations. 

Multiplying the subspace $\cW$ by the function $g(z)$ defines an action
on the Grassmannian, that leaves the differential operator $P$
fixed. The flow parametrized by $t_n$ is equivalent to the $n$th KP
flow (\ref{eqn:KP}). Geometrically it leaves the spectral curve
$\Sigma$ invariant and multiplies $\cL$ by a flat line bundle, so that
it corresponds to a straight line motion on the Jacobian of $\Sigma$. 


\subsection{Topological strings at 1-loop} 

The formalism of the last section can be easily extended to more
general geometric configurations, consisting of a genus $g$ Riemann
surface $\Sigma$ with an arbitrary number $k$ of punctures. A free
fermionic field $\psi(z)$ coupled to a line bundle $\cL$ over $\Sigma$
determines a state $|\cW \rangle$ in the $k$th fold tensor product $\cH^{\otimes
  k}$. This has been exploited in the late eighties to understand the
Polyakov string in the operator formalism (see
e.g. \cite{Ishibashi:1986bd, Vafa:1987es,AlvarezGaume:1988bg}). The KP
formalism determines that the state $|\cW \rangle$ is a Bogoliubov transformation 
\begin{align}\label{eqn:Bogstate}
 | \cW \rangle = \exp \left( \sum_{i,j=1}^k \sum_{r,s>0} a_{rs}^{ij}
   \psi_{-r}^i \psi^{j*}_{-s} \right) |0\rangle 
\end{align}
of the vacuum. 

The most elegant feature of this formalism is that the Riemann surface
can be split up in elementary building stones. The total partition
function can be reconstructed by glueing these using a natural inner
product on $\cH$. One of such elementary pieces is the three-punctured
sphere  
\begin{align}\label{eqn:CFTvertex}
|V \rangle \in \cH^{\otimes 3}.  
\end{align}
Its coefficients may be determined by the CFT equality
$$ \langle \phi_1, \phi_2, \phi_3 |V \rangle = \langle
\phi_1(0)| \phi_2(1) | \phi_3(\infty) \rangle, $$
that follows since the three punctures on the sphere may be fixed at
$0,1, \infty$. 

Returning to the topological B-model on $X_{\Sigma}$, the
$\cW_{\infty}$-constraints seem to suggest that also the dual topological string
partition function $Z^D_{\mathrm{top}}$ can be written as a Bogoliubov
state~(\ref{eqn:Bogstate}) \cite{Aganagic:2003db,
  Aganagic:2003qj}. This implies that also the topological string is
governed by the KP hierarchy. However, it is known that
$Z^D_{\mathrm{top}}$ is generically
\emph{not} of the  
Krichever form, \emph{i.e.} it does not correspond to regular free fermions on a
Riemann surface.  The goal of this chapter is to show how the B-model
partition function should be interpreted geometrically. We will be
motivated by the fact that the fermion fields 
that play a role in the B-model do not obey the usual transformation
rules.    

But before going into this, let us remark that in the limit $\la \to 0$
the B-model fermions do transform semi-classically.  To see this we expand
the fermion field $\psi(x)$  
\begin{align*}
 \psi(x) = e^{\phi_{cl}/\la} \psi_{\textrm{qu}}(x) 
\end{align*}
in a classical piece plus quantum corrections. It is shown in
\cite{Aganagic:2003db,Aganagic:2003qj} that in the limit of small
quantum corrections the Fourier-like transformation
(\ref{eqn:fourierfermion}) reduces to the transformation property
\begin{align*}
 \psi_{\textrm{qu}}(x_j) dx_j^{1/2} = \psi_{\textrm{qu}}(x_i) dx_i^{1/2}
\end{align*}
for the quantum field $\psi_{\textrm{qu}}$.
In other words, the semiclassical approximation of the fermionic field
$\psi(x)$ transforms in the classic way as a spin $1/2$ field on
$\Sigma$.   

Relatedly, let us turn to the Ray-Singer determinant
(\ref{eqn:Ray-Singer}) that describes the (worldsheet) genus one part
of the free energy $\cF$. When the Calabi-Yau threefold
is non-compact and modeled on a Riemann surface $\Sigma$, the
Ray-Singer torsion reduces to the bosonic determinant 
\begin{align*}
\cF_1 = -\frac{1}{2} \mathrm{log\, det}\, \Delta_{\Sigma}.
\end{align*}
By the boson-fermion correspondence this determinant is equal to
the logarithm of the determinant of the $\delbar$-operator on $\Sigma$. In
other words, the KP tau function computes the 1-loop semi-classical partition
function exp$(\cF_1)$ on a non-compact Calabi-Yau background
$X_{\Sigma}$. In particular, the topological vertex reduces to
the CFT vertex (\ref{eqn:CFTvertex}) in the limit that $\la \to 0$ \cite{Aganagic:2003db}. 

We will see this explicitly in some examples in
Chapter~\ref{chapter7}. In these examples we compute exp$(\cF_1)$ for two
threefolds $X_{\Sigma}$ in which $\Sigma$ is a compact genus
one and a compact genus two Riemann surface. Other instances can be found in 
the correspondence between the B-model and matrix models \cite{Dijkgraaf:2002yn}.

\section{$\cD$-modules and quantum curves}\label{sec:b-field}

Let us now release the constraint $\la =0$ and turn our attention to
the full free energy 
$$
\cF = \sum_g \la^{2g-2} \cF_g. 
$$
As we discussed in Chapter~\ref{chapter3+4} the topological string
coupling constant $\la$ is embedded in a type IIA Calabi-Yau compactification
as the graviphoton field strength $C_1$. Let us therefore chase the
graviphoton field 
through the duality web to find out its implications on the
I-brane.

\subsection{Chasing the string coupling constant}\label{sec:chasinglambda}

\begin{figure}[h]
\begin{center}   
\includegraphics[width=7cm]{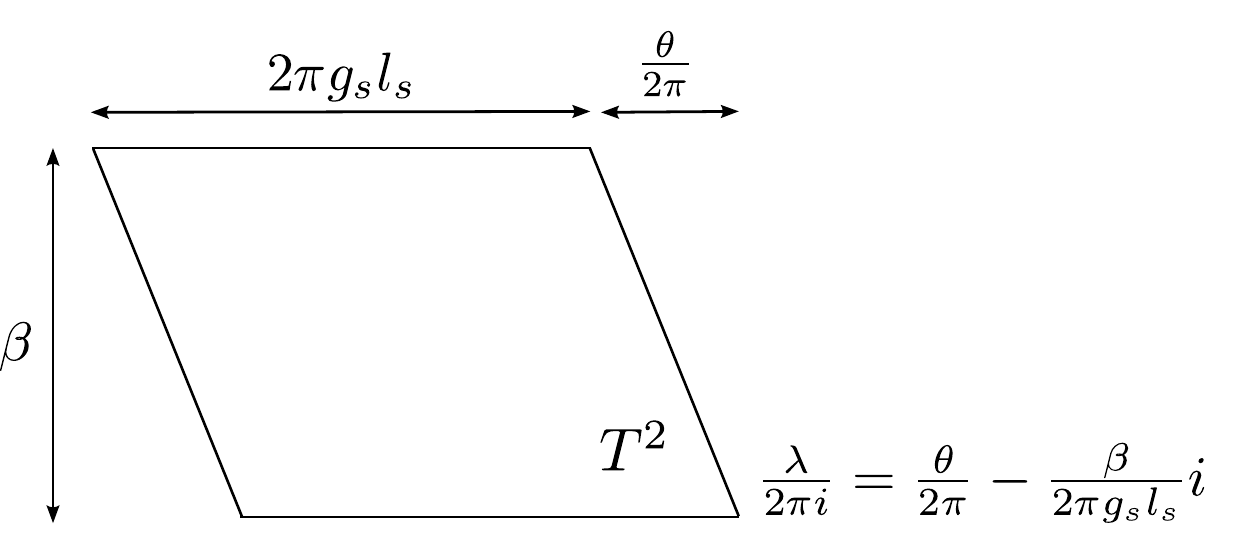}
\caption{The above torus $T^2$ is present asymptotically in the M-theory
  geometry. In a type IIA reduction over the Taub-NUT circle
  $S^1_{\textrm{TN}}$, of size $R_{9} = \beta$, the real part of the
  $T^2$ complex modulus $\la/2 \pi i$ is proportional to the graviphoton
   field $C_1$ integrated over $S^1$.}\label{fig:lambda} 
\end{center} 
\end{figure}

We start with the purely geometric M-theory compactification on $TN \times S^1
\times \wt X$. Asymptotically this geometry has a $T^2$ fibration over
$\R^3$.  The ratio of the radii of the two circles of this two-torus
is given by
$$
{\beta \over 2\pi g_s\ell_s}= {R_9 \over R_{11}},
$$
whereas the complex modulus of this $T^2$ is  
$${\lambda \over 2\pi i} =  \frac{\theta}{2 \pi} -
\frac{\beta}{2 \pi  g_s l_s} i.$$
(The last equality sign follows from equation~(\ref{lambda}).)  
Indeed, remember that in general when we compactify an M-theory circle,
the off-diagonal components $g_{11,i}$ combine into 
the type IIA graviphoton field $C_1$.
In the M-theory background described above, the only off-diagonal
components in the $T^2$-metric are parametrized by the real part of
the $T^2$ complex structure. The real part of the complex modulus
should therefore be proportional to the graviphoton $\theta =
\oint_{S^1} C_1$. The asymptotic torus $T^2$ is illustrated in 
Fig.~\ref{fig:lambda}.

When we exchange the roles of the 9th and 11th dimension, we will perform a
modular transformation or S-duality
$$
\lambda \to 4\pi^2/\lambda.
$$ 
In the dual IIA compactification on $TN \times \wt X$
the asymptotic values for the radius of the circle fibration and
the graviphoton Wilson line are thus proportional to $-1/\lambda$. In
the somewhat singular limit $\beta \to 0$  the complex modulus
$\lambda$ limits to $i\theta$. So after the flip we find that the new
graviphoton field becomes proportional to $1/\lambda$. 

After the flip the Wilson loop only makes sense asymptotically,
since $S^1_{\beta}$ is contractible in $TN$. In fact, the graviphoton
gauge field is given by
$$
C_1 = {1 \over \lambda} \eta,
$$
where the one-form $\eta$ is our friend (\ref{eta}). This field
is not flat but has curvature
\be
\label{Gtwo}
G_2 =dC_1 = {1\over \lambda} \omega.  
\ee 
Here $\omega$ is the unique harmonic 2-form that is
invariant under the tri-holomorphic circle action (\ref{omega}). Note that this is a
natural choice, since in the case of $TN_1$ this form reduces to the
usual constant self-dual two-form in the center $\R^4$ (up to
hyper-K\"ahler rotations).

Summarizing, the topological string partition function will be
reproduced by a type IIA compactification on $TN \times \wt X$ with
graviphoton flux given by (\ref{Gtwo}). Note that in this case the
graviphoton is {\it inversely} proportional to the topological string
coupling!

It is now straightforward to follow this flux further through the
duality chain. In the type IIB compactification on $TN \times X$ the
self-dual 5-form RR field $G_5$ is given by
$$
G_5 = {1\over \la} \omega \wedge \Omega.
$$
with $\Omega$ the holomorphic $(3,0)$ form on the Calabi-Yau
$X$. 

We will now T-dualize this configuration to the IIA background that
includes a NS5-brane. In that case there is 4-form RR-flux
\be
\label{Gflux}
G_4 = {1 \over \la} \omega \wedge dx \wedge dy.
\ee
Here $dx \wedge dy$ is the $(2,0)$ form on the complex surface $\B$.
This 4-form flux can be directly lifted to $M$ theory, where we have
the geometry
$$
TN \times \cB \times \R^3.
$$

\subsubsection{B-field}

Now we have to discuss what the interpretation of this flux is, if we
reduce to IIA theory along the $S^1$ inside $TN$ to produce our system
of intersecting branes. In that case we will have an extra set of
D6-branes with geometry $\B \times \R^3$. We want to argue that the
$G_4$ flux becomes a constant NS $B$-field on their world-volumes.

As a preparation, let us recall again how the world-volume fields of
the D6-branes are related to the $TN$ geometry in the M-theory
compactification. First of all, the centers $\vec{x}_a$ of the $TN$
manifold are given by the vev's of the three scalar Higgs fields of
the 6+1 dimensional gauge theory on the D6-brane.  In a similar
fashion the $U(1)$ gauge fields $A_a$ on the D6-branes are obtained
from the 3-form $C_3$ field in M-theory. More precisely, if $\w_a$ are
the $k$ harmonic two-forms on $TN_k$ introduced in section 2, we have
a decomposition
$$
C_3 = \sum_a \w_a \wedge A_a.
$$
We recall that the forms $\w_a$ are localized around the centers
$\vec{x}_a$ of the $TN$ geometry. So in this fashion a bulk field gets
replaced by a brane field. This relation also holds for a single
D6-brane, because $\w$ is normalizable in $TN_1$.

As a direct consequence of this, the reduction of the 4-form field
strength $G_4=dC_3$ can be identified with the curvature of the gauge
field
$$
G_4 = \sum_a \w_a \wedge F_a.
$$
Combining this relation with the presence of the flux (\ref{Gflux}), we
find that in the I-brane configuration the D6-branes support a
constant flux
$$
\sum_a F_a = {1\ \over \la} dx \wedge dy.
$$

There is simple and equivalent way to induce such a constant magnetic
field on all of the D6-branes: turn on a NS $B$-field in the IIA
background\footnote{This can be most easily understood from the
  worldsheet point of view. The coupling of the B-field to the
  fundamental string in equation (\ref{eqn:worldsheetBfield}) is not
  invariant under gauge transformations $B \mapsto B + d\Lambda$ when
the worldsheet has boundaries. This can be repaired by a simultaneous
gauge transformation $A \mapsto A - \Lambda/ 2\pi \alpha'$ of the
$U(1)$ worldvolume gauge field $A$ on the 
brane which the string is ending on. This
implies that only $B + 2 \pi \alpha' F$ is a gauge-invariant
combination on the brane.}. 
We can therefore conclude that in the presence of the
background flux (\ref{Gflux}) translates into a constant $B_{NS}$
field\footnote{Note that the above B-field is holomorphic
  because we started 
out with a holomorphic graviphoton field strength. This is
unconventional: string theory forces the B-field to be real. More
precisely, the full topological content of 
the intersecting  
brane configuration is captured by a holomorphic and an
anti-holomorphic piece that couple to the real B-field $B +
\bar{B}$.}  induced on the surface $\B$
\begin{align}\label{eqn:B-field}
B_{NS} = {1\over \la} dx \wedge dy.
\end{align}
In the next section we will discuss the full consequences of this.

\subsubsection{Classical term $\cF_0$}

But at this
point we first want to point out one immediate and more elementary
consequence of the $B$-field. The presence of the flux induces a
$U(1)$ gauge field on the D6-brane
\be
\label{A}
A = {1\over \la} y dx.
\ee
For the 4-6 strings we have to restrict $A$ to $\Sigma$. Therefore the
chiral fermions are coupled to a non-zero $U(1)$ gauge field. This
background gauge field gives a contribution to the effective action on
$\Sigma$ of the form
$$
\cF = {1\over 2} \sum_i \oint_{a^i}A \oint_{b_i}A.
$$
Here $(a^i,b_i)$ is a canonical basis of $H_1(\Sigma,\Z)$. Plugging
the expression for $A$ we obtain precisely the (genus zero)
prepotential of topological string theory
\be
\label{Fzero}
\cF = {1\over \la^2} \cF_0,
\ee
where, as always,
$$
X^i = \oint_{a^i} ydx,\qquad \lotjesd_i \cF_0 = \oint_{b_i} ydx.
$$
In fact, we can also include a non-trivial flux $p^i$ through the
cycles $a^i$. This will give a second contribution to the free energy
given by
\be
p^i \oint_{b_i} A = {1\over \la}p^i \lotjesd_i \cF_0.
\label{p-F} 
\ee
We recognize the contributions (\ref{Fzero}) and (\ref{p-F}) as the
genus zero contributions in the expansion for small $\la$ of the
general expression (\ref{pert}).

\subsection{D-Modules} 
\index{$\cD$-module}
\index{quantum curve}

\begin{figure}[b]
\begin{center}
\includegraphics[width=10cm]{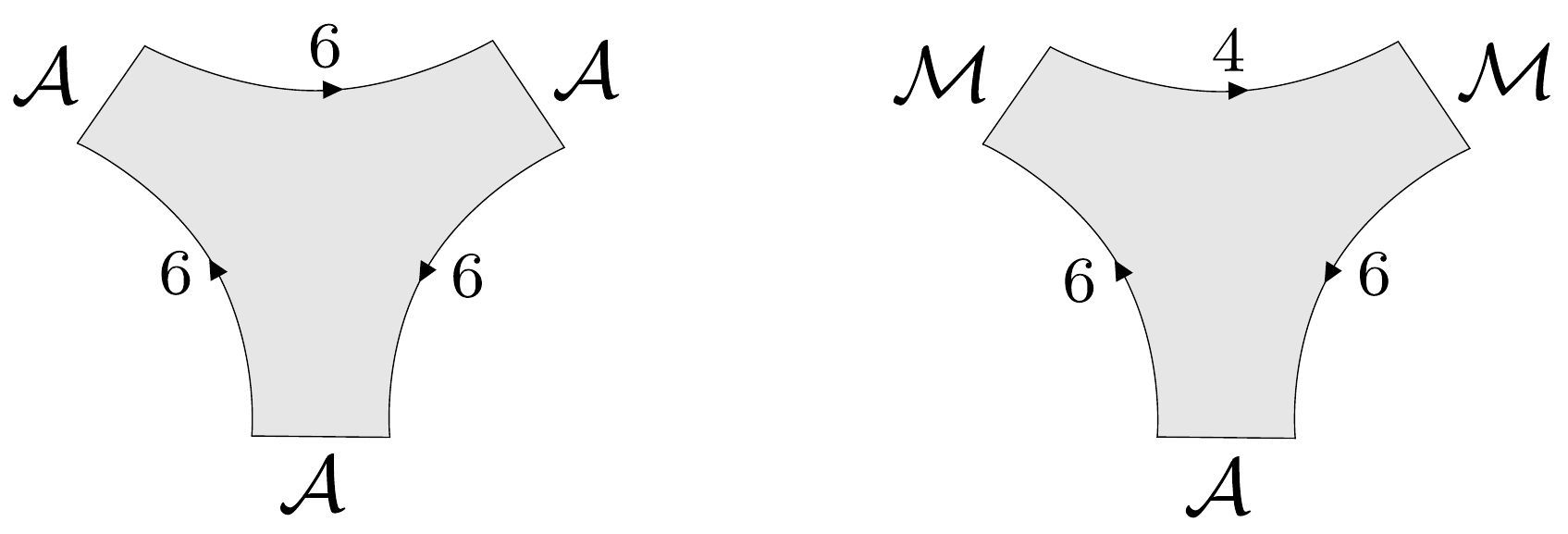}
\caption{The 6-6 strings naturally form a
  non-commutative algebra $\cA$: glueing the inner endpoints of two 6-6
  yields another 6-6 string. On the other hand, 4-6 strings are a module
  $\cM$ for the algebra $\cA$: glueing the right endpoint 4-6 to the
  left one of a 6-6 string yields another 4-6 string.}\label{fig:module}
\end{center} 
\end{figure}

Let us now explain how $\cD$-modules naturally appear in the I-brane
set-up. (See also \cite{Kapustin:2006pk} for a much more involved setting.)
First of all, by very general arguments the algebra $\cal A$ of open
string fields on the D6-brane is naturally non-commutative. This is a
consequence of the fact that the Riemann surface that describes the
interaction
$$
\cA \otimes \cA \to \cA
$$
only has a cyclic symmetry (see Fig. \ref{fig:module}).

This non-commutativity is particularly clear if one includes a
$B$-field as in (\ref{eqn:B-field}). One simple way to see this is, that in the
presence of such a $B$-field a gauge field $A$ is induced that couples
to the open strings. The gauge field satisfies $dA=B$ and can be
chosen as \index{B-field}
$$
A = {1\over \la} y dx.
$$
Therefore, on the 6-6 strings there is a boundary term
$$
\int A = \int dt\ {1\over \la} y \dot x.
$$
If we reduce this term to the zero-modes of the strings we get the
usual term of quantum mechanics, where $\la$ plays the role of
$\hbar$. (Indeed, notice that the momentum coordinate is given by $\partial
L/\partial \dot{x} = y$.)   
Therefore the coordinates $x$ and $y$ become non-commutative
operators
$$
[\hat x,\hat y] = \la.
$$
So, the zero-slope limit of the algebra $\cA$ becomes the
Heisenberg algebra, generated by the variables $\hat x, \hat y$. This
is also known as the non-commutative or quantum plane. In the case
where $\cB = \C^2$ we can identify this algebra with the algebra of
differential operators on $\C$
$$
\cA \cong \cD_\C.
$$
The algebra $\cD_\C$ consists of all operators
$$
D = \sum_i a_i(x) \lotjesd^i,\qquad \lotjesd = {\lotjesd \over \lotjesd x}.
$$
Here we have identified $\hat y = - \la \lotjesd$.

Now suppose that we add some other brane to this system, in our case a
D4-brane localized on $\Sigma$. We pick $x$ as our local coordinate,
so that, once restricted to the curve, the variable $y$ is given by a
function $y=p(x)$. Now the space of 4-6 open strings, that we will
denote as $\cM$, is by definition a module for the algebra $\cA$ of
6-6 strings. This is a simple consequence of the fact that glueing a 6-6
string to a 4-6 string produces again a 4-6 string, as illustrated
in Fig.~\ref{fig:module}. Therefore there is an action
$$
\cA \times \cM \to \cM.
$$
(More completely, $\cM$ is a $\cA$-$\cB$ bimodule, where $\cB$ is the
algebra of 4-4 open strings.)

Modules $\cM$ for the algebra of differential operators are called
\emph{$\cD$-modules}. In this case we are interested in $\cD$-modules that in
the semi-classical limit reduce to curves or equivalently
Lagrangians. Such $\cD$-modules are called holonomic.

So we can draw the following conclusion: in the presence of 
a background flux, the chiral fermions on the I-brane should no
longer be regarded as local fields or sections of the spin bundle
$K^{1/2}$. Instead they should be interpreted as sections of a
non-commutative $\cD$-module. 

Notice that if $\Sigma$ is a non-compact curve, having marked points
at infinity, the symplectic form $dx \wedge dy$ becomes very large at
the asymptotic legs. This can be see by using the appropriate
variables at infinity $x'=1/x$ and $y'=1/y$. In these variables the
$B$-field becomes singular which means that the non-commutativity goes
to zero. This explains why it makes sense to speak about the usual
free chiral fermions at infinity, and to discuss their nontrivial
transformation properties from patch to patch as in \cite{Aganagic:2003qj}.
Considering compact spectral curves seems to be much more involved
from this perspective.

In the context of string theory it is worth stressing the range of parameters $\alpha'$ and $\la$ in which $\cD$-module description is valid. The string coupling $\la$, which enters in the B-field flux as $B = \frac{1}{\la} dz \wedge dw$, plays an important role as quantization parameter. From the $\cD$-module point of view there seems to be no restriction on $\la$, so one might hope that the $\cD$-module even captures non-perturbative information.  However, in a particular system under consideration some restrictions on the values of $\la$ could arise that are related to the radius of convergence of the partition function. Although we do make some additional remarks in Chapter~\ref{chapter6}, we do not study these issues in this thesis.

On the other hand, the string scale $\alpha'$ does not play a fundamental role in the $\cD$-module. The $\cD$-module describes the topological sector of the intersecting brane configuration, which is realized in terms of massless modes of the I-brane system. Therefore the $\cD$-module description is valid only in the regime where $\alpha'$ is small (so that no massive modes interfere with our description). The most interesting case is of course when it is non-zero, as it provides a normalization factor for the worldsheet instanton contributions to the open 4-6 strings in the I-brane partition function (\ref{fermionpartfunction}). Section~\ref{sec:SW} clarifies this with an example.

\subsubsection{$\cD$-modules and differential operators}
\index{$\cD$-module}

The theory of $\cD$-modules was introduced and developed, among  others, by
I. Bernstein, M. Kashiwara, T. Kawai and M. Sato, to study linear
partial differential equations from an algebraic perspective \cite{coutinho,bjork,kashiwara,bernstein}.
Currently this is a very active field, with connections and applications to many other branches
of mathematics.

As already mentioned above, $\cD$-modules are defined as modules for the
algebra of differential operators $\cD$. In general, in a local $\C^n$
patch with complex  
coordinates $(z_1,\ldots,z_n)$, the operators $z_i$ and
$\partial_{z_i}$ represent the $n$th Weyl algebra.
The operators $P\in\cD$ are of the form
\begin{align*}
P = \sum_{i_1,\ldots,i_n}
a_{i_1,\ldots,i_n} \partial_{z_{i_1}}\cdots \partial_{z_{i_n}}. 
\end{align*}
With a set of operators $P_1,\ldots,P_m\in\cD$ one can associate a
system of differential equations 
\be
P_1\Psi = \ldots = P_m\Psi = 0,  \label{PPsi}
\ee 
where $\Psi$ takes values in some function space $\cV$.
An algebraic description of solutions to such a system can be given in terms
of a $\cD$-module $\cM$ determined by the ideal generated by
$P_1,\ldots,P_m\in\cD$ 
\be
\cM = \frac{\cD}{\cD \cdot \langle P_1,\ldots,P_m \rangle}. \label{DDP}
\ee
The advantage of considering such a $\cD$-module is, firstly, that it
captures the solutions to the above system 
of differential equations independently of the form in which this system is
written. Secondly, it is also independent of the function space $\cV$ -- be it 
the space of square-integrable functions, the space of distributions,
the space of holomorphic functions, etc.  

Suppose that one would want to extract solutions $\Psi$ from $\cM$
that take value 
in the function space $\cV$. Such a space $\cV$ is itself a
$\cD$-module. Namely, $\hat x$ and $\hat y$ are realized as
multiplication by $x$ and the differential $-\la \lotjesd_x$. One of the
important properties of $\cD$-modules is that the space of solutions
to the system of differential equations (\ref{PPsi}) in $\cV$ is given
by algebra homomorphism
\begin{align}\label{eqn:dmodhom}
{\rm Hom}_\cD(\cM, \cV).
\end{align}

An important notion is a dimension of a $\cD$-module. The so-called
Bernstein inequality asserts  
that a non-zero $\cD$-module $\cM$ over the $n^{th}$ Weyl algebra 
has a dimension $2n\geq \mathrm{dim} \, \cM \geq n$. In particular, $\cD$
considered itself as 
a $\cD$-module has a dimension $2n$. On the other hand,
$\mathrm{dim}\, \C[x_1,\ldots,x_n] = n$. 
For a non-zero $P\in \cD$, $\mathrm{dim} \, \cD/\cD P = 2n-1$.

A special role in theory of $\cD$-modules is played by the so-called
holonomic $\cD$-modules,  
which by definition have a minimal dimension $n$.
In particular they are cyclic, which means they are determined 
 by a single element $\Psi \in\cM$ called a generator.

In the context of the I-brane in $\C^2$ we are just interested in the
$1$st Weyl  
algebra $\langle x, \la \partial_x \rangle$ of dimension 2. In this
case we immediately 
conclude that the module $\cD/\cD P$ has a dimension $n=1$ for any
non-zero $P$, and is thus holonomic and hence cyclic. It can be
realized as 
\be
\cM = \{D\Psi : D\in\cD\},   \label{DPsi}
\ee
where the generator $\Psi$ is a solution to the differential equation
$P\Psi=0$. 
Such $\cD$-modules reduce to curves or equivalently complex
Lagrangians in the semi-classical limit,  
which is of course the main reason to study them in the context of
I-branes.

\subsubsection{Rank 1 example} 

Let us explain this structure in more detail with a simple example of a
$\cD$-module. We start with the commuting case, in which the
algebra $\cA$ is given by the ring $\cO$ of functions on the plane. If
the spectral curve $\Sigma$ is given by $P=0$, then we can write $\cM
= \cO_\Sigma$ as the quotient
\begin{align*}
\cM = \cO / {\cal I}_\Sigma,
\end{align*}
where ${\cal I}_\Sigma =\cO \cdot P$ is the ideal of functions vanishing on
$\Sigma$.

Now suppose that $P$ is not a polynomial, but a differential operator
\begin{align*}
P \in \cD.
\end{align*} 
Then we can similarly define a $\cD$-module as an equivalence class
of differential operators
\begin{align*}
\cM = \frac{\cD}{\cD \cdot P}.
\end{align*}
One way to think about such a module is in terms of a formal solution
to the equation
\begin{align*}
P \Psi =0.
\end{align*}

Mathematically, such a solution $\Psi \in \cV$ is captured by the
$\cD$-module homomorphism (\ref{eqn:dmodhom}) 
\begin{align*}
\cM =  \frac{\cD}{\cD  \cdot P} ~ \rightarrow~ \cV,
\end{align*}
Indeed, the map that sends the element 
\begin{align*}
[1] \in \cM ~\mapsto~  \Psi(z) \in
\cV
\end{align*}
is well-defined because every element $P' \in \cD P$ is mapped to
zero (remember that $\Psi$ fulfills 
$P\Psi=0$), and it is a bijection; conversely, any map $\cM$ to
$\cV$ is determined by  
a holomorphic solution to the differential equation $P\Psi=0$.
So $\cM$ can also be realized as the vector space of expressions of the form
\begin{align*}
\cM = \{D\Psi;\ D\in \cD\}.
\end{align*}

The $\cD$-module $\cM$ and the corresponding differential operator $P$
should be considered as the non-commutative generalization of the
classical curve $\Sigma$. This is the \emph{quantum spectral curve} from
the theory of quantization of integrable systems, as is known from the
geometric Langlands perspective \cite{Kapustin:2006pk}. 

 Within the context
of string theory it is clear that there should be a {\it unique}
$\cD$-module that corresponds to the curve $\Sigma$. This prescription
should fix possible normal ordering ambiguities in $P$. It would be
interesting to understand this directly from the mathematical
formalism.

\subsubsection{$\cD$-modules and flat connections}
\index{$\cD$-module}

More generally, $\cD$-modules are defined as differential sheaves on any
variety $X$. The sections of the sheaf $\cD_X$ over an open
neighbourhood $U$ are 
given by linear differential operators on $U$. Therefore, both the
structure sheaf $\cO_X$ (of holomorphic functions) as well as the
tangent sheaf $T_X$ (whose local sections are vector fields) may be
embedded in $\cD_X$.
\begin{align*}
\cO_X \hookrightarrow \cD_X \hookleftarrow T_X
\end{align*}  
In fact, $\cD_X$ is generated by these inclusions.

A sheaf $\cM$ on $X$ is defined to be a left module for $\cD_X$ when
$v \cdot s \in \cM$, for any  $v \in \cD_X$ and $s\in
\cM$. Furthermore, it has to fulfill  
\begin{align*}
v \cdot (fs) = v(f) s + f(v \cdot s)\\ 
[ v,w ] \cdot s = v \cdot (w \cdot s) - w \cdot (v \cdot s) 
\end{align*}
for any $v \in \cD_X$, $f\in \cO_X$ and $s\in \cM$. Suppose that $\cM$
is a left $\cD_X$-module whose sections are the local sections of some
vector bundle $V$ (this encomprises all $\cD_X$-modules that are
finitely generated as $\cO_X$-modules). Then the action of $\cD_X$ defines
a connection on $V$ as
\begin{align}\label{eq:connectioncorrD-mod}
\nabla_v (s) &= v \cdot s, 
\end{align}
whose curvature is zero. So a $\cD$-module structure on the sheaf of
sections of a vector bundle $V$ defines a flat connection on this
vector bundle. And conversely, any module consisting of
sections of a vector bundle $V$ with flat connection $\nabla_A$, has
an interpretation as a $\cD$-module defined through the action of the
flat connection. Therefore, a $\cD$-module is in general just a 
system of linear differential equations, changing from patch to patch on $X$. This is known as a local system. In the main part of this paper $X$ is just $\C$ or $\C^*$.

\subsubsection{Examples}
\vspace{2mm}

\noindent 1)~
Take a linear partial differential operator on $\C$, for example 
\begin{align}
P = \la z \partial_z - 1.   \label{Pexmp}
\end{align}
The differential equation $P \psi=0$ is solved by
$\psi(z) = z^{1/\la}$, so according to (\ref{DPsi}) the corresponding
$\cD$-module can be represented as
\begin{align*}
\cM = \langle z,\la\partial_z \rangle \, z^{1/\la}.
\end{align*}

There are many equivalent ways of writing this module. For example,
introducing $\wt \Psi = z\Psi$, 
the above differential equation is transformed into ${\wt P} {\wt
  \Psi} = 0$ with  
\begin{align*}
{\wt P} = \la z \partial_z - \lambda - 1,
\end{align*}
This follows simply from the relation $(\la z\partial_z - \la -1)x =
x(\la z\partial_z -1)$.  
This new operator, as well as the solution to the new equation ${\wt
  \Psi} = z^{1+1/\la}$ look 
different than before. Nonetheless, they represent the same $\cD$-module
\begin{align*}
\cM = \langle z,\la \partial_z \rangle \, z^{1+1/\la} = \langle
z,\la \partial_z \rangle \, z^{1/\la}.
\end{align*}
This simple example indeed shows that the formlism of $\cD$-modules
allows to study solutions to partial differential equations  
independently of the way in which the differential equation is
written. 

The flat connection corresponding to $P$ is determined by the action
of $\partial_z$ on the elements of the $\cD$-module, as in
(\ref{eq:connectioncorrD-mod}). It is given by
 %
%
%
\begin{align*}
\nabla_A = \partial_z dz - \frac{1}{\la z} dz, 
\end{align*}
determining $\psi(z)$ as a local flat section.

\vspace{2mm}

\noindent 2)~
All the modules that we will study in this paper are over $\C$ or
$\C^*$. It is important that they may be of any rank though. Let us
therefore also give a rank two example on the complex plane $\C$. The 
second order differential equation
\begin{align}
P \psi= (\la^2 \partial_z^2 - z)\psi
\end{align}
can be written equivalently as a rank two differential system  
\begin{align*}
P_{ij} \psi_j = 0, \quad \mbox{with}~ P_{ij}=  \left(\begin{array}{cc} 
   \la \partial_z & 0 \\ 0  & 
   \la \partial_z \end{array} \right) -  \left(\begin{array}{cc} 
  0 & 1 \\ z   & 
  0 \end{array} \right). 
\end{align*}
Holomorphic solutions of this linear system are captured by the map 
\begin{align*}
\cM = \frac{\cD^{\oplus 2}}{\cD^{\oplus 2} \, P_{ij}}
\rightarrow \cO_{\C}^{\oplus 2}
\end{align*}
that sends the two generators $[(1,0)^t]$ and $[(0,1)^t]$ to two
independent (2-vector) solutions of $P \psi = 0$. The corresponding
flat connection reads  
 %
%
%
\begin{align*}
\nabla_A = 
\partial_z dz - \frac{1}{\la}
\left( \begin{array}{cc} 0 & 1 \\ z &   0 \end{array} \right)  dz 
\end{align*}
and turns the two solutions into a locally flat frame.

\subsection{Quantum curve}\label{sec:defquant} 
\index{quantum curve}

The $B$-field quantizes the I-brane configuration into a
$\cD$-module. Last subsection we introduced these objects and
saw that they represent solutions to a linear system of
differential equations. However, the I-brane setup doesn't provide us
with just any $\cD$-module: this $\cD$-module must represent a
quantization of the
curve $\Sigma$ 
together with a meromorphic 1-form $\tau$, obeying $d \tau = \omega$,
on it. In this article we focus 
on smooth curves that are given by an equation of the form  
\begin{align}\label{eqn:spectralcurve}
\Sigma: \quad H(z,w) = w^n + u_{n-1}(z) w^{n-1} + \ldots + u_0(z) = 0, 
\end{align}
where $z \in \C$ (or $\C^*$) and $w \in \C$. These play a prominent role
in integrable systems as spectral curves. We will describe
$\cD$-modules corresponding to these curves.

\subsubsection{Semi-classical geometry}

The spectral curve $\Sigma$ in (\ref{eqn:spectralcurve}) is a degree
$n$ cover over $\C$ (or $\C^*$)
\begin{eqnarray*}
\Sigma &\subset& T^*\C \\
 \downarrow& \hspace*{-6mm}\pi\\
\C&& \notag
\end{eqnarray*}
with possible branch points (from now on we restrict to $z \in \C$ for
simplicity in notation). The curve $\Sigma$ is imbedded in $\C^2$ and  
equipped with the (meromorphic) 1-form
%
\begin{align*}
\tau = \frac{1}{\la} w dz|_{\Sigma}. 
\end{align*}

Furthermore, fermions on $\Sigma$ transform as holomorphic sections of
a line bundle $\cL 
\otimes K^{1/2}_{\Sigma}$, provided by the D6-brane. The tuple
$(\cL, \tau)$ on $\Sigma$ pushes forward to a couple 
\begin{align}\label{eqn:Higgspair}
\pi_*:~ (\cL, \tau) \mapsto (V = \pi_*\cL, \phi = \pi_*\tau)
\end{align}
on $\C$ under the projection map $\pi: \Sigma \to \C$. So $V$ is a
rank $n$ vector bundle on $\C$, whereas $\phi \in H^0(\C, K_C \otimes
\mbox{End}V )$ is a meromorphic 1-form valued in $gl(n)$.
 Such an object is called a Higgs field. It endows $V$
with the structure of a Higgs bundle. Setting the characteristic polynomial 
\begin{align*}
\mathrm{det}(\tau - \phi(z))=0
\end{align*}
returns the equation for the spectral curve. The push-forward map
$\pi_*$ sets up a bijection between spectral data and 
(stable) Higgs pairs
\begin{align}\label{eqn:Higgsequivalence}
(\Sigma, \cL) ~\leftrightarrow ~ (V, \phi). 
\end{align}
%

When the base $\C$ is a compact curve $C$ instead, 
the moduli space of stable Higgs pairs forms an algebraically
completely integrable system, the Hitchin integrable system.
The Hitchin map 
\begin{align*}
H^0(C, \mathrm{End}V \otimes K_C) &\to \oplus_{i=1}^n
H^0(C,K_C^{i})\\
\phi &\mapsto \mathrm{det}(w - \phi(z)),
\end{align*}
provides this moduli space with the structure of a Lagrangian fibration, whose
fibers are generically Lagrangian tori. Any point in the image of the
Hitchin map can be 
identified with a spectral curve $\Sigma$, whereas the fiber of the
Hitchin map equals the moduli space of line bundles on $\Sigma$
of a certain degree, which is isomorphic to the Jacobian
$\mathrm{Jac}(\Sigma)$. This verifies that the Hitchin fiber is generically given by a 
torus. Multiplying $\cL$ by a flat bundle on $\Sigma$ defines a linear
flow over the Hitchin fiber, exactly as in our discussion of the
Krichever correspondence in \Cref{sec:KP}.

\subsubsection{Quantum geometry} 

The $\cD$-modules generated by the $B$-field (\ref{eqn:B-field}), as well
as those  
considered in the last subsection, depend on $\la$ (where $\la$ is a
formal variable $\la \in \C[[\la]]$). These are known as
$\cD_{\la}$-modules \cite{Arinkin:2004du}. As a linear differential
system such a 
$\cD_{\la}$-module corresponds to a $\la$-connection~$\nabla_{\la}$ 
\begin{align}
\nabla_{\la} = \la \partial_z - A(z).
\end{align}
that is defined through the Leibnitz rule $\nabla_{\la} (fs) = f
\nabla_{\la}(s) + \la s 
\otimes df$ for any function $f$ and section $s$. Since all the
$\cD$-modules and connections we consider 
are $\cD_{\la}$ and $\la$-connections, we often omit the subscript $\la$. 

Semi-classically, a $\la$-connection $\nabla_{\la}$
reduces to a 1-form $\nabla_0(z)$ with values in $gl(n)$ 
\begin{align}
\nabla_{\la} \mapsto \nabla_0, \quad (\la \to 0).
\end{align}
We just encountered this object as a Higgs field $\phi$. Moreover, we
explained with (\ref{eqn:Higgspair}) that a Higgs $(V,\phi)$ and
spectral data $(\Sigma,\cL)$ 
provide equivalent information. In particular, the spectral curve can
be recovered by the determinant of the Higgs field. This implies that
$\la$-connections quantize spectral data.\footnote{These
  $\la$-connections are also known as $\la$-opers, and play an important
  role in the quantum integrable system of Beilinson and Drinfeld
  \cite{beilinsondrinfeld,talkarinkin}. } 

It tells us exactly which requirements a $\cD$-module quantizing the
I-brane configuration has to satisfy. Fermions on 
a degree $n$ spectral curve have to transform under a rank $n$
$\la$-connection $\nabla_{\la}$ on $\C$, whose semi-classical $\la \to 0$
limit is given by the Higgs field 
\begin{align*}
\nabla_0 = \pi_* (\tau).
\end{align*}
A simple
example of a $\la$-connection is given by  
\begin{align*}
\nabla = \la \partial_z - A(z),
\end{align*}
with $A(z) = \pi_* (\tau)$. Its determinant is a degree $n$
differential equation that canonically quantizes the defining equation
for $\Sigma$. 

%
%

\subsubsection{Examples}

In the last subsection we gave two examples of $\cD_{\la}$-modules: one
of rank 1 and another one of rank 2. Let us examine them here in the
light of deformation quantization. 

\vspace{2mm}
 
\noindent 1)~ In the first example we found the
$\la$-connection 
\begin{align}
\nabla_{\la} =  \la \partial_z - \frac{1}{z}
\end{align}
on the $z$-plane. The semi-classical spectral data is given by the
degree 1 spectral cover
\begin{align*}
\Sigma: \quad  w = \frac{1}{z},  
\end{align*}
with $z,w \in \C^*$, together with
the (meromorphic) 1-form  
\begin{align*}
A = \frac{1}{z} dz.
\end{align*}
Equivalently, the $\cD$-module $\cM$ corresponding to $\nabla_{\la}$ is
generated by
\begin{align*}
\cM = \langle z, \la \partial_z \rangle  z^{1/\la} ,
\end{align*}
which is clearly invariant under the algebra $\cA$  
%
%
of functions on $\Sigma$ at $\la=0$. It is therefore a quantization of
$\Sigma$. This example enters string theory as the deformed
conifold geometry describing the $c=1$ string. It is described as a
$\cD$-module in \cite{Dijkgraaf:2007sw}. We will come back to it in
\Cref{ssec:quantuminvariant}. 

\vspace*{2mm}

\begin{figure}[h!]
\begin{center}   
\includegraphics[width=4.5cm]{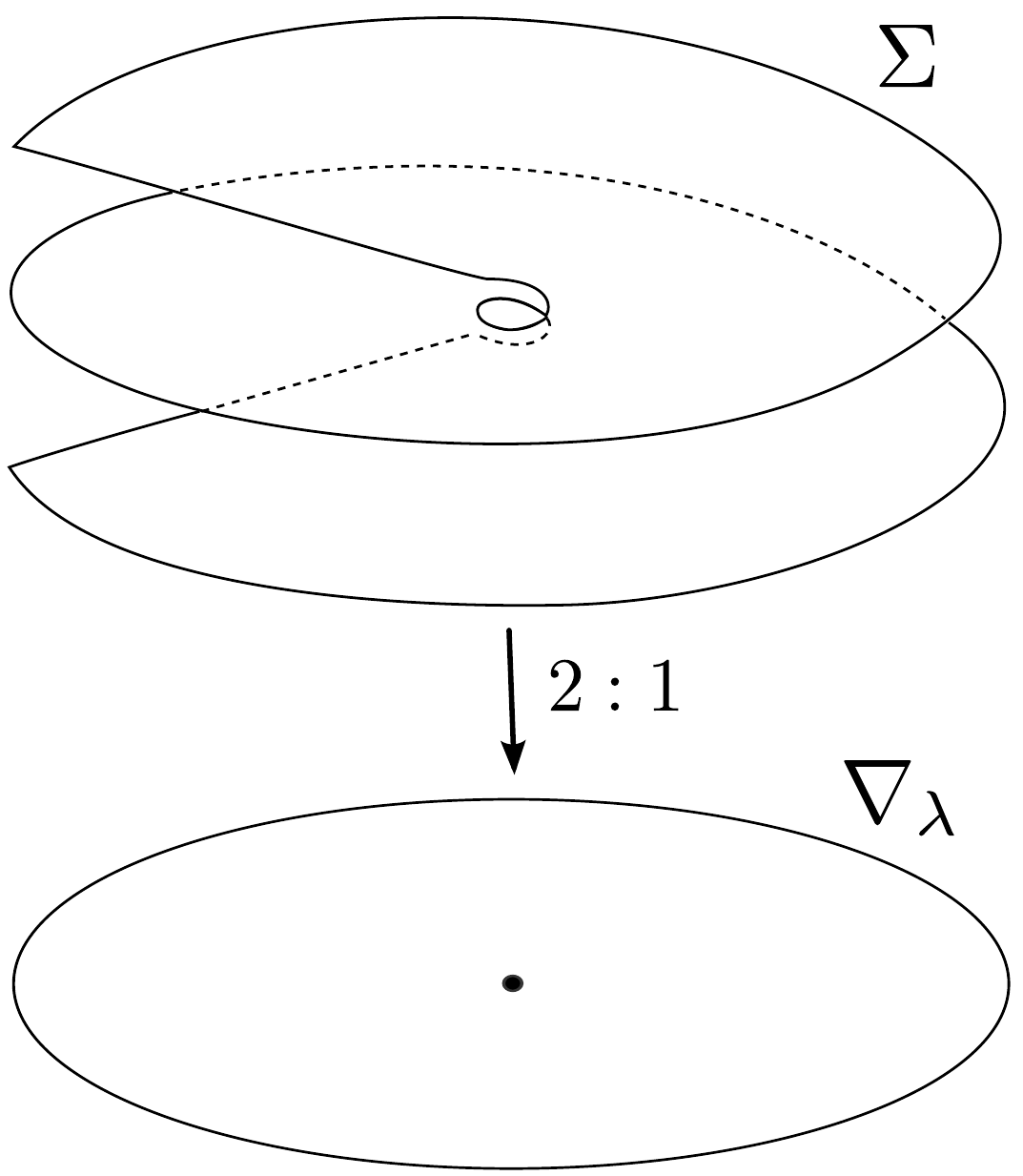}
\caption{A second order differential
  operator $P$ in $\la \partial_z$ defines a rank 2 $\la$-connection
  $\nabla_{\la}$. The determinant of $\nabla_0$ determines a degree 2
  cover over $\C$ which is called the spectral curve
  $\Sigma$.} \label{fig:n=1cover} 
\end{center} 
\end{figure}

\noindent 2)~ In the second example we considered a
$\cD_{\la}$-module generated by a single second order generator
\begin{align}
P = \la^2 \partial_z^2 - z.
\end{align}
The corresponding rank 2 $\la$-connection 
\begin{align*}
\nabla_{\la} =  \left(\begin{array}{cc} 
   \la \partial_z & 0 \\ 0  & 
   \la \partial_z \end{array} \right) -  \left(\begin{array}{cc} 
  0 & 1 \\  z   & 
  0 \end{array} \right) 
\end{align*}
is a $\la$-deformation of the degree 2 spectral cover (illustrated in
Fig.~\ref{fig:n=1cover})  
\begin{align*}
\Sigma: ~w^2 =  z,
\end{align*}
with meromorphic 1-form $\tau =  w dz|_{\Sigma}$. Note that this
one-form pushes forward to the connection 1-form, or Higgs
field,
\begin{align*}
A =  \left(\begin{array}{cc} 
  0 & 1 \\  z   & 
  0 \end{array} \right) dz 
\end{align*}
in the basis $\{ dz, wdz \}$ of ramification 1-forms on the
$z$-plane. Indeed, local sections of $\cL$ push forward to local
sections generated by $1$ and $w$ on the $z$-plane. Now, $wdz \cdot 1
= w dz$ and $ w dz \cdot w = w^2 dz = z dz$ on $\Sigma$.

We discuss the string theory interpretation of this $\cD$-module
 in \Cref{sec:matrixmodels}. 

\section{Fermionic states and quantum
  invariants}\label{sec:quantumpartfunction} 
\index{fermionic state}
\index{tau-function}

In the last section of this chapter we quantify the quantum integrable
system that we described in the last section. Just like the Krichever
map embeds a geometric set-up consisting of a line bundle over a curve
into the KP Grassmannian, there is a straightforward way in which a
quantum curve or $\cD$-module $\cM$ gives
rise to a solution of the KP-hierarchy. By definition $\cM$ carries an 
action of both $x$ and $\partial_x$. However we are free to ignore the
second action, which leaves us with the structure of an $\cO$-module,
$\cO$ being the algebra of functions in $x$. By applying the
infinite-wedge construction to the $\cO$-module $\cW$ we obtain in the usual
way a state $|\cW\rangle$ in the fermion Fock space. Roughly speaking,
$\cW$ can be considered as the space of local sections that can be
continued as sections of a (non-commutative) $\cD$-module, instead
of sections of a line bundle over a curve. We explain this in \Cref{sec:fermstate}. In
\Cref{ssec:quantumstate} and \Cref{ssec:quantuminvariant} we
continue to discuss how to assign quantum 
invariants to quantum curves. Also this is inspired on the
KP-hierarchy. 

\subsection{Fermionic state}\label{sec:fermstate} 

In this section we introduce an infinite dimensional Grassmannian and
its description in terms of the second quantized fermion field
(we learned this material e.g. from
\cite{segalwilson,sato1989,Dijkgraaf:1991qh,mulaseKP,Kac:1993ji}).   
Later on we will extend this formalism in a way which associates
fermionic states to $\cD$-modules, thereby providing a fermionic
description of the I-brane system. This technology is required as
pre-knowledge to Chapter~\ref{chapter6}. 

\subsubsection{Grassmannian and second quantized fermions}

The space $\cH = \C((z^{-1}))$ of all formal Laurent series in
$z^{-1}$ can be given an interpretation of a Hilbert space. Basis
vectors $z^n$, for $n\in \mathbb{Z}$, correspond to one particle
states of energy $n$ associated to the Hamiltonian $z\partial_z$.  
This Hilbert space has a decomposition 
\begin{align*}
\cH = \cH_+ \oplus \cH_-,     
\end{align*}
such that the first factor $\cH_+=\C[z]$ is a subspace generated by
$z^0$, $z^1$, $z^2$, $\ldots$, while $\cH_-$ is generated by negative
powers $z^{-1},z^{-2},\ldots$. Consider now a subspace $\cW$ of $\cH$
with a basis 
$\{ w_k(z) \}_{k\in \mathbb{N}}$. 
We say it is comparable to
$\cH_+$, if in the projection onto positive and negative modes 
\begin{align*}
w_k = \sum_{j\geq 0} (w_+)_{ij}z^j + \sum_{j>0}(w_-)_{ij}z^{-j}
\end{align*}
the matrix $w_+$ is invertible. The Grassmannian $Gr_0$ is the set of all subspaces $\cW \subset \C((z^{-1}))$ which are comparable to $\cH_+$.

In what follows we take much advantage of the
correspondence between $Gr_0$ and the charge zero sector of the second
quantized fermion Fock space $\cF_0$. In this correspondence the
subspace $\cH_+$ is quantized as the Dirac vacuum
\begin{align}
|0 \rangle = z^0 \wedge z^1 \wedge z^2 \wedge \ldots,    \label{wedge-vac}
\end{align}
with all positive energy states filled. The fermionic state associated to the
subspace $\cW$ with basis $w_0(z)$, $w_1(z)$, $w_2(z)$, $\ldots$ is
represented by the semi-infinite wedge\footnote{Actually, we have to
  tensor with $\sqrt{z}$ to make the state fermionic.} 
\begin{align}
|\cW \rangle = w_0 \wedge w_1 \wedge w_2 \wedge \ldots
\end{align}
which is an element of the fiber of a determinant line bundle over the
element $\cW \in Gr$ (and therefore determined up a complex scalar $c$)

To make contact with the usual formulation of the second quantized
fermion Fock space, we can identify the differentiation and wedging
operators with the fermionic modes  
\begin{align*}
\psi_{n+\hf} = \frac{\partial}{\partial z^{-n}} \qquad  \psi_{n+\hf}^*
= z^{n} \wedge.  
\end{align*}
These half-integer modes are annihilation and creation operators which
arise from a decomposition of the fermion field $\psi(z)$ and its
conjugate $\psi^*(z)$ 
\begin{align}
\psi(z) = \sum_{r \in\Z + \hf} \psi_{r} z^{-r-\hf} \qquad \psi^*(z) =
\sum_{r\in\Z+\hf} \psi^*_{r} z^{-r-\hf},   \label{fermion-NS} 
\end{align}
and they obey the anti-commutation relations 
$
\{\psi_{r},\psi^*_{-s}\} = \delta_{r,s}.
$

For subspaces $\cW \in Gr_0$ the determinant of the projection onto
$\cH_+$ is well defined and can be expressed as 
\begin{align*}
\det w_+ = \langle 0 | \cW \rangle. 
\end{align*} 

More generally, one can consider the Fock space $\cF$ which splits
into subspaces of charge $p$   
\begin{align*}
\cF = \bigoplus_{p\in\Z}\, \cF_p.
\end{align*}
Each subspace $\cF_p$ is built by acting with creation and
annihilation operators on a vacuum 
\begin{align*}
|p \rangle = z^p\wedge z^{p+1}\wedge z^{p+2}\wedge\ldots,
\end{align*}
with the property
\begin{eqnarray*}
\psi_{r} |p\rangle & = & 0 \qquad \textrm{for}\ r > p,  \\
\psi^*_{r} |p\rangle & = & 0 \qquad \textrm{for} \ r > -p.
\end{eqnarray*}
The Fermi level of the vacuum $|p \rangle$ is shifted by $p$ units
with respect to the Dirac vacuum $|0 \rangle$. 
This fermion charge is measured by the $U(1)$ current
$$
J(z) = :\psi(z)\psi^*(z): = \sum_n \alpha_n z^{-n-1},
$$
whose components $\alpha_n = \sum_k : \psi_r \psi^*_{n-r}$ satisfy the
bosonic commutation relations  
\begin{align*}
[\alpha_m, \alpha_{-n}] = m \delta_{m,n}.
\end{align*}
With each subspace  $\cW \subset \C((z))$ comparable to the one generated by $(z^k)_{k\geq p}$ one can associate a state $|\cW\rangle \in \cF$ of charge $p$. This charge is equal to the index of the projection operator $pr_+ : \cW \to \cH_+$.

\begin{figure}[b!]
\begin{center}  
\includegraphics[width=6.5cm]{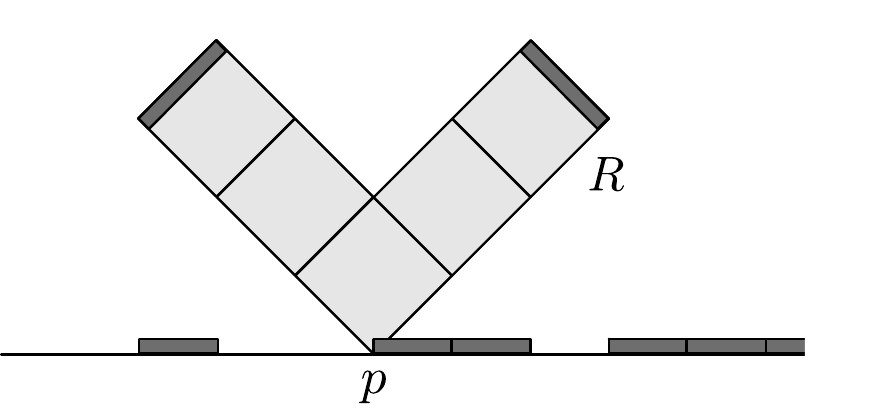}
\caption{Elements of the Fock space
  $\mathcal{F}$  are in a bijective correspondence
  with Maya diagrams. The bottom line represent a Maya diagram
  corresponding to a fermionic state with charge $p$.
  As illustrated it is characterized by a two-dimensional partitions $R$ located
  at position $p$. We therefore denote the state as $|p,R\rangle \in
  \mathcal{F}$.} \label{fig:maya} 
\end{center} 
\end{figure}

A state in the Fock space $\cF$ has also a simple representation in terms of the so-called Maya diagram (see Fig.~\ref{fig:maya}). Black boxes in such a diagram represent excitations, whereas white boxes are gaps in the energy spectrum of the fermion. The charge of a state is given by the number of excitations minus the number of gaps. Fermionic states or Maya diagrams of a fixed charge $p$ can also be associated to two-dimensional partitions. In particular in $p=0$ sector the state
$$
|R\rangle = \prod_{i=1}^{d} \psi^*_{-a_i-\hf} \psi_{-b_i-\hf}|0\rangle
$$
corresponds to the partition $R=(R_1,\ldots,R_l)$ such that
$$
a_i = R_i-i,\qquad b_i = R_i^t-i.
$$
In what follows a state corresponding to a partition $R$ of charge $p$ is denoted as $|p,R\rangle$.

\subsubsection{Flow on the Grassmannian}

Multiplying a basis vector $w_k(z)$ of $\cW$ by a power series $f(z)=
\sum_{n > 0} f_n z^n$, that vanishes at $z=0$, defines an action on the
 Grassmannian:
\begin{align*}
f(z) |\cW \rangle = \sum_k w_0 \wedge \ldots \w_{k-1} \wedge
f \cdot w_k
\wedge w_{k+1} \ldots. 
\end{align*}
When we write $w_k(z)$ in terms of the basis $(z^l)_{l \in \Z}$ this
action is encoded by the multiplication by an infinite matrix in
$gl_{\infty}$, whose $(i,j)^{th}$ entry is given by $f_{i-j}$. 
On the fermionic state $|\cW \rangle$ a multiplication by $z^n$ translates
into a commutator with the bosonic mode $\alpha_n$, since $\alpha_n$
increases the fermionic mode number by $$[\alpha_n, \psi_r] =
\psi_{r+n}.$$ Multiplication by a power series $f(z)$ therefore
translates to the operator 
\begin{align*}
f = \sum_n  f_n  [ \alpha_n, \bullet ]~ \in gl_{\infty}.
\end{align*}
on the Fock space.  

Exponentiating the action of $gl_{\infty}$ yields the group
$Gl_{\infty}$. An element $g(z) = \exp(f(z))$ of this group acts on
$|\cW 
\rangle$ by multiplying all its basis vectors   
\begin{align*}
g(z) |\cW \rangle = g \cdot w_0 \wedge \ldots \wedge g \cdot  w_k \wedge \ldots 
\end{align*}
From the fermionic point of view this action is given by conjugating
each basis vector $w_k$ with the element   
\begin{align*}
 g = \exp\left(\sum f_n \alpha_n \right)= \exp \left( \oint dz ~f(z) J(z) \right)
 \in Gl_{\infty}.
\end{align*}

We call $\Gamma$ the group of exponentials $g(z) : S^1 \to \C^*$. 
An important subgroup of $\Gamma$ is the group $\Gamma_+$ of functions
$g_0: S^1 \to \C^*$ that extend holomorphically over the disk $D_0 = \{z : |z| \le 1 \}$: 
\begin{align*}
\Gamma_+ = \{ g_0: D_0 \to \C^* : g_0(0) =1 \}.
\end{align*}
Another subgroup is the group $\Gamma_-$ of
functions $g_{\infty}: S^1 \to \C^*$ that extend over the disk
$D_{\infty} = 
\{z \in \C \cup \{ \infty \}: |z| \le 1 \}$: 
\begin{align*}
\Gamma_- = \{ g_{\infty}: D_{\infty} \to \C^* : g_{\infty}(\infty) =1 \}.
\end{align*}
Any $g \in \Gamma$ can
be written as an exponential $\exp(f)$. When $g \in \Gamma_+$ the
function $f$ vanishes at $z=0$, and when $g \in \Gamma_-$ it
vanishes at $z=\infty$. 
 
$\Gamma_+$ and $\Gamma_-$ have different properties when acting on
Grassmannian.  
The action of $\Gamma_-$ is free, since any $\cW
\in Gr$ has only a finite number of excitations. On the contrary,
$\Gamma_+$ acts trivially on a vacuum state $|p \rangle$. 
Although the action of the groups $\Gamma_+$ and $\Gamma_-$ on a
subspace $\cW$  
is commutative, as it is just given by multiplication, as operators
on the fermionic state $|\cW \rangle$ it matters which element is
applied first. This introduces normal ordering ambiguities. 

An element  
\begin{align}
g(t,z) = \exp \left( \sum_{k \ge 1} t_k z^k \right) = \exp \left(
  f(t,z) \right) \in \Gamma_+,    \label{gexp}
\end{align}
defines a linear flow over the Grassmannian $Gr$. On the Fock space it
acts as an evolution operator 
\begin{align*}
U(t) = \exp \left(\oint\frac{dz}{2\pi
    i} f(t,z)J(z)\right).
\end{align*}
The determinant
det$(\cW)_+$ is not equivariant with respect to the action of $\Gamma_+$.
The difference is measured by the so-called tau-function 
\begin{align}\label{eqn:tauferm}
\tau_{\cW}(g)  = \frac{\det\,(g^{-1}w)_+} {g^{-1} \det\, w_+} =
\frac{\langle 0 | U(t) | \cW \rangle } {g^{-1} \langle
  0 | \cW \rangle},
\end{align}
which yields a holomorphic function $\tau: \Gamma_+ \to \C$.
This can be regarded as a wave function of $|\cW\rangle$.


%
%

%
%
%
%

\subsubsection{Blending}

So far we considered the Hilbert space $\cH\equiv \cH^{(1)}$ of
functions with values in $\C$. More generally, one can consider a
Hilbert space $\cH^{(n)}$ of functions with values in $\C^n$. Let
$(\epsilon_i)_{i=1,\ldots,n}$ denote a basis of $\C^n$. For each $n$
there is an isomorphism between $\cH^{(n)}$ and $\cH$ given by the
lexicographical identification of the basis
\begin{align*}
\epsilon_i z^k \mapsto z^{nk+i-1}.
\end{align*}
This isomorphism is called \emph{blending}.

In the fermionic language the Hilbert space $\cH^{(n)}$ lifts to the
Fock space of $n$ fermions $\psi^{(i)}$, $i=1,\ldots,n$, each one with
the expansion (\ref{fermion-NS}) and such that 
\begin{align*}
\{ \psi^{(i)}_r, \psi^{*\,(j)}_s \} = \delta_{i,j} \delta_{r,-s}.
\end{align*}
Now blending translates to the following redefinitions of these $n$
fermions into a single fermion $\psi$ 
$$
\psi_{n(r+\rho_i)} = \psi^{(i)}_r,\qquad\qquad \psi^*_{n(r-\rho_i)} = \psi^{*\,(i)}_r,
$$
where
\be
\rho_i = \frac{2i-n-1}{2n}.  \label{rho-def}
\ee

Blending can also be expressed in terms of two-dimensional partitions
introduced above. 
Consider $n$ partitions  $R_{(i)}$ of charges $p_i$, with $\sum_i p_i
= p$, corresponding to states in $n$ independent Hilbert spaces of
fermions $\psi^{(i)}$. Associating with each such partition a state of
a chiral fermion $|p_i, R_{(i)} \rangle$, we have a decomposition
$$
|p,{\bf R}\rangle = \bigotimes_{i=1}^n |p_i, R_{(i)}  \rangle,
$$ 
and the blended partition ${\bf R}$ of charge $p$, corresponding to a
state in the Hilbert space of the blended fermion $\Psi$, is defined
as 
\be 
\{ n(p_i+R_{(i),m} - m) +i-1\ |\ m\in\mathbb{N} \} = \{ p+{\bf
  R}_K -K \ | \ K\in\mathbb{N} \}. \label{blend} 
\ee

\vspace*{.5mm}

\subsubsection{A simple example: rational curves}

Let us illustrate how this formalism can be used to describe quantum
curves. Take a very simple example: the curve
$y=0$. Topologically $\Sigma=\C$ can be viewed as a disc and in the
fermion CFT it will correspond to a state in the Fock space $\cF$.  In
this case the corresponding $\cD$-module just consists of the
polynomials in $x$
$$
\cM = \C[x]
$$
and $\hat y$ is realized as $-\la \lotjesd_x$. The one-form $ydx$ vanishes
identically. The free fermion theory based on this module consists of
the usual vacuum state $|0\rangle$. 

But now we can make a small variation, by picking the curve
$$
y = p(x),
$$
with $p(x)$ some function.  In this case the (meromorphic) one-form is
$p(x)dx$. The corresponding $\cD$-module is still isomorphic to
$\C[x]$ as a vector space, but now the operator $\hat y$ is
represented as
$$
\hat y = -\la \lotjesd_x + p(x).
$$
Of course, there is an obvious map between these two modules: we
simply multiply the functions $\psi(x) \in \C[x]$ as
$$
\psi(x) \to e^{-S(x)/\la} \psi(x),\qquad \lotjesd S(x)=p(x)dx.
$$
In the quantum field theory, where $\psi(x)$ becomes an operator
acting on the Fock space $\cF$, this correspondence is represented by
a linear map $U$ such that
$$
U(t) \cdot \psi(x) \cdot U(t)^{-1} = e^{-S(x)/\la} \psi(x).
$$
If $S(x) = \sum_k t_k x^k$ such a map is given by
$$
U(t) = \exp \sum_k {1\over \la} t_k \alpha_{k}
$$
where $\sum_k \alpha_k x^{-k-1} = \lotjesd\phi(x) = :\psi^\dagger \psi(x):$
is the usual mode expansion of the $U(1)$ current. Here we use that
$[\alpha_k,\psi(x)]= x^k \psi(x)$. The corresponding state in the Sato
Grassmannian is given by $U(t)|0\rangle$. Since $\alpha_k |0\rangle=0$
for $k\geq 0$, this state is only different from the vacuum if the
function $S(x)$ (or $p(x)$) has poles.

\subsection{Second quantizing $\cD$-modules}\label{ssec:quantumstate}

Finally, we have all the ingredients to associate a fermionic state to a
given $\cD$-module on $\C$. The $\cD$-module $\cM$ encodes
holomorphic solutions to a system of differential equations. In
particular it is 
an $\cO_{\C}$-module, and forms a subspace of $\C((z^{-1}))$ that we will
name $\cW$. Second quantization 
turns this subspace into a fermionic state $|\cW \rangle$. Since the
I-brane configuration in fact provides a $\cD_{\la}$-module (in
contrast to a $\cD$-module) the resulting 
I-brane fermionic state $|\cW \rangle$ is a $\la$-deformation as well. 

In \Cref{sec:defquant} we learned that a $\cD_{\la}$-module on $\C$
when $\la \to 0$ reduces to a spectral tuple $(\Sigma,\tau)$. We will
argue here that the fermionic 
state $|\cW \rangle$, associated to the $\cD$-module $\cW$, reduces
to the semi-classical Krichever state corresponding to this spectral data  
in the same limit. The 
tau-function (\ref{eqn:tauferm}) corresponding to this Krichever state
computes the determinant of the $\delbar$-operator, which 
is just the genus one part $\cF_1$ of the all-genus free energy. The
$\cD_{\la}$-module gives a $\la$-deformation of this solution that
computes the full I-brane partition function (\ref{fermionpartfunction}).

In this section we  motivate and explain the second quantization of the
$\cD_{\la}$-module. But first we summarize the
semi-classical correspondence.  

\subsubsection{Semi-classical state}\label{sec:Krichever}

Let us remind ourselves shortly how we associate a semi-classical
fermionic state to a pointed curve $\Sigma- z_{\infty}$, together with
a line bundle $\cL$. As we explained in \Cref{sec:semi-clas} the
traditional way \cite{segalwilson} is to form the space of global holomorphic
sections of $\cL$ on $\Sigma-z_{\infty}$
\begin{align*}
\cW = H^0(\Sigma - z_{\infty}, \cL). 
\end{align*}
Recall that $\cW$ is a an $\cA$-module for  
\begin{align*}
\cA = H^0(\Sigma - z_{\infty}, \cO_{\Sigma}). 
\end{align*}
Tensoring the line bundle $\cL$ with a square root $K^{1/2}_{\Sigma}$
turns sections of $\cL$ into fermionic sections: $\cW$ thus
corresponds to a subspace that is being swept out in time by a free
fermion field $\psi(z)$ living on $\Sigma - z_{\infty} $.

When we denote the semi-infinite set of generators $\cW$ by $w_k(z)$, a second
quantization turns the subspace $\cW$ into the fermionic state  
\begin{align*}
| \cW \rangle  =  w_0 \wedge w_1 \wedge w_2 \wedge \ldots. 
\end{align*}
The map that assigns a subspace $\cW$ to the couple $(\Sigma, \cL)$ is
the inverse of the Krichever map. It yields a geometric solution in the
Grassmannian $Gr$. 

When $\pi$ is an $(n:1)$ projection of $\Sigma$ onto some other curve
$C$, while $\cL$ 
pushes forward to a rank $n$ vector bundle $V = \pi_* \cL$ over $C$, it is
equivalent to look at the subspace of global sections of $V$ on $C -
 \pi(z_{\infty})$. This yields a fermionic state in the
$n$-component Grassmannian $Gr_n$ \cite{Adams:1993,li-mulase}. As we
described in the previous 
subsection blending (or the lexiographical ordening) recovers the state $|\cW
\rangle$ that is part of $Gr_1$. 

The flow over the Grassmannian, generated by $\Gamma_+$, obtains a
geometric interpretation as a linear flow over the Jacobian of
$\Sigma$, and therefore relates such geometric solutions to the
Hitchin integrable system in \Cref{sec:defquant}
\cite{hodge-mulase}.       

The tau-function associated to a fermionic state $|\cW \rangle$,
that is found as a Krichever solution, equals the determinant of the
$\delbar$-operator on $\Sigma$, and thus computes the partition function of
free fermions on $\Sigma$. This is exactly the contribution to the
free energy that we expect from the I-brane set-up when $\la \to 0$,
so that commutativity is restored.  

For all spectral curves in $\C^2$ or $\C^* \times \C$ that we consider later in this thesis, the
line bundle $\cL$ is almost trivial, so that 
the choice of trivialization of $\cL$ at $z_{\infty}$ is the 
only non-trivial piece of data. Picking such a frame is equivalent to
choosing a flat connection on $\Sigma - z_{\infty} $. This
corresponds to an element of $\Gamma_+$. However, we noticed in the
previous subsection that this acts trivially on the fermionic
state. So the Krichever map just yields the Dirac vacuum.


\subsubsection{Rank 1 quantum state}

To find a non-trivial fermionic state $|\cW \rangle$, as an expansion in $\la$,
we start with a $\cD_{\la}$-module. Let us first explain the rank 1
case, with a $\cD_{\la}$-module on $\C$ specified by the (meromorphic)
connection 
\begin{align*}
\nabla_{A} = \partial_z - \frac{1}{\la} A(z)
\end{align*}
that may be trivialized as 
\begin{align*}
\nabla_A =  \partial_z - g_{\la}(z)^{-1}
(\partial_z g_{\la}(z)). 
\end{align*}
When $g_{\la}(z) \in \Gamma_+$ this represents a pure gauge
transformation on the disk (so that $\nabla_A$ corresponds to a
regular flat
connection on $\C$). 

For any $g_{\la}(z)$ a fermionic section 
$\psi(z)$ of $\cL \otimes K^{1/2}$ may be written as  
\begin{align*}
\psi(z) = g_{\la}(z) \xi(z),
\end{align*}
where $\xi(z)$ is a section of $\cL \otimes K^{1/2}$ with trivial
connection $\partial_z$. Flat sections $\Psi(z)$ are defined by the
differential equation  
\begin{align*}
\left( \partial_z - \frac{1}{\la} A(z) \right) \Psi(z) = 0.
\end{align*}
They define a local trivialization of the bundle $\cL$ with
connection $\nabla_A$, and we will use them to translate the geometric
configuration into a quantum state. 

A flat section for the
trivial connection $\partial_z$ is given by $\Xi(z) = 1$. We 
associate the tuple $(\C,\partial_z)$ to the ground state 
\begin{align*}
|0 \rangle = z^0 \wedge z^1 \wedge z^2 \wedge \ldots.
\end{align*}
The gauge transformation $g_{\la}(z)$ maps the trivial solution 
$\Xi(z) = 1$ to $\Psi(z) = g_{\la}(z)$, which transforms
the vacuum into the fermionic state  
\begin{align*}
|\cW \rangle = g_{\la} |0 \rangle,
\end{align*}
where $g_{\la}$ acts as a $Gl_{\infty}$ transformation on the fermionic
modes
\begin{align*}
 |\cW \rangle &= g_{\la}(z) z^0 \wedge g_{\la}(z) z^1 \wedge
g_{\la}(z) z^2 \wedge \ldots.  
\end{align*}
as we explained in the last subsection. 

In other words, we build the quantum state by acting with the
$\cD$-module generator $\Psi(z) = g_{\la}(z)$ on the vacuum
\begin{align*}
\cW = \cD_{\la} \cdot \Psi(z).
\end{align*}
The state $|\cW \rangle$ is just second quantization of the
the $\cD_{\la}$-module $\cW$. Notice that 
this state is non-trivial only when $g_{\la}(z)$ is not a pure 
gauge transformation (which would correspond to a Krichever solution). This
implies that the flat sections will diverge near $z=0$, corresponding
to a distorted geometry in this region.

\subsubsection{Rank $n$ quantum state}

A degree $n$ spectral curve $\Sigma$ is quantized as a
$\la$-connection of rank
$n$. This is equivalent to a $\cD_{\la}$-module $\cM$
that is generated by a single degree $n$ differential operator
$P$.   
%
%
As an $\cO_{\C}$-module, though, $\cM$ is generated by an $n$-tuple
\begin{align*}
(\Psi(z), \del_z \Psi(z), \ldots, \del_z^{n-1} \Psi(z) ),  
\end{align*}
where $\Psi(z)$ is a solution of the differential equation $P \Psi =
0$. In other words, this blends an $n$-vector of solutions
to the linear differential system that the $\la$-connection defines. We
will name this $\cO_{\C}$-module 
\begin{align*}
\cW = \cO_{\C} \cdot (\Psi(z), \del_z \Psi(z), \ldots, \del_z^{n-1}
\Psi(z) ) ~\subset~ \C((z^{-1})).  
\end{align*}
(of course it contains
the same elements as $\cM$)
This is the subspace we want to second quantize into a fermionic state $|\cW \rangle$. 

Now $P$ has $n$ independent solutions $\Psi_i$, that differ in their behaviour
at infinity. These solutions have an asymptotic expansion around $z =
\infty$ that contains a WKB-piece plus an asymptotic
expansion in $\la$, and should thus be interpreted as
perturbative solutions that live on the spectral cover. We suggest
that the asymptotic expansion of any solution can be turned into a
fermionic state that captures the all-genus I-brane partition
function. This partition function thus depends on the choice of
boundary conditions near $z_{\infty}$. 

Some of the WKB-factors will be exponentially suppressed near
$z_{\infty}$, while
others exponentially grow. This depends on the  
specific region in this neighbourhood. The lines that characterize the
changing behaviour of the solutions $\Psi_i$ are called Stokes and
anti-Stokes rays. 
Boundary conditions at infinity specify the solution up to a Stokes
matrix: a solution that decays in that region can be added at no cost. 

This implies that the perturbative fermionic state we assign to a $\cD$-module
depends on the choice of boundary conditions. On the other hand, the
$\cD$-module itself is independent of 
any of these choices and thus in some sense contains non-perturbative
information and goes beyond the all-genus I-brane partition
function. This agrees with the discussion in
\cite{Maldacena:2004sn}. Nonentheless, the focus in this paper is on the
perturbative information a $\cD$-module provides.

\subsection{Quantum invariants and $\la$-deformed CFT's}\label{ssec:quantuminvariant}

Up to here we have explained how a spectral curve may be quantized
into a $\cD_{\la}$-module, and how this translates into a fermionic
state $|\cW \rangle$. 
Most importantly, this allows us to associate a quantum invariant to a
spectral curve. 

Spectral curves embedded in $\C^2$ are
characterized by a single region near infinity. The all-genus I-brane partition function 
is (like in the semiclassical Krichever setting) given by the
determinant of $|\cW \rangle$. When $\Sigma$ is part of $\C^* \times
\C$, it contains two regions at infinity. In this situation the
I-brane partition function is computed as a correlation 
function, by contracting two fermionic states.    

Likewise, we expect that it is important for any I-brane curve
to consider all regions where the curve approaches infinity. 
In such a patch the $B$-field becomes singular so that the
non-commutativity parameter $\la$ tends to 
zero. Hence we can associate a perturbative fermionic
state $| \cW \rangle$ to this neighbourhood, once a
choice of local 
symplectic coordinates $(z,w)$ is made. The result is a quantized module that
is invariant under the action of the Weyl algebra $\cD_{\la}$. 
Any other choice of local coordinates amounts to a combination of the
following two transformations
\begin{align}\label{eq:cxsympltrans}
z \mapsto z + f(w) \quad &\mbox{and} \quad w \mapsto w,\\
z \mapsto w \quad &\mbox{and} \quad w \mapsto -z. \notag
\end{align}
These generate automorphisms of the Weyl algebra, but they
change the $\cD_{\la}$-module associated to the asymptotic neighbourhood.  

Moving from one patch at infinity to
another changes the canonical coordinates by such a complex symplectic
transformation as well. Correspondingly the $\cD_{\la}$-module, and
thus the quantum fermion field, transforms in the
metaplectic representation \cite{Aganagic:2003qj}. Indeed, in the $\la
\to 0$ limit one recovers a semi-classical fermionic section of
$K^{1/2}$. 

This suggests that the 
complete I-brane partition function may be found by writing down the
correct fermionic state for each 
point at infinity and then glueing them by using a symplectic
identification of coordinates. Although the $\cD_{\la}$-modules associated
to the regions near infinity depend on the choice of local
coordinates, this quantum invariant should be independent of the
chosen parametrizations. This is a claim that we cannot yet justify,
except that from the philosophy of string theory there should be a
unique such quantum invariant. A simple supporting example is found
in the Chapter~\ref{chapter6}.   

One of the motivations of Chapter~\ref{chapter6} is to see in practise
how quantum
curves lead to I-brane partition functions. We study several
well-known examples of spectral curves in string theory, and determine
the $\cD_{\la}$-module that underlies their partition function. 
The first set of examples treats matrix model spectral curves embedded in $\C^2$
with just one region at infinity. In the second set we study
Seiberg-Witten geometries embedded in $\C^* \times \C$.


\chapter{Quantum Curves in Matrix Models and Gauge
  Theory}\label{chapter6} 

This chapter illustrates the $\cD$-module
formalism of Chapter~\ref{chapter5} from a string
theory perspective, with examples from the theory of random matrices,
minimal (non-critical) string theory, supersymmetric gauge theory and
topological strings. As a result we connect these familiar 
ingredients in a common framework centered around $\cD$-modules.    
String theory provides solutions to integrable hierarchies of the KP type. This
was first noted in the context of non-critical ($c \neq 26$) bosonic
string theory, which has a dual formulation in terms of Hermitean random
matrices. The matrix model partition function \index{matrix model}
\begin{align}\label{eqn:mmpartfunc}
 Z_{\textrm{mm}}(\la) =  \frac{1}{\mbox{vol}(U(N))} \int D M~ e^{ -
   \frac{1}{\la} \Tr \, W(M) }  
\end{align}
is known  
to be a tau-function of the KP integrable system. Although an
algebraic curve $\Sigma$ emerges in the limit that the size $N$ of the
square 
matrix $M$ tends to infinity, these
matrix model solutions do not correspond to geometric Krichever
solutions. In particular, the relevant Fock space state 
$|\cW\rangle$ does not have a purely geometric interpretation as
being swept out by regular free fermions living on the matrix model
spectral curve $\Sigma$. 
   
The matrix model
partition function admits a formal expansion  $$Z_{\textrm{mm}}(\la) = \exp \sum_g
\la^{2g-2} \cF_g 
$$ in the string coupling constant $\la$. In the  
't Hooft limit $N$ is sent to infinity while the product of $N$ with $\la$ is held
fixed, so that the geometric curve $\Sigma$
equivalently emerges in 
the classical limit $\la 
\to 0$. This suggests that $\lambda$ should be interpreted as some form of
non-commutative deformation of the underlying algebraic curve. In
fact, there have been many indications that this is the right point of
view.

In the simplest matrix models $\Sigma$ appears as an affine
rational curve given by a relation of the form
$$
H(x,y)=0
$$
in the complex two-plane $\C^2$, with a (local) parametrization
$
x=p(z)$ and $ y= q(z),
$
with $p$, $q$ polynomials. Of course, $p$ and $q$ commute: $[p,q]=0$.
However, the string-type solutions with $\lambda \neq 0$ are
characterized by quantities $P$ and $Q$ that no longer commute but instead
satisfy the canonical commutation relation
$$
[P,Q]=\lambda.
$$
In this case clearly $P$, $Q$ cannot be polynomials, but are
represented as differential operators, {\emph i.e.} polynomials in $z$
and $\partial_z$.

As we will point out in \Cref{sec:matrixmodels} these
solutions are naturally captured by a $\cD$-module. Instead of
classical curve in 
the $(x,y)$-plane, 
we should think of a  quantum curve as its analogue in the
non-commutative plane $[x,y]=\lambda$. If we interpret
$$
y = - \lambda {\partial \over \partial x},
$$
one can identify such a quantum curve as a holonomic $\cD$-module
$\cW$ for the algebra $\cD$ of differential operators in $x$.
Roughly speaking,
$\cW$ can be considered as the space of local sections that can be
continued as sections of a (non-commutative) $\cD$-module, instead
of sections of a line bundle over a curve.

One other important instance of integrable hierarchies
in string theory 
is in four-dimensional $\cN=2$ supersymmetric gauge theories. In
Chapter~\ref{chapter3+4} 
we have seen that the low energy effective description of Seiberg-Witten
theories is determined by a twice-punctured algebraic curve
$\Sigma_{SW}$, defined by an equation of the form
\begin{align*}
 H(t,v)=0
\end{align*}
with $t \in \C^*$ and $v \in \C$, that appears as a spectral curve of a Hitchin integrable system. 
In Chapter~\ref{chapter3+4} we geometrically
engineered Seiberg-Witten theory as a Calabi-Yau compactification and
encountered additional gravitational corrections $\cF_g$ to the effective
action.

Like in the matrix model setting these $\cF_g$-terms are
multiplied by some power of the string coupling constant $\la$,
suggesting that the full genus free energy $\cF = \sum_g \la^{2g-2} \cF_g
$ has an interpretation 
in terms of a quantum Seiberg-Witten curve.  
This motivates us to quantize the Seiberg-Witten
surface $\Sigma_{SW}$ in \Cref{sec:SW}. Again, we see that a
 $\cD$-module underlies the structure of the
total partition function.

\section{Matrix model geometries}\label{sec:matrixmodels}

Hermitian one-matrix models with an algebraic potential $$W(M)=
\sum_{j=0}^{d+1} u_j M^j$$  are defined through the matrix integral
(\ref{eqn:mmpartfunc}).  
In the large $N$ limit the distribution of the
eigenvalues $\la_i$ of $M$ on the real axis becomes continuous and
defines a hyperelliptic curve \index{spectral curve}
\begin{align}\label{eq:DVgeometry}
\Sigma_{\textrm{mm}}: \quad y^2 - W'(x)^2 + f(x) =0,
\end{align}
called the (matrix model) spectral curve. The polynomial $f(x) = 4 \mu
\sum_{j=0}^{d-1} b_j x^j$ is determined as
\begin{align*}
f(x) = \frac{4 \mu}{N} \sum_{i=1}^N
\frac{W'(x)-W'(\lambda_i)}{x-\lambda_i},
\end{align*}
with $\mu = N \la$. 
The potential $W(x)$ determines the positions of the cuts of the
hyperelliptic curve, and contains
the non-normalizable moduli. On the other hand, the size of the cuts
is determined by the polynomial $f(x)$, that comprises the
normalizable moduli $b_0, \ldots, b_{d-2}$ and the log-normalizable
modulus $b_{d-1}$. 

R.~Dijkgraaf and C.~Vafa discovered that this matrix model has a dual
description in string theory. In the 't Hooft
limit $N \to \infty$ (with $\mu$ fixed) it is equivalent to 
the topological B-model on a Calabi-Yau geometry $X_{\Sigma}$ modeled on the
matrix model spectral curve $\Sigma_{\textrm{mm}}$
\cite{Dijkgraaf:2002fc,Dijkgraaf:2002vw,Dijkgraaf:2002dh}. 
A good review is \cite{Marino:2004eq}.  
This duality may be generalized by starting with multi-matrix models,
whose spectral curve is a generic (in contrast to hyperelliptic)
algebraic curve in the variables $x$ and $y$.   
\index{Dijkgraaf-Vafa correspondence}

The I-brane picture suggests that the full B-model partition function on these Calabi-Yau geometries can be understood in terms of $\cD$-modules. Even better, we will find that finite $N$ matrix models are determined by an underlying $\cD$-module structure. 

In the past, as well as recently, Hermitean matrix models have been studied in great detail in many contexts.
Already in \cite{Moore:1990mg,Moore:1990cn} an attempt has been made
to understand the string equation $[P,Q] = \la$ in terms of a quantum curve in
terms of the expansion in the parameter $\la$. In G.~Moore's approach
this surface seemed to emerge from an interpretation of the string
equation as isomonodromy equations.  

More recently, quantum curves have appeared in the description of
branes in a dual string model.  
In topological string theory as well as non-critical string theory a
dominant role is played by holomorphic branes: either topological
B-branes \cite{Aganagic:2003qj} or FZZT branes
\cite{Seiberg:2003nm,Kazakov:2004du,Kutasov:2004fg,Maldacena:2004sn}. Their
moduli space equals the spectral curve, whereas the branes themselves may be
interpreted as fermions on the quantized spectral curve.  
As was reviewed in Chapter~\ref{chapter3+4}, in these 
 theories  it is possible to compute correlation functions using a
$W_{1+\infty}$-algebra  that
implements complex symplectomorphisms of the complex
plane~$\mathcal{B}$ -- as in (\ref{eq:cxsympltrans}) -- in quantum
theory as Ward identities \cite{Aganagic:2003qj,Fukuma:1990yk,Fukuma:2006qq}.   

These advances strongly hint at a fundamental appearance of
$\cD$-modules in the theory of matrix models. 
Indeed, this section unifies recent developments in matrix models in
the framework of Chapter~\ref{chapter5}. Firstly, after a
self-contained introduction in double scaled models we uncover the
$\cD$-module underlying the double-scaled $(p,1)$-models. In the second part of this
section we shift our focus to general Hermitian multi-matrix models,
and unravel their $\cD$-module structure.

\subsection{Double scaled matrix models and the KdV hierarchy}

\index{double scaling limit}
Our first goal is to find the $\cD$-modules that explain the quantum
structure of double scaled Hermitean matrix models. This double
scaling limit is a large $N$ limit in which one also fine-tunes the
parameters to find the right critical behaviour of the multi-matrix model
potential. Geometrically the double scaling limit zooms in on some
branch points  
of the spectral curve that move close together.
Spectral curves of double scaled matrix models are therefore of genus
zero and parametrized as 
\begin{align*}
\Sigma_{p,q}: \quad y^p + x^{q} + \ldots =0.
\end{align*} 
The one-matrix model only generates hyperelliptic spectral curves,
whereas the two-matrix model includes all possible combinations of $p$
and $q$.  
 These double scaled multi-matrix models are known
to describe non-critical ($c < 1$) bosonic string theory based on the
$(p,q)$ minimal model coupled to two-dimensional gravity
\cite{Douglas:1989dd,Douglas:1989ve,Gross:1989vs,Brezin:1990rb,Dijkgraaf:1990rs,
 Daul:1993bg} (reviewed extensively in \emph{e.g.}
\cite{Ginsparg:1993is,DiFrancesco:1993nw}).  This field is called
minimal string theory. \index{minimal/non-critical string theory} 


Zooming in on a single branch point yields the geometry
\begin{align*}
\Sigma_{p,1}: \quad y^p = x, 
\end{align*}
corresponding to the $(p,1)$ topological minimal model. This model is strictly 
not a well-defined conformal field theory, but does make sense as
2d topological field theory. For $p=2$ it is known as topological
gravity \cite{Witten:1989ig,Witten:1990hr,Kontsevich:1992ti,Dijkgraaf:1991qh}. 
%
%
%
%

\index{KdV hierarchy}\index{Lax operator}\index{tau function}
All $(p,q)$ minimal models turn out to be governed by two
differential operators 
\begin{align*} 
P &= (\la \partial_x)^p + u_{p-2}(x) (\la \partial_x)^{p-2} + \ldots +
u_0(x), \\
Q &= (\la \partial_x)^q + v_{q-2}(x) (\la \partial_x)^{q-2} + \ldots +
v_0(x),
\end{align*}
of degree $p$ and $q$ respectively, which obey the string (or Douglas)
equation 
\begin{align*}
[P,Q]= \la.
\end{align*}
%
%
%
%
$P$ and $Q$ depend on an infinite set of times
$t= (t_1, t_2, t_3, \ldots )$, which are closed string
couplings in minimal string theory, and evolve in these times as  
\begin{align*}
\la \frac{\partial}{\partial t_j } P &= [(P^{j/p})_+,P],\\
\la \frac{\partial}{\partial t_j } Q &= [(P^{j/p})_+,Q],
\end{align*} 
The fractional powers of
$P$ define a basis of commuting Hamiltonians.
This integrable system defines the $p$-th KdV hierarchy and the
above evolution equations are the KdV flows.

The differential operator $Q$ is completely determined as a function of
fractional powers of the Lax operator $P$ and the times $t$
\begin{align*}
Q = - \sum_{\substack{ j \ge 1 \\  j \neq 0 \mod p }}
 \left( 1+ \frac{j}{p} \right) t_{j+p} P^{j/p}_+, 
\end{align*}
This implies that when we turn off all the KdV times except for
$t_1=x$ and fix $t_{p+1}$ to be constant we find $Q =
\la \partial_x$. This defines the $(p,1)$-models  
\begin{align} \label{eq:(p,1)-model}
P = (\la \partial_x)^p - x, \quad Q = \la \partial_x.
\end{align}
One can reach any other $(p,q)$ model by flowing in the
times $t$. 

The partition function of the $p$-th KdV hierarchy is a
tau-function as in equation~(\ref{eqn:tauferm}). The associated
subspace $\cW \in Gr$ may be found by studying the eigenfunctions
$\psi(t,z)$ of the Lax operator $P$
\begin{align*}
P \psi(t,z) = z^p \psi(t,z).
\end{align*}
The Baker function $\psi_{\la}(t,z)$ represents the fermionic field that
sweeps out the subspace $\cW$ in the times $t$. 

If we restrict to the $(p,1)$-models the Baker function $\psi(x,z)$
can be expanded in a Taylor series
\begin{align*}
\psi(x,z) = \sum_{k=0}^{\infty} v_k(z) \frac{x^k}{k!}.
\end{align*}
Since $\psi(x,z)$ is an element of $\cW$ for all times, this defines a
basis $\{ v_k(z)\}_{k \ge 0}$ of the subspace $\cW$. 
In fact, it is not hard to see that the $(p,1)$ Baker function is
given by the generalized Airy function \index{Airy function}
\begin{align}\label{eqn:Airyfunction}
\psi(x,z) =  e^{\frac{p z^{p+1}}{(p+1) \la}} \sqrt{z^{p-1}} \int dw
~e^{\frac{(-1)^{1/p+1} (x+z^p)w}{\la^{p/p+1}} +
  \frac{w^{p+1}}{p+1}}, 
\end{align}
which is normalized such that its Taylor components $v_k(z)$ can
be expanded as 
\begin{align*}
v_k(z) = z^k (1 + \cO(\la/z^{p+1})). 
\end{align*}
The $(p,1)$ model thus determines the fermionic state
\begin{align}\label{eq:(p,1)state}
| \cW \rangle = v_0 \wedge v_1 \wedge v_2 \wedge \ldots,
\end{align}
where the $v_k(z)$ can be written explicitly in terms of Airy-like integrals
(see  \cite{Dijkgraaf:1991qh} for a nice review). The invariance under
\begin{align*}
z^p \cdot \cW \subset \cW 
\end{align*}
characterizes this state as coming from a $p$-th KdV hierarchy. In the other
direction, the state $|\cW \rangle$ determines the Baker function (and
thus the Lax operator) as
the one-point function
\begin{align*}
\psi(t,z) = \langle t | \psi(z) |\cW \rangle.
\end{align*}

In the dispersionless limit $\la \to 0$ the derivative
$\la \partial_x$ is replaced by a variable $d$, and the Dirac
commutators by Poisson brackets in $x$ and $d$. The leading order
contribution to the string equation is given by the Poisson bracket
\begin{align*}
\{ P_0 , Q_0 \} =1,
\end{align*}
where $P_0$ and $Q_0$ equal $P$ and $Q$ at $\la =0$. The solution to
this equation is 
\begin{align*}
P_0(d;t) &= x \\
Q_0(d;t) &= y(x;t) 
\end{align*}
and recovers the genus zero spectral curve $\Sigma_{p,q}$ of the double
scaled matrix model, parametrized by $d$. The KdV flows \emph{deform} this
surface such a way that its singularities are preserved. (See the
appendix of \cite{Maldacena:2004sn} for a detailed discussion.) 

Note that $\Sigma_{p,q}$ is not a spectral curve for the
Krichever map. The Krichever curve is instead found as the
space of simultaneous eigenvalues of the differential operators 
\begin{align*}
[P,Q]=0,
\end{align*}
that is \emph{preserved} by the KdV flow as a straight-line flow along its
Jacobian. In fact, there is no such Krichever spectral curve corresponding to
the doubled scaled matrix model solutions. 

Wrapping an I-brane around $\Sigma_{p,q}$
quantizes the semi-classical fermions on the spectral curve
$\Sigma_{p,q}$. 
The only point at infinity on $\Sigma_{p,q}$ is given by $x \to \infty$. 
The KdV tau-function should thus be the
fermionic determinant of the quantum state $| \cW \rangle$
that corresponds to this $\cD$-module. In the next subsection we write
down the 
$\cD$-module describing the $(p,1)$ model and show precisely how
this reproduces the tau-function using the formalism developed in
Chapter~\ref{chapter5}.

\subsection{$\cD$-module for topological
  gravity}\label{sec:minimalmodels}

\index{topological gravity}
We are ready to reconstruct the $\cD$-module that yields the
fermionic state $|\cW \rangle$ in equation~(\ref{eq:(p,1)state}).  
For simplicity we study the $(2,1)$-model, associated to an I-brane
wrapping the curve  
\begin{align*}
\Sigma_{(2,1)}: \qquad y^2 = x \qquad \mbox{with}~x,y \in \C.
\end{align*}
Notice that this is an $2:1$ cover over the $x$-plane. It contains
just one asymptotic region, where $x \to \infty$.
Fermions on this 
cover will therefore sweep out a subspace $\cW$ in the Hilbert space
\begin{align*}
\cW  ~\subset~ \cH(S^1) = \C((y^{-1})), 
\end{align*}
the space of
formal Laurent series in $y^{-1}$. The fermionic vacuum $|0 \rangle \subset \cH(S^1)$
corresponds to the subspace 
\begin{align*}
|0 \rangle =  y^{1/2} \wedge y^{3/2} \wedge y^{5/2} \wedge \ldots,
\end{align*}
which encodes the algebra of functions on the disk parametrized by $y$
and with boundary at $y= \infty$. Exponentials in $y^{-1}$
represent non-trivial behaviour near the origin and therefore act
non-trivially on the vacuum state. In contrast, exponentials in $y$ are 
holomorphic on the disk and thus act trivially on the vacuum. 

The $B$-field $B = \frac{1}{\la} dx \wedge dy$ quantizes the algebra of
functions on $\C^2$ into the differential algebra 
\begin{align*}
\cD_{\la} = \langle x, \la \partial_x \rangle.
\end{align*}
Furthermore, it introduces a holomorphic connection 1-form
$A  = \frac{1}{\la} y dx$
on $\Sigma_{(2,1)}$, which pushes forward to the rank two $\la$-connection 
\begin{align}\label{eq:(2,1)connection}
\nabla_A = \la \partial_x-  \left( \begin{array}{cc} 0
    & 1 \\ x & 0  \end{array} \right) 
\end{align}
on the base $\C$, parametrized by $x$. We claim that the
corresponding $\cD_{\la}$-module $\cM$, generated by  
%
 %
\begin{align*}
P= (\la \partial_x)^2 -x, 
\end{align*}
describes the $(2,1)$ model. Let us
verify this. 

Trivializing the $\la$-connection $\nabla_A$ in (\ref{eq:(2,1)connection}) implies
finding a rank two matrix $g(x)$ such that
\begin{align*}
\nabla_A =  \la \partial_x - g'(x) \circ g^{-1}(x).
\end{align*}
The columns of $g$ define a basis of solutions $\Psi(x)$ to the
differential equation $\nabla_A \Psi(x) = 0.$
They are meromorphic flat sections for
$\nabla_A$ that determine a trivialization of the bundle near $x =
\infty$. As the connection $\nabla_A$ is pushed forward from the
cover, $\Psi(x)$ is of the form
\begin{align*}
\Psi(x) = \left( \begin{array}{c} \psi(x) \\ 
    \psi'(x) \end{array} \right). 
\end{align*}

Independent solutions have different asymptotics in the semi-classical
regime where $x \to \infty$. In the $(2,1)$-model the two independent solutions $\psi_{\pm}(x)$ solve the differential equation 
\begin{align*}
P \psi_{\pm}(x) = ((\la \partial_x)^2 - x) \psi_{\pm}(x) =0.
\end{align*}
Hence these are the functions $\psi_+(x) =$ Ai$(x)$ and $\psi_-(x) =$
Bi$(x)$, that correspond semi-classically to the two saddles 
\begin{align*}
w_{\pm} = \pm \sqrt{x} / \la^{1/3}
\end{align*}
of the Airy integral  
\begin{align*}
\psi(x) = \frac{1}{2 \pi i} \int dw~e^{-\frac{xw}{\la^{2/3}} +
  \frac{w^3}{3}}.
\end{align*}

The $\cD$-module $\cM$  can be quantized
into a fermionic state for any choice of boundary
conditions. Depending on this choice we find an  $\cO(x)$-module $\cW_{\pm}$
spanned by linear combinations of $\psi_{\pm}(x)$ and of $\psi_{\pm}'(x)$.  
The fermionic state is generated by asymptotic
expansions in the parameter $\la$ of these elements.

The saddle-point approximation around the saddle $w_{\pm} = \pm
\sqrt{x}/\la^{1/3}$ yields
\begin{align*}
\hspace*{1.3cm} \psi_{\pm}(x) & \sim   y^{-1/2}~ e^{\mp \frac{2  
    y^{3}}{3 \la}} \left( 1 + \sum_{n \ge 1} c_n \la^n (\pm y)^{-3n}
\right) \notag \\
&\sim   y^{-1/2}~ e^{\mp \frac{2 
    y^{3}}{3 \la}}~ v_0(\pm y).
\end{align*} 
To see the last step just recall the definition of 
$v_0(z)$ as being equal to the Baker function $\psi(x,z)$ evaluated at
$x=0$.\footnote{Remark that $x$ and $z^2$ appear equivalently in
$\psi(x,z)$ in equation (\ref{eqn:Airyfunction}), while $\psi(x)$ and
$\psi(x,z)$ only differ in the 
normalization term in $z$.}
A similar expansion can be made for $ \psi'(x)$ with the result 
\begin{align*}
\hspace*{-1.3cm} \psi_{\pm}'(x) &\sim y^{1/2}~
e^{\mp \frac{ 2  y^{3}}{3 \la}}~ v_1(\pm y).
\end{align*} 

Note that both expansions in $\lambda$ are functions in
the coordinate $y$ on the cover. They contain a classical term (the
exponential in $1/\la$), a 1-loop piece and a quantum expansion in $\la
y^{-3}$. When we restrict to the saddle $w = \sqrt{x}/\la^{1/3}$,
these series blend the into the fermionic state
\begin{align*}
| \cW_+ \rangle = \psi_+(y) \wedge \psi_+'(y) \wedge y^2 \psi_+(y)
  \wedge y^2 \psi_+'(y) \wedge \ldots. 
\end{align*} 
Does this agree with the well-known result~(\ref{eq:(p,1)state})?  

First of all, notice that the basis vectors $x^k \psi(x)$ and $x^k
\psi'(x)$, with $k>0$, contain in their expansions the function
$v_k(y)$ plus a sum 
of lower order terms in $v_l(y)$ (with $l <k$). The wedge
product obviously eliminates all these lower order terms. Secondly, the
extra factor $y^{-1/2}$ factors just reminds us that we have written 
down a fermionic state. 

Furthermore, 
the WKB exponentials are exponentials in $y$ and thus elements of
$\Gamma_+$, whereas the expansions $v_k(y)/y^k$ are part of $\Gamma_-$. 
Up to normal ordening ambiguities this shows that the WKB part gives a
trivial contribution. In fact, the tau-function even 
cancels these ambiguities. 

This shows that 
\begin{align*}
| \cW_+ \rangle = v_0(y) \wedge v_1(y) \wedge v_2(y) \wedge \ldots, 
\end{align*} 
which is indeed the same as in equation (\ref{eq:(p,1)state}), when we change
variables from $z$ to $y$ in that equation. Of course, this doesn't change the
tau-function. 

\index{$\cD$-module}\index{quantum curve}
So our conclusion is that the $\cD$-module underlying
topological gravity is the canonical $\cD$-module 
\begin{align*}
\cM = \frac{\cD_{\la}}{\cD_{\la}((\la \partial_x)^2 -x)}. 
\end{align*}
This $\cD$-module gives the definition of the quantum curve
corresponding to the $(2,1)$ model and defines its quantum partition
function in an expansion around $\la$. Exactly the same reasoning
holds for the $(p,1)$-model, where we find 
a canonical rank $p$ connection on the base.  It would be good to be
able to write down a $\cD$-module for general $(p,q)$-models as
well.

\subsection{$\cD$-module for Hermitean matrix models}

$\cD$-modules continue to play an important role in
any Hermitean matrix model. In this subsection we are guided by 
\cite{eynard-isomonodromic} and \cite{Bertola:2001hq,2-matrixdmodule}
of Bertola, Eynard and Harnad. 

We first summarize 
how the partition function for a 1-matrix model defines a tau-function for
the KP hierarchy. As we saw before, such a tau-function corresponds
to a fermionic state $|\cW \rangle$, whose basis elements we will write
down. Following \cite{eynard-isomonodromic} we discover a rank two
differential structure in this basis, whose determinant reduces to the
spectral curve in the semi-classical limit. This $\cD$-module
structure is somewhat more complicated then the $\cD$-module we just
found describing double scaled matrix models.

We continue with 2-matrix models, based on
\cite{2-matrixdmodule}. Instead of one differential equation, these
models determine a group of four differential equations,
that characterize the $\cD$-module in the local coordinates $z$ and
$w$ at infinity. The matrix model partition function may of course be
computed in either frame.

\subsubsection*{1-matrix model}

Let us start with the 1-matrix model partition function
\begin{align}
Z_N = \frac{1}{\mbox{vol}(U(N))} \int D M ~ e^{-\frac{1}{\la}
  \Tr~ W(M)}.   
\end{align}
By diagonalizing the matrix $M$ the matrix integral
may be reduced to an integral over the eigenvalues~$\la_i$  
\begin{align*}
Z_N = \frac{1}{N! } \int \prod_{i=1}^N \frac{d \la_i}{2 \pi} ~ \Delta(\la)^2 ~ e^{- \frac{1}{\la} \sum_i
  W(\la_i)},  
\end{align*}
with the Vandermonde determinant $\Delta(\la) = \prod_{i<j} (\la_i -
\la_j) = \det(\la^{j-1}_i)$. The method of orthogonal polynomials
solves this integral by introducing an infinite set of polynomials
$p_k(x)$, defined by the properties \index{orthogonal polynomials}
\begin{align*}
& \quad p_k(x) = x^k (1 + \cO(x^{-1})),\\
& \int dx ~ p_k(x)~ p_l(x)~ e^{-\frac{1}{\la} W(x)} = 2 \pi h_k \delta_{k,l}.
\end{align*} 
The normalization of their leading term determines the coefficients
$h_n \in \C$. Since the Vandermonde determinant $\Delta(x)$ is not
sensitive to exchanging its entries $x^{j-1}_i$ for $p_{j-1}(x_i)$,
substituting $\Delta(x) = \det(p_{j-1}(x_i))$ turns the partition
function into a product of coefficients
\begin{align*}
Z_N = \prod_{k=0}^{N-1} h_k.   
\end{align*}
With the help of orthogonal polynomials the large $N$ behaviour of
$Z_N$ may be studied, while keeping track of $1/N$ corrections.

The orthogonal polynomials are crucial since they
build up a basis for 
the fermionic KP state. \hyphenation{fer-mi-o-nic} In an appendix of
\cite{eynard-isomonodromic} 
it is shown that one should start at $t=0$  with a state $|\cW_{0}
\rangle$ generated by the polynomials $p_k(x)$ for $k \ge N$
\begin{align*}
|\cW_{0} \rangle = p_N(x) \wedge p_{N+1}(x) \wedge p_{N+2}(x) \wedge \ldots.
\end{align*}
Notice that the vector $p_{N}(x)$ thus corresponds to
the Fermi level and defines the Baker function in the double scaling
limit. Acting on them with the commuting flow generated by  
\begin{align*}
\Gamma_{+} = \left\{ g(t) = e^{\sum_{n \ge 1} \frac{1}{n} t_n x^n} \right\}   
\end{align*}
defines a state  $|\cW_{t} \rangle = | g(t) \cW_{0} \rangle$ at time
$t$, which allows to compute a tau-function at time $t$. If the
coefficients $u_j$ in the potential $W(x)$ are taken to be $u_j =
u_j^{(0)} + t_j$, this $\tau$-function equals the ratio of the matrix
model partition function $Z_N$ at time $t$ divided by that at
$t=0$.

Multiplying the orthogonal polynomials by $\exp(-\frac{1}{2
  \la}W(x))$ doesn't change the fermionic state $\cW = \cW_0$ in a
relevant way, since this
factor is an 
element of $\Gamma_{+}$. To find the right $\cD$-module structure, it is
necessary to proceed with the quasi-polynomials 
\begin{align*}
\psi_k(x) = \frac{1}{\sqrt{h_k}} p_k e^{- \frac{1}{2\la} W(x)}, 
\end{align*}
which form an orthonormal basis with respect to the bilinear form
\begin{align}\label{eq:mmbilinearform}
(\psi_k, \psi_l) = \int dx~\psi_k \psi_l .
\end{align}
 
It is possible to express both multiplication by $x$ and
differentiation with respect to $x$ in terms of the basis of
$\psi_m$'s. The Weyl algebra $\langle x, \la \partial_x \rangle$ acts on
these (quasi)-polynomials by two matrices $Q$ and $P$
\begin{align*}
x \psi_k(x) &= \sum_{l=0}^{\infty} Q_{kl} \psi_l \\
\la \partial_x \psi_k(x) &= \sum_{l=0}^{\infty} P_{kl} \psi_l(x),
 \end{align*}
and the space of quasi-polynomials $\psi_k$ is thus a
$\cD_{\la}$-module.

Notice that we anticipate that the $\cD$-module possesses a rank two structure, since we started with a flat connection $A
=\frac{1}{\la} ydx$ on 
an I-brane wrapped on a hyperelliptic curve. 

Now, the matrices $Q$
and $P$ only contain non-zero entries in a finite band around the
diagonal. The action
of $\partial_x$ on the semi-infinite set of $\psi_k(x)$'s can
therefore indeed be summarized in a rank two  
differential system (\cite{eynard-isomonodromic} and references
therein)   
\begin{align}\label{eq:MMDmodule}
\la \partial_x \left[ \begin{array}{c} \psi_{N}(x) \\
    \psi_{N-1}(x) \end{array} \right] = A_N(x)
\left[ \begin{array}{c} \psi_{N}(x) \\ \psi_{N-1}(x) \end{array}
\right], 
\end{align}
where $A_N(x)$ is a rather complicated $2 \times
2$-matrix involving 
the derivative $W'$ of the potential and the infinite matrix $Q$:
\begin{align*}
A_N(x) =  \frac{1}{2} W'(x) \left[ \begin{array}{cc} -1 & 0 \\ 0 & 1 \end{array} \right] &
+ \gamma_N  
\left[ \begin{array}{ll} -\widetilde{W}'(Q,x)_{N,N-1} &
    \widetilde{W}'(Q,x)_{N,N}\\  -\widetilde{W}'(Q,x)_{N-1,N-1} &
    \widetilde{W}'(Q,x)_{N-1,N} \end{array} \right], 
\end{align*}
with \index{$\cD$-module} \index{quantum curve}
\begin{align*} 
\widetilde{W}'(Q,x) = \left( \frac{W'(Q)-W'(x)}{Q-x} \right)\quad
\mbox{and} \quad 
\gamma_N = \sqrt{\frac{h_N}{h_{N-1}}}.
\end{align*}
Equation (\ref{eq:MMDmodule}) is thus the rank two $\la$-connection
defining the $\cD_{\la}$-module structure on $\cW$ that we were
searching for!  As a check, the determinant of this connection reduces
to the spectral 
curve in the semiclassical, or dispersionless, limit
\cite{eynard-isomonodromic}:  
\begin{align*}
  \Sigma_N: \quad 0 &= \mbox{det} \left( y 1_{2 \times 2} - A_N(x)
  \right) \\& = y^2 - W'(x)^2 + 4 \la \sum_{j=0}^{N-1} \left( \frac{W'(Q) - W'(x)}{Q-x} \right)_{jj}
\end{align*}
(To make the coefficients in the above equation agree with
(\ref{eq:DVgeometry}), we rescaled $y \mapsto y/2$.)
In conclusion we found the $\cD$-module structure underlying Hermitean
1-matrix models. 

Remark that in the $N \to \infty$ limit we expect that the hyperelliptic
curve defining the B-model Calabi-Yau (\ref{eq:DVgeometry}) emerges
from $\Sigma_N$. Indeed, in the 't Hooft limit $Q$
corresponds classically to the
coordinate $x$ on the curve, whereas quantum-mechanically it is an
operator whose spectrum is described by the eigenvalues $\la_i$ of
the infinite matrix $M$. In the large $N$ limit we can therefore
replace the matrix $Q_{ij}$ in the definition for $\Sigma_N$ by
$\lambda_i \delta_{ij}$. 

We can rewrite the rank two connection for
the vector $(\psi_N, \psi_N')^t$ as 
\begin{align*}
\la \partial_x \left[ \begin{array}{c} \psi_{N}(x) \\
    \psi'_{N}(x) \end{array} \right] = \left[ \begin{array}{ll}
   0 & 
    1 \\  -\det(A_N(x)) + \la Y  &
    \la Z  \end{array} \right]
\left[ \begin{array}{c} \psi_{N}(x) \\ \psi'_{N}(x) \end{array}
\right],
\end{align*}
at least when tr$(A_N(x))=0$,
with $Y$ and $Z$ some derivatives of entries of $A_N(x)$.
This brings the $\la$-connection in the familiar form
of Chapter~\ref{chapter5}. In the next subsection we clarify the
differential structure in a simple example.
  
\subsubsection*{2-matrix model}

Let us first say a few words on the $\cD$-module structure underlying
multi-matrix models, which capture spectral curves of any degree in
$x$ and $y$ \cite{Bertola:2001hq, 2-matrixdmodule}. The partition function
for a two-matrix model, with two rank $N$ matrices $M_1$ and $M_2$, is 
\begin{align*}
Z_N =  \frac{1}{\mathrm{Vol}(U(N))^2} \int DM_1 DM_2 ~e^{- \frac{1}{\la}  \Tr \left( W_1(M_1) +
    W_2(M_2) - M_1 M_2 
  \right) }, 
\end{align*} 
where $W_1$ and $W_2$ are two potentials of degree $d_1+1$ and
$d_2+1$. Choosing $W_2$ to be
Gaussian reduces the 2-matrix model to a 1-matrix model. The 2-matrix
model is solved by introducing two sets of orthogonal polynomials
$\pi_k(x)$ 
and $\sigma_k(y)$. Again it is convenient to turn them into quasi-polynomials 
\begin{align*}
\psi_k(x) =  \pi_k(x) e^{- \frac{1}{\la} W_1(x)} , \quad
\phi_k(y) = \sigma_k(y) e^{- \frac{1}{\la} W_2(y)}.
\end{align*}
obeying the orthogonality relations 
\begin{align}\label{eq:2-matrixortho}
\int dx dy ~\psi_k(x) \phi_l(y) e^{\frac{xy}{\la}} = h_k \delta_{kl}. 
\end{align}

Multiplying with or taking a derivative with respect to either $x$ or
$y$ yields (just) two operators $Q$ and $P$ (and their transposes
because of (\ref{eq:2-matrixortho})), that form a representation of
string equation $[P,Q]=0$. Since $Q$ is only non-zero in a band around
the diagonal of size $d_2+1$ and $P$ of size $d_1+1$, the
quasi-polynomials may be folded into the vectors 
\begin{align*}
\vec{\psi} = [ \psi_{N}, \ldots, \psi_{N-d_2} ]^t, \quad
\vec{\phi} = [ \phi_{N}, \ldots, \phi_{N-d_1} ]^t.
\end{align*}
Any other quasi-polynomial can be expressed as a sum of entrees of
these vectors, with coefficients in the polynomials in $x$ and $y$. 
These vectors are called windows. The differential operators
$\la \partial_x$ 
and $\la \partial_y$ respect them, so that their action is
summarized in a rank $d_2+1$ resp. rank $d_1+1$ $\la$-connection
\begin{align}\label{eq:2matrixconnection}
\la \partial_x \vec{\psi}(x) = A_1(x) \vec{\psi}(x) , \quad
\la \partial_y \vec{\phi}(y) =  A_2(y) \vec{\phi}(x).
\end{align}
This we interpret as two representations of the $\cD_{\la}$-module
underlying 2-matrix 
models. Indeed, \cite{Bertola:2001hq} proves that the determinant of both
differential systems equals the same spectral curve $\Sigma$, in the
limit $\la \to 0$ when we 
replace $\la \partial_x \to y$ and  $\la \partial_y \to x$. The
defining equation of $\Sigma$ is of degree $d_1+1$ in $x$ and of
degree $d_2+1$ in 
$y$.  

In fact, it is useful to introduce two more semi-infinite sets of
quasi-polynomials $\underline{\psi}_k(y)$ and $\underline{\phi}_k(x)$,
as the Fourier transforms of $\psi_k(x)$ and $\phi_k(y)$
respectively. The action of the Weyl algebra on them may be encoded as
the transpose of the above linear systems. The full system can
therefore be summarized by (compare to (\ref{eq:c=1module}))
\begin{align*}
x\mbox{-axis}: \quad \{ \psi_k(x),~\underline{\phi}_k(x)\}, \quad
& \nabla_{\la} = \la \partial_x - A_1(x), \\
y\mbox{-axis}: \quad \{\phi_k(y),~\underline{\psi}_k(y)\}, \quad &
\nabla_{\la} = \la \partial_y - A_2(y). \notag
\end{align*}
Moreover, the matrix model partition function can be rewritten as a
fermionic correlator in either local coordinate
\begin{align*}
Z_N &\propto \frac{1}{N!} \int \prod_i d \la_i^{1} d \la_i^2~ \Delta(\la^1) \Delta(\la^2)~
e^{- \frac{1}{\la} \sum_i  W_1(\la^1_i) + W_2(\la^2_i) - \la^1_i
  \la^2_i} \\
&= \prod_{k=0}^{N-1} \langle \psi_k(x)  |  \underline{\phi}_k(x) \rangle 
=  \prod_{k=0}^{N-1} \langle \phi_k(y)  |  \underline{\psi}_k(y)
\rangle \notag 
\end{align*}
with respect to the bilinear form in (\ref{eq:mmbilinearform}).

Furthermore, Bertola, Eynard and Harnad study the dependence on the
parameters $u^{(1)}_j$ and $u^{(2)}_j$ appearing in the potentials
$W_1$ and $W_2$. Deformations in these parameters leave the
two sets of quasi-polynomials invariant as well. On
$\vec{\psi}$ and $\vec{\phi}$  they act as matrices $U^{(1)}_j$ and
$U^{(2)}_j$. This yields the 2-Toda system \index{Toda hierarchy}
\begin{align*}
\partial_{u^{(1)}_j} Q = - [Q,U^{(1)}_j] \qquad \partial_{u^{(1)}_j}
P = - [P,U^{(1)}_j] \\
\partial_{u^{(2)}_j} Q =  [Q,U^{(2)}_j] \qquad \partial_{u^{(2)}_j}
P = [P,U^{(1)}_j]. 
\end{align*}
In \cite{Bertola:2001hq} it is proved that the linear differential systems
(\ref{eq:2matrixconnection}) are compatible with these
deformations, so that the parameters $u^{(1)}_j$ and $u^{(2)}_j$
in fact generate isomonodromic deformations. This shows precisely how the
non-normalizable parameters in the potential respect the
central role of the $\cD_{\la}$-module (\ref{eq:2matrixconnection}) in the 
2-matrix model. \index{isomonodromy}

\subsection{Gaussian example}

\index{Gaussian matrix model}
Let us consider the Gaussian 1-matrix model with quadratic potential 
\begin{align}
 W(x) =  \frac{x^2}{2}, 
\end{align}
that is associated to the spectral curve
\begin{align}\label{eq:spectralcurvegaussian}
y^2 = x^2 - 4 \mu
\end{align}
in the large $N$ limit. In the Dijkgraaf-Vafa correspondence this
matrix model is thus dual to the topological 
B-model on the deformed conifold geometry (see Fig.~\ref{fig:defconifold}). 
 
The Hermite functions 
\begin{align*}
 \psi^{\la}_k(x) &=  \frac{1}{\sqrt{h_k}} e^{-\frac{x^2}{4 \la}}
 H^{\la}_k(x), \quad \mbox{with}\\ 
 H^{\la}_k(x)&=  \la^{k/2} H_k\left( \frac{x}{\sqrt{\la}} \right) =
 x^k \left( 1 + \cO\left( \frac{\sqrt{\la}}{x} \right) \right),
\end{align*}
form an orthogonal basis for this model. Their inner product is given
by
\begin{align*}
\int \frac{dx}{2 \pi} ~ \psi^{\la}_k(x)  \psi^{\la}_l(x) =
 \la^k k! \sqrt{\frac{\la}{2 \pi}} \delta_{kl} \quad \Rightarrow \quad
 h_k =    \la^k k! \sqrt{\frac{\la}{2 \pi}}.  
\end{align*}
The partition function of the Gaussian matrix model can be computed as
a product of the normalization constants $h_k$. Using the asymptotic
expansion of the Barnes function $G_2(z)$, that is defined by
$G_2(z+1) = \Gamma(z) G_2(z)$, the free energy can be expanded in powers 
of $\la$ 
\begin{align}\label{eqn:Gaussianpartfunc}
 \cF_N &=  \log \prod_{k=1}^{N-1} h_k = \log \left(  G_2(N+1)
   \frac{\la^{N^2/2}}{(2  \pi)^{N/2}} \right) \\
& = \frac{1}{2} \left( \frac{\mu}{\la} \right)^2 \left( \log  \mu - \frac{3}{2} \right) -
\frac{1}{12} \log \mu + \zeta'(-1) +
\sum_{g=2}^{\infty} 
\frac{B_{2g}}{2g(2g-2)} \left( \frac{\la}{\mu} \right)^{2g-2} \notag,
\end{align}
where $B_{2g}$ are the Bernoulli numbers and $\mu = N \la$.  

The derivatives of the Hermite functions are related as 
\begin{align*}
\la \frac{d}{dx} \left[ \begin{array}{cc} \psi^{\la}_{k}(x) \\
    \psi^{\la}_{k-1}(x) \end{array} \right] = 
\left[ \begin{array}{cc} -x/2 & \sqrt{k \la} \\ -\sqrt{k \la} &
    x/2\end{array} \right] \left[ \begin{array}{cc} \psi^{\la}_{k}(x) \\
    \psi^{\la}_{k-1}(x) \end{array} \right]. 
\end{align*}
So, according to the previous discussion, the $\cD_{\la}$-module
connection is given by \index{$\cD$-module}\index{quantum curve}
\begin{align}\label{eq:Dmodulegaussian}
\la \frac{d}{dx} - A_N(x) = \la \frac{d}{dx} +
\left[ \begin{array}{cc} x/2 & -\sqrt{N \la} \\ \sqrt{N \la} &
   - x/2 \end{array} \right]. 
\end{align}
Here we choose $\vec{\psi} = [ \psi_N, \psi_{N-1}]^t$ as window.
In the large $N$ limit the determinant of this rank two
differential system indeed yields the spectral curve
(\ref{eq:spectralcurvegaussian}) with $\mu = N \la$.  

Instead of using $\psi_{k}^{\la}$ and $\psi_{k-1}^{\la}$ as a basis,
we can also write down the differential system for $\psi_{k}^{\la}$
and its derivative ${\psi'}_{k}^{\la}(x)= \la \partial_x
\psi_{k}^{\la}(x)$. Since this 
derivative is a linear combination of $\psi_{k-1}^{\la}$ and $x
\psi^{\la}_{k}(x)$ (as we saw above), it is equivalent to use this basis to generate
the fermionic state $\cW $. We compute that 
\begin{align*}
\la \frac{d}{dx} \left[ \begin{array}{cc} \psi^{\la}_{N}(x) \\
    {\psi'}^{\la}_{N}(x) \end{array} \right] = 
\left[ \begin{array}{cc} 0 & 1 \\ x^2 - \la N - \la/2 &
    0 \end{array} \right] \left[ \begin{array}{cc} \psi^{\la}_{N}(x) \\
    {\psi'}^{\la}_{N}(x) \end{array} \right]. 
\end{align*}
The spectral curve in the large $N$ limit hasn't changed. Notice that
in this form it is clear that the rank 2 connection is the
push-forward of the connection $A = \frac{1}{\la} y dx$ on the
spectral curve $y^2 = x^2 - 4 \mu$ to the $\C$-plane, up to some $\la$-corrections.

In the double scaling limit the limits $N \to \infty$ and $\la \to 0$
are not independent as in the 't Hooft limit, but correlated, such
that the higher genus contributions to the partition function are
taken into account. In terms of the Gaussian spectral curve this limit implies
that one zooms in onto one of the endpoints of the cuts. The
orthogonal function $\psi^{\la}_N(x)$ turns into the Baker function
$\psi(x)$ of the double scaled state $\cW$.
 
In the Gaussian matrix model this is implemented by letting $x \to 
\sqrt{\mu} + \epsilon x$, where $\epsilon$ is a small parameter. So
the double scaled spectral curve reads  
\begin{align*}
y^2 = x,
\end{align*}
while the differential system reduces to 
\begin{align*}
\la \frac{d}{dx} \left[ \begin{array}{cc} \psi(x) \\
    \psi'(x) \end{array} \right] = 
\left[ \begin{array}{cc} 0 & 1 \\ x &
    0 \end{array} \right] \left[ \begin{array}{cc} \psi(x) \\
    \psi'(x) \end{array} \right].
\end{align*}
This is indeed the $\cD$-module corresponding to the $(2,1)$-model.

\section{Conifold and $c=1$ string}\label{sec:c=1}


\index{$c=1$ string theory}
The free energy~(\ref{eqn:Gaussianpartfunc}) of the Gaussian matrix
model pops up in the theory of bosonic $c=1$ strings. This $c=1$
string theory 
is formulated in terms of a single bosonic coordinate $X$, that is
compactified on a circle of radius $r$ in the Euclidean theory. A
critical bosonic string theory (with 
$c = 26$) is obtained by coupling the above CFT to a Liouville
field $\phi$. The Liouville field corresponds to the non-decoupled conformal mode of
the worldsheet metric. The local worldsheet action reads 
\begin{align*}
\frac{1}{4 \pi} \int d^2 \sigma \left(\frac{1}{2} (\lotjesd X)^2 +  (\lotjesd \phi)^2
   +  \mu e^{\sqrt{2} \phi} + \sqrt{2} \phi R  \right),
\end{align*}
where the coupling $\mu$ is seen as the worldsheet cosmological
constant. In the Euclidean model there are only two sets of operators, that
describe the winding and momenta modes of the field $X$. These vertex
and vortex operators can be added to the action as marginal
deformations with coefficients $t_n$ and $\tilde{t}_n$.  

Just like in $c<1$ minimal string theories (the
$(p,q)$-models of last section), the partition 
function of the $c=1$ string is first computed using a dual matrix
model description \cite{Gross:1990ub}.
At the self-dual radius $r=1$ it agrees with the Gaussian matrix
model partition function in equation~(\ref{eqn:Gaussianpartfunc}),
where $\la$ now plays the role of the $c=1$ string coupling constant.  

The matrix model dual to the $c=1$ string is called matrix quantum
mechanics. This duality is reviewed in much detail in \emph{e.g.}
\cite{Klebanov:1991qa,Polchinski:1994mb,Alexandrov:2003ut}. 
Matrix quantum mechanics is described by a gauge
field $A$ and a scalar 
field $M$ that are both $N \times N$ Hermitean matrices. The momentum
modes of the $c=1$ string correspond to excitations of $M$, whereas
the winding modes are excitations of $A$. If we focus on the momentum
modes, the (double scaled) matrix model is governed by the Hamiltonian
\begin{align*}
H = \frac{1}{2} \, \Tr \left( - \la^2 \frac{\lotjesd^2}{\lotjesd M^2} - M^2 \right).
\end{align*}

Let us focus on solutions that depend purely on the eigenvalues
$\la_i$ of $M$. The Hamiltonian may be rewritten in terms of the
eigenvalues as 
\begin{align*}
H = \frac{1}{2} \,  \Delta^{-1}(\la) \sum_i \left( - \la^2
    \frac{\lotjesd^2}{\lotjesd \la_i^2} - \la_i^2 \right) \Delta(\la),
\end{align*}
where $\Delta(\la)$ is Vandermonde determinant. It is convenient to
absorb the factor $\Delta$ in the wavefunction solutions, making them
anti-symmetric. Hence, the singlet sector of matrix quantum mechanics
describes a system of $N$ free fermions in an upside-down Gaussian
potential.   

To describe the partition function of the $c=1$ model it is convenient to move over to light-cone
coordinates $\la_{\pm} = \la \pm p$,
so that elementary excitations of the $c=1$ model are
represented as collective excitations of free fermions near the Fermi
level 
\begin{align}\label{eqn:fermilevel}
 \la_+ \la_- = \mu.
\end{align}
When we restrict to $\la_{\pm}>0$, scattering amplitudes can be
computed by preparing asymptotic free fermionic states $\langle \tilde{t} |$
and $| t \rangle$ at the regions where
one of $\la_{\pm}$ becomes very large.

In this picture the generating function of scattering amplitudes has a
particularly simple form. It can be formulated as a fermionic
correlator \cite{Dijkgraaf:1992hk}  
\begin{align}\label{c=1:toda}
Z =  \langle t |  S  | \tilde{t} \rangle, 
\end{align}
where the fermionic scattering matrix $S \in GL(\infty,\C)$ was first
computed in \cite{Moore:1991zv}.  Moreover, in
\cite{Alexandrov:2002fh} (see also Chapter V of 
\cite{Alexandrov:2003ut}) and later in \cite{Aganagic:2003qj} it is noticed that $S$ just equals the Fourier
transformation
\begin{align}
(S \psi) (\la_-) = \int d\la_+ \, e^{ \frac{1}{\la} \la_- \la_+} \psi(\la_+).\label{eqn:c=1fourier} 
\end{align}
In the next section we show that
this follows naturally from the perspective of $\cD$-modules. 

The result (\ref{c=1:toda}) shows that $c=1$ string theory is an integrable
system, just like the 
$(p,q)$-models in the last section. Since it depends on two
sets of times this integrable system is not a KP system. Instead, the
above expression defines a tau function of a 2-Toda
hierarchy. \index{Toda hierarchy}

%
%

Notice that the Fermi level (\ref{eqn:fermilevel}) is a real cycle on
the complex curve  
\begin{align}\label{conifold}
\Sigma: \quad  zw = \mu,
\end{align}
which is a different parametrization of the spectral curve $y^2 = x^2 -
\mu$ of the Gaussian 1-matrix model. 
In the revival of this subject a few years ago, a number of other
matrix model interpretations have been found. This includes a duality with the
Hermitean 2-matrix model, which makes the 2-Toda structure manifest
\cite{Bonora:1994ka}, a Kontsevich-type 
model \cite{Distler:1990mt,Imbimbo:1995yv} at the self-dual radius,
and a so-called normal matrix model 
\cite{Alexandrov:2003qk,Mukherjee:2005aq}, that parametrizes the dual
real cycle on the complex curve $\Sigma$. Let 
us also mention that the well-known duality of the $c=1$ string with
the topological B-model on the deformed conifold
\cite{Ghoshal:1995wm}, that follows, with a detour, from the more
general Dijkgraaf-Vafa correspondence.



\subsubsection{$\cD$-module description of the $c=1$ string}

This paragraph reproduces the $c=1$ partition
function (\ref{c=1:toda}) from a
$\cD$-module point of view. The discussion
continues the line of thought in Section 5.5 of \cite{Aganagic:2003qj}. 

As we have just seen, the $c=1$ string is geometrically characterized
by the presence of a holomorphic curve in $\C \times \C$ defined by 
\begin{align*}
\Sigma_{c=1}: \quad zw = \mu.
\end{align*}
Let us consider an I-brane wrapping the curve $\Sigma_{c=1}$. 
When we assume $z$ as local coordinate the curve quantizes into the
differential operator  
\begin{align}
P = - \la z\lotjesd_z - \mu.
\end{align}
It is amusing that the differential operator $P$ appears as a canonical
example in the theory of $\cD$-modules (see {\it e.g.}
\cite{kashiwara}) in the same way as the $c=1$ string is an elementary
example of a string theory. 

We recognize this example from Chapter~\ref{chapter5}, where a
$\cD$-module was associated to the differential operator $P$. However, now it is important not to
forget that there are \emph{two} asymptotic points $z_{\infty}$ and 
$w_{\infty}$. Let us call their local neighbourhoods $U_z$ and
$U_w$, as local coordinates are  $z$ and $w$ respectively.  
At both asymptotic points the I-brane fermions will sweep out an
asymptotic state. The quantum partition function should therefore be
constructed from two quantum states. 

Before constructing these states for general $\la$, let us 
first consider the semi-classical limit $\la \to 0$. In this limit the
I-brane degrees of freedom are just conventional chiral fermions
on $\Sigma_{c=1}$.
The genus 1 part $\cF_1$ of the free energy is obtained as the
partition function of these semi-classical fermions. It can be
computed by assigning the Dirac vacuum $$| 0 \rangle_z = z^{1/2} \wedge
z^{3/2} \wedge z^{5/2} \wedge \ldots$$ to $U_z$ and likewise the conjugate 
state $$ |0 \rangle_w = w^{1/2} \wedge w^{3/2} \wedge w^{5/2} \wedge
\ldots$$ to $U_w$. To compare these states, we need an operator $S$
that relates $z$ to $1/z$. The semi-classical partition can then be
computed as a fermionic correlator 
$
_w \hspace{-.1mm} \langle 0 |S|0 \rangle_z,
$ 
with the result that 
\begin{align}\label{eq:F1c=1}
e^{\cF_1} =\, _w \hspace{-.2mm} \langle 0 |S|0 \rangle_z =  
\prod_{k\ge 0} \mu^{k+1/2}. 
\end{align}
Using $\zeta$-function regularization we find that this expression
yields the familiar answer $\cF_1 = -{1\over 12} \log \mu$.

In order to go beyond 1-loop, we should think in terms of
$\cD$-modules. Let us for a moment not represent their elements in
terms of differential operators yet. In both asymptotic regions we then
find the $\cD$-modules 
\begin{align*}
U_z: \quad &\cM = \cD / \cD P, \quad \textrm{with} \quad  P= \hat{z}
\hat{w} - \mu, \\ 
U_w: \quad &\underline{\cM} = \cD / \cD \underline{P}, \quad
\textrm{with} \quad 
\underline{P} =  \hat{w} \hat{z} - \mu + \la.
\end{align*}

Notice that the Weyl algebra $\cD = \langle \hat{z}, \hat{w} \rangle$,
with the relation $[\hat{z}, \hat{w}] = \la$, acts 
on monomials $z^k$ and $w^k$ in the module $\cM$ as   
\begin{eqnarray*}
 \hat{z} (z^k) = z^{k+1} && \hat{z} (w^k) =  \left( \la \lotjesd_w + \frac{\mu -
    \la}{w} \right) w^{k} \\ 
 \hat{w} (z^k) =  \left(- \la \lotjesd_z + \frac{\mu}{z}
\right)  z^{k}
&& \hat{w} (w^k) = w^{k+1}.  
\end{eqnarray*}

Here, we just used the relation $\cD P \equiv 0$ and wrote the elements
in the basis $\{ z^k, w^k \, | \, k \in \Z \}$ of $\cM$. A basis of a
representation of $\cM$ on which $\hat{z}$ and 
$\hat{w}$ just act by multiplication by $z$ resp. differentiation
with respect to $z$ is given by
\begin{align*}
&v^z_{k} (z) = z^k \cdot z^{-\mu/\la}, \\
&v^w_{k}(z) =  \int dw~e^{-zw/\la} ~ w^{k-1} \cdot w^{\mu/\la}.  
\end{align*}
Indeed, differentiation with respect to $z$ clearly gives the same
result as applying $\hat{w}$. Moreover, multiplying $v^w_{k}$ by
$z$ gives
\begin{align*}
z \cdot v^w_{k} (z) &= \la \int dw~ e^{-zw/\la}
\frac{\partial}{\partial w} \left( w^{k-1+ \mu/\la} \right) = (\mu
+\la( k -1)) v^w_{k-1}.
\end{align*}

Similarly, in the module $\underline{M}$ one can verify that 
\begin{eqnarray*}
 \hat{w} (w^k) = w^{k+1} && \hat{w} (z^k) =  \left( -\la \lotjesd_w  +
  \frac{\mu}{w} \right) w^{k} \\ 
 \hat{z} (w^m) =  \left(\la \lotjesd_z +
  \frac{\mu-\la}{z} \right)  z^{k}
&& \hat{z} (z^k) = z^{k+1}.  
\end{eqnarray*}
Hence in the
representation of $\underline{\cM}$ defined by
\begin{align*}
&\underline{v}^w_{k} (w) = w^{k-1} \cdot w^{\mu/\la}, \\
&\underline{v}^z_{k}(w) =  \int dz~e^{zw/\la} ~ z^k \cdot z^{-\mu/\la},
\end{align*}
$w$ and $\partial_w$ act in the usual way. 

Since we moved
over to representations of the $\cD$-module where the differential
operator acts as we are used to, the $S$ transformation, that connects
the $U_z$ and the $U_w$ patch and thereby exchanges $\hat{z}$ and
$\hat{w}$, must be a Fourier transformation. This is clear from the
expressions for the basis elements $w$ and $\tilde{w}$: $S$
interchanges $v^z_{k}(z)$ with
$\underline{v}^z_{k}(w)$, and $v^w_{k}(z)$ with $\underline{v}^w_{k}(w)$.
In total we thus find the $\cD$-module elements
\begin{align}\label{eq:c=1module}
U_z: \quad &v_k^{z}, ~{v}_k^{w} \\
U_w: \quad &\underline{v}_k^{w},~ \underline{v}_k^{z} \notag
\end{align}

Representing the $\cD$-module in terms of differential operators
of course gives the same result. A fundamental solution of $P \Psi(z)=0$ is
$\Psi(z) = z^{-\mu/\la}$, so that acting with $\cD =
\langle z, \partial_z \rangle$ on $\Psi(z)$ gives the elements $v^z_k$
in $\cM$. Likewise, we reconstruct the elements $\underline{v}^w_k$
from the fundamental solution of $\underline{P}
\underline{\Psi}(w)=0$. Since $\cD =\langle z, \partial_z \rangle$ and
$\underline{\cD} =\langle w, \partial_w \rangle$ are related by a
Fourier transform, an element $v_k$ of the $\cD$-module in one asymptotic
region is represented by its Fourier transform in the opposite
region. This reproduces all elements in 
(\ref{eq:c=1module}).



A $\la$-expansion of the $\cD$-module element  $\underline{v}^z_{k}$,
using for example the stationary
phase approximation, yields as zeroth order contribution
\begin{align*} 
e^{\mu/\la} \left( \frac{\mu}{w} \right)^{k-\mu/\la},
\end{align*}
while the subdominant contribution is given by 
\begin{align*} 
\sqrt{-\frac{2 \pi \la \mu}{w^2}}.
\end{align*}
So in total we find that 
\begin{align*}
\underline{v}^z_{k}(w) = \sqrt{-2 \pi \la}~(\mu/e)^{-\mu/\la}~ w^{\mu/\la}~
\mu^{k+1/2}~ w^{-k-1}~ \psi_{\textrm{qu}}\left(\frac{\mu}{w}\right).
\end{align*}
This summarizes the contributions that we found
before: the genus zero $w^{\mu/\la}$ and genus one
$\mu^{k+1/2} w^{-k-1}$ results, plus the higher  
order contributions that are collected in $\psi_{\textrm{qu}}$. 

The all-genus partition function $Z$ of this I-brane system can be
easily computed exactly. Schematically it equals the
correlation function     
\begin{align*}
Z_{c=1} =  \langle \cW_w | S_{\mu} |  \cW_z \rangle, 
\end{align*}
where the $S$-matrix implements the Fourier transform between the two
asymptotic patches. Similar to the arguments in (the appendices of)
\cite{Alexandrov:2002fh} and \cite{Aganagic:2003qj}\footnote{The
  argument presented in the appendix of \cite{Aganagic:2003qj} is not
  fully correct. The proper argument (as shown below) recovers a
  slightly different prefactor in front of the Gamma-function, related
  to the doubling in the appendix of \cite{Alexandrov:2002fh}.}  
we find that the
result reproduces the perturbative expansion of the free energy as in 
equation (\ref{eqn:Gaussianpartfunc}).
For completeness let us review the argument by comparing 
$\underline{v}^z_k(w)$ with $\underline{v}^w_k(w)$. 

Notice that $\underline{v}^z_{k}(w)$ almost equals the 
gamma-function $\Gamma(z) = \int_0^{\infty}
dt~e^{-t}~t^{z-1}$. 
Indeed, let us take the integration contour from $-
i\infty$ to $i\infty$ and choose the cut of the logarithm to run from
 $0$ to $\infty$. Then 
\begin{align*}
& \underline{v}^z_{k}(w) =  \left( \frac{\la}{w} \right) \int_{- i\infty}^{i \infty} dz'~e^{z'} ~
\left( \frac{\la z'}{w} \right)^{k-\frac{\mu}{\la}} \\
&= \left( \frac{i \la }{w}
\right)^{k+1-\frac{\mu}{\la}}  \left[ \int_{- \infty}^{0} dz'~e^{iz'}
  ~e^{(k- \frac{\mu}{\la}) \log z'}  + \int_{0}^{\infty} dz'~e^{iz'}
  ~e^{(k-\frac{\mu}{\la}) \log z'} \right] \\
& = \left( \frac{i \la }{w}
\right)^{k+1-\frac{\mu}{\la}}  \left[ \int_{i \infty}^{0} dz'~e^{iz'}
  ~e^{(k- \frac{\mu}{\la}) \log z'}  + \int_{0}^{i\infty} dz'~e^{iz'}
  ~e^{(k- \frac{\mu}{\la}) \log z'} \right],
\end{align*}
where we moved the contour along the positive imaginary axis. A change
of variables and using that
$\log(iz' - \epsilon) = \log z' - 3 i \pi/2$ and $\log(iz' + \epsilon) =
\log z' + i \pi/2$, for $\epsilon$ small and real, then yields   
\begin{align*}
&\underline{v}^z_k(w) =  \left( \frac{i\la }{w}
\right)^{k+ 1-\frac{\mu}{\la}} 
\left[ e^{\pi i (k +1- \frac{\mu}{\la})/2} -  e^{-3 \pi i (k +1 -
   \frac{\mu}{\la})/2} \right]
\Gamma \left( k + 1- \frac{\mu}{\la} \right). 
\end{align*}
which is the same as the theory of type II result in the appendix of
\cite{Alexandrov:2002fh}.\hyphenation{ig-no-ring} 
Ignoring the exponential factor (which will only play a role
non-perturbatively), we find that the free energy $\cF$ equals the sum 
\begin{align*}
 \cF \left( \la,\, \mu \right) &=  \sum_{k \ge 0}  \left(k+1 -{\mu \over
     \la}\right) \log  \la +  \log \Gamma \left( k + 1-
 \frac{\mu}{\la} \right). 
 \end{align*}
It obeys the recursion relation 
 \begin{align}
 \cF \left( \la,\, \mu  + {\la \over 2} \right) - \cF \left( \la, \, \mu
    - {\la \over 2} \right) =   \left( \frac{1}{2}  - {\mu
     \over \la}  \right) \log \la + \log
 \Gamma \left(  {1 \over 2} - {\mu \over \la}  \right). \notag 
 \end{align}
which is known to be fulfilled by the $c=1$ string (see for example
Appendix A in  \cite{Nekrasov:2003rj}), up to a term $ -{1 \over 2}
\log (2 \pi \la)$ that can be taken care of  by normalizing the
functions $\underline{v}_k$. The same result is found when analyzing
the function ${v}_k$.


This concludes our discussion of the $c=1$ string. It is the first
$\cD$-module example 
where we see how to handle curves with two
punctures. The physical interpretation of the I-brane set-up
furthermore provides a check of our formalism. Moreover, this example agrees
with the claim that the $\cD$-module partition function should be
invariant under different parametrizations. Both the representation as
$c=1$ curve, $\Sigma_{c=1}: \,zw = \mu$, and that as a Gaussian matrix
model spectral 
curve, $\Sigma_{mm}: \,y^2  = x^2 + \mu$, yield the same partition function.

\section{Seiberg-Witten geometries}\label{sec:SW}

\index{Seiberg-Witten theory} 
More than once $\mathcal{N}=2$ supersymmetric gauge theories have proved to
provide an important theoretical framework to test new ideas in
physics. The most important advances in
this context are the solution of Seiberg and Witten in terms of a
family of hyperelliptic curves, as well as the explicit solution of
Nekrasov and Okounkov in terms of two-dimensional partitions. 
In what follows we will provide a novel perspective on these
results, by wrapping an I-brane around a Seiberg-Witten curve. The
$B$-field on the I-brane quantizes the curve, and a fermionic state is
obtained from the corresponding $\cD$-module. As we will see,
this state sums over all possible fermion fluxes through the
Seiberg-Witten geometry, and may be interpreted as a sum over
geometries. First we briefly review 
the Seiberg-Witten and Nekrasov-Okounkov approaches.

The solution of the $U(N)$ Seiberg-Witten theory is encoded in its partition
function $Z(a_i,\lambda,\Lambda)$, which is a function of the scale
$\Lambda$, the coupling $\lambda$ and boundary conditions for the
Higgs field denoted by $a_i$ for $i=1,\ldots,N$ (with $\sum_i a_i=0$
for the $SU(N)$ theory). The partition function is related to the free
energy $\cF$ as 
\begin{align*}
Z(a_i,\lambda,\Lambda) = e^{\cF} = e^{\sum_{g=0}^{\infty} \lambda^{2g-2} \cF_g(a_i,\Lambda)}.
\end{align*}  
In the above expansion $\cF_0$ is the prepotential which contains in
particular an instanton expansion in powers of $\Lambda^{2N}$, while
higher $\cF_g$'s encode gravitational corrections. The $U(N)$ Seiberg-Witten
solution identifies the $a_i$'s and the derivatives of the prepotential
$\frac{1}{2\pi i} \frac{\partial \cF_0}{\partial a_i}$ as the $A_i$ and
$B_i$ periods of the meromorphic differential  
\begin{align*}
\eta_{SW} = \frac{1}{2\pi i} v \frac{dt}{t}
\end{align*}  
on the hyperelliptic curve (\ref{eqn:swcurve})
\begin{align}
\Sigma_{SW} : \quad \Lambda^N (t + t^{-1}) = P_N(v) = \prod_{i=1}^N (v-\alpha_i).   \label{SW-curve}
\end{align}  
Despite great conceptual advantages, extracting the instanton
expansion of the prepotential from this description is a non-trivial
task. However, an explicit formula for the partition function,
encoding not only the full prepotential but also entire expansion in
higher $\cF_g$ terms, was postulated by Nekrasov in \cite{Nekrasov:2002qd}.
Subsequently this formula was derived rigorously jointly by him and Okounkov in
\cite{Nekrasov:2003rj} and independently by Nakajima and Yoshioka in
\cite{Nakajima:2003pg,Nakajima:2005fg}.  
For $U(N)$ theory this partition function is given by a sum over $N$
partitions $\vec{ R}=(R_{(1)},\ldots,R_{(N)})$  
\be \label{eqn:U(N)NOpartfunct}
Z(a_i,\lambda,\Lambda) =  Z^{pert}(a_i,\lambda) \sum_{\vec{ R}}
\Lambda^{2N |\vec{R}|} \mu^2_{\vec{R}}(a_i,\la),  
\ee
where
\begin{align}
\mu^2_{\vec{ R}}(a_i,\la) & =  \prod_{(i,m)\neq (j,n)} \frac{a_i - a_j
  + \la(R_{(i),m} - R_{(j),n} + n - m)}{a_i - a_j +
  \la(n-m)}, \label{muR}  
\end{align}
and
\begin{align}
Z^{pert}(a_i,\lambda) & =  \textrm{exp}\,\Big(\sum_{i,j} \gamma_{\lambda}(a_i-a_j,\Lambda)  \Big).   \label{Zpert}
\end{align}
The function $\gamma_{\lambda}(x,\Lambda)$ is related to the free energy of the topological string theory on the conifold, and its various representations and properties are discussed extensively in \cite{Nekrasov:2003rj} in Appendix A.
The vevs $a_i$ are quantized in terms of $\lambda$, so that for $p_i\in\mathbb{Z}$,
\begin{align*}
a_i = \lambda (p_i+\rho_i),\qquad \qquad \rho_i = \frac{2i-N+1}{2N}.
\end{align*} 
The approach of \cite{Nekrasov:2003af} is based on the localization technique in
presence of the so-called $\Omega$-background.
In general this background provides a two-parameter generalization of the
prepotential: the coupling $\lambda$ is replaced by two geometric
parameters $\epsilon_1$ and $\epsilon_2$. The prepotential, as given
above, is recovered for $\lambda=\epsilon_1=-\epsilon_2$. By the
duality web Fig.~\ref{fig:webofdualities} supersymmetric gauge theories
are related to intersecting brane configurations. The
Nekrasov-Okounkov solution must therefore have an interpretation in
terms of a quantum Seiberg-Witten curve, where $\la$ plays the role of
the non-commutativity parameter.

\subsection{Dual partition functions and fermionic correlators}

\index{dual partition function}
For a relation to the I-brane partition
function (\ref{fermionpartfunction}), it is necessary to consider the dual 
of the partition function (\ref{eqn:U(N)NOpartfunct}). This is introduced in
\cite{Nekrasov:2003rj} as the Legendre dual 
\begin{align}
Z^D(\xi,p,\lambda,\Lambda) = \sum_{\sum_i p_i = p} Z(\lambda(p_i+\rho_i),\lambda,\Lambda) \, e^{\frac{i}{\lambda}\sum_j p_j\xi_j}.  \label{Zdual}
\end{align}
An important observation of Nekrasov and Okounkov is that this dual
partition function can be elegantly written as a free fermion
correlator. This is a consequence of the correspondence between fermionic 
states and two-dimensional partitions described in \Cref{sec:fermstate}.  
For $U(1)$ there is no difference between the partition function and its dual and both can be written as
\begin{align} \label{eq:fermcorr-u1}
Z_{U(1)}^{D} (p,\la,\Lambda) & = \langle p | e^{-\frac{1}{\la}
  \alpha_{1}} \Lambda^{2 L_0} 
 e^{\frac{1}{\la} \alpha_{-1}} |p \rangle,
\end{align}
where $|p \rangle$ is the fermionic vacuum whose Fermi level is raised
by $p = a/\la$ units and $L_0$ measures the energy of the state. A version of the
boson-fermi correspondence implies the following decomposition  
\begin{align}
 e^{\frac{1}{\la} \alpha_{-1}} |p \rangle = \sum_R
 \frac{\mu_R}{\la^{|R|}} |p;R \rangle              \label{u1-bos-fer}
\end{align}
in terms of partitions $R$, where $\mu_R$ is the Plancherel measure  
\begin{align*}
\mu_R = \prod_{1\leq m<n <\infty} \frac{R_m - R_n + n - m}{n-m} = \prod_{\square \in R}\frac{1}{h(\square)}
\end{align*}
which can be written equivalently as a product over hook lengths $h(\Box)$. 

For general $N$ the dual partition function (\ref{Zdual}) looks very similar 
\begin{align} \label{eq:fermcorr}
Z_{U(N)}^{D} (\xi_i;p,\la,\Lambda) & = \langle p | e^{-\frac{1}{\tilde{\la}}
  \alpha_{1}} e^{H_{\xi_i}} \Lambda^{2 L_0} 
 e^{\frac{1}{\tilde{\la}} \alpha_{-1}} |p \rangle, 
\end{align}
however, now this expression is obtained by blending $N$ free fermions
$\psi^{(i)}$ 
into a single fermion $\psi$, as explained in \Cref{sec:fermstate}. 
In particular 
\begin{align*}
  H_{\xi_i} = \frac{1}{\la} \sum_r \xi_{(r+1/2) \hspace{-1mm} \mod N}
\psi_r  \psi^{\dag}_{-r},
\end{align*} 
while the bosonic mode $\alpha_{-1}$ arises from the bosonization of
the single blended 
fermion $\psi$. In formula (\ref{u1-bos-fer}) the Plancherel measure
of a blended partition ${\bf R}$ 
can be decomposed into $N$ constituent partitions as
\be
\mu_{\bf R} = \sqrt{Z^{pert}(a_i,\lambda)} \, \mu_{\vec{R}}(a_i,\lambda),   \label{Planch-Rvec}
\ee
with $\mu_{\vec{R}}$ and $Z^{pert}$ given in (\ref{muR}) and (\ref{Zpert}). 
When read in terms of the $N$ twisted fermions $\psi^{(i)}$,  
the correlator (\ref{eq:fermcorr}) involves a sum over the individual fermion charges $p_i$. 


Our aim in this section is to derive the above fermionic expressions
for the dual partition function  
from the $\cD$-module perspective.
In the next subsections we will see how first quantizing the
Seiberg-Witten curve in terms of a $\cD$-module elegantly reproduces
to the fermionic correlators (\ref{eq:fermcorr-u1}) and  
(\ref{eq:fermcorr}).

\subsection{Fermionic correlators as $\cD$-modules}

\begin{figure}[h!]
\begin{center}  
\includegraphics[width=5.7cm]{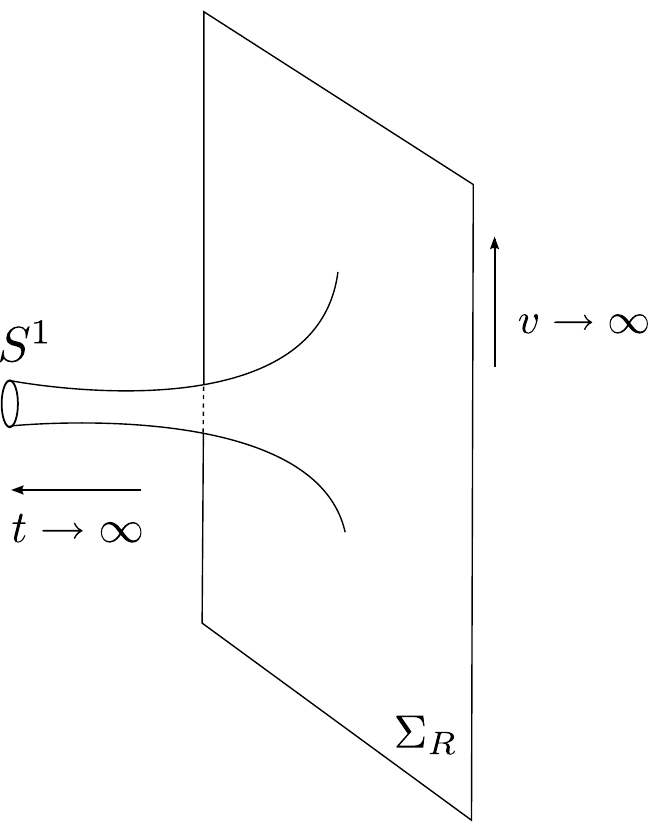}
\caption{The right-half Seiberg-Witten
  geometry is distorted around the asymptotic point $(t \to 0, v
  \to \infty)$. A fermion field on the quantized curve can be
  described as an element of a $\cD$-module, and sweeps out a state
  $|\cW \rangle$ at the $S^1$-boundary where $t \to \infty$.} \label{halfSWgeometry}
\end{center} 
\end{figure}

In this section we compute the I-brane partition function for $U(N)$
Seiberg-Witten geometries. We start with the simpler $U(1)$ and $U(2)$ examples
and then generalize this to $U(N)$. As a first principal step we
notice that the $U(N)$ Seiberg-Witten geometry 
\begin{align}
\Sigma_{SW} : \quad \Lambda^N (t + t^{-1}) = P_N(v) = \prod_{i=1}^N (v-\alpha_i),  \label{uN-SWcurve}
\end{align}  
can be rewritten as 
\begin{align*}
(P_N(v) - \Lambda^N t)(P_N(v) - \Lambda^N t^{-1})= \Lambda^{2N}.
\end{align*}
This shows that the Seiberg-Witten surface may be seen as a transverse
intersection 
of a left and a right half-geometry defined by 
\begin{align*}
\Sigma_{L}:~ \Lambda^N t = P_N(v) \quad
\mbox{resp.} \quad  \Sigma_{R}: ~\Lambda^N t^{-1} = P_N(v), 
\end{align*}
which are connected
by a tube of size $\Lambda^{2N}$. The left geometry  parametrizes
the asymptotic region 
where both $t \to \infty$ and  $v \to \infty$, whereas the right
geometry describes the region where  $v
\to \infty$ while $t \to 0$. This is illustrated in Fig.~\ref{halfSWgeometry}.

\index{$\cD$-module} \index{quantum curve}

Next we wish to associate a subspace in the Grassmannian to both half
Seiberg-Witten 
geometries. This will be swept out by a fermion field on the curve
that couples to the holomorphic part of the $B$-field
\begin{align*}
B = \frac{1}{\la} ds \wedge dv
\end{align*} 
Since this $B$-field quantizes the coordinate $v$ into the differential
operator~$\la \partial_s$, any subspace in this section is a
$\cD$-module for the differential algebra
\begin{align*}  
D_{\C^*} = \langle t, \la \partial_s \rangle.
\end{align*}

The free fermions on the Seiberg-Witten curves couple to the gauge field
$A = \frac{1}{\la} \eta_{SW}.$ This determines their flux through the
$A_i$ cycles of the Seiberg-Witten geometry as 
\begin{align*}
p_i = \frac{1}{\la} \int_{A_i} \eta_{SW}.
\end{align*}
The flux leaking through infinity is $p = \sum_{i=1}^N p_i$, which is
zero for $SU(N)$. A fermion field with fermion flux $p$ at infinity,
will sweep out a 
fermionic state in the $p$th Fock space. 
The parameters $\xi_i = \int_{B_i} \eta_{SW}$ are
dual to the fermion fluxes. Notice that in the perturbative regime $p_i$ can be
written as a $\la$-expansion 
\begin{align*}   
 \la p_i =  \alpha_i + \mathcal{O}(\la).
\end{align*} 

Since both half Seiberg-Witten geometries are distorted near $v =
\infty$ (see Fig.~\ref{halfSWgeometry}), while a 
fermionic subspace can be read off in the neighbourhood where $v$ is
finite, both half-geometries parametrize a subspace of $\C((v))$: 
\begin{align*}
\cW_{L}, ~\cW_{R} \subset \C((v)).
\end{align*}
The trivial geometry corresponds to a disk with
origin at $v=\infty$, whereas its boundary encloses the
point $v=0$. The vacuum state is therefore given by 
\begin{align}\label{eqn:SWvacuum}
|0 \rangle = v^{0} \wedge v^{-1} \wedge v^{-2} \wedge
\ldots.
\end{align}
Exponentials in $v^{-1}$ act trivially (as pure gauge
transformations in $ \Gamma_+$) on this  
state, whereas exponentials in $v$ transform the vacuum into a
non-trivial fermionic state.

Finally, the partition function is recovered by contracting the left
and the right fermionic state. 
Note that $s = - \log t$ is a local spatial coordinate on both half
Seiberg-Witten geometries, which tends to $-\infty$ on the left and to
$+ \infty$ on the right. This makes a huge difference with the $c=1$ geometry
discussed in \Cref{sec:c=1}, where the local coordinate is the
exponentiated coordinate, which on the left
is the inverse of that on the right. While in that example a non-trivial
$S$-matrix is required to identify the left and right half-geometries,
here we can just glue the fermionic states using the classic
Hamiltonian~$L_0$.  
Let us now find these quantum states. 

\subsubsection*{$U(1)$ theory}

The $U(1)$ Seiberg-Witten curve is embedded in $\C^* \times \C$ as 
\begin{align*}
\Lambda (t+t^{-1}) = v - \alpha, \qquad (t = e^s \in \C^*,~ v \in \C)
\end{align*}
where $\alpha \in \C$ is a normalizable mode. This geometry may be
factorized into a left and a right geometry        
\begin{align*}
\Sigma_L: ~ v = \Lambda t +  \alpha \quad & \mbox{and} &  \quad \Sigma_R:
~ v = \Lambda t^{-1} + \alpha,
\end{align*}
that intersect transversely with degeneration parameter $\Lambda^2$.

The symplectic form $B = \frac{1}{\la} ds \wedge dv$ quantizes both
half geometries into $\cD_{\la}$-modules on a punctured disc $\C^*_t$,
parametrized by $t$. We claim that these are characterized by the
$U(1)$ $\la$-connections  
\begin{eqnarray*}
\nabla_L = - \la t \partial_t +  \Lambda t + \la p \quad & \mbox{and} &  \quad 
\nabla_R = \la t \partial_t  + \Lambda t^{-1} + \la p. 
\end{eqnarray*}
These are just the canonical quantizations of the classical Seiberg-Witten
geometries, where additionally $u$ is quantized into $\la p$, with $p \in
\Z$. They yield the linear differential equations   
\begin{align}
P_L \psi^{\la}_{L}(t;p) &=  \left( - \la t \partial_t + \Lambda t +
  \la p  \right) \psi^{\la}_{L}(t;p)
 = 0,  \label{eqn:u(1)diffeqn} \\
P_R \psi^{\la}_{R}(t;p) &=  \left( \la t \partial_t + \Lambda t^{-1} +
  \la p \right)
\psi^{\la}_{R}(t^{-1};p)  = 0.\notag
\end{align}
The $\cD_{\la}$-modules are of the canonical form
\begin{align*}
\cM_{L/R} = \frac{\cD_{\la}}{ \cD_{\la}\cdot P_{L/R}},
\end{align*}
and are generated by the solutions 
\begin{eqnarray*}
\psi_L^{\la}(t;p) = t^p e^{\frac{\Lambda}{\la} t} \quad & \mbox{and} & \quad
\psi_R^{\la}(t;p) = t^{-p} e^{\frac{\Lambda}{\la} t^{-1}}.
\end{eqnarray*}
%

%
%


\begin{figure}[h!]
\begin{center}  
\includegraphics[width=5cm]{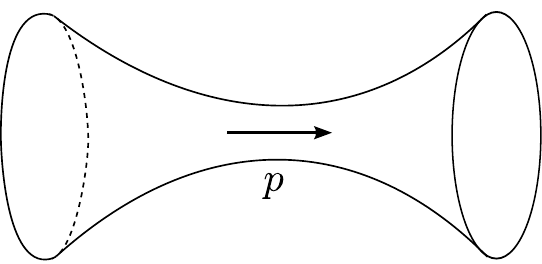}
\caption{Contracting two
  Seiberg-Witten half-geometries yields the Nekrasov-Okounkov
  partition corresponding to a fermion flux $p$ through the surface.} \label{NOwithflux}
\end{center} 
\end{figure}

From the discussion in \Cref{sec:fermstate} it follows that 
the factor $t^{-p}$ acts on the right Dirac vacuum by raising the
Fermi level into $|p\rangle$, while the exponent of $ t^{-1}$
translates to the exponentiated $\alpha_{-1}$ operator. With an
analogous statement for the left state, the modules $\cM_{L/R}$
translate into the Bogoliubov states 
\bea
\langle \cW_L| = \langle p | e^{\frac{\Lambda}{\la}\alpha_1} \qquad &\mbox{and}& \qquad 
| \cW_R \rangle = e^{\frac{\Lambda}{\la}\alpha_{-1}}|p\rangle.    \label{u1-LR}
\eea

The $U(1)$ Nekrasov-Okounkov partition function with fermion flux $p$ (see
Figure~\ref{NOwithflux}) is found by contracting the above fermion states
\begin{align*}
Z^{\la}_{NO}(p;\Lambda) = \langle p | e^{\frac{\Lambda}{\la}\alpha_1}
e^{\frac{\Lambda}{\la}\alpha_{-1}}|p\rangle.  
\end{align*}
The factors $\Lambda$ can be pulled out of the exponentials by using
the commutator $[L_0,\alpha_{\pm 1}] = \alpha_{\pm 1}$. Up to an extra
factor $\Lambda^{-p^2/2}$ we find that 
\begin{align*}
Z^{\la}_{NO}(p;\Lambda) \sim \langle p
|e^{\frac{\alpha_{1}}{\lambda}} 
\Lambda^{2 L_0} e^{\frac{\alpha_{-1}}{\lambda}} |p \rangle.  
\end{align*}
This has a nice geometrical explanation, since the left and right half
geometries are connected by a tube of size
$\Lambda^2$ as in the factorized form of the complete $U(1)$
geometry. The factor $\Lambda^{2 L_0}$ is the Hamiltonian that
describes the propagation of the fermion field along the tube. There
is no need to 
generalize this standard-CFT factor, since both patches are described
by the same space-coordinate $s$.

We also note that, as consistent with \cite{Aganagic:2003qj}, the solution
$\psi^{\la}_{R}(t;u)$ to 
$P_R \psi = 0$ equals the one-point-function  
\begin{align*}
\langle p-1 | \psi(t) |\cW_R \rangle = \sum_n t^{-p-n} \langle
p;R_n|\cW_R \rangle = t^{-p} e^{\frac{\la}{\Lambda} t^{-1}} = 
\psi^{\la}_{R}(t;u), 
\end{align*}
where $R_n$ represents a Young tableau consisting of just one row of $n$
boxes.

\subsubsection*{$U(2)$ theory}

We apply now the above strategy for the $U(2)$ geometry. We split the corresponding curve into a left and
a right half geometry, and for brevity focus just on the right part defined by 
\bea
\Sigma_R: \quad \Lambda^2 t^{-1} = (v - \alpha_2)(v -
\alpha_1).    \label{su2-right}  
\eea  
The $B$-field quantizes this equation into the second order differential equation 
\bea \label{eqn:rank2swdiffeqn} 
P_{R} \psi(t) = \left\{ \la^2 (t \partial_t -  p_2 ) ( t \partial_t -
   p_1 ) -
  \Lambda^2 t^{-1} \right\}
\psi(s) = 0.
\eea
%

A change
of variables $z= 2 t^{-1/2}$
%
%
followed by the ansatz $\psi(z) = z^{-(p_1+p_2)} \phi(z)$ and the
rescaling $z \mapsto 
(\la/\Lambda)z$ transforms this differential equation into the
familiar Bessel equation  
\begin{align*}
\left( z^2 \partial_z^2 + z \partial_z -
  \nu^2 - z^2 \right) \phi(z) = 0, \quad \mbox{with} \quad \nu^2 =
(p_1 - p_2)^2, 
\end{align*}
whose linearly independent solutions are given by modified Bessel
functions $I_{\nu}(z)$ and $K_{\nu}(z)$ of the first kind. The total solution in
the original $t$-coordinate is therefore a linear combination of
\begin{align}\label{eqn:solsu(2)sw}
\psi_R^{\la}(t;p_1,p_2) &= \left\{ \begin{array}{c}
t^{\frac{p}{2}} I_{\nu}\left( \frac{2 \Lambda}{\la \sqrt{t}} \right),\\  
t^{\frac{p}{2}} K_{\nu}\left( 
\frac{2 \Lambda}{\la \sqrt{t}}
\right), 
\end{array} \right.
\end{align}
where $p = p_1 + p_2$. These modified Bessel functions have different
asymptotics at infinity 
and relate to each other by going around the punctured disc $\C^*_t$.  

The second order differential operator $P_R$ defines the
$\cD_{\la}$-module 
\begin{align*}
\cM_{R} = \frac{\cD_{\la}}{\cD_{\la} \cdot P_R},
\end{align*}
which we claim represents fermions on the quantum $SU(2)$
Seiberg-Witten geometry.  
To check this statement, we have to find the fermionic state
corresponding to $\cM_R$. So we asymptotically expand of the
modified Bessel functions around $t=0$ in $\la$:
\begin{align*} 
I_{\nu}\left(\frac{2 \Lambda}{\la \sqrt{t}}\right) & \sim
t^{1/4}\exp \left(\frac{2
    \Lambda}{\la \sqrt{t}}\right)  \Big\{ 1 - 
\frac{(\mu -1)}{8}\, \frac{\la \sqrt{t}}{2
  \Lambda} + \frac{(\mu -1)(\mu-  
  9) }{ 2! \cdot 8^2} \, \frac{\la^2 t}{4
  \Lambda^2} + \ldots  \Big\}  \\
K_{\nu}\left(\frac{2 \Lambda}{\la \sqrt{t}}\right) &  \sim
t^{1/4}\exp \left(-\frac{2
    \Lambda}{\la \sqrt{t}}\right) \Big\{ 1 + 
\frac{(\mu -1)}{8}\, \frac{\la \sqrt{t}}{2
  \Lambda} + \frac{(\mu -1)(\mu-  
  9) }{ 2! \cdot 8^2} \, \frac{\la^2 t}{4
  \Lambda^2} + \ldots  \Big\},
\end{align*}
with $\mu = 4
\nu^2$.   

Recall that equation~(\ref{eqn:SWvacuum})  implies that any exponential
function in the local coordinate 
$v^{-1}=\sqrt{t}$ near the puncture acts trivially on the vacuum 
state. Equivalently, this is true for any asymptotic series in $\sqrt{t}$ that
assumes the value 1 at $\sqrt{t}=0$. In other words, we can
forget about the complete expansion in $\sqrt{t}$! Only the
WKB pieces
\begin{align*}
t^{1/4} \exp \left(\pm {\frac{2 \Lambda}{\la \sqrt{t}}}\right)
\end{align*}
are relevant in writing down the fermionic state. This is 
exactly opposite to the matrix model examples, where the
WKB-piece can be neglected and the perturbative series in $\la$
defines the fermionic state. 
 
The derivatives of the above solutions have one term
proportional to $\psi(s)$ (which we may forget about), and a term
proportional to the derivative of the Bessel functions. The latter may
be expanded as 
\begin{align*} 
\partial_s I_{\nu}(t) & \sim   t^{-1/4} \exp \left(\frac{2
    \Lambda}{\la \sqrt{t}}\right) \Big\{ 1 -
\frac{(\mu+3)}{8}\,  \frac{\la \sqrt{t}}{2
  \Lambda} + \frac{(\mu -1)(\mu+
  15)}{2! \cdot 8^2} \, \frac{\la^2 t}{4
  \Lambda^2}   + \ldots  \Big\}  \\
\partial_s K_{\nu}(t) & \sim t^{-1/4} \exp \left(\frac{2
    \Lambda}{\la \sqrt{t}}\right) \Big\{ 1 +
\frac{(\mu+3)}{8}\,  \frac{\la \sqrt{t}}{2
  \Lambda} + \frac{(\mu -1)(\mu+
  15)}{2! \cdot 8^2} \, \frac{\la^2 t}{4
  \Lambda^2}   + \ldots  \Big\}
\end{align*}
around $\sqrt{t}=0$. Again with the same reasoning only the WKB piece
is necessary to write down the quantum state. Taking into account the
extra factor $t^{\frac{p}{2}}$ in 
(\ref{eqn:solsu(2)sw}) the subspace $\cW^+_{R}$ is thus generated by the
$\cO(t)$-module  
\begin{align*} 
t^{\frac{p}{2}} \left( \begin{array}{cc}  t^{\frac{1}{4}}  \exp \left(\frac{2
    \Lambda}{\la \sqrt{t}}\right) \\ 
      t^{-\frac{1}{4}}  \exp \left(\frac{2
    \Lambda}{\la \sqrt{t}}\right)\end{array} \right) \cO(t),
\end{align*}
and blends (via the lexicographical ordening) into the fermionic  state 
\begin{align*} 
| \cW^+_R \rangle = v^{-p} ~ e^{
      \frac{\Lambda}{\tilde{\la} }v} \left(  v^{0}  \wedge v^{-1}   \wedge   v^{-2} \wedge v^{-3} \wedge \ldots \right)
\end{align*}
on the cover. Here we used a cover coordinate $v^{-1}$ obeying $v^{-2} =
t$, and rescaled the topological string coupling as $\tilde{\la} = \la/2$. $\cW^+_R$ is thus simply generated by a single function 
\begin{align*} 
\psi^{\la}(v) = v^{-p} e^{\frac{\Lambda}{\tilde{\la}} v} 
\end{align*}
Hence the fermions blend into the Bogoliubov state 
\begin{align} 
| \cW^+_{R} \rangle = e^{\frac{\Lambda}{\tilde{\la}} \alpha_{-1}} |p \rangle,   \label{su2-R}
\end{align}
when $p$ is an integer.

Note that the only modulus that appears in this expression is
$p$. This represents the diagonal $U(1)$, denoting the total fermion 
flux through the geometry. The moduli $p_1$ and $p_2$ measures the
fermion flux through an internal cycle and are not visible in the
result, because the final state sums over all internal
momenta. 
In general any $SU(2)$ Seiberg-Witten geometry with the same quantized
$p$ yields the same fermionic state.   

The fermionic (or dual) partition function is found by
contracting the left and the right states,  
similarly as in the $U(1)$ example above. The left state is just the
complex conjugate of the right one, so we find
\begin{align*}
Z_{NO}^{D} (p;\la,\Lambda) = \langle p | e^{\frac{\Lambda}{\tilde{\la}} \alpha_{1}} 
 e^{\frac{\Lambda}{\tilde{\la}} \alpha_{-1}} |p \rangle  \sim \langle
 p | e^{\frac{1}{\tilde{\la}} \alpha_{1}} \Lambda^{2 L_0} 
 e^{\frac{1}{\tilde{\la}} \alpha_{-1}} |p \rangle.
\end{align*}
The result is very similar to the $U(1)$ example, up to the shift $\la
\mapsto \la/2$. But notice that this fermionic state is written in
terms of a single blended fermion. Decomposing this fermion into two
twisted fermions makes it natural to insert an extra operator in the
middle of the correlator, that measures the momenta of the two
fermions through the $A$-cycles of the SW geometry. Weighting these
momenta with a potential $\xi_i$, for $i=1,2$, yields 
\begin{align*} 
Z_{NO}^{D} (\xi_i,p;\la,\Lambda) & \sim \langle p | e^{\frac{1}{\tilde{\la}}
  \alpha_{1}} e^{H_{\xi_i}} \Lambda^{2 L_0} 
 e^{\frac{1}{\tilde{\la}} \alpha_{-1}} |p \rangle,
\end{align*}
where $H_{\xi_i} = \frac{1}{\la} \sum_r \xi_{(r+1/2) \hspace{-1mm} \mod 2} \psi_r
\psi^{\dag}_{-r} = \frac{1}{\la} (p_1 \xi_1 + p_2 \xi_2)$. This is the
answer conjectured by Nekrasov and Okounkov in \cite{Nekrasov:2003rj}.



\subsubsection*{$U(N)$ theory}

It is not difficult to extend this discussion to the $U(N)$ 
theory (\ref{uN-SWcurve}), 
whose corresponding right half geometry we write as
\be
\Sigma_N :~ \Lambda^N t^{-1} = \prod_{i=1}^N
(v- \alpha_i).   \label{uN-SWcurve-bis}
\ee
Canonically quantizing this geometry and changing the coordinates 
$z=  \left( \frac{\Lambda}{\lambda}
\right)^N t^{-1}$,
brings us to the degree $N$ differential equation
\be
P_N \psi(z) = \left( \prod_{i=1}^N (z\partial_z - p_i) - z \right) \psi(z) = 0.  \label{suN-eqn}
\ee

It turns out that a solution to the above equation is given by a
particular Meijer 
G-function, denoted $G^{m,n}_{p,q}(z)$. The Meijer G-function is a complicated special function which was introduced
in order to unify a number of standard special function
\cite{Meijer:1946,Luke:1959,G-Fields},
and is defined in terms of a complex integral
\begin{align*}
G^{m,n}_{p,q}\left( \begin{array}{c}
a_1,\ldots,a_p \\
b_1,\ldots,b_q \end{array} | \, z \right) = \frac{1}{2\pi i} \int_L \frac{\prod_{j=1}^m \Gamma(b_j-t) \prod_{j=1}^n \Gamma(1-a_j+t)\, z^t }{\prod_{j=m+1}^q \Gamma(1-b_j+t) \prod_{j=n+1}^p \Gamma(a_j-t)} \, dt,
\end{align*}
where $L$ is a contour which goes from $-i\infty$ to $+i\infty$ and
separates the poles of $\Gamma(b_j-t)$, for $j=1,\ldots,m $, from
those of $\Gamma(1-a_i+t)$, for $i=1,\ldots,n$.  

It can be shown that the Meijer G-function solves the differential equation
\begin{align}
\left( \prod_{i=1}^q (z\partial_z - b_i) + (-1)^{p-m-n+1}z
  \prod_{j=1}^p(z\partial_z - a_j + 1)\right) \,G(z) =
0. \label{Meijer-diffeq} 
\end{align}
So, indeed
the Seiberg-Witten differential equation (\ref{suN-eqn})  is a special
case of Meijer differential equation~(\ref{Meijer-diffeq})  with
$p=n=0$ and $q=N$. 
Therefore the differential equation (\ref{suN-eqn}) is solved by 
\begin{align*}
\psi(z) =  G^{0,0}_{0,N}\left( \begin{array}{c}
\emptyset \\
p_1,p_2,\ldots,p_N \end{array} | \, z \right).
\end{align*}

Similarly as before we claim that the $\cD$-module corresponding
to $U(N)$ Seiberg-Witten curve is generated by $P_N$. A subspace $\cW$
corresponding to this $\cD$-module is this generated by a solution
$\psi(t)$ and its derivatives in $t \partial_t$. 

For $p<q$ the Meyer differential equation~(\ref{Meijer-diffeq}) has a
regular singularity at 
$z=0$ and an irregular one for $z=\infty$.
To extract the I-brane fermionic state, we are interested in
the behaviour around the irregular singularity, where $t
\to 0$.
It turns out that one of the independent solutions
of the Seiberg-Witten differential equation~(\ref{suN-eqn}) has the asymptotic
expansion \cite{Meijer:1946,Luke:1959,G-Fields} 
\begin{align*}
\psi(v) \sim e^{\frac{\Lambda}{\lambda/N}v} \, v^{\frac{1-N}{2}}\, v^{p} \sum_{j=0}^{\infty} k_j v^{-j},
\end{align*}
around this singularity,  which is conveniently written in
the cover coordinate $(-v)^N=t^{-1}=\left( \frac{\lambda}{\Lambda}   
\right)^N z$. The other solutions are found
by multiplying the coordinate $v$ by $N$-th roots of unity, and thus
behave distinctly at infinity. As before,
$p = \sum_{i=1}^N p_i$.  

To find the fermionic state corresponding to the $U(N)$ Seiberg-Witten curve,
we act with $\psi(v)$ on the Dirac vacuum. 
The positive power of $v$ in the exponent of $\psi(v)$ corresponds in
the operator 
language to $\alpha_{-1}$, whereas $v^p$ lifts the Fermi level.
The remaining series just contains negative powers of $v$ which translate to a
trivial action on the vacuum in the operator formalism. Therefore, the
above asymptotic solution and its derivatives (in $t \partial_t$)
blend into the state  
\be
|\mathcal{W}_R\rangle = e^{\frac{\Lambda}{\tilde{\lambda}}\alpha_{-1}} |p\rangle,     \label{suN-state}
\ee
with rescaled topological string coupling $\tilde{\lambda} = \lambda/N$.
Like for the $U(2)$ Seiberg-Witten geometry the dependence on the
individual moduli $p_{i}$ has dropped out. 

Similarly as in $U(1)$ and $U(2)$, in the present case we also find
the $U(N)$ Nekrasov-Okounkov dual partition function
\be
Z_{NO}^{D} (\xi_i;\la,\Lambda)  = \langle p | e^{\frac{1}{\tilde{\la}}
  \alpha_{1}} e^{H_{\xi_i}} \Lambda^{2 L_0} 
 e^{\frac{1}{\tilde{\la}} \alpha_{-1}} |p \rangle.\label{Z-NO}
\ee   
This
fermionic correlator is indeed the one postulated in
\cite{Nekrasov:2003rj}. For $N=1$ or $N=2$ the Meijer G-function
specializes respectively to the 
exponent and Bessel functions, which reproduces the results derived in
previous subsections.

Although the normalizable moduli $p_i$ disappear in the final I-brane partition
function, they reappear when the state is unblended in terms of $N$
single fermions
\begin{equation}\label{eq:microstates}
e^{\frac{1}{\tilde{\lambda}} \alpha_{-1}}|p\rangle = \sum_{ R} \frac{\mu_{ R}}{\tilde{\lambda}^{|{ R}|}} |p,{
R}\rangle = \sum_{ \sum p_i=p} \sum_{R_{(i)}} \sqrt{Z^{pert}(p)} \, \frac{\mu_{\vec{
R}}(p,\tilde{\lambda})}{\tilde{\lambda}^{|{ R}|}} \bigotimes_{l=1}^N |p_i,R_{(i)} \rangle,
\end{equation}
as may be seen from (\ref{u1-bos-fer}) and (\ref{Planch-Rvec}). The
charges $p_i$ have an interpretation as the fermion fluxes through the
$N$ tubes of  
the Seiberg-Witten geometry we started with. 

Actually, we find the same fermionic state when starting with any other
Seiberg-Witten geometry whose fermion flux at infinity is $p$. Hence
one microstate in the total sum~(\ref{eq:microstates}) can be
interpreted as a fermion flux through an infinite set of geometries. This
gives the state~(\ref{eq:microstates}) as well as the partition function
(\ref{Zdual}) the interpretation of a sum over geometries.

\subsection{Topological string theory and quantum groups}  \label{ssec:sum} 

\index{topological string theory}

Nekrasov and Okounkov also derive a partition function for
the 5-dimensional $U(N)$ Seiberg-Witten theory compactified on the
circle of circumference 
$\beta$ \cite{Nekrasov:2003af,Nekrasov:2003rj,Nakajima:2005fg} . It is
given by a 
$K$-theoretic generalization of the 4-dimensional formula in
equation (\ref{eqn:U(N)NOpartfunct}).

\begin{figure}[h!]
\begin{center}  
\includegraphics[width=10cm]{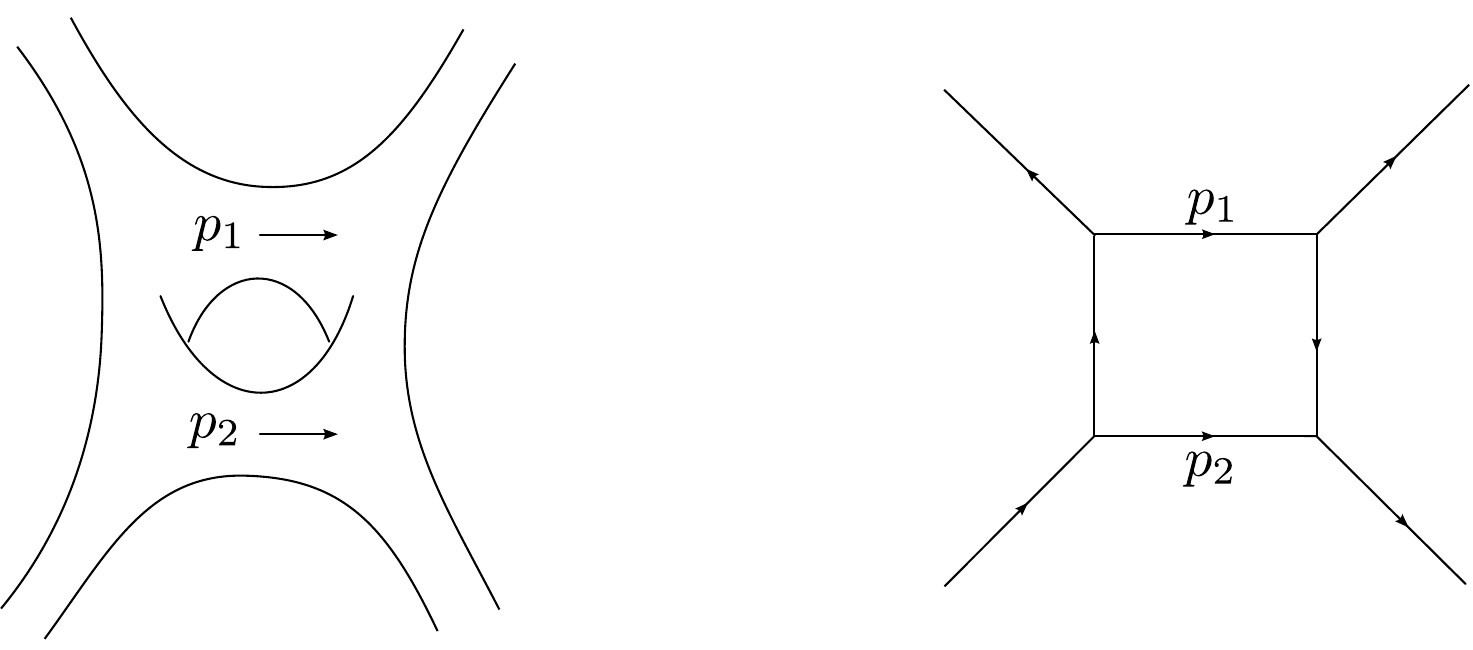}
\caption{On the left we see the
  five-dimensional $U(2)$
  Seiberg-Witten surface with fermion fluxes through its
  $A$-cycles, and on the right a corresponding toric diagram. The
  fermion flux deforms the  K\"ahler lengths of the toric diagram as
  in equation~(\ref{kahlersize}).}  \label{fig:SWdiagramwithflux}
\end{center} 
\end{figure}

%
%

This 5-dimensional theory is closely related to the topological
string theory by geometric engineering on a toric
Calabi-Yau background \cite{Iqbal:2003ix,Iqbal:2003zz}. Namely, the
partition function of the topological string theory on an
$A_N$-singularity fibered over $\P^1$ (whose toric diagram consists of $N-1$
meshes as in Fig.~\ref{fig:SWdiagramwithflux}) is equal to the
partition function of the 
5-dimensional gauge 
theory given above, when the 
K\"ahler sizes of the internal legs are (see \Cref{sec:revisitSW})
\begin{align}\label{kahlersize}
Q_{F_i} = e^{\beta (a_{i+1} - a_{i} )}, \qquad  Q_B = \left( \frac{\beta \Lambda}{2} \right)^{2N}, 
\end{align}
where $F_i$ labels the vertical legs and $B$ the horizontal ones. 
In the so-called gauge theory limit, when $\beta\to 0$, the
topological string partition function reduces to the 4-dimensional
Seiberg-Witten partition function. The corresponding B-model 
mirror geometry is of the form 
\begin{align*}
X_{SW}: \quad xy - H(t,v) = 0,
\end{align*}
where $H(t,v)=0$ represents a Riemann surface
of genus $N-1$. In the gauge theory limit this surface
becomes the Seiberg-Witten curve $\Sigma_{SW}$, parametrized as in the
equation~(\ref{SW-curve}). 

In topological string theory it is natural as well to write down a dual
partition function \cite{Aganagic:2003qj}. In a local B-model this allows the
possibility of arbitrary fermion fluxes through the handles of the Riemann
surface. In this setting it has been argued before that turning on a
fermion flux is equivalent to deforming the geometry. More precisely,
fermion flux parametrized by $\mathcal{P} = p_i B_i$
changes the integral of the holomorphic 3-form over any linking
3-cycle $A^3_i$, 
%
%
and thereby shifts the complex structure moduli $S_i = \int_{A_i}
\Omega$ as
\begin{align*}
S_i \mapsto S_i + \la p_i 
\end{align*}
In the A-model fermion flux translates into wrapping D4 branes 
around 4-cycles, and thereby deforms the K\"ahler moduli. The I-brane
partition function 
thus equals the dual topological string partition function.  

 
Because the Seiberg-Witten surface is embedded in $\C
\times \C^*$, $A^3$ and $B^3$-cycles in the toric threefold
will have topologies $S^1 \times S^2$ and $S^3$, respectively (see
Fig.~\ref{SW3cycles}). In 
particular, a basis of $A^3_i$-cycles can be chosen to reduce to the
surface as the combination of 1-cycles $A^1_i - A^1_{i+1}$. Now notice that the
3-cycle $A^3_i$ with topology $S^1 \times S^2$ is mirror to the
vertical 2-cycle $F_{i}$ that connects the $i$-th and 
the $i+1$-th horizontal leg. So turning on a fermion flux $p_i$
through the $i$-th leg of the Seiberg-Witten geometry changes the
complex structure parameter $S_i$ by an amount proportional to $a_i -
a_{i+1}$. This explains the K\"ahler size $Q_{F_i}$ in (\ref{kahlersize}) 
in terms of fermionic fluxes through the Seiberg-Witten curve, and in
reverse why (\ref{eq:microstates}) may be interpreted as a sum over
Seiberg-Witten geometries, or equivalently toric diagrams.  
So we conclude that the fermionic interpretation in 4d of Nekrasov and
Okounkov is dual in 6d to the fermionic interpretation of the
topological string, and has a deeper interpretation in terms of $\cD$-modules.

\subsubsection{Topological vertex} 

An important step to understand Seiberg-Witten curves (as well as
other local Calabi-Yau geometries) is the
topological vertex, introduced in
\Cref{sec:toricgeometry}. Recall that its mirror is a genus zero 
curve with three punctures 
given by the equation 
\begin{align}\label{eqn:topvertex}
x + y - 1 =0
\end{align}
in $\C^* \times \C^*$. In this case the symplectic form is given by
$du \wedge dv$ where $u,v$ are logarithmic coordinates: $x=e^u$ and $y=
e^{v}$. The corresponding $\cD$-module is now given by the operator 
\cite{Aganagic:2003qj}
\begin{align}\label{Deqn:topvertex}
P =  e^{u} + e^{-\la \lotjesd_u} - 1.
\end{align}
$P$ is actually a difference operator, instead of a differential
operator, so we have to generalize the notion of a $\cD$-module
somewhat. This is a well-known procedure in the field of quantum
groups. These quantum groups appear because in the $\C^*$ case the
operators $\hat x$ and $\hat y$ now satisfy the Weyl algebra or
$q$-commutation relation
\begin{align*}
\hat x \,\hat y = q \,\hat y \,\hat x,\qquad q=e^\la.
\end{align*}
The fundamental solution to $P\Psi=0$ is the quantum dilogarithm
\begin{align*}
\Psi(u)= \prod_{n=1}^{\infty}  \left(1 - e^{u} q^{n} \right).
\end{align*}

The corresponding module $\cM$ for the Weyl algebra can again be
written in terms of the coordinate $u$ or in terms of the dual
variable $v$. There is another unitary map $U$ that implements this
transformation on the free fermion fields. Because of the hidden
cyclic symmetry of the vertex, this can be made transparent by
writing it as 
\begin{align*}
e^{u_1} + e^{u_2} + e^{u_3} =0.
\end{align*}
Up to an overall rescaling of the three variables $u_i$, the map $U$
satisfies $U^3=1$.  This line of reasoning leads one directly to the
formalism of \cite{Aganagic:2003qj}, but we will not pursue this here in more
detail. We reach the important conclusion that the notion of a quantum curve, as
expressed in the concept of a (generalized) $\cD$-module, is the right
framework to derive the complicated transformations of
\cite{Aganagic:2003qj}. We will later use this correspondence in two concrete
examples of compact curves, but first make a few remarks about
five-dimensional $U(1)$ Seiberg-Witten theory. 

\subsubsection{Five-dimensional $U(1)$ theory}

Quantizing any five-dimensional Seiberg-Witten geometry yields
a difference (instead of differential) equation. Working out
$\cD$-modules for these geometries we leave for future work. Let us
treat one example in detail though. The five-dimensional right-half $U(1)$
Seiberg-Witten half-geometry 
\begin{align}\label{eqn:5dU(1)SW}
\Sigma^{5d}_R: \quad \beta \Lambda e^{-\beta \la} t^{-1} +
e^{-\beta v} 
-1 = 0 
\end{align}
is isomorphic to the topological vertex (\ref{eqn:topvertex}) and may
be drawn as a pair of pants. In the field theory limit $\beta \to 0$ 
it reduces to the familiar equation $\Lambda t^{-1} = v$ for the right-half
Seiberg-Witten geometry (with $u=0$). 

\begin{figure}[h!]
\begin{center}  
\includegraphics[width=9.5cm]{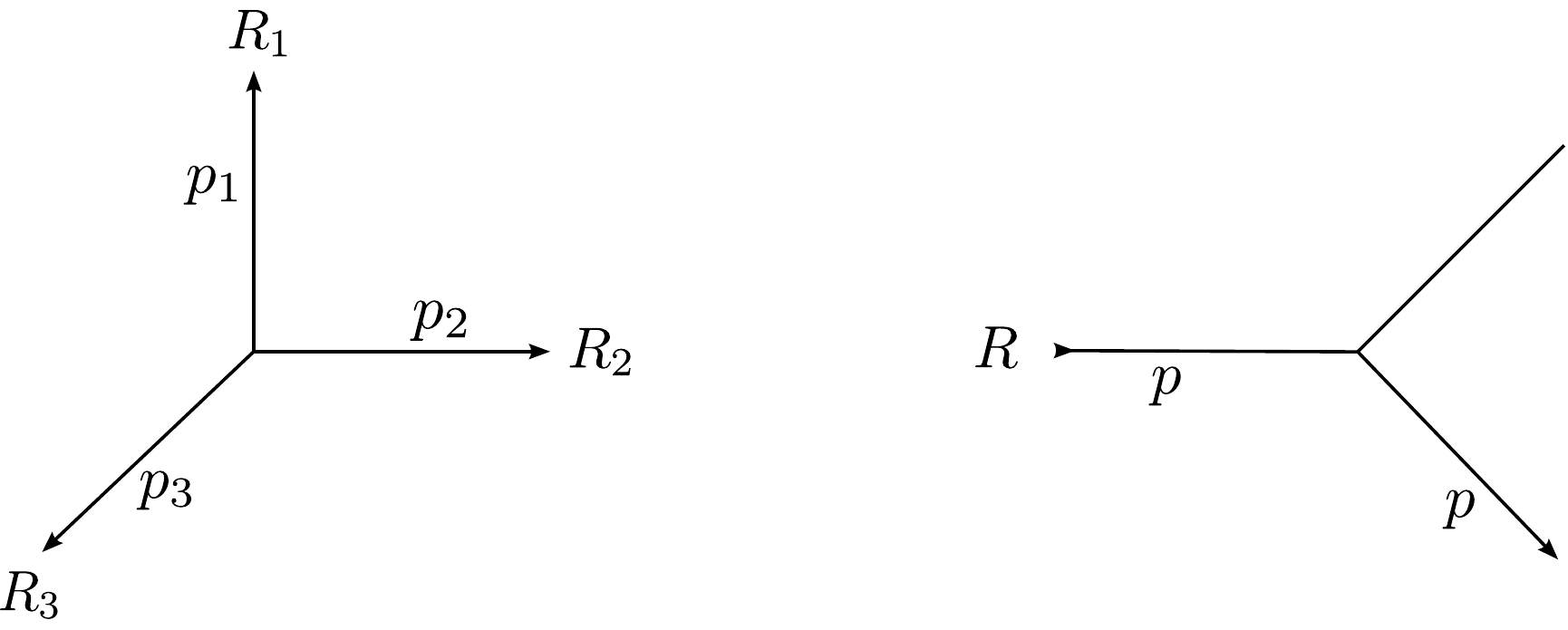}
\caption{The B-model vertex (on the left)
  may be
  expanded as a sum over fermionic states $|p_1, R_1 \rangle \otimes
  |p_2, R_2 \rangle \otimes |p_3, R_3 \rangle$, with $p_1 + p_2 + p_3
  =0$, corresponding to a conserved fermion flux through the pair of
  pants. The five-dimensional right-half Seiberg-Witten geometry (on
  the right) with
  charge $p$ only has one partition $R \neq 0$. }\label{fig:vertices}
\end{center} 
\end{figure}

In the B-model the most general state
assigned to a local pair of pants geometry is given by a Bogoliubov state
\cite{Aganagic:2003qj} 
\be
|\cW \rangle  = \exp \Big[\sum_{i,j} \sum_{m,n=0}^{\infty} a^{ij}_{mn} \psi^i_{-m-1/2} \psi^{* j}_{-n-1/2}\Big]|0\rangle,  \label{V-state}
\ee 
where the index $i=1,2,3$ describes the fermion field on the three
asymptotic regions of the pair of pants, and the coefficients are
determined by a comparison with the A-model topological vertex. This
exponent can be expanded as a sum over states (see Fig.~\ref{fig:vertices})  
\begin{align*}
|p_1, R_1 \rangle \otimes |p_2, R_2 \rangle \otimes |p_3, R_3 \rangle,
\end{align*}
where the fermion flux is conserved: $p_1 + p_2 + p_3 =0$. To describe
the 5d Seiberg-Witten $U(1)$ geometry we won't need this 
state in full generality. 

The B-field quantizes this geometry into the difference equation
\begin{align}\label{eq:difference-eqn}
P(t) \Psi(t)  = \left( \beta \Lambda e^{-\beta \la } t^{-1} + e^{\beta
 \la  t \partial_t} 
-1 \right) \Psi(t) =0.
\end{align}
%
Like for the topological vertex its fundamental solution is the
quantum dilogarithm   
\begin{align*}
\Psi(t) 
= 
\exp \sum_{n>0} \frac{ (\beta \Lambda)^{n} t^{-n} }{ n (1-e^{\beta \la n} ) }. 
\end{align*}

As an intermezzo, notice that quantizing the equation
\begin{align*}
\beta v = - \log \left( 1- \beta \Lambda e^{- \beta \la} t^{-1}
\right), 
\end{align*}
which is just a rewriting of equation (\ref{eqn:5dU(1)SW}) for
$\Sigma^{5d}_R$, we 
find a differential equation which may be interpreted as the WKB
approximation of difference equation (\ref{eq:difference-eqn}). A
fundamental solution of the 
differential equation is given by the genus~0 disc amplitude 
\begin{align*}
\Psi_0(u) 
= \exp \sum_{n>0} \frac{ (\beta \Lambda)^n t^{-n} }{ \la n^2 e^{\beta
    \la n} }. 
\end{align*}

Acting with the five-dimensional dilogarithm on the Dirac vacuum state
yields the fermionic state 
\begin{align*}
| \cW \rangle^{5d}_{U(1)} = \exp  \sum_{n>0}
\frac{ (\beta \Lambda)^{n} \alpha_{-n}}{n(1-e^{\beta \la n})}|0 \rangle.
\end{align*}
This describes a subset of $| \cW
\rangle$ where only the quantum number $R_1$ is non-trivial. 
Summing over all external states of the form 
\begin{align*}
|-p,R \rangle \otimes |p,\bullet \rangle \otimes |0, \bullet \rangle,
\end{align*}
incorporates a fermion flux $p$ through the pair of pants. In the field
theory limit $\beta \to 0$ the resulting state reduces to the familiar
four-dimensional state $$\exp( \alpha_{-1}/\la ) |p \rangle
\otimes|p,\bullet \rangle \otimes |0, \bullet \rangle.$$

The partition function is found as the contraction of the left and
right 5d half-geometries. (Or equivalently in the topological B-model by
inserting a propagator \cite{Aganagic:2003qj}.) This yields the
fermionic correlator  
\begin{align*}
\langle 0| \tilde{\Gamma}_+ \tilde{\Gamma}_-|0 \rangle  
= \langle 0| \Gamma_+ (\beta \Lambda)^{2L_0} \Gamma_-| 0 \rangle, 
\end{align*}
with 
\begin{align*}
\tilde{\Gamma}_{\pm}=\exp  \sum_{\pm n>0}
\frac{ (\beta \Lambda)^{|n|} \alpha_{n}}{|n|(1-e^{\beta \la n})}
\quad \mbox{and} \quad
 \Gamma_{\pm}=\exp\sum_{\pm n>0}
\frac{ \alpha_{n}}{|n|(1-e^{\beta \la n})}.
\end{align*}
Indeed, the result equals the five-dimensional $U(1)$ partition function 
\begin{align*}
Z^{U(1)}_{5d}(\lambda, \Lambda, \beta) = \textrm{exp}\sum_{n=1}^{\infty}
\frac{(\beta \Lambda)^{2n}}{4n\,\textrm{sinh}^2(\beta \la n/2)},
\end{align*}
that was found by Nekrasov and Okounkov in \cite{Nekrasov:2003rj}.

\section{Discussion}\label{sec:ch6discussion}

In this chapter we argued that the fundamental objects underlying
various matters in theoretical physics are chiral fermions
living on quantum curves. In our formulation the quantum curve is
defined, similarly to an affine classical curve, in terms of an equation
$P(z,w)=0$. Its crucial feature, however, is the non-commutative
character of the coordinates $z,w$. It thereby generalizes the
classical curve that comes up in the standard formulation of a given
topic. Examples of such classical curves are spectral curves in
matrix models, $c=1$ string theory, Seiberg-Witten theory, and more
generally in topological string
theory. Semi-classically their (genus one) free energy
is computed as a fermionic determinant on the classical curve. In our
approach chiral 
fermions on the quantum curve 
generate the all-genus expansion of the free energy with respect to the
non-commutativity parameter~$\la$.  

As we explained in Chapter~\ref{chapter5}, fermions on a
non-commutative curve can be realized physically within 
string theory as massless states of open strings on an I-brane
in the presence of the $B$-field. In this chapter we have exploited
this system in a few
important examples. At the   
same time we stressed the fundamental importance of $\cD$-modules,
which are the appropriate mathematical structures describing
non-commutative holomorphic curves. First of all we showed, while
reinterpreting the results in \cite{eynard-isomonodromic}, that
I-branes and $\cD$-modules provide an insightful formulation of matrix
models. This quite general statement is also appealing  when certain
matrix model limits are considered, such as a double scaling
limits. In this case one recovers an I-brane formulation of minimal
string theory, topological gravity and $c=1$ string theory.   

The I-brane configuration can be related to
topological string theory and to Seiberg-Witten theory via a sequence
of string dualities Fig.~\ref{fig:webofdualities}. In the last part of this
chapter we focused on supersymmetric gauge theories. Using the
$\cD$-module formalism we derived the fermionic expression for the
$U(N)$ partition function of 
the pure $\mathcal{N}=2$ gauge theory, reproducing
the dual all-genus 
partition function introduced in \cite{Nekrasov:2003rj}.  
We considered mainly 4-dimensional Seiberg-Witten theories with
unitary gauge groups, though, and explained only the simplest $U(1)$ example of
5-dimensional theory. It would be insightful to extend these results
to other gauge groups and include matter content. It is clear that
this should be possible, as these aspects of the 5-dimensional
Seiberg-Witten theory are captured by topological strings on
toric manifolds. The latter 
system can be solved in a fermionic B-model formulation of the
topological vertex \cite{Aganagic:2003qj} which is equivalent to the I-brane
fermions. Nonetheless, finding the quantum I-brane curve
representing such configurations appears to be a nontrivial task.     

In the process of unraveling the $\cD$-module structure in both sets
of examples, we noticed some crucial differences. While the WKB piece
of the $\cD$-module generator can be ignored in finding the 
matrix model partition function, we discovered that it plays an
eminent role for the Seiberg-Witten geometries. Another distinction is
the difference in (non-)normalizable modes. While the potential $W$
parametrizes non-normalizable modes that appear in the $\cD$-module as
parameters, in contrast, the normalizable modes in the Seiberg-Witten
geometries are eaten by the $\cD$-module, and only visible as a sum
over internal fermion fluxes in the geometry. On the other hand,
varying the $\cD$-module with respect to the non-normalizable modes
yields differential equations which relate to isomonodromy and the
Stokes phenomenon.

While in this chapter our focus has been to associate a
$\la$-perturbative quantum state to a spectral curve, we noticed that
$\cD$-modules in fact contain non-perturbative information. These bits
get lost when we turn the $\cD$-module in a fermionic state by making
an asymptotic expansion of the $\cD$-module generators in $\la$. This
is in line with the discussion on non-perturbative aspects of minimal
string theory in \cite{Maldacena:2004sn}, where it is
argued that non-perturbative effects drastically modify the
non-trivial target space curve into a plain complex plane. 

It also agrees with more recent studies of non-perturbative effects in matrix
models \cite{Eynard:2008he, eynard-2009-0903, Marino:2007te,
  Marino:2008ya}. These articles revealed that a series of
instantons in the matrix model can be summarized in a non-perturbative
partition function that sums over all possible filling fractions $p_i
= \frac{1}{\la} \oint_{a_i} \eta$ as 
\begin{align*}
 Z_{\textrm{non-pert}}(\mu, \nu) = \sum_{p \in \Z^g}
 Z_{\textrm{pert}}(\la(p+ \mu)) e^{2 \pi i p \nu}. 
\end{align*}
In this formula $(\mu,\nu)$ is a choice of characteristics on the
matrix model spectral curve, that encodes the choice of integration
contour in the matrix model. The integer $g$ is the genus of the
spectral curve. Interestingly,
this partition function turns out to have very nice
properties. $Z_{\textrm{non-pert}}(\mu, \nu)$ is 
not only holomorphic, but also transforms in a modular fashion under
the symplectic group $Sp(g,\Z)$. Moreover, it satisfies the Hirota
equations and can thus be interpreted in terms of a twisted fermion
field on $\Sigma$ with twists $(\mu, \nu)$. 

The partition function $Z_{\textrm{non-pert}}(\mu, \nu)$ is obviously
closely related to the I-brane partition function, that is defined in
equation~\ref{fermionpartfunction} and studied from several angles
in this chapter. A choice of saddle in the $\la$-expansion of our
$\cD$-module partition function corresponds to a choice of
characteristics in $Z_{\textrm{non-pert}}(\mu, \nu)$.



How do these latter matrix model results and our $\cD$-module insights fit in 
with other developments that have taken place the last years in the
area of topological string theory?
Let us start with the observation that the (perturbative) topological
string partition function is known to 
suffer from background dependence \cite{Witten:1993ed}. As a result
the free energy does not transform as a proper modular form. The
modularity can be restored, however, but then the resulting function
is not holomorphic anymore \cite{Aganagic:2006wq}. Instead it obeys
the holomorphic anomaly equations \cite{Bershadsky:1993cx}. 
It is natural to suggest that a 
partition function which is both holomorphic and modular, is a
candidate for a non-perturbative completion of topological string
theory. This is argued in \cite{Eynard:2008he}. 

The claim is strengthened by the following discoveries. 
In a sequence of papers \cite{Marino:2006hs,
  Eynard:2007hf, Bouchard:2007ys} the Dijkgraaf-Vafa correspondence
between matrix models and topological string theory has been extended
to arbitrary local Calabi-Yau geometries modeled on a Riemann surface
$\Sigma$. As a result topological string amplitudes can be computed in terms
of a simple recursion relation that originates from the theory of matrix models
\cite{eynard-orantin}.  The Eynard-Orantin formalism is
closely related to the Kodaira-Spencer formulation of the
B-model, and may be viewed as the bosonized version of our fermionic
formulation \cite{Dijkgraaf:2007sx}. It would be valuable to
understand this non-commutative version of the familiar boson/fermion
correspondence and its interpretation in terms of $\cD$-modules in
more detail. 

Moreover, our formalism seems to be closely related to a
non-commutative extension of the Eynard-Orantin formalism, that is
studied in \cite{eynard-latest}. The resulting non-commutative
invariants depend on two deformation parameters: The first deformation
parameter is the usual topological 
string coupling constant $\la$, whereas the second one is an
independent non-commutative deformation. The connection to our
$\cD$-module formalism should arise when we only turn on this second
deformation. Turning on either of the deformation parameters is
possibly equivalent.



Let us also note that while we mainly studied the web of dualities
Fig.~\ref{fig:webofdualities} in the large radius regime, where the
topological string partition function has an expansion in terms of the
usual Gromov-Witten and Donaldson-Thomas invariants, the $\cD$-module
formalism suggests a relation to invariants in other regimes. Since we
need to make a choice of boundary conditions, when we turn
a $\cD$-module into a quantum state, the final state
is troubled by the Stokes effect: Solutions that decay faster can be
added at no cost and the state changes when one crosses certain lines
in the moduli space. This suggests that the $\cD$-modules we studied
 may be helpful in understanding the phenomenon of wall-crossing
in $\mathcal{N}=2$ theories \cite{Denef:2007vg,Gaiotto:2008cd}. We discuss
wall-crossing in $\cN=4$ theories extensively in
Chapter~\ref{chapter7}.

More mathematically, $\cD$-modules play an important role in the geometric
Langlands program
\cite{Frenkel:2005pa,Kapustin:2006pk,Witten:2007td,Gukov:2008ve}. 
In the physical description of this program
$\cD$-modules enter in the  
description of eigenbranes of the magnetic 't Hooft operator in a reduction of
4-dimensional $\cN=4$ gauge theory down to a 2-dimensional sigma
model. However, in this sigma model (which is not yet coupled to
gravity) the $\cD$-modules describe coisotropic A-branes. This is in
contrast to their physical appearance in our intersecting brane
configuration.  

There seems to be a deeper connection of our formalism to quantum
integrable systems as 
they are studied in for example
\cite{beilinsondrinfeld,talkarinkin,Chervov:2006xk}. Quantum curves
feature in these quantum systems as so-called opers, that 
parametrize the base of the integrable system, in the same way that
spectral curves parametrize the base of the Hitchin integrable system. It would
be enlightening to find out whether the fermions on the quantum curve
can be described in a similar way in terms of the quantum integrable system
as holds in the semi-classical limit. Does this lead to a better
description of the quantum fermion CFT on the quantum curve?
Is our set-up related to WZW models based on opers in the geometric
Langlands program \cite{Frenkel:2005pa}?  
Most importantly, we would like to be able to write down a 2-dimensional
action for the quantum fermion theory. In \Cref{sec:compactcurves} we succeed in
writing down the action for a propagator in the I-brane geometry, but
not yet for a 3-vertex.

        \chapter{Dyons and Wall-Crossing}\label{chapter7}

In this chapter we study examples of local Calabi-Yau
threefolds that are modeled on a compact Riemann surface. We start in 
\Cref{sec:compactcurves} with a family of Calabi-Yau's
that are built on a 2-torus, and then consider threefolds based on a genus 2
curve. In both examples we will discover nice automorphic structures
and confirm the relation of the exponent of $\cF_1$ to the fermionic
determinant $\textrm{det}\, \bar{\lotjesd}$.  

Notice that these compact curves have genera $g>0$, and no asymptotic
endpoints at infinity. To compute their I-brane partition
function we thus have to cut the curves in affine pieces. In the
genus 1 example we employ the $\cD$-module techniques to glue
the end-points of a cylinder. In the genus 2
example we use the topological vertex formalism to compute the
contribution of a pair of pants. 

In \Cref{sec:BPSdyons} and \Cref{sec:wall-crossing} we focus on the
semi-classical contribution to the partition function on the genus 2
Calabi-Yau. This 
is summarized in an automorphic invariant that surprisingly turns out to count
the number of non-perturbative BPS dyons in $\cN=4$ theories. We make
this relation explicit in \Cref{sec:BPSdyons}. 
Since the generating function of these BPS
invariants merely corresponds to
$\exp \cF_1$ in topological string theory, it is relatively easy to study
additional structures, that go beyond
properties of the partition function in the large radius regime.  

Wall-crossing is one such topic. 
The charge of a BPS state varies over the moduli space of the
theory. When it aligns with the charge of another BPS state, these BPS
states can form a bound state. This gives a complication in the
counting of BPS states. There are so-called walls of marginal
stability in the moduli space, where the number of BPS states 
may jump. This is an important issue in $\cN=2$ theories, but it also
plays a major role in the counting of the above non-perturbative BPS states in
4-dimensional $\cN=4$ string theory. 

One may wonder what happens to the generating function of
such invariants under a crossing of a wall of marginal stability, and
whether this phenomenon has an interpretation in terms of 
the underlying Riemann surface. In \Cref{sec:wall-crossing} we
will answer these questions for quarter BPS dyons in $\cN=4$
theories.

\section{Compact curves of genus 1 and 2}\label{sec:compactcurves}
\index{automorphic form}

Let us begin with finding Calabi-Yau threefolds that are modeled on
compact curves of genus 1 and 2. The goal of this section is to
analyze their all-genus partition functions in the framework of
I-branes and $\cD$-modules. 

\subsection{Elliptic curve} 

A well-studied example is the geometry mirror to the total space
of a rank two bundle over a 2-torus
\be
\wt X:\ \mathcal{O}(-r)\oplus\mathcal{O}(r)\rightarrow T^2. 
\label{torus}
\ee
The latter has a description in toric geometry as gluing the toric
propagator to itself with a framing factor $r$. This factor changes the
intersection of $[T^2]$ with the 4-cycles that project onto $T^2$ into
$\pm r$ \cite{Vafa:2004qa}. Here we show how one can
use the free fermionic system living on the boundary of the non-commutative
plane to completely solve this model and
recover the existing results for the all genus topological string
amplitudes for this background.

The local $T^2$ model (\ref{torus}) has a simple interpretation in the
B-model obtained after 
mirror symmetry. Note that we can write this geometry as
a global quotient of $\C^* \times \C \times \C$. If we pick toric
coordinates $(e^{u},e^{v},e^{w})$, the identification is
$$
(u, v,w) \sim (u + t, v + ru, w - ru).
$$
This transformation is an affine transformation consisting of a shift
$(t,0,0)$ and a linear map
$$
A = \bem
1 & 0 & 0 \\
r & 1 & 0 \\
-r& 0 & 1 \\
\eem \in SL(3,\Z).
$$
The linear transformation $A$ is the monodromy of the fiber, when we
view this non-compact CY as a $T^3$ fibration. Mirror symmetry will
now replace the torus fibers with their duals, and the monodromy $A$
with the dual monodromy $A^{-T}$. So the B-model can be described as a
quotient of the dual coordinates given by
$$
(u, v,w) \sim (u + t - rv + rw, v, w).
$$

In order to map this B-model to the NS5-brane and finally the
I-brane, we have to perform one more T-duality on the combination
$v+w$. That coordinate is not touched by the action of the framing and
it will be subsequently ignored. If we relabel the coordinates as
$$
x=u,\qquad  y=v-w,
$$
we see that this gives indeed a $T^2$ curve, embedded as the zero
section $y=0$ in the geometry $\cal B$ defined as the quotient of $\C^*
\times \C$ by
\be
\label{toro}
(x,y) \sim (x + t - ry,y).
\ee

The A-model topological string partition function 
is computed as \cite{Vafa:2004qa, Bryan:2004iq}
$$
Z_{\textrm{top}}(t,\la) = e^{-t^3/6 r^2 \la^2}Q^{-1/24} \sum_R Q^{|R|} q^{r \kappa_R/2},
$$
where $Q=e^{-t}$ with $t$ the K\"ahler parameter of the torus and
$q=e^{-\lambda}$, whereas $|R|$ is the number of boxes of the Young
tableau $R$ and $\kappa_R = 2 \sum_{\Box \in R}
i(\Box)-j(\Box)$. After the mirror transformation $t$ becomes the 
modulus of the elliptic curve $T^2$.  The instanton part of $Z_{\textrm{top}}$
can be rewritten in the form 
\begin{align}
\label{T2}
& Z_{\textrm{qu}}(t,\la) = \\ 
& = \oint \frac{dy}{2 \pi i y} \prod_{n=0}^{\infty}
\left(1+y\,Q^{n+1/2}q^{r(n+1/2)^2/2}\right)
\left(1+y^{-1}Q^{n+1/2} q^{-r(n+1/2)^2/2}\right) \notag
\end{align}
which is familiar from \cite{dijkgraaf-mirror,Douglas:1993wy} in the
case $r=1$.  In this model the genus zero answer does not have
instanton contributions and 
so is given entirely by the classical cubic form $\cF_0(t) = -{1\over
  6r^2}t^3$, while at genus one the classical and quantum
contributions combine into 
$$ \cF_1(t) = -\log \eta(Q).
$$
The $g$-loop contributions $\cF_g$, for $g >1$ and $r>0$, incorporate
only quantum effects. They are quasi-modular forms of weight $6g-6$
that can be expressed as polynomials in the Eisenstein series $E_2(q)$,
$E_4(q)$ and $E_6(q)$, where 
\begin{align*}
 E_k(q) = 1 - \frac{2k}{B_k} \sum_{n=1}^{\infty} \frac{n^{k-1} q^n}{1-q^n} 
\end{align*}
and $B_n$ are the Bernoulli numbers.

In fact, it is well-known that this answer is reproduced by a chiral
fermion field with action \cite{Dijkgraaf:1996iy}
\be
\label{fermform}
S = \frac{1}{\pi} \int_{T^2} d^2x  \ \psi^\dagger\left(\bar{\partial} -
r\lambda \partial^2\right) \psi. 
\ee
We will re-derive this same answer from the fermionic perspective
we have developed in this thesis below.  For now note that
this action can be bosonized into \cite{Douglas:1993wy,Gross:1990st}
\be
\label{T2boson}
S = \frac{1}{\pi} \int_{T^2} d^2x \left( \frac{1}{2} \partial \phi
\bar{\partial} \phi  -
\frac{r \lambda}{6} (\partial \phi)^3 \right), 
\ee
which is closely related to the Kodaira-Spencer field theory on the
Calabi-Yau manifold.  This Kodaira-Spencer theory reduces to a free
boson $\phi$ on a cylinder, while the framing quantizes into an action
of the zero mode of the $W^3$ operator \cite{Aganagic:2003qj}
$$
W^3_0 = \oint dx \frac{(\partial \phi)^3}{3}.
$$
This implies that $W^3_0$ defines how to glue the torus
quantum mechanically,
$$
Z_{\textrm{top}}= \Tr \exp \left(-{r \la \over 2}~ W^3_0 \right),
$$
explaining (\ref{T2boson}). The action of $W^3_0$ is quadratic in the
fermions and therefore acts on the single fermion states. 

The topological string partition function (\ref{T2}) is obtained as the
fermion number zero sector. Including a sum over the $U(1)$ flux
gives the full fermion partition function that corresponds to the
I-brane. This can be thought of as a 
generalized Jacobi triple formula \cite{kanekozagier}. Adding the
classical contributions we obtain
\begin{align*}
& Z(v,t,\la) = \\
&= e^{-t^3/6r^2\la^2} Q^{-1/24}\prod_{n=0}^{\infty} (1+yQ^{n+1/2}q^{r(n+1/2)^2/2})(1+y^{-1}Q^{n+1/2} q^{-r(n+1/2)^2/2})  \\
& = \sum_{p=-\infty}^{\infty} y^p e^{-t^3/6r^2\la^2} e^{-pt^2/2 r \la} Q^{p^2/2-1/24} q^{rp^3/6-rp/24} 
  Z_{\textrm{qu}}(t+rp\la,\la) \\
& =  \sum_{p=-\infty}^{\infty} y^p Z_{\textrm{top}}(t+rp\la,\la).
\end{align*}
In the second line we have extracted a factor $e^{-t^2/2 r \la}$ out of $y$. This is the result of turning on flux in the I-brane set-up, and corresponds to the D4-brane tension on the BPS side. Notice that the combination $rp \in r \Z$. This is because $rp$ is the Poincar\'e dual of the four-cycle having intersection number $\pm r$ with $[T^2]$ . 
Hence this indeed reproduces formula (\ref{theta}) with an appropriate
choice of cubic form.  For $r=0$ this result reduces to the standard
Jacobi triple formula
\begin{align}\label{eqn:r=0result}
Z_{r=0}= {\theta_3(y,Q) \over \eta(Q)} = \sum_{n\in\Z} {Q^{n^2/2} y^n
  \over \eta(Q)}.
\end{align}

We now come to deriving (\ref{fermform}) from the perspective of this
thesis.  From the considerations in the last chapters it is clear that we have
a free fermion system living on $T^2$ with the {\it standard} action.
The only subtlety has to do with the fact that $T^2$ is at the boundary
of a non-commutative plane and as we will see this is crucial in
recovering (\ref{fermform}).  From (\ref{toro}) we see that $x\sim
x+t-ry$.  If we treated $y$ as commuting with $x$ we could set it to
$y=0$ and we have a copy of the torus.  But here we know that $y$ does
not commute with $x$.  So we have a free fermion on a torus where the
modulus is changed from
$$t \rightarrow t-ry.$$
The variation of $t$ can be absorbed into the fermionic action by the 
usual
Beltrami differential $\mu_{\bar z}^z =\delta t$:

$$
S = \frac{1}{\pi} \int_{T^2} d^2x  \ \psi^\dagger\left(\bar{\partial} +
\mu \partial\right) \psi.
$$ 
Here we need to substitute $\mu =\delta t = -r y$.  In the classical
case where $y$ is commuting, this would give $\mu =0$ and we get the
same system as the usual fermions.  However, since $x$ and $y$ do not
commute, we should view $y=\lambda \partial_x$ leading to $\mu = -ry = -r
\lambda \partial_x$.  Substituting this operator for $\mu$ in the above
action reproduces (\ref{fermform}).  We have thus re-derived the known
result for the topological string in this background from our
framework.

\subsection{Genus two curve}\label{sec:genustwo} 

An interesting generalization of the elliptic curve example is given
by a local Calabi-Yau geometry containing a genus two curve. Its mirror model can be
constructed using the topological vertex technology of
\Cref{sec:topvertex}. Although the 
vertex technology is able to deal with arbitrary toric curves, it is
instructive to see this explicit case in more detail.

\begin{figure}[t]
\begin{center}
\includegraphics[width=9cm]{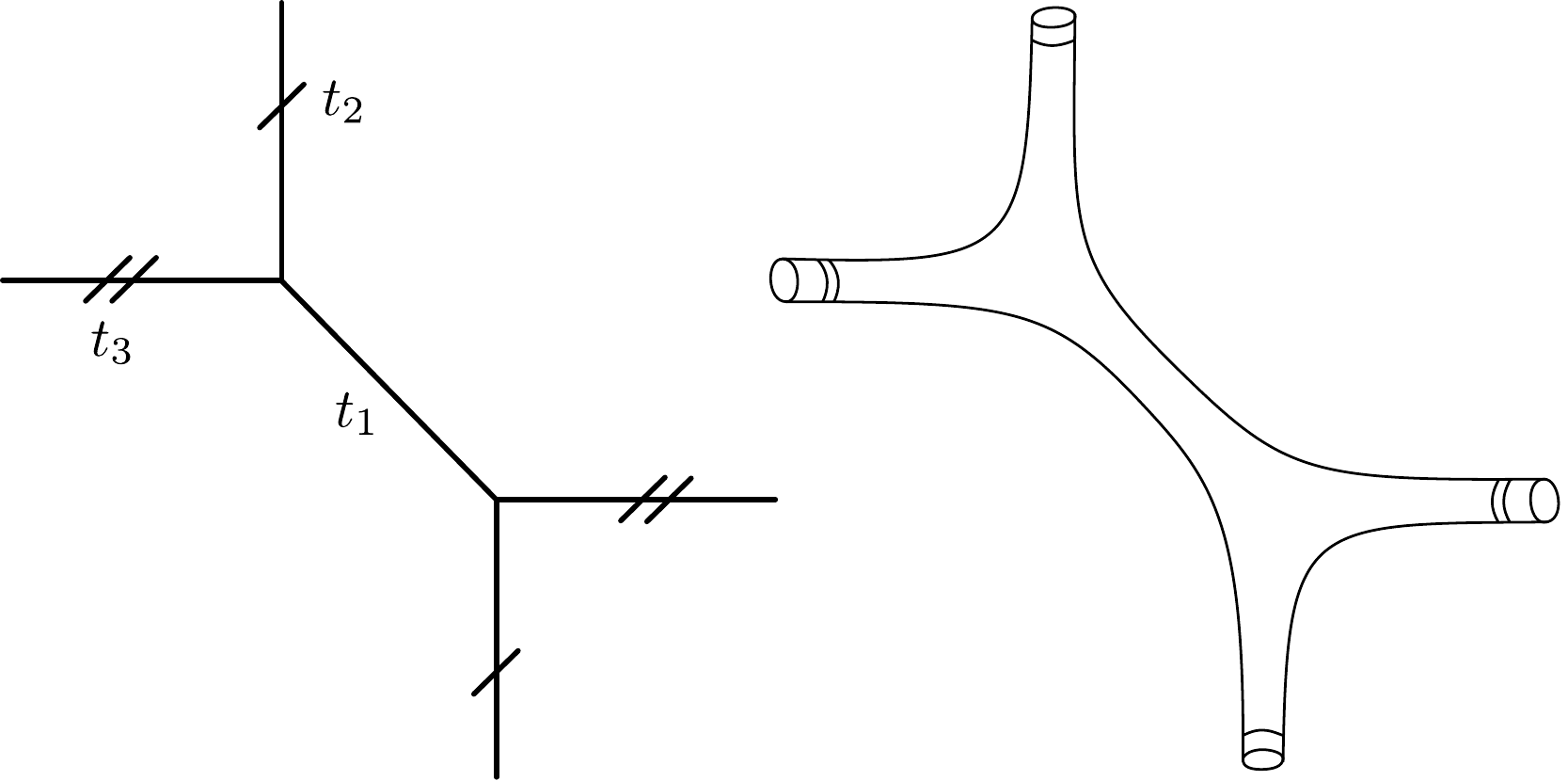}
\caption{The resolved conifold with
  identified legs (left) and its mirror (right). The parameters $t_1$,
  $t_2$ and $t_3$ parametrize the K\"ahler lengths
  of the toric diagram on the left.}
  \label{fig:genustwo}  
\end{center}
\end{figure}

Let us start in the A-model with the toric diagram of the resolved
conifold $\cO(-1) \oplus \cO(-1) \to \P^1$ (see \Cref{sec:resconifold})
and identify the two pairs
of parallel external legs, as shown in Fig. \ref{fig:genustwo}. In
this section we refer to this
geometry as $\tilde{X}$. The B-model geometry corresponding to
$\tilde{X}$ is a locally elliptic Calabi-Yau $X$, described by an
equation of the form $uv = H(x,y)$, where $H$ vanishes on a compact
genus two Riemann surface $\Sigma$.

This B-model geometry is well-studied in \cite{Hollowood:2003cv} as an
example of an  
elliptic threefold geometrically engineering a 6-dimensional gauge theory 
on $\R^4 \times T^2$. The prepotential of this gauge theory is computed  
as the A-model partition function of $\tilde{X}$. Since this is a 
topological vertex calculation, the all-genus partition function is known. 
Moreover, instanton calculus in the 6-dimensional gauge theory shows 
that it can be elegantly rewritten in terms of the equivariant elliptic
genus of an instanton moduli space. The equivariant parameter 
$q$ equals $e^{-\lambda}$ on the A-model side.

Explicitly, the A-model on $\tilde{X}$ can be expressed in topological 
vertices as
\begin{align*}
Z_{\textrm{qu}}(Q_1,Q_2,Q_3) = \sum_{R_1,R_2,R_3} Q_1^{R_1} Q_2^{R_2} 
Q_3^{R_3} (-)^{l_{R_1}+l_{R_2}+l_{R_3}}C_{R_1 R_2 R_3} C_{R_1^t R_2^t R_3^t}, 
\end{align*}
where $Q_i = \exp(-t_i)$ represent the exponentiated K\"ahler classes of the legs
with attached $U(N)$ representations $R_i$, and where $C_{R_1 R_2
R_3}$ is the topological vertex (see \Cref{sec:topvertex}). Notice that $C_{R_1 R_2 R_3}$ is
symmetric under permutations of the $R_i$, while in terms of the toric 
graph it is more natural to use the variables
\begin{align}\label{eqn:toricvariablesgenus2}
Q_{\sigma} := Q_1 Q_3, \qquad Q_{\rho} := Q_1 Q_2, \qquad 
Q_{\nu} = Q_1,
\end{align}
that exhibit the $\mathbb{Z}_2$ symmetry between $Q_{\sigma}$ and 
$Q_{\rho}$. Using these definitions
\begin{align*}
& Z_{\textrm{qu}}(q,\rho,\sigma,\nu) =  \sum_R
Q_{\rho}^{l_R} \prod_{\Box \in R} \frac{(1-Q_{\nu}
  q^{h(\Box)})(1-Q_{\nu}^{-1} q^{h(\Box)})} {(1-q^{h(\Box)})^2} \\
 & \times \prod_{k=1}^{\infty} \frac{(1-Q_{\sigma}^k Q_{\nu} q^{h(\Box)})
  (1-Q_{\sigma}^k Q_{\nu} q^{-h(\Box)}) (1-Q_{\sigma}^k Q_{\nu}^{-1}
  q^{h(\Box)}) (1-Q_{\sigma}^k Q_{\nu}^{-1} q^{-h(\Box)}) }
   {(1-Q_{\sigma}^k q^{h(\Box)})^2 (1-Q_{\sigma}^k q^{-h(\Box)})^2
     (1-Q_{\sigma}^k)}.  \notag
\end{align*}
And this may be rewritten as  \cite{Li:2004ef}
\begin{align}\label{eq:prodAmodel}
&Z_{\textrm{qu}}(q,\rho,\sigma,\nu) \,= \,\sum_{k \ge 0} Q_{\rho}^k \chi
((\mathbb{C}^2)^{[k]};Q_{\sigma},Q_{\nu})(q,q^{-1}) \, = \\
&
\prod_{\substack{ k,a \ge 0,\\ l > 0, \\ 2j \ge 0  }} 
\prod_{\substack{ b=-j, \\ c \in \Z }}^{j}  \left(
\frac{(1-Q_{\rho}^l Q_{\sigma}^a Q_{\nu}^{c-1} q^{2b+k+1})
(1-Q_{\rho}^l Q_{\sigma}^a Q_{\nu}^{c+1} q^{2b+k+1})}
{(1-Q_{\rho}^l Q_{\sigma}^a Q_{\nu}^{c} q^{2b+k+2})
(1-Q_{\rho}^l Q_{\sigma}^a Q_{\nu}^{c} q^{2b+k})}
\right)^{(k+1)C(la,j,c)} \notag
\end{align}
with $b=-j, -j+1, \ldots,j-1,j$ and $q=e^{-\lambda}$, whereas the
coefficients $C(a,j,c)$ are related to the equivariant elliptic genus
of $\C^2$ in the following way
\begin{align*}
\chi(\C^2,y,p,q) =& \prod_{n \ge 1}
  \frac{(1-yp^nq)(1-y^{-1}p^nq^{-1})(1-yp^nq^{-1})(1-y^{-1}p^nq)}
{(1-p^nq)(1-p^nq^{-1})(1-p^nq^{-1})(1-p^nq)} \notag \\
=~& \sum_{a, 2j\ge 0}\sum_{ c\in \mathbb{Z}}
  C(a,j,c) p^a (q^{2j} + q^{2(j-1)} + \ldots + q^{-2j}) y^c. 
\end{align*}
Starting with the IIA background 
$TN_1 \times \tilde{X}$ and going backwards through the duality chain, we find 
ourselves in the I-brane set-up   
on $\R^3 \times T^4 \times \R^2 \times S^1$. The genus two curve $\Sigma$ is 
holomorphically embedded in the abelian surface $T^4$ by the Abel-Jacobi map. 
The I-brane is the intersection of a D4-brane wrapping $\R^3 \times \Sigma$
and a D6-brane wrapping $T^4 \times \R^2 \times S^1$. The aim of this
section is to give an interpretation of the above A-model
result on $\tilde{X}$ in the I-brane picture.

\subsubsection{The case $\lambda=0$}

As a result of the duality chain, we expect that the 1-loop free energy 
$\cF_{1,\textrm{top}}$ of the topological A-model equals the free energy 
$\cF_{1,\textrm{boson}} = - {1\over 2} \log \det \Delta_\Sigma$
of a chiral boson on $\Sigma$. Another sum over the lattice of momenta 
should then result in the chiral fermion determinant. Since not only
the A-model partition function, but also the partition  
function of chiral bosons on a genus two surface is known, we can perform 
an explicit check of these conjectures.

Singling out the $\lambda^0$-part of the  A-model partition function
(\ref{eq:prodAmodel}), yields the sum
\begin{align*}
\cF_{1,\textrm{qu}}(\rho,\sigma,\nu)= \tilde{c}(kl,m)
\prod_{k,l,m} \log \left( 1-e^{2 \pi i (k\rho + l\sigma +
      m\nu)} \right),
\end{align*}    
where the coefficients $\tilde{c}(kl,m)$ are related to the Fourier coefficients $C(a,j,c)$ as
\begin{align}
\tilde{c}(kl,m) = - \sum_{j \in (\mathbb{Z}/2)_{\ge 0}} \sum_{b=-j}^{j}  
\Big[ \big( & 2b^2-\frac{1}{12} \big) \big(
  C(kl,j,m+1) \, + \label{eqn:ctilde} \\
& + 
 C(kl,j,m-1) \big) 
- \left(4b^2 + \frac{5}{6} \right) C(kl,j,m) \Big]. \notag
\end{align}

Remarkably, the same relation can be found by rewriting the
elliptic genus of a $K3$ surface, which is the unique weak Jacobi form
of index 1 and weight 0. This elliptic genus has an expansion
\begin{align*}
\chi(K3,\tau,z) = \sum_{h \ge 0, m \in \mathbb{Z}} 24 \, c(4h-m^2) e^{2 \pi
  i(h\tau+mz)},
\end{align*}
and can be represented as an integral over
the equivariant elliptic genus of $\C^2$: 
\begin{align}\label{eq:ellgenK3}
& \chi(K3,y,p) \notag \\
&= - y^{-1} \int_{K3} x^2 \prod_{n\ge 1} 
\frac{(1-yp^{n-1}q^{-1})(1-yp^{n-1}q)(1-y^{-1}p^{n}q^{-1})
(1-y^{-1}p^{n}q)}
{(1-p^{n-1}q^{-1})(1-p^{n-1}q)(1-p^{n}q^{-1}) (1-p^{n}q)} \notag \\
&= - \int_{K3} x^2 \left( \frac{y+y^{-1}-q-q^{-1}}{A(x)A(-x)} \right) \chi(\C^2,y,p,q),
\end{align}
with $q = e^x$ and $A(x) :=
\sum_{k\ge0}\frac{x^k}{(k+1)!}$. Writing out this equation in terms of
the $K3$ and $\C^2$ coefficients, reveals exactly equation
(\ref{eqn:ctilde}) where $\tilde{c}(kl,m)$ is exchanged with $c(kl,m)
= c(4kl-m^2)$.  

This identification implies that $Z_{1, \textrm{qu}} =
\exp(\cF_{1,\textrm{qu}})$ corresponds to the 24th root of the generating
function of the elliptic genera of  symmetric products of $K3$'s:
\begin{align}
\label{eq:genustwo}
\left( Z_{1, \textrm{qu}}(\rho, \sigma, \nu) \right)^{24} & =  \sum_N  e^{2 \pi i N \sigma}
\chi_{\rho,\nu} \left((K3)^N/S_N  
\right) =\nonumber \\
& =
\prod_{\substack{k >0 ,l \ge 0,\\ m \in \Z}} \!\! \left( 1-e^{2 \pi i (k\rho + l\sigma +
      m\nu)} \right)^{-24 c(4kl-m^2)}.
\end{align}
When the $K3$ surface is realized as an elliptic fibration, with 24
points on the elliptic base where the fibration degenerates in the
simplest possible way, we can think about the $K3$ surface as consisting of 
24 local $TN_1$-spaces. The above result then motivates us to relate
$Z_{1,\textrm{qu}}$ to one such $TN_1$-factor.   

Furthermore, $Z_{1, \textrm{qu}}$ is closely related to the generating
function 
\begin{align}
 e^{-\pi i
  (\rho + \sigma+v)/12} \!\! 
\prod_{(k,l,m)>0} \!\! \left( 1-e^{2 \pi i (k\rho + l\sigma +
      m\nu)} \right)^{-c(4kl-m^2)},
\end{align}
where $(k,l,m) >0$ means $k, l \ge 0$, $ m \in \Z$, but $m<0$ when
$k=l=0$. The first terms on the second line of 
equation~(\ref{eq:genustwo}) have a clear interpretation as classical 
contributions to the genus 1 topological string amplitude
(proportional to the K\"ahler class $t =\rho+ \sigma + \nu$). We
 loosely refer to the above generating function
as the total genus 1 partition function $Z_{1,
  \textrm{top}}$ for the genus 2 Calabi-Yau. 

The 24th power of this topological partition function $$Z_{1,
  \textrm{top}}(\Omega)^{24} = \frac{1}{\Phi_{10} (\Omega)}, \quad
\textrm{with} \quad \Omega =
\bem \rho & \nu \\ \nu & \sigma \eem$$ 
is well-known to both mathematicians and physicists. Mathematically, it is 
characterized by its nice
transformation properties under the symplectic group $Sp(2,\Z)$.  
The form $\Phi_{10}(\Omega)$  transforms as 
\begin{align*}
  \Phi_{10} \left(  g(\Omega) \right) =\left( \textrm{det} (C \Omega +
    D )\right)^{10} \Phi_{10} \left(  \Omega \right) 
\end{align*}
under $Sp(2,\Z)$-transformations. It is thus $Sp(2,\Z)$-automorphic of weight
10. In fact, it is the unique form with this property
\cite{igusa1,igusa2}.  In relation to the topological string
amplitude, we should of course interpret $\Omega$ as the period matrix
of the genus 2 
surface. 

The automorphic form $1/\Phi_{10}$ is familiar in string theory as the
partition function of 24 chiral bosons
\cite{Moore:1986rh,Belavin:1986tv}. More precisely, it appears as 
the holomorphic part of the (worldsheet) genus 2 bosonic string
partition function, or equivalently, as the left-moving
partition function of the heterotic string wrapping a genus 2
surface. Both partition functions describe 26 chiral bosons in the
light-cone gauge. Effectively, their partition function thus captures
24 chiral bosons.\footnote{There is a subtlety here. Performed computations of
  this partition function show that, unlike for the partition function
 $1/\eta^{24}$ of a heterotic string wrapping a 2-torus, the
 ghost determinants and the light-cone directions do not completely
 cancel out in the genus 2 configuration (see the remarks in
 \cite{Dabholkar:2006bj}). However, we expect that there exists a 
 gauge choice in which there is a clean interpretation in
 terms of 24 bosons. We thank E. Verlinde for a discussion on this point.}


So after adding the classical contributions to
$\cF_{1,\textrm{qu}}(\tilde{X})$, we conclude that the total 1-loop partition
function $Z_{1, \textrm{top}} = \exp (\cF_{1, \textrm{top}})$ of the B-model topological string on
$X$ equals  the  
partition function $Z_{\textrm{boson}}$ of a single boson: 
\begin{align*}
Z_{\textrm{boson}}(\rho,\sigma,\nu) = e^{\cF_{1,\textrm{top}}(\rho,\sigma,\nu)}.
\end{align*}

In order to find the contribution to the all-genus partition function
for small $\lambda$, we have
to consider $\cF_{0,\textrm{top}}$ as well. In the B-model on $X$ its
second derivative has a simple interpretation: it is just the period
matrix $\Omega_{ij}$ of the genus two curve $\Sigma$. In terms of the
mirror map, 
these periods will have classical contributions linear in $\rho$,
$\sigma$ and $\tau$, and quantum corrections determined by
$Z_{\textrm{qu}}(\tilde{X})$.  
We will write these down in the next paragraph. Right now, let us
conclude with 
\begin{align*}
Z_{\textrm{fermion}}(\rho,\sigma,\nu)=\sum_{p_1,p_2 \in \Z} e^{\pi i p_i
  \Omega_{ij} p_j} e^{\cF_{1,\textrm{top}}}(\rho,\sigma,\nu).
\end{align*}

\subsubsection{Automorphic properties}

Knowing the full instanton partition function (\ref{eq:prodAmodel})
makes it possible to examine the $\lambda$-corrections to $\cF_{1,\textrm{top}}$
explicitly. In fact, let us start more generally with the
Gopakumar-Vafa partition function (\ref{eqn:GVpartitionfunction})
\begin{align*}
Z_{\textrm{qu}} = \prod_{d \in H_2} \prod_{m \in \Z} \prod_{k=-m}^m
\prod_{n=1}^{\infty} (1-q^{k+n} Q^{d})^{(-1)^{m+1}n \textrm{GV}^{m}_d}.
\end{align*}
In order to get the $g$-loop free energies we note that
\begin{align}\label{eq:expanding}
&\log \prod_{k \ge 1} (1- Y q^{k+l})^k =\notag \\
 &\quad = - \frac{1}{\lambda^2}\mbox{Li}_3(Y) + {\frac{1}{2}(l^2 -
  \frac{1}{6})} \log(1-Y) - \lambda^2 \left( \frac{1}{240} - 
\frac{l^2}{24} + \frac{l^4}{24} \right) \mbox{Li}_{-1}(Y)
 - \ldots \notag \\
&\quad =: - \sum_{g\ge 0} \lambda^{2g-2} P_{2g}(l) \sum_{n \ge 1} n^{2g-3} Y^n,  \notag
\end{align}
where the degree $2g$ polynomials $P_{2g}(l)$ are defined through the last
equality. Hence
\begin{align*}
\cF_{\textrm{qu}} = - \sum_{g \ge 0}  \lambda^{2g-2} \sum_{d \in
  H_2} \sum_{m \in \Z} \sum_{k=-m}^m 
 (-1)^{m+1}  P_{2g}(k)\, \textrm{GV}^{m}_{d} \, \sum_{n=1}^{\infty} n^{2g-3}
 \left(Q^{d} \right)^n.
\end{align*}
Making this expansion for the genus two Calabi-Yau $X$ reveals that
the coefficients $$ c_g = \sum_{m \in \Z}\sum_{k=-m}^m
(-1)^{m+1} P_{2g}(k)\, \textrm{GV}^{m}_{d} $$ are the Fourier coefficients
of Jacobi forms $J_g(q,y) = \sum_{k,l} c_g(k,l) q^k y^l$ of weight
$2g-2$ and index 1. More precisely, we can write the $\cF_{g}$'s as
\begin{align*} 
\cF_{g}(\lambda;Q_{\rho},Q_{\sigma},Q_{\nu}) &= -
\sum_{k,l,m} c_g(kl,m) \sum_{n\ge 1} n^{2g-3} (Q_{\rho}^k
Q_{\sigma}^l Q_{\nu}^m)^n \\
&=- \sum_{N > 0} Q_{\rho}^N \sum_{kn =N} n^{2g-3}
\sum_{l \ge 0, m} c_g(kl,m) Q_{\sigma}^{ln} Q_{\nu}^{mn}  \\
&= - \sum_{N > 0} Q_{\rho}^N \sum_{kn=N} N^{2g-3} \sum_{b=0}^{k-1} k^{2-2g} \, J_g
\left(\frac{n \sigma+b}{k},n \nu \right) \\
&= - \sum_{N > 0} Q_{\rho}^N T_{g,N} (J_g),
\end{align*}
where $T_{g,N}$ are Hecke operators acting on Jacobi forms of weight
$2g-2$. This implies  that all $\cF_{g,\textrm{top}}$'s are lifts of
Jacobi forms, and 
therefore almost automorphic forms of $O(3,2,\Z)=Sp(4,\Z)$ 
\cite{Borcherds:1995}.
\index{Jacobi form}

\subsubsection{Interpretation in the duality chain}
\index{M5 brane}

First of all, notice that the partition function of $\tilde{X}$ can be
build out of topological vertices, and as such is known to have an
interpretation in terms of chiral bosons and fermions
\cite{Aganagic:2003db,Aganagic:2003qj, Bouchard:2007ys}. The duality chain
elucidates these  
observations: the chiral fermions can be
identified with the intersecting brane fermions. Moreover, the $B$-field
on the D6-brane makes it necessary to treat these fermions as
non-commutative objects, which gives an explanation for the nontrivial  
transformation properties in \cite{Aganagic:2003qj}.

In terms of the gauge theory picture we can just refer to 
\cite{Hollowood:2003cv}. Here it is shown that the six dimensional gauge theory on $TN_1
\times T^2$ can be engineered with matrix model techniques, revealing
$\Sigma$ as the Seiberg-Witten curve, whose period matrix equals the
second derivative of $\cF_{0,\textrm{top}}$. 

Finally, the automorphic properties of $\cF_{\textrm{top}}$ seems to fit in
best in the M5-brane frame of the duality chain. Recall that 
S-duality relates IIB on $TN_1 \times X$ to a NS5-brane wrapping
around $TN_1 \times \Sigma$ in the background $TN_1 \times T^4 \times
\R^2$. This lifts to a M5-brane in M-theory on
$TN_1 \times T^4 \times \R^2 \times S^1$. Since the M5-brane partition
function is expected to be an automorphic form of $O(3,2,\Z)$
\cite{Dijkgraaf:1998xr}, this perspective offers a physical reason for
the automorpic properties.
 
Actually, we know exactly which Jacobi forms enter: $J_0 =
\phi_{-2,1}$ is the unique Jacobi form of weight $-2$ and index 1,
$J_1 = - \frac{1}{12} \phi_{0,1} = - \frac{1}{24} \chi(K3,q,y)$ as we
encountered before, $J_2 = \frac{1}{240} E_4 \phi_{-2,1}$, $J_3 =
-\frac{1}{6048} E_6 \phi_{-2,1}$ and $J_4 =\frac{1}{172800} E_4^2
\phi_{-2,1}$ etc. Interestingly, these can all be defined as twisted
elliptic genera of $TN_1$ in the sense that (compare with
(\ref{eq:ellgenK3})) 
\begin{align*}
&J_g(q,y) \, =\\
& -y^{-1} \int_{TN_1} x^{4-2g} \prod_{n\ge 1} 
\frac{(1-yp^{n-1}q^{-1})(1-yp^{n-1}q)(1-y^{-1}p^{n}q^{-1})
(1-y^{-1}p^{n}q)}
{(1-p^{n-1}q^{-1})(1-p^{n-1}q)(1-p^{n}q^{-1}) (1-p^{n}q)},
\end{align*}
coinciding with the M5-brane point of view. This results longs for a
two dimensional conformal field theory interpretation.

\section{Quarter BPS dyons} \label{sec:BPSdyons}
\index{BPS dyon}

The next few sections center on the weight 10 Igusa cusp form
\begin{align*}
\Phi_{10}(\rho, \sigma, \nu) = pqy
\prod_{(k,l,m)>0} \left(1-p^k q^l y^m \right)^{24 c(4kl-m^2)},
\end{align*}
with $p = \exp( 2 \pi i \rho)$, $q = \exp( 2 \pi i \sigma)$ and $y =
\exp( 2 \pi i\nu)$, that we
encountered in equation~(\ref{eq:genustwo}) when analyzing the
semi-classical contribution to the topological partition function on
the genus 2 Calabi-Yau. Let us explain why it plays an important
role in the counting of quarter BPS states in an $\cN=4$
compactification of string theory down to 4 dimensions.

One perspective to study 4-dimensional $\cN=4$ string theory is in the
framework of type II 
theory as a $K3\times T^2$ compactification. Another 
duality frame is heterotic string theory compactified on a
6-torus. The heterotic compactification is 
described by an $\cN=4$ supergravity multiplet coupled to 22
additional vector multiplets. In this compactification a
vector multiplet 
consists of a single vector field and six real scalars, while the
supergravity multiplet contains the metric, the heterotic axion-dilaton field
and six graviphotons. Altogether there are 28 $U(1)$
gauge fields, and hence 28 electric and 28 magnetic conserved
charges. Both the magnetic charge vector $P$ and the electric charge
vector $Q$ are elements in the lattice $\Gamma^{22,6}$. They may be
combined into a charge matrix  
\begin{align}
\bem Q \\ P \eem \in \Gamma^{22,6}
\oplus \Gamma^{22,6},   
\end{align}
on which heterotic S-duality acts as a $SL(2,\Z)$ transformation. From
the type IIB point of view the heterotic S-duality group is visible as a
geometric T-duality group that acts as the mapping class group on the
$T^2$-factor.  

Half BPS states in $\cN=4$ string theory are relatively easy to
describe. They either carry a purely electric or 
magnetic charge, so that they can be counted in the perturbative
regime of heterotic string theory in terms of 24 bosonic oscillators
in the left-moving sector. This yields   
\begin{align}\label{eqn:half-BPS}
 d(Q) = \oint d \sigma \frac{e^{-\pi i Q^2 \sigma}}{\eta^{24} (\sigma)} 
\end{align}
states with electric charge $Q$, and an analogous formula for the
number $d(P)$ of magnetic states with charge $P$ (now using $\rho$ as
integration variable). These formulas suggest  
that counting half BPS states in $\cN=4$ string theory is in some way related to
24 free bosons on a genus 1 surface (compare for example with
formula~(\ref{eqn:r=0result}) and with the $K3$ and the $\R^4$ example in
Chapter~\ref{chapter2}).

Quarter BPS states are more complicated to analyze. As observed in
\cite{Dijkgraaf:1996it}, it is natural to introduce the genus 2
period matrix 
\begin{align*}
  \Omega = \bem \sigma & \nu \\ \nu &
      \rho \eem 
\end{align*}
as this decouples into two genus 1 period matrices when $\nu \to
0$. The complex structure parameters of the resulting tori describe the
half-BPS electric and magnetic states. The degeneracy formula  
\begin{align}\label{eqn:quarter-BPS}
  d(Q,P) = (-1)^{Q \cdot P+1} \oint d \Omega \frac{e^{ -\pi i ( Q^2
      \sigma + 2 Q\cdot P \nu + P^2 \rho)}}{\Phi_{10} (\Omega)}  
\end{align}
indeed reduces to a product of the electric with the magnetic half-BPS
formula (\ref{eqn:half-BPS}) when the limit $\nu \to 0$ is taken. Notice that the degeneracies $d(Q,P)$ should be invariant under the symmetries of the charge-lattice. Indeed, they are functions $d( 1/2\, Q^2, Q \cdot P, 1/2 \, P^2)$ in terms of the integer invariants $1/2 \, Q^2$, $1/2 \,P^2$ and $Q \cdot P$ on the even lattice $\Gamma^{22,6} \oplus \Gamma^{22,6}$.  The
physical relation of quarter BPS states to the genus 2 surface with
period matrix $\Omega$ is not
immediately clear at all, though. As we explain in a little bit, this relation
can be understood through a number of string dualities.  

Recent years have seen a lot of
progress in the understanding of the BPS spectrum of 4-dimensional
$\cN=4$ string theory. For 
example, the above proposal for the  
generating function of BPS states has been extended to so-called
CHL orbifolds \cite{Jatkar:2005bh}. In
the heterotic string perspective such a CHL compactification is
obtained by orbifolding $T^4 \times S^1 \times
\tilde{S}^1$ by a $\Z_N$ symmetry, that is generated by a product of
an internal symmetry together with an order $N$ translation along the
circle $\tilde{S}^1$. The original Dijkgraaf-Verlinde-Verlinde partition
function as well as its CHL extension are now fairly well-understood. 
They have passed all consistency  checks performed so far, such as
$SL(2, \Z)$ invariance (on which we come back in \Cref{sec:wall-crossing})
and agreement with black hole physics
\cite{LopesCardoso:2004xf,David:2006yn,Banerjee:2008pu}.\footnote{See also \cite{Shih:2005uc, Gaiotto:2005hc, 
 David:2006ji, David:2006ru,David:2006ud, Dabholkar:2006bj, Sen:2007vb, Dabholkar:2007vk,
 Cheng:2007ch, Mukherjee:2007nc, Dabholkar:2008zy,
 Cheng:2008fc, Banerjee:2008yu, Cheng:2008kt, Banerjee:2008ky,
 Murthy:2009dq}.}

\begin{figure}[h!]
\begin{center}
\includegraphics[width=10cm]{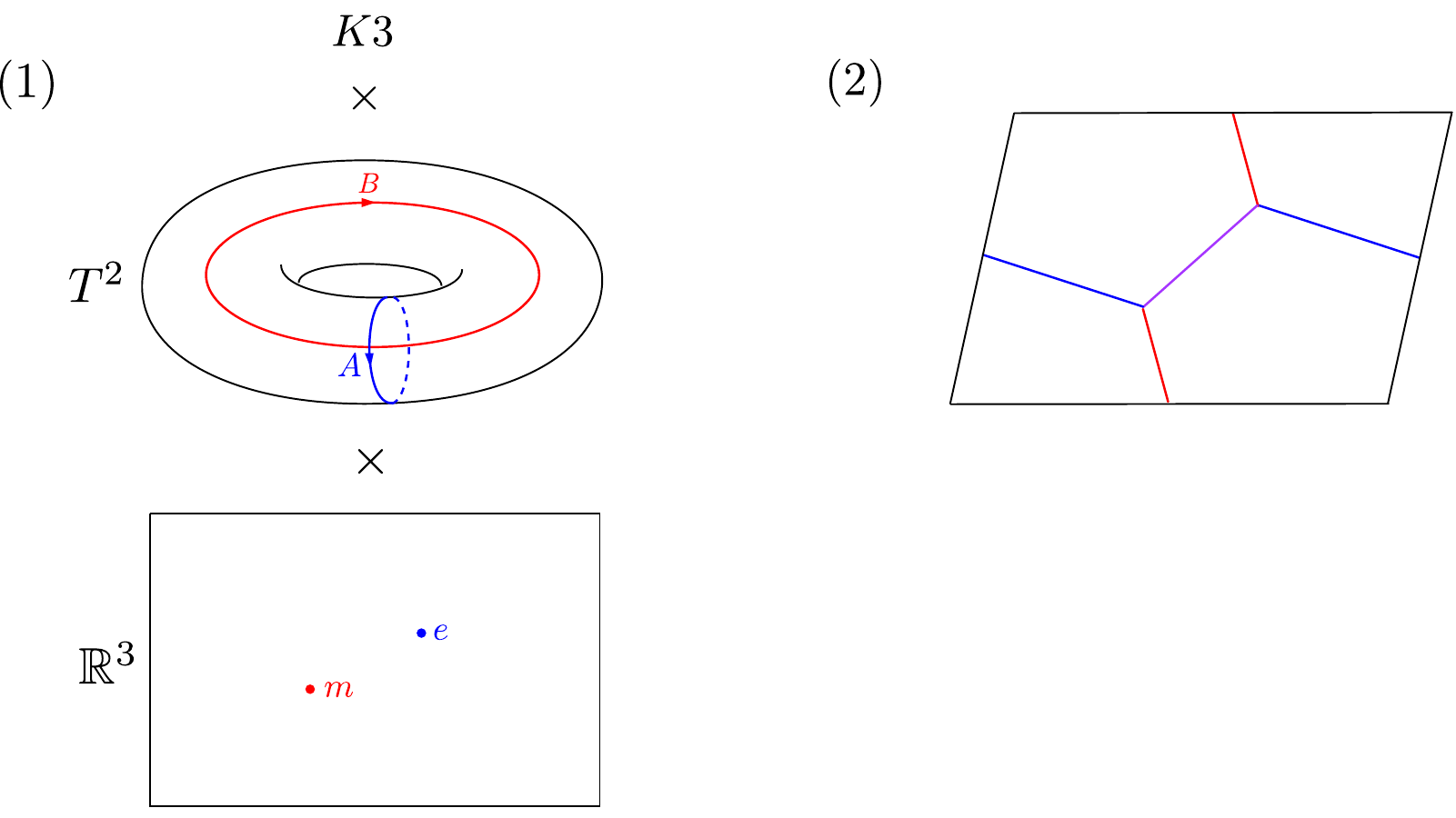}
\caption{(1): BPS dyons in an $\cN=4$ compactification of string
  theory on $K3 \times T^2$ correspond to D4/NS5 brane wrappings over the internal
K3-manifold times a circle in $T^2$. Depending on which
1-cycle in the $T^2$ is wrapped the resulting 4-dimensional BPS
particle will carry electric or magnetic charge. (2): When both 1-cycles
in the $T^2$ are wrapped it is more efficient for the branes to
recombine into a network wrapping the $T^2$.}
  \label{fig:dyons}  
\end{center}
\end{figure}

Schematically, the genus two surface occurs in the dyon counting problem in the
following way, illustrated in Fig. \ref{fig:dyons}
\cite{Gaiotto:2005hc,Dabholkar:2006bj}. Consider type 
IIB string theory compactified on $K3\times T^2$. Half BPS states
(that are pointlike in the non-compactified directions in $\R^4$) in
this frame are represented as bound states of D5/NS5-branes
that wrap the $K3$-fold, with D3-branes that wrap some 2-cycles in the $K3$
and with D1/F1-branes. The total bound states furthermore wraps one
cycle of the $T^2$, depending on whether the half BPS state is
electric or magnetic. A quarter BPS state carries both electric and
magnetic charges. It can be represented by a network of D5 and NS5-brane bound
states that is embedded in the 2-torus. An example of such a network is shown
in Fig.~\ref{fig:dyons}. It consists of two three-vertices
that are known as three-string junctions. Compactifying the time
direction, in order
to calculate a partition function, and lifting into M-theory we obtain
Euclidean M-theory compactified on $K3\times T^4$. The BPS dyon is now represented as an M5 brane wrapping $K3$
times a genus two Riemann surface $\Sigma$ that is holomorphically embedded in
$T^4$. Electric BPS states wrap one A-cycle on $\Sigma$ while magnetic
BPS states wrap the other one.  
Since an M5-brane that wraps a $K3$-surface is dual to a heterotic string,
we furthermore recover the partition function $1/\Phi_{10}$ of 24
chiral bosons on a
genus two surface (although there are some unsolved subtleties
\cite{Dabholkar:2006bj,Banerjee:2008yu}).

This is very similar to the M5-brane description of topological string
theory on a non-compact Calabi-Yau threefold as alluded to in
 \Cref{sec:compactcurves}.  
In fact, when we view the $K3$-surface 
as 24 copies of a $TN_1$-space and the M5-brane wrapping this
$K3$-surface as 24 M5-branes each wrapping a copy of $TN_1$, the set-up reduces
in type IIA to the intersecting brane background
\begin{align}\label{eqn:genus2Ibrane}
\textrm{IIA:} \quad  \R^3 \times \left( \Sigma \subset T^4 \right) \times \R^2
\end{align}
with 24 D4-branes wrapping $\R^3 \times \Sigma$ and a
D6-brane wrapping $T^4\times \R^2$, that intersect on the genus 2
surface $\Sigma$. This relates the appearance of the automorphic form
$\Phi_{10}$ as the semi-classical amplitude $\cF_1$ in topological
string theory and as a dyon index in 4-dimensional $\cN=4$ string
theory physically. 
   

In this first section about quarter BPS dyons we explain the duality
between the $\cN=4$ dyons, string webs and the genus two surface in
detail.
We pay extra attention to the dependence on
background moduli, as this will be important in the description of
wall-crossing in the next section.
In \Cref{The Five-brane Network} we review and extend the
results of A.~Sen in \cite{Sen:1997xi} in detail and
discuss how the $1/4$-BPS dyons are realized as a periodic
network of effective strings in type IIB frame at arbitrary moduli. In
\Cref{The Riemann Surface} we review and extend the results
of S.~Banerjee et al. in \cite{Banerjee:2008yu} by going to Euclidean
M-theory and analyze 
the Riemann surface wrapped by the M5 brane that makes up the
$1/4$-BPS dyon. In particular we analyze the complex structure of the
surface and its relation to the stability of the dyon states. On
general grounds and from earlier results we expect the 
Riemann surface to degenerate in a certain way when the moduli cross a
wall of marginal stability \cite{Sen:2007vb,Cheng:2007ch}. We analyze
wall-crossing in great detail in \Cref{sec:wall-crossing}.

\subsection{The five-brane network}
\label{The Five-brane Network}
\index{string network}

Following S.~Banerjee, A.~Sen and Y.~Srivastava \cite{Banerjee:2008yu}, in this
subsection we consider $1/4$-BPS dyons made up from a type IIB
$T^2$-compactified network of effective strings which are bound states
of $(p,q)$ strings and  $K3$-wrapped five-branes. Working in the limit
of large $K3$ and thus heavy five-branes and using the supersymmetry
condition of the network \cite{Sen:1997xi}, we will write down
explicit expressions for the shape and size of the network with a
given range of values of the IIB axion-dilaton field $\la$ and the
complex structure $\tau$ of the 2-torus. After that we briefly discuss
how the network is 
realized at generic values of moduli, while leaving the details to
\Cref{Moduli Space as the Dual Graph}.

First consider type IIB string theory compactified on the product of a $K3$ manifold and a 2-torus which we shall call $T^2_{\mathrm{(IIB)}}
$, and two effective strings wrapping the two homological cycles of
the torus. Each  
effective string is a bound state of F1 and D1 string together with NS5 
and D5 branes wrapped on $K3$. 

To be more specific, let's consider the following charges. Suppose 
we have the $Q$ effective string, which is a bound string of a $
(n_1,n_2)$ string together with a $K3$-wrapped $(q_1,q_2)$ five-brane, wrapping the $A$-cycle of the $T^2_{\mathrm{(IIB)}}$. Wrapping the 
$B$-cycle is what we call the $P$  effective string, which is a bound 
string 
of a $(m_1,m_2)$ string together with a $K3$-wrapped $(p_1,p_2)$ five-brane. The three T-duality invariants corresponding to this charge 
configuration are given by
\be \label{charges_initial}
Q^2 = 2\sum_{i=1}^2 n_i q_i,\;\;P^2 =  2\sum_{i=1}^2 m_i p_i,\;\;Q
\cdot P = \sum_{i=1}^2 ( m_i q_i + n_i p_i)\;.
\ee

In the limit of large $K3$, the tension of the $Q$- and $P$- string are given by
\be\label{central_charge_vector}
T_Q =  \, q_1-\bl q_2,\quad T_P = p_1 - \bl p_2
\ee
rescaled by a factor of the volume of $K3$ in 10-dimensional Planck unit $V_{K3}^{\mathrm{(P)}}= V_{K3}\la_2$.
Here $V_{K3}$ denotes the the volume of $K3$ in string unit, and $-\bl = -\la_1 + i\la_2 $ is the axion-dilaton of the type IIB theory. In particular, the string coupling is given by $g_s = \la_2^{-1}$.  Similarly, we will denote by $-\bt = -\tau_1 + i \tau_2 $ and $R_B^2 \tau_2 $ the complex structure 
and the area of the type IIB torus $T^2_{\mathrm{(IIB)}}$ respectively. 

Using the above convention,  the $SL(2,\Z)\times SL(2,\Z)$ symmetry of the 
theory acts as
\begin{align*}
\tau \to \frac{a\tau+b}{c\tau+d},\;\; \bem Q \\ P \eem \to \bem a &b \\ c & 
d\eem \bem Q\\ P \eem,\;\; \g =\bem a &b \\ c & d\eem \in SL(2,\Z)
\;,
\end{align*}
and independently
\be\label{IIB_S_dual}
\la \to \frac{a'\la+b'}{c'\la+d'},\;\; \bem \G_1 \\ \G_2 \eem \to \bem a' &b' \\ c'& d'\eem \bem \G_1\\ \G_2 \eem,\;\; \g'=\bem a' &b' \\ c' & d'\eem\in 
SL(2,\Z)
\ee
for all $(\G_1,\G_2)$ strings or five-branes.
The second symmetry is the type IIB S-duality, while the first symmetry 
is the modular transformation of the type IIB torus $T^2_{\mathrm{(IIB)}}$, 
which is mapped to the S-duality of the heterotic string under string duality.

It will turn out to be useful to organize the above complex structure of the  torus $T^2_{\mathrm{(IIB)}}$ and the type IIB axion-dilaton field in terms of 
the following $2\times 2$ symmetric real matrices
\begin{align*}
{\cal M}_\tau = \frac{1}{\tau_2} \bem|\tau|^2 & \tau_1 \\ \tau_1 & 1\eem,\quad
{\cal M}_{\la} = \frac{1}{\la_2} \bem|\la|^2 & \la_1 \\ \la_1 & 1\eem\;,
\end{align*}
which transforms as ${\cal M}_\tau \to \g{\cal M}_\tau \g^T$ and ${\cal M}_{\la} \to \g'{\cal M}_{\la} \g'^T$ under the above $SL(2,\Z)\times SL(2,\Z)$ transformation. Furthermore, we will use the following standard metric on the space of $2\times 2$ 
symmetric real matrices $X$ 
\be\label{Lorenztian_metric}
\Vert X \Vert^2  ={\textrm{\small det}X}\;,
\ee
such that both ${\cal M}_{\tau},{\cal M}_{\la} $ have unit space-like length.

To make the analysis more explicit, let us assume a certain orientation of the string network, given by
$q_1 p_2 - p_1 q_2 >0$. To ensure the irreducibility of the string network made of 
the $(q_1,q_2)$ and the $(p_1,p_2)$ five-branes, we will further 
require $q_1p_2 -q_2 p_1 = 1$, namely that the corresponding $2\times 2$ matrix
$$
\G =\bem q_1&q_2 \\ p_1 & p_2\eem
$$
is an $SL(2,\Z)$ matrix \cite{Dabholkar:2006bj}.
The generalization to the charges with $\G \in GL(2,\Z)$, including 
the opposite orientation of the string network with $q_1 p_2 - p_1 q_2 =-1$, is a straightforward modification of the following discussion and will not be separately discussed here\footnote{
 It simply involves exchanging $\tau$ and $\bt$ in equations
 (\ref{eq_kaehler_parameter}) - (\ref{theta}),
 (\ref{mass1}), (\ref{coord_1}), (\ref{J_integral}), (\ref{eq_Omega}),
 (\ref{area_J_integral}).}
. 

\begin{figure}
\centering
\includegraphics[width=\textwidth]{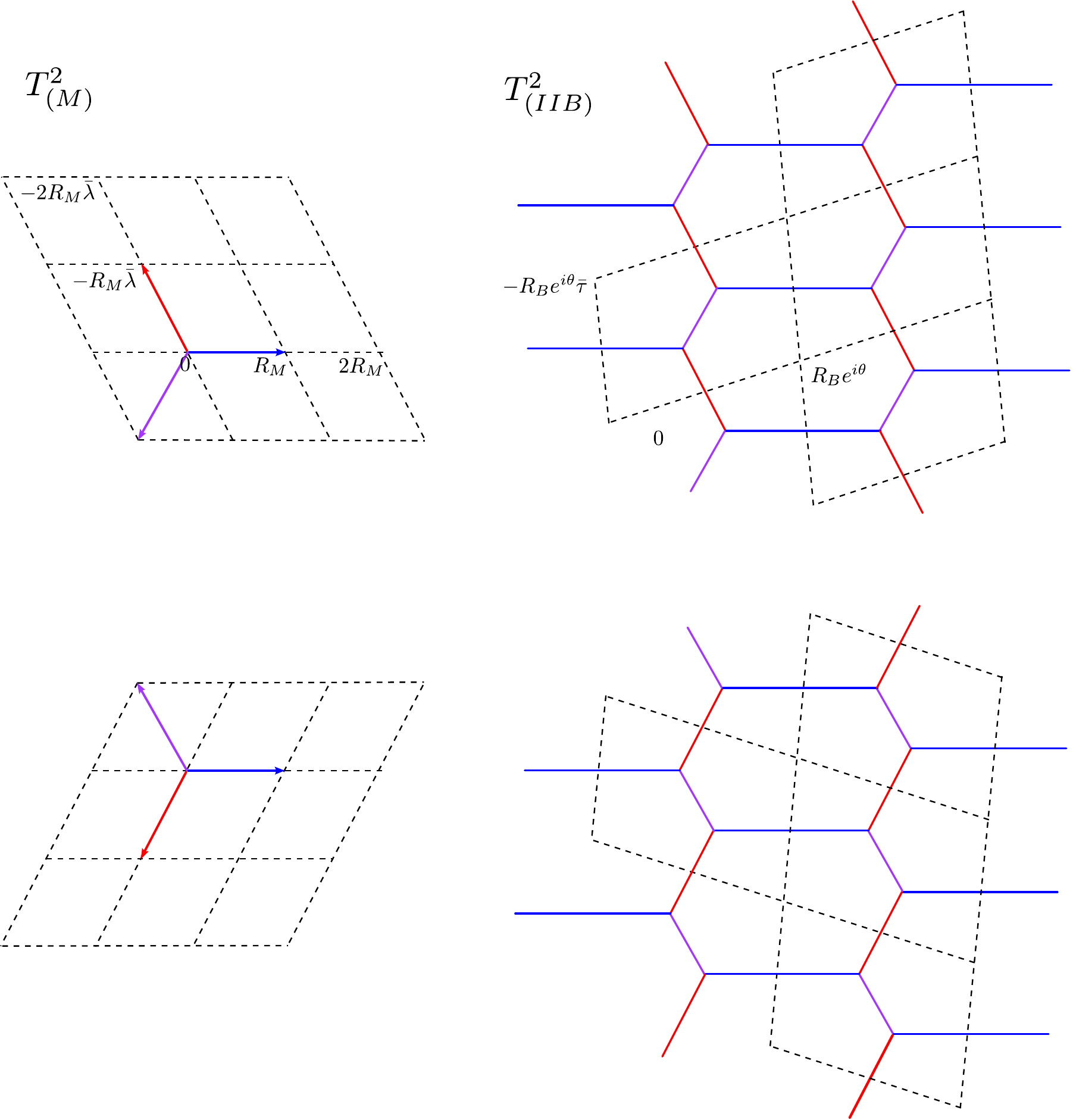} 
\caption{\label{network_fig}An example of the effective string network, with $(q_1,q_2)=(1,0)$ and $(p_1,p_2)=(0,1)$. Which one of the network is realized depends on the sign of the off-diagonal component of the moduli vector ${Z}$.}
\end{figure}

Simple kinematic consideration, or relatedly supersymmetry, requires 
that the three lines meeting at a vertex satisfy the following constraints \cite{Sen:1997xi}. The angles formed by the three legs meeting at a vertex in the periodic string network must be the same as the angles formed by the three tension vectors (\ref{central_charge_vector}) of the corresponding charges in a complex plane. 
Two examples are shown in Fig.~\ref{network_fig}. 

As we shall see shortly, how the supersymmetric network will be realized depends on the background moduli of the theory.  For the time being, let us focus on the one specific case depicted in the first figure in Fig.~\ref{network_fig}. In this case the statement about the angles simply means the following. If we view the compactification torus as $\C/R_B(\Z-\bt\Z)$ and draw the network on the same complex plane, the three vectors  
$\ell_{1,2,3}\in \C$ in this periodic network are given by 
\be\label{network_1}
\ell_1 = t_1 (T_Q+ T_P),\quad\ell_2 =t_2 T_P,\quad\ell_3 = \,t_3 T_Q,
\ee
where the tension vectors $T_{Q,P}$ are given in (\ref{central_charge_vector}) and $t_{1,2,3}\in\R_+$ are the length parameters given by the background moduli in a way we will now describe.

The fact that this network fits in the geometric torus $T^2_{(IIB)}$
means the length parameters satisfy 
\be\label{eq_kaehler_parameter}
\bem t_1+ t_3 & t_1 \\ t_1 & t_1+ t_2\eem\bem T_Q \\ T_P \eem = 
\bem t_1+ t_3 & t_1 \\ t_1 & t_1+ t_2\eem\G \bem 1 \\ -\bar{\la}\eem =
e^{i\theta} R_B \bem 1 \\ -\bt\eem
\ee
for some angle $\theta$ as shown in Fig.~\ref{network_fig}. The obvious
fact that $$ (T_Q+ T_P) \bar\ell_1 + T_P \bar\ell_2+ T_Q \bar\ell_3
\in \R_+$$ then gives 
\be\label{theta}
\theta = \text{Arg} 
(T_Q-\tau T_P).
\ee
The mass of the string network, which is given by the sum of the product of the length of the legs in the type IIB torus and their respective tension, is then given by 
\be
\label{mass1}
M_{\mathrm{IIB}} =(T_Q+ T_P) \bar\ell_1+T_Q \bar\ell_2 + T_P \bar\ell_3  =  R_B V_{K3}\la_2\,\rvert \,T_Q-\tau T_P\,
  \lvert.
\ee

Furthermore, by first solving (\ref{eq_kaehler_parameter}) for the simplest case with $\G = \mathds 1_{2\times 2}$ and considering other solutions related to it by a type IIB S-duality (\ref{IIB_S_dual}), we obtain the expression for the lengths of the three different legs in the string network
\begin{align*}
\bem t_1+ t_3 & t_1 \\ t_1 & t_1+ t_2\eem = \sqrt{\frac{R_B^2\tau_2}{\la_2}}\,\frac{{\cal M}_\tau^{-1} + (\G^{-1})^T {\cal M}_{\la} \G^{-1}}{\Vert {\cal M}_\tau^{-1} + (\G^{-1})^T {\cal M}_{\la} \G^{-1}\Vert}.
\end{align*}
Like in equation~(\ref{eqn:toricvariablesgenus2}) this shows that the
lengths of the string network are naturally expressed in terms of variables
$t_1$, $t_1 + t_2$ and $t_1+t_3$. As will become clear in
equation~(\ref{period_matrix}), the above matrix corresponds as well
to the period matrix of the genus 2 surface. The string network is thus
equivalent to the toric graph in \Cref{sec:genustwo}.  

While the quantity on the right-hand side depends on our specific choice among charges lying on the same T-duality orbit and furthermore its derivation is only valid in the part of the moduli space with $V_{K3} \gg 1$, in what follows we shall see how this quantity can naturally be written as an T-duality invariant expression which is well-defined for general values of moduli.  

From the 4-dimensional macroscopic analysis we know the BPS mass of a dyon should be expressed in 
terms of the charges and the moduli in a specific way \cite{Cvetic:1995bj,Cvetic:1995uj,Cheng:2008fc}. 
Especially, in the heterotic frame it depends on the right-moving 
charges only, which can be combined into the following T-duality invariant matrix
\begin{align*}
\Lambda_{Q_R,P_R}  = \bem Q_R \cdot Q_R & Q_R \cdot P_R
\\Q_R \cdot P_R & P_R \cdot P_R
\eem 
\end{align*}
and further combined with the heterotic  axion-dilaton into the 
matrix
\begin{align}
\label{mod_vec}
{Z} &= 
 \frac{1}{\tau_2} \bem 1 & -\tau_1 \\ -\tau_1 & 
|\tau|^2\eem + \frac{1}{\Vert\Lambda_{Q_R,P_R} \Vert}\bem P_R \cdot P_R& -Q_R \cdot P_R
\\-Q_R \cdot P_R & Q_R \cdot Q_R \eem,
\end{align}
which is again invariant under T-duality transformation.

In terms of these $2\times 2$ matrices, the mass in string frame is 
given by 
\begin{align}\notag
M^2_{\mathrm{IIB}} &= V_{K3}\, R_B^2\, \la_2^2 \bigg( |Q_R -\bt P_R|^2 +2 
\tau_2\,\Vert\Lambda_{Q_R,P_R}\Vert\bigg)\\ \label{mass_sugrav}
&=  V_{K3}\, R_B^2\, \tau_2 \la_2^2\, \Vert \Lambda_{Q_R,P_R}\Vert\, \Vert{Z}
\Vert^2.
\end{align}

Comparing with the mass formula for the string network (\ref{mass1}), 
we can read out the expression for $Q_R,P_R$ 
\begin{align*}
\Vert\Lambda_{Q_R,P_R}\Vert  = V_{K3} \la_2 = V_{K3}^{\mathrm{(P)}},\quad\frac{\Lambda_{Q_R,P_R} }{\Vert\Lambda_{Q_R,P_R}\Vert }= \G {\cal M}_\la^{-1} \G^T,
\end{align*}
and thus
\begin{align*}
{Z} 
={\cal M}_\tau^{-1} + (\G^{-1})^T {\cal M}_{\la} \G^{-1}.
\end{align*}

From this we see that the moduli vector ${Z}$ has the following two physical roles in the type IIB supersymmetric string network. First its length gives the mass of the network as in (\ref{mass_sugrav}). Furthermore its direction dictates the relation between the lengths of various legs of the network by
\be\label{length_parameters_1}
\bem t_1+ t_3 & t_1 \\ t_1 & t_1+ t_2\eem  = \sqrt{\frac{R_B^2\tau_2}{\la_2}}\, \frac{{Z}}{\Vert { Z}\Vert}.
\ee

But there is clearly a problem with this formula. As the reader might have noticed, the above formula is devoid of a geometric meaning when one or more of the length parameters $t_i$ is negative. To take the simplest example, while the diagonal terms of the matrix $Z$ are manifestly positive (\ref{mod_vec}), the off-diagonal term can be of either sign. It means that when the entries of $Z$ fail to be all positive, for example, the network we have just described cannot exist. 

The solution to this problem is the following. As we have mentioned earlier, there are more than just one possible way to realize a supersymmetric string network with given $4D$ charges. For illustration let's now consider the following example. 
Writing 
$
Z =\big(\begin{smallmatrix} z_{1} & z \\ z & z_2 \end{smallmatrix} \big)
$
and assume $z_1,z_2 > -z >0$ such that the network we discussed above does not exist, we will now see that the network is realized as a periodic honeycomb network with three legs given by
\be\label{network_2}
\ell_1 = t_1 (T_Q- T_P),\quad\ell_2 =-t_2 T_P,\quad\ell_3 = \,t_3 T_Q
\;.
\ee
Repeating the same analysis as before we obtain the same expression for the angle $\theta$ which measures the ``tilt'' of  the network (\ref{theta}) and the mass of the network (\ref{mass1}), but now the length parameters are given instead by 
 \be\label{eq_kaehler_parameter_2}
\bem t_1+ t_3 & -t_1 \\-t_1 & t_1+ t_2\eem  =
\sqrt{\frac{R_B^2\tau_2}{\la_2}}\, \frac{{Z}}{\Vert {\cal
    Z}\Vert}.
\ee
It is then easy to see that the above network (\ref{network_2}), shown in the second figure in Fig.~\ref{network_fig},  does exist for the range of moduli space $z_1,z_2 > -z >0$. 

In general, as will be discussed in detail in \Cref{Moduli Space as the Dual Graph}, for any arbitrary point in the moduli space, exactly one network which is given by effective strings with charges $aQ+bP$ and $cQ+dP$ wrapping the cycles $dA-cB$ and $-bA+aB$, will be realized. Here we again use $A$ and $B$ to denote the $A$- and $B$-cycle of the compactification torus $T^2_{\mathrm{(IIB)}}$. 
And the integers
$$
\g= \bem a & b \\ c&d\eem \in GL(2,\Z) 
$$
are determined by the value of moduli, which is given by the values of $\la,\tau$ in the five-brane system we consider. Recall that the requirement that the inverse of an element in $GL(2,\Z)$ is again an element of the same group means that the matrix $\g$ must have determinant $\pm 1$.

In more details, the periodic network will consist of three legs given by
\begin{align*}
\ell_1 =t_1\,\big(  (a+c)T_Q + (b+d)T_P \big),\quad\ell_2 =t_2\, (cT_Q+dT_P),\quad\ell_3 = \,t_3\, (a T_Q+ bT_P)
\end{align*}
 with length parameters given by
\be\label{length_general}
\bem t_1+ t_3 & t_1 \\tau_1 & t_1+ t_2\eem  = \sqrt{\frac{R_B^2\tau_2}{\la_2}}\, \frac{(\g^{-1})^T{Z}\g^{-1}}{\Vert {Z}\Vert}.
\ee
As will be explained in more details in \Cref{Moduli Space as the Dual Graph}, for a given point in the moduli space, the integral matrix $\g$ has to satisfy the requirement that the above equation has a  solution with $t_{1,2,3} \in \R_+$.

\subsection{The Riemann surface}
\label{The Riemann Surface}
\index{hyper-K\"ahler structure}

Following the idea of \cite{Gaiotto:2005hc} and adopting the approach of \cite{Banerjee:2008yu}, in this subsection we study the holomorphic embedding of a Riemann surface wrapped by the M5 brane in Euclidean M-theory which makes up the $1/4$-BPS dyons of the theory.  In particular, following \cite{Banerjee:2008yu} we write down the period matrix of such a surface for generic values of the moduli of the theory, and discuss the relationship between the degeneration of the surface and the crossing of walls of marginal stability where some dyon states might become unstable.

In order to compute the dyon partition function of the compactified type IIB theory discussed in the previous Subsection, it is necessary to go to the Euclidean spacetime with a Euclidean time circle. 
Now recall that type IIB compactified on a circle is equivalent to M-theory 
compactified on a torus, which we will refer to as the ``M-theory torus'' $T^2_{\mathrm{(M)}}$, by a T-duality transformation followed by a lift to eleven dimensions. In particular, letting the eleventh-dimension circle to have asymptotic radius $R_M$, the complex moduli and the area of the M-theory torus $T^2_{\mathrm{(M)}}$ are given by the type IIB axion-dilaton as $-\bl$ and $R_M^2 \la_2$.

In other words, in order to discuss the dyon partition function we consider M-theory compactified down to $\R^3$ on the internal manifold $K3 \times T^2_{\mathrm{M}}\times T^2_{(\mathrm{IIB})}$. 
Since the configuration we will be considering is the M5 brane wrapping the whole $K3$, 
we will now focus on the $ T^2_{\mathrm{(M)}} \times T^2_{(\mathrm{IIB})}$ 
factor whose moduli play the most important role in the rest of the Chapter. Clearly, it can 
be thought of as a space of the form $\C^2/{\mathbf{\Lambda}}$, where the two complex planes can be taken to be the complex planes associated with the tori  $ T^2_{\mathrm{(M)}} $ and $ T^2_{(\mathrm{IIB})}$ respectively. Writing the coordinate 
of the two complex planes as $z_1=x_1+i y_1$ and $z_2=x_2+i y_2$, 
the lattice ${\mathbf{\Lambda}}$ is generated by the following four vectors in 
\(\R^4\) parametrized by $(x_1,y_1,x_2,y_2)$: 
\begin{align}\notag
e_1 &= R_M\,(1,0,0,0)\quad\\ \notag e_2 &= R_M\,(-\re\bl,-\im\bl,0,0)\\ \notag
e_3 &=R_B\,(0,0,\re \,e^{i\theta},\im \,e^{i\theta})\\
\label{coord_1}
e_4 &=
R_B\,(0,0,-\re \,e^{i\theta}\bt ,-\im \,e^{i\theta}\bt )
.
\end{align}
For convenience we have chosen the coordinates of $\R^4$ such that the $Q$-string lies along the $x_2$-axis. See Fig.~\ref{network_fig}.

A priori there is no reason to require the two tori $T^2_{\mathrm{(M)}}$ and $T^2_{(\mathrm{IIB})}$  
be orthogonal to each other. A non-zero inner product in $\R^4$ between the vectors $\{e_1,e_2\}$ and $\{e_3,e_4\}$(\ref{coord_1})
corresponds to turning on time-like Wilson lines for the $B$- and $C$- 
two-form fields along the $A$- and $B$-cycles of of compactification torus $T^2_{(\mathrm{IIB})}$ in 
the original type IIB theory. But since they are absent in the Lorentzian type IIB theory we started with, in most of the following discussion we will assume that such a cross-term is absent.

\begin{figure}[h]
\centering
\includegraphics[width=7cm]{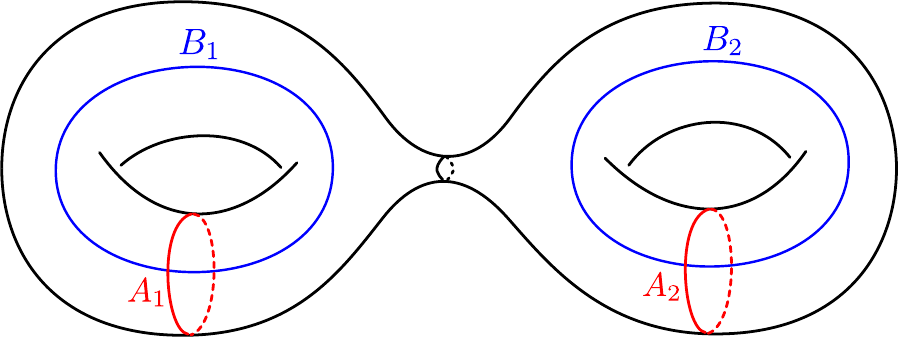} 
\caption{The degeneration of the genus two surface described in
  equation (\ref{degenerate_area}).}\label{genus2degenerating} 
\end{figure}

After describing the M-theory set-up we now turn to 
the dyons in the theory.  The type IIB effective string network 
discussed in \Cref{charges_initial} now becomes a genus two Riemann 
surface $\S$ inside $T^4$ upon compactifying the temporal direction and going to the M-theory frame, which has the effect of fattening the network in Fig.~\ref{network_fig}.  As usual, we would like to choose a canonical basis for the homology cycles of the Riemann surface $\S$ such
 that the $A$- and $B$-cycles have the following canonical
intersections:
\be\label{intersection_A_B}
A_a \cap B_b = \delta_{ab},\quad A_a \cap A_b = B_a \cap B_b =0,\quad a,b=1,2.
\ee
We now choose the basis cycles $A_{1,2}$ and $B_{1,2}$ as shown in 
Fig.~\ref{genus2degenerating}. Beware that they are not directly related to the $A$- and $B$-cycles of the tori $T^2_{\mathrm{(IIB)}}$ and $T^2_{\mathrm(M)}$.

From the charges of the network, which translate in the geometry into the homology classes of the two-cycle  in $T^4$ wrapped by the M5 brane, we see that the Riemann surface $\S$ defines a lattice inside ${\mathbb R}^4$, with generators related to those of $\Lambda$ in the following way

\be\label{jacobian}
\bem \oint_{A_1} dX \\ \oint_{A_2} dX  \eem  =  \G \bem e_1 \\ e_2 \eem,\quad
\bem \oint_{B_1} dX \\ \oint_{B_2} dX  \eem = \bem e_3 \\ e_4 \eem
.
\ee
In the above formula, $dX=(dx_1,dy_1,dx_2,dy_2)$ is the pullback on the Riemann surface $
\S$ of the one-forms on $\mathbb R^4$ in which $\S$ is embedded\footnote{For convenience and given that there's little room for confusion, here and elsewhere in this section we will not distinguish in our notation for a form in $\R^4$ and its pullback along the embedding map (\ref{embedding_map_jac}) onto the Riemann surface. }. It is easy to see that the this lattice is identical to the lattice $\Lambda$ (\ref{coord_1}) generated by $e_{1,\dotsi,4}$ which defines the spacetime four-torus in $\R^4$, as long as we restrict to the M5 brane charges with  $|\text{\small det}\G|= g.c.d.(Q\wedge P) =1$. We shall say more about the role of this lattice for the Riemann surface $\S$ shortly, but for that we will first need to discuss the complex structure of this surface.

The spacetime supersymmetry requires that the genus two Riemann surface to be holomorphically embedded in the spacetime $T^4$. To find the period matrix of the Riemann surface, we are interested in finding the complex structure of $\R^4$ which is compatible with the holomorphicity of $\S$. By definition this complex structure will then determine the complex structure of the Riemann surface. 
Using the natural flat metric on $\R^4$, its volume form is given by 
$$
\textrm{vol}=dx_1 \wedge dx_2 \wedge dy_1 \wedge dy_2, 
$$
and the space of self-dual two-forms in $\R^4$ will then be spanned by the following three two-forms
\begin{align*}
f_1 &= 
dx_1 \wedge dy_1-dx_2 \wedge dy_2\\ 
f_2 &= 
 dx_1 \wedge dy_2+dx_2 \wedge dy_1\\ 
f_3 &=
  dx_1 \wedge dx_2+dy_1 \wedge dy_2.
\end{align*}
Recall that this 3-dimensional space corresponds to the $S^2$ worth of complex structures of the hyper-K\"ahler space $\R^4$ in the following way. For a given complex structure two-form $\Upsilon$, the space of self-dual two-forms is spanned by the $(2,0)$, $(1,1)$ and $(0,2)$ form $\Upsilon=\Upsilon_1 + i\Upsilon_2 $, $J$ and $\bar{\Upsilon}=\Upsilon_1 - i\Upsilon_2 $, where $J$ is the K\"ahler form. From 
\begin{align*}
\Upsilon\wedge \bar{\Upsilon} = J\wedge J = {\textrm{vol}} \\
\Upsilon\wedge \Upsilon = \Upsilon\wedge J =0,
\end{align*}
we conclude that $J,\Upsilon_1,\Upsilon_2$ are mutually perpendicular in the pairing $\frac{\cdot \wedge \cdot }{\textrm{vol}}$ for two-forms and 
$\Upsilon_1\wedge\Upsilon_1=\Upsilon_2  \wedge\Upsilon_2 =\frac{1}{2}
J\wedge J$. 

If the Riemann surface $\S$ is holomorphically embedded in $\R^4$ with respect to the complex structure $\Upsilon$, the following condition is satisfied
\begin{align*}
\int_\S \Upsilon = 0.
\end{align*} 
To find the complex structure $\Upsilon$ compatible with the holomorphicity of $\S$ we therefore have to find a vector $J$ in the 3-dimensional space of self-dual two-forms, such that the plane normal to it is the plane of all two-forms $f$ satisfying $\int_\S f =0$. This plane will then be the plane spanned by $\Upsilon_1$ and $\Upsilon_2$. From (\ref{jacobian}) we can compute the value of $f_{1,2,3}$ integrated over the surface $\S$ using the Riemann bilinear relation. From the results
\begin{align}
\int_\S {f_1} &= 0  \notag \\
\int_\S{f_2} &= -R_BR_M\; \im\left( e^{-i\theta} \big((q_1-\bl q_2)
  -\tau (p_1 -\bl p_2 )\big)\right)=0 \notag \\
\int_\S{f_3} &= R_BR_M\; \re\left( e^{-i\theta} \big((q_1-\bl q_2) -\tau
  (p_1 -\bl p_2 )\big)\right) \label{J_integral} \\
&= R_BR_M\;\vert (q_1-\bl q_2) -\tau (p_1 -\bl p_2 )\vert, \notag
\end{align}
 we see that the correct complex structure of $\R^4$ that gives the holomorphic embedding of the surface $\S$ is as follows 
\begin{align} \notag
\Upsilon&= f_1 + i f_2 = w_1 \wedge w_2,\quad w_1 = dx_1 + idx_2\;,\; w_2 = dy_1 + idy_2 \\ \label{holom_one_form}
J &= f_3\;.
\end{align}
In particular, the above one-forms $w_{1},\,w_2$ form a basis of the holomorphic one-forms on the Riemann surface when pulled back along the embedding map. 
Notice that, although the above expression for the complex structure $\Upsilon$ seems to be independent of the charges and moduli, this is not quite true since we have hidden the dependence in our choice of coordinates $x_{1,2},y_{1,2}$ of $\R^4$ (\ref{coord_1}). More explicitly, one can view the complex structure as charge- and moduli-dependent through our definition of the angle $\theta$
(\ref{theta}).

Now we are ready to discuss the embedding of $\S$ into the spacetime tori $ T^2_{\mathrm{(M)}}\times T^2_{(\mathrm{IIB})}$.
Recall that the Jacobian variety of a genus $g$ Riemann surface $\S^{\mathrm{(g)}}$ is given by the complex torus ${\cal J}(\S^{\mathrm{(g)}})=\C^g/\Lambda(\S^{\mathrm{(g)}})$, where $\Lambda(\S^{\mathrm{(g)}})$ is the lattice generated by the $2g$ vectors 
\be\label{jacobian_lattice}
\begin{array}{ccc}
&( \oint_{A_1} w_1, \dotsi, \oint_{A_1} w_g)&\\
&\vdots&\\
&( \oint_{A_g} w_1, \dotsi, \oint_{A_g} w_g)&\\
&( \oint_{B_1} w_1, \dotsi, \oint_{B_1} w_g)&\\&\vdots&\\
&( \oint_{B_g} w_1, \dotsi, \oint_{B_g} w_g)&
\end{array}
\ee
and $\{w_1,\dotsi,w_g\}$ is a basis of one-forms on the Riemann surface which are holomorphic with respect to its given complex structure. The following map, the so-called Abel-Jacobi map, then gives a holomorphic embedding of the Riemann surface $\S^{\mathrm{(g)}}$ into its Jacobian $\Lambda(\S^{\mathrm{(g)}})$:
\be\label{embedding_map_jac}
\varphi: \S^{\mathrm{(g)}} \to {\cal J}(\S^{\mathrm{(g)}}),\quad \varphi(P)= \bem \int_{P_0}^P w_1, & \dotsi& ,\int_{P_0}^P w_g\eem,
\ee
where $P_0$ is a given arbitrary point on $\S^{\mathrm{(g)}}$. Notice that the Jacobian is defined in such a way that the above map is well-defined, namely that the images are independent of the path of integration.
In the case of our genus two surface $\S$, using the holomorphic one-forms $w_1,w_2$ given in (\ref{holom_one_form}),  from (\ref{jacobian}) we see that $\Lambda = \Lambda(\S)$, and therefore the Jacobian of the surface ${\cal J}(\S)$ is naturally identified with the spacetime $T^4$. The Abel-Jacobi map (\ref{embedding_map_jac}) therefore provides us with an explicit holomorphic embedding of the M5 brane Riemann surface $\S$ into the spacetime torus, as was suggested in \cite{Gaiotto:2005hc}.  

After discussing the complex structure and the embedding of the surface, now we are ready to compute its normalized period matrix $\W$. Consider two holomorphic one-forms 
$(\hat w_1\;\hat w_2 ) = (w_1\;w_2) V$, where $V$ is a real $2\times 2$ matrix,
such that 
\be\label{solve_V_eq1}
\bem \oint_{A_1} \hat w_1 &  \oint_{A_1} \hat w_2 \\ \oint_{A_2} \hat w_1 &   \oint_{A_2} \hat w_2 \eem =
\bem 1 &0\\ 0&1 \eem.
\ee
The (normalized) period matrix $\W= \re\W + i\, \im\W$ is then the symmetric $2\times 2$ matrix given by
\begin{align*}
\W
=\bem \oint_{B_1} \hat{w}_1 &  \oint_{B_1} \hat{w}_2  \\ \oint_{B_2}
\hat{w}_1& \oint_{B_2} \hat{w}_2   \eem =  \bem \rho & \n \\ \n & \s
\eem, \quad \rho,\s,\n \in \C \;.
\end{align*}

Comparing (\ref{solve_V_eq1}) and the first part of (\ref{jacobian}) one can easily obtain the explicit solution for the real matrix $V$. Integrating the resulting $\hat{w}_{1,2}$ over the B-cycles then gives $\re\W$ satisfies
\be\label{eq_Omega}
\im\W\, \G\, \bem 1 \\ -\bar{\la}\eem = e^{i\theta} \frac{R_B}{R_M} \bem 1 \\ -\bt\eem\;.
\ee
Up to a multiplicative factor involving the M-theory radius, this is exactly the same equation (\ref{eq_kaehler_parameter}) that the matrix of the length parameters $t_{1,2,3}$ of the type IIB string network satisfies.
We therefore conclude that the period matrix of the genus two curve wrapped by the supersymmetric M5 brane configuration is given by
\be\label{period_matrix}
\im \W =\sqrt{\frac{R_B^2\tau_2}{R_M^2\la_2}}\, \frac{{Z}}{\Vert { Z}\Vert},\quad\re\W=0.
\ee

Note that the direction of the above vector in $\R^{2,1}$ is given by the 
moduli vector $Z$ (\ref{mod_vec}), while the length is given  by 
the ratio of the area of the two spacetime tori. And the requirement $
\Vert\im\W\Vert\gg 1$ for rapid convergence of the partition function is 
the physical requirement that we work in the low temperature limit in the type IIB frame in which $R_B^2 \tau_2 \gg 
R_M^2 \la_2$.

The fact that the period matrix is purely imaginary is really a consequence of the fact that 
our two spacetime tori $T^2_{\mathrm{(M)}}$ and 
$T^2_{\mathrm{(IIB)}}$ are orthogonal to each other, which in turn reflects the 
absence of temporal Wilson lines in the original type IIB setup. If these 
Wilson lines are turned on, the real part of the period matrix will 
instead be
\be\label{re_omega}
\re\,\W=  \bem C_{t1} & B_{t1} \\ C_{t2} & B_{t2}
\eem\G^{-1}=(\G^{-1})^{T} \bem C_{t1} & C_{t2} \\ B_{t1} & B_{t2}\eem,
\ee
where $B_{t1}$,$B_{t2}$,$C_{t1}$,$C_{t2}$ denote the background two-form $B$- and $C$-fields along the $A$- and $B$-cycles of the torus $T^2_{\mathrm{(IIB)}}$ and the temporal circle in type IIB. The extra condition on these Wilson lines $\mathrm{Re}\W= (\mathrm{Re}\W)^T$ could be thought of as a part of the supersymmetry condition, since if the Wilson lines do not satisfy this condition, the holomorphic embedding of the M5 brane world volume into the spacetime four-torus is not possible with respect to the given complex structure $\Upsilon$ (\ref{holom_one_form}). Put in another way, turning on the temporal Wilson lines for the two-form fields will generically change the complex structure of the surface $\S$, with exception when (\ref{re_omega}) is satisfied. But as mentioned before, we will not consider this possibility further.
 
 Finally we would like to comment on the fact that
 the surface area of the holomorphically embedded genus two 
surface $\S$ is simply given by  
 \be\label{area_J_integral}
 A_{\mathrm\S} =   \int_{\S} J = \frac{i}{2}\int_{\S} (w_1\wedge \bar w_1+w_2\wedge \bar w_2) = R_M R_B\,\big\vert T_Q-\tau T_P  \big\vert
 \ee
 as already computed in (\ref{J_integral}). 
 As expected, the surface area is related to the mass of the BPS object 
in the following simple way
\begin{align*}
 A_{\mathrm{\S}}= \frac{R_M}{V_{K3} \la_2} M_{\mathrm{IIB}} = \frac{R_M}{V_{K3}^{\mathrm{(M)}}} 
M^{\mathrm{(M)}}
\end{align*}
where the quantities with the superscript $\scriptstyle{(M)}$ denote the 
quantities in the M-theory unit. 

This relation between the mass and the area of the corresponding 
Riemann surface suggests a geometric way of understanding the walls of 
marginal stability, defined as the subspace in the moduli space where the BPS masses of the components of a potential bound state sum up to the BPS mass of the total charges. 
When the Riemann surface degenerates in such a 
way that it falls apart into different component surfaces which are simultaneously holomorphic, the 
area of the combined surface clearly equals to the sum of the 
area of each component surface. Upon using the above relation between the area and the BPS mass, this then directly
translates into an expected correspondence between the wall of marginal stability and wall of degeneration of the surface $\S$.
 
One simplest example of the above-mentioned phenomenon is when the genus two curve $\S$ degenerates in
such  a way that it splits from the middle and falls apart into two tori as shown in Fig.~\ref{genus2degenerating}. In this simple case, one can indeed check explicitly that the criterion 
on the period matrix for such a degeneration to happen 
is exactly the criterion that the mass, or the surface area, becomes the 
sum of the contribution of the two components
\begin{align}
\label{degenerate_area}
\W = \bem \rho & 0 \\ 0& \s\eem \quad \Leftrightarrow \quad A_{\mathrm{\S_1}} + A_{\mathrm{\S_2}} = A_{\mathrm{\S}}
\end{align}
where $ A_{\mathrm{\S_1}} = \vert q_1 -\bl q_2 \vert $, $ A_{\mathrm{\S_2}} = \vert -\tau(p_1 -\bl p_2) \vert $.
In other words, the above wall of marginal stability is the co-dimension one subspace of the moduli space such that the two tori defined by $\oint_{A_1} dX$, $\oint_{B_1} dX$ and $\oint_{A_2} dX$, $\oint_{B_2} dX$ respectively (\ref{jacobian}), are simultaneously holomorphic with respect to the complex structure $\Upsilon$.

To obtain a geometric understanding of the physics of crossing the walls 
of marginal stability, in the following subsection we will study the degeneration of the genus two 
Riemann surfaces of this kind in detail. As we shall see, this geometric consideration will lead to 
a construction of a group of crossing the walls of marginal stability and therefore provides a geometric derivation of the group of dyon wall-crossing observed in \cite{Cheng:2008fc}.

\section{Wall-crossing in $\cN=4$ theory}\label{sec:wall-crossing}
\index{Weyl group}

The $Sp(2,\Z)$ automorphic form $\Phi_{10}$ is part of an even richer
algebraic structure than we have described so far (see e.g.
\cite{gritsenko-1995,borcherds-1998-132} for a mathematical treatment and 
\cite{Dijkgraaf:1996it,Cheng:2008fc} for its importance in 4-dimensional $\cN=4$
string theory). Rewriting
it in the form  
\begin{align}\label{eqn:denominator}
\Phi_{10} (\Omega) = e^{- 2  \pi i ( \rho, \Omega' )} \prod_{\alpha \in
  R_+} \left( 1- e^{- \pi i ( \alpha, \Omega')} \right)^{c( - ||
  \alpha ||^2/2)},
\end{align}
where $\Omega'$ has coefficient $-\nu$ instead of $\nu$, makes an
underlying generalized Lie algebra structure apparent, that we reveal while
explaining the symbols used 
in this formula. The lattice 
\begin{align*}
R_+ = \{ \Z_+ \alpha_1 + \Z_+ \alpha_2 + \Z_+ \alpha_3 \}  
\end{align*}
is spanned by three elements, that may be represented as the $2 \times 2$
matrices 
\begin{align}\label{eqn:simpleroots}
\alpha_1 = \bem0&-1\\-1&0\eem,\quad \alpha_2 = \bem2&1\\1&0\eem,\quad \alpha_3 =
\bem0&1\\1&2\eem. 
\end{align}
These matrices can be interpreted as the three simple roots of a Lie
algebra, with Weyl vector
$$\rho = \frac{1}{2} \sum_{i=1}^3 \alpha_i.$$ 

The bilinear form $(.,.)$ on this Lie algebra determines
the equivalence of $\Phi_{10}$ with the right-hand side of
(\ref{eqn:denominator}). For any two symmetric real $2 \times 2$
matrices $X$ and $Y$ it is given by 
\begin{align*}
 (X,Y)  = - \det Y \, \Tr \, (XY^{-1} ).
\end{align*} 
Notice the factor $-2$ difference with
equation~(\ref{Lorenztian_metric}); this is needed to bring the Cartan
matrix in a standard form. Indeed, we now find the Cartan matrix
\begin{align}\label{eqn:cartanmatrix}
 ( \alpha_i, \alpha_j) = \bem \,\,\,2 & -2 & -2 \\ -2 &
     \,2 & -2 \\ -2 & -2 & \,\,\, 2 \eem.
\end{align}
This matrix is not positive-definite, as it has one negative
eigenvalue. Therefore, the simple roots
$\alpha_i$ are part of a Kac-Moody algebra. 

The Weyl group $W$ of this algebra is generated by the
reflections 
\begin{align*}
 s_i: \, X \, \mapsto \, X - 2\frac{(X, \alpha_i)}{(\alpha_i, \alpha_i)}
 \alpha_i \, =: \, w_i (X)
\end{align*}
with respect to the simple roots $\alpha_i$, where $w_i(X) = w_i X
w_i^T$ and $w_i$ is a symmetric real $2 \times 2$ matrix. Acting with
the Weyl group on the simple roots $\alpha_i$ generates all the roots
of the Kac-Moody algebra. Half of these roots are positive; by
definition these are positive combinations of the simple roots and thus part of
$R_+$. Remark that these positive roots have length $2$, they are 
\emph{real} or \emph{space-like}, and thus contribute to the product formula
(\ref{eqn:denominator}) with a power $c(-1) = 2$.   

Note that no other space-like elements in $R_+$ show up in the
infinite product (\ref{eqn:denominator}).  However, there are many
more contributions from vectors in $R_+$ that have
a non-positive length, for instance $2 \rho$ whose length is
$-6$.
To describe the full algebraic structure underlying 
$\Phi_{10}$ we should therefore extend our notion of a Kac-Moody
algebra. An extended Kac-Moody algebra that incorporates \emph{imaginary}
roots, which are either \emph{light-like} or \emph{time-like}, is
called a Borcherds Kac-Moody algebra.  

Finally, the coefficients $c(-|| \alpha||^2/2)$ can be either
positive or negative. This property hints at the presence of both bosonic and
fermionic roots, that contribute respectively to the numerator and
denominator of an index formula. The full algebraic structure behind
(\ref{eqn:denominator}) is thus a \emph{Borcherds Kac-Moody
superalgebra}. 
\index{Borcherds Kac-Moody superalgebra}
Indeed, the subset $\Delta_+ \subset R_+$ that
contributes to $\Phi_{10}$ may be interpreted as the total set of roots of
such an ``automorphic form corrected'' superalgebra. The infinite product
(\ref{eqn:denominator}) is then called a \emph{denominator formula} for
this algebra, where the powers $c(-||
\alpha||^2/2)$ determine the multiplicities of the real and
imaginary roots.  
\index{denominator formula}

As a side-remark we notice that the half-BPS partition function
$\eta^{24}(\sigma)$ can likewise be written in the algebraic form 
\begin{align*}
 \eta^{24}( \sigma) = e^{- \pi i (\beta, \Omega')} \prod_{\tilde{R}_+} \left( 1- e^{-
     \pi i (\beta, \Omega')} \right)^{c(-||\beta||^2/2)} 
\end{align*}
where the degenerated lattice $\tilde{R}_+$ is one-dimensional and
generated by $$\beta = \bem 0 & 0 \\ 0 & 2 \eem. $$


In \cite{Cheng:2008fc} it was found that the Weyl group $W$ has an
elegant interpretation in terms of the wall-crossing of quarter BPS
dyons. Using the $\cN=4$ central charge matrix
\begin{align*}
  \cZ = \frac{1}{\sqrt{\tau_2}} (P_R - \tau Q_R)^m \G_m,  
\end{align*}
one finds that the central charges of two BPS dyons can align on all
$PGL(2,\Z)$-images of the wall 
\begin{align*}
\frac{\tau_{1}}{\tau_{2}} + \frac{(P_R \cdot Q_R)}{| P_R
\wedge Q_R|} = 0
\end{align*}
in the moduli space parametrized by $\tau$ and the left-moving charges
$P_L$, $Q_L$. In terms of the moduli vector $Z$ in
equation~(\ref{mod_vec}) and the roots $\alpha_i$ these walls are
parametrized by  
\begin{align*}
 \left( \frac{Z}{||Z||}\,,\, \alpha \right)=0,   
\end{align*}
where $\alpha$ can be any $PGL(2,\Z)$-image of $\alpha_1$, i.e. any
positive real root. These walls
are straight lines in the upper-half plane, and become arcs of circles
on the Poincar\'e disc. The simple roots $\alpha_i$ form a triangle in
the Poincar\'e disc, whose vertices lie on the boundary of the
Poincar\'e disc. All the other $PGL(2,\Z)$ images can be obtained from
this fundamental domain by a Weyl reflection. This yields a
tessellation of the Poincar\'e disc in triangles, whose vertices all end at
infinity, and realizes each fundamental domain as a hyperbolic Weyl
chamber (see \emph{e.g.} the cover of \cite{Cheng:2008gx}).

Due to the presence of walls of marginal stability, 
where BPS-states could (dis)ap-pear,  the
graded degeneracies of the BPS 
states are generically only piecewise constant functions of the values of
moduli at spatial infinity. In particular they typically jump when a
wall of marginal stability is crossed. 

In the $\cN=4$ set-up wall-crossing has a geometrical interpretation
of the underlying genus two surface with period matrix $\Omega \sim
Z/||Z||$. At any wall of marginal stability this surface
degenerates into a transverse intersection of two tori, so that
$\Phi_{10}(\Omega)$ develops a pole.\footnote{Note that there are more walls of
marginal stability than the ones we just described, since a
degeneration of the genus 2 surface is labeled by an element of $Sp(2,
\Z) \supset SL(2,\Z)$. Recently, these $Sp(2,\Z)$-poles have been
given a supergravity interpretation as well \cite{Murthy:2009dq}.}  
This raises the question whether $\Phi_{10}$ really gives a
good description of the quarter BPS dyons in any chamber of the moduli
space. This has been analyzed in two (equivalent) ways
\cite{Sen:2007vb,Dabholkar:2007vk,Cheng:2007ch}. 

First of
all, the BPS degeneracies are dependent on a choice of contour
\begin{align*}
  d(P,Q) = (-1)^{P \cdot Q+1} \oint_C d \Omega \frac{e^{ - \pi i  (\rho
      P^2 + \s Q^2 + 2\n P\cdot Q)}}{\Phi_{10} (\Omega)}. 
\end{align*}
An $SL(2,\Z)$ transformation of
the moduli and charges in $d(P,Q)$ encodes this degeneracy in 
another Weyl chamber. Since all factors in the integrand are invariant under
such an $SL(2,\Z)$ transformation, exactly same answer would be obtained
when the transformed contour $C'$ is equivalent to $C$. However, the
poles of $\Phi_{10}$ prevent this. They give an extra contribution
to the degeneracy corresponding to half-BPS states on the intersecting 
tori. Still, it is possible to avoid this factor by changing the
charges along.  

Secondly, the expansion of the partition function in the degeneracies 
\begin{align*}
\frac{1}{ \Phi_{10}(\Omega)} = \sum_{P^2,Q^2,P\cdot Q} (-1)^{P\cdot Q +1} d(P,Q)
 e^{\pi i (\rho P^2 + \s Q^2 + 2\n P\cdot Q)}, 
\end{align*}
is dependent on the choice of expansion parameters. Instead of the above
representation one could for example have exchanged $\nu$ on the right-hand side
for $- \nu$, corresponding to a crossing of the wall at $\nu =0$. This changes the degeneracies $d(P,Q)$. Only when one transforms the
charges $P$ and $Q$ accordingly the resulting degeneracies remain the
same. So if one expands the automorphic form using 
moduli-dependent expansion parameters that are appropriate for a
specific Weyl chamber, the partition function $1/\Phi_{10}$ does encode  
the BPS degeneracies at all points in the moduli space.   

These unexpected properties of the BPS degeneracies certainly hint at
deeper structures of the theories yet to be fully
uncovered. Specifically, while the properties pertaining to the
intricate moduli dependence of the BPS index mentioned above have been
observed within the framework of ${\cal N}=4$, $d=4$ supergravity, a
microscopic understanding of these properties is clearly desirable. In
particular, we would like to understand why the different indices at
different points in the moduli space can be extracted from the same
generating function.  
More explicitly, from the fact that  the group of wall-crossing is a
subgroup of the ($\Z_2$-parity-extended) S-duality group, when the
moduli cross a wall of marginal stability, the change of the BPS index
can be summarized by a change of the ``effective charges'' by a Weyl
reflection \cite{Cheng:2008fc}. We would like to understand why the
index should change in such a simple way.

Carrying the analysis of the previous section one step further, we
study the change of the 
surface when it goes through such a degeneration, and find that it is
equivalent to a particular change of the homological cycles of the
surface.  
Using the relation between the homology class in the spacetime $T^4$
of the Riemann surface wrapped by the M5 brane and the conserved
charges, we see how the change of the BPS index when crossing the wall
of marginal stability under consideration amounts to a change of the
``effective charges'' by acting by a certain element of the hyperbolic
reflection group $W$. Following such a strategy and using essentially
only the supersymmetry condition, we derive the specific group
structure underlying the wall-crossing of the theory, and the fact
that the BPS degeneracies at different moduli are given by the same
partition function.  
In particular, we see how the moduli space and its partitioning by the
walls of marginal stability can be identified with the dual graph of
the type IIB $(p,q)$ 5-brane network compactified on the spacetime
torus, with the symmetry group of the network identified with the
symmetry group of the fundamental domain of the group of
wall-crossing. We hope that this microscopic derivation of the Weyl
group will be a first step towards an understanding of the microscopic
origin of the Borcherds-Kac-Moody algebra in the dyon spectrum. 

In \Cref{The First Degeneration} we focus on one specific
degeneration of 
the surface and analyze the change of the homology cycles under such
degeneration by using a hyperelliptic model of the genus two
surface. In this way we derive one of the elements of the reflection
group $W$. In \Cref{The Symmetry} we study the symmetry of
the hyperelliptic surface, or equivalently the symmetry of the
periodic network of effective strings in the type IIB frame. In this
way we obtain the other generators of the group $W$. Using these
results, in \Cref{Moduli Space as the Dual Graph} we discuss
how the moduli space and its partitioning by the walls of marginal
stability, or equivalently the walls of degenerations of the Riemann
surface, can be understood simply as being the dual graph of the
periodic effective string network. We also discuss the implication of
these results for the counting of BPS dyonic states, and in particular
why the index simply changes by an appropriate change of the
``effective charges'' when the moduli cross a wall of marginal
stability. In \Cref{Conclusion} we conclude by a discussion,
in particular we discuss what we cannot derive by such a simple
analysis and sketch an analogous treatment for the case of the CHL
models.


\subsection{Deriving the group of discrete attractor flow}
\label{Deriving the Group of Discrete Attractor Flow}

In this subsection we first study a
specific degeneration of  the 
Riemann surface and show how the effect of going through such a
degeneration boils down to a change of the homology cycles. This then
in turn gets translated into a change of the ``effective charges'' of
the system under the identification between the homology classes of the
cycles of the surface in the internal space and the conserved charges
of the system. Secondly we study the symmetry of
the system and thereby recover the full hyperbolic reflection group
underlying the structure of wall-crossing of the present
theory. Thirdly, we discuss the implication of these results to
the problem of enumerating supersymmetric dyonic states, and show how
it leads to the prescription proposed in \cite{Cheng:2008fc} of
retrieving BPS indices at different points in the moduli space from
the same partition function (see also \cite{Sen:2007vb,Cheng:2007ch}
for earlier discussions).

\subsubsection{The First Degeneration}
\label{The First Degeneration}
\begin{figure}[h]
\centering  
\includegraphics[width=6cm]{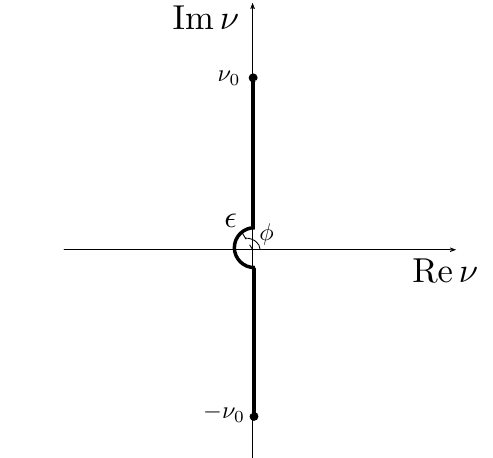}
\caption{In \Cref{The First
      Degeneration} we study the change of the Riemann surface $\S$
    when its period matrix changes as (\ref{change_1}) 
following the above path, where $\e\to 0_+$ and $\rho$ and $\s$ are held fixed at values satisfying $\im\rho\,\im\s\gg (\im\n_0)^2$
.}\label{path}
 \end{figure}

First we will study what happens to the Riemann surface when the 
moduli change such that the surface goes through a degeneration mentioned at the end of the previous Subsection. To remain in the open moduli space of the genus two Riemann surface, we study 
the change of the Riemann surface $\S$ when its period matrix $\W$ changes as 
\begin{align} \label{change_1}
\left( \begin{array}{cc} \rho & -\nu \\ -\nu & \sigma \end{array}
\right) \to \left( \begin{array}{cc} \rho & \nu \\ \nu &
    \sigma \end{array} \right)
\end{align}
 following the path depicted in Fig.~\ref{path}. Clearly, the two end
 points of the path are on the different sides of the wall of marginal
 stability  (\ref{degenerate_area}) considered earlier.

 To focus on what happens to the surface when the wall is crossed, we will further zoom into the part of the path in Fig.~\ref{path} that is a half-circle with vanishing size:
\begin{align} \label{half_circle}
\W=\bem\rho & \e e^{i\f} \\ \e e^{i\f} &\s\eem, \quad  \e\to 0_+,\quad
\f\in \left[-\frac{\pi}{2},\frac{\pi}{2} \right]\;.\end{align}

\begin{figure}[h]
\centering  
\includegraphics[width=9cm]{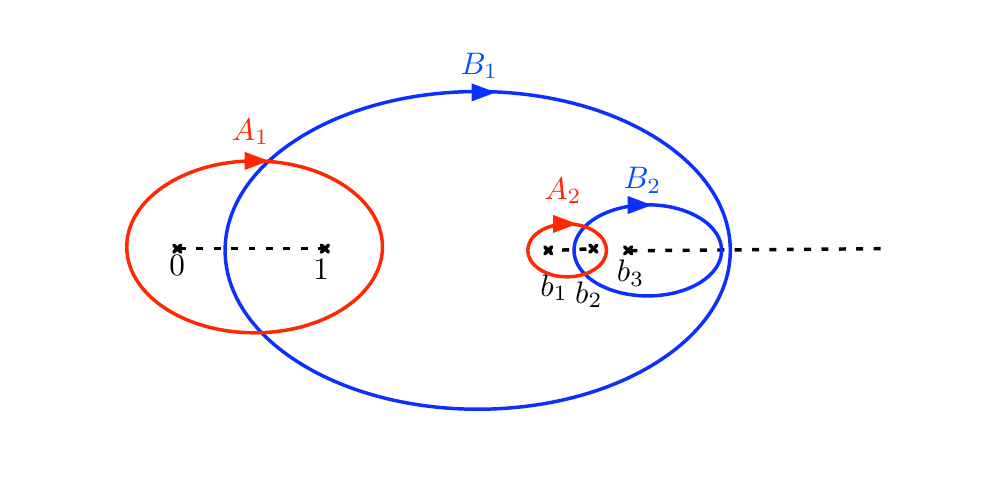}
\caption{Hyperelliptic representation of the genus two surface 
  $\S$ together with a choice of its $A_i$ and $B_i$-cycles. A
  degeneration corresponding to the one shown in Fig.~\ref{genus2degenerating} corresponds to coalescing the branch points $b_1$, $b_2$ and $b_3$. Note that
  when we set the background two-form fields $B$ and $C$ along the time-like direction to zero, so that
  $\re\Omega=0$ (\ref{re_omega}), all branch points are collinear.}\label{hyperellipticgenus2}
 \end{figure}

First recall that, every Riemann 
surface of genus two can be represented as a hyperelliptic surface with six branch points $b_{1,\dotsi,6}$
\be\label{hyperelliptic_eq}
y^2 = x (x- 1)(x- b_1)(x- b_2)(x- b_3),
\ee
where we have used the conformal invariance to fix $b_{4},b_5,b_6$ to be $\infty, 0,1$ respectively.  In other words, we represent the genus two Riemann surface $\S$ as a two-sheet cover of $\C\mathbb P^1$ with six branch points $b_{1,\dotsi,6}$  and three branch cuts between $b_{2i-1}$ and $b_{2i}$ for all $i=1,2,3$, as shown in Fig.~\ref{hyperellipticgenus2}.

To analyze the change of the surface, in particular the homology cycles of the surface, after the imaginary part of $\nu$ changes sign, we would like to determine the normalized basis $\hat{w}_{1,2}$, satisfying (\ref{solve_V_eq1}), in terms of the local coordinate $x$ of $\C\mathbb P^1$.

It is a familiar fact about hyperelliptic curves that the two one-forms 
$$
\frac{dx}{y},~ \frac{x\,dx}{y}
$$
form a basis of the holomorphic one-forms on the genus two surface $\S$ given by (\ref{hyperelliptic_eq}), see for example \cite{tata_theta}.
To achieve our goal we need to compute the integral of the above one-forms along the $A_1$, $A_2$ cycles.  First we observe that, with the choice of cycles as in Fig.~\ref{hyperellipticgenus2}, the integrals of a holomorphic one-form $w$ along the $A$-cycles are given by the so-called ``half-period''
\be\label{half_period}
\frac{1}{2}\oint_{A_1} w = \int_{0}^1 w,\quad\frac{1}{2}\oint_{A_2} w = \int_{b_1}^{b_2} w
\ee
on the upper sheet of the hyperelliptic surface. 

To obtain an expression for these quantities in terms of the period matrix $\W$ and in particular in terms of the angle $\f$ (\ref{half_circle}), we recall that the locations of the branch points $b_{1,2,3}$ are uniquely determined by the genus two Riemann theta functions up to theta function identities \cite{tata_theta}. Explicitly, we have \cite{lebo_degeneration}
 \begin{align*}
 b_1 &= \frac{\theta^2[\begin{smallmatrix} 0 & 0 \\
     0&0 \end{smallmatrix}]\theta^2[\begin{smallmatrix} 0 & 1 \\
     0&0 \end{smallmatrix}]}{\theta^2[\begin{smallmatrix} 1 & 0 \\
     0&0 \end{smallmatrix}]\theta^2[\begin{smallmatrix} 1 & 1 \\
     0&0 \end{smallmatrix}]}(0,\W)\\
b_2 &= \frac{\theta^2[\begin{smallmatrix} 0 & 1 \\
    0&0 \end{smallmatrix}]\theta^2[\begin{smallmatrix} 0 & 0 \\
    0&1\end{smallmatrix}]}{\theta^2[\begin{smallmatrix} 1 & 1 \\
    0&0 \end{smallmatrix}]\theta^2[\begin{smallmatrix} 1 & 0 \\
    0&1\end{smallmatrix}]}(0,\W)
\end{align*}
\begin{align*}
 b_3 &= \frac{\theta^2[\begin{smallmatrix} 0 & 0 \\ 0&0 \end{smallmatrix}]\theta^2[\begin{smallmatrix} 0 & 0 \\ 0&1 \end{smallmatrix}]}{\theta^2[\begin{smallmatrix} 1 & 0 \\ 0&0 \end{smallmatrix}]\theta^2[\begin{smallmatrix} 1 & 0 \\ 0&1 \end{smallmatrix}]}(0,\W),
\end{align*}
where $\theta[\begin{smallmatrix} \varepsilon_1 & \varepsilon_2 \\ \varepsilon'_1 & \varepsilon'_2 \end{smallmatrix}](\zeta,\W)$ is the genus-two Riemann theta functions, defined as
$$
\theta[\begin{smallmatrix} \varepsilon_1 & \varepsilon_2 \\ \varepsilon'_1 & \varepsilon'_2 \end{smallmatrix}](\zeta,\W)  = 
\sum_{n_1,n_2\in \Z}e^{2\pi i \big(\frac{1}{2}(n+\frac{1}{2} \varepsilon)^T\cdot \W\,\cdot (n+\frac{1}{2} \varepsilon)+(n+\frac{1}{2} \varepsilon)^T\cdot (\zeta+ \frac{1}{2}\varepsilon') \big)},
$$
where the ``$\cdot$'' denotes matrix multiplication.

While the details of these formulas are not so important for us, there are a few important immediate consequences of these expressions that we can draw. First of all, due to the fact that the genus two theta functions are a product of two genus one theta functions at leading order in $\n$ when $\n \to 0$:
$$
\theta[\begin{smallmatrix} \varepsilon_1 & \varepsilon_2 \\
  \varepsilon_1'&\varepsilon_2'\end{smallmatrix}]
(0,\big(\begin{smallmatrix}\rho&\n\\ \n & \s\end{smallmatrix}\big) )=
\theta[\begin{smallmatrix} \varepsilon_1 \\
  \varepsilon_1' \end{smallmatrix}](0,\rho)
\theta[\begin{smallmatrix} \varepsilon_2 \\
  \varepsilon_2' \end{smallmatrix}](0,\s) \, \big(1 + {\cal
  O}(\n^2)\big),
$$
the three branch points coalesce when $\n\to 0$
\be
\label{coalesce}
b_1,b_2,b_3 \to b_0 = \Big(\frac{\theta[\begin{smallmatrix}0\\0\end{smallmatrix}](0,\rho)}{\theta[\begin{smallmatrix}1\\0\end{smallmatrix}](0,\rho)}\Big)^4.
\ee

Furthermore, from the definition of the genus two theta functions we see that 
\begin{align*}
\frac{\pa}{\pa \n} b_i\big\lvert_{\n=0} = 0,\quad  i=1,2,3.
\end{align*}
Therefore, for the period matrix on the half-circle given by (\ref{half_circle}) and in Fig.~\ref{path}, we have 
\begin{align*}
b_i = b_0 + \e^2 e^{2i\f} k_i + {\cal O}(\e^4), \quad k_i = \frac{1}{2}\frac{\pa^2}{\pa \n^2} b_i\big\lvert_{\n=0} \,\in \C,\;\; i=1,2,3\;.
\end{align*}
In particular, the branch points go through a $2\pi$ rotation under a change $\f \to \f+ \pi$. In other words, the branch points return to themselves while the period matrix undergoes a change $\n\to -\n$. 

Now we can use the above expression for the branch points near the degeneration point and (\ref{half_period}) to compute the periods along the $A_i$-cycles of the holomorphic one-forms $\frac{dx}{y}$, and $\frac{xdx}{y}$, and 
obtain the following expression for the normalized holomorphic one-forms satisfying (\ref{solve_V_eq1})
\begin{align}\label{normalized_one_forms_hyperelliptic}
\hat w_1 &= \frac{-1}{2(\alpha b_0 -\b)}\frac{(x-b_0) dx}{y} \,\big(1+ {\cal O}(\e^2) \big)\\
\hat w_2 &=\e e^{i\f}\frac{1}{2\g(\alpha b_0 -\b)}\frac{(\alpha x-\b) dx}{y} \,\big(1+ {\cal O}(\e^2) \big),
\end{align}
where $\alpha,\beta,\gamma$ are $\f$-independent, order one constants
\begin{align*}
\alpha &= \int_0^1 \frac{dx}{\sqrt{x(x-1)(x-b_0)^3}} \\ 
\beta &= \int_0^1 \frac{xdx}{\sqrt{x(x-1)(x-b_0)^3}}\\ \
\g& =\frac{1}{\sqrt{b_0(b_0-1)(k_2-k_1)}}\,\int_{0}^1\,\frac{dx}{\sqrt{x(x-1)(x-\frac{k_3-k_1}{k_2-k_1})}}.
\end{align*}

While the precise values of these constants are not important for us, the above expression (\ref{normalized_one_forms_hyperelliptic}) immediately shows that, when  $\im\n$ changes sign by a $\f$ to $\f+\pi$ rotation, the normalized holomorphic one-forms change like
\begin{align*}
\bem \hat{w}_1\\\hat{w}_2 \eem \to \bem \hat{w}_1\\-\hat{w}_2 \eem
\end{align*}
as linear combinations of the holomorphic one-forms $\frac{dx}{y}$ and $\frac{xdx}{y}$,
despite of the fact that the three coalescing branch points $b_{1,2,3}$ simply return to the original locations after a $2\pi$ rotation.

This suggests that, in a representation of the hyperelliptic surface in which the holomorphic one-forms are held fixed, the homology cycles go through the following transformation 
\be\label{basis_transformation_1}
\bem A_1 \\ A_2 \eem \to \bem A_1 \\ -A_2 \eem,\quad\bem B_1 \\ B_2 \eem \to \bem B_1 \\ -B_2 \eem\;.
 \ee
Indeed, it is not difficult to check  that the periods of any holomorphic one-form $w$ along $A_2$ and $B_2$ cycles
$$
\frac{1}{2}\oint_{A_2}w=  \int_{b_1}^{b_2}w,\quad
\frac{1}{2}\oint_{B_2}w=  \int_{b_2}^{b_3}w 
$$
change sign under $\f\to \f+\pi$. 

Another way to understand this change of homology basis is the following. From the expression of the normalized holomorphic one-forms (\ref{normalized_one_forms_hyperelliptic}) we see that, to the leading order in $\e$ we have the two separated genus one surfaces described by
\be\label{two_tori_split}
y'^2 = x(x-1)(x-b_0),\quad y''^2= (x-b_1)(x-b_2)(x-b_3).
\ee
Indeed, from the following relationship between the cross-ratio of the four branch points ${\mathfrak b}_{1,2,3,4}$ of a genus one surface and the torus  complex moduli $\tilde{\tau}$ \cite{tata_theta}
\begin{align*}
\frac{({\mathfrak b}_3-{\mathfrak b}_1)({\mathfrak b}_4-{\mathfrak b}_2)}{({\mathfrak b}_2-{\mathfrak b}_1)({\mathfrak b}_4-{\mathfrak b}_3)}= \Big(\frac{\theta[\begin{smallmatrix}0\\0\end{smallmatrix}](0,\tilde{\theta})}{\theta[\begin{smallmatrix}1\\0\end{smallmatrix}](0,\tilde{\theta})}\Big)^4
\end{align*}
one can check that the following two genus-one curves have complex moduli equal to $\rho$ and $\s$ respectively.
From the above expression (\ref{two_tori_split}) it is manifest that, when $b_i$'s go through a $2\pi$ rotation around their common converging point $b_0$, nothing happens to the first genus one surface while the second one goes through a sheet exchange (or ``hyperelliptic involution") $y'' \to -y''$ corresponding to the monodromy
\begin{align*}
\bem A_2 \\ B_2 \eem \to \bem -A_2 \\ -B_2 \eem.
\end{align*}
The latter can be explicitly seen by substituting
\begin{align*}
x = b_0 + e^{2i\f} \tilde{x},\quad y'' = e^{3i\f} \tilde{y} 
\end{align*}
in the second equation of (\ref{two_tori_split}).

In general, when we change the basis such that the $A_i$-cycles are changed to 
\be\label{change_basis_A}
\bem A_1 \\ A_2 \eem \to \bem a & b \\ c& d \eem \bem A_1 \\ A_2 \eem,\quad \g = \bem a & b \\ c& d \eem \in GL(2,\Z),
\ee
the corresponding change of the $B_i$-cycles is then fixed by the canonical intersection (\ref{intersection_A_B}) to be  
\begin{align*}
\bem B_1 \\ B_2 \eem \to \pm \bem d& -c\\ -b & a\eem\bem B_1 \\ B_2 \eem= (\g^T)^{-1} \bem B_1 \\ B_2 \eem,
\end{align*}
where the $\pm$ signs are taken when $ad-bc=\pm1$. 
Under this transformation, the period matrix transforms as  
\be\label{shorthand}
\W \to \g(\W)\equiv(\g^{-1})^T \W \g^{-1}.\ee

Without changing the Riemann surface, such a change of basis has an interpretation as performing a physical S-duality in the heterotic frame, extended with the $\Z_2$ spacetime parity exchange. To see this, first inspect the expression (\ref{jacobian}) for the vectors defining the Jacobian of the surface. The effect of the above change of basis 
on these vectors is equivalent to the following heterotic S-duality transformation, or equivalently the modular transformation of the torus in the type IIB frame
\begin{align}\label{S_dual}
 \bem Q \\ P \eem \to \bem a & b \\ c & d \eem  \bem Q \\ P \eem, \quad \tau \to \frac{a\tau + b}{c\tau+d} \, \text{  or    } \, \frac{a\bar\tau + b}{c\bar\tau+d} \,
\textrm{   for   } \, ad-bc = \pm 1 
\end{align}
with the corresponding change of $R_B$ such that the area of the type IIB torus remains invariant. In particular, the fact that the moduli vector ${Z}$ transforms as ${Z} \to (\g^{-1})^T {Z} \g^{-1}$ under the above S-duality transformation is then consistent with the transformation of the period matrix under a change of homology basis.

Now let's go back to the evolution (\ref{half_circle}) of the Riemann
surface through the degeneration wall $\n=0$, due to the corresponding
change of the moduli vector ${Z}$ (\ref{period_matrix}). 

What we have seen can be summarized as follows: when the moduli change
across the wall of marginal stability following the path corresponding
to an angle-$\pi$ rotation of the phase of $\n$ (\ref{half_circle}),
the holomorphic one-forms of the surface $\S$ change in such a way
that all their $A_2$, $B_2$ periods change signs while their periods
along the $A_1$, $B_1$ cycles remain the same. 
This is equivalent to keeping the surface unchanged but change the basis for the homology cycles in the way (\ref{change_basis_A}) 
given by the following element in $GL(2,\Z)$
\be\label{def_R}
R = \bem  1 & 0 \\ 0&-1 \eem.
\ee

In other words, the process of keeping the charge fixed while varying the moduli across the wall of marginal stability following (\ref{half_circle}) is equivalent to keeping the moduli vector $Z$ unchanged but changing the charges 
\be\label{new_charge_1}
\bem Q \\ P \eem \to R \bem Q \\ P \eem. 
\ee

 This observation has the following implication for the counting of the BPS states. Consider the partition function of the theory
 $$ Z(\W) = \sum_{P^2,Q^2,P\cdot Q} (-1)^{P\cdot Q +1} d(P,Q) e^{i\pi (\rho P^2 + \s Q^2 + 2\n P\cdot Q)}, $$
 which is a path integral computed on the Riemann surface $\S$. It clearly depends on its period matrix $\W$ and therefore on the moduli vector ${Z}$ through its relation to the period matrix (\ref{period_matrix}). When the moduli change in such a way that the Riemann surface goes through a degeneration described in  (\ref{degenerate_area}), from the above reasoning we see that the partition function remains unchanged while a transformation of ``effective charges'' given in (\ref{new_charge_1}) has to be performed. This corresponds to the change of the highest weight of the Verma module as described in \cite{Cheng:2008fc}. 
  
 It is also easy to understand the nature of this degeneration in the  type IIB five-brane network picture.
From the relation between the period matrix of the genus two curve in M-theory and the length parameters for the periodic string network (\ref{length_general},\ref{period_matrix}), we see that the degeneration of the Riemann surface characterized by $\n=0$ corresponds to the degeneration of the string network characterized by $t_1=0$. For example, starting from a region in the moduli space with ${Z} = \big(\begin{smallmatrix}z_1 & z\\ z&z_2\end{smallmatrix}\big)$ with $z_1,z_2>z>0$,
what happens when $t_1=0$ is a transition from the network described
by (\ref{network_1}), or the first figure in Fig.~\ref{network_fig}, to
the network described by (\ref{network_2}), or the second figure in
Fig.~\ref{network_fig}. The above claim that the final surface has the
same period matrix under a change of homology basis corresponding to
(\ref{new_charge_1}), is then reflected by the fact that the two
defining equations (\ref{network_1}), (\ref{network_2}) transform into each other under the transformation of the charges (\ref{new_charge_1}).

\subsubsection{The symmetry of the Weyl chamber}
 \label{The Symmetry}
 
 In the previous paragraph we have studied in detail a particular degeneration of the Riemann surface and what it implies for the index counting the BPS states under the crossing of the corresponding wall of marginal stability. In this paragraph we will turn to studying the symmetry of the system and see how it will help us to uncover the full structure of the group of wall-crossing of the theory. 
 
 First we note that, under our convention that the two $A$-cycles of the surface are chosen to circle two of the three pairs of branch points $\{b_{2i-1},b_{2i}\}$, the choice shown in Fig.~\ref{hyperellipticgenus2} is not quite unique. In other words, from all the possible change of basis of the form (\ref{change_basis_A}), the exchange and permutation of the cycles $A_1,A_2, -A_1-A_2$ correspond to a symmetry of our hyperelliptic model (\ref{hyperelliptic_eq}) in that we do not need to change the set of branch points $\{b_1,\dotsi,b_6\}$ in order for the new $A_i$-cycles to again circle the cuts joining the pairs  $\{b'_{2i-1},b'_{2i}\}$ of the new branch points. 
We therefore conclude that there is a symmetry group with six elements acting on the hyperelliptic surface (Fig.~\ref{hyperellipticgenus2}), corresponding to six ways of associating the three cuts joining $\{b_{2i-1},b_{2i}\}, i=1,2,3$,  to the three homology cycles $(A_1,A_2, -A_1-A_2)$. From the above discussion we see that this group $D_3\subset GL(2,\Z)$ is the same as the symmetry group of a regular triangle, generated by the order two element which acts on the period matrix as
\be\label{action_D3_1}
\W \to RS(\W)
\ee
and which corresponds to $(A_1,A_2, -A_1-A_2)\to(A_2,A_1, -A_1-A_2)$, together with the order-three element which acts as
 \be\label{action_D3_2}
\W \to ST(\W)
\ee
and corresponds to $(A_1,A_2, -A_1-A_2)\to(A_2,-A_1-A_2,A_1 )$, where $T$ and $S$ denotes the usual T- and S- transformation matrix $\big(\begin{smallmatrix}1&1\\0&1\end{smallmatrix}\big)$ and $\big(\begin{smallmatrix}0&1\\-1&0\end{smallmatrix}\big)$, while $R$ was already given in (\ref{def_R}). 
Also here we have used the shorthand notation introduced in (\ref{shorthand}).

\begin{figure}[h]
\centering
\includegraphics[width=\textwidth]{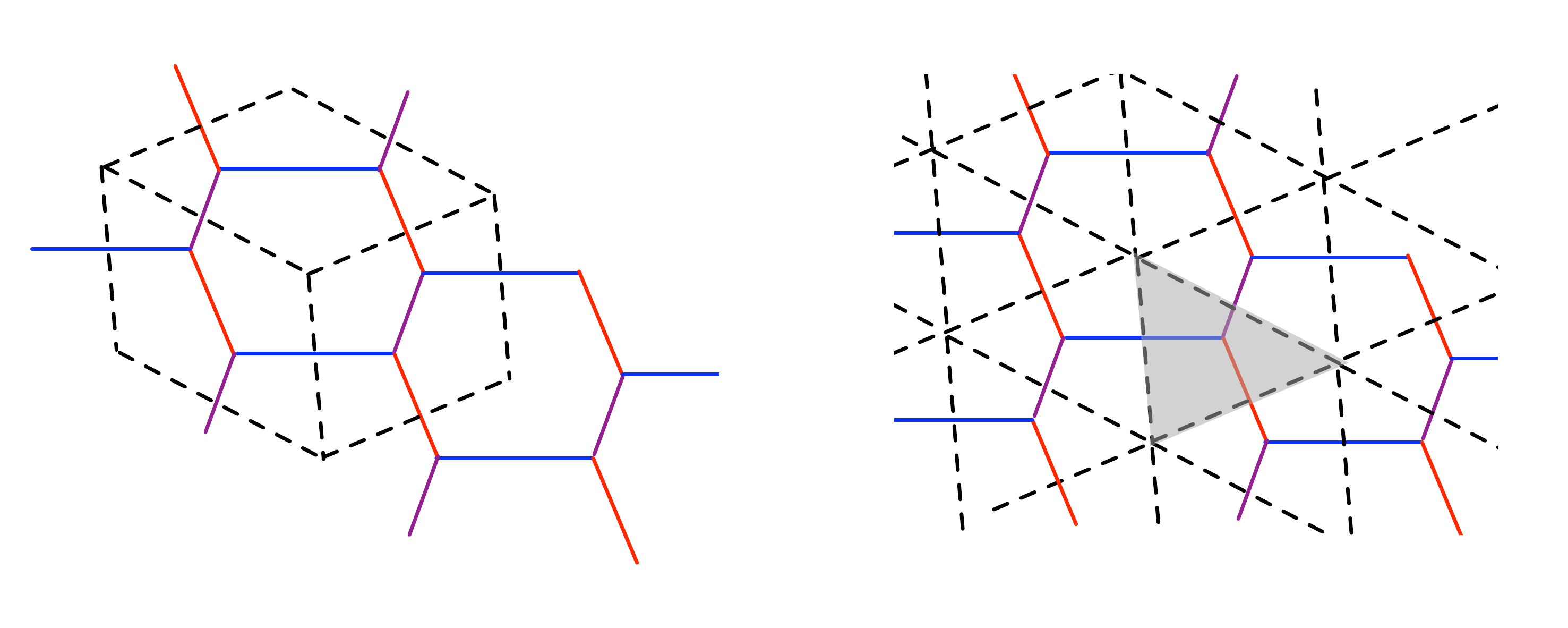}
\caption{{\bf(i)} Different possible ways of compactifying the periodic network on a torus. {\bf(ii)} The moduli space as the dual graph of the five-brane network. }\label{dual_graph}
\end{figure}

The existence of this symmetry is even more apparent in the type IIB picture of five-brane network. For concreteness of the discussion we will now assume that $Z = \big(\begin{smallmatrix}z_1&z\\z&z_2\end{smallmatrix}\big)$ satisfies $z_1,z_2 > z >0$, so that the network  shown in the first figure in Fig.~\ref{network_fig} is realized. But, suppose that we are given this periodic network given by (\ref{network_1}), there is actually more than one way to fit it into a parallelogram tessellation of the plane. In other words,  there is in fact more than one torus compactification of the network possible. From the fact that the vertices of the parallelogram lie at the center of the honeycomb lattice, we conclude that there are three such parallelogram tessellations possible, as shown in Fig.~\ref{dual_graph} as resembling the three sides of a 3-dimensional cube. These three parallelograms then give six possible tori (for each parallelogram we have two ways of choosing the $A$- and $B$-cycle), corresponding to $3!=6$ ways of placing the charge labels $(Q,P,-Q-P)$ to the three legs of the network. This singles out a six-element subgroup of the extended type IIB modular group (or the heterotic S-duality group) $GL(2,\Z)$ (\ref{S_dual}). Not surprisingly, this is exactly the same $D_3$ we discovered earlier as the symmetry group of the hyperelliptic representation of the Riemann surface $\S$.  This correspondence is to be expected from the identification between the change of basis of the homology cycles on $\S$ and the heterotic S-duality discussed in the previous sub-subsection (\ref{change_basis_A}-\ref{S_dual}).

More explicitly, from (\ref{action_D3_1},\ref{action_D3_2}) and the relation between the period matrix and the length parameters of the network  (\ref{length_parameters_1},\ref{period_matrix}) we see that the length parameters indeed transform as
\begin{align*}
RS: (t_1,t_2,t_3)\to(t_1,t_3,t_2),\quad ST:(t_1,t_2,t_3)\to(t_2,t_3,t_1)
\end{align*}
under the action of the order two and three generators of the symmetry group $D_3$. 

As mentioned earlier, this symmetry group is the symmetry group of a equilateral triangle. Geometrically, the relevant equilateral triangle here cannot be literally the triangle in the dual graph of the five-brane network as shown in Fig.~\ref{dual_graph}, since in general there is no reason to expect them to be equilateral using the flat metric on the plane. This inspires us to take a closer look into the matrix of length parameters, which can be written in a way which makes the  $D_3$ symmetry manifest
\begin{align*}
R_M\,\im \W=\bem t_1+ t_3 & t_1 \\ t_1 & t_1+ t_2\eem= \frac{t_2+t_3}{2}\alpha_1+\frac{t_1+t_3}{2}\alpha_2+\frac{t_1+t_2}{2}\alpha_3
\end{align*}
where
\be\label{alphas}
\alpha_1 = \bem0&-1\\-1&0\eem,\quad \alpha_2 = \bem2&1\\1&0\eem,\quad \alpha_3 = \bem0&1\\1&2\eem\;.
\ee
Remember that this natural basis $\{\alpha_{1,2,3}\}$ has the following
matrix of inner products using the standard $GL(2,\Z)$-invariant
Lorentzian metric (\ref{Lorenztian_metric}) 
\be\label{cartan}
-2\, (\alpha_i,\alpha_j) = \bem 2 &-2 &-2\\-2 &2 &-2\\-2 &-2 &2 \eem
\ee    
and therefore forms a equilateral triangle in the hyperbolic space $\R^{2,1}$. 
The group $D_3$ which permutes $\alpha_{1,2,3}$ can therefore be thought
of the symmetry group of this equilateral triangle. We note that the
basis $\{\alpha_{1,2,3}\}$ we used above is exactly the
basis (\ref{eqn:simpleroots}) for the roots of the Borcherds-Kac-Moody
algebra adopted in \cite{Cheng:2008fc}, and in particular the matrix
of inner products in (\ref{cartan}) is simply the real part of the
Cartan matrix (\ref{eqn:cartanmatrix}) of the algebra.

To summarize, we have found a six-element symmetry group $D_3$ of the
dyon system at a given point in the moduli space which is evident in
both the M-theory Riemann surface picture as well as the type IIB
network picture. In the following paragraph we will use this symmetry
to find all the generators of the hyperbolic reflection group $W$
playing the role of the group of wall-crossing in the $\cN=4$ theory
we discussed, and subsequently derive the full group structure of the
dyon BPS index.

\subsubsection{Moduli space as the dual graph}
\label{Moduli Space as the Dual Graph}

In \Cref{The First Degeneration} we have studied in detail the change of the Riemann surface across a degeneration point (\ref{half_circle}) where the surface falls into two separate tori as depicted in Fig.~\ref{genus2degenerating}. 
From the above discussion about the symmetry of the system, we see
that there are two more natural degenerations of the genus two Riemann
surface $\S$ we should consider. The corresponding transformation of
the period matrix is simply the transformation (\ref{change_1}) of the
first degeneration we have studied in \Cref{The First
  Degeneration}, now conjugated with elements of the symmetry group
$D_3$.  
From (\ref{action_D3_1},\ref{action_D3_2}), we see that apart from the group generator $w_1=R$  (\ref{new_charge_1}), we should also consider the generators $$w_2=(ST)^{-2}R\, (ST)^2
 \quad \text{and}  \quad w_3 = (ST)^{-1}R\,(ST)
 \;. $$ Together they generate a non-compact reflection group, which we will denote by $W$. Furthermore, it is not difficult to show  \cite{Feingold-Frenekel,Cheng:2008fc} that the extended S-duality group $PGL(2,\Z)$\footnote{Recall that $PGL(2,\Z)$ is obtained from $GL(2,\Z)$ by identifying the elements $\g$ and $-\mathds{1}\cdot \g \in GL(2,\Z)$. In the heterotic frame this corresponds to identifying two systems with the same value of axion-dilaton $\tau$ and charges which are related to each other by a charge conjugation $\big( \begin{smallmatrix}Q\\P \end{smallmatrix}\big) \to \big( \begin{smallmatrix}-Q\\-P\end{smallmatrix}\big) $. In the type IIB frame this corresponds to a trivial change of basis for the homology cycles $\big( \begin{smallmatrix}A\\beta \end{smallmatrix}\big) \to \big( \begin{smallmatrix}-A\\-B\end{smallmatrix}\big) $ of the compactification torus $T^2_{\mathrm{(IIB)}}$.} is a semi-direct product
\begin{align*}
PGL(2,\Z) = W\rtimes D_3\;
\end{align*}
of the reflection group $W$ and the symmetry group $D_3$.

It is clear what these three degenerations correspond to in the type IIB and as well as in the  M-theory picture. In the former case they are the three ways in which the five-brane network can disintegrate, namely letting one of the three legs having vanishing length. In the latter case, on the other hand, they correspond to coalescing the branch points $b_{3,4,5}$, $b_{5,6,1}$ or   $b_{1,2,3}$. Remember that coalescing three out of the total of six branch points is equivalent to coalescing the complementary set of three branch points.

More explicitly, these three generators $w_{1,2,3}$ of the group $W$ correspond to the following change of the network
\begin{align*}
w_i : t_i\to -t_i,\quad t_i+t_j \;\text{invariant  for   }j\neq i.
\end{align*}
Equivalently, they can also be represented as the following reflections in the $(2+1)$-dimensional Minkowski space in which the period matrix $\im\W$ takes its value
\begin{align*}
w_i :\, \W \to\W - 2\frac{(\alpha_i,\W)}{(\alpha_i,\alpha_i)}\alpha_i,
\end{align*}
where the basis vectors $\alpha_i$'s are defined in (\ref{alphas}).
Recall that we have chosen the M-theory background such that the period matrix $\W$ is purely imaginary and therefore directly related to the length parameters of the five-brane network.

Applying the same analysis as in \Cref{The First Degeneration} to the other two degenerations corresponding to $w_2$ and $w_3$, one can conclude that going through such a degeneration wall has the following effect on the counting of BPS states. Upon applying a suitable change of basis analogous to (\ref{basis_transformation_1}), after the degeneration we regain the original partition function but now with a different effective charges related to the original charges by 
\be\label{group_charges}
\bem Q \\ P \eem\to w_i \bem Q \\ P \eem.
\ee

From the above consideration, we arrive at a picture of the moduli space with its partitioning by the walls of marginal stability 
given by the the dual graph of the honeycomb lattice representing the five-brane network. This is shown in Fig.~\ref{dual_graph}. 
To understand this better, let's start in one of the dual triangles, let's say the gray triangle which denotes the part of the moduli space with $Z = \big(\begin{smallmatrix}z_1&z\\z&z_2\end{smallmatrix}\big)$, $z_1,z_2 > z >0$, such that the network (\ref{network_1}) as shown in the first figure in Fig.~\ref{network_fig} is realized.
We shall choose it to be our ``fundamental domain'' ${\cal W}$, a name that will be justified shortly.
 
A degeneration happens when one of the length parameters $t_i$'s goes
to zero. As we discussed at the end of \Cref{The Riemann
  Surface}, this corresponds to crossing a physical wall of marginal
stability. When this happens we move to the neighboring triangle,
divided from the fundamental triangle ${\cal W}$ by the side of  the
triangle intersecting the leg of the network whose length parameter
has just goes through a zero. In this new triangle, the effective
charges are related to the original one by the corresponding group
element (\ref{group_charges}). For instance, associated to the
triangle that shares one side with ${\cal W}$ which intersects the leg
whose length parameter is denoted by $t_1$ are the effective charges
$(Q,-P)$ and the region in the moduli space with $Z =
\big(\begin{smallmatrix}z_1&z\\z&z_2\end{smallmatrix}\big)$, where
$z_1,z_2 > -z >0$. 

Given the original charges, this procedure can then be iterated. We thus conclude that each triangle of the dual graph has the effective charges $(Q_v,P_v)$ associated to it, where $v$ labels the vertices in the hexagonal lattice, or equivalently the triangles (the faces) of the dual graph, which represent the corresponding regions in the moduli space. Furthermore, in this way each of the triangles can be identified with the fundamental domain of the group $W$, generated by the three elements $w_{1,2,3}$ (\ref{group_charges}), which by construction plays the role of crossing the walls of marginal stability of the theory. 

Now that each triangle has a set of charges $(Q_v,P_v)$ associated to it, while the period matrix of the genus two Riemann surface $\S$ and therefore the partition function remains the same for each triangle, generically we conclude that there is also a different BPS index $D(Q,P)\vert_v= D(Q_v,P_v)$ associated with each triangle. The difference between $D(Q_v,P_v)$ with different $v$ has been calculated in \cite{Sen:2007vb,Cheng:2008fc} for the present theory and was shown to be consistent with the macroscopic wall-crossing formula. 

To sum up, we have derived the following one-to-one correspondence
\begin{align} \label{conclusion}
&\text{vertex  }v  \text{  of string network } \leftrightarrow \text{a triangle in dual graph} \\ \notag
&\leftrightarrow \text{effective charges } (Q_v,P_v) \leftrightarrow \text{BPS index  }D_v =D(Q_v,P_v)   \\ 
&\leftrightarrow \text{an element  }w_v\in W  \leftrightarrow\text{a
  region in the moduli space  } {Z} \in w_v({\cal W}). \notag
\end{align}
The property of the BPS dyon index of the present $\cN=4$ theory that the indices in different parts of the moduli space are given by the same partition function and have the form $D(Q,P)\vert_v= D(Q_v,P_v)$ was observed in \cite{Sen:2007vb,Cheng:2007ch} based on the macroscopic prediction for the change of index upon crossing a wall of marginal stability. And the fact that these different regions of the moduli space with different BPS indices are in one-to-one correspondence with elements of a hyperbolic reflection group $W$ is later observed in \cite{Cheng:2008fc} based on a 4-dimensional macroscopic analysis. What we have seen is how these properties can be understood as the consequence of the simple consideration of the supersymmetry of the effective string network, or equivalently the holomorphicity of the M5 brane world-volume, when the limit of decoupled 4D gravity is taken.

\subsection{Discussion}
\label{Conclusion}

In this section we worked in the decompactification limit
$(V_{K3}\!\gg\! 1)$ and  
showed how the group structure underlying the moduli dependence of the
dyon BPS index of the $\cN=4$ $K3 \times T^2$ compactification of type
II theory can be understood as simply a consequence of the
supersymmetry of the dyonic states. From the other point of view, this
group structure is simply the consequence of the fact that the BPS
spectrum of the theory is given by the appropriate representation of a
Borcherds-Kac-Moody algebra. The Weyl group of the algebra, which is a
symmetry group of the root system of the algebra, then plays the role
of the group of wall-crossing for the physical degeneracy of the
dyonic states \cite{Cheng:2008fc}. Therefore, we hope that the
microscopic derivation of the Weyl group presented in this section
will be the first step towards an understanding of the 
microscopic origin of the Borcherds-Kac-Moody algebra in the dyon
spectrum.  

For this purpose, it is important to be clear about what we do not
derive from the simple analysis of this section. First of all, while
we assume that the partition function is a functional integral on the
genus two Riemann surface $\S$ and therefore depends only on the
period matrix of the surface, justified by the fact that an M5 brane
wrapping the surface and the $K3$ manifold in the Euclidean spacetime
is equivalent to a fundamental heterotic string whose world volume is
the genus two surface \cite{Witten:1995ex}, we have not derived the
partition function itself from our simple consideration. A discussion
about the subtleties of computing the partition function in a very
similar context can be found in  \cite{Banerjee:2008yu} and will
therefore not be repeated here. Relatedly, the presence of the
Borcherds-Kac-Moody algebra \cite{Dijkgraaf:1996it,Cheng:2008fc} is
far from evident from our simple analysis. It seems likely that the physical interpretation of this complete algebra can be found in terms of the chiral fermions on the genus 2 surface in type IIA, or equivalently, in terms of fluctuations on the M5-brane wrapping the genus 2 surface in M-theory.

Furthermore, we have not commented on the role of the group $W$ as the group of a discretized version of attractor flows. 
As discussed in detail in \cite{Cheng:2008fc}, this interpretation naturally arises due to  the existence of a natural ordering among the elements of the hyperbolic reflection group $W$, and the fact that for given total charges, there is a unique endpoint of this ordering, corresponding to the attractor point of these charges. From the point of view of the Borcherds-Kac-Moody algebra, the Verma module relevant for the BPS index is the smallest one when the moduli are at their attractor value. By working in the limit that the type IIB five-branes are all much heavier than all the $(p,q)$ strings, we have no way of telling which of the triangles in the dual graph in Fig.~\ref{dual_graph} contains the attractor point. This is of course consistent with the fact that our analysis in the text is independent of the values of the T-duality invariants $Q^2, P^2, Q\cdot P$ (\ref{charges_initial}), due to the decompactification limit we are taking. 
But this can easily be cured by going to the next leading order in ${\cal O}(V_{K3}^{-1})$. See also \cite{Banerjee:2008yu}. By minimizing the surface area of the genus two surface (\ref{area_J_integral}) with the volume of the two tori $R_B^2\tau_2,R_M^2\la_2$ held fixed, with now the next leading corrections included, one indeed obtains the attractor equation 
$$
{\cal M}_\tau = \frac{\Lambda_{Q_R,P_R}}{\Vert \Lambda_{Q_R,P_R} \Vert} = \frac{1}{\sqrt{Q^2 P^2 -(Q\cdot P)^2}} \bem Q\cdot Q & Q\cdot P \\ Q\cdot P & P\cdot P\eem\;,
$$
as expected. This extra piece of information will then single out a
triangle in the dual graph as the attractor region and completes the
interpretation of the hyperbolic reflection group derived in this section as the group underlying the macroscopic attractor flow of the theory.

Moreover, we would like to comment on the cases of other $\cN=4$ string
theories.  Here we have focused on the $\cN =4$
theory of $K3\times T^2$ compactified type II theory, while from the
analysis in \cite{Cheng:2008kt} we expect very similar group
structures to be present also in the $\Z_n$-orbifolded theories, the
so-called CHL models \cite{Chaudhuri:1995fk,Chaudhuri:1995bf,Chaudhuri:1995dj}, for $n<4$. For the $\Z_2$-orbifold theory, considering the double cover of the genus two Riemann surface relevant for the computation of the partition function \cite{Dabholkar:2006bj,Dijkgraaf:1987vp} and the corresponding string network, a similar analysis can be employed to understand the group structure in that case. The situation of other orbifolded theories is much less clear. 
In particular, in \cite{Cheng:2008kt} it was discovered that a similar
group structure and an underlying Borcherds-Kac-Moody algebra cease to
exist when $n> 4$. In particular, macroscopic analysis 
showed that the symmetry group of a fundamental region bounded by walls
of marginal stability has infinitely many elements when $n>4$. From
the analysis of this section, this symmetry group is expected to be the symmetry group of the dyonic string network/Riemann surface. One might thus suspect that the corresponding dyon network does not exist in the $\Z_{n>4}$ theories.
 It would be interesting to understand the group structure of the
 dyon degeneracies of other orbifolded $\cN=4$ theories.


Finally, we would like to see our analysis of 1/4 BPS states in
$\cN=4$ theories as a starting point
to study wall-crossing of 1/2 BPS states on local Calabi-Yau manifolds
in $\cN=2$ theories. That these topics are closely related follows
from the duality of the dyon background with the 
intersecting brane background (\ref{eqn:genus2Ibrane}). Do
notice that this 
duality only maps the dyon generating function to the
semi-classical piece $\cF_1$ of the topological string partition
function, since the I-brane background doesn't include a graviphoton
field strength. To describe $\cN=2$ wall-crossing fully we should thus go
beyond deformations of the Riemann surface $\Sigma$. This naturally
raises the question whether we can formulate $\cN=2$ wall-crossing in
terms of the underlying quantum curve that we studied in
Chapter~\ref{chapter5} and Chapter~\ref{chapter6}. We leave this for
future work. 

Another relation between $\cN=2$ and $\cN=4$ string theories
is that both the Gopakumar-Vafa partition function
(\ref{eqn:GVpartitionfunction}) and the dyon
partition function (\ref{eq:genustwo}) can be formulated as an infinite product
expansion. The wall-crossing examples that are worked out in $\cN=2$ context, can actually be formulated most elegantly in this representation: depending on the region in the moduli space additional factors have to be added to the partition
function\footnote{Recent developments  
can be found in e.g. \cite{Denef:2007vg, Szendroi:2007nu,
  kontsevich-2008, Jafferis:2008uf, Chuang:2008aw, Mozgovoy:2008fd,
  Gaiotto:2008cd, Dimofte:2009bv}.}.
One might wonder whether also in
this setting wall-crossing can be described in terms of a
universal generating function.
  
\begin{figure}[h]
\begin{center}
\includegraphics[width=3.6cm]{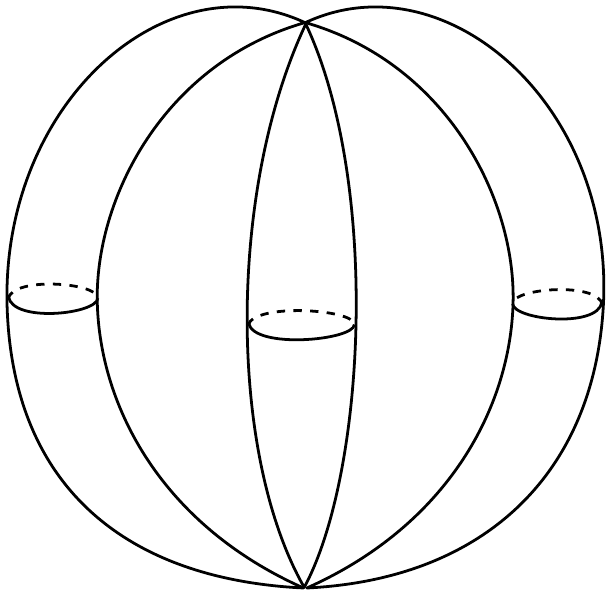}
\caption{The intersection matrix of a configuration of three 2-spheres, that have self-intersection number -2 and that intersect each other 2-sphere transversely in two points, equals the Cartan matrix (\ref{eqn:cartanmatrix}).}
  \label{fig:dyonmckay}  
\end{center}
\end{figure}

Let us point out one interesting observation in this respect. The 
quantum McKay correspondence formulated by J.~Bryan and A.~Gholampour
in \cite{bryan-2008, bryan-2009} relates the
positive roots of an ADE Lie algebra $\lieg$ to 1/2 BPS states of a
Calabi-Yau resolution of the quotient singularity $\C^3/\widetilde{\Gamma}$,
where $\widetilde{\Gamma}$ is the corresponding finite subgroup of $SO(3)$. (These
Calabi-Yau singularities are closely related to the hyper-K\"ahler
surface singularities $\C^2/\Gamma$ discussed in
Chapter~\ref{chapter2} through the double cover of $SU(2)$
over $SO(3)$ \cite{boissiere-sarti}.) More precisely, they prove that 
\begin{align*}
 Z_{GW}'(t. \la) = \prod_{\alpha \in R_+} \prod_{k=1}^{\infty} (1 -
 e^{t \cdot \alpha} q^k)^{k/2}. 
\end{align*}
Here, $\alpha$ is a positive real root that is mapped to a curve in $Y$
by the complex 3-dimensional McKay correspondence (roughly it is the
push-forward of the map $\C^2/\Gamma \to
\C^3/\widetilde{\Gamma}$), where the sum doesn't include positive
roots that are mapped to zero in homology. Up to the factor $1/2$ in
the overall power (which is due to non-compactness), this product equals the part of the Gopakumar-Vafa partition function (\ref{eqn:GVpartitionfunction})
corresponding to the genus 0 curves in the resolution of the
singularity. In particular, each curve contributes a factor  
\begin{align*}
\prod_{\alpha \in R_+} (1 - e^{t \cdot \alpha} )^{1/2}
\end{align*}
semi-classically. From this point of view the dyon counting formula
(\ref{eqn:denominator}) seems to be a
generalization of the (quantum) McKay correspondence, where the Cartan matrix
(\ref{eqn:cartanmatrix}) corresponds geometrically to the transverse
intersection of three genus 0 curves in two points. This is illustrated in Fig.~\ref{fig:dyonmckay}. Indeed, the
corresponding toric diagram (or string web) exactly represents this
configuration of curves. It is interesting to find out whether this locus can indeed appear as a resolved Calabi-Yau singularity. Of course, one might also wonder whether such an algebra can be extracted from other toric Calabi-Yau's that are modeled on a number of compact spheres, and about the relation to wall-crossing.

        \chapter{Fluxes and Metastability}\label{chapter8}


Over the last years much progress has been made in studying Calabi-Yau 
compactifications \index{flux compactification}
\index{Calabi-Yau compactification} 
where one turns on extra fluxes, so-called flux
compactifications. Turning on fluxes  is a method  to lift
part of the large degeneracy in the four-dimensional   
scalar moduli fields. By now there is strong evidence
that there is a huge number of supersymmetric vacua with negative
cosmological constant in which all scalar moduli are stabilized, the
so-called landscape of string theory.  Typical constructions start
with a warped Calabi-Yau compactification of type IIB string theory to
four dimensions. (In such a compactification the
four-dimensional scale is dependent on the coordinates of the internal
space.)  Some of the scalar moduli are stabilized by the
addition of fluxes through the compact cycles of the internal manifold
and others by various quantum effects.  

Since supersymmetry is broken in the real world, it is necessary 
to extend the previous constructions to non-supersymmetric
(meta)stable vacua with small positive cosmological constant to make
contact with phenomenology. 
\index{supersymmetry breaking} 
For this we need to understand the mechanism of
supersymmetry breaking in 
string theory. So far several methods of supersymmetry breaking for
string vacua have been proposed, such as the introduction of
anti-branes and the existence of
metastable points of the flux-induced 
potential. The main drawback of these
constructions is that, in most cases, they are not under complete
quantitative control.

While the question of supersymmetry breaking should be ultimately
understood in an honest compactification, that is in a theory including
gravity in four dimensions, it is technically easier to study simpler
systems where the gravitational dynamics has been decoupled from the
gauge theory degrees of freedom. This typically happens in the limit
where a local singularity develops in the Calabi-Yau manifold. In such a
situation all the interesting dynamics related to the degrees of freedom
of the singularity takes place at energy scales much lower than the four
dimensional Planck scale. Assuming that supersymmetry breaking is
related to these light degrees of freedom, it is then possible to zoom
in towards the singularity and forget about the rest of the
Calabi-Yau.

Recently K.~Intriligator, N.~Seiberg and D.~Shih discovered that even simple
supersymmetric gauge theories can exhibit dynamical supersymmetry
breaking in metastable vacua \cite{Intriligator:2006dd}. From a
phenomenological point of view this possibility is quite  
attractive.  A certain class of gauge theories where
supersymmetry breaking in metastable vacua can be studied with good
control is that of ${\cal N}=2$ gauge theories perturbed by a small
superpotential, initiated by H.~Ooguri, Y.~Ookouchi
and C.-S.~ Park in \cite{Ooguri:2007iu}.
In such theories the exact K\"ahler metric on the moduli space is known.
This makes it possible to 
compute the scalar potential that is produced by the perturbation of the theory
by a small superpotential exactly to first order in the perturbation. It
was shown that generically there are metastable supersymmetry breaking
vacua generated by appropriate superpotentials. We will refer to this as
the \Index{OOP mechanism} for supersymmetry breaking in ${\cal N}=2$ theories.

String theory in a local Calabi-Yau singularity realizes geometric 
aspects of supersymmetric gauge theories, see
Chapter~\ref{chapter3+4}. So also supersymmetry 
breaking in these two systems should be related. The first 
goal in this chapter is to make this connection more
precise by proposing a 
geometric realization of the OOP supersymmetry breaking mechanism in
type II theory on a local Calabi-Yau singularity. Starting from a
geometrically engineered type II compactification, in
\Cref{sec:generalcase} we  
introduce the appropriate superpotential as a non-conventional 3-form
flux in the non-compact Calabi-Yau threefold. This flux does not pierce the
compact cycles of the local geometry, but instead has support at
infinity.\footnote{In \cite{Aganagic:2008qa} a related set-up is
  studied from a different perspective.} 
In \Cref{sec:OOPlocalCY} we exemplify this with the study of
local Calabi-Yau geometries modeled on a Riemann surface.

In \Cref{sec:factorization} we turn to the second goal of this
chapter: Finding a ``natural'' way to generate
the supersymmetry breaking flux configurations described above while starting
from a more standard setup. In this process, we also clarify the meaning
of flux which has support at infinity and the various subtleties related
to it. The natural interpretation of the
flux described in the previous paragraph emerges once we embed the
previous supersymmetry-breaking local singularity into a bigger IIB
compactification with standard flux of compact support. \Cref{sec:extwo}
supplements this general discussion by  
providing an explicit demonstration of
the factorization limit in the class of local geometries studied in
\Cref{sec:OOPlocalCY}. Matrix model techniques can be
used to compute the prepotential in the factorization limit.\footnote{This chapter is a 
   shortened version of 
 \cite{Hollands:2008cs}. In particular, it doesn't include the
   Appendices 
 of \cite{Hollands:2008cs} where the prepotential for
 the Cachazo-Intriligator-Vafa/Dijkgraaf-Vafa geometry is studied using
 the dual matrix model.
 }  
Finally, we finish in \Cref{sec:conclusions} with some concluding
remarks concerning the 
generalization of our story to other $\cN=2$ contexts, such as $M$ and
$F$-theory compactifications.

Let us start by introducing superpotentials, scalar potentials, metastable vacua
and the OOP mechanism.

\section{Ooguri-Ookouchi-Park formalism}\label{sec:OOP}

$\cN=2$ supersymmetry in four-dimensional gauge theories, as introduced in
Section~\ref{sec:SWcurves}, can be broken to $\cN=1$ supersymmetry by
adding a \Index{superpotential} $W(\phi)$ to the low energy effective
action. The superpotential generates a \Index{scalar potential}
$V(\phi)$ on the complex structure moduli space $\cM_q$ of the $\cN=2$ gauge
theory. When the gauge theory is perturbed by just a small
superpotential, the scalar potential is computed by
\begin{align*}
 V(\phi) = G^{i \bar{j}} \lotjesd_i W \overline{\lotjesd_j W},    
\end{align*}
to lowest order in the perturbation. As an implication not all the
vacua parametri-zed by $\cM_q$ are equivalent  
anymore. Furthermore, the value of $\phi$ will be stabilized in a local minimum of
the scalar potential. So adding a superpotential lifts the degeneracy
in the moduli of $\phi$.

Notice that since the quantum metric $G_{i \bar{j}}$ is positive
definite, the scalar potential $V(\phi) \ge 0$.  
Now, since the Hamiltonian  of a supersymmetric theory is
an anti-commutator of the supersymmetry generators,  $H = \{Q^{\dag},
Q \}$,  the potential $V$ should vanish for a four-dimensional vacuum
that preserves  supersymmetry. This implies that other local
minima of $V$ correspond to non-supersymmetric four-dimensional
vacua.  When the potential has more than one (local)
minimum, there is a probability of quantum tunneling.   
Such non-supersymmetric vacua are therefore \index{metastable vacuum}
\emph{metastable}.   

For phenomenological reasons it is very interesting to study the
above perturbed $\cN=2$ gauge theories. In particular, it is
important to find out whether there are choices for the superpotential
$W$ such that the scalar potential $V$ admits non-supersymmetric vacua.  
This was addressed in
\cite{Intriligator:2007py,Pastras:2007qr,Ooguri:2007iu} and affirmed
in the latter two articles. H.~Ooguri, Y.~Ookouchi
and C.-S.~ Park observed in \cite{Ooguri:2007iu} that it is even possible
to create a metastable 
vacuum at any generic point of the complex structure moduli space of an $\cN=2$
supersymmetric gauge theory by turning on a suitable small
superpotential deformation. Their proof just
depends on the sectional curvature of the quantum moduli space. To
create a metastable vacuum at $p \in \cM_q$ the sectional curvature
at $p$ should be positive semi-definite. This means that for any holomorphic
vector field $w \in  T \cM_q$ the curvature $R$ satisfies $$ \langle \,
w, \,R(v,v) \, w \,\rangle \ge 0,$$ for all $v \in (T \cM_q)_p$. The
curvature of special K\"ahler manifolds, of which the complex 
structure moduli space is an example, is indeed semi-positive
\cite{Ooguri:2007iu}.   

If we want to realize a nonsupersymmetric minimum at some
point $p$ in the moduli space, the OOP procedure tells us to
first construct holomorphic normal coordinates around $p$. Choose any
local coordinates $X^i$ near $p$. Then the holomorphic K\"ahler normal coordinates
are defined as 
\begin{equation}
Z^i =
X^{'i}+\tilde{G}^{i\bar{j}}\sum_{n=2}^{\infty}\frac{1}{n!}\partial_{i_3}\ldots\partial_{i_n}\tilde{\Gamma}_{j
  i_1 i_2} X^{' i_1} X^{' i_2}\ldots 
X^{' i_n},\label{KNCdef}
\end{equation}
where $X^{'i}=
X^i-X_0^{i}$ and $X^i_0 = X^i(p)$
\cite{AlvarezGaume:1981hn,Hull:1985pq, Higashijima:2000wz}. 
Furthermore, $\Gamma$ is a Christoffel symbol and the tilde $\tilde{\
}$ means evaluation at $X=X_0$.  We 
then choose the superpotential 
\begin{equation}
W=k_i Z^i,\label{WeffKNC}
\end{equation}
with $k_i \in \C$, consisting of a linear combination of the $Z^i$. 
Stability can be demonstrated by expanding $V$ near $p$
\begin{equation*}
V(Z^i)=k_i\bar{k}_{\bar{j}}\tilde{G}^{i\bar{j}}+k_i\bar{k}_{\bar{j}}\tilde{R}^{i\bar{j}}_{\,\,\,\,k\bar{l}}Z^k\bar{Z}^{\bar{l}}+{\cal{O}}(Z^3). 
\end{equation*}
Since $R$ is positive definite at generic points $p \in \cM_q$, the
quadratic term in the above expansion is generically positive. In that
case the vacuum at $Z^i=0$ is naturally metastable.

As a result, any potential that agrees with \eqref{WeffKNC}
near $X_0^{i}$ to cubic order will engineer a nontrivial vacuum at
$X_0^{i}$. For non-generic $X_0$, the curvature may have a zero
eigenvalue in which case higher order agreement with \eqref{WeffKNC}
is required. More remarks about the OOP formalism can be found in
\cite{Marsano:2007mt}.

\section{Geometrically engineering the OOP formalism}\label{sec:generalcase}

In string compactifications one can generate scalar potentials by
for example turning on higher dimensional gauge fields across cycles of the
internal manifold, inserting D-branes and/or non-perturbative
effects. Recent reviews include \cite{Denef:2008wq,Douglas:2006es}.
In this section we geometrically engineer the OOP formalism by turning
on fluxes in a type IIB local Calabi-Yau compactification. These fluxes
will however be non-conventionally supported at infinity.     

\subsection{Flux at infinity}\label{sec:fluxinfinity}

Before introducing these fluxes that grow large at infinity, let us
review some general aspects of type IIB flux compactifications. 

In \Cref{sec:CYcomp} we summarized how a type IIB Calabi-Yau
compactification leads to an ${\cal N}=2$ supergravity theory in 4d
coupled to $h^{2,1}$ vector multiplets and $h^{1,1}+1$
hypermultiplets. We discussed how the vector multiplets play an
important role in the geometrical engineering of $\cN=2$
supersymmetric gauge theories. In this chapter we concentrate as well
on the dynamics of the vector multiplets.   

Denote a symplectic basis of 3-cycles on the
compact Calabi-Yau threefold $X$ by $\{{A}^i,{ B}_j\}$ with
$i,j=0,1,\dots,h^{2,1}$, and denote the periods of the nowhere
vanishing holomorphic $(3,0)$-form $\Omega$ by $X^i$ and $F_j$. Remember
that the metric on the complex structure moduli space is special
  K\"ahler  and the K\"ahler potential is given by
\begin{equation}
  K = -\log\left( i \int \Omega \wedge \overline{\Omega}\right),
\label{Kcomplex}
\end{equation}
which is an exact result which does not receive any $\alpha'$ or $g_s$
corrections.

The easiest way to lift the phenomenologically unrealistic moduli space
of these Calabi-Yau compactifications is to turn on fluxes through the compact
cycles of the Calabi-Yau. In type IIB theory we can turn on RR and NS-NS
3-form flux $F_3$ and $H_3$ through the 3-cycles of the threefold. This
generates a Gukov-Vafa-Witten superpotential for the complex structure moduli
\cite{Michelson:1996pn, Gukov:1999ya, Taylor:1999ii} given by
\begin{equation}
  W = \int G_3 \wedge \Omega,
\label{gvw}
\end{equation}
where $G_3 = F_3 -\tau H_3$ and $\tau = C_0 + i/g_s$. The scalar
potential is computed by the standard ${\cal N}=1$ 
supergravity expression\footnote{In this expression the indices $a,b$
  run over complex structure moduli, K\"ahler moduli and the
  axion-dilaton. We denote by $\widetilde{K}$ the total K\"ahler
  potential for all moduli and by $K$, as in \eqref{Kcomplex}, the one for the
  complex structure moduli alone.}
\begin{equation*}
  V= e^{\widetilde{K}} \left( G^{a\overline{b}}D_a W \overline{D_bW}
-3|W|^2\right),
\label{potsugra}
\end{equation*}
where $G_{a\overline{b}}$ is the metric on the moduli space derived
from the K\"ahler potential $\widetilde{K}$, and where we have introduced the
K\"ahler covariant derivative $D_a W = \partial_a W +
( \partial_a\widetilde{K} ) W$.

The $F_3$ and $H_3$ fluxes generate charge for the $F_5$-form via a
Chern-Simons coupling in the 10d IIB supergravity action. The $F_5$
flux has nowhere to end, so we are lead to the tadpole cancellation
condition for IIB compactifications
\begin{equation*}
  {1\over l_s^4}\int F_3 \wedge H_3 + Q_{D3} = 0,
\label{tadcanc}
\end{equation*}
where $Q_{D3}$ receives positive contribution from probe D3 branes and
negative contribution from induced charge on D7 and orientifold
planes.

\subsubsection{Local limit and flux at infinity}
\label{subsec:locallimit}

In the limit in which we zoom in to some singularity locus of
a compact Calabi-Yau (see \Cref{sec:decouplegravity}) the structure of special geometry described above
reduces to rigid special geometry, which is relevant for the low
energy dynamics of ${\cal N}=2$ gauge theories. In this case the
K\"ahler potential reduces to
\begin{equation}
  K = i\int \Omega \wedge \overline{\Omega},
\label{krigid}
\end{equation}
and the K\"ahler covariant derivative $D_i$ reduces to the ordinary
derivative $\partial_i$.

As in the compact case, the addition of fluxes to the local Calabi-Yau
introduces a superpotential for the moduli. The dynamics of the
K\"ahler moduli and the dilaton decouple, and we can concentrate on
the normalizable complex structure moduli. The superpotential is still
given by \eqref{gvw}, but now the scalar potential is computed by the
rigid ${\cal N}=2$ expression
\begin{equation*}
  V = G^{i\overline{\jmath}}\partial_i W \overline{\partial_j W}=\int G_3
  \wedge * \overline{G_3}.
\label{potrigid}
\end{equation*}

Since we are in a noncompact Calabi-Yau it is not necessary to impose
the tadpole cancellation condition. Instead, the quantity
\begin{equation*}
  \int F_3 \wedge H_3
\label{tadrigid}
\end{equation*}
represents the $F_5$ flux going off to infinity and remains constant
as we vary the moduli. We will use this to simplify the potential in the
next section.

In most treatments of fluxes in noncompact Calabi-Yau manifolds the
assumption is made that the flux is threading the compact cycles of the
singularity and is going to zero at infinity.  As we explained in the
introduction the goal of this chapter is to study the dynamics in the case
where the flux is actually coming in from infinity and is not supported
on the compact three-cycles. Of course, in a local singularity inside a
bigger compact Calabi-Yau, what is meant by infinity is the rest of the
Calabi-Yau and we should think of flux coming from infinity as flux
leaking towards the singularity from the other compact cycles.

More precisely, in a noncompact Calabi-Yau threefold we consider the vector space
$H^3(X)$ of harmonic 3-forms which do not necessarily have compact
support, so they can grow at infinity. The harmonic 3-forms of compact
support form a linear subspace $H^3_{\text{cpct}}(X)\subset H^3(X)$. There is a
natural way to define the complement subspace $H^3_{\infty}(X)\subset
H^3(X)$ as the harmonic forms with vanishing integrals on the compact
3-cycles\footnote{We should clarify that we are not interested in the
most general harmonic 3-form with noncompact support, but only in a
restricted subset characterized by 3-forms which grow in a
``controlled'' way at infinity. This means that we want to consider
forms which have at most a ``pole'' of finite order at infinity, and not
essential singularities.  This statement has a nice interpretation in
the example where we have a local Calabi-Yau based on a Riemann surface
that we will study later. Another way to state this restriction is that
we will consider harmonic 3-forms on a local Calabi-Yau which do have a
lift to the original Calabi-Yau that we started with before we took the
local limit near its singularity.\label{footnote:bura}}.  Then we have
the decomposition
\begin{equation}
  H^3(X) = H^3_{\infty}(X) \oplus H^3_{\text{cpct}}(X).
\label{formdec}
\end{equation}
We will also refer to the forms in $H^3_{\text{cpct}}(X)$ as harmonic 3-forms with
compact support and to those in $H^3_{\infty}(X)$ as 3-forms with support
at infinity.

Now we want to consider the case where the 3-form field strength that
we have turned on has support at infinity
\begin{equation*}
  G_3 \in H_{\infty}^3(X),
\label{g3infinity}
\end{equation*}
which means that $G_3$ has zero flux through the compact cycles
\begin{equation*}
  \int_{A^i}G_3 = \int_{B_i}G_3 =0.
\label{g3zerocomp}
\end{equation*}
The intuitive picture that one should keep in mind, is that this flux at
infinity represents usual flux piercing other 3-cycles which are very 
far away from the singularity in the big Calabi-Yau. As we will see in more
detail in the next section, in this case and if one zooms into the
local singularity it is a good approximation to treat the flux from the
distant 3-cycles as flux which ``diverges'' at infinity. In other words
both $H^3_\infty (X)$ and $H^3_{\text{cpct}}(X)$ correspond to the usual
$H^3_{\text{cpct}}(\widetilde{X})$ of the bigger Calabi-Yau $\widetilde{X}$ in
which the singularity $X$ develops.

\subsubsection{Flux potential}

What is maybe more surprising is that the 3-form flux $G_3$ with
support at infinity generates a potential for the complex structure
moduli of the singularity $X$, 
even though it is not directly piercing the compact cycles of $X$, as can
be seen from \eqref{g3zerocomp}. 
Our starting point for the computation of this potential is the
energy stored in the 3-form field
\begin{equation}
  \widetilde{V} = \int G_3 \wedge * \overline{G_3}.
\label{vinf}
\end{equation}
Since $G_3$ has noncompact support, this is a divergent integral
meaning that the energy of the flux is infinite. This was to be
expected and is not really a problem, since we are interested in the
\emph{changes} of this energy as we vary the sizes of the 3-cycles in
the neighborhood of the singularity. We would like to throw away the
divergent, moduli independent piece of this quantity and keep the
finite, moduli dependent one. A nice way to achieve this is to use the
fact that the net $F_5$ form flux leaking off at infinity, being a
topological quantity, has to be kept constant as we vary the
moduli. It is easy to show that we can write
\begin{equation*}
  \int G_3 \wedge \overline{G_3}= (\tau - \overline{\tau}) \int F_3 \wedge H_3,
\label{fhg}
\end{equation*}
and the left hand side must be constant for the reason we explained. Since it is a constant we can subtract it from the potential and define
\begin{equation*}
  V \equiv \int G_3 \wedge * \overline{G_3} - \int G_3 \wedge  \overline{G_3} .
\label{vregul}
\end{equation*}
It is easy to show that this is equal to
\begin{equation}
  V = \int G_3^- \wedge * \overline{G_3^-},
\label{potf}
\end{equation}
where $G^-_3$ is the imaginary anti-self dual part of the $G_3$ flux
\begin{equation*}
  * G^-_3 = -i G^-_3.
  \label{g3asd}
\end{equation*}
The expression \eqref{potf} is the finite and moduli dependent piece
of the potential \eqref{vinf}.

\subsubsection{Simplifying the potential}
\label{subsec:potsimpl}
In this subsection we simplify the expression \eqref{potf} for the
potential. In general we have the following relation between
the Hodge decomposition and the $*$ operator on a threefold 
\begin{equation*}\begin{split}
  & * H^{3,0} = -i H^{3,0}, \, \qquad * H^{1,2} = -i H^{1,2},\\
  & * H^{2,1} = i H^{2,1}, \quad \qquad * H^{0,3}= i H^{0,3}.
\end{split}
\label{hodgestar}
\end{equation*}
Before we proceed we would like to analyze the relation between
the decomposition \eqref{formdec} and the Hodge decomposition. In general
we have the following decomposition\footnote{Again, we are only considering
a certain subset of all harmonic 3-forms with noncompact support, as 
explained in footnote \ref{footnote:bura}.}
\begin{equation*}
  H^3(X) =  H^{3,0}_{\infty}\oplus H^{3,0}_{\text{cpct}}\oplus
H^{2,1}_{\infty}\oplus
H^{2,1}_{\text{cpct}}\oplus \{ c.c.\}.
\label{hodgedec}
\end{equation*}
Harmonic forms in $H^{p,q}_{\text{cpct}}$ have compact support, while those in
$H^{p,q}_{\infty}$ do not, and are chosen to have vanishing
$\mathcal{A}$-periods on the compact cycles. (Notice that a harmonic
$(p,q)$-form cannot have vanishing periods on all compact cycles unless
it is identically zero.)  Since we do not want to break supersymmetry
explicitly by the boundary conditions of the system, we want our
configuration to be supersymmetric at infinity, which means that the
flux at infinity has to be imaginary self dual so
\begin{equation}
  G_3 \in H^{2,1}_\infty \oplus H^{2,1}_{\text{cpct}} \oplus H^{1,2}_{\text{cpct}}.
\label{g3atinf}
\end{equation}
where the subscript $\infty$ means that we have to consider the 
elements of the cohomology with noncompact support. We pick a basis
\begin{equation*}
  \Xi_m \in H^{2,1}_\infty,\qquad \Omega_i \in H^{2,1}_{\text{cpct}}
\label{formbasis}
\end{equation*}
with the following periods
\begin{equation}\begin{split}
 & \int_{A^i} \Xi_m = 0, \qquad\qquad \int_{A^i} \Omega_j =
    \delta^i_{j},\\ &\int_{B_i}\Xi_m = K_{im},\,\,\,\quad
    \,\,\,\,\,\int_{B_i} \Omega_j = \tau_{ij},
\end{split}\label{periodbasis}\end{equation}
where $\tau_{ij}$ is the period matrix of the Calabi-Yau, and $K_{im}$ are
holomorphic functions of the normalizable-complex structure moduli.

The flux has an expansion of the form
\begin{equation}
  G_3 = T^m \Xi_m + h^i \Omega_i + \overline{l^i}\, \overline{\Omega_i}.
\label{formflux}
\end{equation}
The parameters $T^m$ are fixed by the boundary conditions and have
to be kept constant as we vary the normalizable moduli. 
We have also assumed that
\begin{equation}
  \int_{A^i}G_3 = \int_{B_i} G_3 =0.
\label{zerocompflux}
\end{equation}
which means
\begin{equation}\begin{split}
 & T^m\int_{A^i} \Xi_m + h^j \int_{A^i} \Omega_j +
    \overline{l^j}\int_{A^i} \overline{\Omega_j} =0\\ &
    T^m\int_{B_i} \Xi_m + h^j \int_{B_i} \Omega_j +
    \overline{l^j}\int_{B_i} \overline{\Omega_j} =0.
\label{compactfluxzero}
\end{split}\end{equation}
The first equation of \eqref{compactfluxzero} implies that
\begin{equation*}
  \overline{l^j} = - h^j.
\label{blabla1}
\end{equation*}
and the second
\begin{equation*}
  h^i = - {1\over 2 i}\left({1\over \im\tau}\right)^{ij}\left(K_{jm}T^m \right).
\label{blabla2}
\end{equation*}
As we explained before, only the imaginary anti-self dual part of the
flux $ G_3^- = \overline{l^i}\,\overline{ \Omega_i}$ contributes to the
regularized potential and we have
\begin{align}
  V &= \int G_3^-\wedge \overline{G_3^-} \nonumber \\
&= {1\over 4}~ \overline{\left(K_{im} T^m \right)}\left({1\over \im \tau}\right)^{ij} \left(
  K_{jn}T^n\right).
\label{finalpot}
\end{align}
In this final expression the period matrix $\tau^{ij}$ and $K_{im}$ are
functions of the normalizable complex structure moduli, while $T^m$'s
have to be considered as constants which play the role of external
parameters.  This potential is in general very complicated and can have
local nonsupersymmetric minima for appropriate choices of the parameters
$T^m$ as we will explain later. 

Although we do not discuss this
here, from the viewpoint of flux compactification it is
a natural generalization to consider fluxes through the compact
3-cycles, relaxing the condition \eqref{zerocompflux}.  Such flux will
make additional contribution to the superpotential of the form $N^i
F_i-\alpha_i X^i$, $\alpha_i=\int_{B_i}\Omega$, which cannot be
controlled by external parameters and makes realization of OOP-like
vacua more difficult.

\subsubsection{Recovering the OOP potential}
\label{subsec:potprops}

The potential \eqref{finalpot} looks familiar. It
shares the same basic structure as the scalar potential that arises when
one adds a small superpotential to Seiberg-Witten theory.  This
connection can be made even more transparent by noting that $K_{im}$ can
in general be written as a total derivative with respect to the special
coordinates $X^i$.
\begin{equation*}K_{im}=\frac{\partial}{\partial
    X^i}\kappa_m(X^j),\qquad\qquad X^i=\oint_{{{A}}^i}\,\Omega. \end{equation*}
 One quick way to see this is to use the
identity $\int_X\,\Xi_m\wedge \partial_i\overline{\Omega}=0$ to derive
$$K_{im}\sim \int_{\partial X} \Lambda_m\wedge
\partial_i\overline{\Omega}$$ for a 2-form $\Lambda_m$ satisfying
$d\Lambda_m=\Xi_m$ on the boundary (at infinity) of $X$.  Because the
divergent contributions to $\Lambda_m$ at infinity can be chosen
independent of the dynamical moduli, we can pull the derivative outside
of everything.

With this notation, \eqref{finalpot} takes the standard form
\begin{equation}V=\frac{1}{4}\overline{\left(\frac{\partial W_{\text{eff}}(X^k)}{\partial X^i}\right)}
\left({1\over\im\tau}\right)^{ij}\left(\frac{\partial W_{\text{eff}}(X^k)}{\partial X^j}\right),\label{finalpotOOP}\end{equation}
where
\begin{equation*}W_{\text{eff}}(X^k)=T^m\kappa_m(X^k)\end{equation*}
is in fact proportional to the Gukov-Vafa-Witten superpotential
induced by the flux $G_3$. These equations make manifest the relation
between our flux-induced potential \eqref{finalpot} and that which arises in
deformed Seiberg-Witten theory and allows us to utilize the OOP technology
of \Cref{sec:OOP} to engineer supersymmetry-breaking vacua.  
\index{OOP potential}

\subsubsection{Lifetime of supersymmetry-breaking vacua}

Because we have managed to achieve supersymmetry-breaking vacua while
freezing all non-normalizable moduli, the energies $V_0$ will in general
be finite and independent of the cutoff scale $\Lambda_0$ that we use to
regulate the local geometry.  This means that our vacua are truly
metastable, even within this local model, and can decay to any of the
supersymmetric vacua that exist in these models.  Because the number the
supersymmetric vacua is potentially large and their properties quite
model-dependent, it is difficult to make general statements about the
lifetime of our OOP vacua.  Nevertheless, we recall here one observation
from \cite{Ooguri:2007iu}, namely that the decay rates will in general
scale like
\begin{equation}e^{-S} \quad \textrm{with} \quad  S\sim \frac{(\Delta Z)^4}{V_+},\end{equation}
where $\Delta Z$ is the distance in field space between the initial and
final vacuum state and $V_+$ is the difference in their energies.  By
simultaneously scaling all $T^m$ by a common factor,
$T^m\rightarrow\epsilon\, T^m$, we can retain our supersymmetry-breaking
vacua while decreasing $V_+$ by the same factor, $V_+\rightarrow\epsilon
V_+$.  In this manner, we see that, just as with OOP vacua in deformed
Seiberg-Witten theory, these OOP flux vacua can be made arbitrarily
long-lived. Because we should really think of the local Calabi-Yau as
sitting inside some larger compact geometry, one important caveat to
this statement of longevity is that the noncompact fluxes $T^m$ in
reality derive from a suitable set of compact fluxes in the full
Calabi-Yau.  This means that there will be a series of quantization
conditions that must be imposed that may affect the degree to which they
may be tuned.

\subsection{Local Calabi-Yau examples}\label{sec:OOPlocalCY} 

In the previous section, we saw that, starting from a compact Calabi-Yau and
taking a decoupling limit, one ends up with a local Calabi-Yau with noncompact
flux with support at infinity, which is nothing but the flux leaking
from the rest of the full Calabi-Yau that have been decoupled, towards ``our''
local Calabi-Yau\@.  Furthermore, this noncompact flux induces potential
\eqref{finalpot} for the complex structure moduli in the local Calabi-Yau.
Depending on the noncompact flux, this potential can be very complicated
and create nonsupersymmetric metastable vacua in the local Calabi-Yau; the OOP
mechanism \cite{Ooguri:2007iu} reviewed in
\Cref{sec:OOP} tells us 
exactly how this can be done.

In this section we study specific examples of our formalism. 
Let us
start with a non-compact Calabi-Yau modeled on a Riemann surface,
defined by
\begin{align}
X_{\Sigma}: \quad  uv-H(x,y)=0,
 \label{generalRSCY}
\end{align}
where $x,y$ can both be variables in $\C$ or $\C^*$. Recall that
the holomorphic 3-form of $X_{\Sigma}$ is given, e.g. for $x,y\in\C$, by
\begin{align*}
 \Omega = {du\wedge dx\wedge dy\over \partial H/\partial v}
 =\frac{du}{u} \wedge dx \wedge dy.
\end{align*}

Many important properties of the noncompact Calabi-Yau threefold
$X_{\Sigma}$ have an interpretation in terms of the underlying Riemann
surface $\Sigma$. For example, the compact 3-cycles $\{ A^i,
B_j \}$ in
$X_\Sigma$ are lifts of compact 1-cycles on $\Sigma$, which we 
denote by $\{a^i,b_j\}$ here. This one-to-one correspondence between 3- and
1-cycles shows an equivalence between the complex structure moduli on
$X_\Sigma$ and $\Sigma$.
 
A basis of (2,1)-forms with compact support on $X_\Sigma$ is given by
derivatives of $\Omega$ with respect to the normalizable complex
structure moduli: $\{\Omega_i=\partial_i \Omega\}$.  If $X_{\Sigma}$
were compact, these derivatives $\partial_i$ would be K\"ahler covariant
derivatives $D_i$ on the moduli space. Being noncompact instead, the
moduli space is described by rigid special geometry and, as we saw
before, the covariant derivatives simplify into partial derivatives.
Another reduction over the compact 3-cycles in the Calabi-Yau shows that
all these compactly supported $(2,1)$-forms $\Omega_i$ reduce to a basis
of holomorphic 1-forms $\omega_i$ on $\Sigma$.  Similarly, $(1,2)$-forms
$\overline{\partial_i\Omega}$ in $X_\Sigma$ reduce to antiholomorphic
1-forms $\overline{\omega}_i$ on $\Sigma$. The $\omega_i$ satisfy the
relations
\begin{equation}
 {1\over 2\pi i}\int_{a^i} \omega_j = \delta^i_{j},\qquad
 {1\over 2\pi i}\int_{b_i} \omega_j = \tau_{ij},
\label{prdomegaRS}
\end{equation}
where $\tau_{ij}$ is the period matrix of $\Sigma$.

The relation between the 3-cycles/3-forms on $X_{\Sigma}$ and the
1-cycles/1-forms on $\Sigma$ through the trivial $uv$-fibration being
understood, we can rewrite the various relations in
\Cref{sec:generalcase} in terms of the Riemann surface $\Sigma$.
First 
of all, the holomorphic 3-form $\Omega$ of $X_\Sigma$ is easily seen to
reduce to a meromorphic 1-form $\eta = y\,dx$ on the Riemann surface
in this case \cite{Klemm:1996bj, Aganagic:2003qj}. The special coordinates
parametrizing complex structure moduli are
\begin{align}
 X^i={1\over 2\pi i}\int_{a^i}\eta,\qquad
 F_i={1\over 2\pi i}\int_{b_i}\eta,
 \label{spclcoordRS}
\end{align}
and the K\"ahler potential \eqref{krigid} is given by
\begin{align}
 K=i\int_\Sigma \eta \wedge\overline{\eta}.
 \label{KahlerpotRSbla}
\end{align}
Recall that, in the special coordinates $\{X^i\}$, the moduli space
metric takes a particularly simple form:
\begin{align}
 ds^2=\left({\partial^2 K\over \partial X^i\partial\overline{X^j}}\right)
 dX^i d\overline{X^j}
 =(\im \tau)_{ij}\, dX^i d\overline{X^j},
 \label{spclmtrcRS}
\end{align}
as can be shown using ${\partial_i \eta}=\omega_i$ and
the Riemann bilinear relation. 

Let us now consider a very small deformation of the system breaking
supersymmetry to $\cN=1$, thus generating a potential $V$ for the
moduli. As we saw before, this can be accomplished by turning on 3-form
flux $G_3$ with support at infinity in the local Calabi-Yau.  This flux can be
thought of as leaking from the other part of the full compact Calabi-Yau, which
has been frozen in the decoupling limit.
We assume that the decoupling limit is taken consistently with the
elliptic fibration structure; namely, we assume that the noncompact flux
is supported at the asymptotic infinities of $\Sigma$, while being
compact in the direction of the $uv$-fibers.

The basis of (2,1)-forms with noncompact support, $\{\Xi_m\}$, in the
Calabi-Yau $X_\Sigma$ descend to meromorphic 1-forms $\{\xi_m\}$ on the
Riemann surface $\Sigma$, satisfying the relations 
\begin{align}
\int_{a^i}\xi_m=0,\qquad
\int_{b_i}\xi_m=K_{im},\label{KIM_lclCY}
\end{align}
which are reductions of \eqref{periodbasis}. Therefore, the
3-form flux $G_3$ with noncompact support on $X_\Sigma$, as given in
\eqref{formflux}, descends to a harmonic 1-form flux
\begin{equation}
 \begin{split}
 g&=g_H+\overline{g_A},\\
 g_H&= T^m \xi_m + h^i \omega_i,\qquad
 \overline{g_A}=\overline{l^i}\overline{\omega_i},
\end{split}
\label{1formRS}
\end{equation}
which will have poles at the punctures (or asymptotic legs) of
$\Sigma$. 
%
%
The 3-form flux $G_3$ in $X_\Sigma$ induces 
superpotential
\eqref{gvw}, which reduces to an integral on $\Sigma$:
\begin{equation*}
  W = \int_{\Sigma} g \wedge \eta,
\end{equation*}
while the associated scalar potential \eqref{potf} reduces to an
integral on $\Sigma$:
\begin{equation}
 V = \int_{\Sigma} g_A \wedge \overline{g_A}.
\label{scalarpotRS}
\end{equation}

If we require the condition \eqref{zerocompflux} that the flux
\eqref{1formRS} is zero through compact 3-cycles of $X_{\text{SW}}$, which
translates into
\begin{align}
 \int_{a^i}g=\int_{b_i}g=0,\label{0cptflxRS}
\end{align}
then by reducing the argument we made for general Calabi-Yau's in the
previous section to the Riemann surface $\Sigma$ (or simply
by borrowing the result \eqref{finalpot}), we can rewrite
\eqref{scalarpotRS} in terms of periods on $\Sigma$:
\begin{equation}
  V = {1\over 4} \overline{(K_{im}T^m)}
   \left({1\over \im \tau}\right)^{ij} 
   K_{jn}T^n.\label{scalarpotRS2}  
\end{equation}

Let us study this potential in more detail for Seiberg-Witten and
Dijkgraaf-Vafa geometries, and make remarks on the gauge theory
interpretation of the physics of these geometries. 



\subsubsection{Differentials on a hyperelliptic surface}

When the underlying Riemann surface is hyperelliptic, say
\begin{align*}
 \Sigma: \quad y^2 = f(x) 
\end{align*}
where $f(x)$ is a polynomial of degree $2N$, there are convenient
representations for $\{\xi_m\}$ and $\{\omega_i\}$. Since we will need
them later, let us briefly review them here.  

A basis of holomorphic differentials $\omega_i$ can be constructed by
\begin{align}
 \omega_i&={Q_i(x)\over y}dx={Q_i(x)\over \sqrt{f(x)}}dx,
\label{hypellomegaI}
\end{align}
where $Q_i(x)$ is a polynomial of degree up to $N-2$ chosen so that
\eqref{prdomegaRS} holds.  Note that this $\omega_i$ asymptots to $
\cO(x^{-2})dx$ when $x\to\infty$, $\widetilde\infty$. This means that it
is regular at $x=\infty,\widetilde\infty$.

On a hyperelliptic surface it is convenient to take the meromorphic
differentials of the second kind, $\xi_m$, as
\begin{align}
 \xi_m&={R_m(x)\over y}dx={R_m(x)\over \sqrt{f(x)}}dx,\qquad m\ge 1.
\label{hypellxiM}
\end{align}
Here, $R_m(x)=mx^{m+N-1}+\dots$ is a polynomial and the coefficients of
$x^{m+N-2}$, $\ldots$, $x^{N-1}$ are chosen so that
\begin{align}
 \xi_m&=\pm \left[mx^{m-1}+\cO(x^{-2})\right]dx,\qquad
 x \sim \infty,\widetilde\infty
\label{hypellxiMasym}
\end{align}
is satisfied.  Note that this
$\xi_m$ has poles at two points, $x=\infty,\widetilde\infty$, instead of
one.  The coefficients of $x^{N-2},\dots,x^0$ are chosen so that
\eqref{KIM_lclCY} is satisfied.

The meromorphic differential of the third kind, $\xi_0$, can be defined
likewise using a polynomial $R_0(x)=x^{n-1}+\dots$, where the
coefficients are chosen so that
\begin{align*}
 \xi_0&={R_0(x)\over y}dx=\pm \left[{1\over x}+\cO(x^{-2})\right]dx,\qquad
 x\sim\infty,\widetilde\infty
\end{align*}
holds and \eqref{KIM_lclCY} is satisfied.

Let us derive a formula that will be useful.  By
expanding the right hand side of the trivial identity $0=\int_{\Sigma}
\omega_i\wedge \xi_m$ by the Riemann bilinear identity, one finds
\begin{align*}
 0&=\sum_j \left(\int _{a^j}\omega_i \int_{b_j}\xi_m
 -\int _{a^j}\xi_m \int_{b_j}\omega_i
 \right)+\sum_{p=\infty,\widetilde\infty}\oint_{p}\omega_i\,d^{-1}\xi_m
 \notag\\
 &=K_{im}+ \sum_{p=\infty,\widetilde\infty}\oint_{p}\omega_i\,d^{-1}\xi_m.
\end{align*}
Because the behaviors of $\omega_i$ and $\xi_m$ at $x=\infty$ is the same as
those at $x=\widetilde\infty$ up to a sign, we find that
\begin{align}
 K_{im}
 &=-\sum_{p=\infty,\widetilde\infty}\oint_{p}\omega_i\,d^{-1}\xi_m
 =-2\oint_{\infty}\omega_i\,d^{-1}\xi_m
 =-2\oint_{\infty}x^m\omega_i.
\label{KIMidentity}
\end{align}

\subsubsection{Seiberg-Witten geometries}
\index{Seiberg-Witten geometry}

The $SU(N)$ Seiberg-Witten geometry 
\begin{equation}
X_{\text{SW}}:\quad
uv - H_{\text{SW}}(v,t) =0,\label{SWCY}
\end{equation}
with $v \in \C$ and $t \in \C^*$,
is an illustrative example of a local
Calabi-Yau threefold. The underlying Riemann surface $\Sigma_{\text{SW}}$ is a hyperelliptic curve 
\begin{equation}
\Sigma_{\text{SW}}:\quad
H_{\text{SW}}(v,t) = \Lambda^N \left( t + \frac{1}{t} \right) - P_N(v)=0
\label{SWcurve}
\end{equation}
where $P_N(v)=\prod_{i=1}^N (v- \alpha_i)$ is a polynomial of degree $N$
with the coefficient of $v^{N-1}$ being zero.
The coefficients of $P_N(v)$ are
normalizable moduli, while $\Lambda$ is a fixed parameter. 

 The
holomorphic 3-form on $X_{\text{SW}}$ is $\Omega_{\text{SW}} = \frac{du}{u} \wedge dv
\wedge \frac{dt}{t}$ and reduces to $ \eta_{\text{SW}} = v \frac{dt}{t}$
on the Riemann surface $\Sigma_{\text{SW}}$. For a description of the
bijection between 3-cycles on the local Calabi-Yau and 1-cycles on the
Seiberg-Witten curve, we refer to \Cref{sec:revisitSW}. 

The complex structure moduli space is conveniently parametrized by the
special coordinates \eqref{spclcoordRS}, which are conventionally
denoted by $a_i$, $i=1,\dots, N-1$ in the Seiberg-Witten
case
\begin{align}
 a_i
 ={1\over 2\pi i}\int_{A^1_i}\eta_{SW}
 ={1\over 2\pi i}\int_{A^1_i}v{dt\over t}.
 \label{spclcoordSW}
\end{align}
(To avoid confusion we denote the 1-cyles in the geometry
  with capital letters.)
As in \eqref{spclmtrcRS}, the moduli space metric takes the special form
for these:
\begin{align}
 ds^2=(\textrm{Im}\tau_{ij})\,da_i d\overline{a_j}.
 \label{SWmetrica}
\end{align}

Using $a_i$, the normalized basis of holomorphic 1-forms $\omega_i$ can
be obtained as follows. Differentiating \eqref{spclcoordSW} with respect
to $a^j$,
\begin{align*}
 \delta^i_j&={1\over 2\pi i}{\partial \over \partial a_j}\int_{A_1^i}v{dt\over t}.
\end{align*}
Comparing with the first equation in \eqref{prdomegaRS}, this means that
\begin{align}
 \omega_i= {\partial \over \partial a_i}\left(v{dt\over t}+d\sigma\right),
 \label{omegaIasader}
\end{align}
where the total derivative term $d\sigma$ is fixed by requiring that
$\omega_i=\cO(v^{-2})dv$ as $v\to\infty$.  Specifically, this leads to
$d\sigma= d(-v\log t)$ and $\omega_i$ is given by
\begin{align*}
 \omega_i
 &
 ={\partial \over \partial a_i}(-\log t\,dv)
 =-{{\partial P_N(v)/ \partial a_i}\over\sqrt{P_N(v)^2-4\Lambda^{2N}}}dv.
\end{align*}
Although $\log t$ may appear problematic because it is not single-valued
on the Riemann surface, its $a_i$ derivative is single-valued and does
not cause any problem.

As we discussed in \Cref{sec:fluxinfinity}, turning on
noncompact flux breaks $\cN=2$ supersymmetry to $\cN=1$ by inducing a
superpotential. As in \eqref{1formRS}, the 
3-form flux in $X_{\text{SW}}$ reduces to a harmonic 1-form
\begin{equation}
g = \sum_{m\ge 1} T^m  \xi_m 
 + \sum_{i=1}^{N-1} h^i \omega_i 
 + \sum_{i=1}^{N-1} \overline{l^i}\overline{\omega_i}.\label{1formgSW}
\end{equation}
 on $\Sigma_{\text{SW}}$.
Under the condition that the compact flux vanishes (\ref{0cptflxRS}),
this leads to the scalar potential \eqref{scalarpotRS2}.

We can write the superpotential we are adding to the system in a form
that will be useful later.  By manipulating the quantity $K_{jn}T^n$
appearing in \eqref{scalarpotRS2},
\begin{align*}
 K_{jn}T^n
 &=T^n\oint_{B^1_j} \xi_n
 =-2T^n\oint_{\infty} v^n \omega_j
 =2T^n{\partial \over \partial a_j }
 \left(\oint_{\infty} v^n \log t\, dv\right)
 \notag
\end{align*}
\begin{align*}
 &=-{2T^n\over N+1}{\partial \over \partial a_j }
 \left(\oint_{\infty} v^{n+1} {dt\over t}\right).
\end{align*}
Here we used \eqref{KIM_lclCY}, \eqref{KIMidentity} and \eqref{omegaIasader}.
By examining \eqref{scalarpotRS2},
and \eqref{SWmetrica}
one sees that the superpotential is given by
\begin{align}
 W_{\text{SW}}=\sum_{m} T^m u_{m+1},
 \label{supotRSgen}
\end{align}
where we defined
\begin{align}
 u_m\equiv {1\over 2\pi i m}\oint_{\infty} v^{m-1} \eta_{\text{SW}}
 ={1\over 2\pi i m}\oint_{\infty} v^m {dt\over t}.
 \label{defuMgeom}
\end{align}

So far everything was about geometry.  Now let us turn to the gauge
theory interpretation of these.  As we mentioned above, the local CY
geometry \eqref{SWCY} without flux realizes $\cN=2$ Seiberg-Witten
theory, with the hyperelliptic curve \eqref{SWcurve} identified with the
$\cN=2$ curve of gauge theory.  The special coordinates $a_i$ defined in
\eqref{spclcoordSW} correspond to the $U(1)$ adjoint scalars in the IR
and parametrize the Coulomb moduli space.  The superpotential
\eqref{supotRSgen} also has a simple gauge theory interpretation.  To
see it, we need the relation between the vev of the adjoint scalar
$\Phi$ and the curve $\Sigma_{SW}$, given by \cite{Cachazo:2002ry,
Cachazo:2002zk}:
\begin{align*}
 \langle \, \Tr \, {dv\over v-\Phi} \, \rangle
 ={dt\over t}=
 {P'_N(v)\over\sqrt{P_N(v)^2-4\Lambda^{2N}}}dv.
\end{align*}
In other words, $u_m$ defined geometrically in \eqref{defuMgeom} has an
interpretation in gauge theory as follows:
\begin{align*}
 u_m= {1\over m}\langle \,\Tr \, \Phi^m \, \rangle.
\end{align*}
From this, one immediately sees that  the superpotential
\eqref{supotRSgen} can be written as
\begin{align}
 W_{SW}
 =\sum_{m\ge 1}  {T^m\over m+1} \Tr \,\Phi^{m+1}
 =\Tr \,[W(\Phi)],\label{spotpertSW}
\end{align}
where we defined
\begin{align*}
  W(x)=\sum_m {T^m\over m+1} v^{m+1}.
\end{align*}
In \eqref{spotpertSW} $\Phi$ is understood as the chiral superfield
 whose lowest component is the adjoint scalar.

Therefore, the $\cN=2$ gauge theory perturbed by the single-trace
superpotential \eqref{spotpertSW} corresponds to the geometry
\eqref{SWCY} with the flux $g$ obeying the following asymptotic boundary
condition:
\begin{align}
 g\sim \sum_m m T^m v^{m-1}\,dv = W''(v)dv,
\end{align}
where we used \eqref{hypellxiMasym}.    The
perturbed $\cN=2$ theory is precisely the system which was shown in
\cite{Ooguri:2007iu, Pastras:2007qr} to have nonsupersymmetric
metastable vacua if the superpotential is chosen
appropriately. (It was shown in \cite{Ooguri:2007iu} to be
possible to create metastable vacua by a single-trace superpotential of
the form \eqref{spotpertSW} at any point in the Coulomb moduli space for
$SU(2)$ and at least at the origin of the moduli space for $SU(N)$.)
Therefore, it tautologically follows that the IIB Seiberg-Witten
geometry with flux at infinity also has metastable vacua, if we tune the
parameters $T^m$ appropriately.

As we mentioned above, the IIB Seiberg-Witten geometry is dual to a IIA
brane configuration of NS5-branes and D4-branes which can be lifted to
an M5-brane configuration.  In \cite{Marsano:2008ts} it was shown that
a superpotential perturbation corresponds in the M-theory setup to
``curving'' the $\cN=2$ configuration of the M5-brane at infinity in a
way specified by the superpotential.  The metastable gauge theory
configuration of \cite{Ooguri:2007iu, Pastras:2007qr} was realized as a
metastable M5-brane configuration and its local stability was given a
geometrical interpretation.  The above proof of \eqref{spotpertSW} is
exactly in parallel to the one given in \cite{Marsano:2008ts} for the
M-theory system.
In passing, it is also worth mentioning that the M-theory analysis of
\cite{Marsano:2008ts} revealed that at strong coupling the
nonsupersymmetric configuration ``backreacts'' on the boundary condition
and it is no longer consistent to impose a holomorphic boundary
condition specified by a holomorphic superpotential, which is in accord
with \cite{Bena:2006rg}.  Therefore, also in the IIB flux setting, it is
expected that if we go beyond the approximation that the flux does not
backreact on the background metric, nonsupersymmetric flux
configurations will backreact and it will be impossible to impose a
holomorphic boundary condition of the type \eqref{1formgSW}.

Although we do not discuss this here, from the viewpoint of
flux compactification, it is a natural generalization to consider fluxes
through the compact 3-cycles.  Such flux will make additional
contribution to the superpotential of the form $e_i a^i+m^i F_i$.
On the gauge theory side, in the Seiberg-Witten theory, this can be
interpreted as perturbation one adds at the far IR, but its UV
interpretation is not clear \cite{Marsano:2007mt}.

\subsubsection{Dijkgraaf-Vafa geometries}
\index{Dijkgraaf-Vafa geometry}

Another example of geometries of the type \eqref{generalRSCY} is type
IIB on
\begin{equation}
  X_{\text{DV}}:\qquad uv - H_{\text{DV}}(x,y)=0, \qquad x,y \in \C,
   \label{DVCY}
\end{equation}
where the underlying Riemann surface $\Sigma_{\text{DV}}$ is a hyperelliptic curve
\begin{equation}
   \Sigma_{\text{DV}}: \qquad
    H_{\text{DV}}(x,y)\equiv y^2-\left[P_n(x)^2- f_{n-1}(x)\right]=0
\label{DVcurve}
\end{equation}
and $P_n(x)$ and $f_{n-1}(x)$ are polynomials of degree $n$ and $n-1$,
respectively. If we write
\begin{align*}
 f_{n-1}(x)=\sum_{i=1}^{n-1} b_{i} x^i,
\end{align*}
then the coefficients of $P_n(x)$ as well as $b_{n-1}$ are
non-normalizable and fixed,
while $b_i$, $i=0,\dots,n-2$ are normalizable complex structure moduli.
The holomorphic 3-form is $\Omega_{\text{DV}}={du\over u}\wedge dx\wedge dy$
which reduces to
\begin{align*}
 \eta_{\text{DV}}=x\,dy  
\end{align*} 
on the Riemann surface $\Sigma_{\text{DV}}$.  The geometry \eqref{DVCY} was
studied by Cachazo, Intriligator and Vafa (CIV) \cite{Cachazo:2001jy}
(see also \cite{Cachazo:2002pr}) in the context of large $N$ transition
\cite{Gopakumar:1998ki, Vafa:2000wi} and further generalized in
\cite{Cachazo:2001gh, Cachazo:2001sg}. The Dijkgraaf-Vafa (DV)
conjecture \cite{Dijkgraaf:2002fc, Dijkgraaf:2002vw, Dijkgraaf:2002dh}
was also based on the same geometry.  We will refer to this geometry as
the CIV-DV geometry \eqref{DVCY} or as the Dijkgraaf-Vafa geometry
henceforth.

The structure of the underlying hyperelliptic Riemann surface
$\Sigma_{\text{DV}}$ \eqref{DVcurve} is similar to the Seiberg-Witten case
\eqref{SWcurve}; $\Sigma_{\text{DV}}$ is a genus $n-1$ surface with two punctures
at infinity.  If we represent $\Sigma_{\text{DV}}$ as a two-sheeted $x$-plane
branched over $2n$ points, those infinities correspond to $x=\infty$ on
the two sheets.  The coefficients of $P_n(x)$, which are
nonnormalizable, determine the position of the $n$ cuts on the
$x$-plane, while the coefficients of $f_{n-1}(x)$, which are
normalizable, are related to the sizes of the cuts.  

The first homology
$H_1(\Sigma_{\text{DV}})$ is spanned by $n-1$ pairs of compact $a$- and
$b$-cycles $(a^i,b_j)$, $i,j=1,\dots,n-1$ with in addition a closed
cycle $a^\infty$ around one of the infinities which is dual to the
noncompact $b$-cycle $b_\infty$ connecting two infinities.  Because
$x,y\in\C$, compact $a$- and $b$-cycles on $\Sigma_{\text{DV}}$ are all
contractible in the $x,y$-plane and hence all compact 1-cycles on
$\Sigma_{\text{DV}}$ lifts to 3-cycles in $X_{\text{DV}}$ with $S^3$ topology.

The special coordinates \eqref{spclcoordRS} in this case is conventionally
denoted by $S^i$, $\Pi_i$:
\begin{align}
 S^i
 ={1\over 2\pi i}\int_{a^i}\eta_{\text{DV}}
 , \qquad
 \Pi_i
 ={1\over 2\pi i}\int_{b_i}\eta_{\text{DV}}
 ,
 \label{spclcoordDV}
\end{align}
for which, as in \eqref{spclmtrcRS}, the moduli space metric takes the
special form:
\begin{align*}
 ds^2=(\textrm{Im}\tau_{ij})\,dS^i d\overline{S^j}.
\end{align*}
We can write the basis of holomorphic
1-forms $\omega_i$ using $S^i$ as:
\begin{align*}
 \omega_i
 &
 ={\partial \over \partial S^i}(-y\,dx)
 ={\partial f_{n-1}(x)/ \partial S^i \over 2\sqrt{P_n(x)^2-f_{n-1}(x)}}\,dx.
\end{align*}

Adding flux at infinity  works
the same as in the Seiberg-Witten case.  The Riemann surface $\Sigma_{\text{DV}}$
is hyperelliptic and we take $\{\xi_m\}$ and $\{\omega_i\}$ to be the
ones given in \eqref{hypellomegaI} and \eqref{hypellxiM}.  Just like
\eqref{1formRS} and \eqref{1formgSW}, the 3-form flux in $X_{\text{DV}}$ reduces
to a harmonic 1-form on $\Sigma_{\text{DV}}$
\begin{equation*}
g = \sum_{m\ge 1} T^m  \xi_m 
 + \sum_{i=1}^{N-1} h^i \omega_i 
 + \sum_{i=1}^{N-1} \overline{l^i}\overline{\omega_i}.
\end{equation*}
Under the condition that the compact flux vanishes
\eqref{0cptflxRS}, the 1-form $g$ leads to the scalar
potential \eqref{scalarpotRS2} which, just as we derived
\eqref{supotRSgen}, can be shown to correspond to the following
superpotential:
\begin{align}
 W_{\text{DV}}=\sum_m T^m \Sigma_{m+1},
 \label{supotDVgeo}
\end{align}
where we defined
\begin{align}
 \Sigma_m
 \equiv {1\over 2\pi i m}\oint_{\infty} x^{m-1} \eta_{\text{DV}}
 = {1\over 2\pi i m}\oint_{\infty} x^{m}\, dy.
 \label{defSigmageo}
\end{align}
The 1-form $\eta_{\text{DV}}$ depends on the complex structure moduli $S^i$
of the Riemann surface \eqref{DVcurve}. Therefore, by changing the
parameters $T^m$, we can generate a superpotential which is a quite
general function of $S^i$'s.
The OOP mechanism \cite{Ooguri:2007iu} states that, if one tunes
superpotential appropriately, one can create a metastable vacuum at any
point of the $\cN=2$ moduli space.  Therefore, also for this
Dijkgraaf-Vafa geometry, we expect to be able to create metastable vacua
by appropriately tuning $T^m$, {\it i.e.}, flux at infinity.  Indeed,
at the end of this section 
we will demonstrate the existence of metastable
vacua in a simple example.

We have been focusing on the case where there is flux at infinity but
there is \emph{no} flux through \emph{compact} cycles.  However, let us
digress a little while and think about the case where there \emph{is}
flux through compact cycles but there is \emph{no} flux at infinity.  In
this case, the IIB system has a standard interpretation
\cite{Cachazo:2001jy, Cachazo:2002pr, Dijkgraaf:2002fc,
Dijkgraaf:2002vw, Dijkgraaf:2002dh} as describing the IR dynamics of
$\cN=2$ $SU(N)$ theory broken to $\cN=1$ by a superpotential
$W=\Tr[W_n(\Phi)]$, $W'_n(x)=P_n(x)$, with the moduli $S^i$ identified
with glueball fields.  More precisely, if there are $N^i$ units of flux
through the cycle $a^i$, where $N=\sum_i N^i$, then the system
corresponds to the supersymmetric ground state of $SU(N)$ gauge theory
broken to $\bigl[\prod_i SU(N^i)\bigr]\times U(1)^{n-1}$.
It is important to note that this equivalence between the Dijkgraaf-Vafa
flux geometry and gauge theory is guaranteed to work only for
holomorphic dynamics, or for the $F$-term.  On the geometry side, one is
considering the underlying geometry \eqref{DVCY} determined by $P_n(x)$
and small flux perturbation on it.  On the gauge theory side, this
corresponds to the limit of large superpotential, where one has no
control of the $D$-term.  Therefore, there is no {\it a priori\/} reason
to expect that the $D$-term of the Dijkgraaf-Vafa geometry, which
governs {\it e.g.}\ existence of nonsupersymmetric vacua, and that of
gauge theory are the same, even qualitatively.  After all, two systems
are different theories and it is only the holomorphic dynamics that is
shared by the two.

Despite such subtlety, it is interesting to ask what is the gauge theory
interpretation of adding flux at infinity, in addition to flux through
compact cycles.
It is known that the curve \eqref{DVcurve} is related to the vev in
gauge theory as \cite{Dijkgraaf:2002fc, Dijkgraaf:2002vw,
Dijkgraaf:2002dh, Cachazo:2002ry, deBoer:2004he}:
\begin{align*}
  -{1\over 32\pi^2} \langle \, \Tr \, \frac{\cW^2}{x-\Phi} \,\rangle dx
  ={y\,dx}
  =\sqrt{P_n(x)^2-f_{n-1}(x)}\,dx.
 \end{align*}
where $\cW^2=\cW_\alpha \cW^\alpha$ and $\cW_\alpha$ is the gaugino
field.  Comparing this with \eqref{defSigmageo}, one finds that the
quantity $\Sigma_m$ defined geometrically in \eqref{defSigmageo} has the
following interpretation:
\begin{align*}
 \Sigma_m= \frac{1}{32\pi^2}\langle \, \Tr \, \cW^2\Phi^{m-1} \, \rangle.
\end{align*}
Therefore the superpotential \eqref{supotDVgeo} can be written as
\begin{align}
 W_{\text{DV}}
 =\frac{1}{32\pi^2} \sum_m T^m \Tr \, [\cW^2\Phi^m]
 =\frac{1}{32\pi^2}\Tr \, [\cW^2 M(\Phi)],
 \label{WDVgt}
\end{align}
where we defined
\begin{align*}
 M(x)=\sum_m T^m x^{m}.
\end{align*}
So flux at infinity of the 
asymptotic form
\begin{align*}
 g\sim \sum_m m T^m x^{m-1}\,dx = M'(x)dx,
\end{align*}
corresponds in gauge theory to adding a novel superpotential of the form
\eqref{WDVgt}.  Again, this correspondence must be taken with a grain of
salt, since it holds only for holomorphic physics.

Note also that flux through compact cycles will induce glueball
superpotential \cite{Cachazo:2001jy} of the form $\alpha_i S^i + N^i
\Pi_i(S)$ added to \eqref{supotDVgeo}.  Because this part does not contain
tunable parameters such as $T^m$ that can be made very small, it is
difficult, if not possible, to use the OOP mechanism to produce
metastable vacua in that case.

\begin{figure}[h!]
 \begin{center}
\includegraphics[width=7cm]{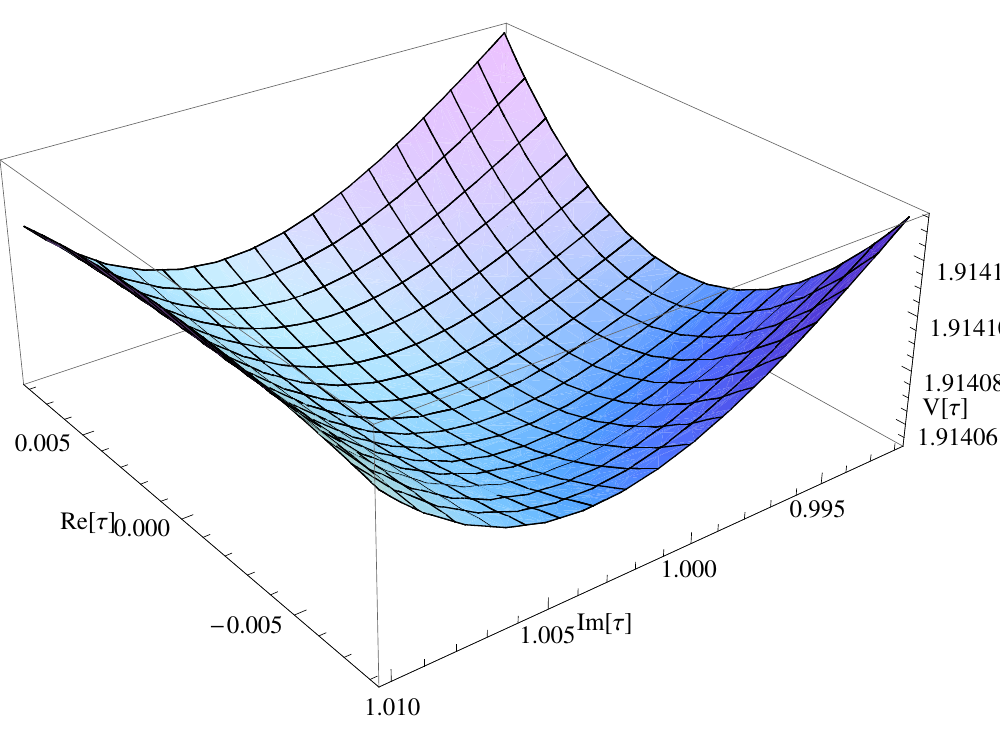}
 \caption{Plot of $V(\tau)$ in the neighborhood of our engineered OOP
   minimum at $\tau=i$.} \label{fig:potexample} 
 \end{center}
\end{figure}

In the situation where there is no flux through compact cycles, we do not
have an interpretation of the system as such an $SU(N)$ theory described
above, simply because $N=\sum_i N^i=0$.  

In Section 3.2 of
\cite{Hollands:2008cs} we work out a simple example, based on the
$SU(2)$ Dijkgraaf-Vafa geometry
\begin{align*}
H_{DV}(x,y)=y^2- (P_2(x)^2-b_0 )= 0, \quad \textrm{with} \quad P_2(x) =
x^2 - \frac{\Delta^2}{4}, 
\end{align*}
with flux at infinity and no flux
through compact cycles, where we demonstrate that we can truly realize
metastable vacua in type IIB using
the OOP mechanism outlined in the previous section by simply adjusting the
parameters $T^m$. Choosing $\tau=i$ for convenience, we find a
metastable vacuum by turning on
\begin{align*}
T^2=\frac{885}{8},\qquad T^6=-\frac{5832}{5\Delta^4},\qquad T^{10}=\frac{2400}{\Delta^8}.\end{align*}
The corresponding local minimum of the potential is plotted in Fig.~\ref{fig:potexample}.

\section{Embedding in a larger Calabi-Yau}\label{sec:factorization} 

In the previous sections we described how we can generate a
supersymmetry breaking potential for the complex structure moduli of a
local Calabi-Yau singularity by the introduction of 3-form flux which
has support at infinity. Allowing flux with noncompact support may
lead to various conceptual difficulties, such as the divergence of the
total energy density. To clarify these difficulties we would like to
sketch how such a 
system can be interpreted as an approximation of a larger Calabi-Yau
threefold with flux of compact support in a certain factorization limit.

\begin{figure}[h!]
 \begin{center}
\includegraphics[width=8.5cm]{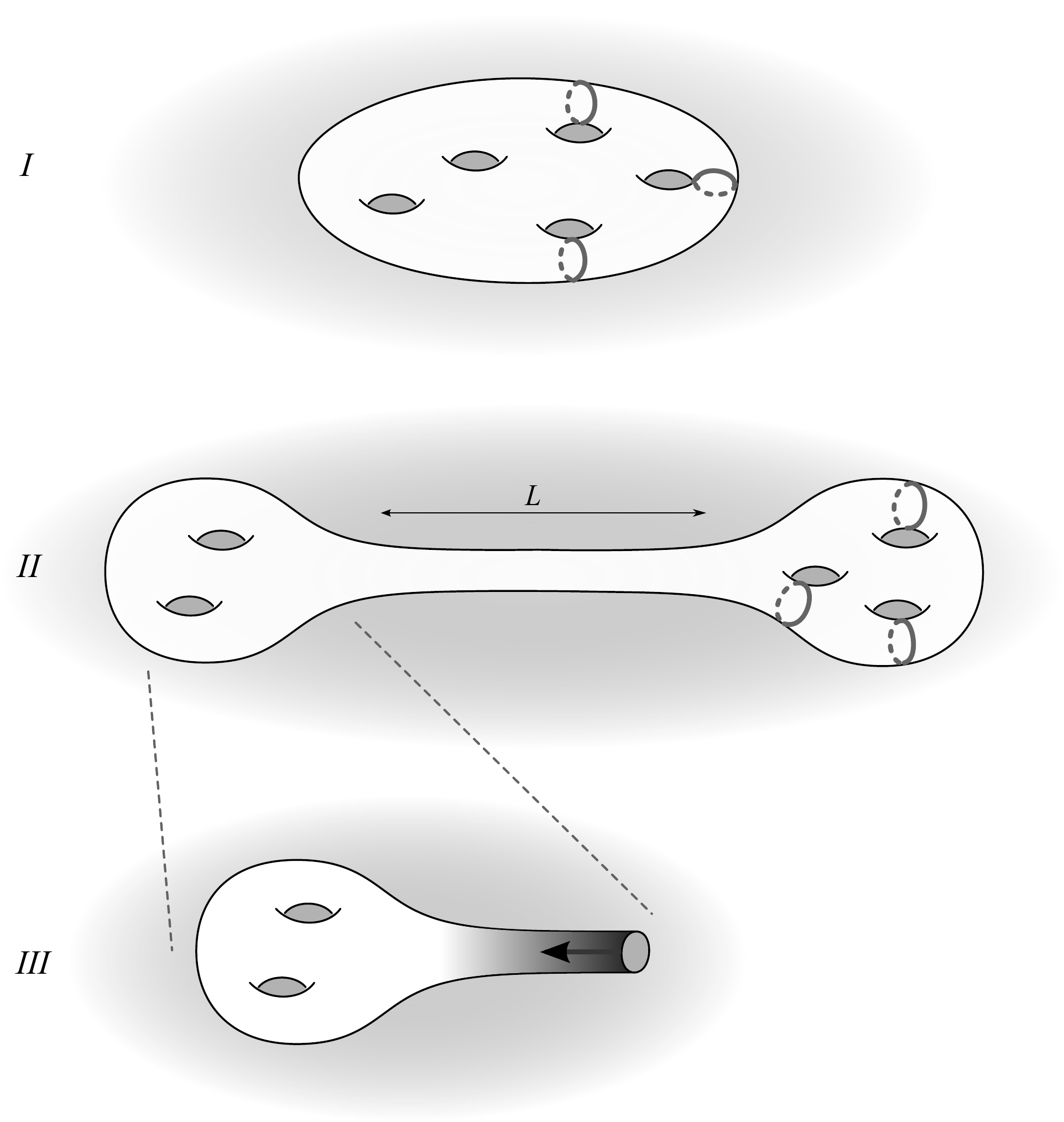}
 \caption{Factorization idea: In I we start with a generic
  Calabi-Yau with flux piercing through some of its 3-cycles, while making the
  distance between the cycles with and without flux very large in 
  II. This is seen as flux from infinity in the left sector without compact
  flux in III, and generates an OOP-like potential in that sector. 
  }
 \label{fig:overview}
 \end{center}
\end{figure}

 As shown in
figure~\ref{fig:overview}, the physical idea is to start with a
Calabi-Yau manifold with a set of three-cycles which are isolated from
the other three-cycles by a large distance. We turn on 3-form flux on
all cycles except for the isolated set. While the flux that we have
turned on is not piercing the isolated cycles, it does leak into their
region. (This means that the 3-form field strength is nonzero in
the region around the isolated set of 3-cycles, but once integrated over
one of these 3-cycles the integral is zero.)  It produces a
potential for their complex structure moduli. 

In the limit where the
distance between the two sets of cycles of the Calabi-Yau becomes very
large, which we will refer to as the \emph{factorization limit}, the
flux leaking towards the isolated set will start to look like the flux
coming from ``infinity''. In this sense, we manage to embed the
scenario considered in the previous section as a small part of a
larger Calabi-Yau with compactly supported flux.

\subsection{Factorization}

In this section we would like to understand this embedding into a bigger
Calabi-Yau in more detail. Our goal is to see how the potential
\eqref{finalpot} arises starting from the standard Gukov-Vafa-Witten
superpotential for 3-form flux in the larger Calabi-Yau. 
For simplicity we will work with a noncompact Calabi-Yau $X$,
\begin{equation*}
X:\quad  uv -H(x,y) =0,
\label{cyfact}
\end{equation*}
which is based on a Riemann surface $\Sigma$ given by $H(x,y)=0$.  

\vspace*{1mm}

As we
explained before the complex parameters entering the defining equation
of the Riemann surface correspond to complex structure moduli of the
Calabi-Yau. Some of them are non-normalizable and can be considered as
external parameters. We want to tune these parameters to approach the
limit where the surface $\Sigma$ factorizes into two surfaces $\Sigma_L$
and $\Sigma_R$ connected by long tubes. This factorization lifts to the
entire Calabi-Yau $X$ and divides it into two regions $X_L$ and $X_R$
that are widely separated. We introduce 3-form flux $G_3$ of compact
support on the 3-cycles of $X_R$. The superpotential and scalar
potential are given by
\begin{equation}  
W = \int G_3 \wedge \Omega
 \qquad 
 \mbox{and} \qquad
 V = G^{I\overline{J}}\partial_I W
\overline{\partial_{J} W},
\label{pottwo}
\end{equation}
where the indices $I,J$ run over all complex structure moduli 
of the total threefold $X$. Using the properties of the K\"ahler
metric $G_{I\overline{J}}$ in the factorization limit we show
that the part of the potential \eqref{pottwo} which depends on the
complex structure moduli of $_L$ is of the form
\eqref{finalpot}. Furthermore, we find an understanding of the
effective value of the parameters $T^m$.

\subsubsection{Geometry of factorization}

\begin{figure}[h!]
 \begin{center}
\includegraphics[width=11.5cm]{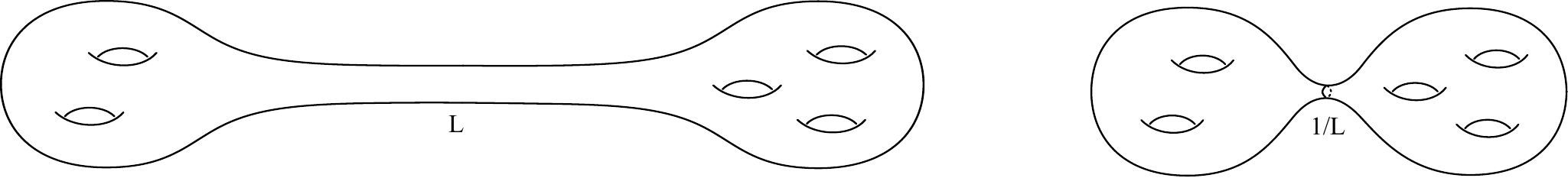} 
\caption{Two
 conformally equivalent ways of viewing the factorization of a Riemann
 surface into two parts. Physically though, we should distinguish both
 points of view, since particle masses depend on the size of the
 cycles. Because in our situation no new massless appears in the
 factorization limit, the left diagram represents our point of view
 best. } \label{fig:section6plumb}
 \end{center}
\end{figure}

Let us study the degeneration of a
Riemann surface $\Sigma$ into two components $\Sigma_L$ and
$\Sigma_R$, depicted in figure~\ref{fig:section6plumb}.\footnote{In
  general these components  could be connected  in a non-trivial
  way. We restrict our computations in this section to the case in 
  which they are linked by just one long tube. These should be
  easily extendible to more general cases.}  In this factorization
data of the full Riemann surface is expressed in terms of the complex 
structure of the individual surfaces. It is well known that in the
limit where the length of the tubes $L= 1/\epsilon$ goes to
infinity the period matrix of the full surface becomes block diagonal
\begin{equation}
  \tau = \left(\begin{array}{cc} \tau^{LL} & 0\\ 0& \tau^{RR}
    \\ \end{array}\right) + {\cal O}\left(\epsilon\right).
\label{periodfactorized}
\end{equation}
While the off-diagonal components $\tau^{LR}$ go to zero in the
factorization limit, their subleading behavior is quite important
in our analysis since it expresses the weak interaction between the
two sectors. The period matrix $\tau^{LR}$ can be computed
systematically in an expansion in $\epsilon$ from data on each of
the two surfaces as we explain below.

Technically, we describe the factorization of the Riemann surface with
the \emph{plumbing fixture} method \cite{Vafa:1987ea}. So consider two
Riemann surfaces $\Sigma_L$ and $\Sigma_R$ of genus $g_L$ and $g_R$
respectively.  On the left surface $\Sigma_L$ we have $g_L$ holomorphic
differentials $\omega_i$, while on the right surface $\Sigma_R$
similarly $g_R$ holomorphic differentials $\omega_{i'}$. The complex
structure of the left surface is determined by the periods of the
holomorphic differentials
\begin{equation*}
  {1\over 2\pi i}\int_{a^i}\omega_j = \delta^i_{j},\qquad {1\over 2\pi
    i}\int_{b_i}\omega_j = \tau^{LL}_{ij}, 
\label{periodholdif}
\end{equation*}
where $\tau^{LL}_{ij}$ is the period matrix of $\Sigma_L$, and we
choose our definitions similarly for the right surface.

The plumbing fixture method works after choosing a puncture $P$ on
$\Sigma_L$ and $P'$ on $\Sigma_R$. It
connects the two surfaces by a long tube of length $L$ 
which is glued onto neighborhoods of the punctures $P$ and $P'$. More
precisely, we pick a local holomorphic coordinate $z$ 
around the puncture $P$ such that $z(P)=0$ and a holomorphic coordinate
$z'$ near $P'$ with $z'(P')=0$. Then we identify points in these
neighborhoods as
\begin{equation*}
  z z' = \epsilon.
\label{plumbing}
\end{equation*}

Now we want to compute the period matrix of the full Riemann surface in
terms of complex structure data of the two surfaces. For this we need
to understand how the differentials $\omega_i$ and $\omega_{i'}$ extend to
well-defined holomorphic differentials on the full surface $\Sigma=
\Sigma_L \cup \Sigma_R/\sim$, where $\sim$ is the above
identification. Let us first consider how to lift
the differential $\omega_i$. Around the puncture $P$ it may be
expanded as
\begin{equation*}
  \omega_i = \sum_{m=1}^\infty K^{P}_{im} z^{m-1} dz,
\label{bla1}
\end{equation*}
where the functions $K_{im}^P$ are defined in equation~(\ref{KIM_lclCY}).
Once we write  this in terms of $z'$ we observe that, as seen from the
right surface, the differential has a Laurent expansion. So $\omega_i$ will 
be written as a linear combination of the meromorphic differentials 
$\xi_{m}^{P'}$ of the right surface. A meromorphic differential
has the following expansion around the puncture
\begin{equation*}
  \xi_m^P = \left({m\over z^{m+1}} + \sum_{n=1}^\infty h^P_{mn}
  z^{n-1}\right)dz.
\label{bla2}
\end{equation*}
Here we have introduced the functions $h_{mn}^P$, which depend on the
complex structure moduli of the surface and the position of $P$.
So in general the differential $\omega_i$ will lift to a differential 
$\widetilde{\omega_i}$ on the full surface which can be written as
\begin{equation*} 
\widetilde{\omega_i} = \left\{ \begin{array}{ll} \omega_i +
    \displaystyle\sum_{m=1}^\infty x_{i m} \xi^P_m &
\text{on}\quad \Sigma_L,\\[1ex]
  \displaystyle\sum_{m=1}^\infty y_{i m} \xi^{P'}_{m} &
\text{on}\quad \Sigma_R. \end{array} \right.
\label{bla3}\end{equation*}
for some coefficients $x_{i m}$ and $y_{i m}$. Matching the differential on the two
sides we find the following conditions
\begin{equation*} \begin{split}
 x_{im} &= - {\epsilon^m\over m} \sum_{n=1}^\infty  y_{in} h^{P'}_{nm},
		  \qquad
 y_{im}  = - {\epsilon^m\over m} \left( K^P_{im} + \sum_{n=1}^\infty x_{in}
  h^{P}_{nm} \right).
\end{split}
\end{equation*}
This allows us to compute the cross-period matrix as
\begin{equation}
\begin{split}
  \tau^{LR}_{ij'} = \int_{b_{j'}} \omega_i 
 &= \sum_{m=1}^\infty  K^{P'}_{j'm} y_{im}  
 = -\sum_{m,n=1}^\infty {\epsilon^n\over n} K_{im}^P G^{-1}_{mn} K^{P'}_{j'n},\\
 G_{mn}&\equiv\delta_{mn}-\sum_{l=1}^\infty {\epsilon^{n+l}\over n l}h'_{ml}h_{ln}.
\end{split}
 \label{crossperiod}
\end{equation}
From this equation we can
read off all order $\epsilon$-corrections to the 
off-diagonal piece of the period matrix when a surface $\Sigma$ degenerates. 

Also, this procedure gives a clear understanding of the term ``flux at
infinity''. We see that the flux at infinity is generated by regular
forms on the degenerated surface, and therefore will at most have finite
order poles at the punctures.

Notice that for a Calabi-Yau threefold that is based on a Riemann surface,
the factorization region is described by the
deformed conifold geometry
\begin{equation*}
 uv+ x^2 + y^2 = \epsilon, \quad \mbox{or equivalently}  \quad uv + zz' = \epsilon. 
\end{equation*}
Usually, this is described as a 3-sphere shrinking to zero-size when
$\epsilon \to 0$. However, as for the complex 1-dimensional plumbing
fixture case we want the two sectors to be far apart from each other. Therefore
we consider the conformally  
equivalent setup where the 3-sphere is scaled to be of
finite size, while the transverse directions are made very
large. The finite size three-sphere reduces to the cross-section of
the tube on the left in figure~\ref{fig:section6plumb}, whereas the
transverse directions reduce to the tube-length.  

To describe the left and right neighborhoods of the degeneration, we
can fix $x= \sqrt{\epsilon - y^2 -uv}$ on the left and $x=-
\sqrt{\epsilon - y^2 -uv}$ on the right. In the limit that $\epsilon
\to 0$ these neighborhoods will not just intersect in a point, but in
the divisor $uv + y^2 = 0$. This is the region where regular forms on
the total threefold will develop poles when the degeneration
starts.



\begin{figure}[h!]
\begin{center}
\includegraphics[width=7.5cm]{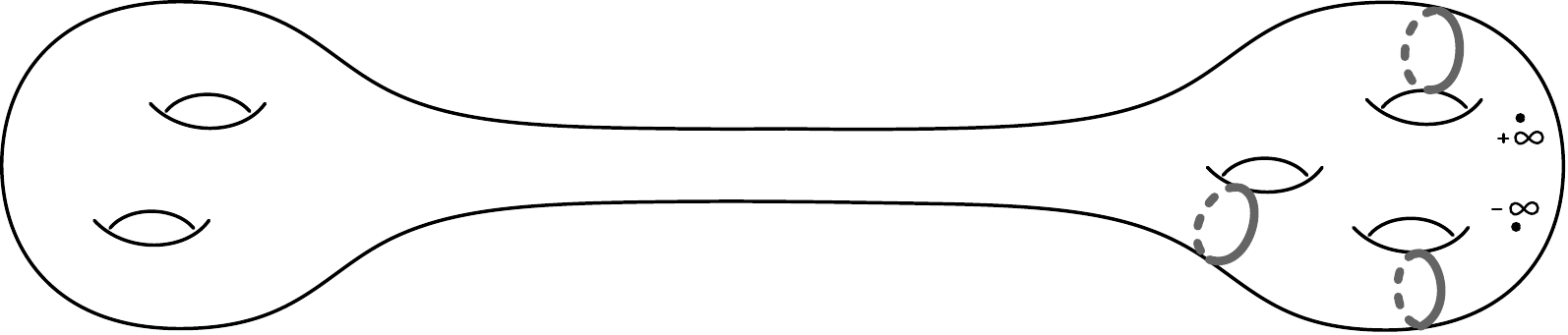}
\caption{Turning on flux on the right part of the factorized Calabi-Yau.}
\label{fig:section5degflux}
\end{center}
\end{figure}

\subsubsection{Dynamics}
\label{subsec:dynamics}

Now we consider turning on flux on the threefold. For simplicity we
again take a Calabi-Yau \eqref{cyfact} that is
based on a factorized Riemann surface. We turn on 3-form flux $G_3 =
F_3 - \tau H_3$ which is only piercing the set of A-cycles
corresponding to $\Sigma_R$, as can be seen in
figure~\ref{fig:section5degflux}, and write down the corresponding
(super) potential. For regularization issues later, we take
two more punctures on the right surface labeled by 
$\pm\infty$ and turn on some flux $\alpha$ through the noncompact ${
  B}_{\infty}$ cycle running from $+\infty$ to $-\infty$.

A basis of $A$ and $B$ cycles is given by the compact
3-cycles on the left and the right, together with the lift
$A^{\infty}$ of the $A$-cycle enclosing $+\infty$ and
$B_{\infty}$. So the flux is determined by
\begin{equation*} \begin{split}
&\int_{A^{i}} G_3 = 0, \qquad  \int_{A^{i'}} G_3 = N^{i'}, \qquad
\int_{A^{\infty}} G_3 = 0, \\
& \int_{B_i} G_3= 0,\qquad \int_{B_i'} G_3 = 0, \qquad \quad
\int_{B_{\infty}} G_3 
  = \alpha. \end{split}
\end{equation*}

Let us denote the complex structure moduli and their duals by $X^I$ and
$F_I$, which are the $A^I$ resp.\ $B_I$ periods of
the holomorphic 3-form $\Omega$.  Here we use the capital indices
$I=\{i,i', \infty\}$ to run over both the left and the right sides.  Then
the GVW superpotential for the complex structure moduli is given by
\begin{equation*}
  W = \int G_3 \wedge \Omega = \alpha  X^{\infty}  + \sum_{i'} N^{i'} F^R_{i'},
\label{gvwfact}
\end{equation*}
and the corresponding scalar potential by
\begin{equation*}
  V = \sum_{I,J} G^{I\overline{J}} \partial_I W \overline{\partial_J W}.
\label{scalarfact}
\end{equation*}
Since $X^{\infty}$ corresponds to a log-normalizable period and the
derivatives in the above potential just correspond to normalizable
modes, the $\alpha$-factor decouples. This shows that
\begin{align}
  V = &
  \sum_{i,j,k',l'}  
  \left(N^{k'}\tau_{k'i}^{LR}\right)
  \left({1\over \textrm{Im}\tau}\right)_{\!LL}^{ij} 
  \overline{\left(N^{l'}\tau_{l'j}^{LR}\right)} \notag \\
  &\qquad +\sum_{i,j',k',l'}
  \textrm{Re}\left[ \left( N^{k'}\tau_{k'i}^{LR} \right)
  \left({1\over \textrm{Im}\tau}\right)_{\!LR}^{ij'} 
  \overline{\left(N^{l'}\tau_{l'j'}^{RR}\right) }\right] \label{longpot}\\
  &\qquad
  + \sum_{i',j',k',l'}
  \left( N^{k'}\tau_{k'i'}^{RR}\right)
  \left({1\over \textrm{Im}\tau}\right)_{\!RR}^{i'j'} 
  \overline{ \left(N^{l'}\tau_{l'j'}^{RR}\right)}.\notag
\end{align}
Thus the total potential is the sum of three terms, which we
denote in the obvious way by $V=V_1+V_2+V_3$. 

Next we consider what happens in the limit where the distance $L$
between the two sets of 3-cycles gets very large. As explained
before the period matrices $\tau^{LL}$ and $\tau^{RR}$
remain of order one in this limit and become almost independent of the
moduli $X_R$ 
and $X_L$, respectively. 

On the other hand, $\tau^{LR}$ goes to zero which would make the first term
$V_1$ in the potential vanish in the limit that
$\epsilon \to 0$, at least if we don't scale the fluxes $N^{i'}$  appropriately.
Since $V_1$ describes the interaction between the
two sides of the Calabi-Yau, we really want to scale the fluxes
$N^{i'}$ to go to infinity in such a way that the term $V_1$
remains finite. 

Then it becomes clear that the term $V_3$ of the
potential dominates over the other two contribution to $V$. 
This implies that in the limit $\epsilon \to 0$ the term $V_3$ should
be minimized first, {\it i.e.},
\begin{equation*}
  \sum_{k'} N^{k'}\tau_{k'i'}^{RR} =0, \quad \forall i',
\label{susyright}
\end{equation*}
which is a set of $n_R$ equations for the $n_R$ moduli $x^{j'}$. The solutions
of this system correspond to supersymmetric vacua for the 3-cycles on
the right side. 
Once we have fixed all $X^{j'}$ to their supersymmetric values $\widehat{X}^{j'}$, 
we can consider the effect of the backreaction of the right side to the
left. This is purely expressed through the potential $V_1$, since
the term $V_2$ vanishes as well at the supersymmetric point. 

So effectively the potential for the complex structure
moduli $X_L^i$ of the left surface is 
\begin{equation}
  V_1 = \sum_{i,j,k',l'}  
   \left(N^{k'}\tau_{k'i}^{LR}\right) 
   \left({1\over \textrm{Im}\tau}\right)_{\!LL}^{ij} 
   \overline{\left(N^{l'}\tau_{l'j}^{LR}\right)}.
   \label{v1ff}
\end{equation}
This may be written as
$  V_1 =\sum_{i,j} \partial_i W_{\text{eff}} (1/\textrm{Im}\tau)_{LL}^{ij} \overline{\partial_j W_{\text{eff}}},$
where we define the effective ``superpotential'' for the left complex
structure moduli as 
\begin{equation*}
 \partial_i W_{\text{eff}} \equiv \sum_{k'} N^{k'} \tau_{k'i}^{LR}.
\label{effpotential}
\end{equation*}

Comparing with expression \eqref{crossperiod} it is clear that the
fluxes on the right should be scaled in such a way that the coefficients
\begin{equation}
T^m =  \epsilon^m   \sum_{k'} N^{k'} K'_{k'm}
\label{tms}
\end{equation}
remain constant. In that situation the effective superpotential is
\begin{equation}
  \partial_i W_{\text{eff}}  = \sum_m  T^m K_{im}
\label{finalsup}
\end{equation}
to leading order in $\epsilon$, which is precisely of the form \eqref{finalpot}.

\subsubsection{Genericity of Potential and Metastable Vacua}
\label{subsec:fivepointthree}

Let us summarize what we have demonstrated so far. We started with a
large Calabi-Yau that consists of two parts $X_L$ and $X_R$ separated
by a large distance, and turned on a large 3-form flux on one of the
sides, say $X_R$. This flux generates a large potential for the
complex structure moduli of $X_R$, which are therefore set to their
supersymmetric minima. The flux on $X_R$ is also weakly backreacting
to the other side $X_L$, inducing a small superpotential for the
complex structure moduli of $X_L$. We computed this superpotential in
equations \eqref{effpotential} and \eqref{finalsup} and found that it
is of the form \eqref{finalpot}. The main point is that the side $X_L$
only knows about $X_R$ via the parameters $T^m$ given by \eqref{tms}.

In this section we discuss two questions. The first is to which 
degree we can tune the parameters $T^m$ independently. And the
second is whether these $T^m$'s can be chosen to
realize an OOP supersymmetry breaking superpotential.

As we can see from \eqref{tms}, the values of the parameters $T^m$
depend on the fluxes $N^{l'}$ on the cycles of $X_R$ and also on the
value of the (generalized) period matrix $K'_{l'm}$. The last one
depends on the choice of the supersymmetric vacuum $\widehat{X}^{j'}$ on the
right side. For given large fluxes $N^{l'}$ there is a huge number of
supersymmetric vacua, or solutions of equation \eqref{susyright}, with different
values of $\widehat{X}^{j'}$ and consequently of $K'_{l'm}$. The density of
such supersymmetric vacua over the complex structure moduli space of
$X_R$ has been studied before \cite{Douglas:2003um, Ashok:2003gk,
Denef:2004cf, Denef:2004ze, Torroba:2006kt}, and it is believed that the
vacua become dense in the moduli space in the limit where the fluxes are
very large.

The coefficients $K'_{l'm}$ are holomorphic functions over the complex
structure moduli space of $X_R$. So naively one would conclude that
when the dimension of this moduli space is large enough, meaning that
the number of 3-cycles in $X_R$ is large, we can always find
supersymmetric points where the $K'_{l'm}$'s have the desired
values. However the functions $K'_{l'm}$ are not ``generic'' and there
may be relations between them which affect the naive counting. We have
not analyzed this problem in detail but we think the following statement
is true. Any number of the $T^m$'s in the superpotential
\eqref{finalsup} can be tuned by considering a Calabi-Yau whose right
side $X_R$ has a sufficiently large number of 3-cycles, and there will
be some supersymmetric vacua with right values of $K'_{l'm}$ to
reproduce the desired $T^m$'s to good accuracy.

This claim is made more intuitive by the following physical
interpretation of equation \eqref{tms}. Start by turning on fluxes
$N^{l'}$ on the cycles of $X_R$, which is based on the Riemann surface
$\Sigma_R$. When reduced on the Riemann surface the flux looks like the
electric field produced by a charge in two dimensions. The set of fluxes
$N^{l'}$ resembles a charge distribution on the cycles of the Riemann
surface.  To compute the field produced by these charges in the distant
region of the other set of cycles $\Sigma_L$, one has to consider a
multipole expansion. Since the matrix $K'_{l'm}$ computes the $m$th
multipole expansion of a charge distributed along the $l'$th cycle, the
coefficients $T^m$ are exactly the multipole moments of the charge
distribution. In this formulation our first question reads whether we
can arrange a charged distribution to have the desired multipole moments
given by the coefficients $T^m$. We expect that the answer is positive.

The second question is more subtle. To realize a metastable
nonsupersymmetric vacuum via the OOP mechanism, one has to tune the
superpotential in a way which is determined by properties of the
K\"ahler metric at that point. As we saw in \Cref{sec:OOP} one
has to tune the coefficients of the effective superpotential only up to
cubic order in an expansion around the candidate metastable point. Since
we have a very large number of parameters $T^m$ at our disposal it seems
that generically we should be able to tune them to generate metastable
vacua at most points on the moduli space.  However we do not have a
proof of this statement and it is possible that various relations
between the period matrices and the K\"ahler metric invalidate the naive
counting\footnote{This question is similar to whether one can realize
the OOP mechanism with a single trace superpotential for the adjoint
scalar in an $SU(N)$ gauge theory. In \cite{Ooguri:2007iu} it was
demonstrated that for $SU(2)$ a metastable vacuum can be generated
anywhere on the moduli space by a single trace superpotential, and for
$SU(N)$ at the center of the moduli space. It was not fully analyzed
whether this is possible in generality.}.

\subsection{Example of factorization}
\label{sec:extwo}

In the previous section, we argued, based on the factorization of the
Riemann surface and the Calabi-Yau, that it is possible to embed the
nonsupersymmetric metastable vacua we found in
\Cref{sec:OOPlocalCY} in a ``larger'' Calabi-Yau, the idea being
that the flux threading compact cycles on one side of the Calabi-Yau
looks like flux coming from infinity from the viewpoint of the other
side of the Calabi-Yau\@.
In this section, we will discuss the Dijkgraaf-Vafa geometries
\begin{align} 
\Sigma_{\text{DV}}: \quad y^2 &= P_n(x)^2 - f_{n-1}(x),\qquad
P_n(x) = \prod_{I=1}^n (x- \alpha_I),\label{DVgeo_factn}
\end{align}
as an example where our proposal can in principle be implemented, and
make some steps towards actually confirming our proposal.



\hyphenation{nor-ma-li-za-ble}

\subsubsection{Factorization in practice}

Remember that the $\alpha_I$'s are non-normalizable
parameters which represent the positions of the cuts on the $x$-plane,
while the coefficients in $f_{n-1}(x)$, or equivalently variables $S^I$
defined in \eqref{spclcoordDV}, are normalizable (or at least
log-normalizable) and hence are dynamical variables describing the size
of those cuts.  Therefore, in this Dijkgraaf-Vafa case
\eqref{DVgeo_factn}, $\alpha_I$ are the parameters we want to adjust in
order to approach the factorization limit where $\Sigma_{\text{DV}}$
degenerates into two subsectors.

So, what we should do is clear: we divide the $n$ cuts into two parts as
$n= n_L + n_R$, the ones on the left indexed by $i$ and on the right by
$i'$, and send these two groups apart from each other by a large factor
$L= 1/\epsilon$ so that
\begin{align*}
\alpha_i - \alpha_{i'} = \mathcal{O}(L) \qquad \mbox{(when}~
L \to \infty).  
\end{align*}
In the $L\to\infty$ limit, the left and right sides will be very far
apart and the factorization we discussed in the previous section must be
achieved.  For example, the period matrix of the total Riemann surface
must diagonalize as in \eqref{periodfactorized} up to $1/L$ correction.

There is one thing we should be careful about when taking the
$L\to\infty$ limit.  If we try to separate the two sets of cuts by
naively taking the typical difference between $\alpha_i$ and
$\alpha_{i'}$ to be of order $L$ while keeping the size of the cuts
fixed, then a simple estimate of the scaling of $S_i^L,S_{i'}^R$ using
\eqref{spclcoordDV} shows that the physical size of the 3-cycles in the
Calabi-Yau blows up.  What we want instead is to end up with two sets of
3-cycles of finite size, separated by a large distance, so that we are
left with nontrivial dynamics of $S_i^L,S_{i'}^R$.  To achieve this we
must also scale the size of the cuts, as we send $L\to\infty$. Let $x_L$
and $x_R$ be local coordinates in the left and right sectors,
respectively, and set
\begin{equation}
  \widetilde x_L = L^{r} x_L,\qquad  \widetilde x_R = L^{r'} x_R, 
   \label{scalecuts}
\end{equation}
where
\begin{equation*}
  r = {n_R \over n_L+1} ,\qquad r' = {n_L \over n_R+1}.
\end{equation*}
Then, from \eqref{spclcoordDV}, it is not difficult to see that we can
keep $S_i^L,S_{i'}^R$ finite if we keep $\widetilde x_L$, $\widetilde
x_R$ finite while taking the $L\to\infty$ limit.
A similar rescaling of local coordinates must be also necessary when
taking a factorization limit in any other examples than
\eqref{DVgeo_factn}.

\subsubsection{Computation of Period Matrix}
\label{subsec:comptau}

In the Dijkgraaf-Vafa geometry \eqref{DVgeo_factn}, the period matrix is
given by
\begin{equation}
 \tau_{IJ} = 
  {\partial^2 \mathcal{F}_0 \over \partial S^I \partial S^J},\label{tauIJF0}
\end{equation}
Here, $\cF_0$ is the B-model prepotential, which by the Dijkgraaf-Vafa
relation \cite{Dijkgraaf:2002fc, Dijkgraaf:2002pp} is related to matrix
models.
The precise way to scale various quantities to take the factorization
limit being understood, it is
in principle possible to confirm our proposal for the Dijkgraaf-Vafa
geometry using \eqref{tauIJF0}.
For doing that, it is important to be able to compute the prepotential
$\cF_0$ for a large number of cuts $n$.  The results from Section
\ref{sec:OOPlocalCY} show that generating a metastable vacuum requires
quite a lot of coefficients $T^m$.  Since we roughly need the same
number of cuts on the right as the number of tuned $\Sigma_m$'s on the
left, the total Riemann surface must have quite a large number of cuts.
So, in this subsection we will explain the way to compute $\cF_0$ and
thus $\tau_{IJ}$ for an arbitrary $n$.

For Dijkgraaf-Vafa geometries \eqref{DVgeo_factn} the prepotential
$\cF_0$ may in fact be computed for any number of cuts $n$ in a number
of ways. The most direct way is evaluating the period integrals on the
hyperelliptic curve. This has been done up to cubic order in $S^I$ in
\cite{Itoyama:2002rk}.  Duality with a $U(N)$ matrix model
\cite{Dijkgraaf:2002fc, Dijkgraaf:2002pp}
\begin{equation*}
 Z =  \exp \left[\sum_{g=0}^\infty
  g_s^{2g-2} \mathcal{F}_g(S) \right] =\int d^{N^2}\! \Phi \,\exp \left[
  \frac{1}{g_s} \Tr W(\Phi) \right], 
\end{equation*}
where the matrix model action is given by 
\begin{equation*}
  W'(x) = P_n(x) = \prod_{I=1}^n (x-\alpha_I)
\end{equation*}
makes this computation quite a bit simpler. Let us quickly show this
argument \cite{Dijkgraaf:2002pp}.

The field $\Phi$ is an $N \times N$ matrix. Say $N^I$ eigenvalues of
$\Phi$ are placed at the critical point $x=\alpha_I$ and divide the
matrix $\Phi$ into $N^I \times N^J$ blocks $\Phi_{IJ}$, where
$\sum_{I=1}^n N^I=N$. One can go to the gauge $\Phi_{IJ}=0$ for $I \neq
J$ by introducing fermionic ghosts in the matrix model action. This
produces the following extra term in the action, where $\Phi_I\equiv
\Phi_{II}$:
\begin{align*}
 W_{\text{ghost}}=\sum_{I \neq J}\Tr(B_{JI}\Phi_{I}C_{IJ}+C_{JI}\Phi_{I}B_{IJ}).
\end{align*}

To write down Feynman diagrams, we expand $\Phi_I$ around
$x=\alpha_I$ as $\Phi_{I}=\alpha_{I}+\phi_{I}.$ A Taylor series of
$W(\Phi_{I}) = W(\alpha_{I} + \phi_{I})$ around $\alpha_{I}$ yields
the propagator and $p$-vertices for $\phi_{I}$. In particular, this
shows that the propagator for $\phi_{I}$ is given by
\begin{align*}
 \langle \phi_{I}\phi_{I} \rangle={1\over W''(\alpha_I)}={1\over \Delta_I},
\end{align*}
where $\Delta_I =W''(\alpha_I)=\prod_{J\neq I}^n
\alpha_{IJ}$. Moreover, expanding the ghost action determines the ghost
propagator to be 
\begin{align*}
 \langle B_{JI}C_{IJ} \rangle= \frac{1}{\alpha_{IJ}},
\end{align*}
and gives the Yukawa interactions between $\phi_{I}$,
$B_{JI}$ and $C_{IJ}$.

\begin{figure}
  \begin{center}
\includegraphics[width=11cm]{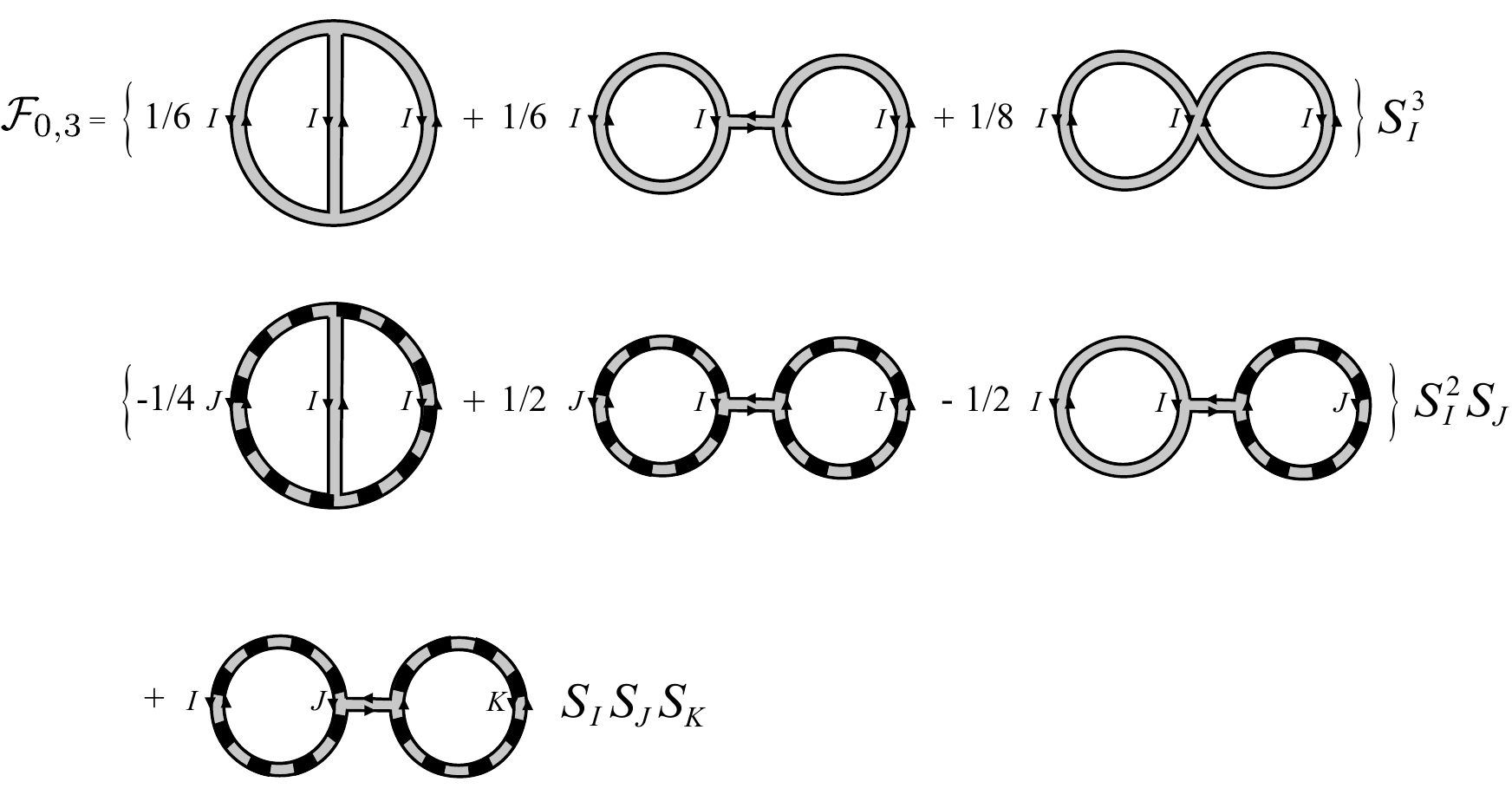} 
 \caption{The contribution to $\mathcal{F}_{0,3}$ given in terms of
 matrix diagrams.  Gray double lines represent $\phi_I$ fields, while
 black-and-gray double lines represent $BC$ ghosts.  \label{F03}}
 \end{center}
\end{figure}

The contribution to the prepotential $\mathcal{F}_0$ of order three in
the $S^I$'s is given by planar diagrams with three holes, see Fig.~\ref{F03}.
 Writing down the expressions $g_{I,3}$ and $g_{I,4}$ in terms of
$\alpha$'s and $\Delta$'s shows that
\begin{equation*}
 \cF_{0,3}=\sum_{I=1}^n u_I S_I^3+\sum_{I\neq J}^n u_{I;J}S_I^2S_J
 +\sum_{I< J< K}^n u_{IJK} S_I S_J S_K,
\end{equation*}
where
\begin{align*}
 u_I&={2\over 3}\biggl(
 -\sum_{J\neq I}{1\over \alpha_{IJ}^2\Delta_J}
 +{1\over 4\Delta_I}\sum_{J<K\atop J,K\neq i}{1\over \alpha_{IJ}\alpha_{IK}}
 \biggr), \\
 u_{I;J}&=
 -{3\over \alpha_{IJ}^2\Delta_I}+{2\over \alpha_{IJ}^2\Delta_J}
 -{2\over \alpha_{IJ}\Delta_I}\sum_{K\neq I,J}{1\over \alpha_{IK}}
 \quad  \mbox{and}\\
 u_{IJK}&=4\left({1\over \alpha_{IJ}\alpha_{IK}\Delta_I}
 +{1\over \alpha_{JI}\alpha_{JK}\Delta_J}
 +{1\over \alpha_{KI}\alpha_{KJ}\Delta_K}
 \right).
\end{align*}

In appendix D of \cite{Hollands:2008cs} we discuss the generalization of this
result to higher order in $S^I$.  In particular, we compute $\cF_0$ up
to $S^5$ terms.

\subsubsection{Scaling of Period Matrix} 

The method explained in \Cref{subsec:comptau} allows one in
principle to compute the period matrix to any order in $S^I$ for general
Dijkgraaf-Vafa curves \eqref{DVgeo_factn}.  Then the factorization limit
can be achieved simply by taking the $L \to \infty$ limit of the result
and one can start looking for metastable vacua.
In this subsection, as a step towards it, let us pursue a more modest
goal of seeing the factorized behavior of the period matrix,
\eqref{periodfactorized}.

The form of the scaling can be elegantly derived for any possible
contributing matrix model diagram to $\mathcal{F}_0$. First 
note that $\Delta_{i}$ scales as $L^{2r}$ as $L \to \infty$, and
$\Delta_{i'}$ as $L^{2r'}$. All propagators with indices from
either side of the surface have an expansion in terms of
$\alpha_{IJ}$'s and $\Delta_I$'s, and thus a scaling in $L$ which is
easy to determine. The total scaling of a planar diagram with an
arbitrary number of these elements turns out to depend just on the
number of ghost vertices that connect the left side to the right
side. It is given by
\begin{equation}
\frac{1}{L^{(1+r)N_{ii'}+(1+r')N_{i'i}}},
\end{equation}
where $N_{ii'}$ is the number of ghost vertices with external ghost
lines indexed by $(i,i')$ and the external $\phi$-line by $(i,i)$.  Note
that in deriving this we assumed the scaling \eqref{scalecuts} and thus
$S^i_L,S^{i'}_R$ are of order one.

This shows that a diagram with only indices on the left (or on the
right) will be of order 1 in $L$. Since such diagrams contribute to the
period matrix $\tau_{ij}$ (or $\tau_{i'j'}$), this shows that the
period matrix is of order 1 in $L$, with corrections in $1/L$ from
diagrams that contain at least two loops indexed by $i$ and $j$. On the
other hand, the off-diagonal pieces of the period matrix $\tau_{ii'}$
and $\tau_{i'i}$ contain at least one ghost cross-vertex with indices
$i$ and $i'$. These parts will therefore scale at least as $1/L$. In
particular, for large $L$ the properties of the full Riemann surface
$\Sigma$ are determined by those of the two factors $\Sigma_L,\Sigma_R$,
and the period matrix $\tau_{IJ}$ indeed diagonalizes as in
\eqref{periodfactorized}.
%

Having checked the diagonalization \eqref{periodfactorized}, the problem
of actually finding an example of a metastable vacuum then just amounts
to solving equation (\ref{tms}) together with (\ref{susyright}) using
the data from matrix model, for $T^m$ giving a metastable vacuum. \hyphenation{me-ta-sta-ble}
Solving these equations is nontrivial, since the relation between the
flux parameters $N^{i'}$ on the right and the coefficients in the
superpotential $T^m$ we want on the left are non-linear, although we
expect that the solutions do exist by the multipole argument we gave in
\Cref{sec:factorization}. We leave matrix model computations up
to requisite orders as well as finding the actual metastable vacua by
solving those equations for the future work.

\section{Concluding remarks}\label{sec:conclusions}


Summarizing, we found that turning on flux with support at infinity in
local Calabi-Yau in type IIB induces a superpotential for the moduli in the local
Calabi-Yau, thus breaking $\cN=2$ of the Calabi-Yau compactification down to $\cN=2$.
Then we demonstrated that one can create metastable vacua by tuning the
flux at infinity using the OOP mechanism, using a Dijkgraaf-Vafa
(CIV-DV) geometry as a primary example.  The metastable vacua known to
exist \cite{Ooguri:2007iu, Pastras:2007qr} in perturbed Seiberg-Witten
theory can also be understood in terms of metastable flux configuration.


Flux diverging at infinity may appear problematic, but in reality a
local Calabi-Yau must be regarded as a local approximation of a larger compact Calabi-Yau
and the flux at infinity has a natural interpretation there; there is
flux floating around in the rest of the Calabi-Yau, which ``leaks'' into our
local Calabi-Yau and just appear to be coming in from infinity.  This,
furthermore, motivates a more natural setting to realize metastable flux
vacua: in a part, say on the right side, of the full Calabi-Yau $X$, there are
some 3-cycles threaded by flux (and possibly O-planes to cancel net
charge if $X$ is compact) and on the left side there are some 3-cycles
without flux through them.  If the distance between the left and right
sectors is large, the full Calabi-Yau $X$ factorizes into an almost decoupled
system of $X_L$ and $X_R$, and the flux in $X_R$ appears to be flux at
infinity from the viewpoint of $X_L$ and induces superpotential in
$X_L$.  By adjusting the number of fluxes in $X_R$, we can tune the
superpotential and generate metastable vacua in $X_L$. This is a very
well controlled setting to analyze flux vacua, which may shed light on
the structure of the nonsupersymmetric landscape of string vacua.
We also made some steps toward actually embedding metastable vacua in a
larger Calabi-Yau as sketched above in the case of Dijkgraaf-Vafa geometry by
computing certain matrix model amplitudes.  Actually finding explicit
vacua along that line is an interesting open problem.

%

Note that we needed just two main ingredients to achieve this
result: Firstly, the OOP mechanism requires that the complex structure moduli
space is special K\"ahler. Secondly, it is important that the generated
Gukov-Vafa-Witten superpotential is very much controllable by tuning
the flux. This means that we can generalize the
above story to any setting which fulfills these two requirements. Other
possibilities therefore include M-theory and F-theory on Calabi-Yau
fourfolds \cite{Becker:1996gj, Gukov:1999ya}. Let us finish by saying a
few words on these two setups.  

%
%

Compactifying M-theory on a Calabi-Yau fourfold $X_4$ with fluxes yields
a three-dimensional low energy theory with 4 supercharges. The complex
structure moduli of the Calabi-Yau are part of the chiral
supermultiplets and are described by variations of the holomorphic
$(4,0)$-form $\Omega$. In the local limit where the fourfold becomes
noncompact, the K\"ahler potential on the moduli space is given by
\begin{equation} \label{eqn:MthKahlerpot}
  K = \int_{X_4} \Omega \wedge \overline{\Omega},
\end{equation}
so that the metric on the moduli space is indeed special
K\"ahler. Moreover, it is well-known that the complex moduli may be
stabilized by turning on 4-form flux $F_4$, which introduces the
superpotential 
\begin{equation*}
  W = \int_{X_4} F_4 \wedge \Omega.
\end{equation*}
The condition for unbroken supersymmetry is $W=dW=0$, so that $F_4$ has
to be a $(2,2)$-form. Stabilizing the K\"ahler moduli as 
well requires that the flux is primitive under the Lefschetz
decomposition (and in particular self-dual). Turning on primitive
$(2,2)$ flux on some compact 4-cycles, we can now follow an equivalent
procedure as in IIB.    
 
M-theory compactified on $X_4$ is equivalent to compactifying
F-theory on $X_4 \times S^1$, at least if $X_4$ is an elliptically
fibered Calabi-Yau. This leads to a four-dimensio-nal space-time with 4
supercharges. So again, the K\"ahler potential is given by
(\ref{eqn:MthKahlerpot}), and the flux $F_4$ is a primitive
$(2,2)$-form. The relation with IIB consistently reduces $F_4$ to a
harmonic $(2,1)$-flux $G_3$. The extra seven-branes that must be inserted
in IIB when reducing over a singular $T^2$ do not contribute to the
superpotential and thus don't play an important role here.   

In particular, consider as an example the local Calabi-Yau fourfold
\begin{equation*}
 u^2 + v^2 +  w^2 + H(x,y) =0,
\end{equation*}
where all variables are $\mathbb{C}$ (or $\mathbb{C}^*$) valued, and
$F(x,y)$ defines a smooth curve in the $x,y$-plane. Its holomorphic
four-form is given by  
\begin{equation*}
\Omega = \frac{du \wedge dv}{w} \wedge dx \wedge dy.
\end{equation*}
The $u,v,w$--fiber defines a two-sphere over each point in the
$x,y$-plane, which shrinks to zero-size over the curve
$H(x,y)=0$. (Like in the Calabi-Yau threefold case, the real
  part of $H(x,y)$ changes sign when crossing the Riemann
  surface. This flop changes the parametrization of the compact $S^2$
  in the $T^*S^2$-fiber from a  ``real'' $S^2$ into an ``imaginary'' $S^2$.)  
Four-cycles can be constructed as an $S^2$ fibration over some disk $D$
ending on the curve and have the topology of a four-sphere (when $x$
and $y \in \mathbb{C}$). Notice that the intersection lattice is
symmetric now and not simply symplectic anymore, so that the bilinear
identity takes a more complicated form. However, like in the threefold
case all relevant quantities reduce to the Riemann surface, and the
analysis is similar as before.


        \appendix 
        
\chapter{Level-rank
  duality}  \label{app-decompose}\index{level-rank duality}

The affine algebras $\widehat{su}(N)_k$ and $\widehat{su}(k)_N$ are
related by the so-called level-rank duality
\cite{frenkel,jimbo-miwa,nakanishi,Naculich:1990pa,Hasegawa}, which maps to
each other orbits of their irreducible integrable representations under
outer automorphism groups. Let us explain this in more detail. The
Dynkin diagram of $\widehat{su}(N)_k$ consists of $N$ nodes permuted
in a cyclic order by the outer automorphism group $\mathbb{Z}_N$. This
also induces an action on affine irreducible integrable
representations. There are
\be
\frac{(N+k-1)!}{(N-1)!\,k!}   \label{nr-weights}
\ee
such representations of $\widehat{su}(N)_k$, which can be identified
in a standard way with Young diagrams $\rho$ with at most $N-1$ rows
and at most $k$ columns. We denote the set of such diagrams by
$\cY_{N-1,k}$. In particular, the generator of the outer automorphism
group $\sigma_N$, the so-called basic outer automorphism, has a simple
realization in terms of a Young diagram
$\rho=(\rho_1,\ldots,\rho_{N-1})$ corresponding to a given integrable
representation. The action of $\sigma_N$ amounts to adding a row of
length $k$ as a first row of $\rho$, and then reducing the diagram,
{\it i.e.} removing $\rho_{N-1}$ columns which acquired a length $N$ (so
that indeed $\sigma_N(\rho)\in \cY_{N-1,k}$),
\be 
\sigma_N (\rho_1,\ldots,\rho_{N-1}) =
(k-\rho_{N-1},\rho_1-\rho_{N-1},\ldots,\rho_{N-2}-\rho_{N-1}).  \label{cyclic}
\ee
It follows that $\sigma_N^N(\rho)=\rho$, as expected for
$\mathbb{Z}_N$ symmetry. All $N$ irreducible integrable
representations related by an action of $\sigma_N$ constitute an orbit
denoted as $[\rho] \subset \cY_{N-1,k}$. As an example, the  $\mathbb{Z}_4$
orbit generated from $\widehat{su}(4)_3$ irreducible integrable
representation corresponding to a diagram $\rho=(2,1)\in \cY_{3,3}$ is
given by
$$ 
\Yvcentermath1 \yng(2,1) \ \to \ \yng(3,2,1) \ \to \ \yng(2,2,1)
\ \to \ \yng(2,1,1)
$$

The number of such $\mathbb{Z}_N$ orbits is given by
(\ref{nr-weights}) divided by $N$. For both $\widehat{su}(N)_k$ and
$\widehat{su}(k)_N$ this number is the same, therefore a bijection
between orbits of their integrable irreducible representations
exists. The level-rank duality is a statement that for
$\widehat{su}(N)_k$ orbit represented by a diagram
$\rho\in\cY_{N-1,k}$ there is a canonical bijection realized as
\bea 
\cY_{N-1,k} \supset [\rho] & =&
\{\sigma^j_N(\rho)\ |\ j=0,\ldots,N-1 \}\ \mapsto \nonumber \\ 
& \mapsto & \{\sigma^a_k(\rho^t)\ |\ a=0,\ldots,k-1\} = [\rho^t]
\subset \cY_{k-1,N}, \label{lr-orbit} \eea
where $^t$ denotes a transposition and a diagram $\rho^t$ should be
reduced (i.e. all columns of length $k$ should be removed if $\rho_1$
was equal to $k$).

The level-rank duality can also be formulated in terms of the
embedding
$$ 
\widehat{u}(1)_{Nk} \times \widehat{su}(N)_k \times
\widehat{su}(k)_N \subset \widehat{u}(Nk)_1.
$$
The $\widehat{u}(Nk)_1$ affine Lie algebra can be realized in terms of
$Nk$ free fermions, so that their total Fock space
$\mathcal{F}^{\otimes Nk}$ decomposes under this embedding as
\be
\cF^{\otimes Nk} = \bigoplus_{\rho} U_{\|\rho\|} \otimes V_{\rho} \otimes
{\wt V}_{\wt\rho},  \label{lr-fock}
\ee
where $U_{\|\rho\|}$, $V_{\rho}$ and ${\wt V}_{\wt\rho}$ denote
irreducible integrable representations of $\widehat{u}(1)_{Nk}$,
$\widehat{su}(k)_N$, and $\widehat{su}(N)_k$ respectively. In the
above decomposition only those pairs $(\rho,\wt\rho)$ arise, which
represent orbits mapped to each other by the duality
(\ref{lr-orbit}). For a given $\widehat{su}(N)_k$ orbit $[\rho]$
represented by $\rho$, these pairs are therefore of the form
$(\sigma^j_N(\rho),\sigma^a_k(\rho^t))$, where $\sigma_N$ and
$\sigma_k$ are appropriate outer automorphism groups. The $U(1)$
charge corresponding to such a pair is $\|\rho\|=(|\rho|+jk +
aN)\ \textrm{mod}\ Nk$, where $|\rho|$ is the number of boxes in the
Young diagram $\rho$. With such identifications, the decomposition
(\ref{lr-fock}) can be written in terms of characters as
\cite{Hasegawa}
\begin{align}   \label{uNk-decompose}
&\chi^{\widehat{u}(Nk)_1}(u,v,\tau) = \\
&\sum_{[\rho]\subset
  \mathcal{Y}_{N-1,k}} \sum_{j=0}^{N-1} \sum_{a=0}^{k-1}
\chi^{\widehat{u}(1)_{Nk}}_{|\rho|+jk+aN}(N |u| + k
|v|,\tau)\ \chi^{\widehat{su}(N)_k}_{\sigma^j_N(\rho)} (\overline{u},\tau)
\chi^{\widehat{su}(k)_N}_{\sigma^a_k(\rho^t)}
(\overline{v},\tau). \notag
\end{align}
Here $u=(u_j)_{j=1...N}$ are elements of the Cartan subalgebra of $u(N)$,
$|u|=\sum_j u_j$ and $\overline{u}$ denotes the traceless
part. Similarly $v=(v_a)_{a=1...k}$ are elements of Cartan subalgebra
of $u(k)$.  $\chi^{\widehat{su}(N)_k}_{\rho} (\overline{u},\tau)$ are
characters of $\widehat{su}(N)_k$ at level $k$ for an integrable
irreducible representation specified by a Young diagram $\rho$, and
$\chi^{\widehat{u}(1)_N}_j$ characters are defined as
$$ 
\chi^{\widehat{u}(1)_N}_j(z,\tau) =
\frac{1}{\eta(q)}\sum_{n\in\mathbb{Z}} q^{\frac{N}{2}(n+j/N)^2}
e^{2\pi i z (n+j/N)}
$$
for $q=e^{2\pi i \tau}$. 

As an example of a decomposition (\ref{lr-fock}) let us consider the
case of $\widehat{u}(1)_{12} \times \widehat{su}(4)_3 \times
\widehat{su}(3)_4 \subset \widehat{u}(12)_1$, with $N=4$ and
$k=3$. From (\ref{nr-weights}) we deduce there are 5 orbits of outer
automorphism groups $\mathbb{Z}_4$ and $\mathbb{Z}_3$. Let us consider
$\widehat{su}(4)_3$ integrable representation related to a diagram
$\rho=\square$, and the corresponding $\widehat{su}(3)_4$ diagram
$\rho^t = \square$. The two orbits under $\sigma_4$ and $\sigma_3$ are
shown respectively in the first row and column of a table below. All
12 pairs of representations appear in the decomposition
(\ref{lr-fock}) with $\widehat{u}(1)_{12}$ charges given in the
table. Note that acting with $\sigma_4$ takes us to another pair of
weights given by a step to the right in the table, and increases
$\widehat{u}(1)_{12}$ charge by 3 (modulo 12). The action of
$\sigma_3$ takes us a step to the bottom in the table and increases
$\widehat{u}(1)_{12}$ charge by 4 (modulo 12). Of course the same
table is generated when we build it starting from any other
element of these two orbits.
\medskip

$$
\Yvcentermath1
\begin{array}{c|ccccccc}
   & \yng(1) & \to & \yng(3,1) & \to & \yng(3,3,1) & \to & \yng(2,2,2) \\ 
\\ \hline
\\
\yng(1) & 1 & & 4 & & 7 & & 10  \\ 
\downarrow & & & & & \\
\yng(4,1) & 5 & & 8 & & 11 & & 2 \\
\downarrow & & & & &\\
\yng(3,3) & 9 & & 0 & & 3 & & 6
\end{array}
$$
\medskip

Pairs of $\widehat{su}(4)_3 \times \widehat{su}(3)_4$ integrable
weights with the same fixed $\widehat{u}(1)_{12}$ charge, arising in
the decomposition of $\widehat{u}(12)_1$, are easily found if all 5
such tables of orbits are drawn. For example for charge 0 we then get
$$
\Yvcentermath1
\bullet\otimes\bullet\ +\ \yng(3,1)\otimes\yng(3,3)\ +\ \yng(2,2)\otimes\yng(4,2)\ + 
$$
$$
\Yvcentermath1
+\ \yng(3,3,2)\otimes\yng(3)\ + \ \yng(2,1,1)\otimes\yng(2,1).
$$
%


    \backmatter
        \phantomsection
        \addcontentsline{toc}{chapter}{Bibliography}
        \begin{small}
        \begin{singlespace}
  	  \bibliographystyle{JHEP}
	  \bibliography{thesis}

\providecommand{\href}[2]{#2}\begingroup\raggedright\begin{thebibliography}{10%
0}

\bibitem{Dijkgraaf:2007sw}
R.~Dijkgraaf, L.~Hollands, P.~Sulkowski, and C.~Vafa, {\it {Supersymmetric
  Gauge Theories, Intersecting Branes and Free Fermions}},  {\em JHEP} {\bf 02}
  (2008) 106, [\href{http://xxx.lanl.gov/abs/0709.4446}{{\tt 0709.4446}}].

\bibitem{Dijkgraaf:2008fh}
R.~Dijkgraaf, L.~Hollands, and P.~Sulkowski, {\it {Quantum Curves and
  D-Modules}},  \href{http://xxx.lanl.gov/abs/0810.4157}{{\tt 0810.4157}}.

\bibitem{Cheng:2009hm}
M.~C.~N. Cheng and L.~Hollands, {\it {A Geometric Derivation of the Dyon
  Wall-Crossing Group}},  {\em JHEP} {\bf 04} (2009) 067,
  [\href{http://xxx.lanl.gov/abs/0901.1758}{{\tt 0901.1758}}].

\bibitem{Hollands:2008cs}
L.~Hollands, J.~Marsano, K.~Papadodimas, and M.~Shigemori, {\it
  {Nonsupersymmetric Flux Vacua and Perturbed N=2 Systems}},  {\em JHEP} {\bf
  10} (2008) 102, [\href{http://xxx.lanl.gov/abs/0804.4006}{{\tt 0804.4006}}].

\bibitem{Strominger:1996it}
A.~Strominger, S.-T. Yau, and E.~Zaslow, {\it {Mirror symmetry is T-duality}},
  {\em Nucl. Phys.} {\bf B479} (1996) 243--259,
  [\href{http://xxx.lanl.gov/abs/hep-th/9606040}{{\tt hep-th/9606040}}].

\bibitem{Thomas:2005xf}
R.~P. Thomas, {\it {The geometry of mirror symmetry}},
  \href{http://xxx.lanl.gov/abs/math/0512412}{{\tt math/0512412}}.

\bibitem{gross-2008}
M.~Gross, {\it {The Strominger-Yau-Zaslow conjecture: From torus fibrations to
  degenerations}},  \href{http://xxx.lanl.gov/abs/0802.3407}{{\tt 0802.3407}}.

\bibitem{Joyce-slag}
{Joyce, Dominic}, {\em {Calabi-Yau manifolds and Related Geometries}},
  ch.~{Lectures on special Lagrangian geometry}.
\newblock {Springer}, 2003.
\newblock \href{http://xxx.lanl.gov/abs/math/0108088}{{\tt math/0108088}}.

\bibitem{Gross:1999hc}
M.~Gross, {\it {Topological mirror symmetry}},
  \href{http://xxx.lanl.gov/abs/math/9909015}{{\tt math/9909015}}.

\bibitem{kontsevich-2004}
{Kontsevich, Maxim and Soibelman, Yan}, {\it {Affine structures and
  non-archimedean analytic spaces}},
  \href{http://xxx.lanl.gov/abs/math/0406564}{{\tt math/0406564}}.

\bibitem{gross-2000}
{Gross, Mark and Wilson, P.~M.~H.}, {\it {Large Complex Structure Limits of K3
  Surfaces}},  \href{http://xxx.lanl.gov/abs/math/0008018}{{\tt math/0008018}}.

\bibitem{ruan}
{Wei-Dong Ruan}, {\it {Lagrangian torus fibration of quintic {C}alabi-{Y}au
  hypersurfaces I: Fermat quintic case}},  in {\em Winter school in mirror
  symmetry, vector bundles and Lagrangian submanifolds} (S.~Yau and C.~Vafa,
  eds.), AMS and International Press, 1999.
\newblock \href{http://xxx.lanl.gov/abs/math.DG/9904012}{{\tt
  math.DG/9904012}}.

\bibitem{Greene:1990ud}
B.~R. Greene and M.~R. Plesser, {\it {Duality in Calabi-Yau moduli space}},
  {\em Nucl. Phys.} {\bf B338} (1990) 15--37.

\bibitem{Candelas:1990rm}
P.~Candelas, X.~C. De~La~Ossa, P.~S. Green, and L.~Parkes, {\it {A pair of
  Calabi-Yau manifolds as an exactly soluble superconformal theory}},  {\em
  Nucl. Phys.} {\bf B359} (1991) 21--74.

\bibitem{Huang:2006hq}
M.-x. Huang, A.~Klemm, and S.~Quackenbush, {\it {Topological String Theory on
  Compact Calabi-Yau: Modularity and Boundary Conditions}},  {\em Lect. Notes
  Phys.} {\bf 757} (2009) 45--102,
  [\href{http://xxx.lanl.gov/abs/hep-th/0612125}{{\tt hep-th/0612125}}].

\bibitem{Harvey:1982xk}
R.~Harvey and J.~Lawson, H.~B., {\it {Calibrated geometries}},  {\em Acta
  Math.} {\bf 148} (1982) 47.

\bibitem{Aganagic:2003db}
M.~Aganagic, A.~Klemm, M.~Marino, and C.~Vafa, {\it {The topological vertex}},
  {\em Commun. Math. Phys.} {\bf 254} (2005) 425--478,
  [\href{http://xxx.lanl.gov/abs/hep-th/0305132}{{\tt hep-th/0305132}}].

\bibitem{Gopakumar:1998ki}
R.~Gopakumar and C.~Vafa, {\it {On the gauge theory/geometry correspondence}},
  {\em Adv. Theor. Math. Phys.} {\bf 3} (1999) 1415--1443,
  [\href{http://xxx.lanl.gov/abs/hep-th/9811131}{{\tt hep-th/9811131}}].

\bibitem{mnop1}
{D. Maulik and N. Nekrasov and A. Okounkov and R.~Pandharipande}, {\it
  {Gromov-Witten theory and Donaldson-Thomas theory I}},
  \href{http://xxx.lanl.gov/abs/math.AG/0312059}{{\tt math.AG/0312059}}.

\bibitem{mnop2}
{D. Maulik and N. Nekrasov and A. Okounkov and R.~Pandharipande}, {\it
  {Gromov-Witten theory and Donaldson-Thomas theory II}},
  \href{http://xxx.lanl.gov/abs/math.AG/0406092}{{\tt math.AG/0406092}}.

\bibitem{Szendroi:2007nu}
B.~Szendroi, {\it {Non-commutative Donaldson-Thomas theory and the conifold}},
  {\em Geom. Topol.} {\bf 12} (2008) 1171--1202,
  [\href{http://xxx.lanl.gov/abs/0705.3419}{{\tt 0705.3419}}].

\bibitem{Gopakumar:1998ii}
R.~Gopakumar and C.~Vafa, {\it {M-theory and topological strings. I}},
  \href{http://xxx.lanl.gov/abs/hep-th/9809187}{{\tt hep-th/9809187}}.

\bibitem{Gopakumar:1998jq}
R.~Gopakumar and C.~Vafa, {\it {M-theory and topological strings. II}},
  \href{http://xxx.lanl.gov/abs/hep-th/9812127}{{\tt hep-th/9812127}}.

\bibitem{Labastida:2000yw}
J.~M.~F. Labastida, M.~Marino, and C.~Vafa, {\it {Knots, links and branes at
  large N}},  {\em JHEP} {\bf 11} (2000) 007,
  [\href{http://xxx.lanl.gov/abs/hep-th/0010102}{{\tt hep-th/0010102}}].

\bibitem{Gukov:2004hz}
S.~Gukov, A.~S. Schwarz, and C.~Vafa, {\it {Khovanov-Rozansky homology and
  topological strings}},  {\em Lett. Math. Phys.} {\bf 74} (2005) 53--74,
  [\href{http://xxx.lanl.gov/abs/hep-th/0412243}{{\tt hep-th/0412243}}].

\bibitem{Gukov:2007tf}
S.~Gukov, A.~Iqbal, C.~Kozcaz, and C.~Vafa, {\it {Link homologies and the
  refined topological vertex}},  \href{http://xxx.lanl.gov/abs/0705.1368}{{\tt
  0705.1368}}.

\bibitem{Dijkgraaf:2009sb}
R.~Dijkgraaf and H.~Fuji, {\it {The Volume Conjecture and Topological
  Strings}},  \href{http://xxx.lanl.gov/abs/0903.2084}{{\tt 0903.2084}}.

\bibitem{Okounkov:2003sp}
A.~Okounkov, N.~Reshetikhin, and C.~Vafa, {\it {Quantum Calabi-Yau and
  classical crystals}},  \href{http://xxx.lanl.gov/abs/hep-th/0309208}{{\tt
  hep-th/0309208}}.

\bibitem{dijkgraaf-2009-811}
{Dijkgraaf, Robbert and Orlando, Domenico and Reffert, Susanne}, {\it {Quantum
  Crystals and Spin Chains}},  {\em Nuclear Physics B} {\bf 811} (2008) 463,
  [\href{http://xxx.lanl.gov/abs/0803.1927}{{\tt 0803.1927}}].

\bibitem{Ooguri:2008yb}
H.~Ooguri and M.~Yamazaki, {\it {Crystal Melting and Toric Calabi-Yau
  Manifolds}},  \href{http://xxx.lanl.gov/abs/0811.2801}{{\tt 0811.2801}}.

\bibitem{Macmahon}
{Percy A. Macmahon}, {\em {Combinatory Analysis}}.
\newblock Cambridge University Press, 1915.

\bibitem{Iqbal:2003ds}
A.~Iqbal, N.~Nekrasov, A.~Okounkov, and C.~Vafa, {\it {Quantum foam and
  topological strings}},  {\em JHEP} {\bf 04} (2008) 011,
  [\href{http://xxx.lanl.gov/abs/hep-th/0312022}{{\tt hep-th/0312022}}].

\bibitem{Ooguri:2009ri}
H.~Ooguri and M.~Yamazaki, {\it {Emergent Calabi-Yau Geometry}},
  \href{http://xxx.lanl.gov/abs/0902.3996}{{\tt 0902.3996}}.

\bibitem{Hori:2000kt}
K.~Hori and C.~Vafa, {\it {Mirror symmetry}},
  \href{http://xxx.lanl.gov/abs/hep-th/0002222}{{\tt hep-th/0002222}}.

\bibitem{calabi}
{Calabi E}, {\it {M\'etriques K\"ahl\'erienes et fibr\'es holomorphes}},  {\em
  {Ann. Sci. de \'Е. N. S.}} (1979), no.~12 266—292.

\bibitem{tianyaulocalCY1}
{G. Tian and S. T. Yau}, {\it {Existence of K\"ahler-Einstein metrics on
  complete K\"ahler manifolds and their applications to algebraic geometry}},
  in {\em Mathematical Aspects of String Theory} (S.~Yau, ed.), World
  Scientific Publishing Co., Singapore, 1987.

\bibitem{tianyaulocalCY2}
{Tian, Gang and Yau, Shing-Tung}, {\it {Complete Kähler manifolds with zero
  Ricci curvature I}},  {\em Amer. Math. Soc. 3 (3)} (1990).

\bibitem{tianyaulocalCY3}
{Tian, Gang and Yau, Shing-Tung}, {\it {Complete Kähler manifolds with zero
  Ricci curvature II}},  {\em Invent. Math. 106 (1)} (1991).

\bibitem{nakajima}
{H.~Nakajima}, {\it {Instantons on ALE spaces, quiver varieties, and Kac-Moody
  algebras}},  {\em {Duke Math.}} {\bf 76} (1994) 365--416.

\bibitem{Vafa:1994tf}
C.~Vafa and E.~Witten, {\it {A strong coupling test of S duality}},  {\em Nucl.
  Phys.} {\bf B431} (1994) 3--77,
  [\href{http://xxx.lanl.gov/abs/hep-th/9408074}{{\tt hep-th/9408074}}].

\bibitem{Montonen:1977sn}
C.~Montonen and D.~I. Olive, {\it {Magnetic Monopoles as Gauge Particles?}},
  {\em Phys. Lett.} {\bf B72} (1977) 117.

\bibitem{Witten:1995gf}
E.~Witten, {\it {On S duality in Abelian gauge theory}},  {\em Selecta Math.}
  {\bf 1} (1995) 383, [\href{http://xxx.lanl.gov/abs/hep-th/9505186}{{\tt
  hep-th/9505186}}].

\bibitem{Verlinde:1995mz}
E.~P. Verlinde, {\it {Global aspects of electric - magnetic duality}},  {\em
  Nucl. Phys.} {\bf B455} (1995) 211--228,
  [\href{http://xxx.lanl.gov/abs/hep-th/9506011}{{\tt hep-th/9506011}}].

\bibitem{Witten:1978mh}
E.~Witten and D.~I. Olive, {\it {Supersymmetry Algebras That Include
  Topological Charges}},  {\em Phys. Lett.} {\bf B78} (1978) 97.

\bibitem{Witten:1994ev}
E.~Witten, {\it {Supersymmetric Yang-Mills theory on a four manifold}},  {\em
  J. Math. Phys.} {\bf 35} (1994) 5101--5135,
  [\href{http://xxx.lanl.gov/abs/hep-th/9403195}{{\tt hep-th/9403195}}].

\bibitem{Witten:1992xu}
E.~Witten, {\it {Two-dimensional gauge theories revisited}},  {\em J. Geom.
  Phys.} {\bf 9} (1992) 303--368,
  [\href{http://xxx.lanl.gov/abs/hep-th/9204083}{{\tt hep-th/9204083}}].

\bibitem{Blau:1995rs}
M.~Blau and G.~Thompson, {\it {Localization and diagonalization: A review of
  functional integral techniques for low dimensional gauge theories and
  topological field theories}},  {\em J. Math. Phys.} {\bf 36} (1995)
  2192--2236, [\href{http://xxx.lanl.gov/abs/hep-th/9501075}{{\tt
  hep-th/9501075}}].

\bibitem{Cordes:1994fc}
S.~Cordes, G.~W. Moore, and S.~Ramgoolam, {\it {Lectures on 2-d Yang-Mills
  theory, equivariant cohomology and topological field theories}},  {\em Nucl.
  Phys. Proc. Suppl.} {\bf 41} (1995) 184--244,
  [\href{http://xxx.lanl.gov/abs/hep-th/9411210}{{\tt hep-th/9411210}}].

\bibitem{harveylawson}
{R.~Harvey and H.B.~Lawson}, {\it {Calibrated geometry}},  {\em {Acta Math.}}
  {\bf 148} (1982) {47--157}.

\bibitem{Bershadsky:1995qy}
M.~Bershadsky, C.~Vafa, and V.~Sadov, {\it {D-Branes and Topological Field
  Theories}},  {\em Nucl. Phys.} {\bf B463} (1996) 420--434,
  [\href{http://xxx.lanl.gov/abs/hep-th/9511222}{{\tt hep-th/9511222}}].

\bibitem{Green:1996dd}
M.~B. Green, J.~A. Harvey, and G.~W. Moore, {\it {I-brane inflow and anomalous
  couplings on D-branes}},  {\em Class. Quant. Grav.} {\bf 14} (1997) 47--52,
  [\href{http://xxx.lanl.gov/abs/hep-th/9605033}{{\tt hep-th/9605033}}].

\bibitem{Cheung:1997az}
Y.-K.~E. Cheung and Z.~Yin, {\it {Anomalies, branes, and currents}},  {\em
  Nucl. Phys.} {\bf B517} (1998) 69--91,
  [\href{http://xxx.lanl.gov/abs/hep-th/9710206}{{\tt hep-th/9710206}}].

\bibitem{Strominger:1995ac}
A.~Strominger, {\it {Open p-branes}},  {\em Phys. Lett.} {\bf B383} (1996)
  44--47, [\href{http://xxx.lanl.gov/abs/hep-th/9512059}{{\tt
  hep-th/9512059}}].

\bibitem{Rocek:1991ps}
M.~Rocek and E.~P. Verlinde, {\it {Duality, quotients, and currents}},  {\em
  Nucl. Phys.} {\bf B373} (1992) 630--646,
  [\href{http://xxx.lanl.gov/abs/hep-th/9110053}{{\tt hep-th/9110053}}].

\bibitem{Buscher:1987qj}
T.~H. Buscher, {\it {Path Integral Derivation of Quantum Duality in Nonlinear
  Sigma Models}},  {\em Phys. Lett.} {\bf B201} (1988) 466.

\bibitem{Obers:1998fb}
N.~A. Obers and B.~Pioline, {\it {U-duality and M-theory}},  {\em Phys. Rept.}
  {\bf 318} (1999) 113--225,
  [\href{http://xxx.lanl.gov/abs/hep-th/9809039}{{\tt hep-th/9809039}}].

\bibitem{Seiberg:1988ur}
N.~Seiberg, {\it {Supersymmetry and Nonperturbative beta Functions}},  {\em
  Phys. Lett.} {\bf B206} (1988) 75.

\bibitem{Dijkgraaf:1996tz}
R.~Dijkgraaf and G.~W. Moore, {\it {Balanced topological field theories}},
  {\em Commun. Math. Phys.} {\bf 185} (1997) 411--440,
  [\href{http://xxx.lanl.gov/abs/hep-th/9608169}{{\tt hep-th/9608169}}].

\bibitem{franscesco}
{di Franscesco, Philippe and Mathieu, Pierre and S\'en\'echal, David}, {\em
  {Conformal Field Theory}}.
\newblock Springer, 1997.

\bibitem{mckay}
{McKay, John}, {\it {Graphs, singularities, and finite groups}},  in {\em {The
  Santa Cruz Conference on Finite Groups}}, vol.~37 of {\em {Proc. Symp. Pure
  Math.}}, pp.~183--186, {Amer. Math. Soc.}, 1980.

\bibitem{mckay-slodowy}
{P.~Slodowy}, {\it {Platonic solids, Kleinian singularities, and Lie groups in
  Algebraic geometry}},  in {\em {Algebraic Geometry}} ({J. Dolgachev}, ed.),
  vol.~1008 of {\em {Lecture Notes in Mathematics}}, pp.~102--138.
\newblock {Springer-Verlag}, 1983.

\bibitem{Nakajima:1995ka}
H.~Nakajima, {\it {Instantons and affine Lie algebras}},  {\em Nucl. Phys.
  Proc. Suppl.} {\bf 46} (1996) 154--161,
  [\href{http://xxx.lanl.gov/abs/alg-geom/9510003}{{\tt alg-geom/9510003}}].

\bibitem{Bianchi:1996zj}
M.~Bianchi, F.~Fucito, G.~Rossi, and M.~Martellini, {\it {Explicit Construction
  of Yang-Mills Instantons on ALE Spaces}},  {\em Nucl. Phys.} {\bf B473}
  (1996) 367--404, [\href{http://xxx.lanl.gov/abs/hep-th/9601162}{{\tt
  hep-th/9601162}}].

\bibitem{Fucito:2004ry}
F.~Fucito, J.~F. Morales, and R.~Poghossian, {\it {Multi instanton calculus on
  ALE spaces}},  {\em Nucl. Phys.} {\bf B703} (2004) 518--536,
  [\href{http://xxx.lanl.gov/abs/hep-th/0406243}{{\tt hep-th/0406243}}].

\bibitem{Fucito:2006kn}
F.~Fucito, J.~F. Morales, and R.~Poghossian, {\it {Instanton on toric
  singularities and black hole countings}},  {\em JHEP} {\bf 12} (2006) 073,
  [\href{http://xxx.lanl.gov/abs/hep-th/0610154}{{\tt hep-th/0610154}}].

\bibitem{frenkel}
{I. Frenkel}, {\em {Lie Algebras and Related Topics}}, vol.~933,
  ch.~{Representations of affine Lie algebras, Hecke modular forms and
  {K}orteweg-{D}e {V}ries type equations}, p.~71.
\newblock Springer, 1982.

\bibitem{licata}
{A. Licata}, {\it {Framed rank $r$ torsion-free sheaves on $CP^2$ and
  representations of the affine {L}ie algebra $\widehat{gl(r)}$}},
  \href{http://xxx.lanl.gov/abs/math.RT/0607690}{{\tt math.RT/0607690}}.

\bibitem{Gaiotto:2005gf}
D.~Gaiotto, A.~Strominger, and X.~Yin, {\it {New Connections Between 4D and 5D
  Black Holes}},  {\em JHEP} {\bf 02} (2006) 024,
  [\href{http://xxx.lanl.gov/abs/hep-th/0503217}{{\tt hep-th/0503217}}].

\bibitem{Gaiotto:2005xt}
D.~Gaiotto, A.~Strominger, and X.~Yin, {\it {5D black rings and 4D black
  holes}},  {\em JHEP} {\bf 02} (2006) 023,
  [\href{http://xxx.lanl.gov/abs/hep-th/0504126}{{\tt hep-th/0504126}}].

\bibitem{Shih:2005uc}
D.~Shih, A.~Strominger, and X.~Yin, {\it {Recounting dyons in N = 4 string
  theory}},  {\em JHEP} {\bf 10} (2006) 087,
  [\href{http://xxx.lanl.gov/abs/hep-th/0505094}{{\tt hep-th/0505094}}].

\bibitem{Dijkgraaf:2006um}
R.~Dijkgraaf, C.~Vafa, and E.~Verlinde, {\it {M-theory and a topological string
  duality}},  \href{http://xxx.lanl.gov/abs/hep-th/0602087}{{\tt
  hep-th/0602087}}.

\bibitem{Bena:2006qm}
I.~Bena, D.-E. Diaconescu, and B.~Florea, {\it {Black string entropy and
  Fourier-Mukai transform}},  {\em JHEP} {\bf 04} (2007) 045,
  [\href{http://xxx.lanl.gov/abs/hep-th/0610068}{{\tt hep-th/0610068}}].

\bibitem{Ooguri:1995wj}
H.~Ooguri and C.~Vafa, {\it {Two-Dimensional Black Hole and Singularities of CY
  Manifolds}},  {\em Nucl. Phys.} {\bf B463} (1996) 55--72,
  [\href{http://xxx.lanl.gov/abs/hep-th/9511164}{{\tt hep-th/9511164}}].

\bibitem{Ruback:1986ag}
P.~J. Ruback, {\it {The motion of Kaluza-Klein monopoles}},  {\em Commun. Math.
  Phys.} {\bf 107} (1986) 93--102.

\bibitem{Sen:1997js}
A.~Sen, {\it {Dynamics of multiple Kaluza-Klein monopoles in M and string
  theory}},  {\em Adv. Theor. Math. Phys.} {\bf 1} (1998) 115--126,
  [\href{http://xxx.lanl.gov/abs/hep-th/9707042}{{\tt hep-th/9707042}}].

\bibitem{Cherkis:2008ip}
S.~A. Cherkis, {\it {Moduli Spaces of Instantons on the Taub-NUT Space}},
  \href{http://xxx.lanl.gov/abs/0805.1245}{{\tt 0805.1245}}.

\bibitem{Witten:2009xu}
E.~Witten, {\it {Branes, Instantons, And Taub-NUT Spaces}},
  \href{http://xxx.lanl.gov/abs/0902.0948}{{\tt 0902.0948}}.

\bibitem{Cherkis:2009jm}
S.~A. Cherkis, {\it {Instantons on the Taub-NUT Space}},
  \href{http://xxx.lanl.gov/abs/0902.4724}{{\tt 0902.4724}}.

\bibitem{Witten:2009at}
E.~Witten, {\it {Geometric Langlands From Six Dimensions}},
  \href{http://xxx.lanl.gov/abs/0905.2720}{{\tt 0905.2720}}.

\bibitem{Bachas:1997kn}
C.~P. Bachas, M.~B. Green, and A.~Schwimmer, {\it {(8,0) quantum mechanics and
  symmetry enhancement in type I' superstrings}},  {\em JHEP} {\bf 01} (1998)
  006, [\href{http://xxx.lanl.gov/abs/hep-th/9712086}{{\tt hep-th/9712086}}].

\bibitem{Hung:2006nn}
L.-Y. Hung, {\it {Comments on I1-branes}},  {\em JHEP} {\bf 05} (2007) 076,
  [\href{http://xxx.lanl.gov/abs/hep-th/0612207}{{\tt hep-th/0612207}}].

\bibitem{Itzhaki:2005tu}
N.~Itzhaki, D.~Kutasov, and N.~Seiberg, {\it {I-brane dynamics}},  {\em JHEP}
  {\bf 01} (2006) 119, [\href{http://xxx.lanl.gov/abs/hep-th/0508025}{{\tt
  hep-th/0508025}}].

\bibitem{Naculich:1990pa}
S.~G. Naculich, H.~A. Riggs, and H.~J. Schnitzer, {\it Group level duality in
  wzw models and {C}hern-{S}imons theory},  {\em Phys. Lett.} {\bf B246} (1990)
  417--422.

\bibitem{Sen:1997kz}
A.~Sen, {\it {A note on enhanced gauge symmetries in M- and string theory}},
  {\em JHEP} {\bf 09} (1997) 001,
  [\href{http://xxx.lanl.gov/abs/hep-th/9707123}{{\tt hep-th/9707123}}].

\bibitem{Gimon:1996rq}
E.~G. Gimon and J.~Polchinski, {\it {Consistency Conditions for Orientifolds
  and D-Manifolds}},  {\em Phys. Rev.} {\bf D54} (1996) 1667--1676,
  [\href{http://xxx.lanl.gov/abs/hep-th/9601038}{{\tt hep-th/9601038}}].

\bibitem{Evans:1997hk}
N.~J. Evans, C.~V. Johnson, and A.~D. Shapere, {\it {Orientifolds, branes, and
  duality of 4D gauge theories}},  {\em Nucl. Phys.} {\bf B505} (1997)
  251--271, [\href{http://xxx.lanl.gov/abs/hep-th/9703210}{{\tt
  hep-th/9703210}}].

\bibitem{jimbo-miwa}
{M.~Jimbo, T.~Miwa}, {\it {On a duality of branching rules for affine Lie
  algebras}},  {\em {Adv. Stud. Pure Math.}} {\bf 6} (1985) 17--65.

\bibitem{Hasegawa}
K.~Hasegawa, {\it Spin module versions of weyl's reciprocity theorem for
  classical kac-moody lie algebras - an application to branching rule duality},
   {\em RIMS, Kyoto Univ. 25} (1989).

\bibitem{Goddard:1985jp}
P.~Goddard, W.~Nahm, and D.~I. Olive, {\it {Symmetric Spaces, Sugawara's Energy
  Momentum Tensor in Two-Dimensions and Free Fermions}},  {\em Phys. Lett.}
  {\bf B160} (1985) 111.

\bibitem{Morrison:1996na}
D.~R. Morrison and C.~Vafa, {\it {Compactifications of F-Theory on Calabi--Yau
  Threefolds -- I}},  {\em Nucl. Phys.} {\bf B473} (1996) 74--92,
  [\href{http://xxx.lanl.gov/abs/hep-th/9602114}{{\tt hep-th/9602114}}].

\bibitem{Morrison:1996pp}
D.~R. Morrison and C.~Vafa, {\it {Compactifications of F-Theory on Calabi--Yau
  Threefolds -- II}},  {\em Nucl. Phys.} {\bf B476} (1996) 437--469,
  [\href{http://xxx.lanl.gov/abs/hep-th/9603161}{{\tt hep-th/9603161}}].

\bibitem{Harvey:2007ab}
J.~A. Harvey and A.~B. Royston, {\it {Localized Modes at a D-brane--O-plane
  Intersection and Heterotic Alice Strings}},  {\em JHEP} {\bf 04} (2008) 018,
  [\href{http://xxx.lanl.gov/abs/0709.1482}{{\tt 0709.1482}}].

\bibitem{Seiberg:1994rs}
N.~Seiberg and E.~Witten, {\it {Monopole Condensation, And Confinement In N=2
  Supersymmetric Yang-Mills Theory}},  {\em Nucl. Phys.} {\bf B426} (1994)
  19--52, [\href{http://xxx.lanl.gov/abs/hep-th/9407087}{{\tt
  hep-th/9407087}}].

\bibitem{Ferrari:1996sv}
F.~Ferrari and A.~Bilal, {\it {The Strong-Coupling Spectrum of the
  Seiberg-Witten Theory}},  {\em Nucl. Phys.} {\bf B469} (1996) 387--402,
  [\href{http://xxx.lanl.gov/abs/hep-th/9602082}{{\tt hep-th/9602082}}].

\bibitem{Witten:1997sc}
E.~Witten, {\it {Solutions of four-dimensional field theories via M- theory}},
  {\em Nucl. Phys.} {\bf B500} (1997) 3--42,
  [\href{http://xxx.lanl.gov/abs/hep-th/9703166}{{\tt hep-th/9703166}}].

\bibitem{Bershadsky:1995sp}
M.~Bershadsky, C.~Vafa, and V.~Sadov, {\it {D-Strings on D-Manifolds}},  {\em
  Nucl. Phys.} {\bf B463} (1996) 398--414,
  [\href{http://xxx.lanl.gov/abs/hep-th/9510225}{{\tt hep-th/9510225}}].

\bibitem{Katz:1996xe}
S.~H. Katz and C.~Vafa, {\it {Matter from geometry}},  {\em Nucl. Phys.} {\bf
  B497} (1997) 146--154, [\href{http://xxx.lanl.gov/abs/hep-th/9606086}{{\tt
  hep-th/9606086}}].

\bibitem{Katz:1996fh}
S.~H. Katz, A.~Klemm, and C.~Vafa, {\it {Geometric engineering of quantum field
  theories}},  {\em Nucl. Phys.} {\bf B497} (1997) 173--195,
  [\href{http://xxx.lanl.gov/abs/hep-th/9609239}{{\tt hep-th/9609239}}].

\bibitem{Katz:1996th}
S.~H. Katz and C.~Vafa, {\it {Geometric engineering of N = 1 quantum field
  theories}},  {\em Nucl. Phys.} {\bf B497} (1997) 196--204,
  [\href{http://xxx.lanl.gov/abs/hep-th/9611090}{{\tt hep-th/9611090}}].

\bibitem{Katz:1997eq}
S.~Katz, P.~Mayr, and C.~Vafa, {\it {Mirror symmetry and exact solution of 4D N
  = 2 gauge theories. I}},  {\em Adv. Theor. Math. Phys.} {\bf 1} (1998)
  53--114, [\href{http://xxx.lanl.gov/abs/hep-th/9706110}{{\tt
  hep-th/9706110}}].

\bibitem{Klemm:1996bj}
A.~Klemm, W.~Lerche, P.~Mayr, C.~Vafa, and N.~P. Warner, {\it {Self-Dual
  Strings and N=2 Supersymmetric Field Theory}},  {\em Nucl. Phys.} {\bf B477}
  (1996) 746--766, [\href{http://xxx.lanl.gov/abs/hep-th/9604034}{{\tt
  hep-th/9604034}}].

\bibitem{Becker:1995kb}
K.~Becker, M.~Becker, and A.~Strominger, {\it {Five-branes, membranes and
  nonperturbative string theory}},  {\em Nucl. Phys.} {\bf B456} (1995)
  130--152, [\href{http://xxx.lanl.gov/abs/hep-th/9507158}{{\tt
  hep-th/9507158}}].

\bibitem{Tong:2002rq}
D.~Tong, {\it {NS5-branes, T-duality and worldsheet instantons}},  {\em JHEP}
  {\bf 07} (2002) 013, [\href{http://xxx.lanl.gov/abs/hep-th/0204186}{{\tt
  hep-th/0204186}}].

\bibitem{Harvey:2005ab}
J.~A. Harvey and S.~Jensen, {\it {Worldsheet instanton corrections to the
  Kaluza-Klein monopole}},  {\em JHEP} {\bf 10} (2005) 028,
  [\href{http://xxx.lanl.gov/abs/hep-th/0507204}{{\tt hep-th/0507204}}].

\bibitem{kaluza}
{Kaluza, Theodor}, {\it {Zum Unit\"atsproblem in der Physik}},  {\em
  {Sitzungsber. Preuss. Akad. Wiss. Berlin. (Math. Phys.)}} ({1921})
  {966--972}.

\bibitem{klein}
{Klein, Oskar}, {\it {Quantentheorie und f\"unfdimensionale
  Relativit\"atstheorie}},  {\em {Zeitschrift f\"ur Physik a Hadrons and
  Nuclei}} {\bf {37 (12)}} ({1926}) {895–906}.

\bibitem{deWit:1984px}
B.~de~Wit, P.~G. Lauwers, and A.~Van~Proeyen, {\it {Lagrangians of N=2
  Supergravity - Matter Systems}},  {\em Nucl. Phys.} {\bf B255} (1985) 569.

\bibitem{Strominger:1990pd}
A.~Strominger, {\it {Special geometry}},  {\em Commun. Math. Phys.} {\bf 133}
  (1990) 163--180.

\bibitem{Craps:1997gp}
B.~Craps, F.~Roose, W.~Troost, and A.~Van~Proeyen, {\it {What is special
  Kaehler geometry?}},  {\em Nucl. Phys.} {\bf B503} (1997) 565--613,
  [\href{http://xxx.lanl.gov/abs/hep-th/9703082}{{\tt hep-th/9703082}}].

\bibitem{Billo:1998yr}
M.~Billo, F.~Denef, P.~Fre, I.~Pesando, W.~Troost, A.~Van~Proeyen, and D.~Zano,
  {\it {The rigid limit in special Kaehler geometry: From K3- fibrations to
  special Riemann surfaces: A detailed case study}},  {\em Class. Quant. Grav.}
  {\bf 15} (1998) 2083--2152,
  [\href{http://xxx.lanl.gov/abs/hep-th/9803228}{{\tt hep-th/9803228}}].

\bibitem{Witten:1988xj}
E.~Witten, {\it {Topological Sigma Models}},  {\em Commun. Math. Phys.} {\bf
  118} (1988) 411.

\bibitem{Vonk:2005yv}
M.~Vonk, {\it {A mini-course on topological strings}},
  \href{http://xxx.lanl.gov/abs/hep-th/0504147}{{\tt hep-th/0504147}}.

\bibitem{Neitzke:2004ni}
A.~Neitzke and C.~Vafa, {\it {Topological strings and their physical
  applications}},  \href{http://xxx.lanl.gov/abs/hep-th/0410178}{{\tt
  hep-th/0410178}}.

\bibitem{notesKlemm}
{A. Klemm}, ``{Introduction to Topological String Theory on Calabi-Yau
  manifolds}.''
  \texttt{http://www.math.ist.utl.pt/~strings/AGTS/topstrings.pdf}, 2003.

\bibitem{horimirrorsymmetry}
{K. Hori et al.}, {\em {Mirror Symmetry}}, vol.~1 of {\em {Clay Mathematics
  Monographs}}.
\newblock {American Mathematical Society}, 2003.

\bibitem{coxkatz}
{D. A. Cox and S. Katz}, {\em {Mirror symmetry and Algebraic Geometry}},
  vol.~{68} of {\em {Mathematical Surveys and Monographs}}.
\newblock {American Mathematical Society}, {1999}.

\bibitem{Bershadsky:1993cx}
M.~Bershadsky, S.~Cecotti, H.~Ooguri, and C.~Vafa, {\it {Kodaira-Spencer theory
  of gravity and exact results for quantum string amplitudes}},  {\em Commun.
  Math. Phys.} {\bf 165} (1994) 311--428,
  [\href{http://xxx.lanl.gov/abs/hep-th/9309140}{{\tt hep-th/9309140}}].

\bibitem{Antoniadis:1993ze}
I.~Antoniadis, E.~Gava, K.~S. Narain, and T.~R. Taylor, {\it {Topological
  amplitudes in string theory}},  {\em Nucl. Phys.} {\bf B413} (1994) 162--184,
  [\href{http://xxx.lanl.gov/abs/hep-th/9307158}{{\tt hep-th/9307158}}].

\bibitem{Katz:1999xq}
S.~H. Katz, A.~Klemm, and C.~Vafa, {\it {M-theory, topological strings and
  spinning black holes}},  {\em Adv. Theor. Math. Phys.} {\bf 3} (1999)
  1445--1537, [\href{http://xxx.lanl.gov/abs/hep-th/9910181}{{\tt
  hep-th/9910181}}].

\bibitem{Ooguri:2004zv}
H.~Ooguri, A.~Strominger, and C.~Vafa, {\it {Black hole attractors and the
  topological string}},  {\em Phys. Rev.} {\bf D70} (2004) 106007,
  [\href{http://xxx.lanl.gov/abs/hep-th/0405146}{{\tt hep-th/0405146}}].

\bibitem{Vafa:2004qa}
C.~Vafa, {\it {Two dimensional Yang-Mills, black holes and topological
  strings}},  \href{http://xxx.lanl.gov/abs/hep-th/0406058}{{\tt
  hep-th/0406058}}.

\bibitem{Aganagic:2004js}
M.~Aganagic, H.~Ooguri, N.~Saulina, and C.~Vafa, {\it {Black holes, q-deformed
  2d Yang-Mills, and non- perturbative topological strings}},  {\em Nucl.
  Phys.} {\bf B715} (2005) 304--348,
  [\href{http://xxx.lanl.gov/abs/hep-th/0411280}{{\tt hep-th/0411280}}].

\bibitem{Dijkgraaf:2002fc}
R.~Dijkgraaf and C.~Vafa, {\it {Matrix models, topological strings, and
  supersymmetric gauge theories}},  {\em Nucl. Phys.} {\bf B644} (2002) 3--20,
  [\href{http://xxx.lanl.gov/abs/hep-th/0206255}{{\tt hep-th/0206255}}].

\bibitem{Dijkgraaf:2002vw}
R.~Dijkgraaf and C.~Vafa, {\it {On geometry and matrix models}},  {\em Nucl.
  Phys.} {\bf B644} (2002) 21--39,
  [\href{http://xxx.lanl.gov/abs/hep-th/0207106}{{\tt hep-th/0207106}}].

\bibitem{Dijkgraaf:2002dh}
R.~Dijkgraaf and C.~Vafa, {\it {A perturbative window into non-perturbative
  physics}},  \href{http://xxx.lanl.gov/abs/hep-th/0208048}{{\tt
  hep-th/0208048}}.

\bibitem{behrend}
{K. Behrend}, {\it {Gromov-Witten invariants in algebraic geometry}},  {\em
  {Invent. Math.}} {\bf 127} (1997) 601--617.

\bibitem{li-tian}
{J. Li and G. Tian}, {\it {Virtual moduli cycles and Gromov-Witten invariants
  of algebraic varieties}},  {\em {JAMS}} {\bf 11} (1998) 119–174.

\bibitem{Aganagic:2003qj}
M.~Aganagic, R.~Dijkgraaf, A.~Klemm, M.~Marino, and C.~Vafa, {\it {Topological
  strings and integrable hierarchies}},  {\em Commun. Math. Phys.} {\bf 261}
  (2006) 451--516, [\href{http://xxx.lanl.gov/abs/hep-th/0312085}{{\tt
  hep-th/0312085}}].

\bibitem{Witten:1993ed}
E.~Witten, {\it {Quantum background independence in string theory}},
  \href{http://xxx.lanl.gov/abs/hep-th/9306122}{{\tt hep-th/9306122}}.

\bibitem{KashaniPoor:2006nc}
A.-K. Kashani-Poor, {\it {The Wave Function Behavior of the Open Topological
  String Partition Function on the Conifold}},  {\em JHEP} {\bf 04} (2007) 004,
  [\href{http://xxx.lanl.gov/abs/hep-th/0606112}{{\tt hep-th/0606112}}].

\bibitem{Bouchard:2007ys}
V.~Bouchard, A.~Klemm, M.~Marino, and S.~Pasquetti, {\it {Remodeling the
  B-model}},  \href{http://xxx.lanl.gov/abs/0709.1453}{{\tt 0709.1453}}.

\bibitem{donaldson-thomas}
{S. K. Donaldson and R. P. Thomas}, {\it {Gauge theory in higher dimensions}},
  in {\em {The geometric universe; science, geometry and the work of Roger
  Penrose}} ({S. A. Huggett et al.}, ed.), vol.~62, pp.~31--47.
\newblock {Oxford University Press}, 1998.

\bibitem{dt-thomas}
{R. P. Thomas}, {\it {A holomorphic Casson invariant for Calabi-Yau 3-folds,
  and bundles on K3 fibrations}},  {\em {J. Diff. Geom.}} {\bf 54} (2000)
  367--438.

\bibitem{maulik-2003}
{D. Maulik and N. Nekrasov and A. Okounkov and R. Pandharipande}, {\it
  {Gromov-Witten theory and Donaldson-Thomas theory, I}},  {\em Compos. Math.}
  {\bf 142} (2006) 1263--1285,
  [\href{http://xxx.lanl.gov/abs/math/0312059}{{\tt math/0312059}}].

\bibitem{maulik-2004}
{D. Maulik and N. Nekrasov and A. Okounkov and R. Pandharipande}, {\it
  {Gromov-Witten theory and Donaldson-Thomas theory, II}},  {\em Compos. Math.}
  {\bf 142} (2006) 1286--1304,
  [\href{http://xxx.lanl.gov/abs/math/0406092}{{\tt math/0406092}}].

\bibitem{maulik-2008}
{D. Maulik and A. Oblomkov and A. Okounkov and R. Pandharipande}, {\it
  {Gromov-Witten/Donaldson-Thomas correspondence for toric 3-folds}},
  \href{http://xxx.lanl.gov/abs/0809.3976}{{\tt 0809.3976}}.

\bibitem{joyce-2008}
{Joyce, Dominic and Song, Yinan}, {\it {A theory of generalized
  Donaldson-Thomas invariants. I. An invariant counting stable pairs}},
  \href{http://xxx.lanl.gov/abs/0810.5645}{{\tt 0810.5645}}.

\bibitem{katz-2006}
{S. Katz}, {\it {Genus zero Gopakumar-Vafa invariants of contractible curves}},
   {\em {J. Diff. Geom.}} {\bf 79(2)} (2008) 185--195,
  [\href{http://xxx.lanl.gov/abs/0601193}{{\tt 0601193}}].

\bibitem{pandharipande-2007}
{R. Pandharipande and R. P. Thomas}, {\it {Stable pairs and BPS invariants}},
  \href{http://xxx.lanl.gov/abs/07011.3899}{{\tt 07011.3899}}.

\bibitem{Hollowood:2003cv}
T.~J. Hollowood, A.~Iqbal, and C.~Vafa, {\it {Matrix Models, Geometric
  Engineering and Elliptic Genera}},  {\em JHEP} {\bf 03} (2008) 069,
  [\href{http://xxx.lanl.gov/abs/hep-th/0310272}{{\tt hep-th/0310272}}].

\bibitem{katz-2004}
S.~Katz, {\it {Gromov-Witten, Gopakumar-Vafa, and Donaldson-Thomas invariants
  of Calabi-Yau threefolds}},  \href{http://xxx.lanl.gov/abs/math/0408266}{{\tt
  math/0408266}}.

\bibitem{Cirafici:2008sn}
M.~Cirafici, A.~Sinkovics, and R.~J. Szabo, {\it {Cohomological gauge theory,
  quiver matrix models and Donaldson-Thomas theory}},  {\em Nucl. Phys.} {\bf
  B809} (2009) 452--518, [\href{http://xxx.lanl.gov/abs/0803.4188}{{\tt
  0803.4188}}].

\bibitem{Henningson:1997hy}
M.~Henningson and P.~Yi, {\it {Four-dimensional BPS-spectra via M-theory}},
  {\em Phys. Rev.} {\bf D57} (1998) 1291--1298,
  [\href{http://xxx.lanl.gov/abs/hep-th/9707251}{{\tt hep-th/9707251}}].

\bibitem{Lambert:1998wc}
N.~D. Lambert and P.~C. West, {\it {Monopole dynamics from the M-fivebrane}},
  {\em Nucl. Phys.} {\bf B556} (1999) 177--196,
  [\href{http://xxx.lanl.gov/abs/hep-th/9811025}{{\tt hep-th/9811025}}].

\bibitem{Duff:1993ye}
M.~J. Duff and J.~X. Lu, {\it {Black and super p-branes in diverse
  dimensions}},  {\em Nucl. Phys.} {\bf B416} (1994) 301--334,
  [\href{http://xxx.lanl.gov/abs/hep-th/9306052}{{\tt hep-th/9306052}}].

\bibitem{Dijkgraaf:2002ac}
R.~Dijkgraaf, E.~P. Verlinde, and M.~Vonk, {\it {On the partition sum of the NS
  five-brane}},  \href{http://xxx.lanl.gov/abs/hep-th/0205281}{{\tt
  hep-th/0205281}}.

\bibitem{Nekrasov:2003rj}
N.~Nekrasov and A.~Okounkov, {\it {Seiberg-Witten theory and random
  partitions}},  \href{http://xxx.lanl.gov/abs/hep-th/0306238}{{\tt
  hep-th/0306238}}.

\bibitem{Eynard:2008he}
B.~Eynard and M.~Marino, {\it {A holomorphic and background independent
  partition function for matrix models and topological strings}},
  \href{http://xxx.lanl.gov/abs/0810.4273}{{\tt 0810.4273}}.

\bibitem{sato1981}
{M. Sato}, {\it {Soliton equations as dynamical systems on infinite dimensional
  Grassmann manifolds}},  {\em {RIMS Kokyuroku}} {\bf 439} (1981) 30--46.

\bibitem{Date:1982tj}
E.~Date, M.~Jimbo, M.~Kashiwara, and T.~Miwa, {\it {Transformation groups for
  soliton equations}},  in {\em RIMS Symp. Nonlinear integrable systems -
  classical theory and quantum theory}, 1983.
\newblock RIMS-394.

\bibitem{segalwilson}
{G.~Segal and G.~Wilson}, {\it {Loop groups and equations of {KdV} type}},
  {\em {Inst. Hautes \'Etudes Sci. Publ. Math.}} {\bf 61} (1985) 5--65.

\bibitem{Ishibashi:1986bd}
N.~Ishibashi, Y.~Matsuo, and H.~Ooguri, {\it {Soliton equations and free
  fermions on Riemann surfaces}},  {\em Mod. Phys. Lett.} {\bf A2} (1987) 119.

\bibitem{Vafa:1987es}
C.~Vafa, {\it {Operator Formulation on Riemann Surfaces}},  {\em Phys. Lett.}
  {\bf B190} (1987) 47.

\bibitem{AlvarezGaume:1988bg}
L.~Alvarez-Gaume, C.~Gomez, G.~W. Moore, and C.~Vafa, {\it {Strings in the
  Operator Formalism}},  {\em Nucl. Phys.} {\bf B303} (1988) 455.

\bibitem{Dijkgraaf:2002yn}
R.~Dijkgraaf, A.~Sinkovics, and M.~Temurhan, {\it {Matrix models and
  gravitational corrections}},  {\em Adv. Theor. Math. Phys.} {\bf 7} (2004)
  1155--1176, [\href{http://xxx.lanl.gov/abs/hep-th/0211241}{{\tt
  hep-th/0211241}}].

\bibitem{Kapustin:2006pk}
A.~Kapustin and E.~Witten, {\it {Electric-magnetic duality and the geometric
  Langlands program}},  \href{http://xxx.lanl.gov/abs/hep-th/0604151}{{\tt
  hep-th/0604151}}.

\bibitem{coutinho}
{S.C. Coutinho}, {\em {A primer of algebraic D-modules}}.
\newblock {Cambridge University Press}, 1995.

\bibitem{bjork}
{J. Bjork}, {\em {Rings of differential operators}}.
\newblock {North-Holland Publishing Company}, 1979.

\bibitem{kashiwara}
{M. Kashiwara}, {\em {D-modules and micolocal calculus}}.
\newblock No.~217 in {Transl. Math. Monographs}. {AMS}, 2000.

\bibitem{bernstein}
{J. Bernstein}, {\it {Algebraic theory of $\mathcal{D}$-modules}},
  \href{http://xxx.lanl.gov/abs/{http://www.math.uchicago.edu/$\sim$arinkin/la%
nglands/Bernstein/}}{{\tt
  {http://www.math.uchicago.edu/$\sim$arinkin/langlands/Bernstein/}}}.

\bibitem{Arinkin:2004du}
D.~Arinkin, {\it {Moduli of connections with a small parameter on a curve}},
  \href{http://xxx.lanl.gov/abs/math/0409373}{{\tt math/0409373}}.

\bibitem{beilinsondrinfeld}
{A.~Beilinson and V.~Drinfeld}, {\it {Quantization of Hitchin's integrable
  system and Hecke eigensheaves}},
  \href{http://xxx.lanl.gov/abs/{http://www.math.uchicago.edu/$\sim$mitya/lang%
lands.html}}{{\tt {http://www.math.uchicago.edu/$\sim$mitya/langlands.html}}}.

\bibitem{talkarinkin}
{D.~Arinkin}, {\it {On quasiclassical limit of Langlands correspondence}},
  {\em {talk at the KITP workshop ``Gauge theory and Langlands duality''}}
  (2008).

\bibitem{sato1989}
{M.~Sato}, {\it {The {KP} hierarchy and infinite-dimensional {G}rassmann
  manifolds}},  {\em {Theta functions -- Bowdoin 1987 Part 1, Proc. Sympos.
  Pure Math.}} {\bf 49} (1989).

\bibitem{Dijkgraaf:1991qh}
R.~Dijkgraaf, {\it {Intersection theory, integrable hierarchies and topological
  field theory}},  \href{http://xxx.lanl.gov/abs/hep-th/9201003}{{\tt
  hep-th/9201003}}.

\bibitem{mulaseKP}
{M.~Mulase}, {\it {Algebraic theory of KP equations}},  {\em {Perspectives in
  Mathematical Physics}} (1994) 157--223.

\bibitem{Kac:1993ji}
V.~G. Kac and J.~W. van~de Leur, {\it {The n component of KP hierarchy and
  representation theory}},  {\em J. Math. Phys.} {\bf 44} (2003) 3245--3293,
  [\href{http://xxx.lanl.gov/abs/hep-th/9308137}{{\tt hep-th/9308137}}].

\bibitem{Adams:1993}
{M. R.~Adams and M. J. Bergvelt}, {\it {The Krichever map, vector bundles over
  algebraic curves, and Heisenberg algebras}},  {\em {Commun. Math. Phys.}}
  {\bf 154} (1993).

\bibitem{li-mulase}
{Y.~Li and M.~Mulase}, {\it {Prym varieties and integrable systems}},  {\em
  {Commun. Anal. Geom.}} {\bf 5} (1997).

\bibitem{hodge-mulase}
{A. Hodge and M. Mulase}, {\it {Hitchin integrable systems, deformations of
  spectral curves, and KP-type equations}},
  \href{http://xxx.lanl.gov/abs/math.AG/0801.0015}{{\tt math.AG/0801.0015}}.

\bibitem{Maldacena:2004sn}
J.~M. Maldacena, G.~W. Moore, N.~Seiberg, and D.~Shih, {\it {Exact vs.
  semiclassical target space of the minimal string}},  {\em JHEP} {\bf 10}
  (2004) 020, [\href{http://xxx.lanl.gov/abs/hep-th/0408039}{{\tt
  hep-th/0408039}}].

\bibitem{Marino:2004eq}
M.~Marino, {\it {Les Houches lectures on matrix models and topological
  strings}},  \href{http://xxx.lanl.gov/abs/hep-th/0410165}{{\tt
  hep-th/0410165}}.

\bibitem{Moore:1990mg}
G.~W. Moore, {\it {Geometry of the string equations}},  {\em Commun. Math.
  Phys.} {\bf 133} (1990) 261--304.

\bibitem{Moore:1990cn}
G.~W. Moore, {\it {Matrix models of 2-D gravity and isomonodromic
  deformation}},  {\em Prog. Theor. Phys. Suppl.} {\bf 102} (1990) 255--286.

\bibitem{Seiberg:2003nm}
N.~Seiberg and D.~Shih, {\it {Branes, rings and matrix models in minimal
  (super)string theory}},  {\em JHEP} {\bf 02} (2004) 021,
  [\href{http://xxx.lanl.gov/abs/hep-th/0312170}{{\tt hep-th/0312170}}].

\bibitem{Kazakov:2004du}
V.~A. Kazakov and I.~K. Kostov, {\it {Instantons in non-critical strings from
  the two-matrix model}},  \href{http://xxx.lanl.gov/abs/hep-th/0403152}{{\tt
  hep-th/0403152}}.

\bibitem{Kutasov:2004fg}
D.~Kutasov, K.~Okuyama, J.-w. Park, N.~Seiberg, and D.~Shih, {\it {Annulus
  amplitudes and ZZ branes in minimal string theory}},  {\em JHEP} {\bf 08}
  (2004) 026, [\href{http://xxx.lanl.gov/abs/hep-th/0406030}{{\tt
  hep-th/0406030}}].

\bibitem{Fukuma:1990yk}
M.~Fukuma, H.~Kawai, and R.~Nakayama, {\it {Infinite dimensional Grassmannian
  structure of two- dimensional quantum gravity}},  {\em Commun. Math. Phys.}
  {\bf 143} (1992) 371--404.

\bibitem{Fukuma:2006qq}
M.~Fukuma, H.~Irie, and Y.~Matsuo, {\it {Notes on the algebraic curves in (p,q)
  minimal string theory}},  {\em JHEP} {\bf 09} (2006) 075,
  [\href{http://xxx.lanl.gov/abs/hep-th/0602274}{{\tt hep-th/0602274}}].

\bibitem{Douglas:1989dd}
M.~R. Douglas, {\it Strings in less than one dimension and the generalized kdv
  hierarchies},  {\em Phys. Lett.} {\bf B238} (1990) 176.

\bibitem{Douglas:1989ve}
M.~R. Douglas and S.~H. Shenker, {\it {Strings in Less Than One-Dimension}},
  {\em Nucl. Phys.} {\bf B335} (1990) 635.

\bibitem{Gross:1989vs}
D.~J. Gross and A.~A. Migdal, {\it {Nonperturbative Two-Dimensional Quantum
  Gravity}},  {\em Phys. Rev. Lett.} {\bf 64} (1990) 127.

\bibitem{Brezin:1990rb}
E.~Brezin and V.~A. Kazakov, {\it {Exactly solvable field theories of closed
  strings}},  {\em Phys. Lett.} {\bf B236} (1990) 144--150.

\bibitem{Dijkgraaf:1990rs}
R.~Dijkgraaf, H.~L. Verlinde, and E.~P. Verlinde, {\it {Loop equations and
  Virasoro constraints in nonperturbative 2-D quantum gravity}},  {\em Nucl.
  Phys.} {\bf B348} (1991) 435--456.

\bibitem{Daul:1993bg}
J.~M. Daul, V.~A. Kazakov, and I.~K. Kostov, {\it {Rational theories of 2-D
  gravity from the two matrix model}},  {\em Nucl. Phys.} {\bf B409} (1993)
  311--338, [\href{http://xxx.lanl.gov/abs/hep-th/9303093}{{\tt
  hep-th/9303093}}].

\bibitem{Ginsparg:1993is}
P.~H. Ginsparg and G.~W. Moore, {\it {Lectures on 2-D gravity and 2-D string
  theory}},  \href{http://xxx.lanl.gov/abs/hep-th/9304011}{{\tt
  hep-th/9304011}}.

\bibitem{DiFrancesco:1993nw}
P.~Di~Francesco, P.~H. Ginsparg, and J.~Zinn-Justin, {\it {2-D Gravity and
  random matrices}},  {\em Phys. Rept.} {\bf 254} (1995) 1--133,
  [\href{http://xxx.lanl.gov/abs/hep-th/9306153}{{\tt hep-th/9306153}}].

\bibitem{Witten:1989ig}
E.~Witten, {\it {On the structure of the topological phase of two-dimensional
  gravity}},  {\em Nucl. Phys.} {\bf B340} (1990) 281--332.

\bibitem{Witten:1990hr}
E.~Witten, {\it {Two-dimensional gravity and intersection theory on moduli
  space}},  {\em Surveys Diff. Geom.} {\bf 1} (1991) 243--310.

\bibitem{Kontsevich:1992ti}
M.~Kontsevich, {\it {Intersection theory on the moduli space of curves and the
  matrix Airy function}},  {\em Commun. Math. Phys.} {\bf 147} (1992) 1--23.

\bibitem{eynard-isomonodromic}
{M. Bertola and B. Eynard and J. Harnad}, {\it {Partition functions for matrix
  models and isomonodromic tau functions}},  {\em {J. Phys. A}} {\bf {36}}
  ({2003}) {3067}, [\href{http://xxx.lanl.gov/abs/nlin/0204054}{{\tt
  nlin/0204054}}].

\bibitem{Bertola:2001hq}
M.~Bertola, B.~Eynard, and J.~P. Harnad, {\it {Duality, Biorthogonal
  Polynomials and Multi-Matrix Models}},  {\em Commun. Math. Phys.} {\bf 229}
  (2002) 73--120, [\href{http://xxx.lanl.gov/abs/nlin/0108049}{{\tt
  nlin/0108049}}].

\bibitem{2-matrixdmodule}
M.~Bertola, B.~Eynard, and J.~Harnad, {\it {Differential systems for
  biorthogonal polynomials appearing in 2-matrix models and the associated
  Riemann-Hilbert problem}},  {\em Comm. Math. Phys.} {\bf 243} (2003)
  193--240, [\href{http://xxx.lanl.gov/abs/nlin/0208002}{{\tt nlin/0208002}}].

\bibitem{Gross:1990ub}
D.~J. Gross and I.~R. Klebanov, {\it {One-dimensional string theory on a
  circle}},  {\em Nucl. Phys.} {\bf B344} (1990) 475--498.

\bibitem{Klebanov:1991qa}
I.~R. Klebanov, {\it {String theory in two-dimensions}},
  \href{http://xxx.lanl.gov/abs/hep-th/9108019}{{\tt hep-th/9108019}}.

\bibitem{Polchinski:1994mb}
J.~Polchinski, {\it {What is string theory?}},
  \href{http://xxx.lanl.gov/abs/hep-th/9411028}{{\tt hep-th/9411028}}.

\bibitem{Alexandrov:2003ut}
S.~Alexandrov, {\it {Matrix quantum mechanics and two-dimensional string theory
  in non-trivial backgrounds}},
  \href{http://xxx.lanl.gov/abs/hep-th/0311273}{{\tt hep-th/0311273}}.

\bibitem{Dijkgraaf:1992hk}
R.~Dijkgraaf, G.~W. Moore, and R.~Plesser, {\it {The Partition function of 2-D
  string theory}},  {\em Nucl. Phys.} {\bf B394} (1993) 356--382,
  [\href{http://xxx.lanl.gov/abs/hep-th/9208031}{{\tt hep-th/9208031}}].

\bibitem{Moore:1991zv}
G.~W. Moore, M.~R. Plesser, and S.~Ramgoolam, {\it {Exact S matrix for 2-D
  string theory}},  {\em Nucl. Phys.} {\bf B377} (1992) 143--190,
  [\href{http://xxx.lanl.gov/abs/hep-th/9111035}{{\tt hep-th/9111035}}].

\bibitem{Alexandrov:2002fh}
S.~Y. Alexandrov, V.~A. Kazakov, and I.~K. Kostov, {\it {Time-dependent
  backgrounds of 2D string theory}},  {\em Nucl. Phys.} {\bf B640} (2002)
  119--144, [\href{http://xxx.lanl.gov/abs/hep-th/0205079}{{\tt
  hep-th/0205079}}].

\bibitem{Bonora:1994ka}
L.~Bonora and C.~S. Xiong, {\it {Extended Toda lattice hierarchy, extended two
  matrix model and c = 1 string theory}},  {\em Nucl. Phys.} {\bf B434} (1995)
  408--444, [\href{http://xxx.lanl.gov/abs/hep-th/9407141}{{\tt
  hep-th/9407141}}].

\bibitem{Distler:1990mt}
J.~Distler and C.~Vafa, {\it {A Critical matrix model at c = 1}},  {\em Mod.
  Phys. Lett.} {\bf A6} (1991) 259--270.

\bibitem{Imbimbo:1995yv}
C.~Imbimbo and S.~Mukhi, {\it {The Topological matrix model of c = 1 string}},
  {\em Nucl. Phys.} {\bf B449} (1995) 553--568,
  [\href{http://xxx.lanl.gov/abs/hep-th/9505127}{{\tt hep-th/9505127}}].

\bibitem{Alexandrov:2003qk}
S.~Y. Alexandrov, V.~A. Kazakov, and I.~K. Kostov, {\it {2D string theory as
  normal matrix model}},  {\em Nucl. Phys.} {\bf B667} (2003) 90--110,
  [\href{http://xxx.lanl.gov/abs/hep-th/0302106}{{\tt hep-th/0302106}}].

\bibitem{Mukherjee:2005aq}
A.~Mukherjee and S.~Mukhi, {\it {c = 1 matrix models: Equivalences and
  open-closed string duality}},  {\em JHEP} {\bf 10} (2005) 099,
  [\href{http://xxx.lanl.gov/abs/hep-th/0505180}{{\tt hep-th/0505180}}].

\bibitem{Ghoshal:1995wm}
D.~Ghoshal and C.~Vafa, {\it {C = 1 string as the topological theory of the
  conifold}},  {\em Nucl. Phys.} {\bf B453} (1995) 121--128,
  [\href{http://xxx.lanl.gov/abs/hep-th/9506122}{{\tt hep-th/9506122}}].

\bibitem{Nekrasov:2002qd}
N.~A. Nekrasov, {\it {Seiberg-Witten Prepotential From Instanton Counting}},
  {\em Adv. Theor. Math. Phys.} {\bf 7} (2004) 831--864,
  [\href{http://xxx.lanl.gov/abs/hep-th/0206161}{{\tt hep-th/0206161}}].

\bibitem{Nakajima:2003pg}
H.~Nakajima and K.~Yoshioka, {\it {Instanton counting on blowup. I}},
  \href{http://xxx.lanl.gov/abs/math/0306198}{{\tt math/0306198}}.

\bibitem{Nakajima:2005fg}
H.~Nakajima and K.~Yoshioka, {\it {Instanton counting on blowup. II:
  K-theoretic partition function}},
  \href{http://xxx.lanl.gov/abs/math/0505553}{{\tt math/0505553}}.

\bibitem{Nekrasov:2003af}
N.~A. Nekrasov, {\it {Seiberg-Witten prepotential from instanton counting}},
  \href{http://xxx.lanl.gov/abs/hep-th/0306211}{{\tt hep-th/0306211}}.

\bibitem{Meijer:1946}
C.~Meijer, {\it On the g-function},  {\em Nederl. Akad. Wetensch. Proc. Ser. A}
  {\bf 49} (1946) 344--356.

\bibitem{Luke:1959}
{Y. Luke}, {\em {The special functions and their approximations}}, vol.~{I}.
\newblock Academic Press, New York, 1969.

\bibitem{G-Fields}
J.~Fields, {\it The asymptotic expansion of the meijer g-function},  {\em
  Mathematics of Computation} {\bf 26} (1972) 757--765.

\bibitem{Iqbal:2003ix}
A.~Iqbal and A.-K. Kashani-Poor, {\it {Instanton counting and Chern-Simons
  theory}},  {\em Adv. Theor. Math. Phys.} {\bf 7} (2004) 457--497,
  [\href{http://xxx.lanl.gov/abs/hep-th/0212279}{{\tt hep-th/0212279}}].

\bibitem{Iqbal:2003zz}
A.~Iqbal and A.-K. Kashani-Poor, {\it {SU(N) geometries and topological string
  amplitudes}},  {\em Adv. Theor. Math. Phys.} {\bf 10} (2006) 1--32,
  [\href{http://xxx.lanl.gov/abs/hep-th/0306032}{{\tt hep-th/0306032}}].

\bibitem{eynard-2009-0903}
{B. Eynard}, {\it {Large N expansion of convergent matrix integrals,
  holomorphic anomalies, and background independence}},  {\em JHEP} {\bf 0903}
  (2009) 003, [\href{http://xxx.lanl.gov/abs/0802.1788}{{\tt 0802.1788}}].

\bibitem{Marino:2007te}
M.~Marino, R.~Schiappa, and M.~Weiss, {\it {Nonperturbative Effects and the
  Large-Order Behavior of Matrix Models and Topological Strings}},
  \href{http://xxx.lanl.gov/abs/0711.1954}{{\tt 0711.1954}}.

\bibitem{Marino:2008ya}
M.~Marino, {\it {Nonperturbative effects and nonperturbative definitions in
  matrix models and topological strings}},
  \href{http://xxx.lanl.gov/abs/0805.3033}{{\tt 0805.3033}}.

\bibitem{Aganagic:2006wq}
M.~Aganagic, V.~Bouchard, and A.~Klemm, {\it {Topological Strings and (Almost)
  Modular Forms}},  {\em Commun. Math. Phys.} {\bf 277} (2008) 771--819,
  [\href{http://xxx.lanl.gov/abs/hep-th/0607100}{{\tt hep-th/0607100}}].

\bibitem{Marino:2006hs}
M.~Marino, {\it {Open string amplitudes and large order behavior in topological
  string theory}},  {\em JHEP} {\bf 03} (2008) 060,
  [\href{http://xxx.lanl.gov/abs/hep-th/0612127}{{\tt hep-th/0612127}}].

\bibitem{Eynard:2007hf}
B.~Eynard, M.~Marino, and N.~Orantin, {\it {Holomorphic anomaly and matrix
  models}},  {\em JHEP} {\bf 06} (2007) 058,
  [\href{http://xxx.lanl.gov/abs/hep-th/0702110}{{\tt hep-th/0702110}}].

\bibitem{eynard-orantin}
{B.~Eynard and N.~Orantin}, {\it {Invariants of algebraic curves and
  topological expansion}},  \href{http://xxx.lanl.gov/abs/math-ph/0702045}{{\tt
  math-ph/0702045}}.

\bibitem{Dijkgraaf:2007sx}
R.~Dijkgraaf and C.~Vafa, {\it {Two Dimensional Kodaira-Spencer Theory and
  Three Dimensional Chern-Simons Gravity}},
  \href{http://xxx.lanl.gov/abs/0711.1932}{{\tt 0711.1932}}.

\bibitem{eynard-latest}
{B. Eynard and O. Marchal}, {\it {Topological expansion of the Bethe ansatz,
  and non-commutative algebraic geometry}},
  \href{http://xxx.lanl.gov/abs/0809.3367}{{\tt 0809.3367}}.

\bibitem{Denef:2007vg}
F.~Denef and G.~W. Moore, {\it {Split states, entropy enigmas, holes and
  halos}},  \href{http://xxx.lanl.gov/abs/hep-th/0702146}{{\tt
  hep-th/0702146}}.

\bibitem{Gaiotto:2008cd}
D.~Gaiotto, G.~W. Moore, and A.~Neitzke, {\it {Four-dimensional wall-crossing
  via three-dimensional field theory}},
  \href{http://xxx.lanl.gov/abs/0807.4723}{{\tt 0807.4723}}.

\bibitem{Frenkel:2005pa}
E.~Frenkel, {\it {Lectures on the Langlands program and conformal field
  theory}},  \href{http://xxx.lanl.gov/abs/hep-th/0512172}{{\tt
  hep-th/0512172}}.

\bibitem{Witten:2007td}
E.~Witten, {\it {Gauge Theory And Wild Ramification}},
  \href{http://xxx.lanl.gov/abs/0710.0631}{{\tt 0710.0631}}.

\bibitem{Gukov:2008ve}
S.~Gukov and E.~Witten, {\it {Branes and Quantization}},
  \href{http://xxx.lanl.gov/abs/0809.0305}{{\tt 0809.0305}}.

\bibitem{Chervov:2006xk}
A.~Chervov and D.~Talalaev, {\it {Quantum spectral curves, quantum integrable
  systems and the geometric Langlands correspondence}},
  \href{http://xxx.lanl.gov/abs/hep-th/0604128}{{\tt hep-th/0604128}}.

\bibitem{Bryan:2004iq}
J.~Bryan and R.~Pandharipande, {\it {The local Gromov-Witten theory of
  curves}},  \href{http://xxx.lanl.gov/abs/math/0411037}{{\tt math/0411037}}.

\bibitem{dijkgraaf-mirror}
{R. Dijkgraaf}, {\it {Mirror symmetry and elliptic curves}},  in {\em {The
  moduli space of curves}}, vol.~129 of {\em {Progress in Mathematics}},
  {Birkhauser}, 1995.

\bibitem{Douglas:1993wy}
M.~R. Douglas, {\it {Conformal field theory techniques in large N Yang-Mills
  theory}},  \href{http://xxx.lanl.gov/abs/hep-th/9311130}{{\tt
  hep-th/9311130}}.

\bibitem{Dijkgraaf:1996iy}
R.~Dijkgraaf, {\it {Chiral deformations of conformal field theories}},  {\em
  Nucl. Phys.} {\bf B493} (1997) 588--612,
  [\href{http://xxx.lanl.gov/abs/hep-th/9609022}{{\tt hep-th/9609022}}].

\bibitem{Gross:1990st}
D.~J. Gross and I.~R. Klebanov, {\it {Fermionic string field theory of c = 1
  two-dimensional quantum gravity}},  {\em Nucl. Phys.} {\bf B352} (1991)
  671--688.

\bibitem{kanekozagier}
{M. Kaneko and D. Zagier}, {\it {A generalized Jacobi theta function and
  quasimodular forms}},  in {\em {The moduli space of curves}}, vol.~129 of
  {\em {Progress in Mathematics}}, {Birkhauser}, 1995.

\bibitem{Li:2004ef}
J.~Li, K.~Liu, and J.~Zhou, {\it {Topological string partition functions as
  equivariant indices}},  \href{http://xxx.lanl.gov/abs/math/0412089}{{\tt
  math/0412089}}.

\bibitem{igusa1}
{J.-L. Igusa}, {\it {On Siegel modular forms of genus two}},  {\em Amer. J.
  Math} {\bf 84} (1962) 175--200.

\bibitem{igusa2}
{J.-L. Igusa}, {\it {On Siegel modular forms of genus two (2)}},  {\em Amer. J.
  Math} {\bf 86} (1964) 164--412.

\bibitem{Moore:1986rh}
G.~W. Moore, {\it {Modular forms and two loop string physics}},  {\em Phys.
  Lett.} {\bf B176} (1986) 369.

\bibitem{Belavin:1986tv}
A.~A. Belavin, V.~Knizhnik, A.~Morozov, and A.~Perelomov, {\it {Two and three
  loop amplitudes in the bosonic string theory}},  {\em JETP Lett.} {\bf 43}
  (1986) 411.

\bibitem{Dabholkar:2006bj}
A.~Dabholkar and D.~Gaiotto, {\it {Spectrum of CHL dyons from genus-two
  partition function}},  {\em JHEP} {\bf 12} (2007) 087,
  [\href{http://xxx.lanl.gov/abs/hep-th/0612011}{{\tt hep-th/0612011}}].

\bibitem{Borcherds:1995}
{R. Borcherds}, {\it {Automorphic forms on $O_{s+2,2}(R)$ and infinite
  products}},  {\em {Invent. Math.}} {\bf {120}} ({1995}) {161--213}.

\bibitem{Dijkgraaf:1998xr}
R.~Dijkgraaf, {\it {The mathematics of five-branes}},
  \href{http://xxx.lanl.gov/abs/hep-th/9810157}{{\tt hep-th/9810157}}.

\bibitem{Dijkgraaf:1996it}
R.~Dijkgraaf, E.~P. Verlinde, and H.~L. Verlinde, {\it {Counting Dyons in N=4
  String Theory}},  {\em Nucl. Phys.} {\bf B484} (1997) 543--561,
  [\href{http://xxx.lanl.gov/abs/hep-th/9607026}{{\tt hep-th/9607026}}].

\bibitem{Jatkar:2005bh}
D.~P. Jatkar and A.~Sen, {\it {Dyon spectrum in CHL models}},  {\em JHEP} {\bf
  04} (2006) 018, [\href{http://xxx.lanl.gov/abs/hep-th/0510147}{{\tt
  hep-th/0510147}}].

\bibitem{LopesCardoso:2004xf}
G.~Lopes~Cardoso, B.~de~Wit, J.~Kappeli, and T.~Mohaupt, {\it {Asymptotic
  degeneracy of dyonic N = 4 string states and black hole entropy}},  {\em
  JHEP} {\bf 12} (2004) 075,
  [\href{http://xxx.lanl.gov/abs/hep-th/0412287}{{\tt hep-th/0412287}}].

\bibitem{David:2006yn}
J.~R. David and A.~Sen, {\it {CHL dyons and statistical entropy function from
  D1-D5 system}},  {\em JHEP} {\bf 11} (2006) 072,
  [\href{http://xxx.lanl.gov/abs/hep-th/0605210}{{\tt hep-th/0605210}}].

\bibitem{Banerjee:2008pu}
S.~Banerjee, A.~Sen, and Y.~K. Srivastava, {\it {Partition Functions of Torsion
  $>1$ Dyons in Heterotic String Theory on $T^6$}},  {\em JHEP} {\bf 05} (2008)
  098, [\href{http://xxx.lanl.gov/abs/0802.1556}{{\tt 0802.1556}}].

\bibitem{Gaiotto:2005hc}
D.~Gaiotto, {\it {Re-recounting dyons in N = 4 string theory}},
  \href{http://xxx.lanl.gov/abs/hep-th/0506249}{{\tt hep-th/0506249}}.

\bibitem{David:2006ji}
J.~R. David, D.~P. Jatkar, and A.~Sen, {\it {Product representation of dyon
  partition function in CHL models}},  {\em JHEP} {\bf 06} (2006) 064,
  [\href{http://xxx.lanl.gov/abs/hep-th/0602254}{{\tt hep-th/0602254}}].

\bibitem{David:2006ru}
J.~R. David, D.~P. Jatkar, and A.~Sen, {\it {Dyon spectrum in N = 4
  supersymmetric type II string theories}},  {\em JHEP} {\bf 11} (2006) 073,
  [\href{http://xxx.lanl.gov/abs/hep-th/0607155}{{\tt hep-th/0607155}}].

\bibitem{David:2006ud}
J.~R. David, D.~P. Jatkar, and A.~Sen, {\it {Dyon spectrum in generic N = 4
  supersymmetric Z(N) orbifolds}},  {\em JHEP} {\bf 01} (2007) 016,
  [\href{http://xxx.lanl.gov/abs/hep-th/0609109}{{\tt hep-th/0609109}}].

\bibitem{Sen:2007vb}
A.~Sen, {\it {Walls of Marginal Stability and Dyon Spectrum in N=4
  Supersymmetric String Theories}},  {\em JHEP} {\bf 05} (2007) 039,
  [\href{http://xxx.lanl.gov/abs/hep-th/0702141}{{\tt hep-th/0702141}}].

\bibitem{Dabholkar:2007vk}
A.~Dabholkar, D.~Gaiotto, and S.~Nampuri, {\it {Comments on the spectrum of CHL
  dyons}},  {\em JHEP} {\bf 01} (2008) 023,
  [\href{http://xxx.lanl.gov/abs/hep-th/0702150}{{\tt hep-th/0702150}}].

\bibitem{Cheng:2007ch}
M.~C.~N. Cheng and E.~Verlinde, {\it {Dying Dyons Don't Count}},  {\em JHEP}
  {\bf 09} (2007) 070, [\href{http://xxx.lanl.gov/abs/0706.2363}{{\tt
  0706.2363}}].

\bibitem{Mukherjee:2007nc}
A.~Mukherjee, S.~Mukhi, and R.~Nigam, {\it {Dyon Death Eaters}},  {\em JHEP}
  {\bf 10} (2007) 037, [\href{http://xxx.lanl.gov/abs/0707.3035}{{\tt
  0707.3035}}].

\bibitem{Dabholkar:2008zy}
A.~Dabholkar, J.~Gomes, and S.~Murthy, {\it {Counting all dyons in N =4 string
  theory}},  \href{http://xxx.lanl.gov/abs/0803.2692}{{\tt 0803.2692}}.

\bibitem{Cheng:2008fc}
M.~C.~N. Cheng and E.~P. Verlinde, {\it {Wall Crossing, Discrete Attractor
  Flow, and Borcherds Algebra}},  {\em SIGMA} {\bf 4} (2008) 068,
  [\href{http://xxx.lanl.gov/abs/0806.2337}{{\tt 0806.2337}}].

\bibitem{Banerjee:2008yu}
S.~Banerjee, A.~Sen, and Y.~K. Srivastava, {\it {Genus Two Surface and Quarter
  BPS Dyons: The Contour Prescription}},
  \href{http://xxx.lanl.gov/abs/0808.1746}{{\tt 0808.1746}}.

\bibitem{Cheng:2008kt}
M.~C.~N. Cheng and A.~Dabholkar, {\it {Borcherds-Kac-Moody Symmetry of N=4
  Dyons}},  \href{http://xxx.lanl.gov/abs/0809.4258}{{\tt 0809.4258}}.

\bibitem{Banerjee:2008ky}
N.~Banerjee, D.~P. Jatkar, and A.~Sen, {\it {Asymptotic Expansion of the N=4
  Dyon Degeneracy}},  \href{http://xxx.lanl.gov/abs/0810.3472}{{\tt
  0810.3472}}.

\bibitem{Murthy:2009dq}
S.~Murthy and B.~Pioline, {\it {A Farey tale for N=4 dyons}},
  \href{http://xxx.lanl.gov/abs/0904.4253}{{\tt 0904.4253}}.

\bibitem{Sen:1997xi}
A.~Sen, {\it {String network}},  {\em JHEP} {\bf 03} (1998) 005,
  [\href{http://xxx.lanl.gov/abs/hep-th/9711130}{{\tt hep-th/9711130}}].

\bibitem{Cvetic:1995bj}
M.~Cvetic and A.~A. Tseytlin, {\it {Solitonic strings and BPS saturated dyonic
  black holes}},  {\em Phys. Rev.} {\bf D53} (1996) 5619--5633,
  [\href{http://xxx.lanl.gov/abs/hep-th/9512031}{{\tt hep-th/9512031}}].

\bibitem{Cvetic:1995uj}
M.~Cvetic and D.~Youm, {\it {Dyonic BPS saturated black holes of heterotic
  string on a six torus}},  {\em Phys. Rev.} {\bf D53} (1996) 584--588,
  [\href{http://xxx.lanl.gov/abs/hep-th/9507090}{{\tt hep-th/9507090}}].

\bibitem{gritsenko-1995}
V.~A. Gritsenko and V.~V. Nikulin, {\it {Siegel automorphic form corrections of
  some Lorentzian Kac--Moody Lie algebras}},
  \href{http://xxx.lanl.gov/abs/9504006}{{\tt 9504006}}.

\bibitem{borcherds-1998-132}
R.~Borcherds, {\it {Automorphic forms with singularities on Grassmannians}},
  {\em Inventiones mathematicae} {\bf 132} (1998) 491--562,
  [\href{http://xxx.lanl.gov/abs/9609022}{{\tt 9609022}}].

\bibitem{Cheng:2008gx}
M.~C.~N. Cheng, {\it {The Spectra of Supersymmetric States in String Theory}},
  \href{http://xxx.lanl.gov/abs/0807.3099}{{\tt 0807.3099}}.

\bibitem{tata_theta}
{David Mumford}, {\em {Tata Lectures on Theta II -- Jacobian Theta Functions
  and Differential Equations}}.
\newblock {Birkh\"auser}, 1984.

\bibitem{lebo_degeneration}
{Aaron Lebowitz}, {\it {Degeneration of a Compact Riemann Surface of Genus 2}},
   {\em {Israel Journal of Mathematics}} {\bf 12} (1972).

\bibitem{Feingold-Frenekel}
{A.~J.~Feingold and I.~B.~Frenekel}, {\it {A hyperbolic Kac-Moody Algebra and
  the theory of Siegel modular forms of genus 2}},  {\em {Math. Ann.}} {\bf
  263} (1983) 87.

\bibitem{Witten:1995ex}
E.~Witten, {\it {String theory dynamics in various dimensions}},  {\em Nucl.
  Phys.} {\bf B443} (1995) 85--126,
  [\href{http://xxx.lanl.gov/abs/hep-th/9503124}{{\tt hep-th/9503124}}].

\bibitem{Chaudhuri:1995fk}
S.~Chaudhuri, G.~Hockney, and J.~D. Lykken, {\it {Maximally supersymmetric
  string theories in D < 10}},  {\em Phys. Rev. Lett.} {\bf 75} (1995)
  2264--2267, [\href{http://xxx.lanl.gov/abs/hep-th/9505054}{{\tt
  hep-th/9505054}}].

\bibitem{Chaudhuri:1995bf}
S.~Chaudhuri and J.~Polchinski, {\it {Moduli space of CHL strings}},  {\em
  Phys. Rev.} {\bf D52} (1995) 7168--7173,
  [\href{http://xxx.lanl.gov/abs/hep-th/9506048}{{\tt hep-th/9506048}}].

\bibitem{Chaudhuri:1995dj}
S.~Chaudhuri and D.~A. Lowe, {\it {Type IIA heterotic duals with maximal
  supersymmetry}},  {\em Nucl. Phys.} {\bf B459} (1996) 113--124,
  [\href{http://xxx.lanl.gov/abs/hep-th/9508144}{{\tt hep-th/9508144}}].

\bibitem{Dijkgraaf:1987vp}
R.~Dijkgraaf, E.~P. Verlinde, and H.~L. Verlinde, {\it {C = 1 Conformal Field
  Theories on Riemann Surfaces}},  {\em Commun. Math. Phys.} {\bf 115} (1988)
  649--690.

\bibitem{kontsevich-2008}
M.~Kontsevich and Y.~Soibelman, {\it {Stability structures, motivic
  Donaldson-Thomas invariants and cluster transformations}},
  \href{http://xxx.lanl.gov/abs/0811.2435}{{\tt 0811.2435}}.

\bibitem{Jafferis:2008uf}
D.~L. Jafferis and G.~W. Moore, {\it {Wall crossing in local Calabi Yau
  manifolds}},  \href{http://xxx.lanl.gov/abs/0810.4909}{{\tt 0810.4909}}.

\bibitem{Chuang:2008aw}
W.-y. Chuang and D.~L. Jafferis, {\it {Wall Crossing of BPS States on the
  Conifold from Seiberg Duality and Pyramid Partitions}},
  \href{http://xxx.lanl.gov/abs/0810.5072}{{\tt 0810.5072}}.

\bibitem{Mozgovoy:2008fd}
S.~Mozgovoy and M.~Reineke, {\it {On the noncommutative Donaldson-Thomas
  invariants arising from brane tilings}},
  \href{http://xxx.lanl.gov/abs/0809.0117}{{\tt 0809.0117}}.

\bibitem{Dimofte:2009bv}
T.~Dimofte and S.~Gukov, {\it {Refined, Motivic, and Quantum}},
  \href{http://xxx.lanl.gov/abs/0904.1420}{{\tt 0904.1420}}.

\bibitem{bryan-2008}
{J. Bryan and A. Gholampour}, {\it {The Quantum McKay Correspondence for
  polyhedral singularities}},  \href{http://xxx.lanl.gov/abs/0803.3766}{{\tt
  0803.3766}}.

\bibitem{bryan-2009}
{J. Bryan and A. Gholampour}, {\it {BPS invariants for resolutions of
  polyhedral singularities}},  \href{http://xxx.lanl.gov/abs/0905.0537}{{\tt
  0905.0537}}.

\bibitem{boissiere-sarti}
{S. Boissiere and A. Sarti}, {\it {Contraction of excess fibres between the
  McKay correspondences in dimension two and three}},  {\em {Ann. Inst.
  Fourier}} {\bf 57} (2007), no.~6 1839--1861,
  [\href{http://xxx.lanl.gov/abs/0504360}{{\tt 0504360}}].

\bibitem{Intriligator:2006dd}
K.~A. Intriligator, N.~Seiberg, and D.~Shih, {\it {Dynamical SUSY breaking in
  meta-stable vacua}},  {\em JHEP} {\bf 04} (2006) 021,
  [\href{http://xxx.lanl.gov/abs/hep-th/0602239}{{\tt hep-th/0602239}}].

\bibitem{Ooguri:2007iu}
H.~Ooguri, Y.~Ookouchi, and C.-S. Park, {\it {Metastable Vacua in Perturbed
  Seiberg-Witten Theories}},  \href{http://xxx.lanl.gov/abs/0704.3613}{{\tt
  0704.3613}}.

\bibitem{Aganagic:2008qa}
M.~Aganagic, C.~Beem, J.~Seo, and C.~Vafa, {\it {Extended Supersymmetric Moduli
  Space and a SUSY/Non-SUSY Duality}},
  \href{http://xxx.lanl.gov/abs/0804.2489}{{\tt 0804.2489}}.

\bibitem{Intriligator:2007py}
K.~A. Intriligator, N.~Seiberg, and D.~Shih, {\it {Supersymmetry Breaking,
  R-Symmetry Breaking and Metastable Vacua}},  {\em JHEP} {\bf 07} (2007) 017,
  [\href{http://xxx.lanl.gov/abs/hep-th/0703281}{{\tt hep-th/0703281}}].

\bibitem{Pastras:2007qr}
G.~Pastras, {\it {Non supersymmetric metastable vacua in N = 2 SYM softly
  broken to N = 1}},  \href{http://xxx.lanl.gov/abs/0705.0505}{{\tt
  0705.0505}}.

\bibitem{AlvarezGaume:1981hn}
L.~Alvarez-Gaume, D.~Z. Freedman, and S.~Mukhi, {\it {The Background Field
  Method and the Ultraviolet Structure of the Supersymmetric Nonlinear Sigma
  Model}},  {\em Ann. Phys.} {\bf 134} (1981) 85.

\bibitem{Hull:1985pq}
C.~M. Hull, A.~Karlhede, U.~Lindstrom, and M.~Rocek, {\it {Nonlinear sigma
  models and their gauging in and out of superspace}},  {\em Nucl. Phys.} {\bf
  B266} (1986) 1.

\bibitem{Higashijima:2000wz}
K.~Higashijima and M.~Nitta, {\it {Kaehler normal coordinate expansion in
  supersymmetric theories}},  {\em Prog. Theor. Phys.} {\bf 105} (2001)
  243--260, [\href{http://xxx.lanl.gov/abs/hep-th/0006027}{{\tt
  hep-th/0006027}}].

\bibitem{Marsano:2007mt}
J.~Marsano, H.~Ooguri, Y.~Ookouchi, and C.-S. Park, {\it {Metastable Vacua in
  Perturbed Seiberg-Witten Theories, Part 2: Fayet-Iliopoulos Terms and
  K\'ahler Normal Coordinates}},  {\em Nucl. Phys.} {\bf B798} (2008) 17--35,
  [\href{http://xxx.lanl.gov/abs/0712.3305}{{\tt 0712.3305}}].

\bibitem{Denef:2008wq}
F.~Denef, {\it {Les Houches Lectures on Constructing String Vacua}},
  \href{http://xxx.lanl.gov/abs/0803.1194}{{\tt 0803.1194}}.

\bibitem{Douglas:2006es}
M.~R. Douglas and S.~Kachru, {\it {Flux compactification}},  {\em Rev. Mod.
  Phys.} {\bf 79} (2007) 733--796,
  [\href{http://xxx.lanl.gov/abs/hep-th/0610102}{{\tt hep-th/0610102}}].

\bibitem{Michelson:1996pn}
J.~Michelson, {\it {Compactifications of type IIB strings to four dimensions
  with non-trivial classical potential}},  {\em Nucl. Phys.} {\bf B495} (1997)
  127--148, [\href{http://xxx.lanl.gov/abs/hep-th/9610151}{{\tt
  hep-th/9610151}}].

\bibitem{Gukov:1999ya}
S.~Gukov, C.~Vafa, and E.~Witten, {\it {CFT's from Calabi-Yau four-folds}},
  {\em Nucl. Phys.} {\bf B584} (2000) 69--108,
  [\href{http://xxx.lanl.gov/abs/hep-th/9906070}{{\tt hep-th/9906070}}].

\bibitem{Taylor:1999ii}
T.~R. Taylor and C.~Vafa, {\it {RR flux on Calabi-Yau and partial supersymmetry
  breaking}},  {\em Phys. Lett.} {\bf B474} (2000) 130--137,
  [\href{http://xxx.lanl.gov/abs/hep-th/9912152}{{\tt hep-th/9912152}}].

\bibitem{Cachazo:2002ry}
F.~Cachazo, M.~R. Douglas, N.~Seiberg, and E.~Witten, {\it {Chiral Rings and
  Anomalies in Supersymmetric Gauge Theory}},  {\em JHEP} {\bf 12} (2002) 071,
  [\href{http://xxx.lanl.gov/abs/hep-th/0211170}{{\tt hep-th/0211170}}].

\bibitem{Cachazo:2002zk}
F.~Cachazo, N.~Seiberg, and E.~Witten, {\it {Phases of N = 1 supersymmetric
  gauge theories and matrices}},  {\em JHEP} {\bf 02} (2003) 042,
  [\href{http://xxx.lanl.gov/abs/hep-th/0301006}{{\tt hep-th/0301006}}].

\bibitem{Marsano:2008ts}
J.~Marsano, K.~Papadodimas, and M.~Shigemori, {\it {Off-shell M5 Brane,
  Perturbed Seiberg-Witten Theory, and Metastable Vacua}},  {\em Nucl. Phys.}
  {\bf B804} (2008) 19--69, [\href{http://xxx.lanl.gov/abs/0801.2154}{{\tt
  0801.2154}}].

\bibitem{Bena:2006rg}
I.~Bena, E.~Gorbatov, S.~Hellerman, N.~Seiberg, and D.~Shih, {\it {A note on
  (meta)stable brane configurations in MQCD}},  {\em JHEP} {\bf 11} (2006) 088,
  [\href{http://xxx.lanl.gov/abs/hep-th/0608157}{{\tt hep-th/0608157}}].

\bibitem{Cachazo:2001jy}
F.~Cachazo, K.~A. Intriligator, and C.~Vafa, {\it {A large N duality via a
  geometric transition}},  {\em Nucl. Phys.} {\bf B603} (2001) 3--41,
  [\href{http://xxx.lanl.gov/abs/hep-th/0103067}{{\tt hep-th/0103067}}].

\bibitem{Cachazo:2002pr}
F.~Cachazo and C.~Vafa, {\it {N = 1 and N = 2 geometry from fluxes}},
  \href{http://xxx.lanl.gov/abs/hep-th/0206017}{{\tt hep-th/0206017}}.

\bibitem{Vafa:2000wi}
C.~Vafa, {\it {Superstrings and topological strings at large N}},  {\em J.
  Math. Phys.} {\bf 42} (2001) 2798--2817,
  [\href{http://xxx.lanl.gov/abs/hep-th/0008142}{{\tt hep-th/0008142}}].

\bibitem{Cachazo:2001gh}
F.~Cachazo, S.~Katz, and C.~Vafa, {\it {Geometric transitions and N = 1 quiver
  theories}},  \href{http://xxx.lanl.gov/abs/hep-th/0108120}{{\tt
  hep-th/0108120}}.

\bibitem{Cachazo:2001sg}
F.~Cachazo, B.~Fiol, K.~A. Intriligator, S.~Katz, and C.~Vafa, {\it {A
  geometric unification of dualities}},  {\em Nucl. Phys.} {\bf B628} (2002)
  3--78, [\href{http://xxx.lanl.gov/abs/hep-th/0110028}{{\tt hep-th/0110028}}].

\bibitem{deBoer:2004he}
J.~de~Boer and S.~de~Haro, {\it {The off-shell M5-brane and non-perturbative
  gauge theory}},  {\em Nucl. Phys.} {\bf B696} (2004) 174--204,
  [\href{http://xxx.lanl.gov/abs/hep-th/0403035}{{\tt hep-th/0403035}}].

\bibitem{Vafa:1987ea}
C.~Vafa, {\it {Conformal theories and punctured surfaces}},  {\em Phys. Lett.}
  {\bf B199} (1987) 195.

\bibitem{Douglas:2003um}
M.~R. Douglas, {\it {The statistics of string/M theory vacua}},  {\em JHEP}
  {\bf 05} (2003) 046, [\href{http://xxx.lanl.gov/abs/hep-th/0303194}{{\tt
  hep-th/0303194}}].

\bibitem{Ashok:2003gk}
S.~Ashok and M.~R. Douglas, {\it {Counting flux vacua}},  {\em JHEP} {\bf 01}
  (2004) 060, [\href{http://xxx.lanl.gov/abs/hep-th/0307049}{{\tt
  hep-th/0307049}}].

\bibitem{Denef:2004cf}
F.~Denef and M.~R. Douglas, {\it {Distributions of nonsupersymmetric flux
  vacua}},  {\em JHEP} {\bf 03} (2005) 061,
  [\href{http://xxx.lanl.gov/abs/hep-th/0411183}{{\tt hep-th/0411183}}].

\bibitem{Denef:2004ze}
F.~Denef and M.~R. Douglas, {\it {Distributions of flux vacua}},  {\em JHEP}
  {\bf 05} (2004) 072, [\href{http://xxx.lanl.gov/abs/hep-th/0404116}{{\tt
  hep-th/0404116}}].

\bibitem{Torroba:2006kt}
G.~Torroba, {\it {Finiteness of flux vacua from geometric transitions}},  {\em
  JHEP} {\bf 02} (2007) 061,
  [\href{http://xxx.lanl.gov/abs/hep-th/0611002}{{\tt hep-th/0611002}}].

\bibitem{Dijkgraaf:2002pp}
R.~Dijkgraaf, S.~Gukov, V.~A. Kazakov, and C.~Vafa, {\it {Perturbative analysis
  of gauged matrix models}},  {\em Phys. Rev.} {\bf D68} (2003) 045007,
  [\href{http://xxx.lanl.gov/abs/hep-th/0210238}{{\tt hep-th/0210238}}].

\bibitem{Itoyama:2002rk}
H.~Itoyama and A.~Morozov, {\it {Calculating gluino condensate prepotential}},
  {\em Prog. Theor. Phys.} {\bf 109} (2003) 433--463,
  [\href{http://xxx.lanl.gov/abs/hep-th/0212032}{{\tt hep-th/0212032}}].

\bibitem{Becker:1996gj}
K.~Becker and M.~Becker, {\it {M-Theory on Eight-Manifolds}},  {\em Nucl.
  Phys.} {\bf B477} (1996) 155--167,
  [\href{http://xxx.lanl.gov/abs/hep-th/9605053}{{\tt hep-th/9605053}}].

\bibitem{nakanishi}
{T. Nakanishi and A. Tsuchiya}, {\it {Level-rank duality of WZW models in
  conformal field theory}},  {\em {Commun. Math. Phys.}} {\bf 144} (1992)
  351--372.

\end{thebibliography}\endgroup
        \end{singlespace}
        \end{small}


    \selectlanguage{dutch} 
	\chapter{Samenvatting}
\markboth{Samenvatting}{Samenvatting}
\hyphenation{Rie-mann-op-per-vlak}

\begin{figure}[b!]
\begin{center} 
  \includegraphics[width=.42\textwidth]{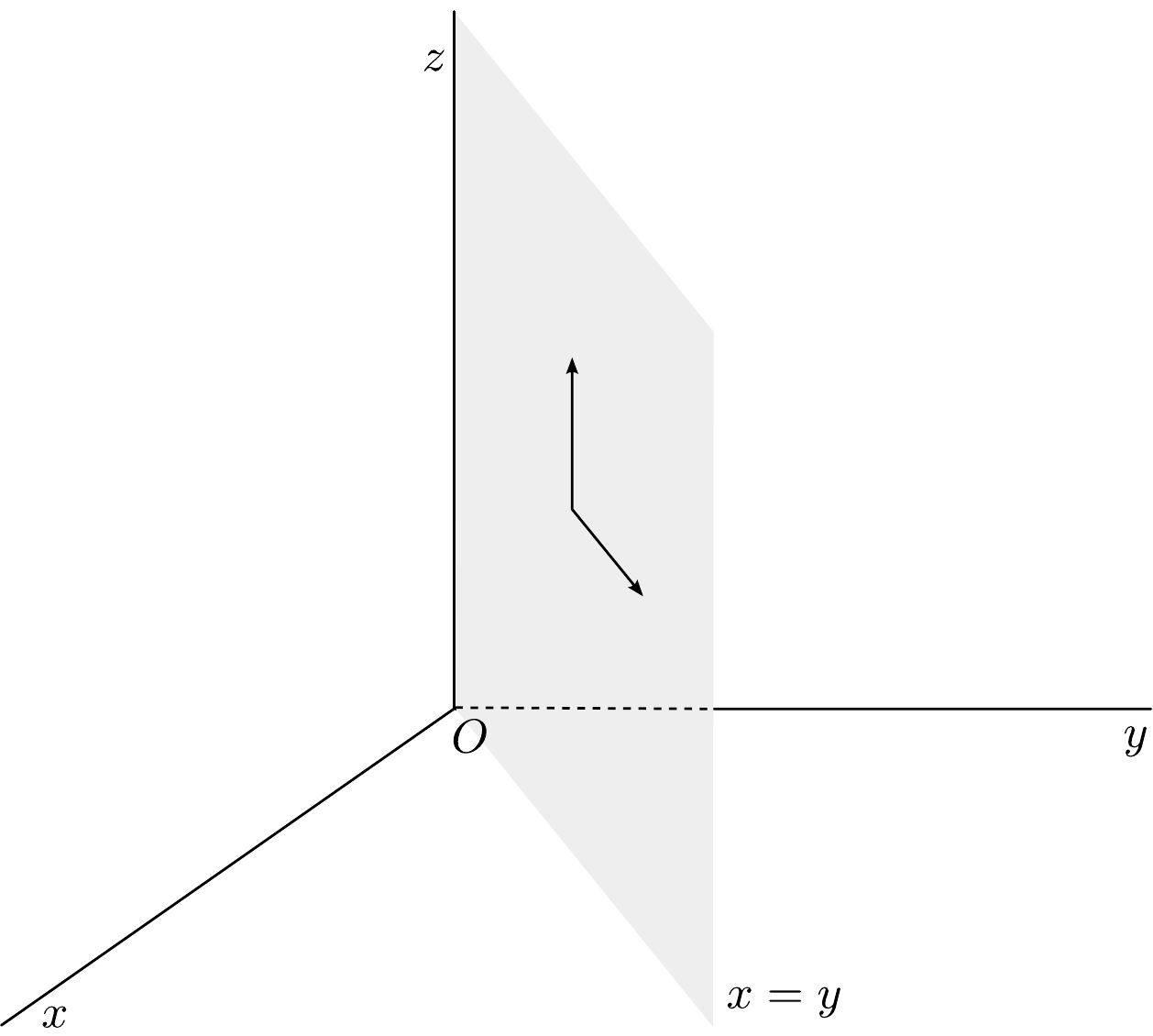}
  \hspace*{.14\textwidth}
  \includegraphics[width=.42\textwidth]{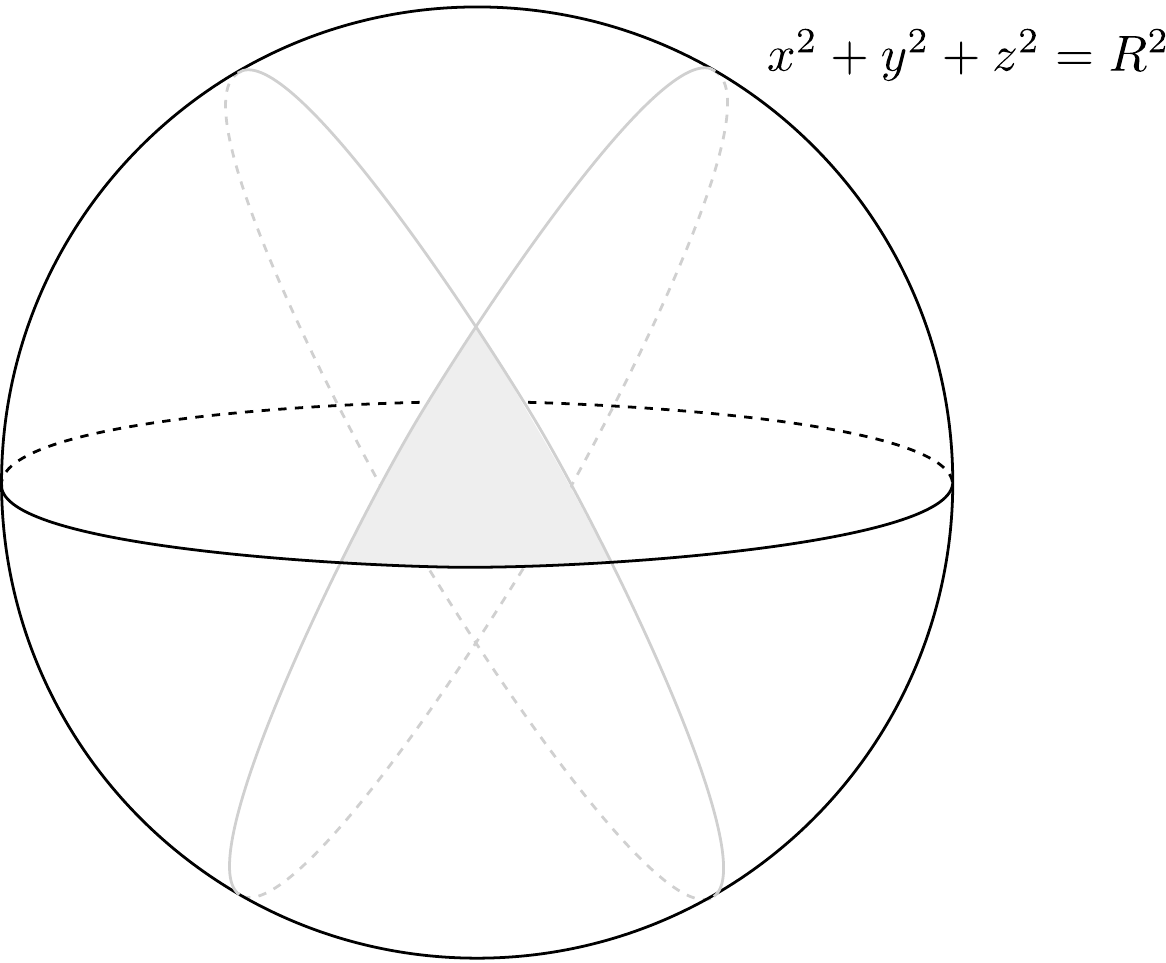}
\caption{Links een meetkundig vlak, gedefinieerd door de lineaire
  vergelijking $x=y$. Op elk punt in dit vlak kun je twee richtingen
  uit. Het vlak is dus een 2-dimensionaal deel van het 3-dimensionale
  $(x,y,z)$-assenstelsel. Bovendien zijn die twee richtingen overal op
  het vlak hetzelfde. \newline 
  Rechts een bol, gedefinieerd door de
  kwadratische vergelijking $x^2 + y^2 + z^2 = R^2$, waarbij $R$ de
  straal van de bol is. De hoeken van de 
  gearceerde driehoek tellen op tot iets meer dan 180 graden. Op ieder punt
  van de bol kun je twee richtingen uit, zodat ook de bol
  2-dimensionaal is. Die twee richtingen veranderen echter 
  als je naar een ander punt op de bol beweegt.}\label{fig:vlakbol}  
\end{center}
\end{figure}

In de Euclidische meetkunde bekijkt men rechte oppervlakken zoals die links
in Fig.~\ref{fig:vlakbol}. Sinds de 19e eeuw bestuderen wiskundigen
echter ook gekromde oppervlakken, zoals de bol, ge{\"i}llustreerd rechts in
deze figuur. Meetkunde op de bol is anders dan op 
een vlak: de hoeken van een driehoek tellen niet op tot 180 graden,
maar meer! Dit is een teken dat de ruimte niet vlak is, maar gekromd. 

\begin{figure}[t]
\begin{center} 
\includegraphics[width=11cm]{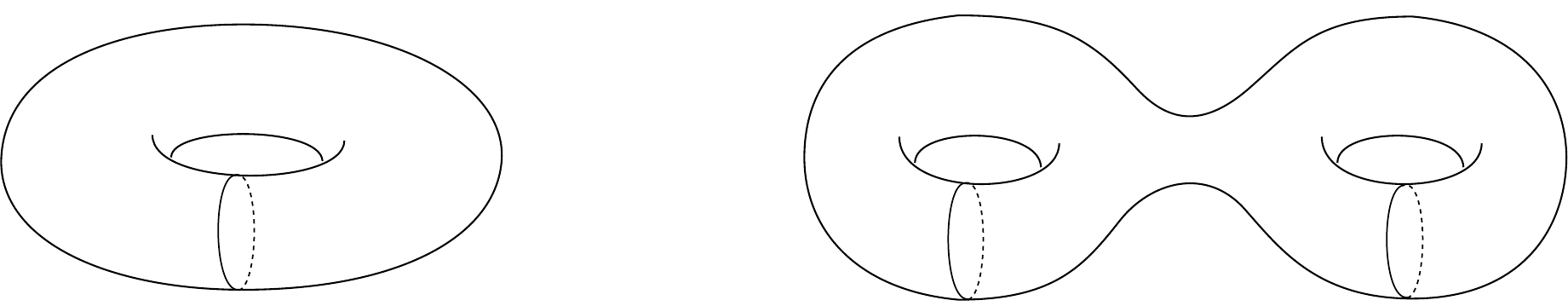}
\caption{Twee voorbeelden van Riemannoppervlakken: links een
  Riemannoppervlak in de vorm van een donut --- dit wordt een torus
  genoemd --- en 
  rechts een oppervlak met twee handvaten.}\label{fig:vbgenustwee}
\end{center}
\end{figure}

Algemenere gekromde oppervlakken zijn als eerste bestudeerd door
B.~Riemann en worden \emph{Riemannoppervlakken} genoemd. 
Door te doen alsof zo'n oppervlak van rubber is, en geleidelijke
vervormingen toe te staan, kan een Riemannoppervlak beschreven worden
door het aantal handvaten in het oppervlak. De bol heeft
bijvoorbeeld geen handvaten. Fig.~\ref{fig:vbgenustwee} toont twee
Riemannoppervlakken met handvaten, die veelvuldig voorkomen in dit
proefschrift.    

\begin{figure}[b!]
\begin{center} 
\includegraphics[width=6cm]{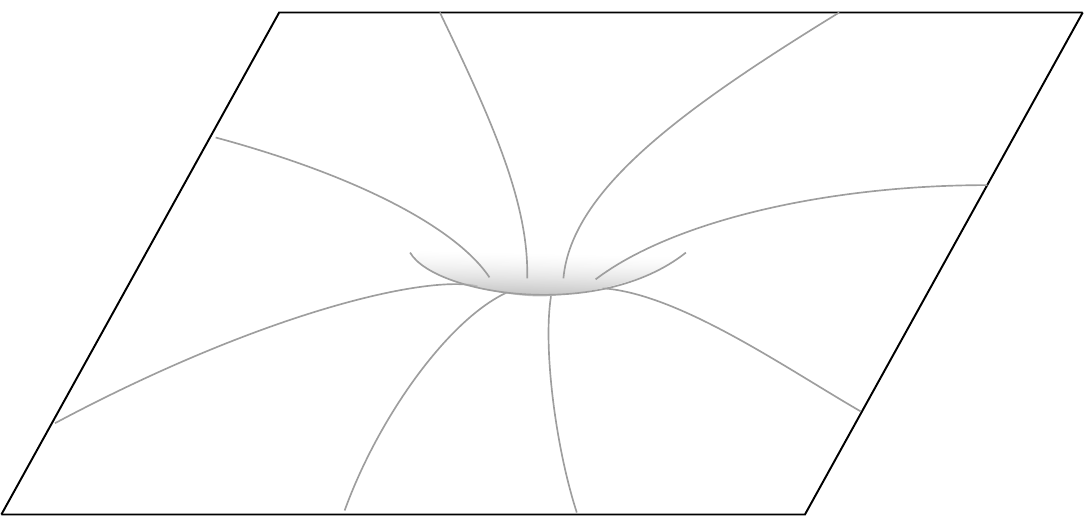}
\caption{Een 2-dimensionale voorstelling van de kromming van de
  ruimte-tijd nabij een zwaar object in de ruimte.}\label{fig:bh}
\end{center}
\end{figure}

In de natuur vinden we veel van zulke gekromde ruimten, zoals het
oppervlak van de aarde. Sterker nog, A.~Einstein ontdekte dat de
ruimte waarin we leven in een diepere zin gekromd is. Om dit te
beschrijven bracht hij ruimte en tijd onder \'e\'en noemer. Grofweg
wordt deze zogenaamde \emph{ruimte-tijd} gegeven door een 4-dimensionaal
assenstelsel dat verandert als we van de ene naar de andere plaats
reizen. Net als op de bol. De kromming is het grootst in de buurt 
van grote massa's zoals sterren en zwarte gaten. 
Fig. \ref{fig:bh} is hier een illustratie van.

\begin{figure}[h]
\begin{center} 
\includegraphics[width=9cm]{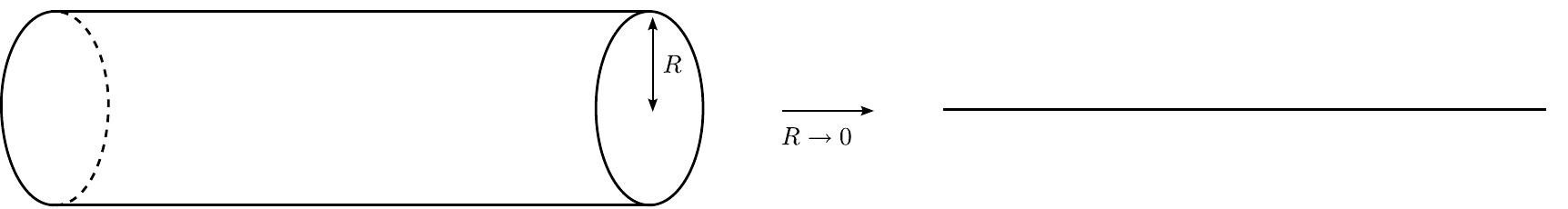}
\caption{Als de straal $R$ van een cylinder heel klein wordt,
  zijn de cylinder en een rechte lijn moeilijk van elkaar te
  onderscheiden.}\label{fig:smalldim} 
\end{center}
\end{figure}

\emph{Snaartheorie} probeert eigenschappen van ons universum te
beschrijven door naar nog hoger-dimensionale gekromde ruimten te
kijken. Die ruimten hebben \hyphenation{mees-tal} meestal 10
dimensies, waarvan 1 
tijd-richting. Oftewel, je kunt in 9 verschillende richtingen reizen
die allemaal loodrecht op elkaar staan. De relatie met onze wereld kan
dan gemaakt worden door zes van die dimensies heel klein te maken,
zodat je ze bijna niet ziet. Zie Fig.~\ref{fig:smalldim} voor een
illustratie. Alhoewel het moeilijk is die kleine dimensies te meten,
zijn ze wel degelijk belangrijk voor de natuurkunde in onze wereld. Ze
worden bestudeerd om oplossingen te vinden voor allerlei raadsels
waarmee natuurkundigen geconfronteerd worden. 

Een van de grootste raadsels is het vinden van een goede beschrijving
van zwaartekracht op heel kleine lengteschaal. Sinds het begin van
de vorige eeuw is bekend dat op afstanden kleiner dan de grootte van
een atoom de kwantummechanica een rol speelt. In deze theorie
kunnen afstand en snelheid niet gelijktijdig exact bepaald
worden,\hyphenation{ge-lijk-tij-dig} en zijn grootheden die continu lijken, zoals energie, opgebouwd uit
discrete pakketjes, de kwanta. Om de zwaartekracht op kleine schaal te
kunnen begrijpen, hebben we een kwantummechanische beschrijving van
Einsteins theorie nodig. Dit blijkt echter erg moeilijk te zijn.  

Een andere formulering van dit probleem is dat we 
een beschrijving van de natuur proberen te vinden waarin we alle vier
de fundamentele krachten---electromagnetische kracht, sterke en zwakke
kernkracht en zwaartekracht---verenigen.
Het zogeheten standaardmodel unificeert de eerste drie van deze
krachten, in het kader van de kwantummechanica. Maar de vierde
kracht, gravitatie, wil niet zo meewerken. Dit staat een beschrijving 
in de weg van de meest fundamentele vraagstukken in het heelal, 
bijvoorbeeld de vraag naar het ontstaan van het heelal.

Snaartheorie is een van de beste kandidaten om inzicht te
verkrijgen in deze fundamentele vraagstukken. Vermoedelijk beschrijft
deze theorie alle vier de krachten. Maar tegelijkertijd is ze 
veelomvattend en ingewikkeld. Hoewel we al ontzettend veel over
snaartheorie weten, is dat nog lang niet genoeg om de hele theorie te
doorgronden. Dit proefschrift zet een paar kleine stapjes in
deze richting. 

Belangrijk om te weten is dat we  nog niet kunnen meten aan
snaartheorie. De theorie is dus volledig gebouwd op fysische
argumenten en een heleboel wiskunde. Dit heeft als voordeel dat er
een actieve interactie is met allerlei takken van de
wiskunde. Snaartheorie blijkt interessante wiskundige vermoedens te
genereren en nieuwe verbindingen tussen verschillende subdisciplines
te leggen. Dit proefschrift speelt ook daar op in.     

In dit proefschrift bestuderen we snaartheorie op een 10-dimensionale
ruimte die we onderverdelen in de 4-dimensionale ruimte-tijd en een
6-dimensionale interne ruimte. Om precies te zijn bestuderen we 
interne ruimten die zogenaamde Calabi-Yau vari\"eteiten vormen, en wel
Calabi-Yau vari\"eteiten die gemodelleerd zijn in termen van een
Riemannoppervlak.\hyphenation{ge-mo-del-leerd} Dit Riemannoppervlak
vormt een rode draad door het proefschrift. 

Eigenlijk hadden we iets nauwkeuriger moeten zijn bij het defini\"eren
van Riemannoppervlakken: het zijn een speciaal soort oppervlakken die er lokaal
uitzien als het complexe vlak, met co\"ordinaten $x + i y$,
waarbij $i^2 = -1$. Een Riemannoppervlak wordt daarom ook wel een
complexe of algebra\"ische kromme genoemd.

Op Calabi-Yau vari\"eteiten die gemodelleerd zijn op een
Riemannoppervlak, kunnen we met behulp van snaartheorie 
partitiefuncties uitrekenen. Een voorbeeld van een relatief
eenvoudige, maar toch interessante partitiefunctie, die we aanduiden
met $Z(\tau)$, is  
\begin{align*}
Z(\tau) = e^{-\pi i \tau/12} \prod_{n \ge 1}^{\infty}
\frac{1}{ (1-e^{2 \pi  i \tau n})}.
\end{align*}
De variabele $\tau$ neemt waarden aan in de bovenste helft van het
complexe vlak. In de wiskunde staat de functie $Z(\tau)$ vooral bekend 
om zijn mooie eigenschappen onder transformaties van zijn argument
$\tau$. De functiewaarde blijft namelijk bijna hetzelfde wanneer
$\tau$ naar $-1/\tau$ gestuurd wordt: 
\begin{align*}
 Z(-1/\tau) = \frac{1}{\sqrt{-i\tau}}~Z(\tau)
\end{align*}

\begin{figure}[h]
\begin{center} 
\includegraphics[width=12 cm]{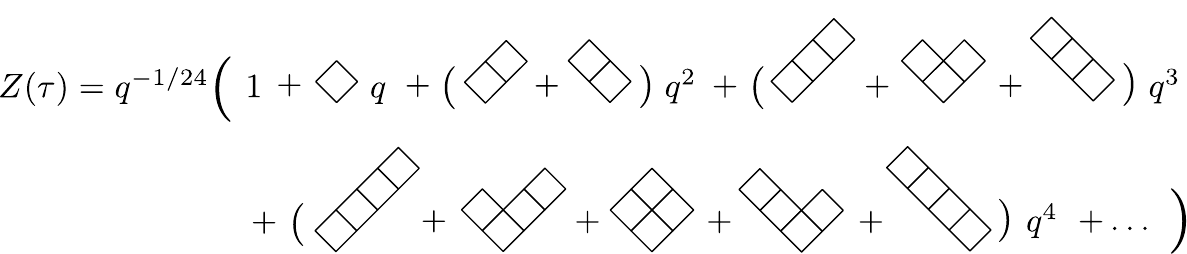}
\caption{In deze reeks voor de partitiefunctie $Z(\tau)$ geeft het
  aantal diagrammen met $k$ vierkantjes de numerieke factor voor de
  term $q^k$ aan. Dat is er dus eentje voor $q$, twee voor $q^2$, drie
  voor $q^3$, vijf voor $q^4$, enzovoorts. }\label{fig:2dpart}
\end{center}
\end{figure}

Deze symmetrie\"en hebben tevens een interessante fysische
interpretatie. Om dit kort uit te leggen, beginnen we met een
reeks-ontwikkeling van de partitiefunctie $Z(\tau)$. Oftewel, we
schrijven het product als een oneindige som:
\begin{align*}
Z(\tau) = q^{-1/24} \left( 1 + q +  
2 q^2 + 3 q^3 + 5 q^4 + \ldots \right),
\end{align*}
Hier hebben we voor de bondigheid $q = e^{2 \pi i \tau}$
\hyphenation{ver-wij-zen}%
gedefinieerd. De puntjes verwijzen naar alle termen $q^k$ met een 
macht $k$ groter dan 4. Het grappige is dat deze reeks 
grafisch ge\"interpreteerd kan worden, zie Fig.~\ref{fig:2dpart}.

\begin{figure}[t]
\begin{center} 
\includegraphics[width=9cm]{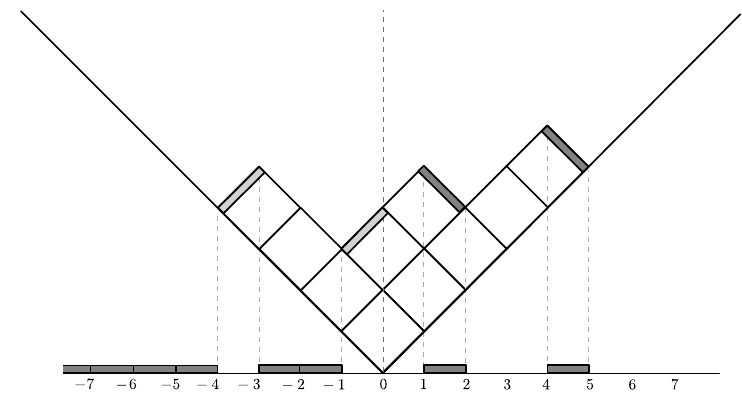}
\caption{Deze figuur laat een iets algemener diagram zien in de
  reeksontwikkeling van $Z(\tau)$ in Fig.~\ref{fig:2dpart}; omdat dit
  diagram uit tien 
  vierkantjes bestaat komt het pas tevoorschijn bij de term
  $q^{10}$. Elk zo'n diagram kun je op een unieke manier 
  afbeelden naar een toestand van fermionen. In de bovenstaande figuur
  bepalen de donkere en lichte rechthoekjes in het diagram welke
  energietoestanden op de getallenlijn wel of resp. niet bezet zijn. Op de 
  getallenlijn indiceren de donker gekleurde rechthoekjes de ingenomen
  energietoestanden.}\label{fig:one2dpart} 
\end{center}
\end{figure}

\indent Deze diagrammen met vierkantjes moeten aan een aantal
regels voldoen. Als je een rechte hoek met het midden op de
grond zet, zoals in Fig. \ref{fig:one2dpart}, moeten ze te verkrijgen
zijn door vierkantjes in deze wig te laten vallen. 

De fysica achter deze reeks heeft te maken met de kleinst mogelijke
deeltjes. Die zijn er namelijk maar in twee soorten:
bosonen en fermionen. Bosonen willen zich graag in dezelfde toestand
bevinden, terwijl de fermionen dit nooit zullen doen. Een van de
simpelste kwantum-systemen beschrijft fermionen op een cirkel. 
Hun energie\"en nemen discrete (oftewel kwantum-) waarden aan.     

Een bepaalde toestand van dit systeem wordt dan beschreven door aan te
geven welke energietoestanden bezet zijn door een fermion. Omdat
de fermionen in dit systeem nooit dezelfde energie zullen hebben, is zo'n
energietoestand \'of onbezet, \'of bezet door een enkel fermion. De totale
toestand van het systeem kan daarom gevisualiseerd worden door een
getallenlijn, waarop bij elk geheel getal wordt aangegeven of er wel
of geen fermion zit.

Nu blijken alle diagrammen in Fig. \ref{fig:2dpart} op een unieke
manier af te beelden naar zo'n fermionische toestand, zoals
ge\"illustreerd in Fig.~\ref{fig:one2dpart}. De partitiefunctie
$Z(\tau)$ codeert dus alle mogelijke toestanden van het fermionische
systeem! 

In de snaartheorie vinden we deze specifieke partitiefunctie $Z(\tau)$
voor een relatief eenvoudige Calabi-Yau vari\"eteit die gebaseerd
is op de torus uit Fig.~\ref{fig:vbgenustwee}. De parameter $\tau$
karakteriseert in die interpretatie de vorm van de torus. De fermionen
leven op de torus: \'e\'en opspannende cirkel van de torus kan gezien
worden als de co\"ordinaat-cirkel, en de ander als de tijd. (Preciezer
gezegd is $Z(\tau)$ de partitiefunctie van chirale, oftewel holomorfe,
fermionen op de torus.)  

\begin{figure}[t]
\begin{center} 
\includegraphics[width=10cm]{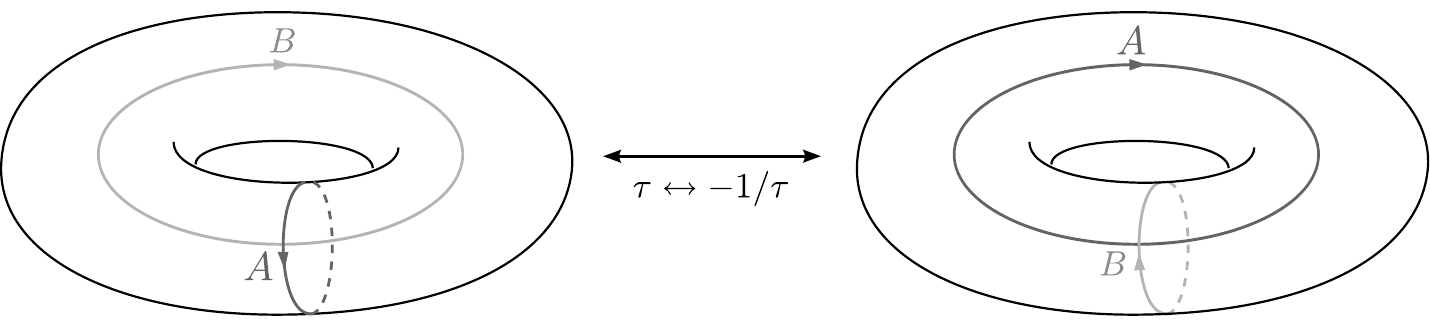}
\caption{De parameter $\tau$ karakteriseert de vorm van de
  torus. Onder de afbeelding $\tau \mapsto -1/\tau$ verwisselen de
  twee opspannende cirkels op de torus.}\label{fig:torusflop} 
\end{center}
\end{figure}

Onder de
afbeelding $\tau \mapsto -1/\tau$ verandert de torus zodat de
ruimte- en tijd-cirkels worden omgewisseld, zie
Fig.~\ref{fig:torusflop}.\hyphenation{ver-an-dert} 
De vorm van de torus blijft echter dezelfde. Dit verklaart waarom de
partitiefunctie $Z(\tau)$ zo goed als invariant blijft onder deze
transformatie.  

Interessant is dat de partitiefunctie $Z(\tau)$ niet alleen een
betekenis heeft in de interne 6-dimensionale ruimte, maar ook in de
4-dimensionale ruimte-tijd. In deze ruimte-tijd is er een duale
theorie die de electromagnetische interacties beschrijft.  
De symmetrie $\tau
\mapsto -1/\tau$ blijkt hierin elektrische en magnetische deeltjes
 te verwisselen. Zo verkrijgen we ook inzicht in de
beschrijving van natuurkunde in onze wereld: er is een dieper liggende
symmetrie tussen de electrische en magnetische krachten! 

Deze wisselwerking tussen enerzijds de wiskunde van interne
6-dimensionale ruimten en anderzijds de fysica in de 4-dimensionale ruimte-tijd
speelt een grote rol in dit proefschrift. Ruwweg heet het vakgebied
waarin die wiskunde van 6-dimensionale Calabi-Yau ruimten wordt ontwikkeld 
\emph{topologische snaartheorie}. In dit proefschrift generalizeren we 
de bovenstaande duale beschrijving van partitiefuncties in termen van
fermionen. 

In het bijzonder zien we dat een algemene partitiefunctie
in de topologische snaartheorie kan worden gezien als een
partitiefunctie van fermionen op een vreemd soort Riemannoppervlak: de
co\"ordinaten van dit Riemannoppervlak (ofwel algebra\"ische 
kromme) gedragen zich niet klassiek, maar\hyphenation{kwan-tum-me-cha-nisch} kwantummechanisch. We
noemen dit dan ook een \emph{kwantum kromme}. Dit verheldert de
naam van het proefschrift: ``Topologische Snaren en Kwantum Krommen''.






        \chapter{Acknowledgments}
\markboth{Acknowledgments}{Acknowledgments}

Let me finally take the opportunity to thank some people who have
been invaluable for the completion of this thesis. 

First and foremost, I would like to express my gratitude to my supervisor
Robbert Dijkgraaf. It must have been at some point in my fourth year in
Utrecht that I stumbled upon his website and was immediately struck by
the elegant combination of mathematics and physics. In these last years
Robbert has guided me through this subject and we have worked together
on several interesting projects. Unfortunately his increasingly busy
schedule didn't always allow time for me. However, our meetings were
always inspiring and I have greatly benefited from his wisdom
and creativity. Most important for this thesis is that Robbert has been
a driving force behind both of our articles. I would like to thank
him very much for this.\hyphenation{bene-fited}     


Secondly, I am indebted to my other co-authors
Miranda Cheng, Kyriakos Papadodimas, Masaki Shigemori, Piotr Su{\l}kowski and Cumrun Vafa
 for sharing their insights and ideas. Only in these
collaborations I really learned what doing 
research encompasses, and I enjoyed it very much. I would also like to
thank many other colleagues I met during the last years
for the stimulating discussions and the pleasant times in Amsterdam
and on all those trips abroad. Special thanks go to Chris Beasley, 
 Jan
de Boer, Sheer El-Showk, Sergei Gukov, Amir Kashani-Poor, Albrecht
Klemm, Johan van de
Leur, Jan Manschot, Marcos
Marino, Ilies Messamah, Andy Neitzke, Hessel Posthuma, Ani Sinkovics, 
Bal\'azs Szendr\"oi, Stefan Vandoren and Erik Verlinde. I'm looking forward
to elaborating on our ideas.

Furthermore I am grateful to all my (former) colleagues in
Amsterdam for the friendly and animate atmosphere in the institute,
in particular to  
Amir, Asad, Ben, Masaki, Tomeu and especially Kyriakos
for being keen to answer all my questions. I also mustn't forget to mention Bianca and Yocklang for
their administrative support.
Moreover, the times that I shared an office with
Jan, Manuela and Piotr 
bring back pleasant memories. Balt, Ilies,
Ingmar, Johannes, Joost, Leo, Meindert, Paul, Sheer, Xerxes and many
others made 
lunches and cookie times, too, very lively with interesting discussions.
 Also,
many thanks to the members of the theoretical physics and mathematics
institutes in Utrecht for welcoming me with open arms on each of my
visits.








In addition to academic pursuits, I have very much enjoyed all
after-hours fun, and would like to thank everyone who contributed to this.
Let me mention here my highschool 
friends Casper \& Sabine, Joost \& Monique, 
Jules, Lieselotte \& Thijs, Martijn, Thijs \& Jiska and Vera \&
Constantijn,  my former study-buddies Gerben, Jaap, Jan \&
Afke, as well as Alex, Erik \& Judith, Hendrik, Josien, Luuk,
Marielle, Rob, Rogier,
Rudy \& Magda, Sheer \& Hanna, Sylvain and  Thijs
\& Hieke. 
Also thanks to Ana \&
Harald and Ichiro \& Kumiko for the 
great holidays in Austria and Japan. I appreciate it particularly that
Gerben and Sheer are willing to be my paranymfs. 

Let me close by adding some words of appreciation to all my
family (in-law). Especially, many thanks   
to my brother Ramon \& Anneke for their frequent company. I am sure they are
going to love our apartment as much as we did. Moreover I am very grateful to my  parents, for always being there
and supporting me. But most of all I would like to thank 
Chris, for just everything.   

\vspace{5mm}    

\flushright{Amsterdam, July 2009}












\end{document}